\documentclass[aps,rmp,reprint,amsmath,amssymb,graphicx,floatfix,longbibliography]{revtex4-1}
\usepackage{braket}
\usepackage{amsmath}
\usepackage{amssymb}
\usepackage{mathtools}
\usepackage{graphicx}
\usepackage{dcolumn}
\usepackage{bm}
\usepackage{multirow}
\usepackage{leftidx}
\usepackage{color}
\usepackage{float}
\usepackage{dsfont}
\usepackage{amsfonts}
\usepackage[english]{babel}
\usepackage{slashed}
\usepackage{verbatim}

\begin{document}

\title{Universal few-body physics and cluster formation}

\author{Chris H. Greene}
 \email{chgreene@purdue.edu}
\author{P. Giannakeas}
\email{pgiannak@purdue.edu}
\author{J. P\'{e}rez-R\'{i}os}
\email{jperezri@purdue.edu}
\affiliation{Department of Physics and Astronomy, Purdue University, West Lafayette, Indiana 47907-2036, USA}
\date{\today}
\begin{abstract}
A recent rejuvenation of experimental and theoretical interest in the physics of few-body systems has 
provided deep, fundamental insights into a broad range of problems.  Few-body physics is a cross-cutting 
discipline not restricted to conventional subject areas such as nuclear physics or atomic or molecular physics.  
To a large degree, the recent explosion of interest in this subject has been sparked by dramatic enhancements
of experimental capabilities in ultracold atomic systems over the past decade, which now permit atoms and 
molecules to be explored deep in the quantum mechanical limit with controllable two-body interactions.  
This control, typically enabled by magnetic or electromagnetically-dressed Fano-Feshbach resonances, 
allows in particular access to the range of universal few-body physics, where two-body scattering lengths 
far exceed all other length scales in the problem. The Efimov effect, where 3 particles experiencing 
short-range interactions can counterintuitively exhibit an infinite number of bound or quasi-bound energy
levels, is the most famous example of universality. Tremendous progress in the field 
of universal Efimov physics has taken off, driven particularly by a combination of experimental and theoretical 
studies in the past decade, and prior to the first observation in 2006, by an extensive set of theoretical 
studies dating back to 1970.  Because experimental observations of Efimov physics have usually relied on
resonances or interference phenomena in three-body recombination, this connects naturally with the processes 
of molecule formation in a low temperature gas of atoms or nucleons, and more generally with N-body recombination
processes. Some other topics not closely related to the Efimov effect are also reviewed in this article, including
confinement-induced resonances for explorations of lower-dimensionality systems, and some chemically interesting 
systems with longer-range forces such as the ion-atom-atom recombination problem.
\end{abstract}


\maketitle

\tableofcontents{}


\section{Introduction and Overview}
\label{Intro}
\begin{figure}[h]
\centering
 \includegraphics[width=6.5 cm]{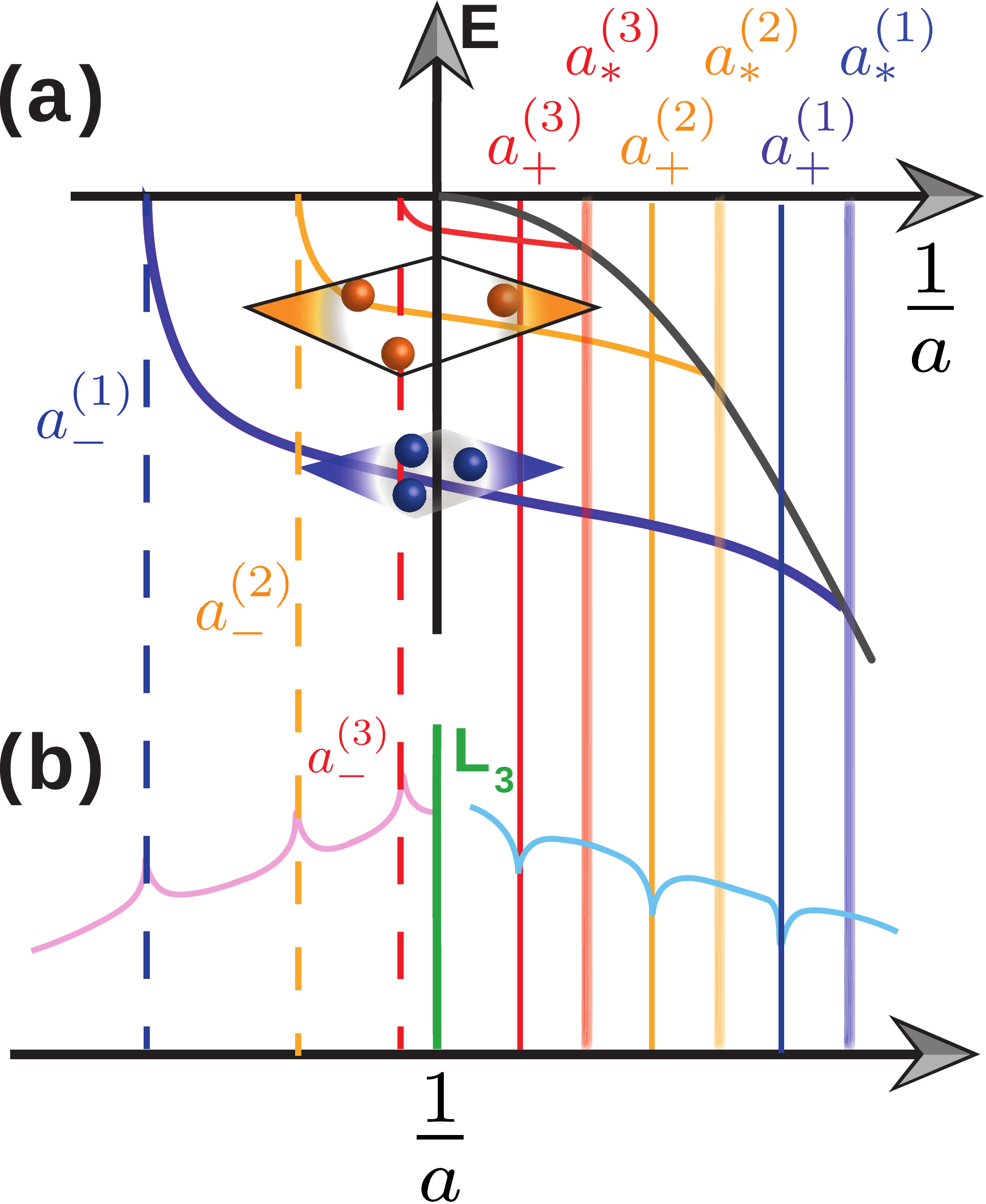}
 \caption{(Color online) Characterization of universal Efimov trimers in ultracold gases. Panel (a) shows 
 the universal trimer energy dependence on the inverse two-body scattering length, i.e. $a$. In particular, the Efimov 
 trimers cross the three-body continuum at $a_{-}^{(n)}$, where $n$ =1 denotes the Efimov ground state, 
 $n$ = 2 the first excited states, etc. Efimov states intersecting the atom-dimer continuum are 
 characterized by $a_{*}^{(n)}$ for the positive two-body scattering length. 
\textcolor{black}{Panel (b) schemmatically illustrates a log-log plot of the recombination rates of three identical bosons at low energies versus the inverse two-body scattering length $a$.} Minima in 3-body recombination
occur at scattering length values denoted here as $a_{+}^{(n)}$.  The negative values of the atom-atom
scattering length marked $a_{-}^{(n)}$ indicate positions of the maxima in $L_{3}$ at ultracold temperatures,
i.e. where Efimov states intersect 
 the atom-atom-atom three-body threshold.}
\label{3B}
\end{figure}

Spectacular recent breakthroughs for the three-body problem with near-resonant two-body interaction, in both experiments and theories, have spawned this review of universal few-body physics, which concentrates on systems with finite-range interactions. Vitaly Efimov's 1970 prediction~\cite{efimov1970plb} that an infinite family of universal three-body states should emerge when two or more two-body scattering lengths are sufficiently large in magnitude first received partial experimental confirmation in 2006 by Rudi Grimm's group in Innsbruck~\cite{kraemer2006NT}. \textcolor{black}{That development was quickly followed by many subsequent experiments~\cite{Pollack-2009,dyke2013PRA,knoop2009NTP,knoop2010PRL,ferlaino2008PRL,ferlaino2009PRL,
ferlaino2011FBS,berninger2011PRL,berni2013pra,zenesini2013NJP,zaccanti2009NTP,gross2009PRL,gross2010PRL,gross2011CRP,machtey2012PRL,machtey2012PRLb,ottenstein2008PRL,wenz2009PRA,wild2012PRL,barontini2009PRL,bloom2013PRL,HuBloom2014,nakajima2010PRL,nakajima2011PRL,lompe2010PRL,roy2013prl,Kunitski-2015, Huang-2014a,Huang-2014b, Huang-2015,Pires-2014,Tung-2014,JohansenChin2016arxiv,Ulmanis-2015b,Ulmanis2016prl,Haefner2017arxiv,Maier-2015,Arlt2016,WangBlumeWang2016} bearing on various aspects of the Efimov effect and universality. See Fig.~\ref{3B}.} This class of phenomena is called {\it universal} because it can occur for systems with vastly different energy and length scales. While it was originally predicted for few-nucleon systems such as the triton, with energy scales of order $10^6$ eV and distance scales of the order of $10^{-14}$ m, all of the convincing demonstrations to date have involved energy and distance scales of order $10^{-12}$ eV and $10^{-7}$ m, respectively.  Some of the few-body physics topics discussed here have already been reviewed elsewhere, and the reader is recommended to explore a large body of literature that can be found in \textcolor{black}{~\cite{NaidonEndoReview2016,braaten2006PRep,nielsen2001PRep,rittenhouse2011JPB,blume2012rpp,wang2013amop,wang2015AnnRev,Yurovsky2008a,Cote2016adv,ohsaki1990PRep,Petrov2012,ZinnerJensen2013jpg,
Frederico2012xh,jensen2004RMP,baranov_condensed_2012,suzuki1998}}.

\subsection{Systems with finite range interactions}
\label{Finite}
It is reasonable to ask why finding a new family of resonances has generated such excitement in few-body physics, excitement that has translated into an exponentially growing rate of citations for the 1970 Efimov paper during the past 15 years.  In fact these resonances are unique and counterintuitive. For every previously known example of a system where infinitely many bound states or resonances exist that converge to a breakup threshold, the forces were infinite in their extent.  The best known example of this is of course the asymptotically attractive Coulomb potential $V(r)\rightarrow -1/r$ which has an energy level formula $E_n \propto -1/n^2$, and a second example is the charge-dipole two-body potential $V(r) \rightarrow -(s^2+1/4)/2r^2$ at $r>r_0$ which has (for a system of units with reduced mass $\mu=1$) an energy level formula $E_n = E_0 \exp(-2 \pi n/s)$.  The presence of a finite versus infinite number of quantized levels below a threshold hinges on the convergence or non-convergence of the zero-energy \textcolor{black}{JWKB} phase integral with Langer correction included. \textcolor{black}{(Vol.II of ~\cite{morse1953} presents a pedagogical derivation of the Langer correction needed for accurate semiclassical calculations, e.g. when the independent coordinate domain is semi-infinite or finite as is true for the radial coordinate in three-dimensional problems.)}  That is, one can deduce the energy level formula relevant to a given two-body potential energy function $V(r)$ by evaluating the zero energy total phase \textcolor{black}{$\phi=\int_{r_0}^\infty \sqrt{-2 \mu V(r)/ \hbar^2-\frac{1}{4r^2} } dr $}.  If this $\phi$ is infinite, then the number of converging energy levels will also be infinite, whereas if $\phi$ is finite then their number is also finite.  \textcolor{black}{This type of analysis also applies to the recently predicted ``super-Efimov effect'' ~\cite{Nishida2013,Volosniev2014JPB,Moroz2014,Gridnev2014JPA,GaoWangYu2015} which is likewise predicted to yield an infinite sequence of bound (or resonant) levels for a system of three fermions in 2 dimensions,(see also ~\cite{EfremovSchleich2013PRL}) with a density of states far smaller than in the original Efimov effect.  }

 A common thread running through this story is the fact that hyperspherical coordinate techniques played a key role in the early theoretical predictions of the Thomas and Efimov effects in pre-1980 studies, and they have played an equally crucial role in showing later that ultracold quantum gases should provide a powerful way to observe universal Efimov physics, by linking the Efimov effect quantitatively to the loss process of three-body recombination.  Hyperspherical studies have shown unusual flexibility, as they have been used on the one hand with zero-range regularized pseudopotential interactions to obtain closed-form analytical results,\cite{efimov1970plb,efimov1971SJNP,efimov1973npa,
efimov1979SJNP,macek1986ZPD, WatanabeEfimov1989, nielsen2001PRep,nielsen1999PRLb,
kartavtsev2002FBS,mehta2008PRA,macek2002FBS,macek2005PRA} and on the other hand as the basis for quantitative numerical solutions using finite-range analytical or numerical three-body Hamiltonians \cite{esry1996JPB,esry1999PRL,suno2002PRA, dincao2005PRL, esry2006NT,wang2012PRL,Wang-2012b}. This flexibility has led to a tremendous deepening of our understanding of three-body recombination and atom-dimer elastic and inelastic scattering over the past two decades, both the quantitative understanding and, equally important, qualitative and semi-quantitative ways to understand the main reaction pathways which govern the corresponding physical mechanisms. Despite this headway, there has not been a comprehensive review or monograph that has presented the full hyperspherical methodology, nor has there been one that covered much of the connections with diverse areas of physics from nuclear systems to cold atoms to exotic species and atomic electron states, and one of the goals of the present review is to bridge this gap in the existing literature.

\begin{figure}[h]
\centering
 \includegraphics[width=4.5 cm]{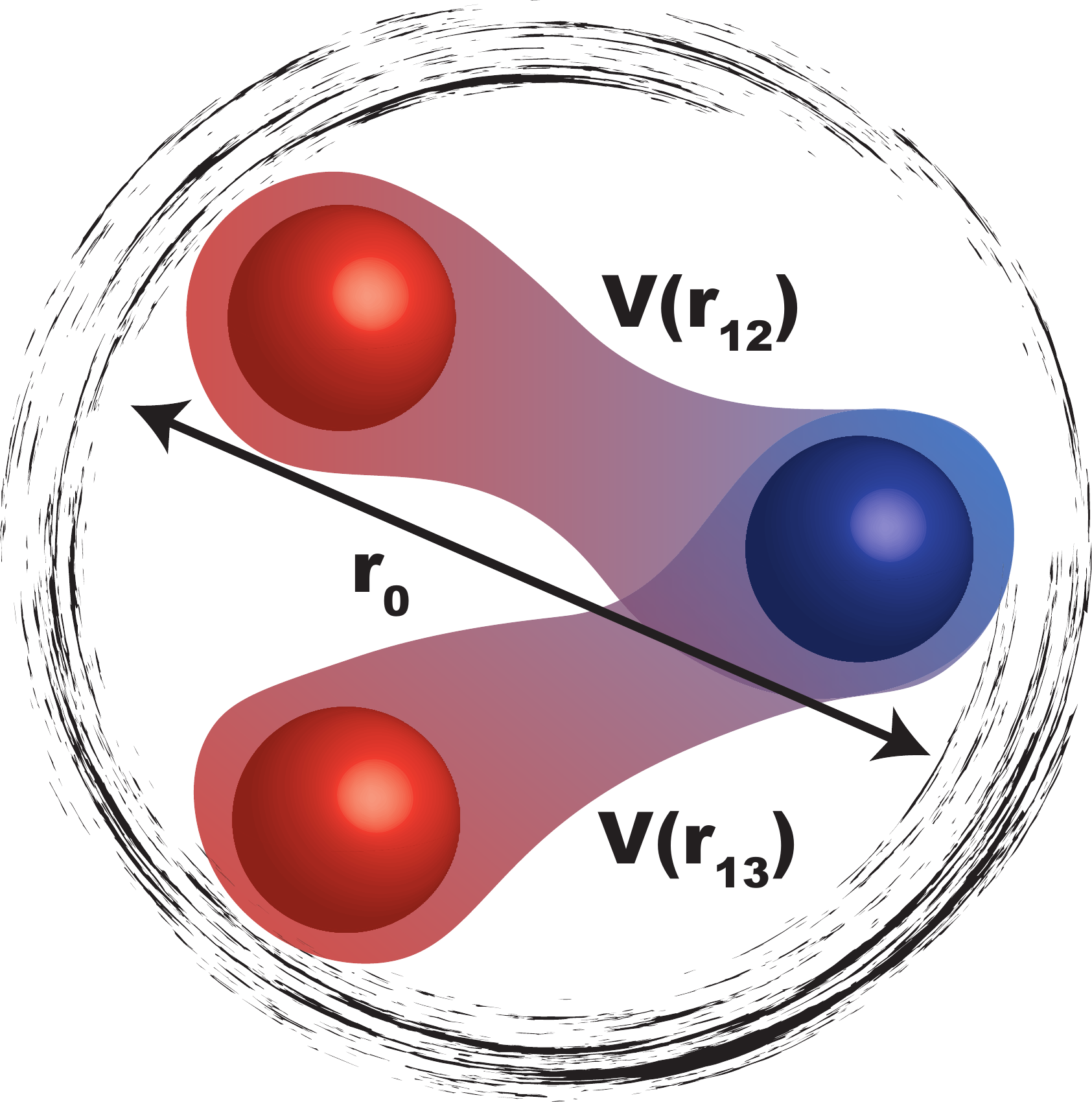}
 \caption{(Color online) Tritium nucleus model assumed by Thomas. The 
 neutron-proton interaction is characterized by a finite-range potential $V(r_{ij})$, 
whose range is given by $r_{0}$, but the neutron-neutron interaction 
is neglected (which is known nowadays to be far from correct). See text for details.}
\label{Fig-Thomas}
\end{figure}

The concepts of the hyperspherical approach are of course far from being a new innovation in few-body theoretical physics.  They go back at least as far as the pioneering work of Llewellen Hilleth Thomas,\cite{thomas1935pr} who realized that three nucleons whose ratio of potential range to scattering length becomes arbitrarily small, $r_0/|a| \rightarrow 0$, must have a ground state energy that ``collapses" to $E\rightarrow -\infty $.  \textcolor{black}{The triton model considered by Thomas is depicted in Fig.\ref{Fig-Thomas}.} This was demonstrated by showing that the effective potential energy of such a system, as a function of the hyperradius $R$ (Thomas denoted this variable as $s$), has the form $-1/R^2$, a potential that exhibits the well-known ``fall to the center" collapse of its ground state energy, as is discussed in quantum mechanics textbooks~\cite{Landau1997}. Another early application of hyperspherical coordinates framework was developed by Julian Schwinger's student at Harvard, R. E. Clapp, in his PhD thesis work on the triton binding energy \cite{CLAPP1949}.  Fock's 1958 study of the analytical nature of the electronic helium atom wavefunction at small hyperradii also utilized hyperspherical coordinates in a fundamental way~\cite{fock1958}.

Some of the deepest insights into the nature of the three-body problem have emerged from Macek's adiabatic hyperspherical methodology~\cite{macek1968JPB}.
The latter consists of a comprehensive theoretical framework in which the Hamiltonian of the system is initially diagonalized at fixed values of the hyperradius $R$, yielding adiabatic curves which represent the energies of the system as functions of $R$.  These give an immediate, dynamics-based representation of the available reaction pathways for any given system, and highlight the emerged structure of the bound and quasi-bound states of the system as well as their excitation and decay mechanisms~\cite{fano1976PT,fano1983RPP,lin1986AMOP,lin1995PRep}. Coupling matrix elements can also be computed which permit, as is shown in this review article, a systematic solution of the full three-body Schr\"odinger equation to the accuracy desired for arbitrary bound state problems as well as two-body inelastic and rearrangement collisions (A+BC), three-body collisions (A+B+C), and photon-assisted collision processes~\cite{fink1985JPB}.   

\subsection{Coulomb Systems}
\label{Coulomb}
The three-body problem in quantum mechanics with Coulomb interactions has generated intense effort throughout the past century. Early in the days of the ``old quantum theory'', it was a major problem to understand the ground and excited states of the helium atom. With Schr\"odinger's wave mechanics, in combination with other tools such as the Ritz variational method, it became possible by the 1930s to calculate properties of such low-lying states in the three-body Coulomb problem to high precision.  For higher excited states lying in the two-body or three-body continua, however, progress was much slower. The ability to nonperturbatively calculate the simple process of electron impact ionization of hydrogen at low energies ($< 1$ eV) above the double escape threshold, for instance, did not emerge until the 1990s~\cite{Bartlett2003,KATO1995,McCurdy1997,Robicheaux1997,Kadyrov2009}, although important theoretical work prior to that had identified the unusual threshold behavior for two-electron escape~\cite{wannier1953pr,rau1984PRep,PETERKOP1971,KLAR1976,fano1983RPP,SELLES1987,greene1982PRL,greene1983JPB,READ1984,watanabe1991JPB}.  Analogous theoretical headway occurred over that same period for other three-body observables, such as double photoionization of He and H$^-$~\cite{Robicheaux1997,meyer1994PRA,meyer1997PRL}.
Of course long before the quantal version of the three-body problem became topical, the Newtonian version with inverse square forces had acquired paramount importance and was singled out by researchers such as Poincar\'e and Hilbert as a crucial bottleneck that had to be solved.  

Early efforts on systems with Coulomb interactions by Macek, Lin, and Fano demonstrated that significant insights into the qualitative and semiquantitative nature of doubly-excited states of He and H$^-$ emerge when an adiabatic hyperspherical approximation is implemented~\cite{macek1968JPB,fano1976PT,lin1986AMOP,lin2000PHYSICSESSAYS}. Surprisingly high doubly-excited states of two-electron atoms can be treated in the adiabatic scheme, as seen for calculations of high states which yielded a simple interpretation of regularities seen in photoabsorption~\cite{sadeghpour1990PRL,domke1991PRL,tang1992PRL,Rau1992science}. Extensions to other atomic systems such as the alkaline earth atoms~\cite{greene1981PRA} and the negative ion of helium~\cite{watanabe1982PRA} were also developed, which showed that nonadiabatic couplings often need to be incorporated in order for the results to be even qualitatively useful~\cite{christensendalsgaard1984PRAb}.  \textcolor{black}{The exploration of near-separability of the two-electron wavefunction in alternative choices of coordinates, which yields nontrivial insights in some cases, was reviewed by ~\cite{TannerRichterRost2000rmp}.}

Another arena where three-body Coulombic interactions have been subjected to intensive study has been in the context of muon-catalyzed fusion~\cite{hino1996PRL}. Interesting studies of the dt$\mu$ reaction of importance for muon-catalyzed fusion were carried out, for instance, using hyperspheroidal coordinates~\cite{hara1988PLA,FUKUDA1990}. Another hyperspheroidal coordinate application was to HD$^+$ by Macek and Jerjian~\cite{macek1986PRA} and by Hara {\it et al.}~\cite{HARA1989} Some of the most suitable systems for an adiabatic representation in hyperspherical coordinates are those with two or more equal mass particles, such as the ion formed from two electrons and one positron, {\it i.e.} the positronium negative ion~\cite{botero1986PRL,botero1985PRA,FabreDeLaRipelle1993FBS}.  Also, not to be overlooked is the fact that this approach can be made quantitatively accurate, in some cases with direct solution of the coupled hyperradial equations in the adiabatic representation~\cite{kadomtsev1987PRA}.  

In fact the adiabatic representation has challenges as the system grows in complexity and in the number of relevant coupled hyperradial equations, and for such systems the clever recasting as a set of diabatic equations, called the ``slow-variable discretization'' (SVD) method proposed by Tolstikhin {\it et al.}~\cite{tolstikhin1996JPB}, improves the efficiency enormously.  When propagation to very large hyperradii is required in order to obtain accurate scattering information, a hybrid method~\cite{wang2011PRA} has proven to be quite efficient and accurate, which uses SVD at small to intermediate hyperradii but solves the direct coupled adiabatic equations at very large hyperradii. One of the most recent applications of the SVD hyperspherical treatment is an investigation of the famous Hoyle triple-$\alpha$ resonance by~\cite{suno2015PRC}.

\begin{figure}[h]
\centering
 \includegraphics[width=8.5 cm]{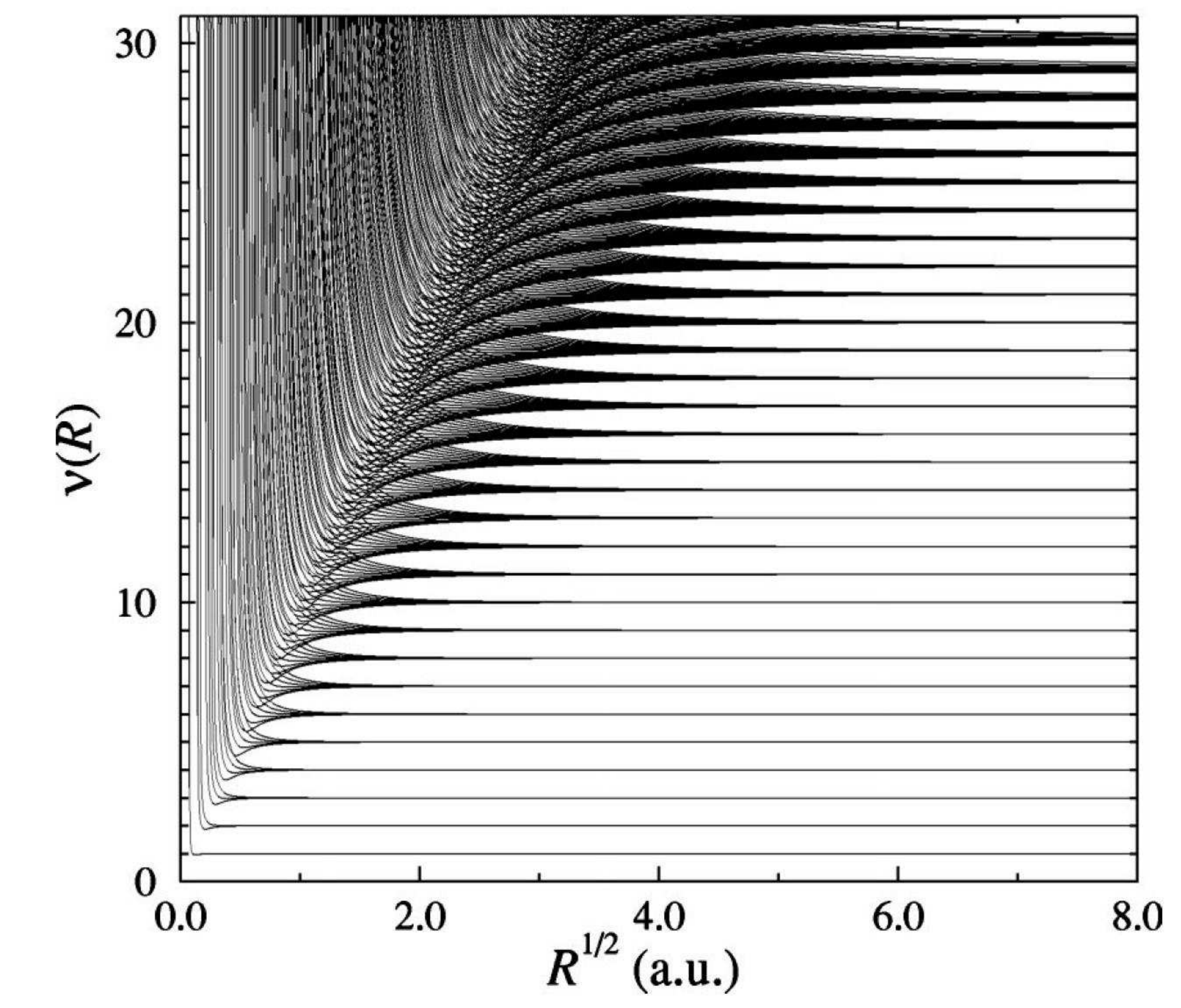}
 \caption{Hyperspherical potential energy curves for the even-parity, zero angular momentum states of the ${\bar p} p e$ system, showing the rich variety of collision channel pathways that exist in this system.  The states shown in the energy range displayed here are mostly of the type ``anti-hydrogen plus electron''.  Note that the energy and hyperradius are displayed respectively here on an ``effective quantum number scale'' $\nu \equiv (-m U(R))^{-1/2}$ with $m$ the proton mass, and square root hyperradial scale $R^{1/2}$, reflecting the usual scaling in a Coulomb potential. For instance $\nu=1$ corresponds to the energy of the ground $1s$ state of the hydrogenic ${\bar p} p$ state on this scale.  Adopted from~\cite{esry2003PRA}.}
\label{Fig-Antiproton}
\end{figure}

\begin{figure}[h]
\centering
 \includegraphics[width=8.5 cm]{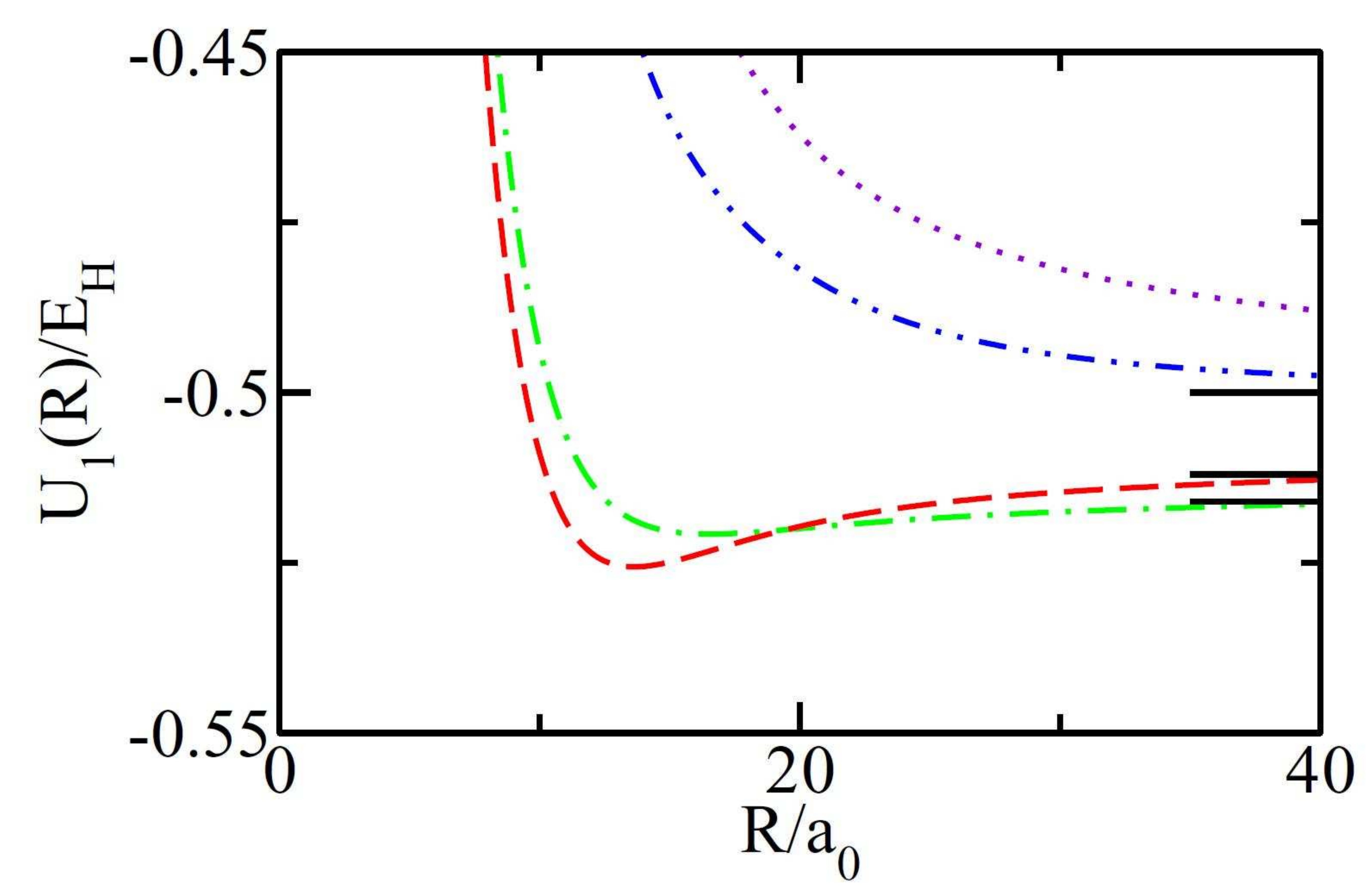}
 \caption{(Color online) The lowest several potential energy curves for zero angular momentum and even total parity are shown versus hyperradius for a 5-body Coulomb system, consisting of three electrons and two positrons. Denoting $(S_+,S_-)$ the separate spin quantum numbers of the positrons and the electrons, these potentials shown as dashed, dash-dotted, dash-dot-dotted, and dotted lines correspond to $(S_+,S_-)= (1,\frac{1}{2}),(0,\frac{1}{2}),(1,\frac{3}{2}),(0,\frac{3}{2}) $, respectively.  The horizontal solid lines ordered from lowest to highest indicate the asymptotic fragmentation threshold energies of Ps$_2$+e$^-$, Ps+Ps$^-$, and 2Ps+e$^-$.  Adopted from \cite{Daily-2014}.}
\label{Fig-5leptons}
\end{figure}

Other exotic examples of three-body Coulombic systems that have been studied include the antiproton+hydrogen atom system, explored by~\cite{esry2003PRA}, which gives an idea of the prototypical hyperspherical potential curves that emerge from applying the adiabatic hyperspherical method to the ${\bar p} p e$ system, as is shown in Fig.~\ref{Fig-Antiproton}.  A huge number of interacting channels are evident, which might initially seem hopelessly daunting in complexity. Closer inspection shows, however, that most of the curve crossings are highly diabatic, and the diabatic potential curves are remarkably simple, suggesting approximately conserved quantum numbers.  Further examples of such simplicity emerging for a seemingly complex system will be demonstrated throughout the present review. Some exciting headway in treating four-body Coulomb systems has also occurred during the last few decades, notably by Morishita, Lin, and collaborators, \cite{morishita1998PRA,morishita1999PRA,morishita1997PRA,dincao2003PRA} in a robust improvement over primitive early studies \cite{clark1980PRA, greene1984PRAc}. A small number of treatments have extended adiabatic hyperspherical ideas to more than four particles, although they are still at a relatively primitive state at this time \cite{morishita2005PRA,bohn1998PRA, blume2000JCP, rittenhouse2006FBS, rittenhouse2008JPB, kim2000JPB, kushibe2004PRA, sogo2005EPL, Daily-2014}. Fig.~\ref{Fig-5leptons} shows an example of the lowest potential energy curves obtained by~\cite{Daily-2014} for a system of 3 electrons and 2 positrons.  These potentials contain bound states of the different symmetries of this 5-body system, and they also describe the lowest energy scattering processes.

Early interest in three-body continuum states in Coulombic systems was mainly triggered by a desire to understand electron impact ionization of atoms, especially in the low energy range, through phenomena such as the Wannier-Rau-Peterkop threshold law for that process \textcolor{black}{derived initially through entirely classical arguments ~\cite{wannier1953pr} and later confirmed through quantum mechanical and semiclassical arguments by ~\cite{RAU1971,PETERKOP1971,PETERKOP1983}. Experimental confirmations of this unusual irrational threshold law for a process with both electrons escaping from a residual particle of positive charge $Ze$, namely $\sigma \propto E^\gamma$ where $\gamma = \frac{1}{4} [(\frac{100Z-9}{4Z-1})^{1/2}-1]$, were measured for electron impact ionization of atomic helium by ~\cite{Cvejanovic1974jpb} and for two-electron photodetachment of H$^-$ ~\cite{Bryant1982prl}(both for $z=1,\gamma=1.127...)$.} Going beyond the double escape threshold law proved to be highly challenging, with some of the first credible absolute cross sections from a theoretical calculation, for the fundamental e + H $\rightarrow$ e + e + p process, emerging first in the numerical ``convergent close-coupling'' studies by \textcolor{black}{ ~\cite{bray1993PRL,Bartlett2003,Kadyrov2009}, performed in ordinary independent electron coordinates. One of the first studies that obtained competitively accurate results within a hyperspherical coordinate framework was that of~\cite{KATO1995,kato1997PRA}, and it was} followed by a subsequent detailed study by \cite{malegat2003PRA,malegat2003PS,malegat2004PRA,malegat2004PSb}. Highly quantitative results are also now obtained for this two-electron escape process by direct solution of the time-independent \cite{McCurdy2004JPB} or time-dependent \cite{pindzola1998PRA,pindzola2007JPB} Schr\"odinger equation. Recent years have seen extensive interest in the time-reversed process: three-body recombination.  For a low-temperature plasma consisting of electrons and protons, this is the reaction e + e +  p $\rightarrow$ H($nl$) + e~\cite{robicheaux2007JPB,robicheaux2010PRL,pohl2008prl} or its antimatter analog with positrons and antiprotons; this mechanism underpins recent exciting progress in the formation of antihydrogen~\cite{andresen2010NT,andresen2011NP}.  

\subsection{Chemical Physics}
\label{Chemical}
Another class of studies that has utilized a 3-body hyperspherical solution to solve a challenging problem in chemical physics is the dissociative recombination (DR) of $H_3^+$~\cite{kokoouline2001NT,kokoouline2003PRL,petrignani2011}.  To describe the DR process where an electron collides with $H_3^+$ and the final state dissociates into $H_2+H$ or $H+H+H$, the use of hyperspherical coordinates has both a practical computational advantage and a qualitative conceptual advantage.  For instance, the theory of DR is much better understood for a diatomic target than for a polyatomic target, so the use of an adiabatic hyperspherical representation of the nuclear positions ultimately maps polyatomic DR theory back in terms of more familiar diatomic DR theory. Those studies also showed that a nontrivial rearrangement collision can be controlled by conical intersection dynamics, more specifically in this case the Jahn-Teller effect.  Part of that solution was a description of the incident electron channels as well as the energetically-closed Rydberg channel pathways using multichannel quantum defect techniques, which will not be discussed in detail here but are summarized elsewhere in the literature~\cite{Kokoouline2011CPL}.

In chemical physics, some of the most impressive theoretical studies of few-atom reactive scattering have been carried out using a hyperspherical coordinate framework. See for instance an early treatment by~\cite{kuppermannJCP1986} of H+H$_2$ scattering. In 1985, experimentalists Neumark and coworkers perform a groundbreaking study\cite{Neumark1985JCP} of the famous F+H$_2 \rightarrow FH+H$ reaction, which required several years before a converged theoretical treatment using hyperspherical coordinates in a variant of Macek's adiabatic representation - the diabatic-by-sector method - was developed by Launay and coworkers~\cite{Launay1990CPL}. Other studies of importance in hyperspherical treatments of reactive scattering were developed by~\cite{pack1987JCP,pack1989JCP}. However, other studies of important rearrangement reactions were carried out using Jacobi or other coordinates, such as the calculation by~\cite{Neuhauser1991CPL}, but our focus in this article is primarily on methodologies that ultimately boil down to solving one or a coupled set of one-dimensional hyperradial Schr\"odinger equations.  \textcolor{black}{A simple and popular method for solving such coupled 1D differential equations is the log-derivative method \cite{Johnson1973jcp, MANOLOPOULOS1993}, while a more advanced technique frequently utilized when there are many close avoided crossings in the potential curves has been developed by \cite{tolstikhin1996JPB} and implemented in various studies such as \cite{wang2011PRA}.}

A handful of studies have even gone beyond three-atom processes and computed scattering cross sections for reactions involving four-atoms using hyperspherical~\cite{Clary1991JCP} or other methods~\cite{Balakrishnan2014JCP}.  These studies can be viewed as solutions to the few-body Schr\"odinger equation, starting from the Born-Oppenheimer potential energy surface as a function of the internuclear coordinates.  Of course a number of important reactive systems have two or more fundamentally coupled potential surfaces, with or without conical intersections, and these require further sophistication even in formulating the basic Born-Oppenheimer Hamiltonian governing the coupled electronic and nuclear degrees of freedom.

\subsection{Fragmentation, recombination, and molecule formation}
\label{Recomb}
The general theory of nuclear reactions was formulated in hyperspherical coordinates in a useful set of papers by \cite{delves1959NP,delves1960NP}, who showed that the usual unitary scattering matrix can be defined in general by inspecting the asymptotic form of the flux-conserving solution at large hyperradii.  Hyperspherical coordinates were picked up by Smirnov and others in the Soviet school of nuclear physics, and that work is reviewed in an extremely practical and general article by~\cite{smirnov1977Sov.J.Part.Nucl.}. The work of that school concentrated on the development of non-interacting solutions in the hyperangular degrees of freedom, the so-called hyperspherical harmonics, including a graphical way to construct the solutions, and the analog of fractional parentage coefficients to achieve their antisymmetrization when applied to several fermionic particles such as nucleons~\cite{smirnov1977Sov.J.Part.Nucl.}. This has been developed further in nuclear collision theory by~\cite{barnea1999JMP,barnea1997AP} and by \cite{nielsen2001PRep} More recently, a model treatment of elastic nucleon scattering of the type $A+A_2$ has shown that there is a significant benefit from adopting adiabatic hyperspherical ideas in the calculation, particularly if the theory is implemented using integral relations for the scattering amplitudes that are developed by \cite{barletta2009PRL,barletta2009FBS,barletta2008FBS}.

The variant of the three-body problem involving short-range forces, particularly relevant in nuclear physics, has served as an independent but equally important testing ground for theory. Whereas in ultracold atomic physics it is a recombination process such as $A+B+C \rightarrow AB+C$ that is of greatest interest, which can form a diatomic molecule in a gas of free atoms, in nuclear physics it is more typically the time reverse of recombination, i.e. $AB+C \rightarrow A+B+C$ whose reaction rates and scattering amplitudes are of interest in laboratory experiments and in astrophysical contexts.  An early study by~\cite{thomas1935pr} showed that the range $r_0$ of two-body nuclear forces cannot be made arbitrarily smaller than the nucleon-nucleon scattering lengths $a_{nn}(S=0)=-18.9$ fm, $a_{np}(S=0)=-23.7$ fm, $a_{np}(S=1)=5.43$ fm, because the \textcolor{black}{three-nucleon} ground state would become {\it arbitrarily deep} and plummet all the way to $-\infty$ in the limit $r_0 \rightarrow 0$, a behavior never observed experimentally, of course, but which is now referred to as the ``Thomas collapse'' effect. Interestingly, however, one sees that the scattering lengths are generally much larger in magnitude than the \textcolor{black}{known range  $r_0\sim 1-2~\rm{fm}$} of the nucleon-nucleon strong force. Another intriguing foray into the behavior of three particles interacting via short range forces came decades later from Efimov, who predicted an effect that bears some connection with the Thomas collapse effect: Efimov predicted that for a system of three particles having infinite two-body scattering lengths, there must be an infinite number of 3-body bound levels that become {\it arbitrarily weak} in their binding.  Efimov's work went on to predict that in the limit where three equal mass particles have common interparticle scattering lengths $a$, the number of such {\emph universal} bound levels becomes finite and is truncated to the approximate value $N \approx \frac{1}{\pi} ln(|a|/r_0)$.  These levels are called universal because they depend only on the dimensionless ratio between the scattering length and the distance $r_0$ beyond which the two-body interactions are negligible, and in some cases an additional parameter is needed, such as the ``three-body parameter'' discussed below in Sec.~\ref{Birth}.

The recombination process that occurs when three ultracold atoms collide, e.g. $A+A+A\rightarrow A_2 + A$ in a Bose-Einstein condensate, became a particularly important topic in the field of degenerate quantum gases in the mid-1990s, when it was increasingly realized that this was the dominant loss process in most experiments.  The reason was that most of the experimental ingenuity had been directed towards turning off inelastic two-body losses by cleverly designing the quantum states of the trapped atoms.  This left little possibility to further turn off inelastic three-body losses, although the gases in real experiments were usually sufficiently dilute that the quantum gas produced could be studied for reasonable periods of time, usually from 0.1 - 100 s.  The process of three-body recombination was studied in a perturbative treatment by Verhaar and collaborators for the case of spin-polarized atomic hydrogen~\cite{Goey-1986}; the rate coefficient for the process is only of order $10^{-38}cm^6/s$, i.e. of extremely low probability because it requires a spin flip via magnetic interactions.  For more typical systems such as alkali atoms that recombine in an ultracold gas, an application of the Verhaar approach~\cite{Moerdijk1996PRA,Moerdijk1996PRA2} predicted that the recombination rate should scale overall as $a^2$, i.e. as the {\it square} of the atom-atom scattering length $a$.  

In fact the growth of the recombination rate \textcolor{black}{coefficient $K_3$} with $a$ was eventually shown to be much faster than quadratic.  The first promising step towards a deeper understanding of three-body recombination emerged from a study by \cite{Fedichev1996PRL} that predicted that the true scaling of $K_3$ should vary much more strongly with scattering length, as $a^4$. Sparked by growing interest throughout the ultracold science community in the need for a deeper understanding of three-body recombination, two nonperturbative treatments of this process at large two-body scattering lengths were published in 1999, one by Nielsen and Macek~\cite{nielsen1999PRLb} and the other by Esry {\it et al.} \cite{esry1999PRL}. While these 1999 Letters confirmed the \cite{fedichev1996PRLb} prediction of an overall $a^4$ scaling of \textcolor{black}{the three-body recombination rate coefficient} $K_3$, they both found an additional Stueckelberg interference modulation with the encouraging potential to cause destructive interference at some very large values of $a$, potentially beneficial for experiments where loss needs to be minimized. In addition, \cite{esry1999PRL} predicted that an infinite number of resonances should periodically enhance the recombination rate at large negative $a$, and that these resonances are Efimov states that have become unbound and merged into the three-body continuum. In the case of homonuclear three-body recombination, those ``zero-energy'' resonances are predicted to have an approximate geometric scaling in the scattering length, with each successive Efimov resonance occurring at a two-body scattering length that is approximately $e^{\pi/s_0} \approx 22.7$ times larger than the preceding one.  

Following these initial predictions, subsequent theoretical studies extended and amplified them, e.g. as reviewed with a focus on the hyperspherical coordinate point of view by several articles \cite{nielsen2001PRep,greene2010PT,rittenhouse2011JPB,wang2013amop,wang2015AnnRev,wang2015AnnRev}.    
Importantly, alternative treatments found largely similar conclusions using methods such as effective field theory \cite{bedaque2000PRL,braaten2001PRL,braaten2003PRA,braaten2006PRep}, a separable interaction application of effective field theory \cite{shepard2007PRA}, two exactly solvable models \cite{macek2006PRA,gogolin2008PRL,MoraGogolinEgger2011} and the treatment by \cite{kohler2002PRL,lee2007PRA} that adopted the early theoretical nuclear physics treatment of \cite{alt1967NPB}.  All of these explorations added tremendously to confidence in the theory community that the Efimov effect should be observable, despite the dearth of experimental confirmation prior to 2006.

Then, however, this field received a tremendous injection of excitement in 2006 when recombination rate measurements for a Cs gas by Grimm's Innsbruck group \cite{kraemer2006NT} observed the aforementioned Efimov resonance in the three-body rate coefficient $K_3$ at a large negative scattering length, in agreement with the 1999 prediction \cite{esry1999PRL}.  That study provided the first experimental confirmation of the Efimov effect.  The scattering length dependence of measured recombination rates in that 2006 experiment closely resembled the predicted shape \cite{esry1999PRL} for a three-body Efimov resonance, but a skeptic might argue that observation of one resonance alone might not be convincing evidence of its Efimov character.  However, subsequent observations of three-body recombination in numerous systems have solidified, confirmed, and extended that interpretation beyond any doubt.  The most dramatic \textcolor{black}{signature} has been observing multiple resonances, separated by the predicted Efimov factor of 22.7 in the scattering length, and multiple predicted interference minima, separated by that same universal factor \cite{nielsen1999PRLb,esry1999PRL,braaten2006PRep,greene2010PT}. 

\begin{figure}[h]
\centering
 \includegraphics[width=6.5 cm]{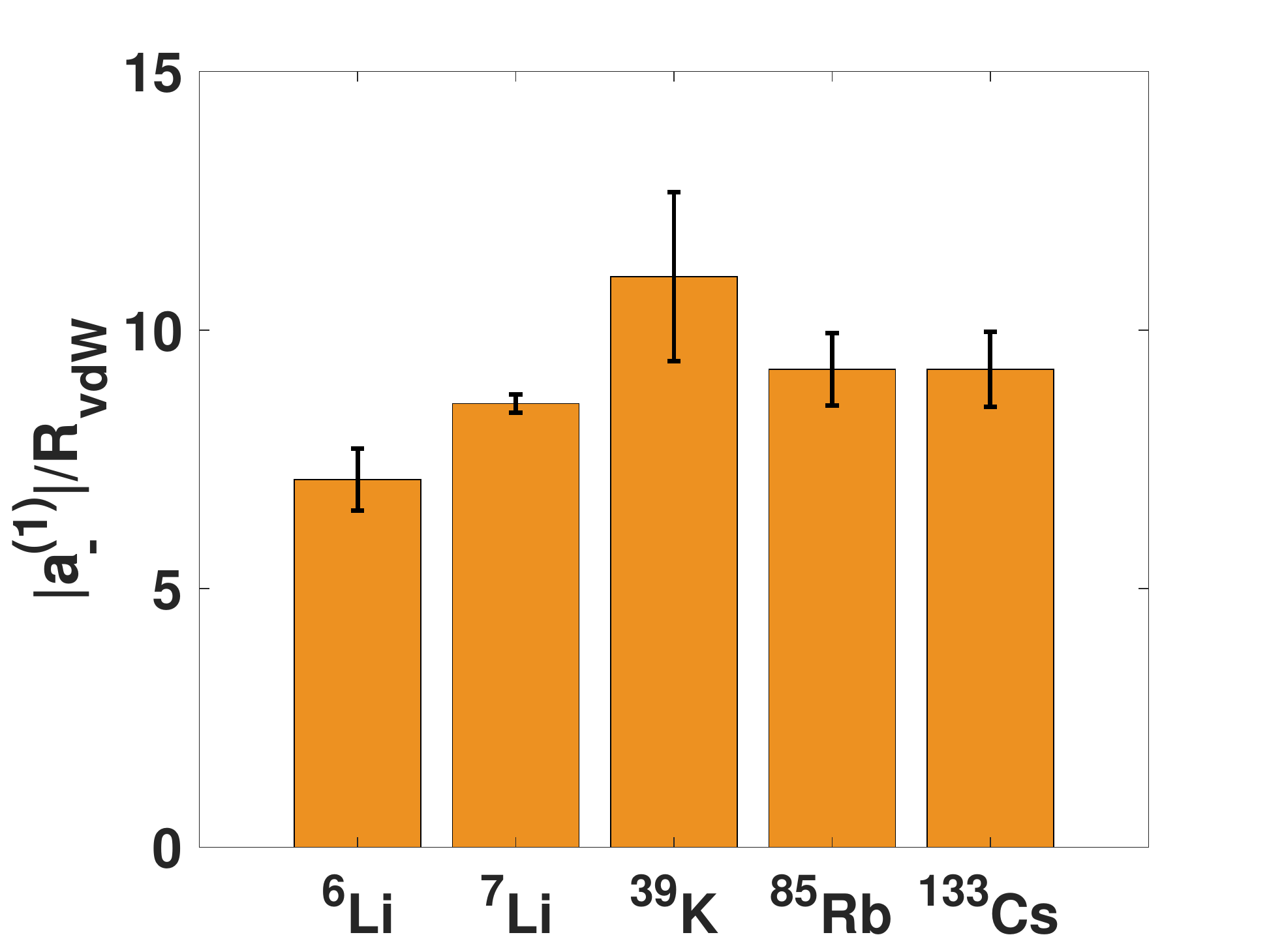}
 \caption{(Color online) Three-body parameter scaled by $\ell_{\rm{vdW}}$ 
 for three equal mass particles. Specifically, this quantity is the value of the (negative) atom-atom scattering length at which the first universal Efimov resonance is observable in a zero energy three-body recombination process. The error bars have been calculated as the weighted mean of the experimental results reported in Sec.\ref{Birth}. }
\label{Universal}
\end{figure}

A further unexpected level of universality emerged from experimental studies with three-atom recombination.  The three-body parameter had been thought by virtually all theorists to occur ``randomly'', and to vary widely from system to system.  The three-body parameter can be viewed as setting the energy $E_0$ of the lowest  Efimov state at $a=\infty$ (unitarity), or alternatively, as the smallest scattering length $a_-^{(1)}$ at which a zero-energy Efimov resonance occurs and thus sets the location of all subsequent resonances through the universal scaling formula, $a_-^{(n)}=a_-^{(1)} e^{(n-1)\pi/s_0}$. 
The remarkable surprise was experimental evidence from the Grimm group \cite{berninger2011PRL} and several others \cite{gross2009PRL,gross2010PRL,gross2011CRP,dyke2013PRA, wild2012PRL,roy2013prl} which showed that for homonuclear three-body systems dominated by van der Waals (vdW) $-C_6 r^{-6}$ two-body interactions at long range, an approximate {\it van der Waals universality} fixes $a_-^{(1)} \approx -10 \ell_{\rm{vdW}}$ in terms of the characteristic length $\ell_{\rm{vdW}} \equiv [m C_6/(16\hbar^2)]^{1/4}$.  As Fig.~\ref{Universal} shows, the three-body parameter is fixed to within approximately 15\% by this simple relation. Shortly after this experimental evidence was published, a theoretical interpretation emerged from \cite{wang2012PRL} which showed that a classical suppression of the two-body probability density whenever two-particles approach to within $r< \ell_{\rm{vdW}}$ produces an effective hyperradial barrier that restricts three-body motion at $R < 2 \ell_{\rm{vdW}}$ and sets the three-body parameter. To clarify, there is a classical suppression because the probability of a classical particle having local velocity $v(r)$ to exist in a region of width $\Delta r$ is proportional to $\Delta r /v(r)$, the time spent by the particle in that region in each traversal.  In the presence of an attractive van der Waals force, the velocity increases suddenly and dramatically when the interparticle distance $r$ decreases to less than the van der Waals length, causing this probability density to plummet in such regions.  The existence of the hyperradial barrier was subsequently confirmed and extended in \textcolor{black}{further studies by \cite{naidon2014prl, naidon2012PRA, naidon2014PRA}} which stressed particularly that a key element of this van der Waals universality is a change from a very floppy equilateral to a roughly linear geometry that occurs near $R \approx 2 \ell_{\rm{vdW}}$; the geometry change then triggers strong non-Born-Oppenheimer repulsion and suppresses the three-body solution at all smaller hyperradii in the relevant potential curve. \textcolor{black}{An alternative toy model addressing the implications of two-body van der Waals forces on the three-body approximate universality has also been published as a preprint by ~\cite{chin2011ARX}.  Other treatments aimed at this issue of three-body parameter universality that start from a two-channel or narrow two-body resonance point of view are presented in ~\cite{wang2014natphys,sorensen2012PRA,schmidt2012EPJB}. }
 
The case of heteronuclear universal Efimov physics appears to be significantly more complicated, e.g. for the particularly interesting case of heavy-heavy-light (HHL) systems that exhibit more favorable Efimov scaling than for the homonuclear three-body systems. But a degree of van der Waals universality has been predicted in \cite{Wang-2012b} to still be relevant \textcolor{black}{for the ``Efimov favored'' HHL case}.  The complexity grows for \textcolor{black}{these heteronuclear systems because more parameters control the universality, namely two van der Waals lengths} and a mass ratio, and the universal energy spectrum now depends on two scattering lengths that are uncorrelated in general.  Nevertheless, early experimental evidence from two different experimental groups \cite{Pires-2014,Tung-2014,Ulmanis2016prl} suggests that this generalized van der Waals universality for HHL systems is at least approximately valid, but still deserves careful study in the future. \textcolor{black}{ A very recent experimental preprint ~\cite{JohansenChin2016arxiv} suggests that for Efimov physics near a {\emph narrow} two-body 
Fano-Feshbach resonance in the $^6$Li-$^{133}$Cs-$^{133}$Cs system, the universal van der Waals theoretical predictions developed for Efimov physics in the HHL system near a broad two-body resonance~\cite{Wang-2012b} will require very substantial modifications, e.g. by implementing a multichannel model for the two-body interaction Hamiltonian as in ~\cite{wang2014natphys,kartavtsev2002FBS,mehta2008PRA}.  For light-light-heavy (LLH) three-body systems, Ref.~\cite{Wang-2012b} stresses that these are ``Efimov-unfavored'', and it is unlikely that a true Efimov state will be observable experimentally.}

\subsection{Recombination processes involving cluster resonances with more than three particles}
\label{More}
\textcolor{black}{A detailed discussion of universal four-boson and five-boson energy levels and recombination resonances is given below in subsection \ref{fourANDfive}, but here we point out some of the basic issues involved in describing cluster resonances in systems of $N>3$ identical bosons having short-range interactions.  Most of these systems have a pairwise attractive long-range interaction and a strong short-range repulsion, as in the case $N$ bosonic helium atoms.  Simple counting then shows that in the relative coordinate system the number of positive terms in the kinetic energy operator  is proportional to $(N-1)$ whereas the number of net attractive terms in the pairwise potential energy is equal to $N(N-1)/2$. Thus, one expects that if one is in a negative region of the two-body scattering length $a$ where 3 particles are not quite attracted strongly enough to bind a universal trimer state, there could be a value of the negative scattering length $a=a_{4B}^-$ where 4 or more particles are able to bind.  Similarly, if one goes to a region where 4 particles are not quite strongly enough to be bound, there should be a negative value of $a = a_{5B}^-$. One can explore this theoretically either by varying the two-body potential strength to modify the scattering length,\cite{stecher2010JPB,yamashita2010pra, stecher2011PRL,gattobigio2012PRA,Nicholson2012prl,YanBlume2015pra} or by artificially changing the particle mass in the calculations for a fixed two-body potential, which also modifies the repulsive/attractive balance in the Hamiltonian~\cite{hanna2006PRA}. This concept has been studied in a number of studies,and some universal aspects have already emerged.  In particular, the most recent careful discussion by ~\cite{YanBlume2015pra} gives evidence that for general short-range two-body interactions, such as gaussians or other short-range potentials, the $N$-body cluster energies at unitarity $a\rightarrow \infty$ are not uniquely specified since they depend on the type of ``three-body regulator'' implemented.  However, there does appear to be a quasi-universality that emerges in the case of van der Waals two-body interactions: the cluster bound state energies at unitarity are then approximately fixed in terms of the van der Waals length scale. }

These and other developments will be addressed in the remainder of this review, including a detailed description of the techniques, while stressing methods of interpretive analysis that have been utilized to study these universal phenomena from a hyperspherical coordinate perspective. \textcolor{black}{A very recent treatment of universal 5-body bound states in a mass-imbalanced fermionic system has been developed by ~\cite{BazakPetrov2016} using alternative (integral equation) techniques~\cite{Pricoupenk2011pra}.}

\section{Adiabatic Hyperspherical Treatment}
\label{Sec.2}

A Schr\"{o}dinger wave equation for $N$ interacting particles, with masses $%
m_{i}$ moving in 3 dimensions per particle, becomes in the absence of
external fields, a $d=3N-3$ dimensional partial differential equation (PDE)
in the relative coordinate system. \ When expressed in hyperspherical
coordinates, a single scalar coordinate, the hyperradius $R$ defined below,
is singled out for special treatment within an adiabatic formulation.\ It is
possible in general to formally transform the $d-$dimensional \textcolor{black}{PDE}, specifically the time-independent Schr\"{o}%
dinger equation $\hat{H}\Psi =E\Psi $ for any potential energy function
dependent on the relative position coordinates only, into an infinite set of 
\textit{ordinary} coupled differential equations in a \textit{single}
adiabatic coordinate $R$. \ Moreover, a conceptual advantage of
hyperspherical coordinates is that every possible fragmentation mode for any
system of particles occurs in the limit $R\rightarrow \infty .$

The basic equations of the adiabatic representation are simple to derive. \
First of all, write the full time-independent Hamiltonian in the form%
\begin{equation}
\hat{H}=\hat{T}_{R}+H_{R=\text{const}},
\label{Eq:Ham}
\end{equation}%
where the term $H_{R=\text{const}}$ depends on $R$ only as a parameter and
is a Hermitian partial differential operator in all other (hyperangular)
coordinates of the system plus spins, denoted collectively here as $\{\varpi
\}.$ \ Next, solve the eigenvalue equation at each value of $R$:%
\begin{equation}
H_{R=\text{const}}\Phi _{\nu }(R;\varpi )=u_{\nu }(R)\Phi _{\nu }(R;\varpi ).
\label{Eq:AdiabaticHam}
\end{equation}%
The exact eigenfunctions of the full $\hat{H}$ can now be expanded into the
complete, orthonormal set of eigenfunctions $\Phi _{\nu }(R;\varpi )$ with $%
R $-dependent coefficients $F_{E\nu }(R),$ as 
\begin{equation}
\Psi _{E}(R;\varpi )=R^{-(d-1)/2}\sum_{\nu }\Phi _{\nu }(R;\varpi )F_{E\nu
}(R),
\label{Eq:WfnAnsatz}
\end{equation}%
giving an infinite set of coupled differential equations for the hyperradial
functions:%
\begin{equation}
(-\frac{\hbar ^{2}}{2\mu }\frac{d^{2}}{dR^{2}}+U_{\nu }(R)-E)F_{E\nu
}(R)=-\sum_{\nu ^{\prime }}\hat{W}_{\nu \nu ^{\prime }}F_{E\nu ^{\prime
}}(R),
\label{Eq:CoupledEqns}
\end{equation}%
where $\mu$ is the $N$-body reduced mass and its explicit form is given in Eq.~(\ref{Eq:NbodyRedMass}).
Observe that for a coordinate space with $d$ dimensions, the hyperradial
kinetic energy operator has the form $\hat{T}_{R}=-\frac{\hbar ^{2}}{2\mu }%
\frac{1}{R^{d-1}}\frac{\partial }{\partial R}R^{d-1}\frac{\partial }{%
\partial R},$ and the rescaling of the radial function eliminates the
first-order derivative of $F_{E\nu }(R)$ on the left-hand side of Eq.(\ref{Eq:CoupledEqns}). \
The rescaling also adds what Fano called a \textquotedblleft
mock-centrifugal term\textquotedblright\ to $u_{\nu }(R),$ giving the full
effective \textcolor{black}{hyperradial} Born-Oppenheimer potential as 
\begin{equation}
U_{\nu }(R)=u_{\nu }(R)+\frac{(d-1)(d-3)\hbar ^{2}}{8\mu R^{2}}
\label{Eq:PotCurve}
\end{equation}%
\ \ \ \ \ The coupling terms on the right-hand side of Eq.(\ref{Eq:CoupledEqns}) which are
responsible for nonadiabatic coupling are given by:%
\begin{equation}
\hat{W}_{\nu \nu ^{\prime }}F_{E\nu ^{\prime }}=-\frac{\hbar ^{2}}{2\mu }%
Q_{\nu \nu ^{\prime }}(R)F_{E\nu ^{\prime }}(R)-\frac{\hbar ^{2}}{\mu }%
P_{\nu \nu ^{\prime }}(R)\frac{dF_{E\nu ^{\prime }}(R)}{dR}.
\label{Eq:CouplingMatrices}
\end{equation}%
Here the two nonadiabatic coupling matrices are given by $Q_{\nu \nu
^{\prime }}(R)\equiv \left\langle \left\langle \Phi _{\nu }(R;\varpi
)\left\vert \frac{\partial ^{2}}{\partial R^{2}}\right\vert \Phi _{\nu
^{\prime }}(R;\varpi )\right\rangle \right\rangle ^{(R)}$ and $P_{\nu \nu
^{\prime }}(R)\equiv \left\langle \left\langle \Phi _{\nu }(R;\varpi
)\left\vert \frac{\partial }{\partial R}\right\vert \Phi _{\nu ^{\prime
}}(R;\varpi )\right\rangle \right\rangle ^{(R)}$ where the double bracket
notation signifies an integral (and spin trace) only over the $\varpi $
degrees of freedom. \ This set of coupled equations is sometimes treated in
the \textit{ \textcolor{black}{hyperradial} Born-Oppenheimer approximation} which neglects the right-hand
side of Eq.~(\ref{Eq:CoupledEqns}). \ In that approximation, the system moves along a single
potential energy curve with no possibility of changing from one potential to
another, and this approximation of course has no possibility of describing
an inelastic collision. \ But in some cases it can give a reasonable
description of energy levels and scattering phaseshifts, although in most 
cases a more accurate result is obtained by retaining 
(except near close avoided crossings) the diagonal terms of 
Eq.~(\ref{Eq:CoupledEqns}) which is usually referred to as the 
\textcolor{black}{hyperspherical}
{\it adiabatic approximation}.\ 

While the \textcolor{black}{hyperradial} Born-Oppenheimer approximation, which considers only a single term
in the expansion for $\Psi _{E}$ in Eq.~(\ref{Eq:CoupledEqns}), is often reasonable, a far richer
set of phenomena emerges when nonadiabatic coupling effects are incorporated,
either by direct solution of the coupled radial equations or else using
semiclassical methods such as Landau-Zener-St\"{u}ckelberg or their
improvements along the lines \textcolor{black}{of \cite{nikitin1970,zhu2001}}. 
This in fact yields a quantitative description
of phenomena such as three-body or four-body recombination, and inelastic
atom-dimer or dimer-dimer scattering.

The following development sketches one explicit version of this recasting of
the Schr\"{o}dinger equation into hyperspherical coordinates for an
N-particle system in 3 dimensions. Note that a similar development for N 2D particles is
presented by \cite{Daily2015PRB}, in the context of the quantum Hall effect. 
\ One first transforms the $N$
laboratory frame position vectors \{$\vec{r}_{i}$\} in terms of a suitable
set of $N-1$ mass-weighted relative Jacobi coordinate vectors \{$\vec{\rho}%
_{i}$\}, plus the center of mass vector which is trivial and is therefore
ignored throughout. \ Extensive arbitrariness and flexibility exists for the
choice of the Jacobi coordinate vectors, but for definiteness, one simple
choice is based on choosing the $j-$th Jacobi vector as the (reduced-mass
weighted) relative vector between particle $(j+1)$ and the center of mass of
the preceding group of particles $1$ through $j$, i.e.:
\begin{equation}
\begin{array}{c}
\vec{\rho}_{1}=\sqrt{\frac{\mu _{12}}{\mu }(}\vec{r}_{2}-\vec{r}_{1}) \\ 
\vec{\rho}_{2}=\sqrt{\frac{\mu _{12,3}}{\mu }}(\vec{r}_{3}-\frac{m_{1}\vec{r}%
_{1}+m_{2}\vec{r}_{2}}{m_{1}+m_{2}})%
\end{array}%
\label{Eq:JacobiVecs}
\end{equation}%
$...$ \ etc., where the $N-1$ Jacobi reduced masses are%
\begin{equation}
\mu _{12}=\frac{m_{1}m_{2}}{m_{1}+m_{2}},\mu _{12,3}=\frac{(m_{1}+m_{2})m_{3}%
}{m_{1}+m_{2}+m_{3}},\text{...etc.,} 
\label{Eq:ReducedMasses}
\end{equation}%
and where the $N$-body reduced mass is%
\begin{equation}
\mu =(\mu _{12}\mu _{12,3}....)^{\frac{1}{N-1}}. 
\label{Eq:NbodyRedMass}
\end{equation}%
Alternative choices for the overall reduced mass $\mu $ are possible and are
sometimes utilized, but this choice in Eq.(\ref{Eq:NbodyRedMass}) is particularly desirable
in many contexts because it preserves the overall volume element. With these
definitions, the nonrelativistic kinetic operator acquires a simple form,
namely%
\begin{equation}
\hat{T}=-\frac{\hbar ^{2}}{2\mu }\sum_{j=1}^{N-1}\vec{\nabla}_{\rho
_{j}}^{2}\equiv -\frac{\hbar ^{2}}{2\mu }\sum_{i=1}^{d}\frac{\partial ^{2}}{%
\partial x_{i}^{2}}
\label{Eq:KineticEnergy}
\end{equation}%
The Cartesian coordinates of all these Jacobi vectors can thus be collected
into a single $d$-dimensional relative vector \ \b{x}$\equiv
\{x_{1},x_{2},x_{3},...x_{d}\},$ and these can in turn be transformed into
hyperspherical coordinates by defining the hyperradius $R$ as the radius of
the $d$-dimensional hypersphere: 
\begin{equation}
R=\sqrt{x_{1}^{2}+x_{2}^{2}+x_{3}^{2}+...x_{d}^{2}}.
\label{Eq:HyperRadius}
\end{equation}%
There are again many possible choices for the $d-1$ hyperangles $\alpha _{k}$%
, but one simple generalization of our usual spherical coordinates is
implied by the chain\cite{avery1989}:%
\begin{equation}
\begin{array}{c}
x_{d}=R\cos \alpha _{d-1} \\ 
x_{d-1}=R\sin \alpha _{d-1}\cos \alpha _{d-2} \\ 
x_{d-2}=R\sin \alpha _{d-1}\sin \alpha _{d-2}\cos \alpha _{d-3} \\ 
...x_{2}=R\prod_{j=1}^{d-1}\sin \alpha _{j},\text{ and }x_{1}=R%
\prod_{j=2}^{d-1}\sin \alpha _{j}\cos \alpha _{1}.%
\end{array}%
\label{Eq:NbodyHyperangles}
\end{equation}%
This easily generalizable choice of the hyperangles is sometimes referred to
as the \textit{canonical} choice. \ The ranges spanned by these hyperangles
\textcolor{black}{are} then

\begin{equation}
0\leq \alpha _{1}\leq 2\pi , \\
0\leq \alpha _{i}\leq \pi ,i=2,...,d-1.
\end{equation}

Now, the nonrelativistic kinetic energy operator in hyperspherical
coordinates can be conveniently written as%
\begin{equation}
\hat{T}=T_{R}+\frac{\hbar ^{2}\boldsymbol{\Lambda} ^{2}}{2\mu R^{2}},
\end{equation}%
where $T_{R}=-\frac{\hbar ^{2}}{2\mu }\frac{1}{R^{d-1}}\frac{\partial }{%
\partial R}R^{d-1}\frac{\partial }{\partial R},$ and where $\boldsymbol{\Lambda} ^{2}$ is
the isotropic Casimir operator for the group O(d), \textcolor{black}{\cite{smirnov1977Sov.J.Part.Nucl., Knirk1974jcp,cavagnero1984pra,cavagnero1986pra}} given explicitly by 
\begin{equation}
\mathbf{\Lambda }^{2}=-\sum_{i>j}\Lambda _{ij}^{2},
\label{Eq:hyperangmoment_def} \ \ \ 
\Lambda _{ij}=x_{i}\dfrac{\partial }{\partial x_{j}}-x_{j}\dfrac{\partial }{%
\partial x_{i}}.  \nonumber
\end{equation}%
The operator $\mathbf{\Lambda }^{2}$ is often referred to as the \textcolor{black}{square of the}
\textquotedblleft grand angular momentum\textquotedblright\ operator of the
system. These equations now show how the physics of this $d$-dimensional
problem can be mapped exactly onto an adiabatic representation in the single
coordinate $R$, with potential energy curves $U_{\nu }(R)$ and nonadiabatic
coupling terms as in standard Born-Oppenheimer theory. \ As is particularly
stressed by \cite{macek1968JPB,fano1983RPP,fano1981PRA}, and as we document 
below, this approach yields tremendous
insights in many physical systems.

Some examples of applying the adiabatic hyperspherical representation to
systems with many particles are summarized below in Sec.(\ref{manybody}). \ But before
turning to examples, we show how far greater symmetry and simplicity emerge
from a clever choice of the hyperangles for three-particle systems, $N=3,$
by adopting a \textquotedblleft body-fixed\textquotedblright\ coordinate
system of the type suggested by \cite{whitten1968JMP}. \ The particular variant
described here adopts the conventions specified by \cite{suno2002PRA}.


Usually we are interested in three-body systems that have exact separability
in the relative and center of mass coordinates, whereby the relative degrees
of freedom can be described by six coordinates, i.e. $d=6$ is the full
dimensionality of this space. Three of these coordinates are conveniently
chosen to be Euler angles $\{\alpha $, $\beta $, $\gamma \}$ that connect
the body-fixed frame to the space-fixed frame. Three remaining coordinates
in this system are the hyperradius $R$ and two hyperangles $\theta $ and $%
\varphi $. Following Refs. \cite{whitten1968JMP,johnson1983JCP,kendrick1999JCP,Lepetit-1990} with only minor
modifications described in \cite{suno2002PRA}, this begins from the
mass-scaled Jacobi coordinates introduced above \cite{delves1960NP} 
\begin{eqnarray}
\vec{\rho}_{1} &=&(\vec{r}_{2}-\vec{r}_{1})/\Delta , \\
\vec{\rho}_{2} &=&\Delta \left[ \vec{r}_{3}-\frac{m_{1}\vec{r}_{1}+m_{2}\vec{%
r}_{2}}{m_{1}+m_{2}}\right] ,
\end{eqnarray}%
with 
\begin{equation}
\Delta ^{2}=\frac{1}{\mu }\frac{m_{3}(m_{1}+m_{2})}{m_{1}+m_{2}+m_{3}}
\label{mass_factor}
\end{equation}%
and $\mu $ is the three-body reduced mass as was defined above, namely 
\begin{equation}
\mu ^{2}=\frac{m_{1}m_{2}m_{3}}{m_{1}+m_{2}+m_{3}}.  \label{reduced_mass}
\end{equation}%
In this expression, particle $i$ with mass $m_{i}$ has position $\vec{r}_{i}$%
. When the three particles have identical mass $m$, the parameters simplify
to $\Delta =(4/3)^{\frac{1}{4}}$ and $\mu =m/\sqrt{3}$. And specializing the
above definition of the hyperradius $R,$ it is given here by: 
\begin{equation}
R^{2}=\rho _{1}^{2}+\rho _{2}^{2},\quad 0\leq R<\infty .
\label{3body_hyperradius}
\end{equation}%
The hyperangles $\theta $ and $\varphi $ are determined by the four nonzero
components of the two Jacobi vectors in the body frame $x-y$ plane by 
\begin{equation}
\begin{array}{l}
(\vec{\rho}_{1})_{x}=R\cos (\theta /2-\pi /4)\sin (\varphi /2+\pi /6), \\ 
(\vec{\rho}_{1})_{y}=R\sin (\theta /2-\pi /4)\cos (\varphi /2+\pi /6), \\ 
(\vec{\rho}_{2})_{x}=R\cos (\theta /2-\pi /4)\cos (\varphi /2+\pi /6), \\ 
(\vec{\rho}_{2})_{y}=-R\sin (\theta /2-\pi /4)\sin (\varphi /2+\pi /6),%
\end{array}
\label{thetaphi}
\end{equation}%
where by definition $\rho _{1z}=0=\rho _{2z}.$\ \ For definiteness, note
that the $x$, $y$, and $z$ right-handed coordinate system of the body-fixed
frame is chosen such that the $z$ axis is parallel to $\vec{\rho}_{1}\times 
\vec{\rho}_{2}$, and the $x$ axis is that with the smallest moment of
inertia. The ranges of the hyperangles are $0\leq \theta \leq \frac{\pi }{2}$
and $0\leq \varphi <2\pi $ \cite{kendrick1999JCP}. \ If the three equal mass
particles are in fact truly identical, then the hyperangle $\varphi $ can be
further restricted to the range $[0,2\pi /3]$. Note that in this case, the
interaction potential is symmetric under the operation $\varphi \rightarrow
\pi /3-\varphi $. Then the bosonic or fermionic symmetry of the Schr\"{o}dinger solutions under exchange of any two particles is particularly simple
to impose as a boundary condition in these coordinates. The volume element
for integrals over $|\Psi |^{2}$ is equal to $d\mathcal{V}\equiv d\varpi
R^{5}dR=2\sin 2\theta d\theta d\varphi d\alpha \sin \beta d\beta d\gamma
R^{5}dR,$ and the Euler angle ranges are $0\leq \alpha <2\pi ,$ $0\leq \beta
<\pi ,$ $0\leq \gamma <\pi $. The full Schr\"{o}dinger equation for the
rescaled wavefunction $\psi _{E}=R^{5/2}\Psi $ describing three identical
particles now takes the form 
\begin{equation}
\left( -\frac{1}{2\mu }\frac{\partial ^{2}}{\partial R^{2}}+\frac{15}{8\mu
R^{2}}+\frac{\boldsymbol{\Lambda} ^{2}}{2\mu R^{2}}+V(R,\theta ,\varphi )\right) \psi
_{E}=E\psi _{E},  \label{SchrodingerEquation}
\end{equation}%
In this expression, $\boldsymbol{\Lambda} ^{2}$ is the squared \textquotedblleft grand
angular momentum operator\textquotedblright\ and is given by \cite%
{kendrick1999JCP,Lepetit-1990} 
\begin{equation}
\frac{\boldsymbol{\Lambda} ^{2}}{2\mu R^{2}}=T_{1}+T_{2}+T_{3},
\end{equation}%
where 
\begin{eqnarray}
T_{1} &=&-\frac{2}{\mu R^{2}\sin 2\theta }\frac{\partial }{\partial \theta }%
\sin 2\theta \frac{\partial }{\partial \theta }, \\
T_{2} &=&\frac{1}{\mu R^{2}\sin ^{2}\theta }\left( i\frac{\partial }{%
\partial \varphi }-\cos \theta \frac{L_{z}}{2}\right) ^{2}, \\
T_{3} &=&\frac{L_{x}^{2}}{\mu R^{2}(1-\sin \theta )}+\frac{L_{y}^{2}}{\mu
R^{2}(1+\sin \theta )}+\frac{L_{z}^{2}}{2\mu R^{2}}.
\end{eqnarray}%
The total orbital angular momentum operator in the body frame is denoted
here as $\vec{L}=\{L_{x},L_{y},L_{z}\}$. For an interacting 3-body system,
one frequently adopts a sum of two-body potential energy functions for $%
V(R,\theta ,\varphi )$, but some explorations are carried out with explicit
non-pairwise additive terms as well. \ That is, most explorations of
universal physics have used an approximate 3-particle $V$ of the form: 
\begin{equation}
V(R,\theta ,\varphi )=v(r_{12})+v(r_{23})+v(r_{31}),
\label{InteractionPotential}
\end{equation}%
where $r_{ij}$ are the interparticle distances. For three equal mass
particles, these distances are expressed in terms of the hyperspherical
coordinates as 
\begin{equation}
\begin{array}{l}
r_{12}=3^{-1/4}R[1+\sin \theta \sin (\varphi -\pi /6)]^{1/2}, \\ 
r_{23}=3^{-1/4}R[1+\sin \theta \sin (\varphi -5\pi /6)]^{1/2}, \\ 
r_{31}=3^{-1/4}R[1+\sin \theta \sin (\varphi +\pi /2)]^{1/2}.%
\end{array}
\label{InterparticleDistances}
\end{equation}

\textcolor{black}{As was indicated above, the 
first step in implementing the adiabatic
representation is to solve the fixed-$R$ adiabatic eigenvalue equation for a
given symmetry $L^{\Pi }$ to obtain the fixed-$R$ adiabatic eigenfunctions ($\Phi
_{\nu },$ sometimes referred to as channel functions) and eigenvalues
(potential energy curves $U_{\nu }(R)$). Here we adopt
an abbreviated notation with $\Omega \equiv (\theta ,\varphi ,\alpha ,\beta
,\gamma )$ and for some systems $\Omega$ includes spin degrees of freedom as well.} For the body-frame choice of hyperangles, it is simplest to
expand the Euler angle dependence of the $\Phi _{\nu }$ in terms of
normalized Wigner D-functions, $\tilde{D}_{MK}^{L}(\alpha \beta \gamma ),$
i.e. as 
\begin{equation}
\Phi _{\nu }^{L\Pi }(R;\Omega )=\sum\limits_{K}\phi _{K\nu}(R;\theta ,\varphi )%
\tilde{D}_{MK}^{L}(\alpha \beta \gamma ).
\end{equation}%
This representation guarantees that $\Phi _{\nu }^{L\Pi }$ is automatically
an eigenfunction of $\vec{L}^{2},$ and it is also an even (odd)
eigenfunction of the parity operator $\hat{\Pi}$ provided $K$ is restricted
to even (odd) values respectively. \ A few more details of this body frame
representation are useful when using this representation to convert the
5-dimensional PDE Eq.(\ref{SchrodingerEquation}) into a set of coupled 2D PDEs in 
$\theta ,\varphi $ only. In this body frame representation of angular
momentum, the raising and lowering operators are defined 
\textcolor{black}{(owing to the anomalous commutation relations of 
body-frame operators)} as:%
\begin{equation}
L_{\pm }=L_{x}\mp iL_{y},
\end{equation}%
where%
\begin{eqnarray*}
L_{\pm }\tilde{D}_{M,K}^{L}(\alpha \beta \gamma ) &=&\sqrt{(L\mp K)(L\pm K+1)%
}\tilde{D}_{M,K\pm 1}^{L}(\alpha \beta \gamma ) \\
L_{z}\tilde{D}_{M,K}^{L}(\alpha \beta \gamma ) &=&K\tilde{D}%
_{M,K}^{L}(\alpha \beta \gamma ).
\end{eqnarray*}%
\ 

\bigskip After inserting the above expressions, one obtains for each value
of $\{L,M,\Pi \}$ a finite number of coupled 2D PDEs in $\theta ,\varphi .$
\ The terms involving $L_{x}^{2}$ and $L_{y}^{2}$ cause couplings between
components $K$ and $K\pm 2.$ \ While these PDEs are for complex solutions,
as written here, it is possible to take linear combinations, e.g. $\phi
_{K}(R;\theta ,\varphi )\pm \phi _{-K}(R;\theta ,\varphi )$ and reformulate
the PDEs in terms of real functions everywhere. 

\subsection{Recombination cross sections and rate coefficients}

It was proven by Delves that the hyperspherical representation preserves 
the usual desired properties of continuum scattering solutions, such as 
flux conservation when the Hamiltonian is Hermitian which ensures unitarity
of the scattering matrix $S$ and symmetry of the $S$-matrix when the 
Hamiltonian is time-reversal invariant.  One simple conceptual aspect of 
Macek's adiabatic hyperspherical representation involving potential energy 
curves and nonadiabatic couplings is that the computation 
of the unitary $S$-matrix can utilize any
of the powerful techniques already developed for treating two-body inelastic
scattering processes.  In other words, just as in standard multichannel
scattering theory \cite{Rodberg1970} or multichannel quantum defect theory
\cite{seaton1983rpp,fano1970,greene1985AMOP,aymar1996RMP,Mies-1984,
burke1998PRLb,Mies-2000,Gao2001,ruzic2013}, one simply propagates solutions 
of the coupled equations in Eq.~(\ref{Eq:CoupledEqns}) out to large distances, fits to linear combination
of energy-normalized regular and irregular radial functions 
$\{f_{E\nu}(R),g_{E\nu}(R)\}$ and in this manner
obtain a real, symmetric reaction matrix $K_{\nu\nu'}(E)$ characterizing solutions
from some large matching radius $R_0$ out to infinity:
\begin{equation}
  \Psi _{E\nu'}(R;\varpi )=\sum_{\nu }
\frac{\Phi _{\nu }(R;\varpi )}{R^{(d-1)/2}}(f_{E\nu}(R)\delta_{\nu\nu'}-g_{E\nu}(R)K_{\nu\nu'})
\end{equation}
Then linear combinations 
of those solutions can be taken to enforce any appropriate boundary conditions at
$R\rightarrow \infty$ for the observable quantities of 
interest \cite{fano1986,aymar1996RMP}.  The usual relations are obtained for 
quantities like $S=(\mathds{1}+iK)(\mathds{1}-iK)^{-1}$ with extra long range phase factors
sometimes needed to satisfy outgoing-wave or incoming-wave boundary conditions.  (See, 
e.g. Sec.II of \cite{aymar1996RMP}.)  Of particular interest in the context 
of ultracold quantum gases is the three-body recombination rate coefficient which was derived
in \cite{esry1999PRL}. The relevant formula for three identical bosonic particles which are in 
a thermal gas rather than a BEC, 
after correcting for a factor of 6 error in the formulas
reported in that paper, is:
\begin{equation}
  K_3(E) = \frac{\hbar k}{\mu} \frac{192 \pi^2}{k^5} \Sigma_{\nu'\nu} |S_{\nu',\nu}|^2
\label{Eq:K3formula}
\end{equation}
Here $k=\sqrt{2 \mu E/\hbar^2}$ and the sum includes all entrance three-body 
continuum channels (A+A+A, $\nu$) for the symmetry of interest, and over all final state two-body bound
channels (A$_2$+A, $\nu'$).  A few words are relevant to explain how this recombination rate
coefficient is to be used in rate equations used to model this reaction in a cold gas.  This quantity 
$K_3(E)$ is the {\it fundamental} coefficient relevant to a single triad of particles in the gas.  
The coefficient in the rate equations for disappearance of atoms or appearance of dimers is another rate 
coefficient, $L_3$, which is determined by the following points.  If one imagines that there are
$N$ atoms in a thermal gas volume $V$, then there are $g_N=N(N-1)(N-2)/3!\approx N^3/6$ distinct triads of the type A+A+A
in the system.  If we define a density as $n \equiv N/V$, then the rate equation for disappearance of atoms from a cold
trapped gas is
\begin{equation}
   \frac{dn}{dt}=-L_3 n^3,
\label{Eq:L3formula}
\end{equation}
where for a thermal trapped gas, $L_3 = 3 K_3 \frac{g_N}{N^3} \approx \frac{K_3}{2}$.  In this last equation, the leading 
factor of 3 in the middle is the number of atoms lost in each recombination event, and the value 3 reflects the fact that for a typical trapped gas of atoms, a recombination event releases so much 
kinetic energy that both the final dimer and the final atom following recombination will be ejected, i.e. all three of the initially free atoms.  If an unusually deep
trap is implemented, or if the binding energy of the dimer produced is far less than the trap depth, then that factor of 3 
would of course be changed to 2 since only the dimer would escape the atom trap, 
though one should then also keep track of the energy deposited into the 
cloud by the remaining hot atom.  As is also well known, \cite{kagan1985JETPL, burt1997PRL, Dalibard1999} if the initial 
atom cloud is in a pure BEC rather than a thermal
gas, then the preceding $K_3$ needs to be reduced by a factor of $3!$.  

Some of the simplest and most important early predictions of the low energy recombination rate behavior
include an expected $a^4$ scaling \cite{fedichev1996PRLb}, which was later seen to be modified in a nontrivial way
that differs depending on whether the atom-atom scattering length $a$ is 
positive \cite{nielsen1999PRLb,esry1999PRL} or negative \cite{esry1999PRL}.  If $a$ is 
large and positive, then there exists a weakly bound dimer state whose energy is approximately $-\hbar^2/m a^2$,
and the recombination rate into that universal dimer channel should have St\"uckelberg 
interference minima at scattering lengths $a_+^{(i)}$ whose
spacings should scale geometrically with the Efimov scaling parameter $a_+^{(i+1)}/a_+^{(i)} = e^{\pi/s_0} \sim 22.7$.
If instead, $a$ is negative, then this implies that there is no weakly-bound universal dimer, and recombination
can occur only into deeper nonuniversal dimer channels.  On this side, even though the attraction is not strong enough
to bind two atoms together into a universal dimer, the Efimov effect can bind trimers at certain values of $a_-^{(i)}<0$.  
Moreover, the successive values of $a$ where a trimer can form at zero energy also obey 
the Efimov scaling, $a_-^{(i+1)}/a_-^{(i)} \sim 22.7$. 
While the first experiments \cite{Pollack-2009,dyke2013PRA,knoop2009NTP,knoop2010PRL,ferlaino2008PRL,ferlaino2009PRL,ferlaino2011FBS,berninger2011PRL,berni2013pra,zenesini2013NJP,zaccanti2009NTP,gross2009PRL,gross2010PRL,gross2011CRP,machtey2012PRL,machtey2012PRLb} were only able to observe a single Efimov resonance ($i=1$) for homonuclear systems, a
recent impressive experiment by \cite{Huang-2014b} has observed the $i=1,2$ resonances and confirmed their approximate ratio to
be close to Efimov's predicted value.  

Much subsequent theory has treated the physics of recombination, and developed compact analytical formulas within the 
framework of zero-range models and/or effective field theory, which are particularly convenient for analyzing experimental
data.  See for instance the following references \cite{braaten2006PRep,macek2006PRA,gogolin2008PRL,MoraGogolinEgger2011}.  A different direction of extending
and generalizing recombination theory has been the treatment of recombination processes for $N>3$ particles.  A generalization 
of Eq.(\ref{Eq:K3formula}) presented above for recombination of $N$ identical bosons into any 
number of bound fragments is derived in \cite{mehta2009PRL}:
\begin{equation}
   K_N(E) =  N!\frac{\hbar k}{\mu} (\frac{2\pi}{k})^{d-1}\frac{\Gamma(d/2)}{2 \pi^{d/2}} \Sigma_{\mu\nu} |S_{\mu,\nu}|^2.
\label{Eq:KNformula}
\end{equation}
Here $d$ is the number of dimensions in the relative coordinate space after 
eliminating the trivial center-of-mass motion, i.e. 
for $N$ particles in 3 dimensions, $d=3N-3$.  This last formula of course reduces 
to the above expression for $K_3$ when $N=3$.



\section{The birth of few-body physics: the effects of Thomas and Efimov}
\label{Birth}

\subsection{The Thomas collapse}
\label{ThomasCollapse}

In the early days of nuclear physics, 
in 1935, a mere three years following the Chadwick discovery of the neutron, L. H. Thomas published a seminal work about the structure 
of the triton, $^3$H \cite{thomas1935pr}. In particular, Thomas studied the existence 
of the triton ground state obtained with different assumptions for the neutron-proton interaction, but 
neglecting neutron-neutron interactions as it is depicted in Fig.~\ref{Fig-Thomas}. 
As a result, \cite{thomas1935pr} found that the neutron-neutron potential energy should have a repulsive character at
short range, and that the neutron-proton interaction cannot be confined to 
a distance very small compared with 1 fm. These findings constitute the 
very first exploration of few-body physics with finite range forces, 
and they sparked the interest of many 
physicists in different fields of physics, especially atomic physics and 
molecular physics in addition to nuclear physics.

The key point of \textcolor{black}{\cite{thomas1935pr} is} 
that it is possible to account for nucleon-nucleon (or atom-atom) interactions having an arbitrary scattering length $a$ with many different two-body interaction models. For a two-body model with arbitrarily short range $r_0$ there must be a corresponding potential depth of order $\hbar^2/2 m r_0^2$ in order to yield a value of $a$ that is independent of potential range and fixed at an experimentally measured value.  For instance, in a spherical square well model having depth $V_0$ and range $r_0$, the zero-energy two-body scattering length for two equal mass particles of mass $m$ and reduced mass $m/2$ is equal to $a=r_0-\tan{q r_0}/q$, where $q=\sqrt{m V_0/\hbar^2}$.  As $r_0$ is decreased to smaller and smaller values, $q$ must increase approximately in proportion to $1/r_0$ in order to maintain any given fixed scattering length.  Thomas then examined the nature of the three-body ground state energy in this limit of decreasing potential range $r_0$ but fixed two-body scattering length.  The qualitative argument is rather simple, namely that when a third particle is brought into the system having equal scattering lengths $a$, this adds two new potential energy terms to the Hamiltonian of the same depth and range, while only adding one new kinetic energy term.  As a result, the three-body system is shown by Thomas to have a ground state energy that must be of order $-\hbar^2/m r_0^2$, which becomes arbitrarily large and negative as $r_0 \rightarrow 0$.

\subsection{Efimov physics and universality in ultracold gases}

Efimov considered an analogous three-body problem which also involved two-body scattering lengths $a$ much larger in magnitude than the potential range, i.e. $|a|/r_0 >>1$, except that Efimov visualized the two-body interaction range $r_0$ to be fixed, and $|a|\rightarrow \infty$.

Three identical particles with resonant two-body interaction will show an 
infinite series of three-body bound states as predicted by Efimov
 \cite{efimov1970plb,efimov1971SJNP,Efimov-1973} more 
than 40 years ago.  This infinity of trimer states follows a 
discrete symmetry scaling, {\it i.e.}, the energy of the $n$-th and $n+1$-th states are related 
through $E_{n+1}=\lambda^2 E_{n}$, where for the particular case of three identical 
bosons $\lambda=e^{\pi/s_{0}}$ with $s_{0}$ = 1.0062 \cite{efimov1970plb,greene2010PT,
ferlaino2011FBS,braaten2006PRep,wang2013amop}, and hence $\lambda \approx$  22.7. Efimov introduced 
the universal theory of three-body collisions thinking in nuclear systems as the preferable 
scenario for the quest of his predictions. However, the first experimental evidence of the 
prediction of V. Efimov came from ultracold gases \cite{kraemer2006NT}, and this early 
evidence has triggered an explosive growth in research into few-body  ultracold physics. 

In ultracold systems the exciting capability to tune two-body atomic scattering length, using magnetic, optical or RF-induced Fano-Feshbach resonances \cite{chin2010RMP,inouye1998,koehler2006,Tscherbul2010pra,Hanna2010njp,Hutson2016rf}. This tunability of ultracold system Hamiltonians makes them perfect candidates to study few-body universality. However, the 
formation of universal trimers must be detected and characterized in such 
systems. The most usual route to such detection is to measure the three-body loss coefficient $L_{3}$ 
as a function of the two-body scattering length $a$, as is schematically shown in Fig.\ref{3B}. 
Specifically, the universal Efimov trimers cause an enhancement of 
$L_{3}$ at a given two-body negative scattering length $a_{-}^{(n)}$, and the Efimov physics exhibits interference minima 
at values of the positive two-body scattering length $a_{+}^{(n)}$, as shown 
in Fig.\ref{3B}. \textcolor{black}{Efimov states can also be studied by radiative or oscillatory field association, as has been achieved in $^6$Li by ~\cite{Lompe-2010b,Nakajima-2011} and in $^7$Li by ~\cite{machtey2012PRLb}.}

\subsection{Faddeev equations for three identical bosons: bound states}
 \subsubsection{Hamiltonian \textcolor{black}{and} Faddeev operator equations}
 In the following three spinless and equal mass particles of bosonic character are considered which interact via short range fields.
 Note that the notation introduced below for deriving the Faddeev equations follows \cite{glockle2012quantum}.
 Then the total Hamiltonian for three $s$-wave interacting bosons obeys the following form:
\begin{equation}
 H=H_0+ \hat{V}_{23}+\hat{V}_{31}+\hat{V}_{12},
 \label{pgeq1}
\end{equation}
where $H_0$ is the three-body kinetic operator, $\hat{V}_{ij}$ indicates the short range potential between the $i-$th and $j-$th particle.
For simplicity the following notation is introduced $\hat{V}_i\equiv \hat{V}_{jk}$ with cyclic permutation of $(i,j,k)$.
The three-body Schr\"odinger equation reads:
\begin{equation}
 (H_0+\sum_i \hat{V}_i)\Psi=E\Psi
\label{pgeq2}
\end{equation}
where $\Psi$ indicates the three-body wave function.
Employing now the three-body non-interaction Green's function, i.e.$\hat{G}_0\equiv[E-H_0]^{-1}$, Eq.~(\ref{pgeq2}) can be recast to the following form:
\begin{equation}
 \Psi= \hat{G}_0\sum_i \hat{V}_i \Psi=\underbrace{\hat{G}_0 \hat{V}_1 \Psi}_{\psi^{(1)}}+\underbrace{\hat{G}_0 \hat{V}_2 \Psi}_{\psi^{(2)}}+\underbrace{\hat{G}_0 \hat{V}_3 \Psi}_{\psi^{(3)}}.
 \label{pgeq3}
\end{equation}
where this holds as long as the Green's function is free of poles.  
In our case this is valid since we focus on the description of three-body bound states, \textcolor{black}{i.e. the energy of any bound state is negative while the zeroth-order Hamiltonian has only kinetic energy and is positive definite.}
As Eq.~(\ref{pgeq3}) illustrates the total three-body wavefunction $\Psi$ can be decomposed in three components, namely $\Psi=\sum_i \psi^{(i)}$ with $i=1\ldots3$ where each $\psi^{(i)}$ indicates the $i$-th Faddeev component of the total three-body wavefunction $\Psi$.
Physically, the $i$-th Faddeev component, i.e.$\psi^{(i)}$ implies that the $i-$th particle is a spectator particle with respect to the interacting pair $(j,k)$.
Employing this decomposition ansatz in Eq.~(\ref{pgeq3}) yields a system of three coupled Faddeev equations which describe the bound state properties of the three-body system. 
\begin{equation}
 \left(\begin{matrix}
	  \psi^{(1)} \\[0.3em]
	  \psi^{(2)} \\[0.3em]
	  \psi^{(3)} \\[0.3em]
      \end{matrix}
\right)= \hat{G}_0\left(\begin{matrix}
	  0 & \hat{t}_1 & \hat{t}_1 \\[0.3em]
	  \hat{t}_2 & 0 & \hat{t}_2 \\[0.3em]
	  \hat{t}_3 & \hat{t}_3 & 0 \\[0.3em]
      \end{matrix}
\right) \left(\begin{matrix}
	  \psi^{(1)} \\[0.3em]
	  \psi^{(2)} \\[0.3em]
	  \psi^{(3)} \\[0.3em]
      \end{matrix}
\right)
\label{pgeq4}
\end{equation}
where the term $\hat{t}_i$ represents the two-body transition operator.
More specifically, $\hat{t}_i$ obeys the following Lippmann-Schwinger equation:
\begin{equation}
\hat{t}_i=\hat{V}_i+\hat{V}_i \hat{G}_0 \hat{t}_i,~{\rm for}~i=(1,2,3)
\label{pgeq5}
\end{equation}
where the term $\hat{G}_0$ denotes the Green's function of three non-interacting bosons.
This implies that the transition operator $\hat{t}_i$ is considered as a two-body operator embedded in a three-body Hilbert space.

The Faddeev equations in Eq.~(\ref{pgeq4}) can be decoupled by taking into account the exchange symmetry between the three particles.
Formally the exchange symmetry can be addressed by a permutation operator $P_{ij}$ which permutes the $i$-th with the $j-th$ particle.
In addition, the considered system consists of \textcolor{black}{three identical bosons} therefore the total wavefunction $\Psi$ is symmetric.
Due to this the exchange operator only permutes the particles in the Faddeev components.
By using the permutation operator, a pair of Faddeev components ($\psi^{(j)}$, $\psi^{(k)}$) can be expressed in terms of $\psi^{(i)}$ and vice versa.
The $\psi^{(i)}$ component of the Faddeev equations in Eq.~(\ref{pgeq4}) then takes the following form:
\begin{equation}
 \psi^{(i)}=\hat{G}_0\hat{t}_i(P_{ij}P_{jk}+P_{ik}P_{jk})\psi^{(i)},~{\rm for} ~(i,j,k=1,2,3),
 \label{pgeq6}
\end{equation}
where the indices $(i,j,k)$ form a cyclic permutation.

Eq.~(\ref{pgeq6}) represents the operator form of the Faddeev equations and in the following Eq.~(\ref{pgeq6}) is expressed in momentum space.
For completeness reasons in the following the Jacobi coordinates and the corresponding momenta are briefly reviewed.

\subsubsection{Faddeev equations in momentum representation}
Consider that the motion of three bosonic particles with masses $m_i$ with $i=1\ldots3$ are described by the lab coordinates $\boldsymbol{x}_i$ whereas their corresponding momentum is $\boldsymbol{k}_i$ with $i=1\ldots3$.
Then in order to describe the relative motion of three particles the following three sets of Jacobi coordinates are introduced:
\begin{equation}
 \boldsymbol{\rho}_i=\boldsymbol{x}_i-\frac{m_j\boldsymbol{x}_j+m_k\boldsymbol{x}_k}{m_j+m_k}~{\rm and}~\boldsymbol{r}_i=\boldsymbol{x}_j-\boldsymbol{x}_k,
 \label{pgeq7}
\end{equation}
where $(i,j,k=1,2,3)$ form a cyclic permutation and the Jacobi vector $\boldsymbol{r}_i$ denotes the relative distance between the $j$-th and $k$-th particles whereas the vectors $\boldsymbol{\rho}_i$ indicate the distance of the $i$-th particle, i.e. the {\it spectator particle}, from the center of mass of the $(j,k)$ pair of atoms.
Note that the coordinate of the center of mass of three particles obeys the simple relation $\boldsymbol{R}=\sum_{i=1}^3 m_i\boldsymbol{x}_i/M$ where $M=\sum_{i=1}^3 m_i$ denotes the total mass of the system.

Similarly, for the Jacobi momenta we obtain the following relations:
\begin{equation}
\boldsymbol{q}_i=\frac{m_k \boldsymbol{k}_j-m_j\boldsymbol{k}_k}{m_j+m_k}~{\rm and}~\boldsymbol{p}_i=\frac{(m_j+m_k)\boldsymbol{k}_i-m_i(\boldsymbol{k}_j+\boldsymbol{k}_k)}{M},
\label{pgeq8}
\end{equation}
where $(i,j,k=1,2,3)$ form a cyclic permutation, the $\boldsymbol{q}_i$ denotes the relative momentum of $(j,k)$ pair and the $\boldsymbol{p}_i$ indicates the momentum of the spectator particle relative to the center of mass of $(j,k)$ pair.
The total momentum is given by the relation $\boldsymbol{P}=\sum_{i=1}^3 \boldsymbol{k}_i$.

According to these definitions the kinetic operator $\hat{H}_0$ in the momentum space takes the following form
\begin{equation}
 H_0=\frac{\boldsymbol{P}^2}{2M}+\frac{\boldsymbol{p}_i^2}{2\bar{\mu}_i}+\frac{\boldsymbol{q}_i^2}{2\mu_i},
 \label{pgeq9}
\end{equation}
where $\mu_i=(m_jm_k)/(m_j+m_k)$ is the reduced mass of the $(j,k)$ pair particles and $\bar{\mu}_i=m_i(m_j+m_k)/M$ denotes the reduced mass of the spectator particle and the center of mass of the $(j,k)$ pair.

In the following is assumed that the collisions occur in the frame of the total center of mass, this means $\boldsymbol{P}=0$.
Therefore, the term $\boldsymbol{P}^2/(2M)$ can be removed from the total Hamiltonian which then takes the form:
\begin{eqnarray}
 H'=\frac{\boldsymbol{p}_i^2}{2\bar{\mu}_i}+\frac{\boldsymbol{q}_i^2}{2\mu_i}+V_i(\boldsymbol{\rho}_i)&+&V_j(\boldsymbol{r}_i+\frac{m_j}{m_j+m_k}\boldsymbol{\rho}_i)\cr
 &+&V_k(\boldsymbol{r}_i-\frac{m_k}{m_j+m_k}\boldsymbol{\rho}_i)
 \label{pgeq10}
\end{eqnarray}

Since the Jacobi coordinates and momenta are introduced, the reduced Faddeev equation (see Eq.~(\ref{pgeq6})) can be transformed into the momentum space.
For this purpose a certain set of Jacobi momenta is chosen, i.e.$(\boldsymbol{p}_1, \boldsymbol{q}_1)$.
This means that in this particular set of Jacobi coordinates the particle 1 is the spectator of the pair (2,3).
Upon introducing a complete set of states $\ket{\boldsymbol{q}_1\boldsymbol{p}_1}$, Eq.~(\ref{pgeq6}) becomes
\begin{eqnarray}
\braket{\boldsymbol{q}_1 \boldsymbol{p}_1|\psi^{(1)}}&=&G_0(\boldsymbol{q}_1,\boldsymbol{p}_1)\int \frac{d \boldsymbol{q}_1'}{(2\pi)^3} \frac{d\boldsymbol{p}_1'}{(2\pi)^3} \braket{\boldsymbol{q}_1 \boldsymbol{p}_1|\hat{t}_1|\boldsymbol{q}_1' \boldsymbol{p}_1'}\cr
&\times& \braket{\boldsymbol{q}_1' \boldsymbol{p}_1'|P_{12}P_{23}+P_{13}P_{23}|\psi^{(1)}},
\label{pgeq11} 
\end{eqnarray}
where the three-body Green's function in momentum space is given by the relation $G_0(\boldsymbol{q}_1,\boldsymbol{p}_1)=[E-\frac{q_1^2}{2 \mu_1} -\frac{p_1^2}{2\bar{\mu}_1}]^{-1}$.
Note that for the bound trimer spectrum the total energy $E$ is negative; thus in this case the $G_0$ Green's function is free of poles.

The matrix elements of the transition operator $\hat{t}_1$ in Eq.~(\ref{pgeq11}) can be evaluated with the help of the corresponding Lippmann-Schwinger equation Eq.~(\ref{pgeq5}):
\begin{equation}
  \braket{\boldsymbol{q}_1 \boldsymbol{p}_1|\hat{t}_1|\boldsymbol{q}_1' \boldsymbol{p}_1'}=\delta(\boldsymbol{p}_1-\boldsymbol{p}_1')\braket{\boldsymbol{q}_1|t(E-\frac{p^2_1}{2\bar{\mu}_1})|\boldsymbol{q}_1' },
  \label{pgeq12}
\end{equation}
where the term $t(E-\frac{p^2_1}{2\bar{\mu}_1})$ is the two-body transition amplitude embedded in the two-body Hilbert space.
This means that the transition amplitude obeys a two-body Lippmann-Schwinger equation of the following form:
\begin{eqnarray}
\braket{\boldsymbol{q}_1|t(\varepsilon)|\boldsymbol{q}_1'}&=& \braket{\boldsymbol{q}_1|\hat{V}_1|\boldsymbol{q}_1'} + \int \frac{d\boldsymbol{q}_1''}{(2\pi)^3}\braket{\boldsymbol{q}_1|\hat{V}_1|\boldsymbol{q}_1''} \times \cr
&\times&\left[\varepsilon-\frac{q_1''^2}{2\mu_1}\right]^{-1}\braket{\boldsymbol{q}_1''|t(\varepsilon)|\boldsymbol{q}_1'}
 \label{pgeq13}
\end{eqnarray}

In addition the exchange operators in Eq.~(\ref{pgeq11}) for equal masses namely $m_1=m_2=m_3=m$ obey the following relation:

\begin{equation}
\begin{split}
  &\braket{\boldsymbol{q}_1' \boldsymbol{p}_1'|P_{12}P_{23}+P_{13}P_{23}|\boldsymbol{q}_1'' \boldsymbol{p}_1''}= \cr
 &= \delta(\boldsymbol{q}_1'+\frac{3}{4}\boldsymbol{p}_1''+\frac{\boldsymbol{q}_1''}{2})
 \delta(\boldsymbol{p}_1'-\boldsymbol{q}_1''+\frac{\boldsymbol{p}_1''}{2}) \cr
 &+\delta(\boldsymbol{q}_1'-\frac{3}{4}\boldsymbol{p}_1''+\frac{\boldsymbol{q}_1''}{2})
 \delta(\boldsymbol{p}_1'+\boldsymbol{q}_1''+\frac{\boldsymbol{p}_1''}{2}).
\end{split}
\label{pgeq14}
\end{equation}

By substituting the Eqs.~(\ref{pgeq12},\ref{pgeq13}) and (\ref{pgeq14}) in the reduced Faddeev equation, namely Eq.~(\ref{pgeq11}) we get the following expression:
\begin{equation}
\begin{split}
& \braket{\boldsymbol{q}_1 \boldsymbol{p}_1|\psi^{(1)}}=\left(E-\frac{q_1^2}{m}-\frac{3 p_1^2}{4 m}\right)^{-1}\times \cr
 & \int \frac{ d  \boldsymbol{p}_1'}{(2\pi)^3}\bigg[\braket{\boldsymbol{q}_1|t(E-\frac{3p^2_1}{4 m})|-\boldsymbol{p}_1'-\frac{\boldsymbol{p}_1}{2}} \braket{\boldsymbol{p}_1+\frac{\boldsymbol{p}_1'}{2};\boldsymbol{p}_1'|\psi^{(1)}}\cr
 &+\braket{\boldsymbol{q}_1|t(E-\frac{3p^2_1}{4 m})|\boldsymbol{p}_1'+\frac{\boldsymbol{p}_1}{2}} \braket{-\boldsymbol{p}_1-\frac{\boldsymbol{p}_1'}{2};\boldsymbol{p}_1'|\psi^{(1)}}\bigg].
\end{split}
 \label{pgeq15}
 \end{equation}

 \subsubsection{Separable potential approximation: two-body transition elements \textcolor{black}{and} the reduced Faddeev equation}
 In the following is considered that the two-body interactions can be modeled by a separable potential, such as the Yamaguchi potential \cite{yamaguchi1954PR}.
 This particular type of potentials simplifies the Faddeev equations [see Eq~(\ref{pgeq15})] into an one-dimensional integral equation.
  Assume that the two particles interact via $s$-wave interactions only through the following {\it non-local} potential:
 \begin{equation}
  \braket{\boldsymbol{q}_1|\hat{V}_1|\boldsymbol{q}'_1}=-\frac{\lambda}{m}\chi(\boldsymbol{q}_1)\chi(\boldsymbol{q}'_1),
  \label{pgeq16}
  \end{equation}
 where $\lambda$ denotes the strength of the two-body interactions, $m$ indicates the mass of the particles and the $\chi(\cdot)$ functions are the so called {\it form~factors}.
 Typically, the $\chi$-form factors  are chosen such that the potential $V$ yields the same scattering length and effective range correction as the {\it real} two body interactions.
 Note that since we are interested in three-body bosonic collisions of neutral atoms in the following subsection we provide the form factors $\chi$ which are derived from a van der Waals potential.
 This particular choice of form factor incorporates in a transparent way the pairwise two-body interactions of the three neutral atoms.
 
After insertion of Eq.~(\ref{pgeq13}), the two-body transition matrix elements for the separable potential in Eq.~(\ref{pgeq16}) obey
 \begin{eqnarray}
\braket{\boldsymbol{q}_1|t(\varepsilon)|\boldsymbol{q}'_1}&=&-\frac{\lambda}{m} \chi(\boldsymbol{q}_1)\tau(\varepsilon)\chi(\boldsymbol{q}'_1),\cr {\rm with~}\tau^{-1}(\varepsilon)&=&1+\frac{\lambda}{m}\int \frac{d\boldsymbol{q}_1}{(2\pi)^3}\frac{|\chi(\boldsymbol{q}_1)|^2}{\varepsilon-\frac{q^2_1}{m}}.
\label{pgeq17}
 \end{eqnarray}

After specializing to states where the three particles have total angular momentum $L=0$ and using the separable potential from Eq.~(\ref{pgeq16}), as well as the two-body transition matrix elements from Eq.~(\ref{pgeq17}), the reduced Faddeev equation in Eq.~(\ref{pgeq15}) reads
\begin{equation}
 \begin{split}
 \braket{\boldsymbol{q}_1 \boldsymbol{p}_1|\psi^{(1)}}&=-2\frac{\lambda}{m}\left(E-\frac{q_1^2}{m}-\frac{3 p_1^2}{4 m}\right)^{-1}\tau(E-\frac{3 p_1^2}{4 m})\times \cr
 & \times \chi(\boldsymbol{q}_1) \int \frac{d  \boldsymbol{p}_1'}{(2\pi)^3}\chi(\boldsymbol{p}_1'+\frac{\boldsymbol{p}_1}{2})\braket{\boldsymbol{p}_1+\frac{\boldsymbol{p}_1'}{2};\boldsymbol{p}_1'|\psi^{(1)}}.
\end{split}
 \label{pgeq18}
 \end{equation}
 
 This integral equation can be further simplified by employing the following ansatz for the Faddeev component $\ket{\psi^{(1)}}$:
 \begin{equation}
 \begin{split}
   \braket{\boldsymbol{q}_1 \boldsymbol{p}_1|\psi^{(1)}}&=\left(E-\frac{q_1^2}{m}-\frac{3 p_1^2}{4 m}\right)^{-1}\chi(\boldsymbol{q}_1) \mathcal{F}(\boldsymbol{p}_1),
   \end{split}
  \label{pgeq18b}
  \end{equation}
  
  Substituting the ansatz of Eq.~(\ref{pgeq18b}) in the reduced Faddeev equation, namely Eq.~(\ref{pgeq17}) an integral equation for the amplitudes $\mathcal{F}$ is obtained where its arguments depend only on the magnitude of the $\boldsymbol{p}_1$ vector states due to the $s$-wave character of the two-body interactions.
  Under these considerations the integral equation of the amplitudes $\mathcal{F}$ reads
  \begin{equation}
  \begin{split}
    \mathcal{F}(p_1) &= -2\frac{\lambda}{m}\tau \left(E-\frac{3 p_1^2}{4 m}\right)\times \cr
    & \times \int \frac{d  \boldsymbol{p}_1'}{(2\pi)^3}\frac{\chi(|\boldsymbol{p}_1'+\frac{\boldsymbol{p}_1}{2}|)\chi(|\boldsymbol{p}_1+\frac{\boldsymbol{p}'_1}{2}|)}{E-\frac{p_1^2}{m} -\frac{p_1'^2}{m}-\frac{\boldsymbol{p}_1 \cdot \boldsymbol{p}'_1}{m}}\mathcal{F}(p_1'),
    \end{split}
   \label{pgeq19}
  \end{equation}
where for a particular choice of $\chi-$form factors the preceding equation is transformed into a matrix equation.
For a given $s$-wave scattering length and effective range parameters, numerically the energy is varied in searching for roots of the corresponding determinantal equation of Eq.~(\ref{pgeq19}).

Finally, it should be noted that replacement of the $\chi$-form factor by $\chi(\boldsymbol{q}_1)\to 1$ in the reduced Faddeev equation in Eq.~(\ref{pgeq18}) one obtains the Skorniakov\--Ter-Martirosian equation \cite{skorniakov1957JETP} for three bosons colliding with zero-range $s$-wave interactions.
The following subsection focuses on deriving a separable potential which is suitable for the two-body interactions of neutral atoms, i.e. van der Waals forces.

  \subsection{Separable potentials for van der Waals pairwise interactions}
  
  In order to study the universal aspects of the three body spectrum of bosonic 
gases it is necessary to focus on the two-body interactions which govern the
collisional behavior of ultracold gaseous matter.
  More specifically, it is well known that neutral bosonic atoms at large 
separation distances experience an attractive  van der Waals type of force 
which asymptotically vanishes as $\sim-1/r^6$.
  Note that we ignore the Casimir-Polder modification due to retardation,\textcolor{black}{\cite{CasimirPolder1948pr}} 
which modifies this at very long range but is largely irrelevant to the 
energy scale of interest here.
  This particular type of interaction potential imprints universal features 
onto the corresponding wavefunction \cite{flambaum1999pra,Gao1998a} which 
becomes manifested in the spectra of three interacting bosons.

   \begin{figure}[h]
 \includegraphics[scale=0.55]{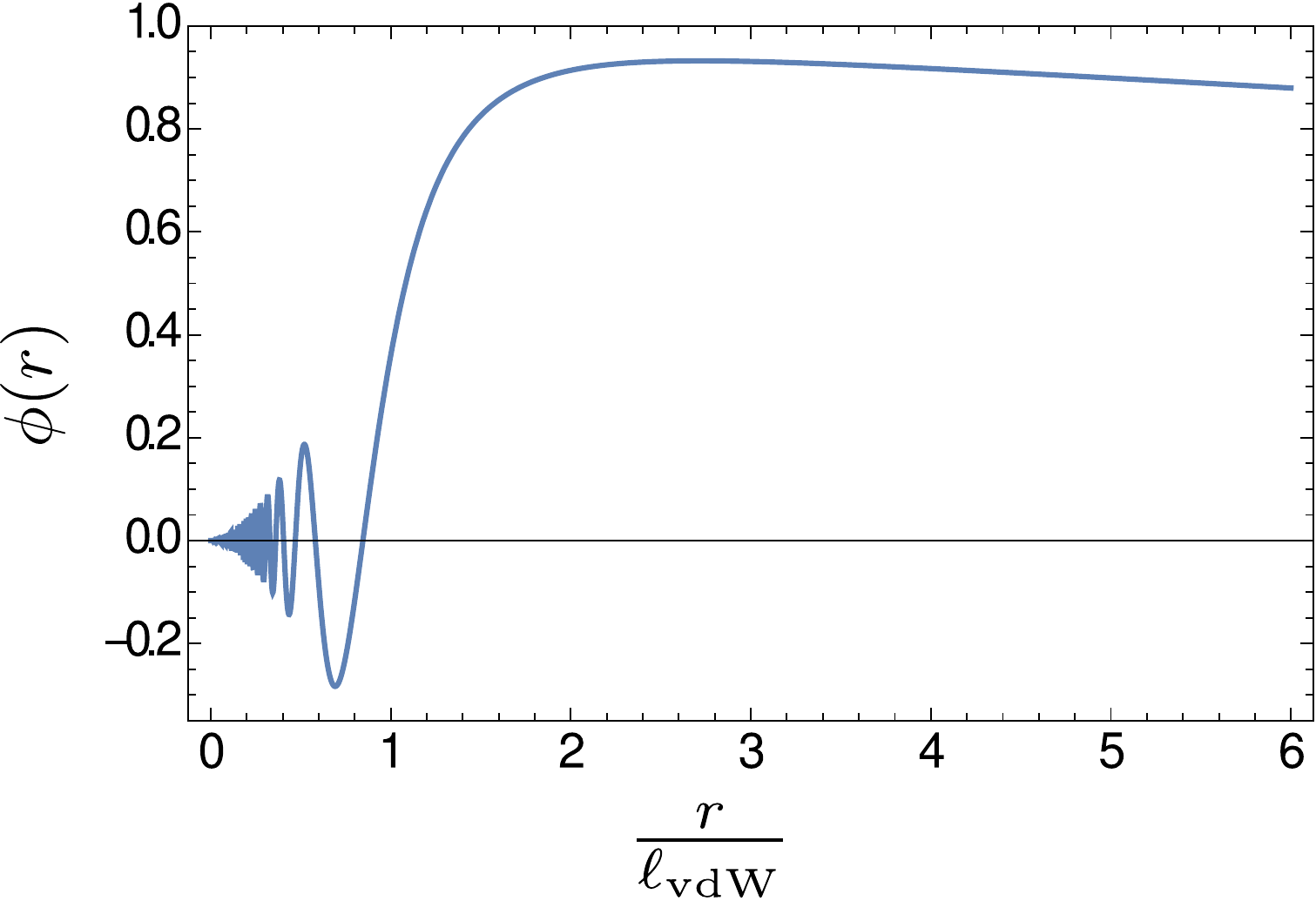}
 \caption{(Color online) The two-body zero energy wavefunction $\phi(r)$ for the van der Waals potential as a function of the scaled interparticle distance $\frac{r}{\ell_{\rm vdW}}$, for an $s$-wave scattering length $a_s=50 \ell_{\rm vdW}$. Note that $\ell_{\rm vdW}$ denotes the van der Waals length scale defined in the text.}
\label{pgfig1}
\end{figure}

  However, as was shown in the previous subsection, the Faddeev equations are best simplified by using the separable potential approach.
 Thus it is of major interest to construct a separable potential which encapsulates the main features of the van der Waals forces.
 \cite{naidon2014PRA,naidon2014prl} show that such a potential can be derived simply by using the analytically known {\it zero-energy} wave function of two particles in the presence of van der Waals potential \cite{flambaum1999pra}.
 Namely, the zero-energy two-body wavefunction for van der Waals interaction is given by the following relation:
 \begin{equation}
 \begin{split}
 \phi(r)&= \Gamma\left(\frac{5}{4}\right)\sqrt{\frac{r}{\ell_{\rm vdW}}} J_{\frac{1}{4}}\left(2 \frac{\ell_{\rm vdW}^2}{r^2}\right)\cr
 &-\frac{\ell_{\rm vdW}}{a_s}\Gamma\left(\frac{3}{4}\right)\sqrt{\frac{r}{\ell_{\rm vdW}}} J_{-\frac{1}{4}}\left(2 \frac{\ell_{\rm vdW}^2}{r^2}\right),
 \end{split}
 \label{pgeq20}
 \end{equation}
 where $a_s$ is the $s$-wave scattering length, $ \ell_{\rm vdW}=\frac{1}{2}(m C_6/\hbar^2)^{1/4}$ is the van der Waals length scale with $C_6$ being the dispersion coefficient.
 The quantities $\Gamma(\cdot)$ and $J_{\pm \frac{1}{4}}(\cdot)$ represent the Gamma and Bessel functions respectively.
 
 Fig.~\ref{pgfig1} depicts the wavefunction in Eq.~(\ref{pgeq20}) for an $s$-wave scattering length $a_s=50 \ell_{\rm vdW}$.
At short distances the two-body wavefunction oscillates fast enough which in essence reflects the fact that the van der Waals potential contains many two-body bound states.
At large distances the wave function of Eq.~(\ref{pgeq20}) obtains the form $\phi(r)\to 1-r/a_s$.
It is evident that a separable potential based on the above mentioned two-body wavefunction contains the correct behavior of the two-body wavefunction as well as effective-range effects due to the short-range oscillatory part of $\phi(r)$.
 The latter is of particular importance since \cite{naidon2014PRA,naidon2014prl} demonstrate that the universality of the three body parameter of the Efimov states relies exactly on the short-range oscillatory part of the two-body wavefunction.
The Yamaguchi potential from  Eq.~(\ref{pgeq16}) is adopted, where the $\chi$-function in the momentum space is defined by:
\begin{equation}
 \chi(q_1)=1-q_1\int_0^\infty d r \left[1-\frac{r}{a_s}-\phi(r)\right]\sin(q_1 r),
 \label{pgeq21}
\end{equation}
where $\phi(r)$ is the zero-energy two-body rescaled radial wavefunction (see Eq.~(\ref{pgeq20})).
Note that the argument of the $\chi$-form factor depends only on the magnitude of the vector $\boldsymbol{q}_1$ due to the $s$-wave character of the wavefunction.
The strength $\lambda$ of the Yamaguchi potential is determined by the following expression:
\begin{equation}
 \lambda=\left[-\frac{1}{4\pi a_s}+\frac{1}{2\pi^2}\int_0^\infty d q_1 |\chi(q_1)|^2 \right]^{-1}.
 \label{pgeq22}
 \end{equation}
  
   \begin{figure}[h]
 \includegraphics[scale=0.55]{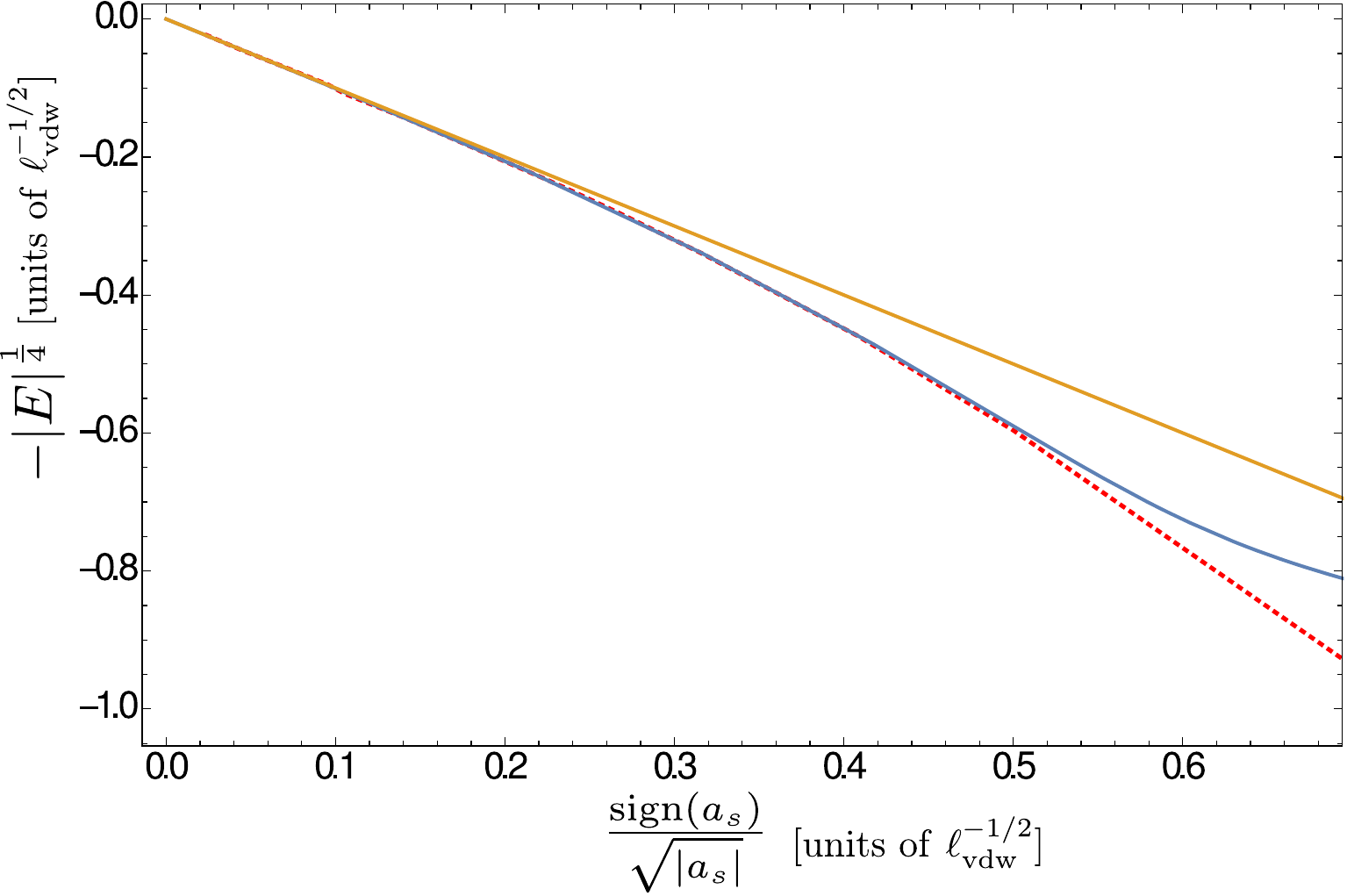}
 \caption{(Color online) The two-body binding energy as a function of the $s$-wave scattering length.
The orange curve refers to the universal dimer energy, the blue solid line indicates the the effective range theory for van der Waals interactions, and the red dotted curve denotes the binding energy within the separable potential approximation.}
\label{pgfig2}
\end{figure}

 \begin{figure}[h]
 \includegraphics[scale=0.55]{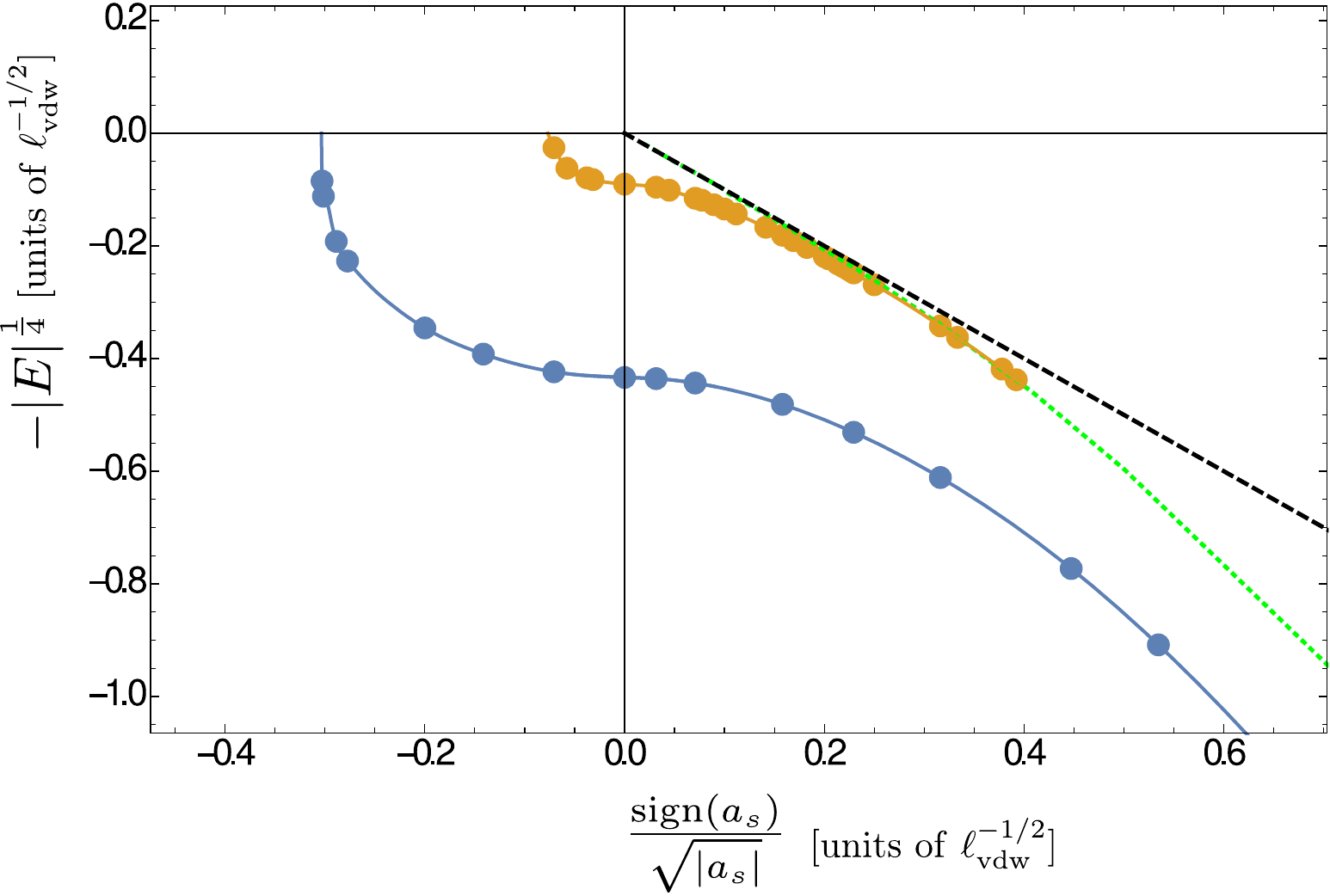}
 \caption{(Color online) The Efimov spectrum for the ground (blue line and dots) and the first excited (orange line and dots) three-body state. The black dashed line is the universal dimer energies. The green dotted curve refers to the dimer binding energies calculated within the separable van der Waals potential approach.}
\label{pgfig3}
\end{figure}

Substitution of Eqs.~(\ref{pgeq21}) and (\ref{pgeq22}) for the non-local interaction in Eq.~(\ref{pgeq16}) specifies a separable potential which mimics a van der Waals interaction between two neutral atoms.
Namely, the van der Waals separable potential in the momentum space reads
\begin{equation}
\braket{\boldsymbol{q}_1|V_1|\boldsymbol{q}_1'}=\frac{\chi(q_1)\chi(q_1')}{\frac{m}{4\pi a_s}-\frac{m}{2\pi^2}\int_0^\infty d q_1 |\chi(q_1)|^2 }.
 \label{pgeq23}
\end{equation}

As an example, Fig.~\ref{pgfig2} illustrates the binding energies versus the $s$-wave scattering length, $a_s$.
The two-body dimer energies within the separable potential approximation, see Eq.~(\ref{pgeq23}), (red dotted line) are compared with the effective range theory of van der Waals interactions given by \cite{flambaum1999pra} (blue solid line). 
The yellow solid line indicates the universal dimer energies.
Evidently, Fig.~\ref{pgfig2} depicts that the separable potential introduced in Eq.~(\ref{pgeq23}) captures the essential two-body physics beyond the effective range approximation.

\subsection{The Efimov spectrum \textcolor{black}{and} its universal aspects}
This subsection focuses on the impact of van der Waals forces on the Efimov spectrum of three identical $s$-wave-interacting bosons, the typical situation for three ultracold atoms but irrelevant for the few-nucleon problem.
In particular, the reduced Faddeev equation in Eq.~(\ref{pgeq11}) is numerically solved within the separable potential approximation.
The separable potential is constructed according to the prescription given in the previous subsection.
Specifically, use of the potential in Eq.~(\ref{pgeq23}) ensures that it contains all the relevant zero-energy information about the van der Waals potential.
Under these considerations, Fig.~\ref{pgfig3} depicts the Efimov spectrum of three neutral atoms as a function of the $s$-wave scattering length.
In particular, the blue curve and dots indicate the ground Efimov trimer state. The orange dots and curve denote the first excited state.
The black dashed curve refers to the universal dimer threshold, {\it i.e.}, $E=-\frac{\hbar^2}{m a_s^2}$, whereas the green dotted curve corresponds to the two-body binding energies given for the potential in Eq.~(\ref{pgeq23}).
Deeply in the regime of unitarity, namely $|a_s|\to \infty$, the trimer energies for the ground and first excited state are $E_1=0.035338~\ell_{\rm vdW}^{-2}$ and $E_2=6.6806\times10^{-5}~\ell_{\rm vdW}^{-2}$ respectively, or in wave vectors we have that $\kappa_1=0.1879~\ell_{\rm vdW}^{-1}$ and $\kappa_2=0.00817~\ell_{\rm vdW}^{-1}$.
For the first two $\kappa$ wave vectors a scaling factor is obtained which is equal to $\kappa_1/\kappa_2=22.9988$.
The latter deviates from the universal scaling law obtained within the zero-range approximation which is given by the relation $\kappa^{\rm ZR}_1/\kappa^{\rm ZR}_2=22.6944$.
This discrepancy between the van der Waals approach and the zero-range approximation can be attributed to the fact that the latter method completely neglects effective range corrections.
Specifically, the ground Efimov state is strongly influenced by finite range effects in the two-body interaction potentials~\cite{platterpra2015}.
Note that the value obtained for $\kappa_0$ is in reasonable agreement within 16$\%$ with the corresponding calculation in~\cite{wang2012PRL}, which was based on a local position space van der Waals interaction.
More specifically, for a hard-core van der Waals potential tail~\cite{wang2012PRL} obtains the value $\kappa_0=0.226(2)~\ell_{\rm vdW}^{-1}$ at unitarity for the ground Efimov state.

Away from unitarity and for negative values of the scattering length, Fig.~\ref{pgfig3} shows that the trimer states cross the three body threshold and become resonances in the 3 particle scattering continuum.
In particular, the ground state crosses the threshold at $a_-^{(1)}=-10.849~ \ell_{\rm vdW}$, whereas the first excited Efimov trimer merges with the three-body continuum at $a_-^{(2)}=-169.199 ~\ell_{\rm vdW}$.
Note that the $a_-^{(1)}$ for the ground Efimov state is in good agreement with the hyperspherical approach employed by Wang {\it et al.} ~\cite{wang2012PRL}. More specifically, \cite{wang2012PRL} for a hard-core van der Waals potential tail obtains the value  $a_-^{(1)}=-9.73 (3)~\ell_{\rm vdW}$for the ground Efimov state.
In addition, the Naidon {\it et al.} model result agrees well with the corresponding experimental values, {\it i.e.} $a_{-}^{(1)}=-9.1~\ell_{\rm vdW}$~\cite{ferlaino2011FBS}.
Remarkably, the separable potential model presented by~\cite{naidon2014PRA,naidon2014prl} reproduces the universal features of the Efimov spectrum without utilizing any auxiliary parameter of the type that is needed within the zero-range approximation. Recall that the three-body spectrum for the Efimov effect is not bounded from below in the zero-range approximation; thus an additional parameter (three-body parameter) is employed in order to define properly the ``ground Efimov state''. In the van der Waals separable potential model the auxiliary parameter becomes unnecessary due to the fact that the potential itself describes not only the asymptotic behavior of the two-body wavefunction but also its behavior at short distances which oscillates rapidly.
Indeed, the fast oscillations of the two-body wavefunction in regions of the three-body configuration space where two particles approach each other translates into an effective repulsive hyperradial barrier, which in return suppresses the probability to find three bosons at distances less than $R\sim 2~ \ell_{\rm vdW}$.
This suppression effect was initially understood by \cite{wang2012PRL} using the hyperspherical approach where the steep attraction of the van der Waals forces leads to an effective three-body potential barrier at this somewhat surprisingly large hyperradius.

 \begin{figure}[h]
 \includegraphics[scale=0.55]{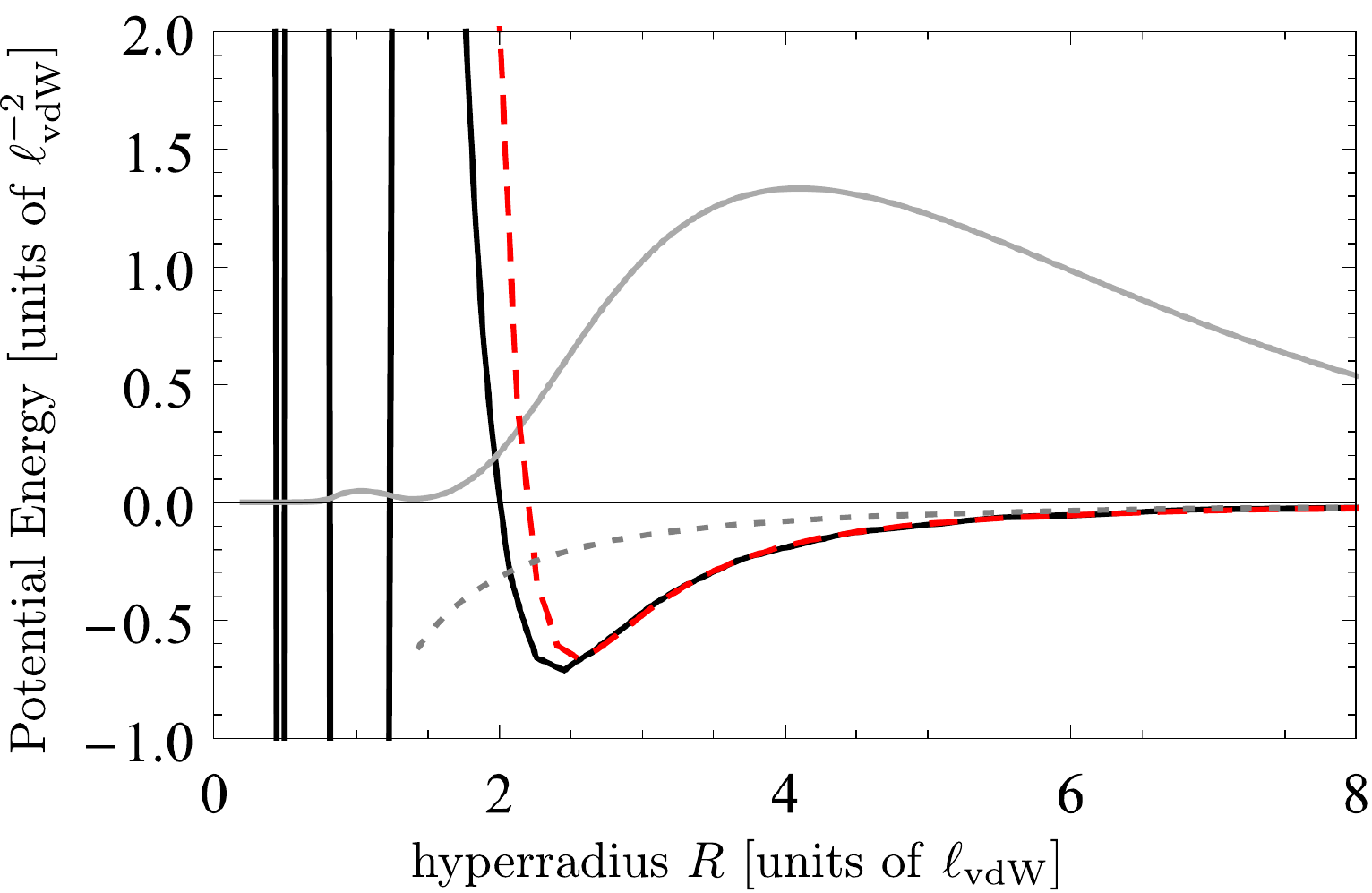}
 \caption{(Color online) The gray solid curve depicts the three-body Faddeev wavefunction in arbitrary units within the separable potential approach  as a function of the hyperradius. The red dashed line corresponds to the hypersherical potential curve including the diagonal adiabatic correction term $Q_{00}$, while the gray dotted curve indicates the asymptotic Efimov potential curve at unitarity. The black solid curve indicates the effective potential implied by the Faddeev equation solution determined within the separable potential approximation.(taken from\cite{naidon2014PRA})}
\label{pgfig4}
\end{figure}

In order to illustrate this point from the reduced Faddeev equation in Eq.~(\ref{pgeq16}) the three-body wavefunction is first obtained in the momentum representation at $a_-^{(1)}$.
%
Then following a Fourier transformation the corresponding configuration space three-body wavefunction is expressed in hyperspherical coordinates.
After integrating the density over all the hyperangles $\alpha$, taking the square root and applying the hyperradius kinetic operator to the resulting hyperradial wavefunction, an effective potential is obtained as a function of the hyperradius $R$.
This effective potential is compared with the corresponding adiabatic potential curve which contains the diagonal correction from the diagonal nonadiabatic coupling term $Q_{00}$. Fig.~\ref{pgfig4} compares the resulting implied hyperradial potential curve from \cite{naidon2014PRA} with the direct adiabatic hyperspherical solution from \cite{wang2012PRL}, showing good general agreement.
The gray dotted curve illustrates the asymptotic $-R^{-2}$ Efimov potential curve at unitarity for comparison. 
While the effective potential curve possesses some additional structure,
\cite{naidon2014PRA} states that this structure is an artifact which mainly arises from the oscillatory behavior of the Faddeev three-body wavefunction.


\subsection{Efimov states in homonuclear systems}

\subsubsection{$^6$Li}

Ultracold gases of fermionic $^6$Li have been the object of different studies about Efimov physics and
universality in three-body physics, e.g. by ~\cite{ottenstein2008PRL,wenz2009PRA,lompe2010PRL,Huckans-2009,
Williams-2009, nakajima2011PRL,nakajima2010PRL}. It should be pointed out that it is a bit of a stretch to include the $^6$Li system in our discussion of the Efimov effect for three identical bosonic atoms.  Owing to the fermionic nature of $^6$Li there is no $s$-wave scattering between atoms in identical spin substates, but atoms in different substates do have an $s$-wave scattering length.  The studies just quoted have in fact considered atoms in three distinguishable substates, but unlike the case of three identical bosons, the three interparticle scattering lengths are in general different in the $^6$Li system.  However, they are all large and negative and therefore the system can be approximately mapped onto and compared with an Efimov system with three identical bosons in identical spin substates. In the following discussion, it should be kept in mind that this mapping is an approximation. It is argued \textcolor{black}{by \cite{wenz2009PRA}} that one conjectured mapping, a definition of an effective ``homonuclear'' scattering length $ a_{\rm ave}$ that applies when all three interspecies scattering lengths are large and negative is:
\begin{equation}
a_{\rm ave}^4 \equiv \frac{1}{3} (a_{12}^2 a_{23}^2+ a_{13}^2 a_{23}^2+ a_{12}^2 a_{13}^2).
\end{equation}
 Nevertheless, quantities in Efimov physics such as the loss to deeply bound dimers and the three body parameter should more rigorously be understood to depend in general on all three separate scattering lengths for a fermionic atom such as $^6$Li, i.e. on $a_{12},a_{23},a_{13}$.   In general, many of the experimental investigations have relied upon radio-frequency (RF) techniques 
for the identification of Efimov trimers. These methods employ RF pulses 
to form different Efimov states, which are detected as atom loss, thus leading
 to the characterization of their binding energies~\cite{lompe2010PRL,nakajima2010PRL,
 nakajima2011PRL,wenz2009PRA}. The measured trimer energies show a clear dependence 
 on the applied magnetic field close to the two-body Feshbach resonances, which has 
 been viewed as evidence for deviations from Efimov's universal three-body physics scenario 
. In particular, the geometric scaling factor $\lambda=$22.7 is not observed between successive resonances, and 
 this has been interpreted as a magnetic field dependence of the three-body parameter. 
 
The apparent non-universality of $^6$Li has been an open question in the last decade, leading 
two different non-universal models beyond non-universal two-body interactions~\cite{nakajima2010PRL}. 
However, \cite{Huang-2014a} have shown that accounting for a realistic two-body 
energy dependent scattering length and taking into account finite temperature effects the three-body 
parameter for $^6$Li turns out to be $a_-^{(1)}/\ell_{\rm{vdW}}=-7.11 \pm 0.6$ 
which is very similar to the 
results obtained for identical bosons Table~\ref{Table-7Li} and Table~\ref{K}. Moreover, the geometric scaling factor shows 
a 10$\%$ deviation with respect to $\lambda$ = 22.7; the universal expected value.

\subsubsection{$^7$Li}

The Efimov physics in bosonic $^7$Li has been extensively studied through 
characterizations of maxima and minima of the three-body loss
coefficient~\cite{gross2009PRL,gross2010PRL,gross2011CRP,Pollack-2009,dyke2013PRA,machtey2012PRL}, as 
well as using radio-frequency fields to measure the binding energies of weakly bound trimers~\cite{machtey2012PRLb}. In 
particular, the Rice group identified the ground Efimov state for $^{7}$Li in 
the $|m_{F} = 1\rangle$ hyperfine state as a resonance in the three-body loss coefficient for 
$a<0$.  An initial suggestion in~\cite{Pollack-2009} that they had also observed the first excited Efimov resonance $a_{-}^{(2)}$ was later attributed to a calibration error.  The recalibration, published in~\cite{dyke2013PRA}, also corrected the position of the first Efimov resonance to $a_{-}^{(1)}$= -252 $\pm 10$. 
Efimov physics was also observed on the $a>0$ branch of the spectrum as 
the expected minima in the three-body loss coefficient~\cite{Pollack-2009}, 
yielding $a_{+}^{(1)}$= 89$\pm$ 4 and $a_{+}^{(2)}$= 1420 $\pm$ 100 when the recalibration of~\cite{dyke2013PRA} was applied. The ratio $a_{+}^{(2)}/ a_{+}^{(1)}=16\pm2$ deviates appreciably from the expected universal ratio of 22.7,\cite{ nielsen1999PRLb, esry1999PRL} but this level of deviation for the first two Efimov features is not unexpected, based on theoretical calculations.

\begin{table}[h]
\centering
\begin{tabular}{c c c c}
\hline
\hline
$m_{F}$ & $a_{+}^{(1)}$ ($a_{0}$)  & $-a_{-}^{(1)}$ ($a_{0}$) & $|a_{-}^{(1)}|$/$\ell_{\rm{vdW}}$  \\
\hline
0 & 243 $\pm$ 35 & 264 $\pm$ 11 & 8.52 $\pm$ 0.35 \\
+1 & 247 $\pm$ 12 & 268 $\pm$ 12 & 8.65 $\pm$ 0.39 \\
\hline
\hline
\end{tabular}
\caption{Fitting parameters to an universal theory obtained by measuring the three-body 
loss coefficient in $^7$Li. Results taken from~\cite{gross2010PRL}.}
\label{Table-7Li}
\end{table}

Similar results for the maxima of the three-body loss rate were obtained by 
Gross {\it et al.}~\cite{gross2009PRL,gross2010PRL,gross2011CRP} for two different hyperfine
 states: $|m_{F} = 1\rangle$ and $|m_{F} = 0\rangle$ as shown in Table\ref{Table-7Li}. 
 However different results for a$_{+}^{(1)}$ in comparison with \cite{Pollack-2009} 
 were obtained as displayed in Table \ref{Table-7Li}. This discrepancy for a$_{+}^{(1)}$ has 
 been explained as a distinct magnetic field-scattering length conversion through a different 
 characterization of the same Feshbach resonance \cite{gross2010PRL}. In Table \ref{Table-7Li} 
 it is also observed the universal character of the three-body parameter $a_{-}^{(1)}$ in terms of 
 the van der Waals length $\ell_{\rm{vdW}}$ for $^7$Li-$^7$Li. In particular, the values obtained 
 for $a_{-}^{(1)}/\ell_{\rm{vdW}}$ are very similar to the values observed in 
 cesium~\cite{berninger2011PRL}, rubidium\cite{wild2012PRL} and potassium \cite{roy2013prl}.

\subsubsection{$^{39}$K}

The study of Efimov states in bosonic $^{39}$K at ultracold temperatures has been 
developed mainly by the LENS group~\cite{zaccanti2009NTP,roy2013prl}. In particular, 
the study of \textcolor{black}{~\cite{roy2013prl}} is a remarkable exploration of the $a_{-}^{(1)}$ three-body parameter 
universality, even including narrow Feshbach 
resonances. This study was carried out by employing different spin states $m_{F}$, as well 
as different Feshbach resonances in an ultracold gas of $^{39}$K, some showing 
open-channel dominance while others are narrower closed-channel-dominated 
resonances.

\begin{table}[h]
\centering
\begin{tabular}{c c c c c c}
\hline
\hline
$m_{F}$ & $R^{*}$($a_{0}$) & $s_{res}$ & $-a_{-}^{(1)}$ ($a_{0}$) & $|a_{-}^{(1)}|$/$\ell_{\rm{vdW}}$ & T (nK) \\
\hline
0 & 22 & 2.8 & 640$\pm$100 & 10.0$\pm$ 1.6 & 50$\pm$5 \\
0 & 456 & 0.14 & 950$\pm$250 & 14.7$\pm$ 3.9  & 330$\pm$30 \\
0 & 556 & 0.11 & 950$\pm$150 & 14.7$\pm$ 2.3 & 400$\pm$80 \\
+1 & 22 & 2.8 & 690$\pm$40 & 10.7$\pm$ 0.6 & 90$\pm$6 \\
-1 & 23 & 2.6 & 830$\pm$140 & 12.9$\pm$2.2 &120$\pm$10 \\
-1 & 24 & 2.5 & 640$\pm$90 & 10.0$\pm$1.4  &20$\pm$7 \\
-1 & 59 & 1.1 & 730$\pm$120 &11.3$\pm$1.9  &40$\pm$5 \\
\hline
\hline
\end{tabular}
\caption{Experimental determined three-body parameter $a_{-}$ for different Feshbach 
resonances and spin states $m_{F}$ in $^{39}$ K taken from \cite{roy2013prl}. 
R$^{*}$ represents the intrinsic length scale and associated with it, the resonance 
strength $s_{res}$. The value for the three-body parameter as a function of the van 
der Waals length $\ell_{\rm{vdW}}$ = 64.49 $a_{0}$ is also reported, as well as the initial 
temperature $T$, which implies a saturation limit of the three-body recombination rate
because the $S$-matrix is unitary.}
\label{K}
\end{table}

Resonances with a small resonance strength $s_{res}$, \cite{chin2010RMP} {i.e.}, narrow 
resonances, have an intrinsic length scale $R^{*}=\hbar^2/(ma_{bg}\delta \mu)$~\cite{chin2010RMP}, where 
$a_{bg}$ represents the background scattering length, $m$ is the reduced mass and 
$\delta \mu$ is the change in the magnetic moment between the initial and 
final states. In such a scenario was predicted that the Efimov physics would be dominated 
by the intrinsic length associated with the resonance $R^{*}$, in particular, 
$a_{-}^{(1)}=-12.90R^*$ 
\cite{Petrov-2004,gogolin2008PRL,MoraGogolinEgger2011}. However, the experimental work
 of \textcolor{black}{~\cite{roy2013prl}} revealed a completely different behavior, as shown in Table~\ref{K}, where 
 the three-body parameter $|a_{-}^{(1)}|/\ell_{\rm{vdW}}\sim 10$, which 
 turns out to be very similar to the \textcolor{black}{experimental and theoretical} values for the case of 
 broad 2-body resonances~\cite{berninger2011PRL,Wang-2012,naidon2014prl}, {\it i.e.}, 
 $|a_{-}^{(1)}|/\ell_{\rm{vdW}}=9.5$. This striking result implies that the intrinsic length scale associated 
 with a narrow resonance apparently plays no role in the determination of the three-body 
 parameter. Thus, for systems with long-range dominant van der Waals interactions, the three-body parameter seems to be universal. 

\subsubsection{$^{85}$Rb}

The study of Tan's contact in an ultracold gas of $^{85}$Rb has been 
realized by~\cite{wild2012PRL}. In particular the two-body and 
three-body contact were determined, as well as the three-body recombination
 rate constant, by varying the two-body scattering length in a sweep of the 
 magnetic field through a Feshbach resonance. The two-body contact is an extensive 
thermodynamic magnitude proportional to the derivative of the internal 
energy of the ultracold gas with respect to the scattering 
length~\cite{tan2008AP,tan2008APb,tan2008APc,WernerTarruellCastin2009epjb,Combescot-2009,Schakel-2010}, {\it i.e}, $C_{2}\propto dE/da$.
The three-body contact $C_{3}$ is defined in terms of the derivative of the internal 
energy with respect to the three-body parameter $C_{3}\propto dE/da_{-}$ ~\cite{Braaten-2011,Castin-2011}. \footnote{Usually C$_3$ is defined in terms of 
the so-called three-body interaction parameter $k_{*}$~\cite{Braaten-2011,Castin-2011}, 
which is related to the three-body parameter by the equation: $a_{-}^{(1)}=(-1.56 \pm 5)/k_{*}$.~\cite{braaten2006PRep}.}

The measurements of the three-body recombination rate were 
performed in dilute, ultracold, non-condensed clouds containing 1.5 $\times$ 10$^5$ 
atoms of $^{85}$Rb at a temperature $T$~=80~nK. Then
the magnetic field was varied through a Feshbach resonance in order 
to explore the region of negative scattering lengths. The obtained three-body 
recombination rate was fitted to the expected form for the Efimov three-body 
rate~\cite{braaten2006PRep}, obtaining $a_{-}^{(1)}$= -759 $\pm$ 6 $a_{0}$. The 
utilized fitting function is only valid at $T$ = 0, and hence the fitting was 
realized for $a<1/k_{\rm thermal}$, where $k_{\rm thermal}=\sqrt{2mk_{B}T}/\hbar$. 
The ratio between the measured three-body parameter and the van der Waals length 
is $a_{-}^{(1)}/\ell_{\rm{vdW}}$ = -9.24 $\pm$ 0.7~\cite{wild2012PRL}. This value is very similar 
to the reported values for $^{133}$Cs~\cite{kraemer2006NT,berninger2011PRL} 
and $^{7}$Li~\cite{gross2009PRL,gross2010PRL,gross2011CRP}.

\subsubsection{$^{133}$Cs}

The first experimental evidence of the Efimov effect was observed in 
an ultracold gas of $^{133}$Cs~\cite{kraemer2006NT} by tuning the 
Cs-Cs scattering length of through a Feshbach resonance, and 
measuring the enhancement and decreases of the three-body loss coefficient 
for negative and positive scattering lengths, respectively. At the same time, this 
pioneering work readily showed the possibility of using ultracold physics in order 
to explore universal physics in few-body physics.\cite{esry2006NT}

\begin{table}[h]
\centering
\begin{tabular}{c c c}
\hline
\hline
B$_{res}$(G) & $|a_{-}^{(1)}| /\ell_{\rm{vdW}}$ & $\eta_{-}$ \\
\hline
 7.56$\pm$0.17 & 8.63 $\pm$ 0.22 & 0.10$\pm$0.03 \\
553.30$\pm$0.4 & 10.19 $\pm$ 0.57  & 0.12$\pm$0.01 \\
554.71$\pm$0.80 & 9.48 $\pm$ 0.79 & 0.19$\pm$0.02 \\
853.07$\pm$0.56 & 9.45 $\pm$ 0.28 & 0.08$\pm$0.01 \\
\hline
\hline
\end{tabular}
\caption{Experimentally determined three-body parameter $a_{-}$ for different Feshbach 
taken from ~\cite{berninger2011PRL}. The position of the 
Feshbach resonances employed are denoted by $B_{res}$, the three-body 
parameter as a function of the van der Waals length ($\ell_{\rm{vdW}}$ = 101 $a_{0}$) is reported, 
and finally $\eta_{-}$ is a nonuniversal quantity that reflects decay into deeply bound 
diatomic states \cite{wenz2009PRA}.} 
\label{Cs}
\end{table}

A few years after the observation of Efimov states in ultracold systems, Berninger {\it et al.}~\cite{berninger2011PRL} employed four different Feshbach resonances to study variations
of the three-body parameter in an ultracold sample of $^{133}$Cs. 
For these four observed Efimov resonances shown in Table~\ref{Cs}, the ratio of the 
three-body parameter 
$a_{-}^{(1)}$ to the van der Waals length $\ell_{\rm{vdW}}$ is approximately equal for all the Feshbach 
resonances analyzed in the experiment, to within only about $\%15$ variations. More recently, 
Huang {\it et al.}~\cite{Huang-2014a} have realized an exhaustive experimental work on the 
negative scattering length branch of the two-body interaction in Cs, confirming the universality 
of the Efimov scaling by seeing for the first time two successive Efimov resonances in 
a homonuclear system. Note, however, that more than one previous experiment has observed the 
expected Efimov scaling between two successive destructive interference 
St\"uckelberg minima $a_+^{(n)}$; it should be remembered that
this behavior of three-body recombination at positive scattering lengths is a 
{\it non-resonant} manifestation of universal Efimov physics.

\subsection{\textcolor{black}{Four-body and five-body} bound states and recombination resonances}
\label{fourANDfive}
\textcolor{black}{Normally one views 3-body recombination 
as a comparatively rare process in
a dilute, ultracold gas. Typical Bose-Einstein condensates, for instance, can have
lifetimes of the order of many seconds.  Thus it may come as a surprise that higher
order processes involving even more than 3 atoms simultaneously colliding in 3D can have
even higher inelastic collision rates in some regimes of scattering length and 
density.  There are theoretical predictions of this resonant $N$-body recombination in ~\cite{wang2009PRL, stecher2009NTP,mehta2009PRL, Rittenhouse-2011JPB, Blume2012prl,wang2012PRLc,BlumeYan2014prl,YanBlume2015pra}, and a few experimental observations ~\cite{Ferlaino-2009,Pollack-2009,dyke2013PRA,zenesini2013NJP,Ulmanis2016prl}. While these usually cause difficulty for applications of interest with quantum degenerate gases or optical lattices, they can be especially interesting and informative to study in their own right, especially from a few-body point of view.}

\textcolor{black}{The 4-body problem has challenged theorists for many years~\cite{lazauskas2006PRA} and is still of fundamental importance and interest.
Extensive attention has been devoted to the question of whether there is an Efimov effect for four or more particles.  For four or more identical particles, an early theoretical study by ~\cite{Amado1973prd} concluded: ``Hence the remarkable Efimov effect seems even more remarkably to be a property of the three-body system only.''
Later, however, a treatment by ~\cite{KrogerPerne1980prc} based on a separable potential model concluded that in certain parameter ranges there is an Efimov effect for four bosons.  To add to this apparent discrepancy between the preceding two references mentioned, the possible existence of an Efimov effect in a 3D four-body system with three heavy particles and one light particle was treated theoretically by ~\cite{AdhikariFonseca1981prd} and later by ~\cite{NausTjon1987fbs}, reaching opposite conclusions (no and yes, respectively).
A subsequent study by ~\cite{Adhikari1995prl} suggests that an additional short-range length (or high-momentum) scale is required for each successively larger number $N$ of particles, in order to pin down the energy even of low-lying states.
Another treatment by ~\cite{Yamashita2006epl} concluded that four-body bound states can exist in the universal regime of large atom-atom scattering lengths, but they will normally not be fixed in energy by 2-body and 3-body physics alone, and will require an independent 4-body parameter.  This conclusion was supported by a later study as well, namely ~\cite{HadizadehTomio2011prl}.
}

\textcolor{black}{Based on 4-identical boson bound state calculations using low-energy effective field theory,} it was conjectured by \cite{platter2004PRA,hammer2007EPJAb} that there should be two 4-boson bound states at unitarity lying at energies between each successive pair of Efimov trimer energies.  These studies suggested, in apparent disagreement with ~\cite{Yamashita2006epl}, that the energies are largely fixed by the three-body parameter, and at least to a good approximation, this would mean that no additional 4-body parameter is needed.  A 4-body hyperspherical calculation \textcolor{black}{was carried out ~\cite{stecher2009NTP} that was based on the use of correlated Gaussian basis functions\cite{suzuki1998} adapted to the adiabatic hyperspherical representation\cite{stecher2009PRA,RakshitBlume2012pra, Mitroy2013rmp, Daily-2014}. Using that method,} \cite{stecher2009NTP, dincao2009PRLb, wang2009PRL, mehta2009PRL} gave supporting evidence to that conjecture, and advanced the theory to the point where detailed predictions could be made of 4-body recombination rate coefficients and resonance positions. In the universal limit, for instance, theory predicted \cite{stecher2009NTP} that the two-body scattering lengths where 4-boson resonances would be observable as zero energy recombination resonances, should be at the following values of the boson-boson scattering length: $a_{4B,1}^- \approx 0.43 _-^{(1)}$ and at $a_{4B,2}^- \approx 0.9 a_-^{(1)}$.  These have since been confirmed in experimental studies \cite{ferlaino2009PRL} of homonuclear recombination processes involving four or more free bosonic atoms, although \cite{stecher2009NTP} \textcolor{black}{pointed out that there was already some evidence for a four body process in \cite{kraemer2006NT}. Exciting theoretical progress in developing a highly quantitative theoretical treatment was subsequently reported for four-body resonances and recombination by \cite{deltuva2012PRA, deltuva2010PRAb, deltuva2011PRA}, in a momentum-space treatment based on a separable two-body interaction, a treatment that does not utilize hyperspherical coordinates. One interesting aspect of those theoretical and experimental efforts is the suggested implication that no additional 4-body parameter is needed to fix the universal behavior of four interacting identical bosons, as it appears to be fixed once the 3-body parameter is known. The extent to which this remains true for interactions of much shorter range than van der Waals potentials remains an active topic of investigation.}

These developments in turn spawn a fundamental question:  Are the universal properties also fixed for 5, 6, 7, and even more bosonic particles once the three-body parameter is known?  If the answer is {\it yes}, this is a crucial point that can greatly simplify the development of realistic many-body theories for interacting bosons.  A number of studies \textcolor{black}{ \cite{blume2000JCP,stecher2010JPB,yamashita2010pra, stecher2011PRL,gattobigio2012PRA} } bear directly on this question.  In particular, \cite{stecher2011PRL} predicts that a universal resonance of 5 identical bosons should occur at zero energy when the two-body scattering length is equal to $a_{5B,1}^- \approx (0.65 \pm 0.01) a_{4B,1}^-$. That prediction was tested and confirmed experimentally by the Innsbruck group\cite{zenesini2013NJP}; this study also compared a detailed theoretical and experimental estimate of the direct 5-body recombination rate, apparently the first time a direct (i.e. non-stepwise) recombination process could be observed experimentally and computed theoretically.  While this is suggestive of a general universality for all N-boson systems, very recent work by \cite{YanBlume2015pra} suggests that this may apply only to systems whose long-range two-body interaction is dominated by van der Waals interactions, as shorter range interacting systems apparently exhibit extensive variability in their N-boson binding energies at unitarity.\textcolor{black}{\cite{yamashita2010pra}}

\begin{figure}[h!]
  \includegraphics[scale=0.16]{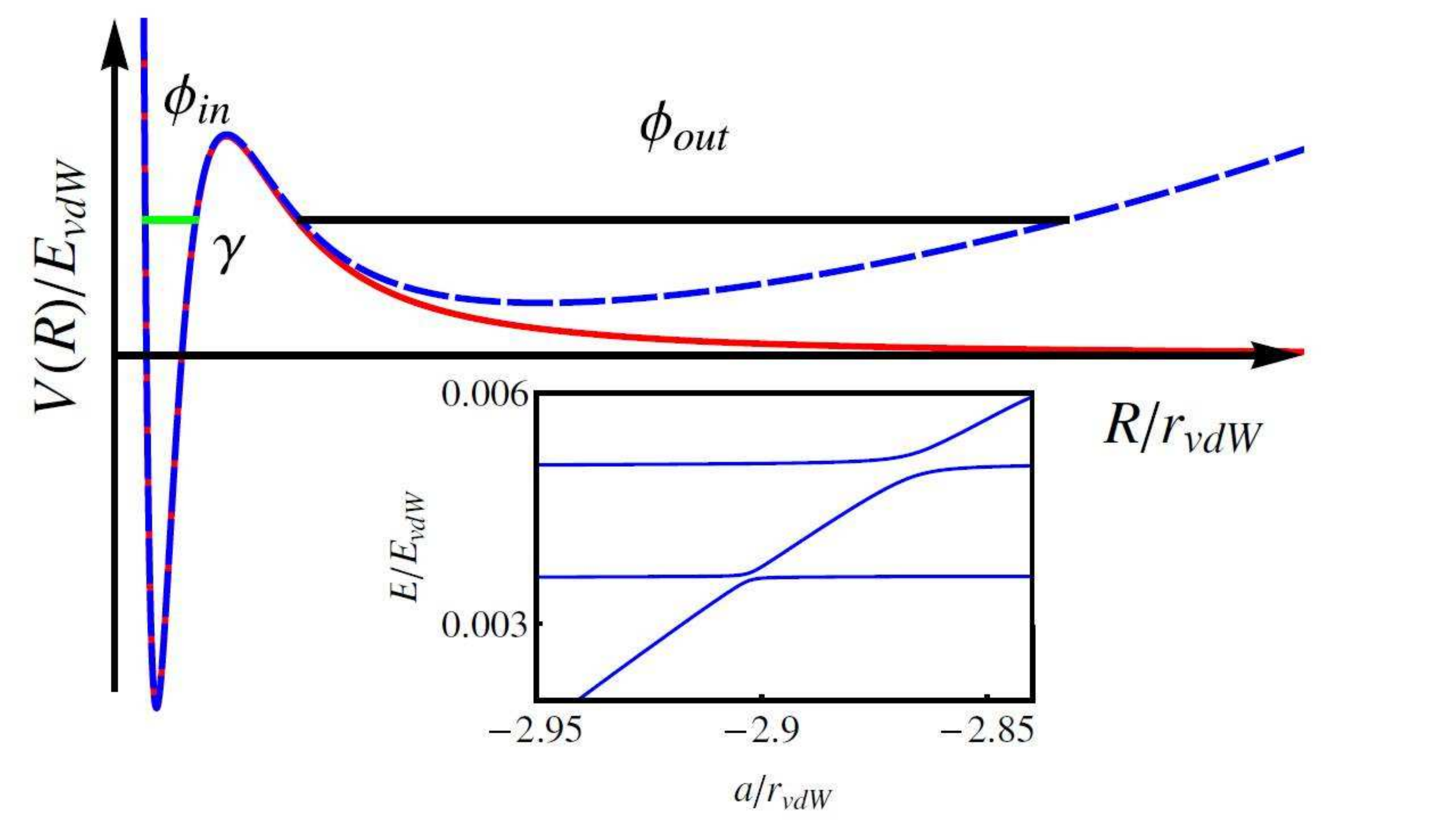}
  \includegraphics[scale=0.16]{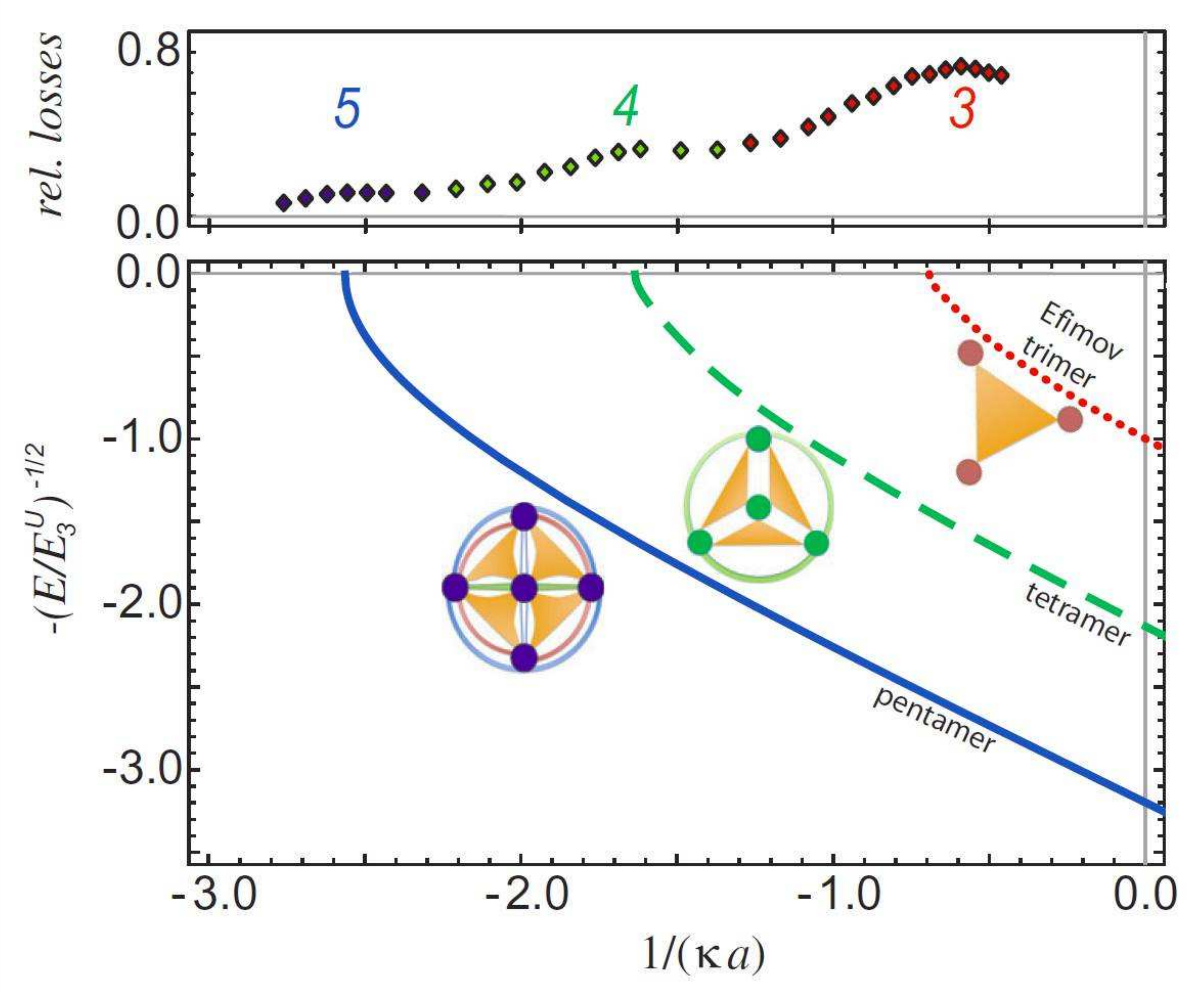}
  \includegraphics[scale=0.16]{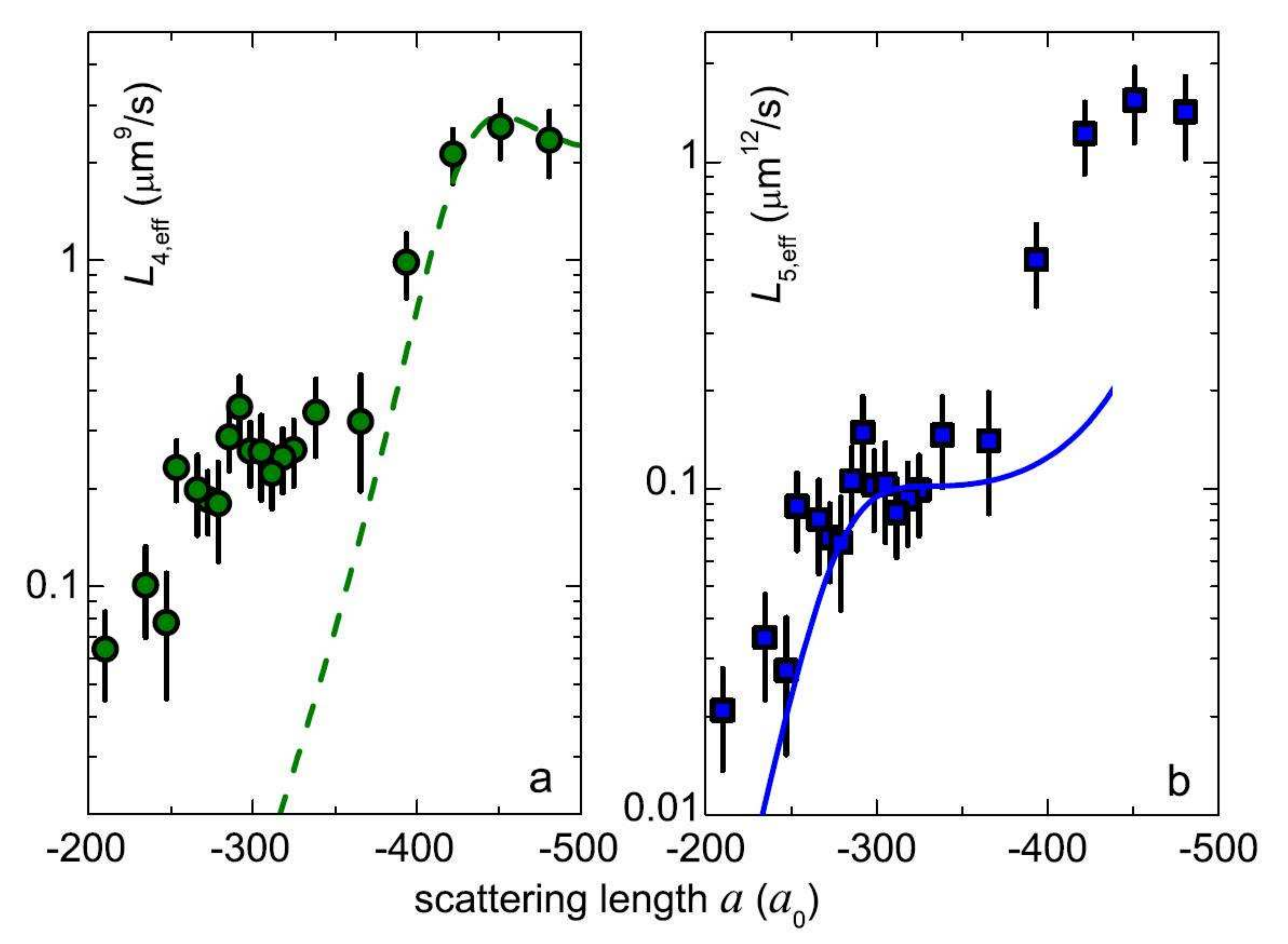}
\caption{(Color online) \textcolor{black}{
{(a)Schematic hyperspherical potential curve relevant to five-boson
recombination when the two-body scattering length is negative, $a<0$.
The curve has also labeled the WKB phases $\phi$ in the two classically allowed
regions or $R$ as well as the tunneling exponent that characterizes
the region of negative incident kinetic energy. The inset shows energy levels 
for this five-boson system in a spherically symmetric harmonic trap, which has
avoided crossings between inner and outer region states whose strengths enable
an estimate of the tunneling exponent $\gamma$ that is important in obtaining
the 5-body recombination rate.
(b) The upper panel shows the experimental atom loss in a Cs
gas is shown as a function of inverse scattering length, rescaled by a 
characteristic wavenumber of the order of $2/\ell_{\rm vdW}$. 
The lower panel shows the energies of trimer, tetramer, and pentamer states and
the points where they merge into the zero-energy continuum. 
(c) Measured loss rate coefficients,
compared with theory that includes either 4-body recombination only 
$L_{4,{\rm eff}}$ (left) or else 5-body recombination only $L_{5,{\rm eff}}$ (right). {\bf Taken from \cite{zenesini2013NJP}}.}}
}
\label{fivebody}
\end{figure}

Following the prediction in~\cite{stecher2010JPB,stecher2011PRL}, the possible existence of a 
universal 5-body recombination resonance was tested and confirmed in \cite{zenesini2013NJP}.  
Fig.\ref{fivebody} shows the comparison between theory and experiment, in a region
that includes both a universal 4-body resonance and a universal 5-body resonance.  
These predictions of universal resonances observable in $N$-body recombination have
been extended in some impressive recent calculations to even larger numbers of 
identical bosons by~\cite{gattobigio2011PRA,gattobigio2012PRA}.

\subsection{Efimov states in heteronuclear mass-imbalanced systems}

The existence of an infinite series of three-body bound states for resonant two-body interaction, 
as predicted by Efimov~\cite{efimov1970plb}, is not only present for homonuclear systems, as 
such universal three-body bound states should appear as well for heteronuclear systems \textcolor{black}{ ~\cite{efimov1973npa, efimov1979SJNP, dincao2006PRA,dincao2006PRAb,helfrich2010PRA,Wang-2012b,Wang-2015erratum,Mikkelsen2015jpb,Petrov2015}}.
In particular, in heteronuclear systems the mass-imbalanced nature of the 
three-body system preserves but modifies in an interesting way the discrete 
symmetry scaling characteristic of Efimov states, {\it i.e.}, 
$a_{-}^{(n)}=\lambda a_{-}^{(n-1)}$, and hence preserving the universality of the three-body bound states. 
Most importantly, it influences the scaling factor $\lambda$, which depends on 
the masses of the three particles involved. In particular, $\lambda$ gets smaller as 
the mass imbalance of the HHL system increases, which has sparked the study of highly mass-imbalanced systems 
as the best possible scenario for studying multiple excited Efimov states, and hence exploring as
deeply and unambiguously as possible the universal characteristics of such states. To date, heteronuclear Efimov states have been 
searched for in 
$^{41}$K-$^{87}$Rb-$^{87}$Rb~\cite{barontini2009PRL,Arlt2016}, $^{39}$K-$^{87}$Rb-$^{87}$Rb~\cite{Arlt2016},
$^{40}$K-$^{87}$Rb-$^{87}$Rb~\cite{bloom2013PRL}, $^{7}$Li-$^{87}$Rb-$^{87}$Rb~\cite{Maier-2015} and 
$^{6}$Li-$^{133}$Cs-$^{133}$Cs~\cite{Ulmanis-2015,Ulmanis-2015b,Ulmanis2016prl,Tung-2014,JohansenChin2016arxiv}.

The study of three-body losses in an ultracold mixture $^{41}$K-$^{87}$Rb performed by 
the LENS group led to the first claimed observation of heteronuclear Efimov states, 
specifically for $^{41}$K-$^{87}$Rb-$^{87}$Rb and $^{41}$K-$^{41}$K-$^{87}$Rb~\cite{barontini2009PRL}. In particular, a three-body 
parameter $a_{-}^{(1)}$ = -246 $\pm$ 14 $a_{0}$ was claimed to be observed for K-Rb-Rb.  However, this claimed observation of 
an Efimov resonance has been questioned in the literature, in part because it is so far from the
expected theoretical range for this system. Owing to the positive value of the Rb-Rb scattering length $a\sim 100$ $a_0$, 
the first Efimov resonance is expected to occur in K-Rb-Rb at around $a_-^{(1)}$(K-Rb) $\le$ -30,000 $a_0$.  In a very recent follow-up by the Aarhus experimental group, they find that there is a two-body $p$-wave feature in the vicinity of the LENS group's claimed Efimov resonance $^{41}$K-$^{87}$Rb-$^{87}$Rb, which adds to doubts about the classification of that loss feature which doesn't fit universal expectations for the 3-body system. 
\cite{Wang-2012b,Ulmanis2016prl} Similarly, the K-K-Rb system is ``Efimov-unfavored (LLH)'' since it has two lighter and one heavier atom, and its first Efimov resonance has been predicted to occur only for $a_-^{(1)}(K-Rb) \le -10^6$ a.u.  Other experiments on potassium mixtures with $^{87}$Rb with either the fermionic isotope $^{40}$K~\cite{bloom2013PRL} or the bosonic isotope $^{39}$K~\cite{Arlt2016} have failed to observe Efimov resonances at a reasonable value of the K$-$Rb scattering length, results which are more consistent with theoretical expectations. In particular, the JILA group~\cite{bloom2013PRL} studied an 
ultracold Bose-Fermi mixture $^{40}$K-$^{87}$Rb, in which only
$^{40}$K-$^{87}$Rb-$^{87}$Rb supports Efimov resonances because  
$^{40}$K-$^{40}$K-$^{87}$Rb is suppressed by spin statistics. The JILA group 
measured the three-body recombination rate with the aim of testing whether 
they could observe approximately the same Efimov resonant 
position as had been seen for $^{41}$K-$^{87}$Rb-$^{87}$Rb by~\cite{barontini2009PRL}. 
This expectation was of course fueled by the assumption of universality of 
the three-body parameter~\cite{Wang-2012b}. However, no trace of any Efimov resonances was 
observed at the expected two-body scattering length~\cite{bloom2013PRL}, consistent with our
current theoretical understanding~\cite{Wang-2012b,Ulmanis2016prl}. 

Very recently an Efimov 
resonance for $^{7}$Li-$^{87}$Rb-$^{87}$Rb was reported, ~\cite{Maier-2015} as a consequence of a 
first exploration of the negative Li-Rb scattering length in an ultracold mixture 
of bosonic Li and $^{87}$Rb. The observed 
resonance is found at $a_-^{(1)}=$-1870 $\pm$ 121 $a_{0}$ in at least approximate agreement with the universal Efimov 
expectation~\cite{Wang-2012b}, and it should be stressed that it is vital to include in the analysis
 the correct heavy-heavy scattering length~\cite{Maier-2015}.  Note that with current experimental
capabilities it is extremely difficult to reliably create and control atom-atom scattering lengths
beyond about $10,000$ $a_0$. in absolute magnitude.  Only one experiment to date, a heroic effort 
by the Innsbruck group in a homonuclear Cs gas \cite{Huang-2014a}, has been able to measure Efimov physics at 
a scattering length as large and negative as $-22,000$ $a_0$.

\begin{table}[h]
\centering
\begin{tabular}{c c c c c}
\hline
\hline
n & $a_{-}^{(n)}$ ($a_{0}$) &$ \lambda$ & $a_{-}^{(n)}$ ($a_{0}$) & $ \lambda$\\
\hline
1 &  -311 $\pm$ 3 &  & -323 $\pm$ 8 & \\
2 & -1710 $\pm$ 70 &  5.48 $\pm$ 0.28 & -1635 $\pm$ 60 & 5.1 $\pm$ 0.2 \\
3 & -8540 $\pm$ 2700 &  5.00 $\pm$ 1.8 & -7850 $\pm $ 1100& 4.8 $\pm$ 0.7 \\
\hline
\hline
\end{tabular}
\caption{Experimental Efimov resonances for $^{6}$Li-$^{133}$Cs-$^{133}$Cs, with the Li-Cs scattering lengths denoted 
here as $a_{-}^{(n)}$. 
 For the spin states utilized in these experiments, the background Cs-Cs scattering length in this region of magnetic field near 843G is approximately in the range $-1600 a_0 < a_{\rm{CsCs}}<-1000 a_0$.
The maxima of the three-body loss rate occur at the indicated valuen $a_{-}^{(n)}$, where 
 $n$ stand for the ordering of the different associated Efimov states. The three-body parameter is 
 denoted here as $a_{-}^{(1)}$. The discrete symmetry scaling factor for two successive 
 Efimov states is denoted by $\lambda$. The \textcolor{black}{Heidelberg group results of} \cite{Ulmanis-2015} are shown in the second 
 and third columns, the results of Chicago group \cite{Tung-2014} in the fourth and fifth columns.
 Note that the Heidelberg group also suggests a re-calibration of the Chicago group's data in Table 2 of \cite{Ulmanis-2015}, but those results are not shown here.}
\label{LiCs}
\end{table}

The most convincing tests of universal Efimov scaling are for the highly mass imbalanced 
case of $^{6}$Li-$^{133}$Cs-$^{133}$Cs, studied independently by the \textcolor{black}{
Heidelberg~\cite{Pires-2014,Ulmanis-2015,Ulmanis-2015b,Ulmanis2016prl,Haefner2017arxiv} and Chicago groups~\cite{Tung-2014,JohansenChin2016arxiv}}. Theoretically, 
the universal Efimov scaling factor for this system should be $\lambda$ = 4.88 ~\cite{Wang-2012b,dincao2006PRAb}, which enables experiments to observe and characterize multiple resonances in a single Efimov series for the first time. 
As in many other approaches to Efimov physics with ultracold atoms, Chicago and Heidelberg groups use a magnetic Fano$-$Feshbach resonance for Li-Cs to vary the two-body scattering length. The results for the observed Efimov resonances, characterized by analysis of the maxima in the three-body loss coefficient, are shown in Table~\ref{LiCs}, where the three-body parameter reported by the Chicago and Heidelberg groups can be seen to agree approximately, to within the error bars. Another interesting difference probed in the experiments by \cite{Ulmanis2016prl} is the contrasting value of the first Efimov resonance depending on the sign of the Cs-Cs scattering length.  For instance, when the Cs-Cs scattering length is large and negative as in the cases reported in Table \ref{LiCs}, the first resonance occurs at a Li-Cs scattering length value equal to $a_{-}^{(1)} \approx -320 a.u.$.  But for a different range of magnetic fields where the Cs-Cs scattering length is positive [$a({\rm Cs-Cs}) \approx 200 a_0$], the first resonance occurs at $a_{-}^{(1)} \approx -2000 a.u.$  As is argued in \cite{Ulmanis2016prl}, this major difference can be understood qualitatively already in the zero-range theory, without invoking van der Waals interactions, although a full model including van der Waals finite-range interactions is needed to make the description quantitatively accurate. These experiments are of course extremely difficult, and we stress the importance of developing highly accurate
two-body scattering models before undertaking the analysis of departures from expected universal behavior. 

The prediction of Efimov and the universality of the three-body 
physics \cite{efimov1970plb,efimov1971SJNP}is strictly true in the case of 
two-body resonant interactions and assuming $T$~=~0, since no 
consideration was given to the kinetic energy of the three-body system. A few
studies have carried out the appropriate Boltzmann average needed to derive
finite temperature predictions of the three-body recombination rate, as in~\cite{Petrov2015,Salomon2013prl}.
Thus, in realistic systems one would expect some deviations \textcolor{black}{from} Efimov's prediction. However, the experimental observations seem to indicate that 
most of the Efimov states are accurately universal.

\subsection{Efimov and universal bound states for fermionic systems}
\textcolor{black}{
It is well known that there is no Efimov effect for homonuclear trimers composed of identical fermions in a single intrinsic spin substate. This is easy to understand because the requirement of antisymmetry adds nodes to the spatial wavefunction and this raises the kinetic energy of the trimer internal degrees of freedom substantially.  For a system of two heavy fermions of mass $M$ in the same spin state and a lighter distinguishable particle of mass $m$, the nodal constraint of antisymmetry is weaker and some interesting predictions for this case have been presented by ~\cite{kartavtsev_low-energy_2007,kartavtsev_recent_2014}.  The Efimov effect emerges for this FFX system with divergent F+X scattering length, provided the mass ratio is sufficiently large, namely $M/m > 13.607$.  For smaller mass ratios than this critical ratio just mentioned, one observes one or two universal states, usually denoted ``Kartavtsev-Malykh universal trimers'', but there is no true Efimov effect and the number of energy levels remains finite.  Some level perturbations that can affect these universal trimer states have been identified by ~\cite{Safavi-Naini2013pra}.
}

\textcolor{black}{The possibility of a four-particle Efimov effect is another intriguing prediction by ~\cite{CastinMoraPricoupenko2010prl}, with 3 heavier identical fermions of mass $M$ and one lighter distinguishable particle of mass $m$.  Specifically, they predict that only in the tiny mass ratio range $13.384 < M/m < 13.607$ should one be able to observe the infinite number of energy levels converging geometrically to zero binding that characterizes Efimov physics.  A very recent preprint from ~\cite{BazakPetrov2016} predicts a pure five-body Efimov effect, for a system of 4 heavy identical fermions and 1 distinguishable particle, again in a small range between $13.279\pm.002 < M/m < 13.384$, based on a stochastic solution of the generalized Skorniakov-Ter-Martirosian (STM) equation~\cite{Pricoupenk2011pra}, based on techniques analogous to diffusion Monte Carlo methods.  The study of ~\cite{BazakPetrov2016} also predicts mass ratios where one expects universal five-body states in regimes where no true Efimov effect exists, and a conjecture that the 5+1 hexamer and higher particle numbers will be qualitatively different rather than simply continuing the trend, a conjecture certainly deserving to be explored. This study agrees and improves on the accuracy of a prediction by ~\cite{Blume2012prl} that a universal 3+1 tetramer should exist at a mass ratio around $M/m \ge 9.5$, with the new and improved computed ratio equal to $M/m \ge 8.862\pm 1$.
}

 \subsection{Naturally occurring Efimov physics in the helium trimer}

The study of helium clusters-- their aggregation, formation and collision dynamics -- has been 
an active research topic in chemical physics, in particular in the field of molecular
 beams~\cite{Campargue}. Molecular beam experiments rely on the supersonic 
 expansion of a chosen gas in vacuum, which induces the cooling of the different 
 molecular degrees of freedom as the gas expands in the chamber. This cooling mechanism 
 is due to inelastic collisions involving electronic, rotational and vibrational degrees of freedom, 
 and hence it strongly depends on the inelastic cross sections as well as the number of collisions 
 through the density of the gas~\cite{Zucrow,Zhdanov,Montero-2014}. The diluteness of the gas as 
 it moves away from the nozzle can be controlled by the initial conditions of the expansion: 
 temperature, pressure and mass flow, through the conservation of enthalpy and mass flow 
 of the fluid. Therefore, any property or process related with the dynamics of the gas, such as 
 cluster formation, could be controlled to some degree in those experiments. Using such methods, Sch\"ollokopf and 
 Toennies~\cite{He-trimer,He-trimerb} experimentally observed the helium dimer and the 
 ground state of helium trimer. 
 
\begin{figure}[h]
\centering
 \includegraphics[scale=0.20]{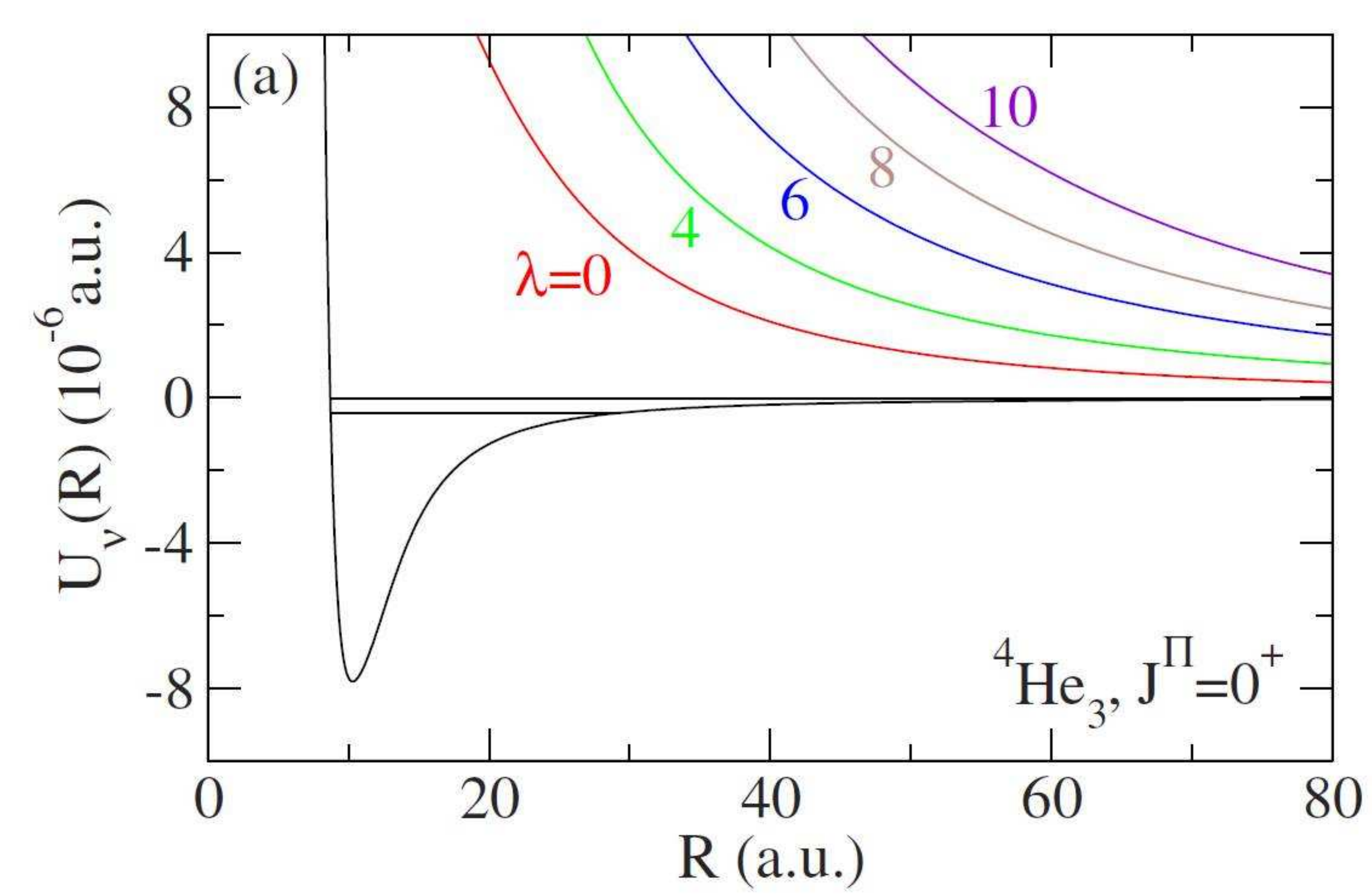}
 \includegraphics[scale=0.14]{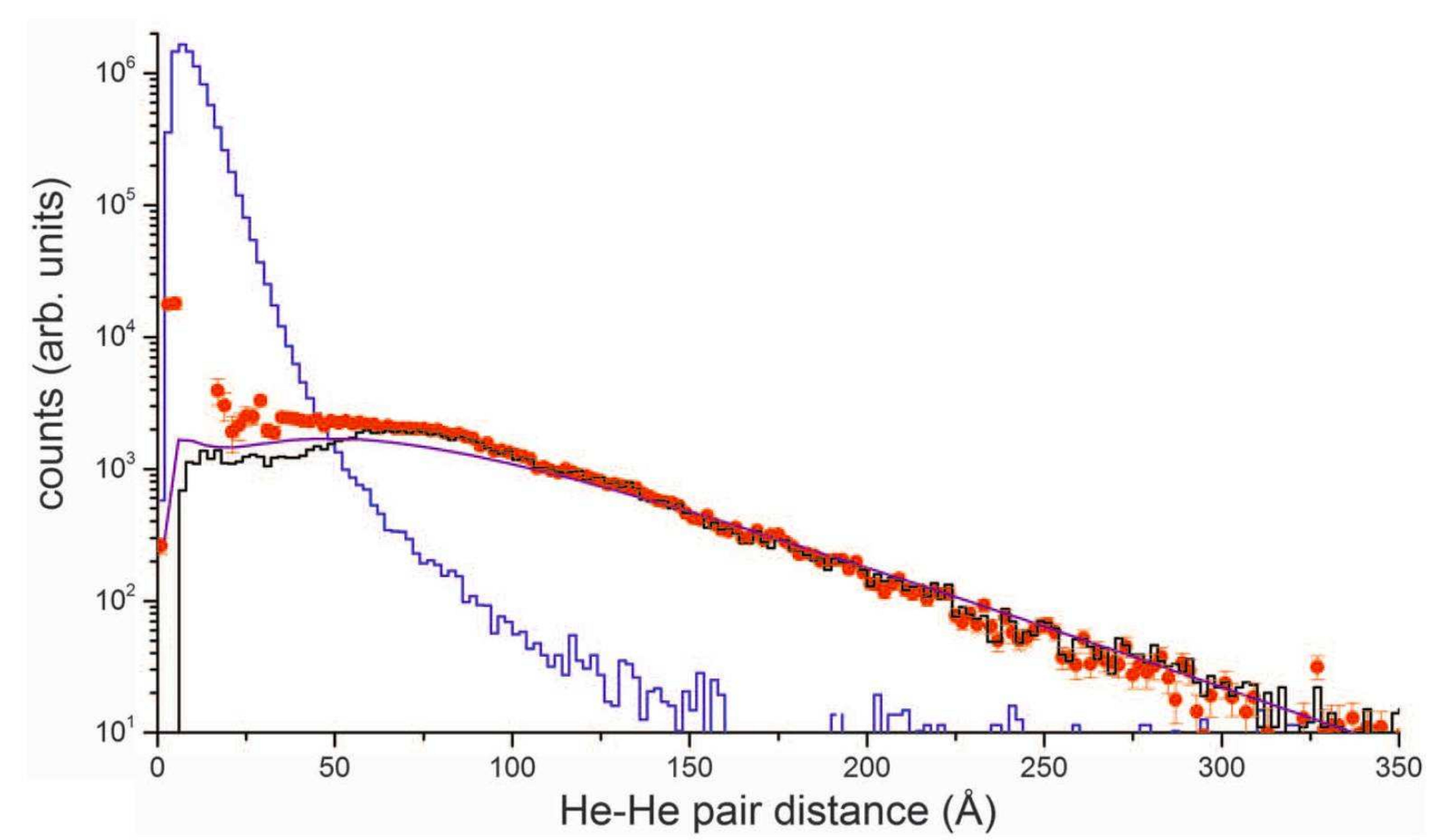}
 \caption{(Color online) (a) Adiabatic $J^\pi=0^+$ hyperspherical potential curves from the calculation of \cite{suno2008PRA}, with the two bound state energies predicted for this symmetry drawn into the potentials. 
 The higher energy of the two, which has an energy so small that it appears to coincide with E=0, was 
 predicted to be an observable Efimov state. (b) \textcolor{black}{Experimentally measured pair distribution functions of the two helium trimer bound states, with a theory comparison for the more diffuse state that is concluded to be an Efimov state. This measurement used laser ionization followed by Coulomb explosion of the three resulting ions, with detection in a COLTRIMS apparatus.  For more details see the combined theoretical and experimental paper published by \cite{Kunitski-2015}.}}
\label{testaaaa}
\end{figure}

 Although the ground state $^4$He$_{3}$ was observed two decades ago, the first excited 
 state of $^4$He$_{3}$, which has Efimov character, was not observed until very 
 recently~\cite{Kunitski-2015}.  \textcolor{black}{Fig.~\ref{testaaaa} summarizes both the key theoretical and experimental results for the system.} This remarkably challenging experiment was performed by 
 joining the technology of molecular beam experiments, atom interferometry, and modern ionization 
 and detection techniques. In particular, a very well controlled nozzle conditions leaded to 
 a supersonic expansion of He, where He trimers were selected by means of matter-wave 
 diffraction through a grating. Then, all three atoms of the trimer are ionized by means of a 
 strong ultrashort pulse, leading to the subsequent Coulomb explosion of the trimer compounds. 
 The momenta of the ions after the Coulomb explosion were detected by cold target recoil 
 ion momentum spectroscopy (COLTRIMS)~\cite{Ullrich-2003,Jagutzki-2002}, which allowed in an analysis the reconstruction of the 
 initial probability distribution of the trimer atom positions, and thereby allowing a deduction of the trimer binding energy. In this way, 
 Kunitski {\it et al.}~\cite{Kunitski-2015} studied the formation of two different He trimer states as functions of the pressure 
 in the nozzle, leading to the first observation of the excited state of $^4$He$_{3}$.
 
The findings of Kunitski {\it et al.}~\cite{Kunitski-2015} revealed the geometry of the the ground 
state and first excited state of helium trimer. In particular, for the ground trimer state it was 
observed a unimodal radial distribution for the atom-atom distance in the trimer, in relation 
with the expected equilateral geometry. However, for the first excited trimer state the radial 
distribution function shows a bimodal character, thus resembling an isosceles triangular 
geometry.  These results demonstrate the Efimov character of the first excited state of 
helium trimer. In particular, the obtained binding energy is 2.60 $\pm$ 0.2 mK in very good 
agreement with some of the most recent theoretical predictions~\cite{Kunitski-2015,Hiyama-2012}, however the bimodal 
radial distribution clearly deviates from what it is expected from the universal Efimov predictions 
for resonant two-body interaction at unitarity, which is not surprising in view of the finite value of the He-He scattering 
length~\cite{blume2000JCPc,Blume2014jcpErr}.

The most recent hyperspherical coordinate calculations of the helium trimer properties in the electronic
ground state appear to be those of~\cite{suno2008PRA}.  Their $J^\pi=0^+$ adiabatic potential energy curves 
obtained with an up to date potential surface, which includes three-body as well as realistic retarded
two-body potential terms, are shown in Fig.~\ref{testaaaa}.  The energies drawn into 
the lowest potential curve are the 
two computed bound state energies, namely -130.86 mK  and -2.5882 mK.  
The more weakly bound of these is the one 
expected to have significant Efimov state character, and it is in fact so weakly bound that its energy
is indistinguishable from $E$~=~0 on the scale of Fig.~\ref{testaaaa}.

\section{Few-body perspectives on many-body systems}
\label{manybody}
\begin{figure}[h!]
  \includegraphics[scale=0.10]{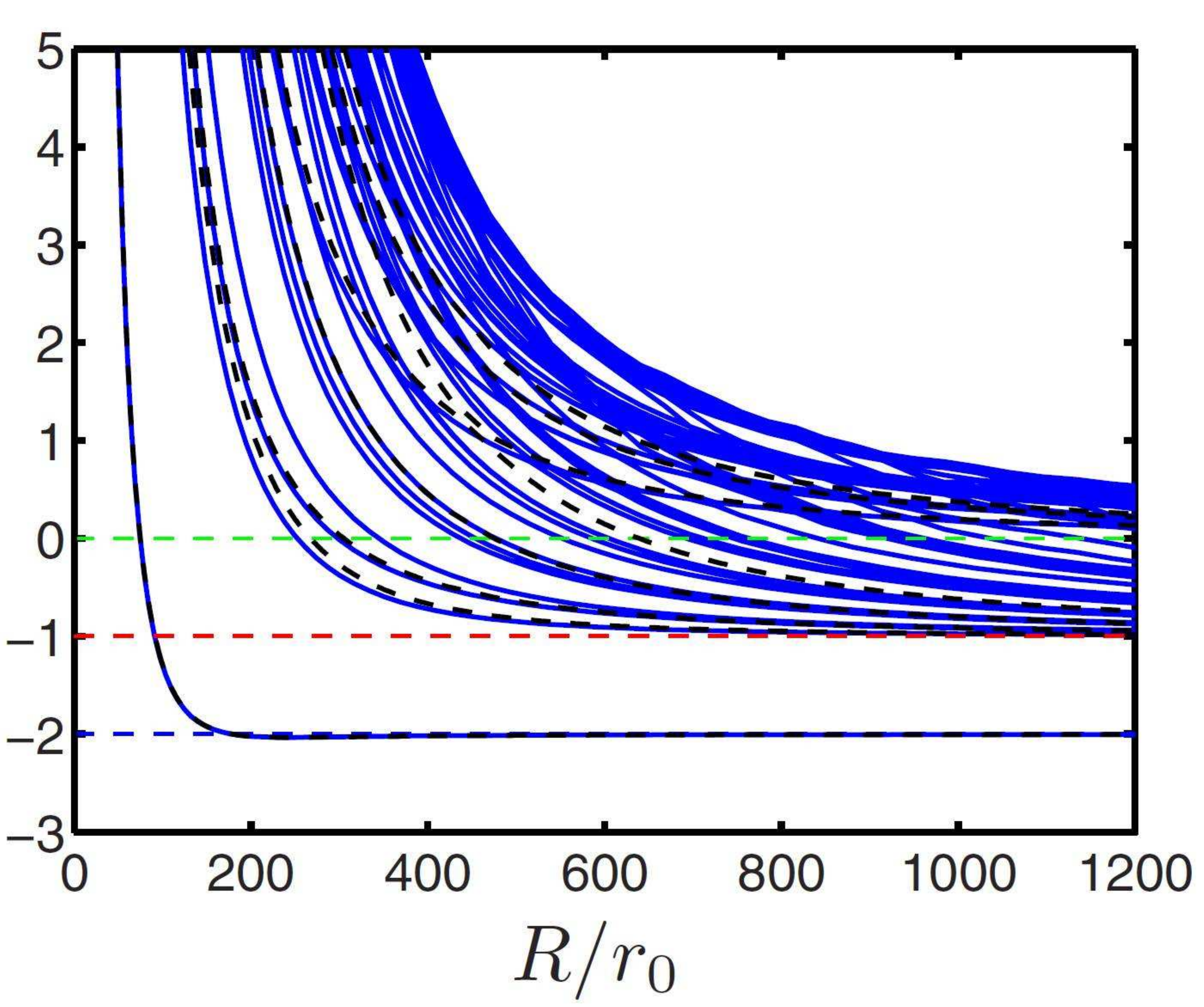}
  \includegraphics[scale=0.10]{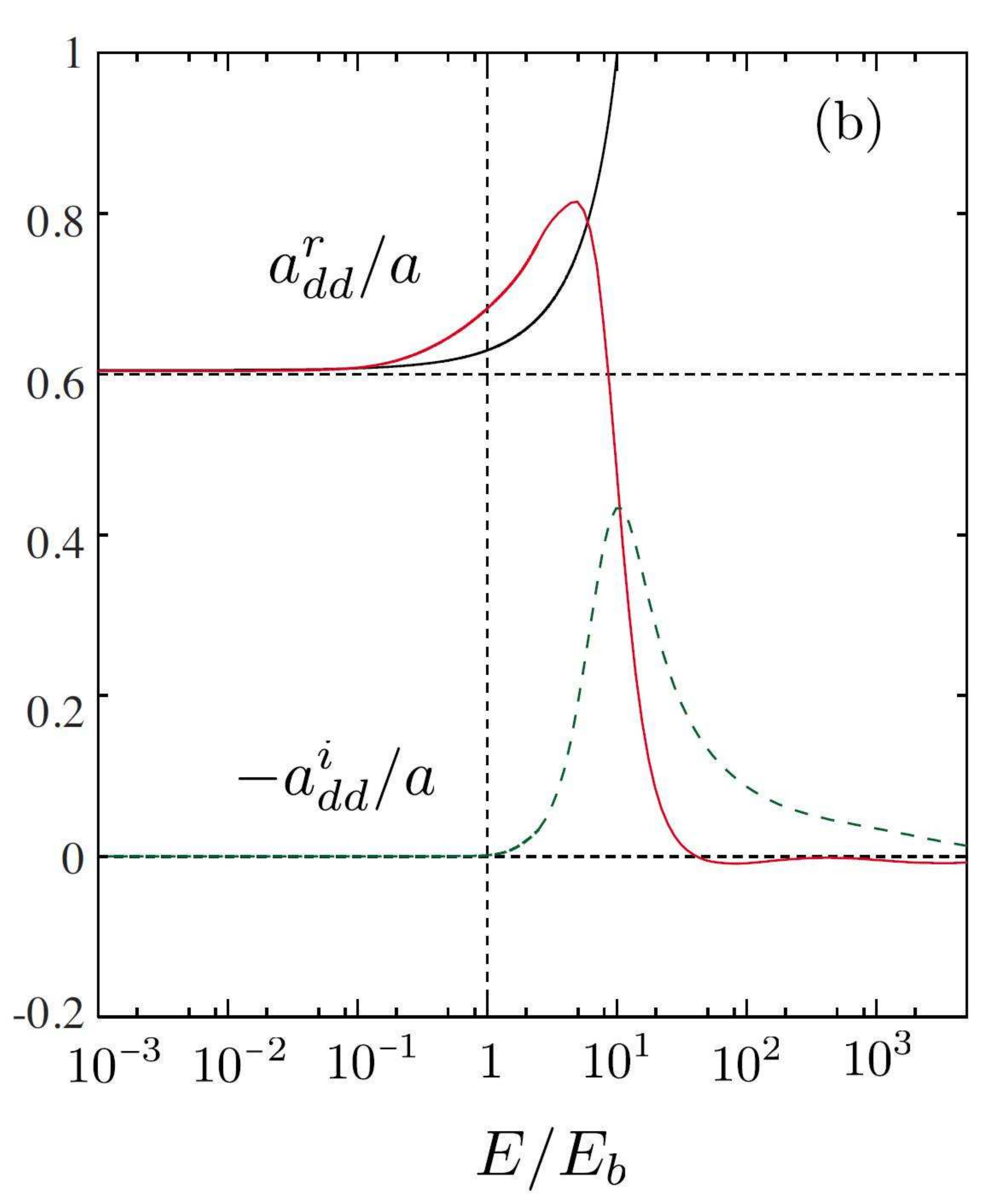}%
\caption{(Color online) 
{(a) Adiabatic $J^\pi=0^+$ four-fermion hyperspherical potential curves from the calculation of~\cite{stecher2009PRA}, for two spin-up and two spin-down identical fermions with a large positive interspecies scattering length, i.e. on the BEC side of the BCS-BEC crossover problem.  The dashed horizontal lines mark the fragmentation \textcolor{black}{thresholds, the lowest of which represents the dissociation of two bound universal dimers ($FF'~+~FF'$), the next highest representing one bound dimer plus two free atoms ($FF'~ + ~F~ +~ F'$), and the highest} which denotes the threshold energy $E=0$ for complete four-body dissociation.
Using these adiabatic potential energy curves and the nonadiabatic couplings, the elastic and inelastic collision properties could be computed for this system. (b) Computed elastic $a_{dd}^r$ and inelastic $a_{dd}^i$ scattering lengths for collisions between two universal dimers, i.e. in an FF'+FF' collision, shown in units of the two-body scattering length $a(F+F')$ as a function of energy measured in units of the dimer binding energy.  Note the smallness of the inelastic (imaginary) scattering length, first understood theoretically by~\cite{petrov2005JPB,petrov2005PRA}, which was crucial for understanding why the two-component Fermi gas has minimal losses close to unitarity\cite{cubizolles2003PRL,strecker2003PRL,regal2004PRL}. These low losses were crucial in enabling the BCS-BEC crossover experiments to be successful and create long-lived quantum gases.}
}
\label{pfig00003}
\end{figure}

There are multiple ways in which few-body physics is useful for understanding, interpreting, 
and predicting new many-body phenomena. The most obvious is through the development of detailed 
theoretical understanding of the microscopic processes involving two, three, four, or some 
cases even a handful more particles within a gas or lattice array of particles.  The detailed 
studies described above, and other review articles~\cite{koehler2006,chin2010RMP,wang2013amop,wang2015AnnRev} have focused largely
on two-body phenomena such as Fano-Feshbach resonances, and on the three-body phenomena that arise
such as Efimov resonances in three-body recombination and related behavior such as St\"uckelberg 
interference minima.  The initially surprising experimental 
result~\cite{cubizolles2003PRL,strecker2003PRL,regal2004PRL} that huge universal fermionic dimers 
have remarkably small losses despite their huge size was understood in an 
important theoretical treatment \textcolor{black}{by ~\cite{petrov2005JPB,petrov2005PRA}. This  
played a key role in stimulating experiments
in the BCS-BEC crossover problem and in triggering explorations of other phenomena in unitary Fermi 
gases.  Fig.~\ref{pfig00003} shows a later theoretical treatment of the two-component four-fermion system in hyperspherical coordinates, including the computed dimer-dimer scattering information.}  For instance, it has now become possible to map out an extremely accurate equation of state
for unitary Fermi gases~\cite{ChevySalomon2011JPConf,KuZwierlein2012science}.

Another way few-body theories have provided some useful perspectives on Bose-Einstein condensates 
and degenerate Fermi gases has simply been through applying the few-body toolkit and ideas - such
as adiabatic hyperspherical potential curves - to the many-particle system directly.  In some cases
this is done by treating only a modest number of particles accurately, while in other cases the
many-particle limit is examined but at a relatively crude level of approximation to estimate the
many-particle potential energy curves.

\begin{figure}[h]
  \includegraphics[scale=0.20]{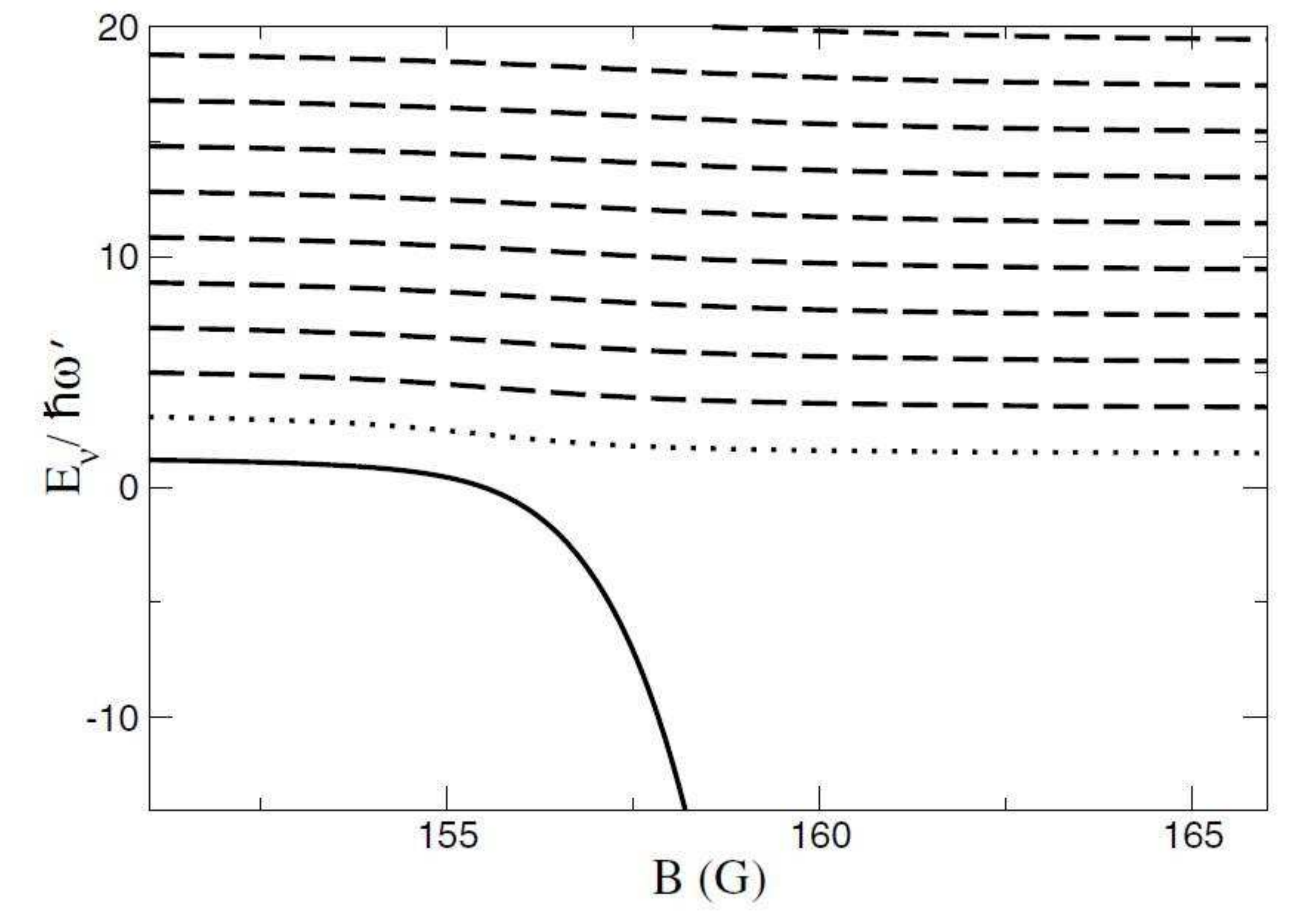}
  \includegraphics[scale=0.18]{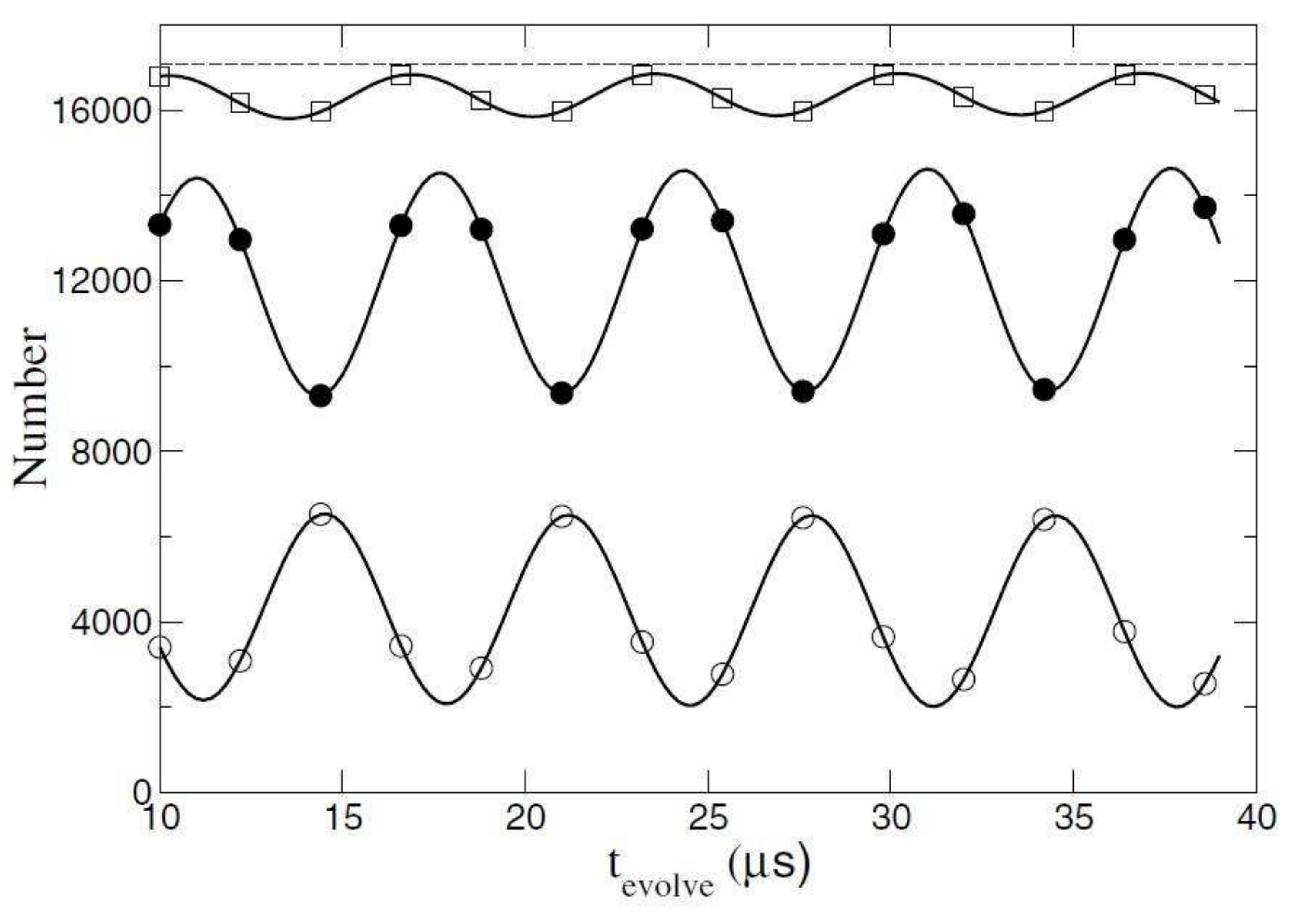}
\caption{(a) Two-body $s$-wave energy levels in $^{85}$Rb are shown 
as a function of magnetic field near the 155~G Fano-Feshbach resonance that was used by \cite{donley2002NT} 
to study quantum beats between atomic states of
a quantum degenerate Rb gas and molecular states. (b) Calculated quantum 
beats reflecting interference between one pathway where a given pair of 
atoms remained atomic and another pathway where that pair of atoms was 
bound into a long-range universal dimer for a delay time $T$. These two-body 
calculations use an artificially tight trap ($\omega' = 2\pi $ kHz whose peak 
density approximately equals the density of 17,100 atoms trapped in the actual 
experiment whose geometric mean trapping frequency was $\omega = 2\pi \times 12 $ Hz. Taken from \cite{borca2003NJP}}
\label{molecule7a}
\end{figure}

Few-body physics also produces insight into systems of trapped atoms through the use of the idea 
of an artificially strong trap.  The premise here is that frequently in a trapped quantum gas with 
thousands or even millions of atoms, the physical trap frequency might only be of order 10 Hz and  
determines only a largely irrelevant energy scale for the system.  More relevant by far are the 
typical scale of interparticle interactions and the average kinetic energy or temperature, reflected in the
separation of atoms $\Delta r$.  The physical content of this separation length scale is sometimes 
referred to as the ``Fermi wavenumber''
$k_F = 2\pi/\Delta r$ even when the gas consists of some or even all bosonic particles.  Then one
can gain insight by treating only 2, 3, or 4 particles in an unphysical artificially tight 
``theoretical trap'' designed to have a high frequency with particle separation $\Delta r$ 
comparable to the $\Delta r$ in the actual many-particle system.  An example
where this strategy enables a simple interpretation~\cite{borca2003NJP}.
 of a complicated many-body problem is the famous 
``atom-molecule'' coherent oscillations or quantum beats observed in an 
$^{85}$Rb experiment\cite{donley2002NT}, \textcolor{black}{depicted in Fig.~\ref{molecule7a}.}

\subsection{\textcolor{black}{Polaron physics attacked from a few-body viewpoint}}

\textcolor{black}{When a slow electron moves inside a bulk material such as a polar crystal or helium liquid, it attracts other particles from the bulk and the entity behaves as a quasi-particle, as described in highly-cited early studies by ~\cite{Froehlich1954AdvPhys,Feynman1955PhysRev}.  Such a quasi-particle was denoted a "polaron", and this term has been generalized to describe a more general situation in which an interaction-dressed minority particle moves in the field of other particles in a medium.  An active field of research to this day, polarons have attracted extensive attention from experimental ~\cite{MichaudSanche1987pra} as well as theoretical studies~\cite{BasakCohen1979prb, FanoStephens1986prb, StephensFano1988pra} in condensed-matter physics.  In recent years polaron physics has become a topic of great interest in the ultracold atomic physics community, owing to the promise of great control and observability.  The few-body side of polaron physics has two major areas of interest.  One aspect is to discern the details of a single quasi-particle in the many-body environment and the more advanced topic of interactions among 2, 3, or 4 quasi-particles, i.e. the effect of a many-body bosonic or fermionic bath of particles on the interactions, energy levels, and recombination of the quasi-particles.~\cite{BellottiZinner2016}  The second area that has received extensive attention is the behavior of few-body analogues of a polaronic system, with small numbers of minority and majority particles, such as the HHL and HHHL and related problems discussed elsewhere in this review.}

\textcolor{black}{Polarons have received attention in ultracold atom experiments and theory over the past decade or so by ~\cite{astrakharchik_motion_2004,kalas_interaction-induced_2006,cucchietti_strong-coupling_2006,tempere_feynman_2009,schirotzek_observation_2009,bei-bing_polaron_2009,koschorreck_attractive_2012,spethmann_dynamics_2012,kohstall_metastability_2012,catani_quantum_2012,rath_field-theoretical_2013,scelle_motional_2013,li_variational_2014,levinsen_impurity_2015,grusdt_renormalization_2015,bruderer_self-trapping_2008} and very recently, as in ~\cite{HuJinCornell2016prl,JorgensenParishArlt2016prl}. Some work has considered an impurity with internal degrees of freedom called an ``angulon'' which is a quasi-particle consisting of a rotating impurity dressed by the quantal many-body environment ~\cite{SchmidtLemeshko2016prx,Lemeshko2017prl}. To date the explorations have concentrated on the behavior of a single impurity in a BEC or DFG, but a future few-body topic that is still in its infancy will be the study of interactions among two or more impurities dressed by the many-body environment.  An initial foray along those lines by ~\cite{Naidon2016arxiv} treats two-body polaron-polaron interactions.  We refer the reader to the excellent recent review of this subject in ~\cite{NaidonEndoReview2016} and references therein.
}

\subsection{Bose-Einstein condensates viewed in hyperspherical coordinates}

\begin{figure}[h]
 \includegraphics[scale=0.25]{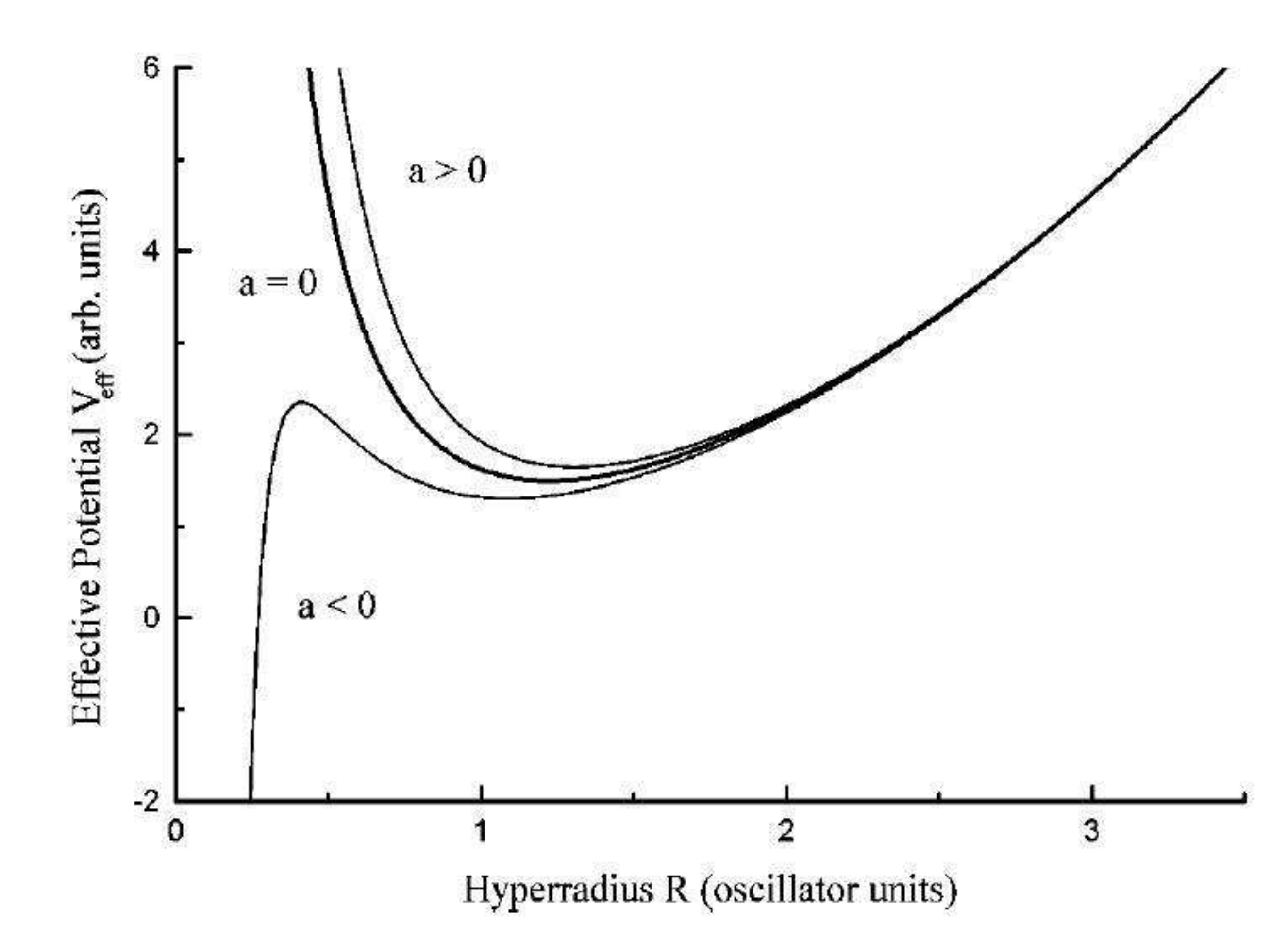}
 \includegraphics[scale=0.18]{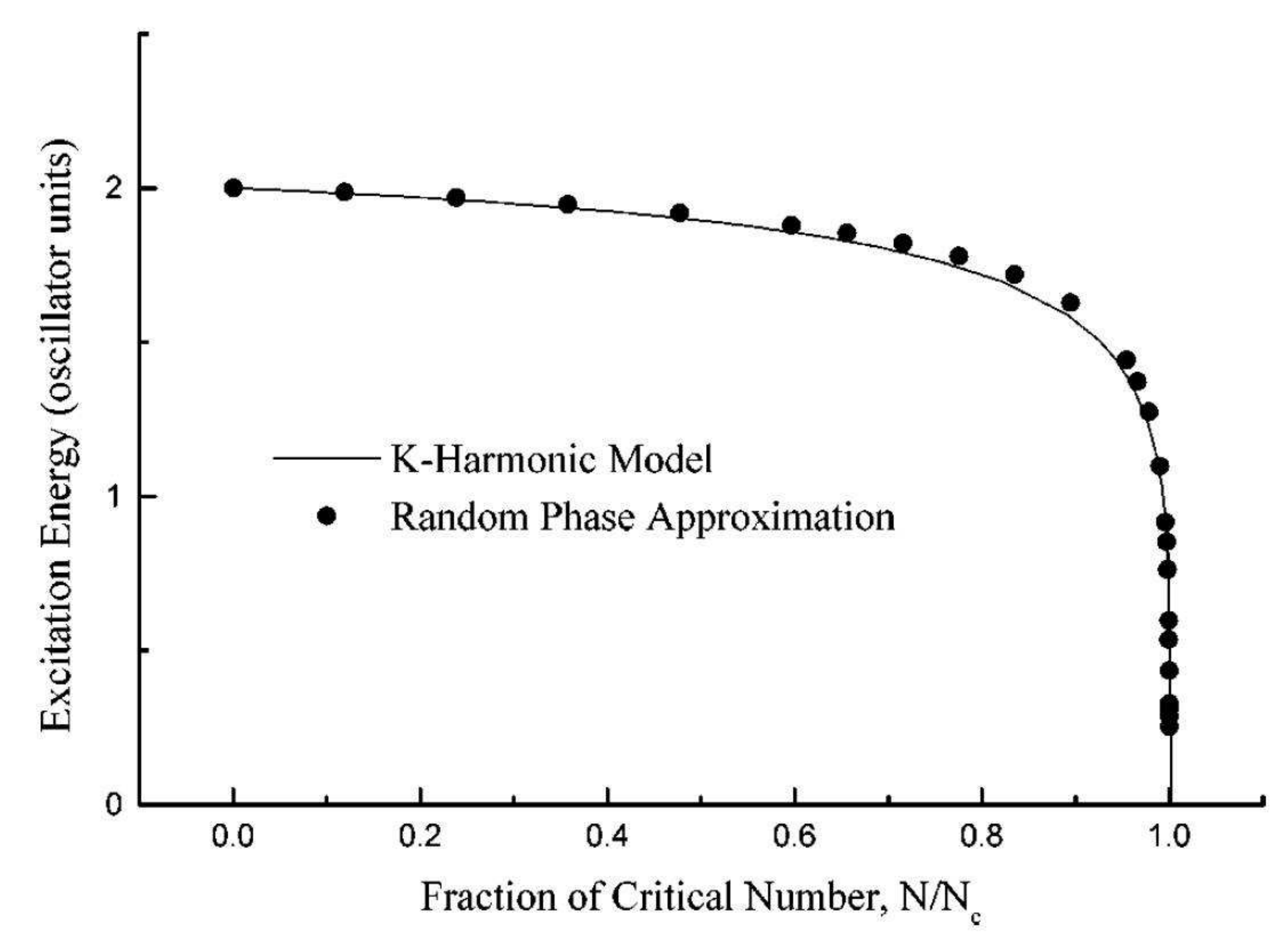}
\caption{(a) Adiabatic hyperspherical potential curves for an attractive BEC, a noninteracting BEC, 
and a repulsively interacting BEC. (b) Excitation energy calculated for an attractive ($a<0$ BEC) 
in the hyperspherical adiabatic theory, compared with the random phase approximation which is 
essentially identical to Bogoliubov theory. Taken from \cite{bohn1998PRA}.}
\label{BEC8a}
\end{figure}

In the Russian nuclear physics literature, a technique evolved during the 1960s and 1970s to treat the 
many-nucleon problem, which was referred
to as ``K-harmonic'' theory.  This treatment was based on a particularly simple approximation 
formulated in hyperspherical coordinates.  The basic idea was to find the lowest grand angular momentum 
state, the ``$K$-harmonic'' $|KQ>$.  This eigenfunction of hyperangular kinetic energy and various 
symmetry operators is usually an appropriately antisymmetrized linear combination of hyperspherical 
harmonics; they are constrained to obey the symmetries of the system such as the Pauli exclusion 
principle, definite parity, and so on.  This approximation then uses this single harmonic to describe
the hyperangular wavefunction of the system. When this technique is applied to the description of a single
component BEC, it is particularly simple because the $K$-harmonic for this system is simply the state of 
vanishing grand angular momentum, $K=0$, which is a constant in the hyperangular space of any $N$-particle 
system.  While the resulting wavefunction is not sufficiently realistic
to describe the true short-range interactions that occur whenever any two particles approach each other, the
use of the Fermi pseudopotential in the many-body Hamiltonian implies that the average energy of interaction
can be approximately represented perturbatively.\cite{bohn1998PRA} In our terminology today, we can view this as 
approximating the lowest adiabatic hyperangular eigenfunction as this $K$-harmonic, after which 
perturbation theory can be applied to the interparticle interaction term in the Hamiltonian to determine the
approximate ground state potential energy curve for the many particle system, namely, $U_0(R)$. \textcolor{black}{See Fig.~\ref{BEC8a}.} This 
treatment does a reasonable job of predicting the approximate maximum number of atoms that can exist in 
a harmonic trap when the scattering length is negative, beyond which the system undergoes a macroscopic
collapse often referred to as the ``Bosenova''~\cite{bradley1997PRL,donley2001NT}.

Since that simple study of \cite{bohn1998PRA}, other studies have considered improvements 
of the hyperspherical BEC treatment, including generalizations to asymmetrical traps and attempts to better include the two-body correlation physics beyond the simplest mean-field approximations.\cite{Watson1999pra, KimZubarev2002pra, kushibe2004PRA, FedorovJensen2004becJPB} 

\subsection{The unitary Bose gas}
Based on what has been learned from studies of dilute Bose gas recombination theory and
experiments, one expects it to be impossible to create a long-lived Bose-Einstein 
condensate in the unitary limit where $a\rightarrow \infty$. This conclusion is
based on the fact that three-body losses are seen to scale with the scattering length 
overall as $a^4$, aside from quantum resonance and interference factors.  Nevertheless 
this is a fascinating limit, in part because it can test whether the recombination loss rates continue
to scale as $a^4$ all the way to $a \rightarrow \infty$ and in part because the many-body 
behavior implied by the Gross-Pitaevskii equation looks so qualitatively different 
for large negative (infinitely attractive and immediate collapse) versus large 
positive scattering lengths (infinitely strong repulsion).  

These questions have long been of theoretical interest, e.g.~\cite{Pethick2002prl,Radzihovsky20082376,LeeLee2010pra}, and recently they have
begun to receive experimental attention. One way to deal with the transient nature of 
any short-lived quantum gas is to perform the experiment as a ``quench'', {\it i.e.} begin 
with a BEC at small positive scattering length and then suddenly ramp to the range of 
unitarity, $a\rightarrow \infty$.  A recent experiment at JILA by~\cite{Makotyn:NaturePhysics:2014}
very quickly stimulated extensive theoretical 
work~\cite{Sykes2014pra,Krauth2014ncomm,JiangZhou2014pra,Smith2014prl,
Levin2014pra,Laurent2014prl,Kira2015ncomm,Ancilotto2015fbs,radzihovsky2016pra,
Jiang2016pra,CorsonBohn2016pra} to understand their main observations,
which were the following. {\it (i)} The $a^4$ scaling of three-body loss is no longer applicable
at very large scattering lengths, {\it i.e.} when $n a^3 >>1$ where $n$ is the density.  
To understand this, Ref.\cite{Makotyn:NaturePhysics:2014}
defines a characteristic wavenumber of the system (analogous to the Fermi wavenumber) as 
$K \equiv (6 \pi^2 n)^{1/3}$, which is comparable to the inverse of the average interparticle
spacing. The authors propose that in any formula involving the scattering length $a$, it
should be viewed as saturating at a constant value of the order of $1/K$ 
as soon as you reach the regime 
$K a \approx 1$. In other words, as soon as $a \rightarrow \infty$, it is no longer
a relevant length scale in the system, and the premise is that the interparticle spacing
becomes the largest relevant scale instead. {\it (ii)} The momentum distribution of the
atom cloud was measured as a function of time after the jump to unitarity, and it settled
down to a quasi-stable distribution, which was the target of various many-body and
few-body theory efforts to try to understand.

One of the main items of interest in this system, from a few-body physics perspective, is 
the saturation of losses at a value far smaller than would be expected from the zero-temperature
$a^4$ scaling. An estimate\cite{Sykes2014pra} has been made of the three-body rate 
coefficient $L_3 \approx 3 \times 10^{-23}\ {\rm cm}^{6}/{\rm sec}$ is reasonably close (around half as large as) the measured value in the JILA experiment of~\cite{Makotyn:NaturePhysics:2014}. This theoretical estimate is made using the few-body ``artificial trap model'' which considers only 3 atoms but places them in an artificially tight trap such that the atom density
approximately matches the experimental average density, which was $\langle n\rangle = 5.5(3) \times 10^{12}\ {\rm cm}^{-3}$. 
As is discussed in greater detail in~\cite{Sykes2014pra}, one obtains somewhat poorer quantitative agreement 
but still the correct order of magnitude for the loss rate at unitarity by replacing the value of $a$ in the
universal 3-body loss rate formulas by $1/K$ as defined above in terms of the average density in the gas.  Other
experimental studies of the unitary Bose gas that have focused on the three-body loss rate include \cite{Salomon2013prl,Hadzibabic2013prl,Eismann2016prx}.  And another intriguing ``universality limit'' where the scattering length is no longer a relevant length scale is the opposite limit $a \approx 0$, recently explored experimentally with some phenomenological conjectures in~\cite{Shotan2014prl}.

\subsection{Two-component Fermi gases and the BCS-BEC crossover problem}

One fascinating type of dilute quantum gas experiment consists of distinguishable fermions, either in two or more spin substates or else composed of two or more types of distinguishable particles.  This system received extensive attention from many theory (e.g.~\cite{YinBlume2015pra,YanBlume2016prl,hu2007NTP,hu2006EPL,hu2004PRL}) and experimental groups,\cite{demarco1999Sci,ketterle2008RIVISTADELNUOVOCIMENTO,KuZwierlein2012science,
zwierlein2004PRL,schunck2005PRA,zwierlein2006Sci,houbiers1997PRAb} with particularly keen interest in the community around a decade ago. Conceptually, one typically starts the experiment by forming a two-component degenerate Fermi gas without interactions, i.e. with vanishing scattering length between the two component atoms.  Interactions between like fermions can usually be ignored, unless one is close to a $p-wave$ Fano-Feshbach resonance.  Now one increases the attraction by making the scattering length $a$ between unlike atoms small and negative, i.e. $-1 << k_f a<0$.  This is the regime usually referred to as the Bardeen-Cooper-Schrieffer (BCS) region, because the weak attraction tends to cause pairing. These BCS-type pairs are sometimes referred to as ``pre-formed pairs'' because the pairing occurs before the attraction is strong enough to form true isolated dimer bound states.  Of course true molecular bound states can be formed only after the attraction has increased beyond $k_f a < -1$, to infinite $a$ and then $a$ is large and positive which allows true universal dimers to form with binding energy $\frac{1}{2 \mu a^2}$.  At that point, when the gas has large and positive $a$, the quantum gas has experienced a ``crossover'' from a degenerate Fermi gas to a BEC of weakly-bound molecules~\cite{demelo1993PRL}. Remarkably, such a system allows an exploration of either Fermi or Bose quantum statistics depending on the range of scattering lengths chosen experimentally. An impressive series of experiments has observed precisely these phenomena~\cite{greiner2003NT,regal2005PRL,regal2004PRLb,loftus2002PRL}, and this crossover physics has been reviewed by \cite{regal2007}.

Of course the single-component Fermi gas is also of interest, with all fermionic atoms in the same intrinsic spin state.  Owing to the absence of $s$-wave collisions in such systems, the cross sections for two-body elastic collisions that are needed for thermalization of the gas are generally quite small at ultracold temperatures.  Use of a $p$-wave Fano-Feshbach resonance can enhance the cross sections, although two-body losses also usually grow in the vicinity of such resonances~\cite{regal2003PRLb}. A two-component Fermi gas has its three-body recombination losses suppressed since the low temperature behavior of the rate coefficient is linear in the temperature $T$, in contrast to a gas of bosons or of 3 distinguishable particles which have a constant low-temperature 3-body recombination rate.  A spin-polarized gas of fermions, on the other hand, has an even stronger suppression of the low-temperature recombination rate, which varies as $T^2$.  This might be expected to give very long-lived Fermi gases when fully polarized, but near a $p$-wave Fano-Feshbach resonance in spin-polarized $^{40}$K (the point where the 
$p$-wave scattering volume $V_p \rightarrow \infty$), the recombination coefficient has been measured~\cite{regal2003PRLb} and calculated~\cite{suno2003PRL,suno2003NJP}, and found to approach within an appreciable fraction of the unitarity limit for the atom loss rate at total relative energy $E$:

\begin{equation}
K_3^{\rm max} = \frac{\hbar^5}{m^3} \frac{144 \sqrt{3} \pi^2}{E^2}.
\end{equation}

\textcolor{black}{In addition to an explosion of effort to understand the many-body physics of the BCS-BEC crossover problem, there has also been extensive fruitful effort directed towards understanding this system from a few-body point of view.  See in particular ~\cite{bulgac2006PRL, werner2006PRL, chang2007PRA, Akkineni2007prb, kestner2007PRA, alhassid2008PRL,ZinnerMolmer2009pra}.  And in fact one limit of the many-body problem, the ``high energy limit'' can be accurately treated using the virial or cluster expansion.  Exciting progress in computing virial coefficients for three particles ~\cite{LiuDrummond2009prl,CastinWerner2013cjp} and for four particles ~\cite{YanBlume2016prl} has been contributed in landmark theoretical papers during the past decade, as well as tested in a few impressive experiments.~\cite{NascimbeneSalomon2010nature}
}

\subsection{A step beyond independent particles: the Tan Contact}
The standard methods used in most many-body calculations usually start with a mean-field
wavefunction ansatz, in some cases going one step farther to the level of Bogoliubov
theory or, what is essentially equivalent, the random phase approximation \textcolor{black}{ ~\cite{fetter2003quantum,esry1997PRA}}.
These approximations have a demonstrated track record of describing gross global properties 
of many body systems, for properties such as chemical potentials, total energies, and excitation 
frequencies~\cite{dalfovo1999RMP,giorgini2008RMP}. One thing that should be kept in mind about 
such treatments is that they are based on ridiculously inaccurate cartoon-level wavefunctions of the
many-body system \textcolor{black}{at interparticle distances less than the van der Waals length. However, the behavior over larger distances of order of the long de Broglie wavelengths in the system, it turns out to be reasonable.    To understand the flaws in the full many-body wavefunction ansatz used to derive the Gross-Pitaevskii equation,
the workhorse equation of Bose-Einstein condensation theory, recall that it is} a product of independent orbitals $\psi$, i.e.
$\Psi(\overrightarrow{r_1},\overrightarrow{r_2},....\overrightarrow{r_N}) = 
\psi(\overrightarrow{r_1})\psi(\overrightarrow{r_2})...\psi(\overrightarrow{r_N}) $.  Why is this a 
patently absurd hypothesis?  Because it says that each particle's probability amplitude is {\it independent} 
of the instantaneous positions of all other particles; but in reality, there simply {\it must} be,
in the ``true'' wavefunction of the system, a two-body character (at the very least) when the distance between
any two particles of the gas gets comparable to the interaction length scale in the potential energy function
between those particles (the van der Waals length in the case of isotropic atoms).  For two $^{87}$Rb atoms in
spin-stretched magnetic substates, for instance, the zero-energy wavefunction of any two approaching atoms in the Rb BEC must have the 39 nodes as the interparticle distance is varied which are guaranteed 
to be there by Levinson's theorem,\cite{Rodberg1970,taylor1972}
since the Rb dimer has 39 triplet bound states in the $L=0$ orbital partial wave.  
(For an ultracold atomic gas, higher partial wave physics is normally suppressed by
the centrifugal barrier, so we will focus only on the $s$-wave physics in this discussion.) This behavior 
is often incorporated into Monte Carlo(MC) calculations by using a ``Jastrow-type'' variational trial function (or a guiding function in
the case of diffusion MC), which includes a product of all zero energy two-body pair wavefunctions~\cite{dalfovo1999RMP,giorgini2008RMP,Carlson2015rmp}.

To see why the cartoon-level approximation gives such a good description of many properties, consider 
the zero-range Fermi pseudopotential representation of the interaction between two
low energy particles, i.e. $V(\overrightarrow{r_{ij}}) = \frac{2 \pi a_{ij} \hbar^2}{\mu_{ij}} \delta(\overrightarrow{r}_{ij})$, 
where $a_{ij}$ is the scattering length between particles $i$ and $j$ and $\mu_{ij}$ is their reduced mass.  \textcolor{black}{ In the important paper by\cite{fermi1934}, he proved that this potential 
gives an accurate interaction energy of two particles with finite range potentials in the zero energy limit,
even when highly inaccurate zeroth-order wavefunctions are utilized; this rescues many-body predictions of energies and other
gross properties of the many-body system.  The success of this ``rescue'' is documented by \cite{Holzmann1999} who demonstrate that the behaviors {\it over large distance scales} of mean field and Bogoliubov wavefunctions are quite reasonable, even though their Hamiltonian does not contain the large number of two-body bound states whose presence would cause rapid, short-range oscillations in any ``exact'' wavefunction of all alkali metal atoms in a many-body gas.}

Nevertheless, some properties of the many-body system go beyond those global properties that are well described
by a separable wavefunction ansatz and its crude improvements at the next level of approximation.  One such property 
identified by Shina Tan in a ground-breaking series of papers~\cite{tan2008AP,tan2008APc,tan2008APb} is the ``high energy'' limit of the 
pair correlation function. This Tan contact parameter has now been measured for a Fermi 
gas~\cite{SagiJin2012prl,SagiJin2013} as well as for a BEC\cite{wild2012PRL}, and 
those experiments have confirmed the basis two-body physics on which Tan's ideas are based.  In brief, one way of looking
at the Tan contact is to acknowledge that there will be a range of distances, as two zero-energy particles begin to 
approach each other at smaller and smaller distances $r_{ij}$, where the wavefunction must be proportional to:
\begin{equation}
  \Psi \propto \frac{1}{a_{ij}} - \frac{1}{r_{ij}}
\end{equation}
In a sense the physics of the Tan contact is just the tip of the iceberg, because at higher momenta one begins to probe 
the full momentum space wavefunction of two-body subsystems, which have a complicated structure that in general depends
on the detailed nature of their short range interactions. \textcolor{black}{For instance, in a gas of Rb or K atoms, one can expect that the momentum space wavefunction above $k_{\rm{vdW}}\sim 1/\ell_{\rm{vdW}}$ or at energies above a few MHz should exhibit deviations from the contact prediction based on the scattering length alone.  Experimental measurements to date appear to be mostly in the 10-100 kHz regime.} Nevertheless, there a wide energy range, high compared to many
body excitation frequencies but low compared to van der Waals energy scales, 
where Tan's contact
and scattering length two-body physics controls the major departure of the atomic quantum 
gas from a description in terms of non-interacting
independent particle wavefunctions~\cite{blume2009PRA,braaten2010PRL,braaten2009LP,Smith2014prl,Sykes2014pra,
YanBlume2013pra,YinBlume2015pra,CorsonBohn2016pra,radzihovsky2016pra}.

\subsection{Towards many-body theory with realistic interactions}
If a mean field separable wavefunction ansatz is attempted in a 
variational calculation that uses realistic atom-atom interactions, 
the results are disastrous and the total energy is overestimated by
many orders of magnitude~\cite{esry1999PRAb}. Basically, the mean-field wavefunction 
is unable to make the wavefunction negligibly small at small
distances where the atom-atom potential is hugely repulsive.  An exact
solution of the Schr\"odinger equation would of course make the wavefunctions
exponentially small in such classically forbidden regions, something that
no independent particle wavefunction is ``smart enough'' to accomplish.  Quantum
Monte Carlo calculations, however, are able to solve for 
ground states of many particle systems and they are 
smart enough to make the wavefunctions exponentially small in regions of strong 
repulsion.  For instance, some of the best calculations of helium cluster energies have been 
obtained using diffusion Monte Carlo calculations~\cite{lewerenz1997JCP, blume2002EPJD}. Obtaining
excited state information from Monte Carlo calculations is notoriously difficult and
limited, however, which makes the technique difficult to use for determining scattering
properties.  

\begin{figure}[h!]
  \includegraphics[scale=0.25]{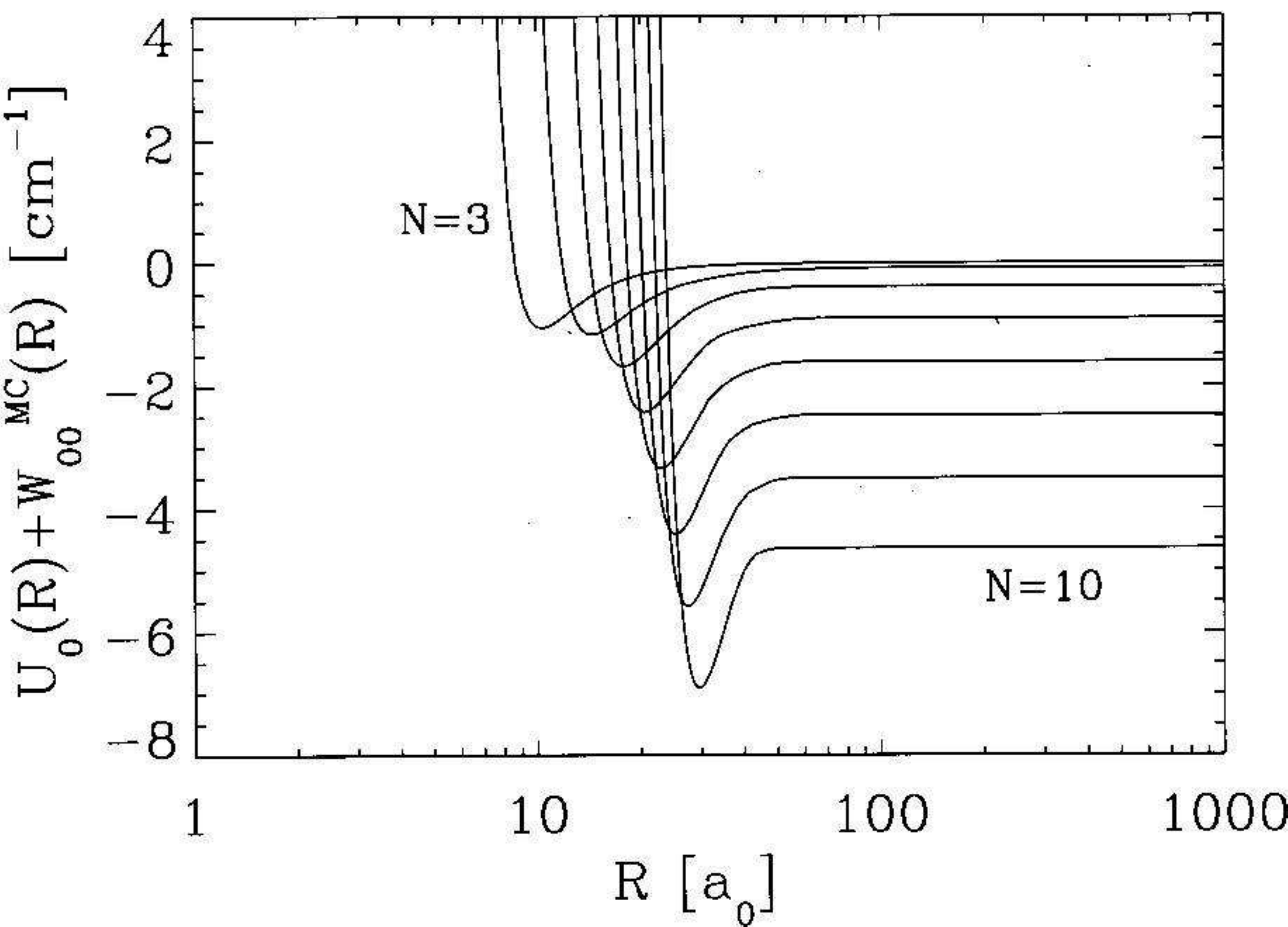}
 \caption{Lowest energy adiabatic hyperspherical potential curves for $N$ $^4$He atoms with $N=3-10$.  
These potential curves are for total angular momentum $L=0$.  They were computed in a hybrid
hyperspherical-diffusion Monte Carlo method.  Based on these potentials, approximate scattering lengths
were computed, for instance $a$(He$_9$+He)=67 $a_0$.  Note that the potential curve for He$_N$ converges 
asymptotically to the ground state energy of the He$_{N-1}$ cluster. Taken from~\cite{blume2000JCP}.
}
\label{BlumeHeliumPotentials}
\end{figure}

One hybrid theory that has shown promise combines Monte Carlo and 
adiabatic hyperspherical ideas~\cite{blume2000JCP}. The basic idea is to carry out a
diffusion Monte Carlo calculation to find the energy of the system at a fixed hyperradius.
By repeating the calculation for many different hyperradii, one maps out the ground state 
potential energy curve of the system. Then, within the adiabatic approximation that neglects
coupling to higher potential curves, at least elastic scattering and \textcolor{black}{a class of} excited bound state 
properties can be computed. That approach has been used to compute hyperspherical 
potential curves (including diagonal adiabatic correction terms) for clusters 
of up to $N=10$ $^4$He atoms, \textcolor{black}{as is shown in Fig.~\ref{BlumeHeliumPotentials},} and observable properties such as binding energies 
and atom-cluster scattering lengths. There appear to be no competing calculations to date of
the atom-cluster scattering lengths, for instance, beyond about $N=6$ helium atoms, although
excellent progress has been achieved up to $N=6$, and in some cases beyond, by using Gaussian
wavefunctions in combination with a Hamiltonian based on soft-core model potentials.
 
\section{Ultracold atoms in low dimensional traps}
The experimental realization of Bose-Einstein condensation in a dilute gas of alkali-atoms in 1995 \cite{anderson1995,bradley1995,davis1995} enables the investigation of pure quantum systems that lie at the interface among atomic, molecular, quantum optical physics and many-body physics.
A key breakthrough emerging from the control of ultracold gaseous matter is the capability to tune interatomic interactions in strength and sign by means of magnetic or optical Fano-Feshbach resonances~\cite{chin2010RMP,inouye1998,koehler2006}. 
Nowadays, this has triggered the next generation of quantum technologies that allow experimental creation and manipulation of low-dimensional ultracold gases~\cite{blochrmp,bongs2004physics,mckay2011cooling,lewenstein2007ultracold,
lewenstein2012ultracold,cazalilla2011one,imambekov2012one,giamarchi2004quantum,
lieb2004one, kolomeisky1996phase} of bosonic or fermionic or mixed symmetry~\cite{giorgini2008RMP}.

Degenerate ultracold atomic gases of reduced dimensionality then serve as a vehicle for experimental realizations and theoretical investigations of exotic quantum phases such as the Tonks-Girardeau gas (TG)~\cite{tonks1936, girardeau1960}. The TG many-body phase consists of a one-dimensional gas of impenetrable bosons with infinite pairwise repulsion. A fundamental property of the TG gas is that it can be viewed as displaying a {\it fermionization} of the bosons.
In this context, {\it fermionization} indicates that the infinite repulsion of the bosons creates a node when any two particles touch, so that the squared wavefunction of the infinitely repelling bosons coincides with that of a noninteracting fermionic gas. When the repulsive 1D interaction coefficient is cranked up beyond the pole and onto the side representing infinite attraction, the bosonic ensemble experiences a metastable many-body state, namely the ``super Tonks-Girardeau'' gas phase \cite{astrakharchik2005beyond}. Clearly the strength and the sign of interactions play an essential role in creating and probing these exotic many-body phases.
It is therefore crucial to study in detail the collisional processes as modified by external confining potentials.
Indeed, in these particular low-dimensional two-body systems, the confinement generates significant modifications to the colliding pair scattering properties. 
Existing theoretical studies on bosonic collisions show that resonant scattering can be induced by the confinement, yielding the so-called confinement-induced resonance (CIR) effect~\cite{Yurovsky2008a,dunjko2011}. 
A CIR occurs when the length scale of the confinement becomes comparable to the $s$-wave scattering length of the colliding bosons and it produces a divergence in the 1D coupling coefficient that is interpreted as a Fano-Feshbach-like resonance. 
The additional control of low dimensional gaseous matter by varying the confinement frequency led to the experimental creation of the TG gas~\cite{kinoshita2004observation,paredes2004tonks} and the Super-TG gas~\cite{haller2009} in cigar-shaped traps.

Evidently, the deepening understanding of collision physics in the low-dimensional ultracold gases translates into an ability to manipulate and even to design new many-body phases by means of the external confinement.

\subsection{Confinement-induced resonances: an interlude}

Early seminal work by \cite{demkovjetp1966} treated the motion 
of an electron in the presence of a uniform magnetic field as well as a zero-range \textcolor{black}{potential}.
That study showed that the motion of the charged particle is bounded in 
the presence of the magnetic and zero-range potentials whereas in the 
magnetic free case the zero-range potential can not bind the electron.
Another physical system which exhibits similar effects is the negative-ion 
photodetachment in a uniform magnetic field~\cite{larsonpra1979,larsonpra1985,
clark1983PRA, greene1987,crawfordpra1988, grozdanovpra1995,robicheauxschwinger2015} 
where in this particular case the electron-atom interaction is treated as a short-range potential.
Note that all these cases are half-collisions, \textcolor{black}{in the sense that they 
arise in photofragmentation processes, and only involve an {\it escape} to infinity, whereas a full 
collision involves both an incoming wave {\it and} an outgoing wave.}

In the realm of ultracold atomic physics, \cite{olshanii1998} showed in his 
seminal work that boson-boson collisions in an axially symmetric waveguide 
relies on virtually identical mathematics as the system treated by Demkov and Drukarev, 
except with the trapping potential playing the role of the transverse diamagnetic 
confinement caused by the magnetic field.
More specifically, Olshanii showed that the confinement not only can create a new 
bound state, but it can also nontrivially enhance the resonant two-body collision 
amplitude.  \textcolor{black}{In many cases the reduced-dimensional 
bound state is not strictly new, but can be viewed as having been shifted from its
position in the 3D or 2D system.  For this reason, the term ``confinement-induced resonance''
is often used synonymously with the term ``confinement-shifted resonance''.}
The following briefly summarizes the two-body scattering and its modification 
under the influence of the trapping potential and highlights the physical implications.

\subsubsection{Two-body collisions in a cigar-shaped trap}
Two bosonic particles in the presence of a quasi-1D waveguide have $s$-wave collisions that can be modeled using a Fermi-Huang regularized pseudopotential.
The waveguide constrains the motion of the particles transversely, they propagate freely in the longitudinal direction.
The quadratic nature of the trapping potential allows separation of the Schr\"odinger equation into center-of-mass and relative degrees of freedom.
The relative coordinate Hamiltonian describes the relevant collisional physics, which in cylindrical coordinates reads
\begin{equation}
 H= -\frac{\hbar^2}{2 \mu}\nabla_{\boldsymbol{r}}^2 + V_{sh}(\boldsymbol{r})+\frac{1}{2}\mu \omega_\perp^2 \rho^2,
 \label{peq1}
\end{equation}
where $\mu$ is the reduced mass of the two bosons, $\omega_\perp$ is the frequency of the confining potential.
As usual, $\rho$ denotes the radial polar coordinate and $V_{sh}$ is the 3D Fermi-Huang pseudopotential operator, defined by:
\begin{equation}
 V_{sh}(\boldsymbol{r})\Psi=\frac{2 \pi \hbar^2 a_s(E)}{\mu} \delta(\boldsymbol{r}) \frac{d}{d r}(r \Psi),
\label{peq2}
 \end{equation}
where $a_s$ indicates the $s$-wave scattering length, $\delta(\boldsymbol{r})$ denotes the three dimensional delta function and the quantity $\frac{d}{d r}(r \cdot~~)$ is the regularization operator.
Note that the $s$-wave scattering length in Eq.~\ref{peq2} depends on the relative collision energy $E$ in the pseudopotential, although in the ultracold limit this energy dependence is often negligible.

The waveguide symmetry implies that the transverse degrees of freedom in the scattering solutions can be expanded in terms of two-dimensional harmonic oscillator eigenstates in the potential $\frac{1}{2}\mu \omega_\perp^2 \rho^2$ and the \textcolor{black}{transverse part of the Laplacian in the Hamiltonian} $H$ (see Eq.~\ref{peq1}). The corresponding transverse eigenenergies are given by $E_{n m}=\hbar \omega_\perp(n +|m| +1)$, with $n=2 n_\rho=0,2,4,...$ being the quantum number associated with the nodes of the wave function in the $\rho$-direction and $m$ the azimuthal angular momentum. \textcolor{black}{This relative Hamiltonian $H$ possesses azimuthal symmetry, so $m$ is a good quantum throughout all the configuration space. Here we concentrate on the case $m=0$.}

\textcolor{black}{Remarkably, the prescription given in~\cite{demkovjetp1966} to regularize a divergent sum that arises in the derivation involves a very similar mathematical analysis as was used in the CIR treatment of ~\cite{olshanii1998} that results in a Hurwitz zeta function that is well-defined.} 

This means that the system interacts strongly at a finite value of the 3D scattering length, whereas in the absence of the confinement the two bosons exhibit a comparatively weak interaction.
This particular phenomenon is the so-called {\it confinement-induced resonance} (CIR).
The main feature of this effect is that the corresponding resonance condition, $a_s(E)/a_\perp=-1/c_1$, can be met either by tuning the trapping frequency or by adjusting the scattering length via a Fano-Feshbach resonance, \textcolor{black}{where $a_{\perp}=\sqrt{\mu \omega_{\perp}/\hbar}$.  }

\begin{figure}[h]
\includegraphics[scale=0.45]{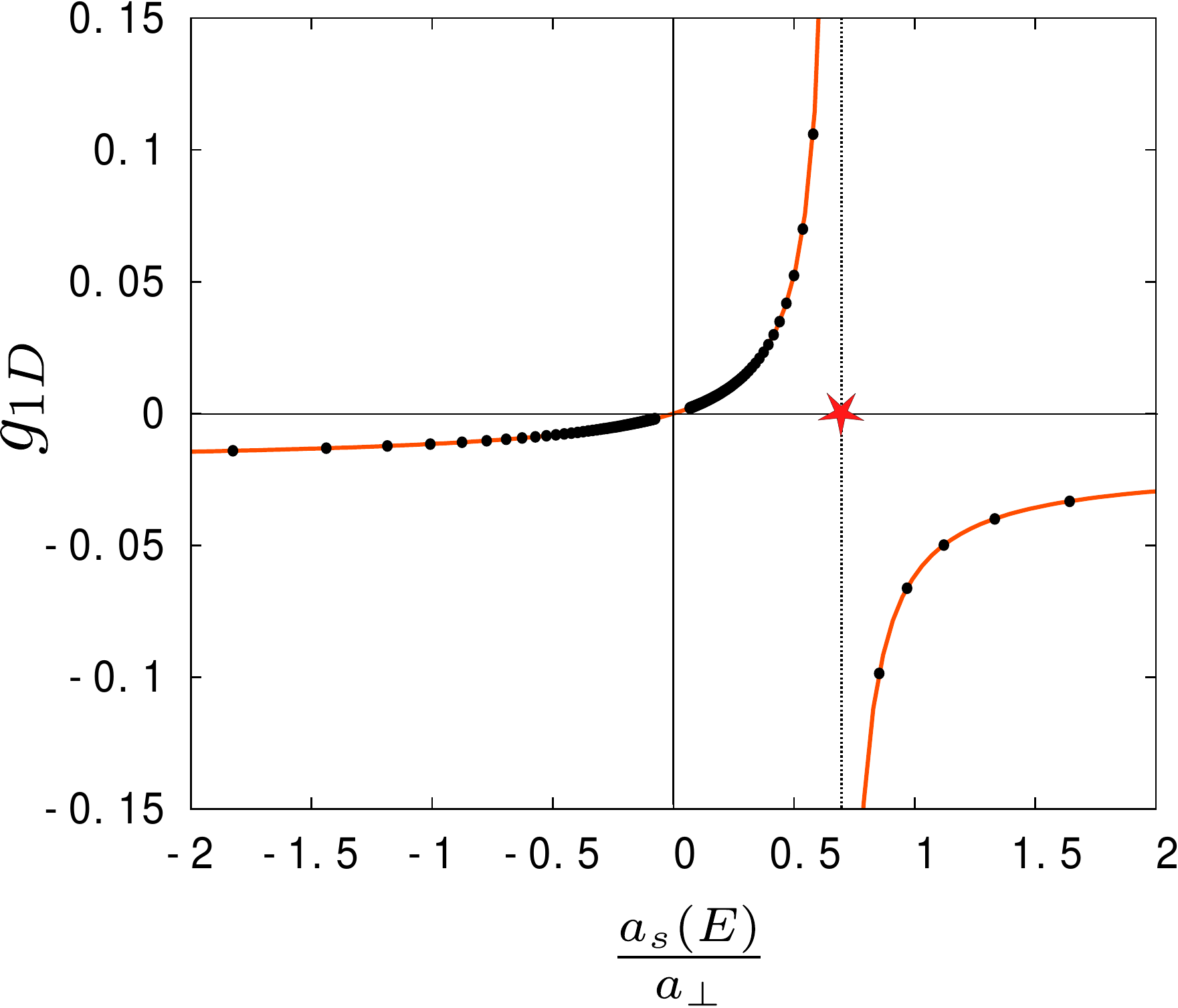}
\caption{(Color online) The one-dimensional coupling strength $g_{1D}$ (in units of $\frac{\hbar^2}{\mu a_\perp}$) as a function of the ratio $a_s(E)/a_\perp$.
The orange solid line corresponds to the analytical expression given in Eq.~(\ref{peq9}). The black dots correspond to full numerical calculations where the two-body interactions are modeled via a 6-10 potential. The red star denotes the position of the confinement-induced resonance.}
\label{pfig1}
\end{figure}

\textcolor{black}{
The treatment by~\cite{olshanii1998} showed that the full Hamiltonian in Eq.~(\ref{peq1}) can be mapped onto an {\it effective} one-dimensional Hamiltonian with a delta function interaction between the two bosons, of strength $g_{1D}$.
\begin{equation}
 H_{eff}=-\frac{\hbar^2}{2 \mu} \frac{d^2}{d z^2}+ g_{1D} \delta(z),~~~g_{1D}=\frac{\hbar^2}{\mu a_\perp}\frac{2a_s(E)}{a_\perp+c_1a_s(E)},
 \label{peq9}
\end{equation}
where the constant $c_1$ is given by 
\begin{equation}
c_1(k)=\zeta \bigg(\frac{1}{2}, 1-\frac{1}{4}\left({ka_{\perp}}\right)^{2}\bigg),
\label{peq7}
\end{equation}
with $\zeta(\cdot,\cdot)$ the Hurwitz zeta function and for $ka_\perp\ll1$ we have the value $c_1\approx-1.46035$. Here $k$ is the wavenumber associated with the total colliding energy $E$. }

\textcolor{black}{This effective two-body interaction derived from first principles  can be utilized directly in a many-body Hamiltonian, which permits an exploration of the underlying physics associated with the Tonks-Girardeau gas.
The effective Hamiltonian based on the coupling strength $g_{1D}$ encapsulates all the relevant scattering information in the full Hamiltonian as well as the non-trivial modifications due to the trap.
More specifically, as the ratio $a_s(E)/a_\perp$ tends to $-1/c_1$ the quantity $g_{1D} \to \infty $.
This particular feature is depicted in Fig.\ref{pfig1} by the vertical dotted line where the position of the CIR is indicated by the red star.
In Fig.~\ref{pfig1} the coupling constant $g_{1D}$ is shown as a function of the ratio $a_s(E)/a_\perp$.
The black dots correspond to the full numerical solution of the Hamiltonian given in Eq.~(\ref{peq1}) where the two body interactions are modeled by a 6-10 potential, {\it i.e.} $V(r)=C_{10}/r^{10}-C_6/r^6$.
The orange solid line denotes the analytical result of the coupling constant $g_{1D}$ given in Eq.~(\ref{peq9}) which agrees accurately with the numerical solution.
Furthermore, Fig.~\ref{pfig1} shows that as the ratio $a_s(E)/a_\perp$ tends to infinity the coupling constant tends asymptotically to a weakly attractive limit in the effective 1D potential energy. 
The resonant  3D free-space two body interactions, on either the repulsive or attractive side, are modified into weakly attractive 1D forces due to the trapping geometry.
This attractive force in the waveguide is determined solely by the $c_1(k)$ constant and the oscillator length $a_\perp$.
Indeed, for $a_s(E)/a_\perp \to \infty$ the coupling strength is $g_{1D}\to \frac{\hbar^2}{2 \mu a_\perp c_1}$.}

\begin{figure}[h]
\includegraphics[scale=0.5]{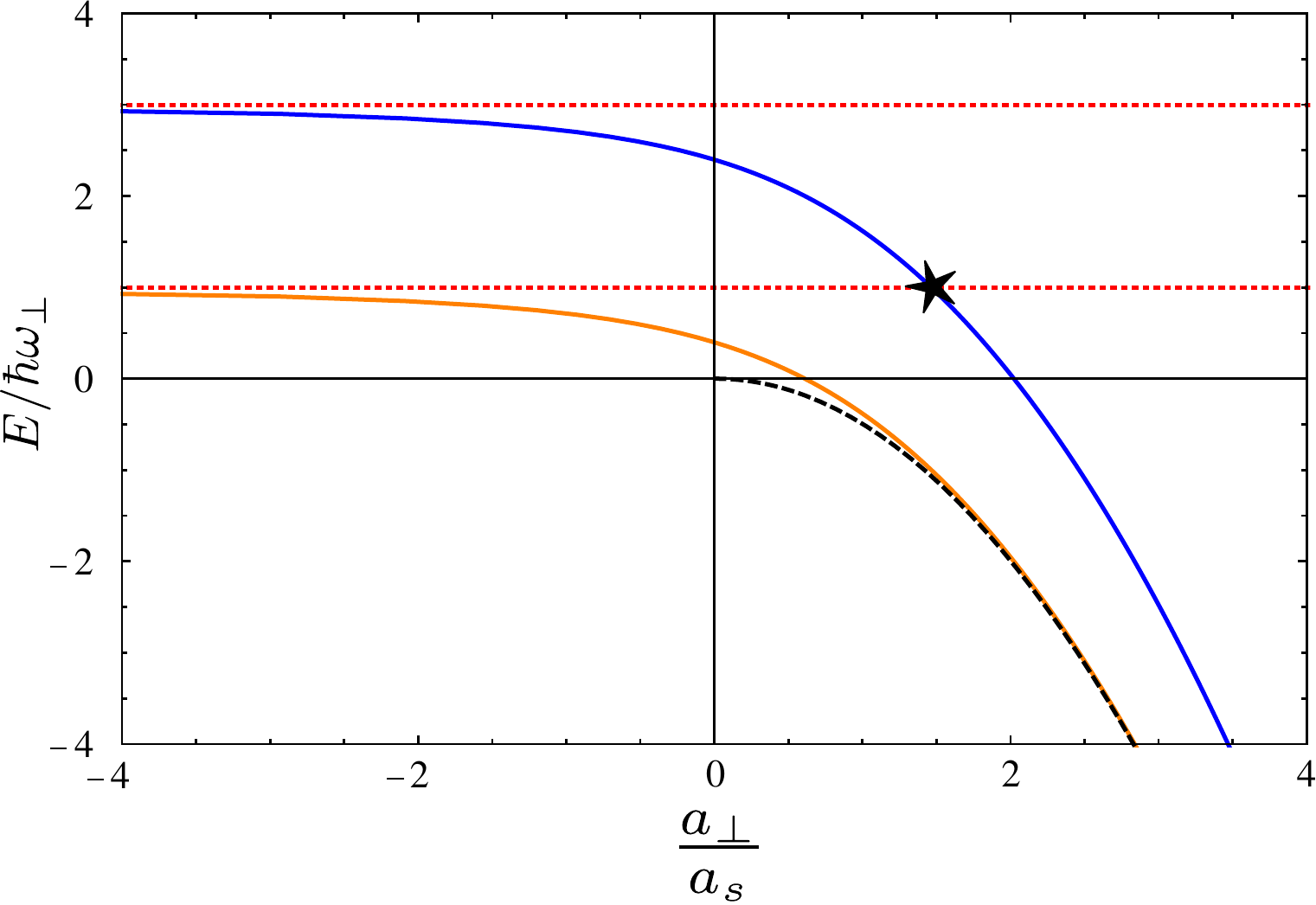}
\caption{(Color online) An illustration of the physical origin of confinement-induced resonances. Bound state energies in units of $\hbar \omega_\perp$, i.e.$E/ \hbar \omega_\perp$ are depicted versus the ratio $a_\perp/a_s$. The red dotted lines indicate the energy levels of the unperturbed two-dimensional harmonic oscillator, which act here as channel threshold energies. The black star denotes the position of the confinement-induced resonance relevant in the ultracold limit. The black dashed curve illustrates the $s$-wave bound state in the absence of the trapping potential. The solid orange curve corresponds to the molecular confinement-induced resonance. The blue curve is the CIR bound state energy supported by all the closed channels. }
\label{pfig2}
\end{figure}

\textcolor{black}{
Qualitatively, the CIR can be viewed as a bound state supported by all the closed channels whose energy coincides with the lowest channel threshold, as in a usual Fano-Feshbach resonance~\cite{bergemann2003}.
This multichannel bound state produces only a single pole in the reactance operator, i.e. the tangent of the 1D even $z$-parity phase shift.
To follow up on this idea, one can obtain the binding energies of the closed channel supported bound state from the following transcendental equation, and its solution is the CIR resonance condition:
\begin{equation}
 \frac{a_\perp}{a_s}= -\zeta(\frac{1}{2}, \frac{3}{2}-\frac{E_r}{2\hbar \omega_\perp}),
 \label{peq16}
\end{equation}
where $\zeta(\cdot,\cdot)$ is the Hurwitz zeta function.
Note that the appropriate value of $E_r$ in the ultracold limit are in the range of few tens of kHz.  On the other hand, the theoretical treatment suggests that at energies below the threshold of the open channel {\it{confinement-induced molecular}} states might be supported.
Eq.(\ref{peq17}) is derived by requiring {\it all} the scattering channels to be closed yielding a wavefunction that vanishes asymptotically.    
In this case the molecular confinement-induced energies obey the following transcendental equation:
\begin{equation}
 \frac{a_\perp}{a_s}=-\sqrt{\frac{2\hbar\omega_\perp}{\hbar\omega_\perp-E_r}}- \zeta(\frac{1}{2}, \frac{3}{2}-\frac{E_r}{2\hbar \omega_\perp}).
 \label{peq17}
\end{equation}
}

Fig.~\ref{pfig2} illustrates the relations in Eqs.~(\ref{peq16}) and (\ref{peq17}), where the corresponding closed channel bound state (blue solid curve) and molecular CIR state (orange solid curve) energies are given as a function of the ratio $a_\perp/a_s$.
The red dashed lines indicate the two dimensional harmonic oscillator eigenenergies in the absence of short range interactions, i.e. the values $E^{(1)}=\hbar \omega_\perp$ and $E^{(2)}=3\hbar \omega_\perp$.
The black dashed curve corresponds to the energy of an $s$-wave molecular pair in the absence of confining trap, i.e. $E_{free}/\hbar \omega_\perp=-\frac{a^2_\perp}{2 a_s^2}$. 
We observe that the closed channel bound state (blue solid curve) tends to $3\hbar \omega_\perp$ as the ratio $a_\perp/a_s\to -\infty$, i.e.in the absence of short range interaction.
As the ratio $a_\perp/a_s$ approaches the value $-c_1=1.46035$ the bound state from the closed channels becomes degenerate with the threshold of the open channel, $E^{(1)}=\hbar \omega_\perp$ (see black star in Fig.\ref{pfig2}), and not coincidentally,
the CIR resonance occurs at $a_\perp/a_s=-c_1=1.46035$.  Note that at $a_{\perp}/a_s>1.46035 $ and for energies less than $E^{(1)}$ the depicted blue line does not have any physical significance and it is just an analytic continuation of Eq.(65). In this energy regime all the channels should be closed and Eq. (65) refers to the case where the system possess a single open channel.

\textcolor{black}{
This analysis illustrates that the CIR has the character of a Fano-Feshbach resonance, where the collective bound state attached to all the closed channels lies in the low-energy continuum of the open channel.
In addition, the confinement-induced molecular state (see orange solid line) exists regardless the strength of the short range interactions. 
This is a manifestation of the impact of the confinement since in free space collisions the bosonic pair forms a weakly bound state only for positive values of the $s$-wave scattering length.
On the positive side of the abscissa in Fig.~\ref{pfig2} as the ratio $a_\perp/a_s$ increases we observe that the energy of free space weakly bound molecule (black dashed line) coincides with the energy of the confinement-induced molecular state (orange solid line).
This occurs at these values of scattering length since the two-body potential is deep enough forcing the wavefunction to vanish before the trapping potential becomes important.
Therefore, in this limit the confinement-induced molecular state behaves as a free space bound state.
Note that similar behavior is observed in the paper by ~\cite{demkovjetp1966}.
}

\begin{figure}[h]
\includegraphics[scale=0.5]{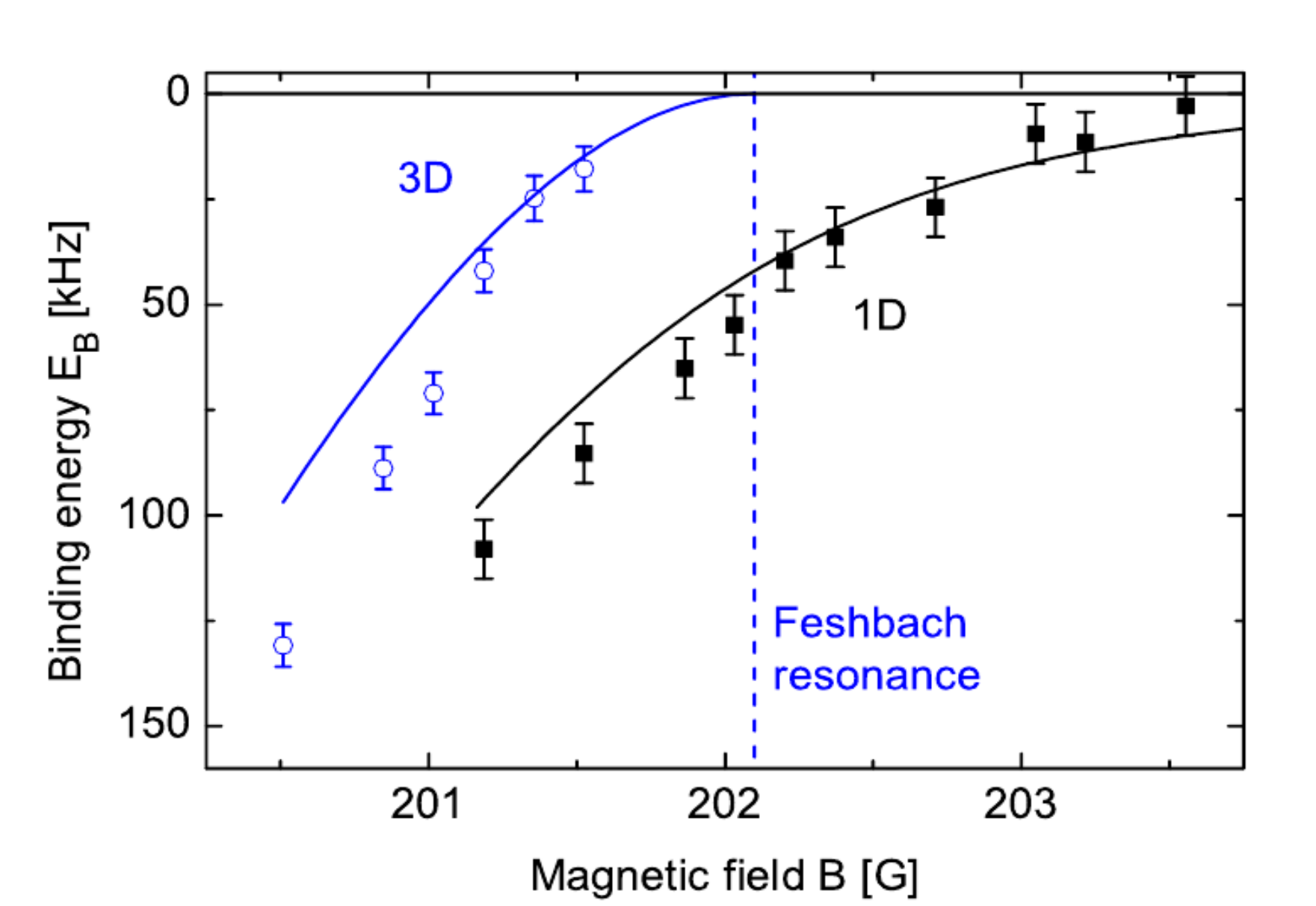}
\caption{(Color online) Confinement-induced (black squares and lines) and Feshbach (blue line and circles) molecules.
The solid lines correspond to the theoretical predictions whereas the circles and squares indicate the experimental measurements.
The vertical dashed line indicates the position of the Feshbach resonance.
{\bf taken from \cite{moritz2005}}
}
\label{pfigmor1}
\end{figure}

\begin{figure}[h]
\includegraphics[scale=0.45]{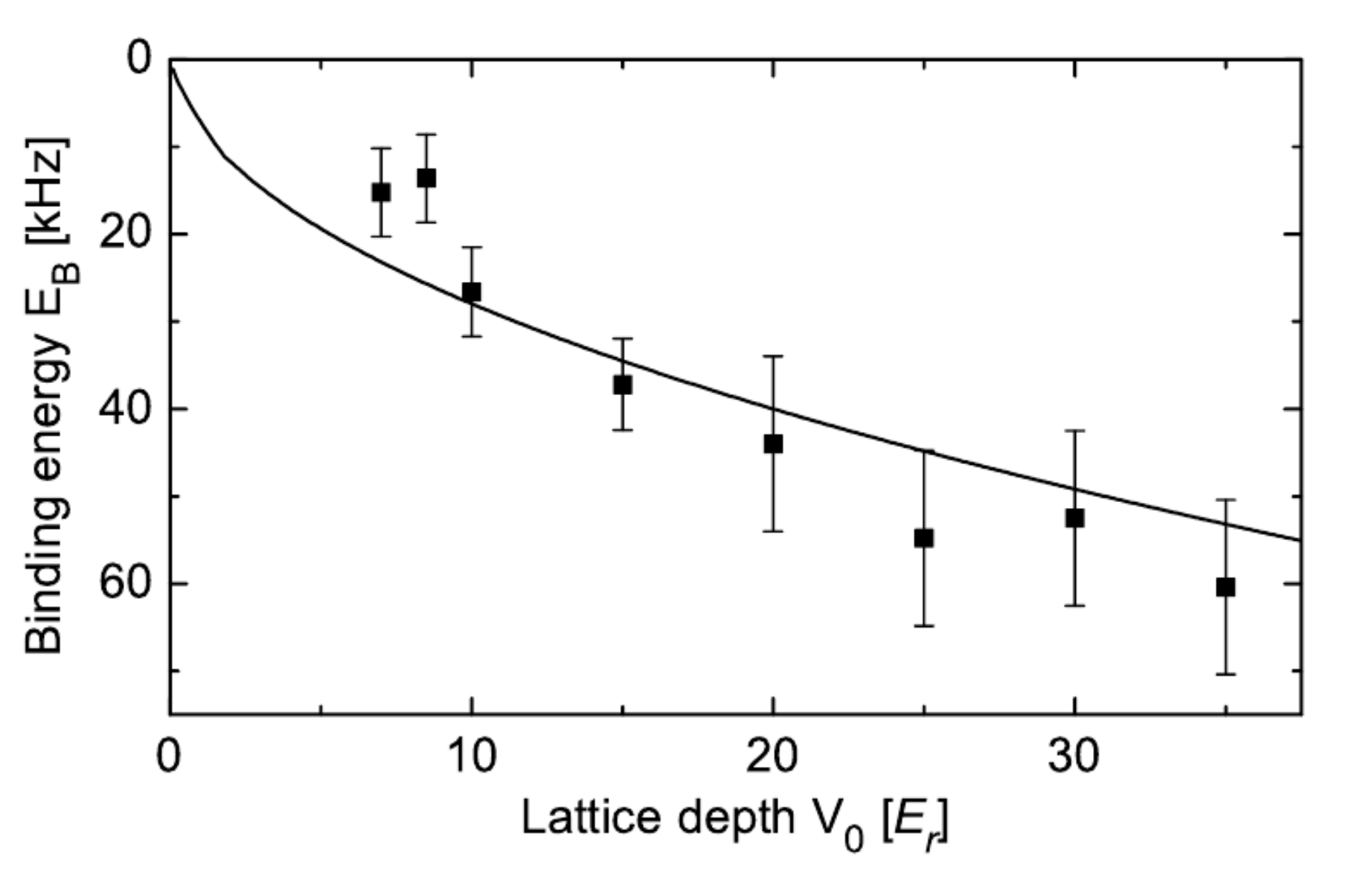}
\caption{The binding energy of the confinement-induced molecules as a function of the lattice depth $V_0$ in recoil units ($E_r$). The spectra are measured very close to the Feshbach resonance at magnetic field $B=202$G.
The black solid line indicates the corresponding theoretical calculations.
Taken from ~\cite{moritz2005}.
}
\label{pfigmor2}
\end{figure}

An experiment by \cite{moritz2005} considered a Fermi gas of $^{40}$K atoms in the presence of harmonic confinement.
The $^{40}$K atoms are prepared in two hyperfine states $\ket{F=9/2,m_F=-9/2}$ and $\ket{F=9/2, m_F=-7/2}$.
Note that the third hyperfine $\ket{F=9/2,m_F=-5/2}$ is not populated initially.
The mutual interactions are tuned via a Feshbach resonance whose position is located at $B=202.1$G and its zero-crossing is at $B=210$G.
Thereafter, by employing radio-frequency spectroscopy confinement-induced molecules are generated and their binding energy is measured as a function of the scattering length and the confinement frequency.
In Fig.~\ref{pfigmor1} the binding energies of the confinement-induced (black solid line and squares) and the Feshbach (blue solid line and circles) molecule are depicted as a function of the magnetic field.
The solid lines denote the theoretical predictions of~\cite{dickerscheid2005feshbach} whereas the scattered data are the experimental measurements.
The blue dashed line indicates the position of the Feshbach resonance where a Feshbach molecule (blue line and circles) is only formed on the positive side of the resonance whereas no measurement occurs on the negative side.
On the other hand, in the case of confinement-induced molecule (black line and squares) we observe that there is always a bound state regardless the sign of the $s$-wave interactions.
This is in accordance with the theoretical predictions.
In addition, we may note that the intersection point of the black line with the blue dashed line, i.e.at the position of infinite scattering length, the binding energy acquires its universal value, {\it i.e.}, $E_B\approx 0.6 \hbar \omega_\perp$.
At this universal value the binding energy of the confinement-induced molecules depends solely on the strength of the confinement.
Note that the strength of the confinement in~\cite{moritz2005} is tuned by changing the lattice depth $V_0$.
Fig.~\ref{pfigmor2} depicts the binding energy of the confinement-induced molecule as a function of the confinement strength in units of recoil energy $E_r$.
The black solid line refers to theoretical calculations and the scatter data indicate the radio-frequency measurements.
Both theory \cite{dickerscheid2005feshbach} and experiment show sufficient agreement.
The minor disagreements between theory and experiment in Figs.~\ref{pfigmor1} and \ref{pfigmor2} associated with the effective range corrections which are not included as was pointed out by\cite{peng2012two}

In the above mentioned analysis of $s$-wave confinement induced resonances it is evident that zero-range approximations are employed.
This means that the short-range part of the Hamiltonian is treated in essence as a single channel.
In experiments, however, (see Fig.~\ref{pfigmor1}) the main toolkit to tune the interactions or the $s$-wave scattering length are the Feshbach resonances.
This particular aspect implies that the short-range part must be treated as a two-channel model in order to obtain a direct comparison with the corresponding experimental advances.
Towards this pathway a tremendous amount of theoretical effort is focused in order to incorporate adequately the two-channel nature of Feshbach resonances in the confinement-induced physics~\cite{yurovsky2005feshbach,yurovsky2006properties,peng2012two,kristensen2015ultracold,saeidian2012}.
All these works pointed out the importance of the effective range corrections particularly on the calculation of the binding energy of confinement-induced molecules.
In addition, it was shown that the effective range corrections become more important for narrow Feshbach resonances.
Note that similar conclusions were drawn also for fermionic species in harmonic waveguides~\cite{saeidian2015shifts}.

Another aspect which was excluded from Refs.~\cite{olshanii1998,bergemann2003,demkovjetp1966} is that the total colliding energy is sufficient for the pair atoms such that no excitation will occur before and after the collision.
Lifting this constraint, {\it i.e.}, going beyond the single mode regime~\cite{moore2004,saeidian2008,hess2015analytical} predicted numerically~\cite{saeidian2008} and analytically~\cite{moore2004,hess2015analytical} the emergent inelastic confinement-induced resonances for bosonic and fermionic exchange symmetries.
In particular,\cite{hess2015analytical} considered also the higher partial wave interactions going beyond $s$- and $p$-wave interactions and obtained universal expressions for the position of all the inelastic confinement-induced resonances.

\textcolor{black}{A study by ~\cite{kim2006}} studied distinguishable particle collisions in the presence of a harmonic waveguide which yield the effect of {\it dual} confinement-induced resonances.
This type of resonances correspond to total transmission due to the destructive interference of $s$- and $p$- partial waves.
The importance of high partial waves on bosonic or fermionic systems in a cigar shaped trap were considered by~\cite{giannakeas2012}.
Due to the anisotropy of the trap all the partial waves associated with either bosonic or fermionic exchange symmetry are coupled yielding in this manner coupled $\ell$-wave confinement-induced resonances.
The analysis of this particular system is based on the idea of the local frame transformation.
This framework was employed for highlighting the underlining physics of fermionic collisions in matter waveguides by~\cite{granger2004PRL} avoiding the complications of zero-range and two-channel models.
\textcolor{black}{The following focuses on a system of spin-polarized fermions in the presence of cigar-shaped traps and the underlying details of the local frame transformation theory.}

\subsubsection{Fermions in a cigar-shaped trap}
It is also of extensive experimental and theoretical interest to explore near-degenerate fermionic gases in low dimensional traps, which requires a detailed understanding of confinement-induced resonances between identical fermions.
In~\cite{granger2004PRL},  a scattering theory was developed to describe collisions between identical spin-polarized fermions in the presence of an axially symmetric harmonic trap.
Owing to the Pauli exclusion principle, ultracold fermions interact in 3D with $p$-wave interactions instead of $s$-wave that was considered in the previous subsection.
The $p$-wave theory requires somewhat different considerations, which we therefore consider in this subsection from the general viewpoint of the $K$-matrix theory; this highlights the behavior imposed on spin-polarized fermionic ensembles by the transverse trapping potential.

Again, the relative Hamiltonian for two spin-polarized fermions expressed in cylindrical coordinates has the same form as in given in Eq.~(\ref{peq1}).
In contrast with the previous $s$-wave treatment, the spherically symmetric two-body interaction is not modeled by a pseudopotential.  Instead the formulation works directly with the short-range phase shift caused by the spherical symmetric two-body atomic interaction.  In the following, azimuthal symmetry is assumed and our analysis is restricted to only 
$m=0$.

In practice the length scales associated with the two potential terms in Hamiltonian $H$ are well separated, with $r_0$ of the short range potential typically orders of magnitude smaller than the waveguide potential oscillator length in a typical ultracold experiment, $a_\perp=\sqrt{\hbar/(\mu \omega_\perp)}$.
At small interparticle distance $r<< a_\perp$, the orbital angular momentum is approximately conserved and therefore the two fermions experience a free-space collision with total colliding energy $E=\hbar^2 k^2/(2 \mu)$.
As usual, $\mu$ denotes the two-body reduced mass.
In this case the $\ell'$-th linearly independent energy eigenstate expressed in spherical coordinates has the following form at $r_0<<r<< a_\perp$:
\begin{equation}
\Psi_{\ell'}(\mathbf{r})=\sum_{\ell}F_{\ell}(r,\theta)\delta_{\ell \ell'}-G_{\ell}(r,\theta)K^{3D}_{\ell\ell'},
\label{peq18}
\end{equation}
where $F_{\ell}(r,\theta)$ ($G_{\ell}(r,\theta)$) is the energy normalized regular (irregular) solution expressed in terms of spherical Bessel $j_{\ell}(r)$ (spherical Neumann $n_{\ell}(r)$) functions multiplied by the corresponding spherical harmonic $Y_{\ell,m=0}(\theta,\phi)$. 
The summation is performed over all odd $\ell$ angular momentum due the Pauli exclusion principle.
The quantity $K^{3D}_{\ell\ell'}$ represents the elements of the reaction matrix $\underline{K}^{3D}$ in three-dimensions.
This $K$-matrix incorporates all the scattering information due to the short-range potential $V_{sh}(\mathbf{r})$, and for a spherically symmetric potential it is diagonal, but for anisotropic interactions such as the dipole-dipole type it could acquire off-diagonal elements in other contexts~\cite{giannakeas2013prl}.

At large distances the waveguide geometry prevails and imposes cylindrical symmetry on the wave function, and 
the total collision energy gets apportioned between the transversal and longitudinal degrees of freedom. 
The energy can be expressed at $|z|>r_0$ as $E= \hbar \omega_\perp(2 n+|m|+1) + \hbar^2 q_n^2/(2 \mu)$, where the term $\hbar \omega_\perp(2 n+|m|+1) $ refers to the energy of the transversal part of the Hamiltonian and $q_n$ is the channel momentum, i.e.is the momentum of the particles in the $z$ direction.
In this region the $n'$-th linearly-independent scattering wave function at energy $E$ can be expressed in cylindrical coordinates as:
\begin{equation}
\Psi_{n'}(\mathbf{r})=\sum_nf_{n}(z,\rho)\delta_{nn'}-g_n(z,\rho)K^{1D}_{nn'},
\label{peq19}
\end{equation}
where the quantity $K^{1D}_{nn'}$ represents the elements of the quasi-1D reaction matrix $\underline{K}^{1D}$ and where $(f_n(z,\rho),~g_n(z,\rho))$ are the energy normalized (regular, irregular) standing wave solutions solely in the presence of the trap. 
The specific form of the regular and irregular solutions which obey the Pauli exclusion principle have odd $z$-parity for $m=0$ and are given by:

\begin{equation}
  \bigg( \begin{array}{l l}
    f_n(z,\rho)\\
    g_n(z,\rho)\\
\end{array}
\bigg)
= (2\pi^2 q_n)^{-1/2} \Phi_n(\rho)\left \{
    \begin{array}{l l}
%

  \bigg(\begin{array}{l l}
      \sin q_nz  \\
      -\frac{z}{|z|}\cos q_nz \\
  \end{array}\bigg)
\end{array}\right.
\label{peq20}
\end{equation}
where $\Phi_n(\rho)$ are the $m=0$ eigenfunctions of the two-dimensional harmonic oscillator and the $z$-dependence describes motion in the unbounded coordinate. For collisions of spin-polarized fermions the factor $z/|z|$ corresponds to the anti-symmetrization operator.

From the Hamiltonian $H$ in Eq.~(\ref{peq1}) it is evident that the corresponding Schr\"odinger equation is non-separable over all the configuration space.
However, as mentioned above there are two distinct subspaces where the resulting Schr\"odinger equation is separable and where all the relevant scattering information can be expressed in terms of reaction matrices, namely $K^{3D}$ [see Eq.~(\ref{peq18})] and $K^{1D}$ [see Eq.~(\ref{peq19})].
The main idea is to define a frame transformation which will permit to express the  $K^{1D}$ reaction matrix in terms of the short range  $K^{3D}$.
Intuitively, the frame transformation $U$ permits us to propagate outwards to the asymptotic region the information of the collisional events occurred close to the origin.

An unusual property of this frame transformation is that it is not {\it{unitary}}. This arises due to the fact that the solutions given in Eq.~(\ref{peq18}) and Eq.~(\ref{peq19}) obey different Schr\"odinger equations.
However, since Hamiltonian $H$ in Eq.~(\ref{peq1}) possesses length scale separation implying the existence of an intermediate region where both potentials are negligible.
This means that in this subspace Eqs.~(\ref{peq18}) and (\ref{peq19}) approximately satisfy the same Schr\"{o}dinger equation, i.e.the Helmholtz equation.
Therefore, in this {\it{Helmholtz}} region one employs {\it{locally}} the above mentioned frame transformation.
The concept of the {\it{local frame transformation}} was introduced by \cite{fano1981,harmin1982,harmin1982prl}
and extended by~\cite{greene1987,wong1988,granger2004PRL,robicheauxschwinger2015,giannakeas2016pra,zhang2013}.

The local frame transformation is derived by matching the energy normalized regular solutions $f_n(\boldsymbol{r})$ and $F_\ell(\boldsymbol{r})$ on a surface $\sigma$ at a finite distance $r_0<r<a_\perp$ inside the Helmholtz region.
Formally, for $r<a_\perp$ $U$ obeys the relation
\begin{equation}
 f_n(\boldsymbol{r})=\sum_\ell F_\ell(\boldsymbol{r})U^T_{\ell n},~~{\rm{with}}~~U^T_{\ell n}=\braket{\braket{F_\ell|f_n}},~~{\rm{for}}~~r<a_\perp,
 \label{peq21}
\end{equation}
where $U^T$ denotes the transpose of the frame transformation matrix $U$, the symbol $\braket{\braket{\cdot|\cdot}}$ indicates that the solutions $F_\ell$ and $f_n$ are integrated only over the solid angle $\Omega=(\theta,\phi)$.
Note that the matrix elements $U$ are independent of the distance $r$.
This occurs since the matching of the regular solutions takes place in the Helmholtz region where both solution possess same $r$ dependence.
In addition, since the set of solutions in Eqs.~(\ref{peq18}) and (\ref{peq19}) are real standing-wave solutions, the matrix $U$ is real. 

It was pointed out by \cite{fano1981} that the irregular parts of Eqs.~(\ref{peq18}) and (\ref{peq19}) can be interconnected by matching in the Helmholtz region the corresponding {\it{principal}} value Green's functions, written in the different coordinate systems.
Formally, the irregular solutions obey:
\begin{equation}
 g_n(\boldsymbol{r})=\sum_\ell G_\ell(\boldsymbol{r})[U^{-1}]_{\ell n},~~{\rm{for}}~~r<a_\perp.
 \label{peq22}
\end{equation}

After inserting Eqs.~(\ref{peq21}) and (\ref{peq22}) into the scattering wave function $\Psi (\boldsymbol{r})$ in Eq.~(\ref{peq18}), the $K^{1D}$ matrix is seen to be expressed in terms of the short range $K^{3D}$ compactly as
\begin{equation}
 K^{1D}=UK^{3D}U^T,
 \label{peq23}
\end{equation}
where the short range $K^{3D}$ includes all the odd partial waves for the spin-polarized fermions.
Also, the $K^{1D}$ matrix depends on the total collision energy $E$.

The $K^{1D}$ matrix contains information about the asymptotically open- and closed-channel components of the wavefunctions.  
To describe the closed-channel components, solutions given in Eq.~(\ref{peq19}) can be analytically continued by setting the channel momentum $q_n$ to be $q_n\to i|q_n|$, in the usual spirit of quantum defect theory.
Then one can derive the local frame transformation $U$ which possesses the same functional form as for the open channels. 
Note that the local frame transformation $U$ in the open channels only for $p$-wave interactions obeys the relation $U_{\ell=1 n}=\frac{\sqrt{2}}{a_{\perp}}\sqrt{\frac{3}{kq_n}}P_{\ell=1}(\frac{q_n}{k})$ where $P_\ell(\cdot)$ indicates the $\ell$-th Legendre polynomial.
This yields the same expression for $K^{1D}$ shown in Eq.~(\ref{peq23}).
The only drawback from these manipulations is that the resulting $K^{1D}$ matrix corresponds to a scattering solution which does not (yet) obey the proper boundary conditions asymptotically.
This is because the closed channel parts of the wave function given in Eq.~(\ref{peq19}) contains exponentially growing pieces at $|z| \rightarrow \infty$.
One sees readily after substituting $q_n\to i |q_n|$ that the regular and irregular solutions in Eq.~(\ref{peq20}) for the closed channel components have both exponentially decaying and growing pieces.
Therefore, in order to enforce the physically accepted asymptotic boundary conditions in Eq.~(\ref{peq19}) concepts from {\it{multichannel quantum defect theory}} are employed~\cite{aymar1996RMP}.

Initially, the scattering wave function in Eq.~(\ref{peq19}) is separated into open (``o``) and closed (''c``) channels.
Then, linear combinations are chosen by demanding that the exponentially growing pieces in the closed channels are canceled.
Formally, we have the following relation for the wave function:

\begin{eqnarray}
\Bigg(\begin{matrix}
	  \Psi_{oo} & \Psi_{oc}  \\[0.3em]
	  \Psi_{co} &  \Psi_{cc} \\[0.3em]
      \end{matrix}
\Bigg)
 \Bigg(\begin{matrix}
	  B_{oo}    \\[0.3em]
	  B_{co}  \\[0.3em]
      \end{matrix}
\Bigg)
&=& \nonumber \Bigg[\Bigg(\begin{matrix}
	  f_{o} & 0           \\[0.3em]
	  0 &  f_{c} \\[0.3em]
      \end{matrix}
\Bigg)
-
 \Bigg(\begin{matrix}
	  g_{o} & 0           \\[0.3em]
	  0 &  g_{c} \\[0.3em]
      \end{matrix}
\Bigg)\cr
&~& \Bigg(\begin{matrix}
	  K_{oo}^{1D} & K_{oc}^{1D}           \\[0.3em]
	  K_{co}^{1D} & K_{cc}^{1D} \\[0.3em]
      \end{matrix}
\Bigg)\Bigg]
 \Bigg(\begin{matrix}
	  B_{oo}    \\[0.3em]
	  B_{co} 	\\[0.3em]
      \end{matrix}
\Bigg),
\label{peq24}
\end{eqnarray}
where the matrices $B_{oo}$ and $B_{co}$ denote the linear combination coefficients.
By eliminating the closed channels the linear combination coefficients acquire the values
\begin{equation}
 B_{oo}= \mathds{1} ~~{\rm{and}}~~ B_{co}=\bigg(\frac{f_c}{g_c}-K^{1D}_{cc}\bigg)^{-1}K^{1D}_{co},
 \label{peq25}
\end{equation}
where the term $f_c/g_c \xrightarrow{|z|\to \infty}-i \mathds{1}$.

The physical wave function, which involves only open channels asymptotically since the closed-channel components decay exponentially, acquires the following form at $|z|\rightarrow \infty$:
\begin{eqnarray}
 \Psi^{phys}&=& f_o-g_o[K_{oo}^{1D}+iK^{1D}_{oc}(\mathds{1}-iK^{1D}_{cc})^{-1}K^{1D}_{co}].
\label{peq26}
\end{eqnarray}
Here the effects of the closed channels on the open channel scattering are included in the corresponding {\it physical} $K$-matrix, given by:
\begin{equation}
K_{oo}^{1D,~phys}\equiv K_{oo}^{1D}+iK^{1D}_{oc}(\mathds{1}-iK^{1D}_{cc})^{-1}K^{1D}_{co}.
\label{peq27}
\end{equation}
The resonances of the collision complex appear as poles of the $K_{oo}^{1D,~phys}$ matrix.
More specifically, the $K_{oo}^{1D,~phys}$ exhibits resonant features at zero eigenvalues of the matrix $(\mathds{1}-iK^{1D}_{cc})$.
Therefore, this argument can be recast into the form of a determinantal equation:

\begin{equation}
 det(\mathds{1}-iK_{cc}^{1D})=0.
\label{peq28}
\end{equation}

Note that despite the appearance of the imaginary unit $i$ in the preceding equations, all of these physical wavefunctions and reaction matrices are real.  When all channels are closed, the roots of Eq.~(\ref{peq28}) yield the bound state energies.  When one or more channels are energetically open, the roots of the above closed-channel determinant approximately identify the real parts of resonance energies.
Eq.~(\ref{peq28}) shows why a confinement-induced resonance can be viewed as a Fano-Feshbach type of resonance.

Consider next the situation where the two fermions collide in the single mode regime, meaning that the relative collision energy lies between the lowest two transverse thresholds.
In addition, these ultracold spin-polarized fermions interact at short distances via $p$-wave interactions only.  Phase shifts associated with higher 3D partial waves are entirely neglected.
Accordingly, the determinantal equation has one nonzero root, and the $K^{1D,~phys}_{oo}$ matrix takes the following form:

\begin{eqnarray}
 K_{oo}^{1D,~phys}&=&K_{oo}^{1D}+iK^{1D}_{oc}\big( \mathds{1}+\frac{i}{1-i\gamma}K^{1D}_{cc}\big)^{-1}K^{1D}_{co} \label{peq30}
\end{eqnarray}
where $\gamma=Tr(K^{1D}_{cc,\ell})$.
Note that the $Tr(K^{1D}_{cc,\ell})$ is an infinite sum which formally diverges and in order to obtain a meaningful answer an auxiliary regularization scheme is employed.
In the particular case a Riemann zeta function regularization scheme is used.
Note that such techniques are totally avoided in the generalized form of the local frame transformation theory \cite{robicheauxschwinger2015,giannakeas2016pra}.
In addition, Eq.~(\ref{peq30}) is further simplified by substituting $(K^{3D})_{\ell' \ell''}=\tan \delta_{\ell=1}(E) \delta_{\ell',1} \delta_{\ell'',1} $, i.e. using the fact that only $p$-wave phase shifts $\delta_{\ell=1}(E)$ are non-zero.
Next, in Eq.~(\ref{peq30}) we substitute the corresponding local frame transformation for the closed channels only $U_{\ell=1 n}=\frac{\sqrt{2}}{a_{\perp}}\sqrt{\frac{3}{ik|q_n|}}P_{\ell=1}(\frac{i |q_n|}{k})$ where $P_\ell(\cdot)$ indicates the $\ell$-th Legendre polynomial.
Eq.(\ref{peq30}) now reads
\begin{equation}
 K_{oo}^{1D,~phys}= - \frac{6 V_p}{a_\perp^3} q_0 a_\perp \bigg [1-12\frac{V_p}{a_\perp^3} \zeta(-\frac{1}{2},\frac{3}{2}-\frac{E}{2\hbar \omega_\perp}) \bigg  ]^{-1},
\label{peq31}
\end{equation}
where the terms inside the square brackets provide the resonance 
condition for the position of the {\it{p-wave confinement-induced resonances}}. 
The term $V_p$ denotes the energy-dependent 3D scattering volume 
which is defined by $V_p(E)=-\tan \delta_{\ell=1}(E)/k^3$ ($k=\sqrt{2 \mu E/\hbar^2})$.

\textcolor{black}{The quantity of greatest experimental relevance is 
the effective interaction strength between two 1D spin-polarized fermions 
which contains the corresponding $p$-wave confinement-induced physics.
As was shown in \cite{grirardeaupra2004,pricoupenko2008prl,kanjilalpra2004} 
this effective 1D interaction is related to the corresponding $K$-matrix 
[see Eq.~(\ref{peq31})] according to the following relation:}
\begin{equation}
 g_{1D}^-=-\frac{\hbar^2 a_\perp}{\mu q_0} K_{oo}^{1D,~phys}.
 \label{pggfer}
 \end{equation}
\textcolor{black}{This coefficient controls the strength of effective 
zero-range pseudopotential that is relevant for describing the interaction 
of identical fermions in 1D in both few-body and many-body contexts, namely:}
\begin{equation}
 V^{\rm pseudo}(z)=  g_{1D}^- \overleftarrow{\frac{d}{dz}} \delta(z) \overrightarrow{\frac{d}{dz}}.
 \label{pseudofermionpot}
 \end{equation}
\textcolor{black}{The left (or right) arrow indicates that the derivative 
operator acts on the bra (or ket) respectively.
In the idealized limit of a zero-range potential, this pseudopotential produces 
a discontinuous wavefunction that obeys the required antisymmetry of the 
identical fermion wavefunction.  This might seem problematical 
since one normally requires wavefunctions in Schr\"odinger wave mechanics 
to be continuous, but it can be accommodated theoretically as is discussed, 
for instance, by ~\cite{Cheon1999prl}.}

\begin{figure}[t!]
\includegraphics[scale=0.55]{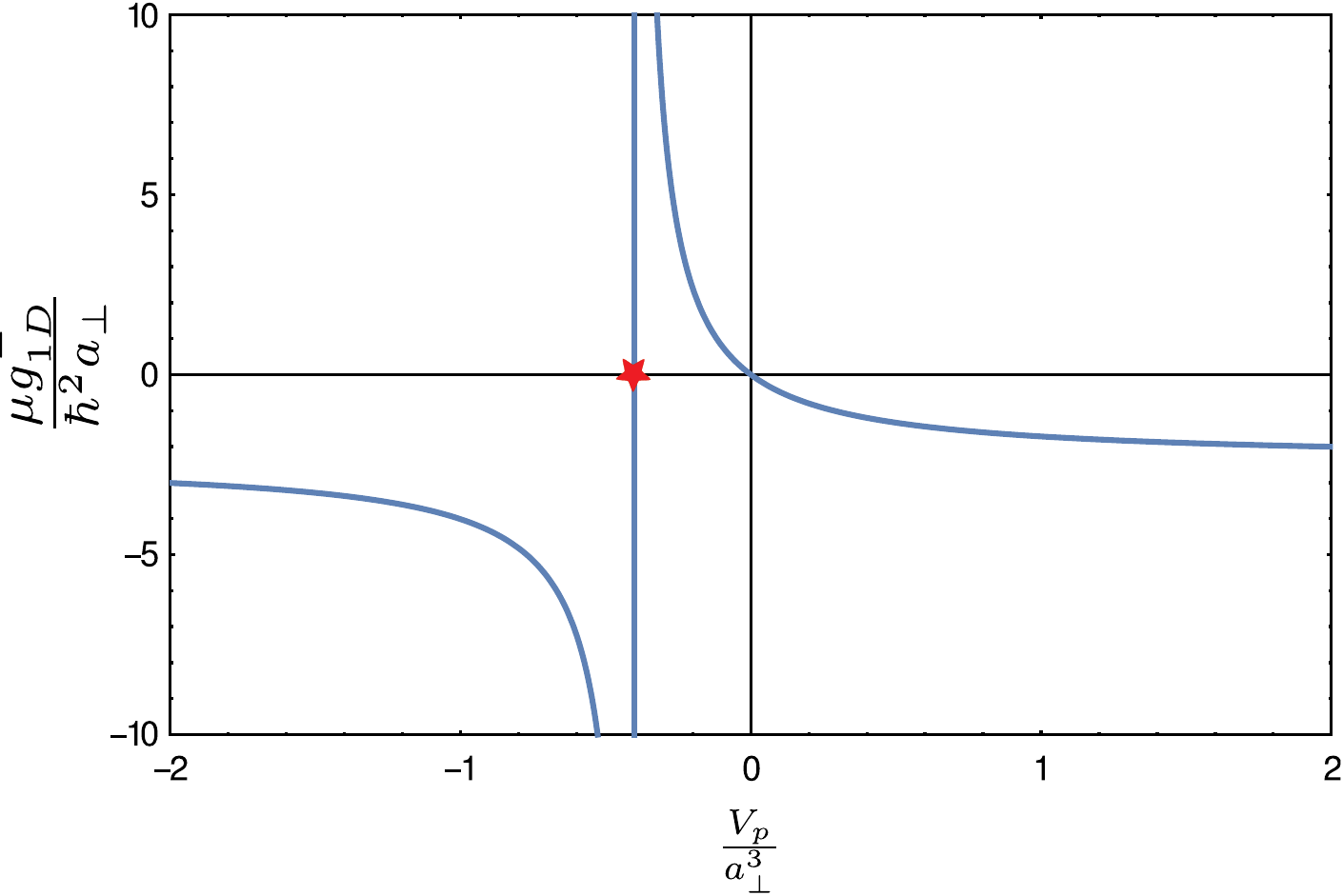}
\caption{(Color online) The low-energy effective interaction $g_{1D}^-$ 
for two spin-polarized fermions in the presence of a harmonic confinement 
as a function of $V_p/a_\perp^3$. Red star denotes the position of the 
$p$-wave confinement-induced resonance.}
\label{pfigg1dfer}
\end{figure}

\textcolor{black}{Fig.\ref{pfigg1dfer} illustrates the dependence of the 
effective coupling constant $g_{1D}^-$ as a function of the $V_p/a_\perp^3$ in the low-energy regime.
 The two fermions interact strongly at the position of $p$-wave 
confinement-induced resonance (see the red star in Fig.\ref{pfigg1dfer}) 
whereas in the limit of $V_p/a_\perp^3\to0$ the corresponding effective 
interaction vanishes.
Interestingly, for strong $p$-wave interactions, i.e. $V_p/a_\perp^3\to\pm \infty$ 
the spin-polarized fermions experience a weak attraction due the transversal 
harmonic confinement. The location of this divergent interaction strength is 
called the $p$-wave confinement-induced resonance, and it occurs where the 
scattering volume of the $p$-wave phase shift is finite and satisfies the 
resonance condition, namely $V_p/a_\perp^3=[12\zeta(-1/2,3/2-E/(2\hbar \omega_\perp))]^{-1}$.}
 
\begin{figure}[h!]
\includegraphics[scale=0.25]{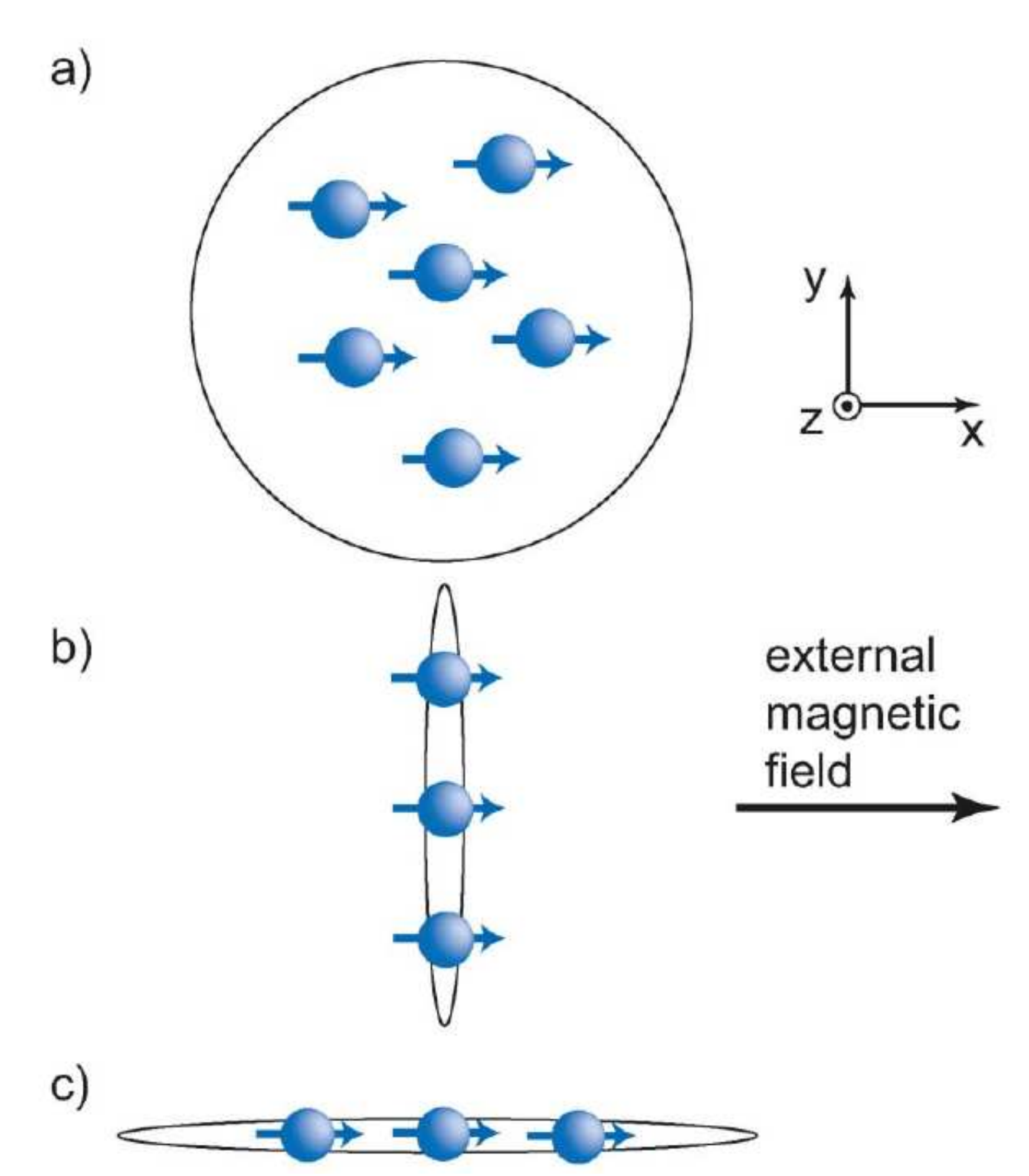}
\caption{(Color online) Schematic illustration of the alignment of the 
spins with respect to pancake and cigar-shaped traps (a) in a pancake 
configuration all the spin alignments are permitted. (b) and (c) refer 
to cigar-shaped traps where in (b) [(c)] only the $|m|=1$ ($m=0$) spin 
configuration of the $p$-wave interactions is allowed. 
(Taken from \cite{gunter2005}.) }
\label{pfig4}
\end{figure}

Apparently, the theoretical scope of confinement-induced resonances is addressed mainly to the elastic collisional aspects of particles with either bosonic or fermionic exchange symmetry.
However, in the experimental advances of \cite{gunter2005,Haller2010a,sala2013,lamporesi2010} the confinement-induced physics is probed via atom loss measurements which inherently are inelastic scattering processes.
These processes mainly emerge due to mechanisms, such as three-body recombination, coupling of the two-body center-of-mass and relative degrees of freedom, spin-flips etc.
Theoretically, the few-body collisions in the presence of external confinement are investigated by \cite{mora2005PRA,mora2004PRL,mora2005PRL,blume2014pra,gharashi2012pra} addressing the three and four-body aspects of the confinement-induced physics where a detailed discussion can be found in the excellent review by \cite{blume2012rpp}.

\cite{gunter2005} experimentally investigated $p$-wave collisions of spin-polarized fermions in the presence of cigar-shaped and pancake traps.
The degenerate gas constitutes of fermionic $^{40}K$ atoms where their mutual interactions are tuned by means of a $p$-wave Feshbach resonance at $198~G$ which possesses a double peaked feature.
As was shown by \cite{ticknor2004multiplet} {\it et al} the doublet structure of a $p$-wave magnetic Feshbach resonance is associated with the different projections of the orbital angular momentum, i.e.$|m|=1$ and $m=0$ for $\ell=1$ and it occurs due to the magnetic dipole-dipole interactions.
\cite{gunter2005} {\it et al. } showed that this feature yields particular signatures also in low-dimensional arrangements.
Qualitatively the impact of the double-peaked Feshbach resonance is depicted in Fig.\ref{pfig4} where three configurations are considered for pancake and cigar-shaped traps.
In particular, panel (a) corresponds in a pancake trap where all the projection alignments, i.e.$|m|=1$ and $m=0$, are considered.
In Fig.\ref{pfig4}(b) a cigar-shaped trap is considered whose longitudinal direction is perpendicular to the magnetic field. 
This implies that the $|m|=1$ configuration of the $p$-wave Feshbach resonance mainly contributes in the scattering process whereas collisional events associated with $m=0$ component of the Feshbach resonance are suppressed. 
Finally, in Fig.\ref{pfig4}(c) the quasi-one dimensional trap is aligned with the magnetic field, whereby the $m=0$ component dominates the $p$-wave collisions.

\begin{figure}[h!]
\includegraphics[scale=0.45]{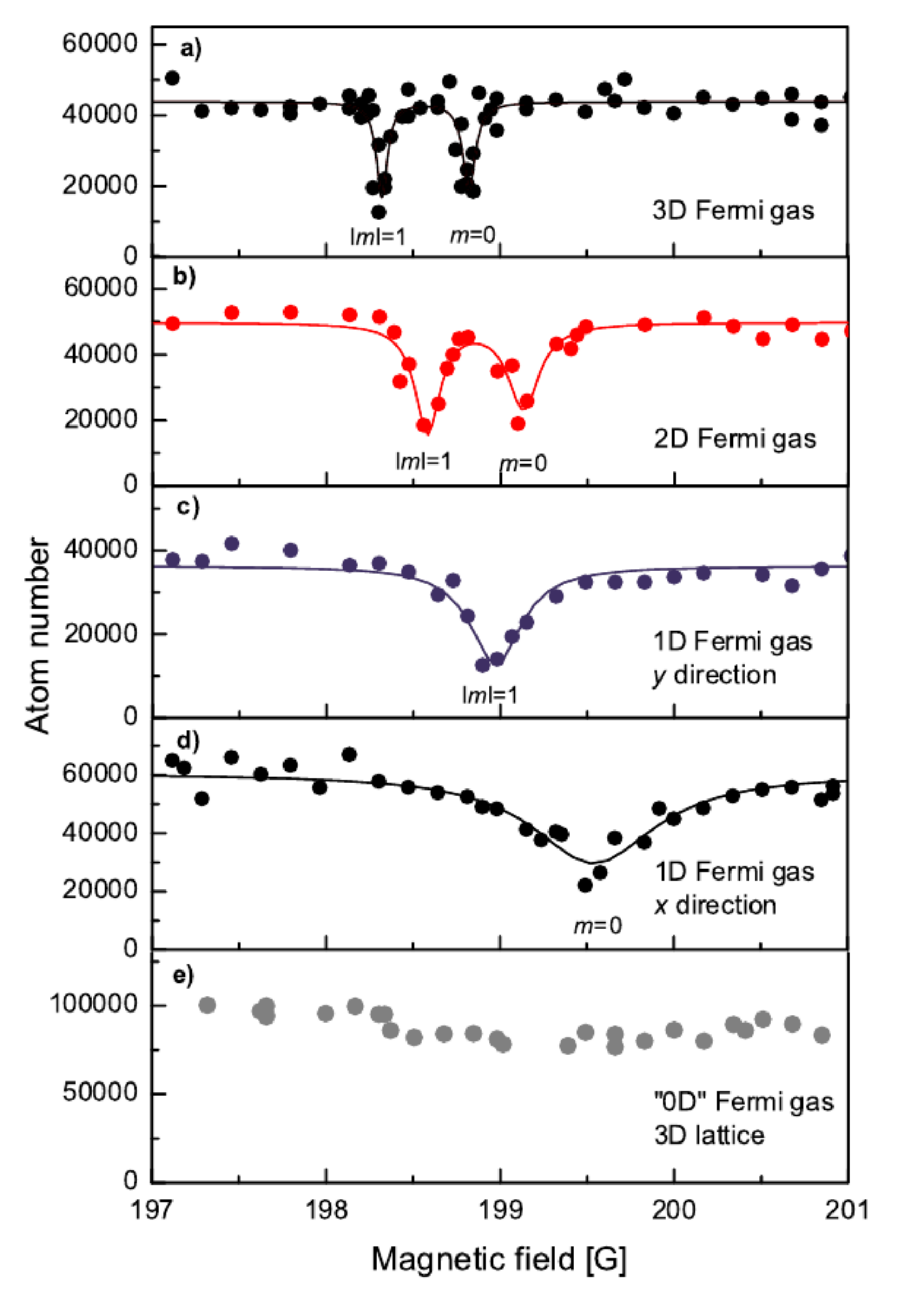}
\caption{(Color online) Atom loss measurements for $^{40}K$ atoms around a multiplet $p$-wave Feshbach resonance for (a) three dimensional dipole trap, (b) a two-dimensional pancake trap, (c) [(d)] a quasi-one dimensional cigar-shaped trap whose longitudinal direction is perpendicular (parallel) to the magnetic field, (e) a three-dimensional optical lattice. ({\bf taken from} \cite{gunter2005}) }
\label{pfig5}
\end{figure}

By measuring the atom loss signal the impact of the multiplet Feshbach resonance is illustrated in Fig.\ref{pfig5} for five different trap configurations.  
In particular Fig.\ref{pfig5} considers (a) a three dimensional optical trap where the double-peaked feature of the p-wave resonance stands out.
In Fig.\ref{pfig5}(b), the potassium atoms are confined in a pancake-shaped trap and the two components of the Feshbach resonance prevail where a confinement induced shift is observed with respect to the measurements of panel (a) is observed.
In Fig.\ref{pfig5}(c) the Fermi gas is confined in a quasi-one dimensional trap with its longitudinal direction positioned perpendicular to the magnetic field.
In contrast to panels (a) and (b) of Fig.\ref{pfig5}, only the $|m|=1$ component of $p$-wave resonance is pronounced whereas the trapping potential induces a shift with respect to the corresponding resonance in Fig.\ref{pfig5}(a).
Similarly, in Fig.\ref{pfig5}(d) the Fermi gas is confined in a cigar-shaped trap with the longitudinal direction being aligned with the magnetic field. 
In this case, solely the $|m|=0$ component of the Feshbach resonance contributes appreciably, and it is shifted towards larger values of field strength with respect to the corresponding measurements of Fig.\ref{pfig5}(a) verifying in this manner the theoretical predictions of \cite{granger2004PRL}.
Finally, a tight three-dimensional optical lattice is considered in Fig.\ref{pfig5}(e) where no pronounced losses are observed since the atoms are effectively confined in ``zero'' dimensions and asymptotic scattering states are restricted. 

From a theoretical viewpoint, \cite{peng2014prl} considered fermionic collisions around a p-wave Feshbach resonance in the presence of quasi-two-and quasi-one-dimensional traps.
In particular, \cite{peng2014prl} takes into account the multiplet structure of the p-wave Feshbach resonance and study within a zero-range model the impact of the relative orientation of the magnetic field with the trapping potentials on the collisional processes.
In this manner, the experimentally, i.e.\cite{gunter2005}, observed spin alignment-dependent confinement-induced resonances for spin-polarized fermions were also verified theoretically by \cite{peng2014prl}.

\subsection{Confinement-induced resonances: delving deeper}
The above discussion has only considered quasi-one dimensional harmonic 
type of confinement  for producing confinement-induced resonances.
The rapid technological advances in laser trapping techniques have 
opened a new avenue that enables ultracold atoms to be confined in 
arbitrary geometries.
From a theoretical viewpoint, this requires extensions of the boundaries 
of our understanding of confinement-induced physics to include more 
general types of trapping potentials. \textcolor{black}{Many of the 
generalizations discussed in the present subsection have 
been reviewed in ~\cite{Zinner2012jpa,blume2012rpp}}

In this direction, \cite{petrov2000prl,petrov2001,pricoupenko2008prl, idziaszek2006} 
considered ultracold collisions in the presence of quasi-two dimensional traps.
In particular, \cite{petrov2001} within the zero range approximation 
studied bosonic collisions in pancake-shaped traps.
The pancake trap modifies the properties of 3D binary collisions yielding 
two-dimensional confinement-induced resonances.
This particular type of resonance also fulfills a Fano-Feshbach scenario 
as do the quasi-one-dimensional resonances.
However, due to the pancake geometry a confinement-induced resonance occurs 
when the two-body potential is not sufficiently deep to produce a true 3D 
universal bound state, i.e. where the corresponding free space scattering length is negative.
Again, this differs from the quasi-one-dimensional confinement-induced 
resonances in the low energy limit, which occur at positive values of the $s$-wave scattering length.
The theoretical predictions of \cite{petrov2001} were experimentally 
confirmed by \cite{froehlich2011}.
\cite{froehlich2011} explored the collisional aspects of a two-component 
Fermi gas in a pancake trap around a free space Fano-Feshbach resonance.
By employing radio-frequency spectroscopic techniques measured molecule 
formation on the negative side of Fano-Feshbach resonance.

\cite{idziaszek2005,peng2010} studied confinement-induced resonances in the presence of an anisotropic waveguide, using a  pseudopotential model of the two-body collisions.
The anisotropy is induced by considering a transverse harmonic potential in the $x-y$ plane, with different frequencies in the $x$ and $y$ directions.
Theory suggests that the system in this type of geometry possesses only one confinement-induced resonance, whose position can be tuned by adjusting the confining frequency aspect ratio \cite{peng2010}.
Moreover, similar conclusions emerged from \cite{zhangwei2011pra} which considered a two-channel model for the short-range 3D interaction.
The simple fact that an anisotropic harmonic waveguide should exhibits only one confinement-induced resonance was not confirmed by the experiment of \cite{Haller2010a}. 
More specifically, \cite{Haller2010a} conducted experiments on Cs atoms confined in quasi-one-to-quasi-two dimensional traps.
In the regime of anisotropic traps through atom loss measurements the corresponding observations showed signatures of {\it two} confinement-induced resonances.

The double peak feature was theoretically resolved by proposing two possible loss mechanisms. One was associated with multichannel inelastic processes \cite{melezhik2011} and the second one was related to the mere fact that trapping potential in the \cite{Haller2010a} exhibit an anharmonicity \cite{peng2011pra, sala2012}.
The anharmonicity of trap induces a coupling between the center-of-mass and relative degrees of freedom of the colliding pair.
This coupling enables the two particles to form a molecule without requiring a third particle since the binding energy can be distributed to the center of mass degrees of freedom.
After implementing these considerations in the theory, two confinement-induced resonances do indeed emerge which confirm the experimental observations of \cite{Haller2010a}.
More recently, in order to pinpoint the physical origin of the double confinement-induced resonances, \cite{sala2013} considered experiments with $^6$Li atoms in an anharmonic waveguide.
More specifically, in that experiment, the trapping potential was loaded with only two $^6$Li atoms in the ground state of the external potential.
Therefore, three-body effects as well as multichannel inelastic multichannel effects were excluded \cite{melezhik2011}.
In this manner, it a double peak structure was observed in atom loss measurements which are attributed to two confinement-induced resonances. 

Dealing with ultracold collisions in arbitrarily-shaped transversal potentials \cite{zhang2013,robicheauxschwinger2015} developed theories based on local frame transformation theory, which can predict a broader class of confinement-induced resonances.
These theoretical treatments also include two-body collisions beyond $s$ or $p$-wave character.
In addition, \cite{robicheauxschwinger2015} using ideas related to the Schwinger variational principle provide infinity-free calculations of scattering observables based on physical grounds, and avoids the need for additional regularization schemes which have been previously utilized in the pseudopotential approaches of \cite{olshanii1998,petrov2001,granger2004PRL}.
However, in the treatments of \cite{zhang2013,robicheauxschwinger2015} the center of mass is coupled with the relative degrees of freedom for particles of finite mass, so they have thus far only been applied in the limit of an infinitely massive particle which is struck by a much lighter one.
\cite{peano2005} considered the coupling of the center of mass in an arbitrary transversal potential, using the Green's function formalism to solve the corresponding Schr\"odinger equation directly in the laboratory frame.
Apart from arbitrary quasi-one-dimensional potentials, \cite{peano2005} considered also the case of two-component ultracold gases in harmonic traps where atoms have different polarizabilities; they therefore experience different harmonic oscillator confining frequencies which results in  coupled center-of-mass and relative degrees of freedom.
In both cases, \cite{peano2005} theoretically predicted that the corresponding scattering observables should exhibit more than one resonant feature associated with the confinement-induced resonances.
Similarly, \cite{kim2005} developed a theory yielding only qualitative predictions since the Hilbert space associated with the closed channel physics was not taken into account.
This aspect, however was taken into account by \cite{melezhik2009} which predicted confinement-induced resonant molecular formation.

\textcolor{black}{ \cite{massigan2006pra} and \cite{nishida2008universal,nishida2010,nishida2011FBS}} focused on 
mixed-dimension collisions in ultracold gases, under the assumption that different atomic 
species experience move in different numbers of spatial dimensions, such as when a 3D gas of atoms
interacts with either different atoms or the same atoms in different internal states that are trapped in an optical lattice.
Again, for this mixed-dimension system the center of mass and relative degrees of freedom 
are inherently coupled.
The concept of mixed dimensionality arises from the fact that in a mixture of 
ultracold gas different particles, i.e. A and B species, experience different confinement frequencies.
The confinement frequencies depend on each atom's polarizability and the laser frequency.
Therefore, by adjusting the laser frequency at a zero of polarizability of one 
atomic species, i.e.A atoms, only the B atoms will experience the trapping potential.
The proposed technique of \cite{massigan2006pra,leblancpra2007} for 
species-selective dipole potentials was first realized by \cite{catani2009prl}.
\cite{nishida2010} studied this idea by considering one atomic species, 
e.g. A atoms, to be totally unconfined, i.e. they move in three dimensions, 
while the B atoms are trapped in a tight spherical trap (a ``zero dimension'' 
configuration), or a cigar-shaped (quasi-one-dimensional configuration) 
trapping potential, or a pancake-shaped trap (quasi-two dimensional configuration).
In addition, particles A and B are assumed to collide at small distances with 
$s$-wave interactions only.
This gives rise to an infinite series of a particular type of confinement-induced 
resonances which possess high orbital angular momentum character despite the fact that the 
two-body collisions are dominated by $s$-wave interactions.~\cite{massigan2006pra}
This effect emerges from the combination of pure $s$-wave interactions and the 
fact that the two-body collisions take place in mixed-dimensions which couples 
the angular momenta of the A and B atomic species.
The theoretical predictions \cite{nishida2008universal,nishida2010} have 
apparently been observed by \cite{lamporesi2010}, who
created a mixed-dimensional confinement of two ultracold atomic species, 
namely $^{41}$K and $^{87}$Rb.
The $^{41}$K atoms are trapped in two-dimensions whereas the $^{87}$Rb 
atoms move in three-dimensions, which is achieved by implementing 
species-selective dipole trapping techniques \cite{massigan2006pra,leblancpra2007,catani2009prl}.
In this mixed-dimensional configuration \cite{lamporesi2010} observed 
up to five resonances by measuring 3-body losses,
in good agreement with theoretical resonance positions.

Pair collisions within a three-dimensional optical lattice were theoretically investigated by \cite{fedichev2004,cui2010}.
\cite{fedichev2004} utilized the tight-binding framework and assumed that the range of the two-body interactions is far smaller than the lattice spacing, i.e. $d$, and the size of the ground state in a lattice side, i.e. $\ell_0$.
Also the curvature within the lattice sites is approximated as a harmonic oscillator, which permits a decoupling of the center of mass and relative degrees of freedom.
Furthermore, in the tight binding model \cite{fedichev2004} assumed that the effective mass of the particles is large enough such that the size of the ground state within the lattice site is small compared to the lattice spacing whereas effects arising from higher Bloch bands were excluded.
Based on these considerations \cite{fedichev2004} predicted a confinement-induced resonance that occurs at negative values of the $s$-wave scattering length.
The resonance condition simply reads $a_s\sim \ell^\ast$, where $\ell^\ast=\ell_0\sqrt{D_0}/(4 \ln{2})$, with $D_0$ the tunneling amplitude to neighboring lattice sides in the lowest Bloch band.
\cite{kohl2005fermionic} conducted experiments on a degenerate Fermi gas in the presence of a three-dimensional optical lattice.
They observed that the Feshbach resonance within the optical lattice exhibits an additional shift from the corresponding Feshbach resonance in the absence of the external potential, which verified the theoretical predictions of \cite{fedichev2004}.
\cite{cui2010} extended the theoretical studies of \cite{fedichev2004} and quantitatively described Bloch wave scattering at different lattice depths.
Also, higher Bloch bands are taken into account as well as intraband effects which occur in the lowest Bloch band.
\cite{cui2010} showed that in the case of true molecular states and at moderate lattice depths the higher Bloch bands effects play crucial role since their neglect overestimates binding energies.

\subsection{\textcolor{black}{Synthetic spin-orbit coupled systems}}
Experimental and theoretical efforts on the confinement-induced physics in low 
dimensional systems consider a regime where the two-body interactions are 
short-ranged and isotropic.
Lifting the latter constraint permits us to generalize the concept of the 
confinement-induced resonances in physical systems which are mainly governed 
by anisotropic binary interactions.

\textcolor{black}{The experimental realization of spin-orbit coupled 
Bose-Einstein condensates opens new avenues to explore the collisional 
aspects of such exotic systems. Developments in this rapidly evolving field 
have been reviewed by ~\cite{HuiZhai2015rpp, williams2012science}.
For example, in free space collisions, spin-orbit coupling yields a 
mixed-partial wave scattering process that alters the corresponding Wigner 
threshold law \cite{xiaoling2012pra, duan2013pra,susocpra}. 
\cite{soc2014,zhang2013effective,Zhang2012pra2d} studied the impact of 
reduced dimensionality on these physical systems in the presence of 
quasi-one and quasi-two dimensional traps. }

In particular, \cite{soc2014} considers resonant collisions of spin-orbit coupled cold atoms with Raman coupling in the presence of an axially symmetric harmonic waveguide.
As a first order approximation, effects due to the coupling of the center-of-mass and relative degrees of freedom are neglected by considering the case of zero center-of-mass momentum.
Then the relative Hamiltonian, apart from the kinetic energy, two-body and confinement potential terms, possesses two additional terms: (i) The spin-orbit coupling term $H_{\rm{SOC}}=\frac{\gamma k}{m} (\sigma_2^x-\sigma_1^x)$ and (ii) Raman coupling term $H_{\rm Raman}=\frac{\Omega}{2}(\sigma_2^z+\sigma_1^z)$.
$\sigma_{1,2}^i$ with $i=x,z$ represents the spin Pauli matrices for each particle, $k$ indicates the relative kinetic energy, $m$ is the atom's mass. 
$\Omega$ represents the strength of the two-photon Raman coupling and $\gamma=2\pi\hbar\sin (\theta/2)/\lambda$ indicates the spin-orbit coupling constant where $\lambda$ is the Raman laser wavelength and $\theta$ denotes the angle between the lasers.

Both spin-orbit and Raman coupling influence inherently the position of the resulting confinement-induced resonance.  
A confinement-induced resonance always exists regardless the sign of the $s$-wave scattering length only in the case where the Raman coupling strength is less than spin-orbit coupling strength, i.e.$\Omega <2 \gamma$.
For strong Raman coupling, i.e.$\Omega \gg 2\gamma$ the position of the confinement-induced resonance occurs  only at smaller values of the ratio $a_s/a_\perp$, namely $ a_s/a_\perp \sim 1/\sqrt{2 \Omega}$.
This provides in essence an extra means to manipulate the position of a CIR without the need of a Fano-Feshbach resonance to tune the magnitude of the 3D scattering length.
Note that similar findings were reported also by \cite{zhang2013effective}.

\subsection{\textcolor{black}{Confined dipoles and dynamical CIR}}
Collisions of magnetic dipolar atoms or of polar molecules pose another 
physical system whose two-body interactions are inherently anisotropic.
The concept of confinement-induced resonances for anisotropic two-body 
interactions has been considered both for quasi-two and quasi-one dimensional 
waveguide geometries.  \textcolor{black}{Such dipolar systems hold particular
interest in the many-body realm for their potential to create novel new 
topological phases of matter,
in addition to quantum information applications.~\cite{baranov_theoretical_2008,baranov_condensed_2012}}
In a numerical study, \cite{hanna2012pra} explored the impact of a pancake geometry on 
nonreactive polar molecules where despite the fact that the confining potential yield broader resonances than in the absence of a trap, the location of resonances are extremely sensitive to the dipole moment strength.

Apart from pancake geometries, the concept of dipolar confinement-induced resonances is also investigated in harmonic waveguides, i.e. in quasi-one dimensional traps \cite{sinha2007,giannakeas2013prl,bartolo2013dipolar} where the dipoles are aligned by an external field in a head-to-tail configuration.
In particular, \cite{giannakeas2013prl,tao2014pra} apply the local frame transformation theory and the pseudopotential techniques, respectively, showed the existence of a broad class of dipolar confinement-induced resonances which are characterized by mixed orbital angular momentum character due to the dipole-dipole interactions and the confinement.

Moreover, for $s$-wave dominated dipolar confinement-induced resonances their \textcolor{black}{position} depends linearly on the ratio of the length scale of dipolar forces over the trapping length scale, i.e.$\sim\frac{l_d}{a_\perp}.$
Note that $(l_d,~a_\perp)= (\mu d^2/\hbar^2,~\sqrt{\hbar/\mu \omega_\perp})$ where $\mu$ indicating the reduced mass of the dipolar, $d$ is the corresponding dipole moment and $\omega_\perp$ indicates the confinement frequency.
This linear dependence of position of the dipolar confinement-induced resonances on the ratio $\frac{l_d}{a_\perp}$ means that the collisional properties of dipoles in the presence of a confinement can be controlled by adjusting the strength of an external field and confining potential frequency in a regime accessible by the experimental advances.

Furthermore, \cite{tao2014pra} showed that by tilting the relative orientation of the external electric field with respect to the longitudinal axis of the harmonic waveguide provides additional means to refine the tuning of dipolar confinement-induced resonance positions.
\cite{simoni2015polar} studied the case of reactive polar molecules in cigar-shaped traps.
In more detail \cite{simoni2015polar} numerically studied the impact of the reduced dimensionality on elastic, inelastic and reaction rates of collision of the reactive molecules in terms of the collisional energy and the strength of the dipole moments.
The full four-body calculations are simplified by employing an asymptotic effective two-body model at large distances where the reactions are suppressed \cite{micheli2010universal}.
The reaction physics is introduced through a WKB-type boundary conditions at short distances that accounts for atom exchange phenomena.
By varying the angle of an external electric field with respect to the longitudinal direction of the trap, i.e.trap axis, it is observed that the reaction rate is greatly suppressed for angles normal to the trap's axis.
For the case of a bosonic KRb molecule the reaction rate can only be efficiently suppressed under strong confinement without yielding any significant advantages over the reaction rate suppression in quasi-two dimensional trapping geometries.

Photon-assisted confinement-induced resonances arise when there is a dynamical mechanism to enhance resonant collisions in the presence of a waveguide.
\cite{leyton2014photon} consider $s$-wave binary collisions in the presence of an RF driven harmonic waveguide whose confining frequency modulation permits the separation of the center of mass degrees freedom. 
The resulting time-dependent Schr\"odinger equation is solved within a zero-range approximation, i.e. using a Bethe-Peierls boundary condition at the origin of the relative degrees of freedom.
The RF modulation of the transversal frequency permits to the two counter-propagating atoms to perform a transition from the continuum state to the confinement-induced molecular state by emitting one or multiple photons.
This dynamical mechanism of photon-assisted confinement-induced resonances can result into a series of resonant features for a given number of photons.

\subsection{\textcolor{black}{Inelastic few-body collisions in 1D}}
\textcolor{black}{An intriguing implication of the exact integrability of the 1D Schr\"odinger equation for $N$ equal mass particles that experience zero-range two-body potentials, and the existence of an analytically known solution due to ~\cite{mcguire1964JMP}, is that there are no inelastic collisions.  One way to understand this is that every time two particles collide, the final state of the two particles in momentum space is kinematically identical to the initial state, since transmission and reflection are indistinguishable.  While the absence of all inelasticity including three-body recombination for this system is straightforwardly clear, given the exact solution of ~\cite{mcguire1964JMP}, it is far from obvious how the inelasticity turns out to vanish when studied in the adiabatic hyperspherical representation.  One way all inelasticity would vanish for a system would be if the hyperradial degree of freedom in the Schr\"odinger equation turns out to be exactly separable, because in that case all nonadiabatic coupling matrices would vanish.  It is not that simple, however, in the case of identical 1D bosons with zero-range interactions, because the nonadiabatic coupling matrices are nonzero. Hence, the different adiabatic hyperspherical channels are coupled, at least locally.  This issue was explored by ~\cite{mehta2005PRA}, which numerically solved the coupled hyperradial equations that describe atom-dimer scattering. }

\textcolor{black}{Later, three-body recombination was calculated in this 1D identical boson system for both the zero range potential case and for a finite-range potential by ~\cite{mehta2007PRA}, again using the adiabatic hyperspherical representation.  That study confirmed as well that there is no inelasticity, i.e. vanishing three-body recombination rate coefficient.  In the hyperspherical representation, the recombination rate and the rates of all other inelastic processes vanish because there is complete destructive interference in the zero-range limit.  For a 1D potential of finite range $L$, however, the study showed how the recombination rate increases for nonzero $L>0$.  The near-threshold behavior of the recombination rate is of interest for 1D and quasi-1D experiments, and it was demonstrated in ~\cite{mehta2007PRA} that three identical particles have the same threshold behavior $K_3^{1D} \propto k^7$ regardless of whether the particles are spin polarized fermions or bosons.  For three identical bosons in particular, this study shows more concretely that in strict 1D, $K_3^{1D} = C(L) (\hbar k/\mu) (ka)^6$, where $k$ is the 1D wavenumber and $a$ is the 1D two-body scattering length.  Experimental evidence is quite limited in this topic, but the results measured to date, e.g. by ~\cite{TolraPhillips1D2004prl} appear not to have reached a regime very close to the strict 1D results, and are probably better viewed as probes of the crossover regime between 1D and 3D or between 1D and 2D.
}

 
%


\subsection{\textcolor{black}{2D, quasi-2D systems, and the super-Efimov effect}}
\textcolor{black}{Two-dimensional systems in both few-body and many-body physics exhibit rich and fascinating behavior, involving logarithmic dependences of nearly all quantities that depend on distance and energy.  In many-particle condensed-matter systems, prototypical phenomena that have generated extensive interest include the Berezinskii-Kosterlitz-Thouless(BKT)-transitionm ~\cite{berezins.vl_destruction_1971,kosterlitz_ordering_1973,
thouless_quantized_1982} relating to the formation and binding of 2D vortices, and of course the fractional quantum Hall effect ~\cite{Stormer1999rmp}. Underlying the theoretical description of striking many-body phenomena in 2D are the effective two-body and three-body interactions that are modified when a three-dimensional gas is squeezed into a pancake-shaped trap geometry.} 

\textcolor{black}{The modification of the 3D atom-atom scattering information into an effective 2D interaction has been addressed by many authors, e.g. \cite{Wodkiewicz1991,KanjilalBlume2006pra}, with a more comprehensive list of references in \cite{Dunjko2011461}.  Implications of 2D confinement for three-body recombination and for the formation of many-body phases were treated by \cite{PetrovHolzmannShlyapnikov2000prl}, for a gas consisting of particles with finite-range interactions.  The three-body problem in 2D for short-range interactions has more recently been examined in a hyperspherical coordinate framework by ~\cite{DIncaoEsry2014pra,DIncaoAnisEsry2015pra}, including a nonperturbative study of 3-body recombination in that geometry. Two-body dipole-interacting particles in a 2D or quasi-2D gas are treated by many publications, two of which are ~\cite{KanjilalBohnBlume2007pra, dincao2011pra}.  The intriguing many-body phenomena that arise in dipolar systems have received extensive theoretical and experimental attention, as has been reviewed in ~\cite{baranov_theoretical_2008,baranov_condensed_2012}.  }

{\textcolor{black}{One class of few-body treatments in two dimensions relates to fractional quantum Hall droplets having modest numbers of particles, typically from 3-10 electrons, or in the ultracold physics context, atoms or polar molecules. A few such explorations in recent years can be found in ~\cite{Daily2015PRB,Rittenhouse2016pra,Wooten2017prb}, which are just a sampling of the work that followed the famous work by ~\cite{Laughlin1983prb} on three 2D electrons in a perpendicular magnetic field.  A degenerate perturbation theory treatment of semiconductor quantum dots in a strong magnetic field, which has numerous cases that can serve as useful benchmark calculations for comparison with few-body theory, can be found in ~\cite{Jeon2007}.
}

{\textcolor{black}{A provocative few-body prediction in recent years has been the super-Efimov effect, which deals with bound states of 3 $p$-wave interacting fermions in 2D.~\cite{Nishida2013, Volosniev2014JPB, Moroz2014, Gridnev2014JPA, GaoWangYu2015}  The interactions are assumed in the derivation of this effect to have a finite range, and each interacting pair in the trimer is assumed to have a zero-energy bound state in the symmetry with $|L_z|=1$, also referred to as a resonant $p$-wave interaction. The resulting trimer energy level formula predicted in this case takes the double-exponential form:  
$E_n \propto \exp[-2 e^{3\pi n/4+\theta}]$ where $\theta$ is a nonuniversal constant defined modulo $3 \pi /4$.Because the size of these super-Efimov states grows so rapidly with $n$, and also the successive binding energies shrink dramatically as $n$ increases, these will be challenging to observe experimentally. Whereas the successive energy levels in the ordinary homonuclear Efimov effect are less bound by a factor of 515, the corresponding ratio in the super-Efimov effect exceeds $10^9$. More promising than the homonuclear systems are heavy-heavy-light heteronuclear trimers, whose super-Efimov states can display a more favorable scaling,~\cite{Moroz2014} as is also the case in the ordinary Efimov effect for heteronuclear trimers.  Another exploration of heavy-heavy-light trimers in 2D with resonant $p$-wave interactions is based on the \textcolor{black}{conventional} Born-Oppenheimer approximation ~\cite{EfremovSchleich2013PRL}.  }

\textcolor{black}{Recently, a theoretical treatment by ~\cite{Nishida2017arxiv} has introduced the ``semi-super Efimov effect''.  This is in a system of four bosons in 2D, which exhibit a different scaling possible for an infinite pattern of energy levels and state sizes, in a scenario where the three-boson interactions are resonant but the two-body interactions are negligible.  In the semi-super Efimov case, the state sizes are predicted to scale with the integer quantum number $n>0$ in proportion to $\exp[(\pi n)^2/27]$.
 }

\textcolor{black}{The theory of the quasi-2D homonuclear three-boson problem 
has been treated in detail by ~\cite{Levinsen2014,Yamashita2015jpb}.  This topic
is also sometimes referred to as the ``crossover'' from 3D to 2D.  
This study demonstrated how the finite number of universal 3-body states in 2D, 
where there is no true Efimov effect,~\cite{BruchTjon1979pra,nielsen2001PRep,nishida2011FBS} 
connect with true Efimov states in 3D as one varies the degree of confinement 
in the transverse dimension.  Three fermions in 2D are also treated theoretically 
in ~\cite{Ngampruetikorn2013epl}, by solving the Skorniakov-Ter-Martirosian(STM) 
integral equation ~\cite{skorniakov1957JETP}.
}

\section{Few-body physics in nuclear and chemical systems}

In previous sections, the current state of the art of few-body ultracold atomic physics has been presented, with an extended discussion of universality 
in three-body and four-body physics. However, the domain of few-body
 physics clearly extends beyond atomic systems, reaching many different branches 
of physics such as chemistry or nuclear or particle physics, among 
others. Indeed, few-body physics was born in nuclear physics motivated 
by the seminal paper of \textcolor{black}{~\cite{thomas1935pr}}, as pointed out above.

The premise of long de Broglie wavelength effective field theory is that
for the low energy physics of a few- or many-body system below a certain
characteristic energy scale, the behavior of the system should not be
sensitive to the details of the Hamiltonian at distances much less than $\lambda$.
 \textcolor{black}{Indeed, 
the physics behind such behavior is closely related with the concept of renormalization 
group theory~\cite{Wilson-1971,Wilson-1974,Wilson-review}}
One of the best pedagogical introductions to the strategy of systematically
building in the correct long wavelength physics has been presented by
 \cite{lepage1997ARX,lepage1989} in the context of nonrelativistic
Schr\"odinger quantum mechanics. \textcolor{black}{In fact, the concept 
of effective field theory was spawned by 
the seminal work of Weinberg that attempted to understand the role of pion-exchange in nuclear forces~\cite{Weinberg-1979,Weinberg-1990,Weinberg-1991}}. 
Example applications to Efimov physics 
and related few-body systems have been developed in detail by 
\cite{braaten2006PRep}.  Other studies that utilize model two-body and three-body
Hamiltonians that produce key information such as the low energy two-body 
scattering length and effective range are making use of the spirit 
of low energy effective field theories even when they proceed via more 
direct solution of the few-body Schr\"odinger equation.  \textcolor{black}{We note as well that trions, excitons, and biexcitons occur as few-body problems in semiconductor physics, as is discussed, e.g. by \cite{Patton2003prb}.}

This section is devoted to the study of
major developments in few-body physics relevant for nuclear physics
and chemistry.
The few-body physics in chemical sciences will be presented in two guises: 
first, few-body physics based on classical trajectory calculations
in hyperspherical coordinates involving neutrals and charged particles. Second, a full 
quantum mechanical treatment revealing the underlying universality of three-body 
collisions involving charged particles.


\subsection{Hyperspherical methods in nuclear physics}

The adiabatic hyperspherical technique has been introduced in previous 
sections  of the present 
review, with examples of its methodology and applications 
in the field of atomic and molecular collisions. Reiterating, the basic idea behind this method is to reduce a complex 
multidimensional problem into set of coupled second-order ordinary differential equations 
in a single variable. The same idea can of course be applied in a field with tremendously 
different energy and length scales: nuclear physics, where the nature of the nucleon-nucleon and related interactions 
exhibit all the complexities of the strong nuclear force. For instance, both exotic nuclear systems and 
the three-nucleon problem, to name two classes of problems, have been studied using the 
adiabatic hyperspherical representation. 

In nuclear physics the adiabatic hyperspherical technique follows the same scheme as is 
presented above in this review: the first step involves solution of the hyperangular 
equation where the hyperradius $R$ is treated as a parameter, thereby giving 
hyperspherical potential energy curves and couplings. Then, these are employed 
to solve a set of coupled ordinary differential equations in the radial coordinate. 
With this strategy one can tackle the few- or many-nucleon interaction problem at different
levels of sophistication.  For instance, the cosmologically important reaction of three alpha particles to form 
$^{12}C$ via the intermediate Hoyle state has been treated within the coupled-channel
adiabatic representation \textcolor{black}{by ~\cite{alvarez-rodriguez2007EPJA, alvarez-rodriguez2008PRC, suno2015PRC}.}  The calculation  of hyperspherical potential curves and couplings \textcolor{black}{by ~\cite{suno2015PRC} }
was based on a binary $\alpha-\alpha$ model Hamiltonian that accurately describes the $^8Be$ resonance
state, and a three-body term was chosen to represent some experimentally-known 
properties of $^{12}C$ such as some particular energy levels.  The final calculation gives
a good energy and width in agreement with experimental values for the $J_n^\pi=0_2^+$ Hoyle resonance state.  \textcolor{black}{The relevant adiabatic hyperspherical potential curves are shown in Fig.~\ref{Hoyle}.}

\begin{figure}[h!]
\centering
 \includegraphics[width=8.5 cm]{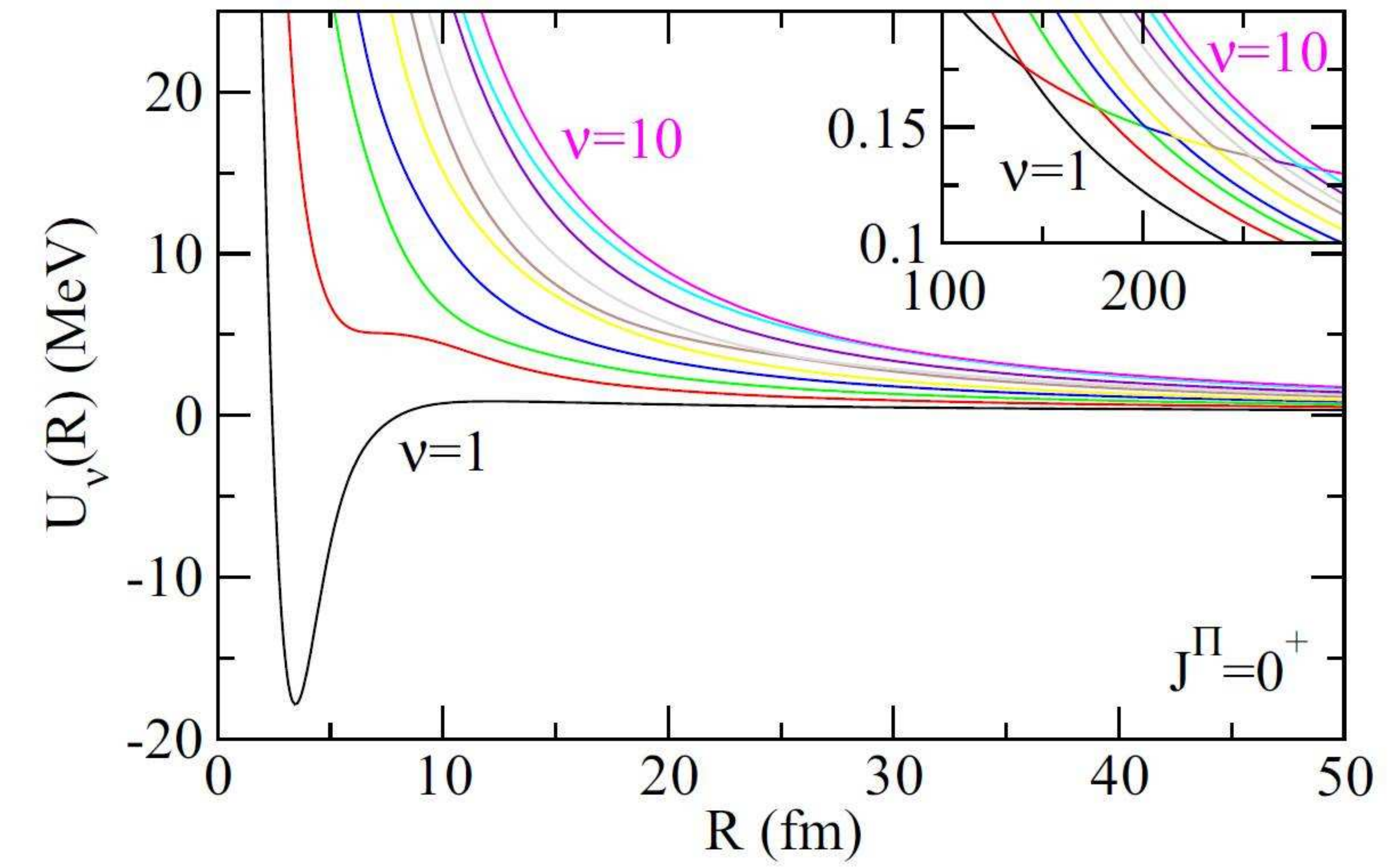}
 \caption{(Color online) Adiabatic hyperspherical potential
curves computed for the triple-$\alpha$ system by~\cite{suno2015PRC} 
for the $J^\pi=0^+$ symmetry which contains the famous Hoyle resonance
thought to be important in nucleosynthesis. Inset: the adiabatic potential curves at large hyper-radius.}
\label{Hoyle}
\end{figure}

Note that when 
this technique is applied to few-nucleon systems, one must keep in mind 
the distinctively different nature of  
the nucleon-nucleon forces compared to the 
atom-atom interactions. This difference comes from the fact that nuclear 
collisions can be understood to first order as resulting from the exchange of virtual pions between 
nucleons. In this sense, pions ($\pi$) can be viewed as the {\it quanta} of the nuclear force, 
and since they represent a massive scalar field their influence is associated with a Yukawa potential 
$e^{-m_{\pi}}r/4\pi r$~\cite{Weinberg-1991}. Indeed, the nucleon-nucleon potential can be 
modeled using an effective field theory based on the exchange of pions. This exchange leads to new and complicated 
interaction terms in the nucleon-nucleon potential, among them, the tensor interaction reads

\begin{equation}
\label{nucl-pot}
V(\bm{r}_{ij})=V_{t}(r_{ij})(\bm{\tau}_{i}\cdot\bm{\tau}_{j})\left[3\frac{(\bm{\sigma}_{i}\cdot 
\bm{r_{ij}})(\bm{\sigma}_{j}\cdot \bm{r}_{ij})}{r_{ij}^2}-\bm{\sigma}_{i}\cdot \bm{\sigma}_{j}\right],
\end{equation}

\noindent
where $\bm{\sigma}_{i}$ and $\bm{\tau}_{i}$ represent the nuclear spin and isospin 
of nucleon $i$, respectively.  \textcolor{black}{For a more detailed modern view of the nucleon-nucleon 
forces, including pion-less theories and chiral effective field theory, we recommend Ref.\cite{epelbaum2009RMP}.}

The very complicated nature of the nucleon-nucleon interaction does not prevent the success of the 
adiabatic hyperspherical technique in nuclear physics, indeed hyperspherical 
coordinates were applied in the context of nuclear physics by \cite{delves1960NP} and 
\cite{Smith-1960} to study three-body nuclear 
systems~\cite{Levinger,Valliers-1976,Fang-1977,Verma-1979}.  Those studies, however, did not implement the adiabatic formulation and thus could not benefit from its insights and accelerated convergence. Hyperspherical methods were also applied to exotic nuclei, such as the hypertriton, and complex nuclei by including realistic 
nucleon-nucleon potentials~\cite{Verma-1979,Verma-1982,Clare-1985}. The potential 
employed included the tensor interaction and many other components of the 
nucleon-nucleon force. In general, many theorists preferred to work with a set of coupled 
integral equations instead of coupled differential equations, which can be regarded as a 
procedure different from the adiabatic hyperspherical machinery most frequently adopted in atomic physics. 
More recently, however, the adiabatic hyperspherical approach has been employed to compute the triton bound state energy~\cite{Daily-2015} in a convergence exploration, using 
realistic nucleon-nucleon potentials with a three-body force as well. 
Other systems that have been considered include exotic species such as kaonic clusters involving three and four particles~\cite{Nuclear-clusters}.

\subsection{Universality in nuclear systems}

As discussed above, the nuclear forces are fundamentally different
 from the interactions in the context of ultracold atomic physics, since they derive from the 
electromagnetic interaction. 
Therefore, due to the very strong and short-ranged nature of the nuclear interactions, some aspects of universality can be expected to differ in nuclear systems compared with atomic and molecular species. In particular, some 
nuclear systems form halo nuclei \textcolor{black}{~\cite{Tanihata-1996,cobis1998PRC,Jensen-2004,Zhukov-1993}}, 
which is a nuclear bound state formed by a tightly bound core and one or two 
valence nucleons. These valence nucleons are characterized to have a very small binding energy in 
comparison with the binding energy of the core nucleons, which 
is reflected in a highly extended bound state wave function.  For this reason, 
halo nuclei exhibit very large radii compared to the core radius. 
The most familiar example of a halo nucleus is the deuteron,  \textcolor{black}{with an average neutron-proton separation of 3.1 fm that is three times larger than both the size of its component nucleons and the range of their interaction potential (both $\approx 1 fm$). While halos also exist in atomic and molecular physics, they are far more prevalent in nuclear systems, with many studies even in
large or medium-sized nuclei.  See for instance ~\cite{Hove2014prc,Hove2016prc} and references therein.  In fact Efimov {\it physics} can be relevant to describing aspects of the wavefunction of a medium-sized nucleus like $^{62}Ca$ with two outlying nucleons,~\cite{Hagen2013prl}.  Nevertheless this type of system exemplifies the {\it Efimov-unfavored scenario} with two light particles and one heavier particle that is discussed in ~\cite{Wang-2012b}; based on the arguments presented there, it is unlikely that a ``true Efimov state'' exists in such systems.}

Halo nuclei with two valence nucleons represent a good playground for 
three-body physics, since these nuclear systems are potential candidates to 
\textcolor{black}{exhibit some properties associated with  
Efimov physics despite being Efimov-unfavored.} These nuclides are found in the bottom of the neutron 
drip line: the line that describes the boundary beyond which the neutron-rich nuclides are 
unstable. Among the different kinds of two nucleon 
halo nuclei, the Borromean~\footnote{The term Borromean 
is associated with the coat of arms of the house of Borromeo family in the north of Italy, which consists 
in three interlayer rings. In particular, the symbol appears in the left escutcheon of the coat 
of arms. However, a similar symbol involving three triangles was already
 used in Norse mythology around the 7th century.} halo nuclei 
have received special attention since these have a three-body 
bound state, \textcolor{black}{despite the fact that} 
none of the two-body subsystems is bound. 
The most studied Borromean halo nuclei to date have been $^{6}$He and
  $^{11}$Li~\cite{Tanihata-1996,Zhukov-1993}.  A schematic representation of one 
such nucleus is shown in the inset of Fig.~\ref{Phillips}, concretely for
$^{6}$He. In this case, the core is \textcolor{black}{an $\alpha$ particle and the two valence nucleons are neutrons}.

\begin{figure}[h!]
\centering
 \includegraphics[width=7.5 cm]{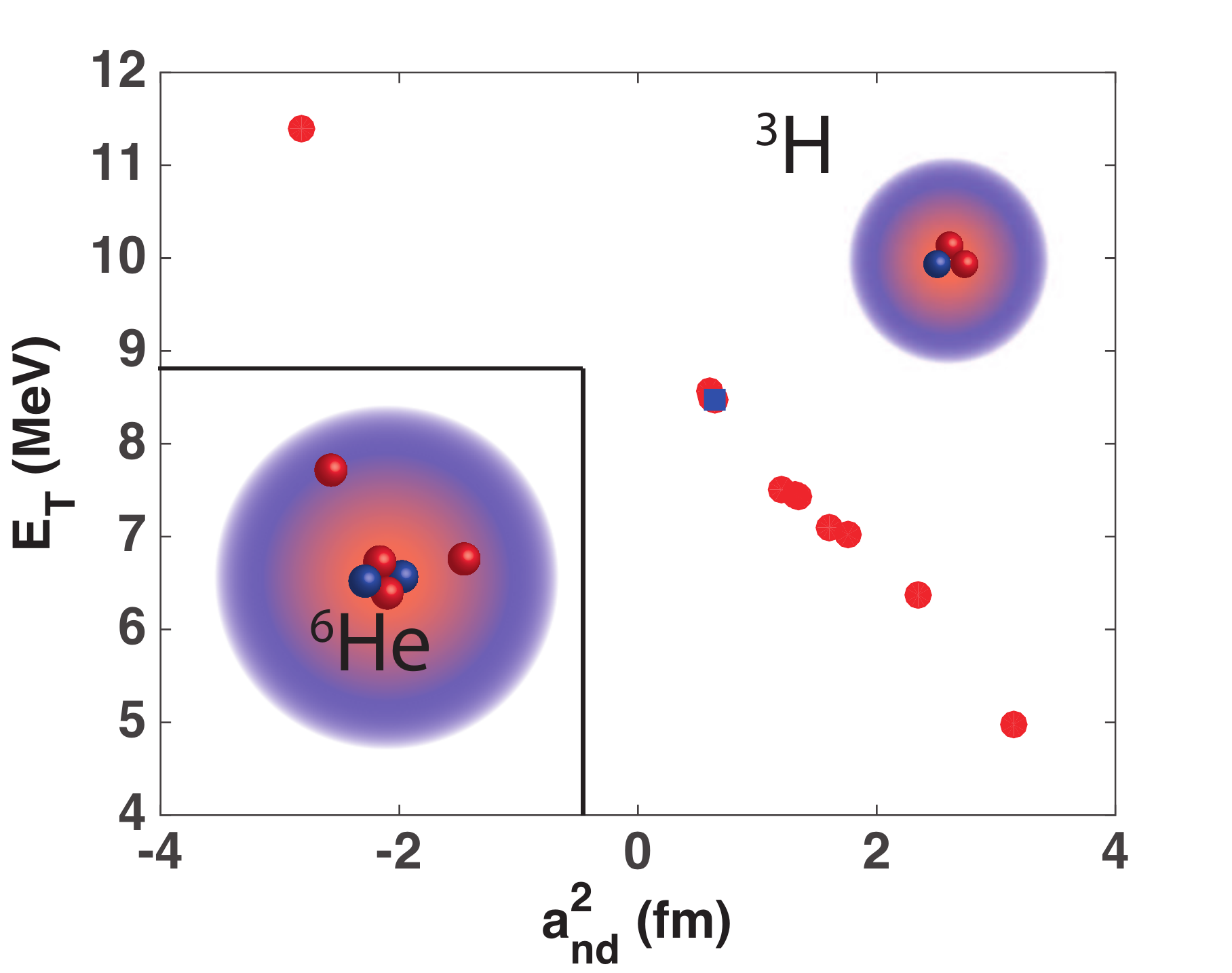}
 \caption{(Color online) Phillips plot for the triton energy $E_{T}$ as a function of the 
 doublet $nd$ scattering length $a^{2}_{nd}$. The red dots stand for different theoretical 
 calculations based on different kinds of two-body and three-body 
 interactions~\cite{Benayoun-1981,Friar-1984,Fedorov-2002}, whereas the blue square 
 represents the experimental result. For a more detailed presentation of the Phillips plot for the
 triton, see the work of Efimov and Tkachenko~\cite{Efimov-1988}. Inset: schematic 
 representation of a halo nuclei with two valence nucleons, namely $^{6}$He.}
\label{Phillips}
\end{figure}

 On the other hand, three-body bound states in some nuclear systems show 
a universal behavior: a correlation between the dimer-nucleon scattering length 
and the nucleon trimer binding energy, which is known as the Phillips 
line~\cite{Phillips-1968}. This correlation is independent of the model employed 
for the calculations: such as a two-body contact interaction, three-body interaction terms, and it should also be present regardless of
 the method employed, i.e. an effective field theory approach, adiabatic hyperspherical treatment, or 
low energy Faddeev equations.  Fig.~\ref{Phillips} presents a Phillips plot 
for the triton including different theoretical results as well as the 
experimental data. This figure exhibits a linear correlation between 
the triton binding energy and the doublet neutron-deuteron scattering 
length $a^{2}_{nd}$. However, other kinds of correlation may 
occur in different three-body nuclear bound states, see {\it e.g.} the work 
of Fedorov and Jensen~\cite{Fedorov-2002}.

At least two halo nuclei have been seen as potential candidates to exhibit 
Efimov universality. However, the universality can only be claimed convincingly if 
the different excited state of the three-body system follow the predicted 
scaling law by Efimov, and those are not experimentally accessible by any currently existing capabilities. For instance in the case of $^{6}$He, 
which is also Borromean, it has been proven that a $p$-wave
 resonance exists in the $J$ = 3/2 channel of $n-\alpha$ scattering which explains 
 the nature of the three-body bound state. However, in the 1990's \cite{Fedorov-1994} and \cite{Amorim-1997} 
explored the Efimov character of several halo nuclei, assuming that the 
ground state is also an Efimov state. \textcolor{black}{Their study suggested that $^{20}$C is 
the only halo nucleus candidate that has appreciable Efimov state character, other than the triton.}

Apart from {\it normal} nuclei, namely those that appear in the table of nuclides, there 
other hypernuclei contain strange quarks, and are the so-called strange nuclei. Some of these} are classified as halo nuclei, and among them the simplest case is the 
hypertriton $^{3}_{\Lambda}$H: a three-body bound state formed by a neutron, a 
proton and the $\Lambda^{0}$. The $\Lambda^{0}$ is the lightest $\Lambda$ hyperon, 
a neutrally charged baryon similar to a neutron but slightly heavier, and 
its quark structure is $uds$; it has strangeness -1. The 
total binding energy of $^{3}_{\Lambda}$H is $\approx$ 2.4 MeV~\cite{Fujiwara-2008}, whereas its 
breakup energy is $\sim$ 0.14 MeV~\cite{Fujiwara-2008}, which is very small in comparison with the 
binding energy of the deuteron 2.22 MeV, and hence it can be considered as a 
two-nucleon halo nucleus. Indeed, it has been extensively studied 
\cite{Gongleton-1992,Cobis-1997,Fedorov-2002}. All of these studies suffer, however, from needing better experimental information concerning the n-$\Lambda$ scattering length, 
so these works might be considered as qualitative or semi-quantitative 
approaches to the Efimov nature of the hypertriton. Nevertheless, new data coming 
from ALICE and STAR may help to understand better the nature of the n-$\Lambda$
 interaction, as well as to yield more accurate measurements of the lifetime of $^{3}_{\Lambda}$H
 ~\cite{ALICE,Zhu-2013,STAR}. On the other hand, a good understanding of the 
 hyperon-nucleon interaction is needed for a proper understanding of high-density 
 matter systems, such as neutron stars~\cite{Weber-2007,Vidana-2013,
 Lonardoni-2014,Lattimer-2004}.

In nuclear systems, universal properties in the four-body sector can be identified. The clearest 
example is the case of the Tjon line~\cite{Tjon-1975}: a correlation between the 
binding energy of the $\alpha$ particle and the triton binding energy \textcolor{black}{that persists across nearly all nucleon interaction models}; in particular, this correlation is approximately linear. 
The origin of the Tjon correlation can be explained as an approximate independence of the four-body energy level spectrum on
any four-body parameter. In other words, \textcolor{black}{the Tjon analysis suggests that
there is no need for a} four-body parameter for the renormalization at leading order in 
the four-body sector~\cite{platter2004PRA,Platter-2005}, for energy levels in the universal regime. However, higher order corrections
 break the expected correlation leading to a band with some scatter depending on the short range physics, instead of a simple, well-defined line~\cite{Nogga-2000}.
 
\subsection{Few-body physics and universality in chemistry}

Traditionally the term {\it few-body physics} has been employed in nuclear 
physics, and subsequently it became adopted in the context of atomic physics, 
especially in ultracold atomic systems \cite{Hess-1983,Hess-1984,Goey-1986,esry1999PRL,burt1997PRL,weber2003PRL,Fedichev1996PRL,suno2003PRL}. 
However, in 
 chemical physics it has not be the case, even though chemical physics studies 
 hinge on our understanding of few-body physics. And of course a deep understanding of fundamental processes in chemical physics is frequently needed 
 in other fields of physics and chemistry, notably  
 in astrophysics, such as the three-body recombination of hydrogen in stellar 
formation\cite{Flower-2007,Forrey-2013a}; in theoretical chemistry: transport 
coefficients in gases\cite{Molecular_transport,Mason-1962,McCourt,
Wang-Chang,Snider-1960,Kohler-1983,Montero-2014}, reactive and non-reactive 
scattering\cite{Levine,Child,Shui-thesis,Truhlar-1975} and three-body 
recombination\cite{Francis-2006,mansbach1969pr,Ermolova-2014}, dissociative 
recombination of H$_{3}^{+}$\cite{kokoouline2001NT,kokoouline2003PRL, petrignani2011};  plasma physics\cite{Zhdanov}; and in cold chemistry\cite{Willitsch-2008,Willitsch-2012,
Hall-2012,harter2012prl,Harter:NaturePhysics:2013,Harter-2014,Ulm}. Some of these 
characteristic and fundamental processes in chemical physics will 
be reviewed from a few-body perspective in the present section, in particular, those 
involving three-body processes, such as three-body recombination and dissociative
 recombination. Special emphasis will be given to universality in three-body 
 recombination processes which are relevant to hybrid trap experiments. 
 
Three-body processes can be viewed as a chemical reaction that converts three free atoms (or molecules or other particles) into diatomic molecules in the absence 
of external fields, {\it i.e.} the reaction A + A + A $\rightarrow$ A$_{2}$ + A. One of the very 
first theoretical treatments of this reaction was developed by Keck~\cite{Keck-1960,Keck-1967} using a 
variational principle following a very early approach proposed by Wigner~\cite{Wigner}. 
In particular, an upper bound for the three-body recombination rate was computed by dividing 
Regions of phase-space by a trial surface that acts as the boundary between reactants 
and products;  this is now denoted the phase-space theory of reaction rates. Almost in parallel,
Smith developed a more microscopic  treatment for three-body recombination~\cite{Smith-1962}, and 
later~\cite{Shui-1970} extended the previous theory of Keck, applying 
this theory to the recombination of nitrogen~\cite{Shui-1970}. An application to the 
recombination of hydrogen~\cite{Shui-1973,Shui-thesis} found
fair agreement with the fairly crude early experiments.

The phase-space theory of reaction rates was introduced by Keck~\cite{Keck-1960} 
and an alternative was presented by Smith~\cite{Smith-1962}. 
Next, the mathematical foundations of a recent approach to calculation of classical 
three-body recombination rates are presented, after which applications to different 
systems involving neutrals as well as charged particles will be reviewed. Classical 
trajectory calculations in hyperspherical approach have been employed to derive
different classical Newtonian threshold laws, which are reviewed here, with special emphasis on their 
universality. The quantum nature of few-body physics in 
chemical systems will be considered at the end, where 
the recombination of hydrogen atoms and the dissociative recombination of H$_{3}^+$ are covered, as fundamental benchmark systems in chemical physics.

\subsubsection{General classical treatment of few-body collisions}

Classically, a two-body collision is envisioned as one particle with 
a definite momentum moving towards a scattering center. The 
cross section is defined as an effective area on the plane perpendicular to the 
initial momentum of the incoming particle which contains the scattering 
center~\cite{Levine}. In classical mechanics, the scattering 
cross section is determined in terms of the scattering probability for a given value of the impact parameter $\bm{b}$. Recall that $\bm{b}$ is defined
 as the component of the position vector which is perpendicular to the momentum vector
 of the incoming particle at infinite distance.

The concept of a two-body collision cross 
section is readily generalized to an arbitrary number ($n$) of dimensions. In particular, 
the cross section is defined as the effective scattering area of the $n-1$ hyperplane 
perpendicular to the initial momentum of the incoming particle, for a given impact parameter $\bm{b}$ and 
initial momentum $\bm{P}_{0}$ 

\begin{equation}
\label{s-1}
\sigma_{process}(\bm{P}_{0})=\int \wp_{process}(\bm{b},\bm{P}_{0})d^{n-1}\bm{b}.
\end{equation} 

\noindent
Here, the opacity function, $\wp_{\rm{process}}(\bm{b},\bm{P}_{0})$ is the
 probability that a trajectory with particular initial conditions leads to the collisional process 
 under investigation, {\it e.g.}, an inelastic collision, or formation of a particular product, etc.

 \begin{figure}[h!]
\centering
 \includegraphics[width=7.5 cm]{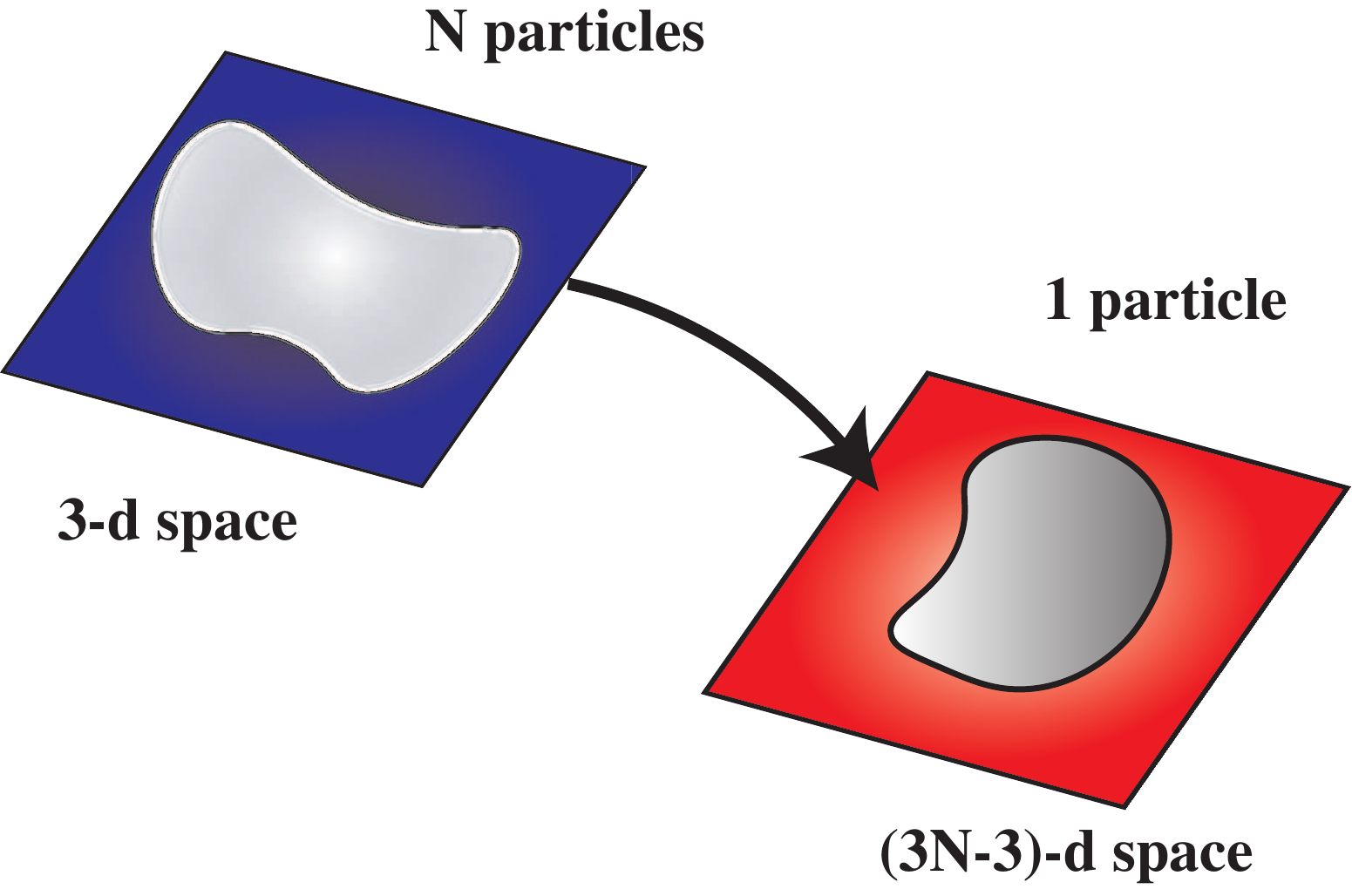}
 \caption{(Color online) Schematic representation of a general classical 
 treatment of few-body collisions. The method is based on the mapping of 
 the degrees of freedom of the system at hand into a problem involving 
 a single effective particle moving in higher dimensional space, in particular 
 the dimension ($n$) is equal to the number of independent relative coordinates of the system. }
\label{mapping}
\end{figure}
 
The classical dynamics of few-body system can be obtained by recasting the 
degrees of freedom of the system at hand $d$ as a two-body 
collision in a $d$-dimensional space, as shown in Fig.~(\ref{mapping}). 
This figure represents the usual case where the center 
of mass motion is decoupled from the relative motion of the interacting 
particles. In this picture the three-body 
recombination cross section is written

 \begin{equation}
\label{s-2}
\sigma_{rec}(P_{0})=\frac{\int \wp_{rec}(\bm{b},\bm{P}_{0})d\Omega_{P_0}^{6}d\Omega_b^5b^4db}{\int d\Omega_{P_0}^6},
\end{equation}  
where the quantity $ d\Omega_{P_0}^6$ represents the differential element associated with the hyperangles of the initial momentum $\bm{P}_0$. In Eq.(\ref{s-2}) the averaging over the degrees of freedom associated with the initial 
momentum is shown explicitly, whereby the final average cross section depends only on the energy.

\subsubsection{Classical trajectory calculations in hyperspherical coordinates}

The classical Hamiltonian for three particles with masses $m_{1}$, $m_{2}$ and $m_{3}$ 
moving in a given potential energy landscape $V(\bm{r}_{1},\bm{r}_{2},\bm{r}_{3})$ is

\begin{equation}
\label{c-1}
H=\frac{\bm{p}_{1}^{2}}{2m_{1}}+\frac{\bm{p}_{2}^{2}}{2m_{2}}+\frac{\bm{p}_{3}^{2}}{2m_{3}}+
V(\bm{r}_{1},\bm{r}_{2},\bm{r}_{3}),
\end{equation}

\noindent
where $\bm{p}_{i}$ and $\bm{r}_{i}$ represent the momentum 
and the vector position of the $i^{th}$ particle, respectively. This 
Hamiltonian can be recast in terms of the Jacobi coordinates 
$(\bm{\rho}_{1},\bm{\rho}_2)$, depicted in Fig.(\ref{Class-method}) 
As in \cite{Karplus-1965}

\begin{equation}
\label{c-2}
H=\frac{\bm{P}_{1}^{2}}{2m_{12}}+\frac{\bm{P}_{2}^{2}}{2m_{3,12}}+\frac{\bm{P}_{\rm{CM}}^{2}}{2M}+
V(\bm{\rho}_{1},\bm{\rho}_{2}).
\end{equation}

\noindent
Here $\frac{1}{m_{12}}=\frac{1}{m_{1}}+\frac{1}{m_{2}}$; 
$\frac{1}{m_{3,12}}=\frac{1}{m_{3}}+\frac{1}{m_{1}+m_{2}}$; $V(\bm{\rho}_{1},\bm{\rho}_{2})$ 
is the potential energy in terms of the relative Jacobi coordinates with the CM 
momentum a separated constant of motion. $\bm{P}_{1}$,  $\bm{P}_{2}$ and $\bm{P}_{\rm{CM}}$ 
represent the canonical momenta conjugate to $\bm{\rho}_{1}$, $\bm{\rho}_{2}$ 
and $\bm{\rho}_{CM}$, respectively. Finally, the relative Hamiltonian is

\begin{equation}
\label{c-3}
H=\frac{\bm{P}_{1}^{2}}{2m_{12}}+\frac{\bm{P}_{2}^{2}}{2m_{3,12}}+V(\bm{\rho}_{1},\bm{\rho}_{2}).
\end{equation}

For three particles, the Hamilton equations of motion can be expressed in terms of
 Jacobi coordinates and momenta as follows:

\begin{subequations}
\begin{eqnarray}
\frac{d\rho_{i,\alpha}}{dt}&=&\frac{\partial H}{\partial P_{i,\alpha}},\label{c-4a} \\
\frac{dP_{i,\alpha}}{dt} &=&-\frac{\partial H}{\partial \rho_{i,\alpha}},\label{c-4b}
\end{eqnarray}
\end{subequations}

\noindent
where $i=1,2$ and $\alpha=x,y,z$ label the Cartesian
 coordinates of each Jacobi vector. Upon adopting the representation of Smith~\cite{Smith-1962}, a 
 6D position vector is constructed from the two mass-weighted Jacobi vectors as 

\begin{equation}
\label{c-5}
\bm{\rho}=\begin{pmatrix} 
   \sqrt{\frac{m_{12}}{\mu}}\bm{\rho}_{1}\\ 
   \sqrt{\frac{m_{3,12}}{\mu}}\bm{\rho}_{2} 
\end{pmatrix},
\end{equation}

\noindent
where $\mu=\sqrt{\frac{m_{1}m_{2}m_{3}}{M}}$. On the other hand, an equivalent 
6D vector position can be expressed in terms of the {\it bare} Jacobi vectors as~\cite{JPR-2014}

\begin{equation}
\label{c-6}
\bm{\rho}_{bare}=\begin{pmatrix} 
  \bm{\rho}_{1}\\ 
  \bm{\rho}_{2} 
\end{pmatrix}.
\end{equation}

\noindent
Similarly the canonical 
momenta are given by

\begin{equation}
\label{c-7}
\bm{P}=\begin{pmatrix} 
   \bm{P}_{1}\\ 
   \bm{P}_{2} 
\end{pmatrix}
\end{equation}

\noindent
and 

\begin{equation}
\label{c-8}
\bm{P}_{bare}=\begin{pmatrix} 
   \sqrt{\frac{\mu}{m_{12}}}\bm{P}_{1}\\ 
   \sqrt{\frac{\mu}{m_{3,12}}}\bm{P}_{2} 
\end{pmatrix},
\end{equation}

\noindent
respectively. It can be shown that the relation between the coordinates 
$(\bm{\rho},\bm{P})$ and $(\bm{\rho}_{bare},\bm{P}_{bare})$ defines 
a canonical transformation\cite{Whittaker-1937,Maslov,Landau-mechanics}, 
and hence both sets of coordinates will describe the same phase-space 
volume. In other words, the scattering observables will be the same for
 either of these sets of coordinates, as one would expect. The 6D position 
 vector $\bm{\rho}$ or $\bm{\rho}_{bare}$ links the three-body
 problem in 3D and the single particle problem in 6D, as is schematically 
 presented in Fig.(\ref{Class-method}).

 \begin{figure}[h!]
\centering
 \includegraphics[width=7.5 cm]{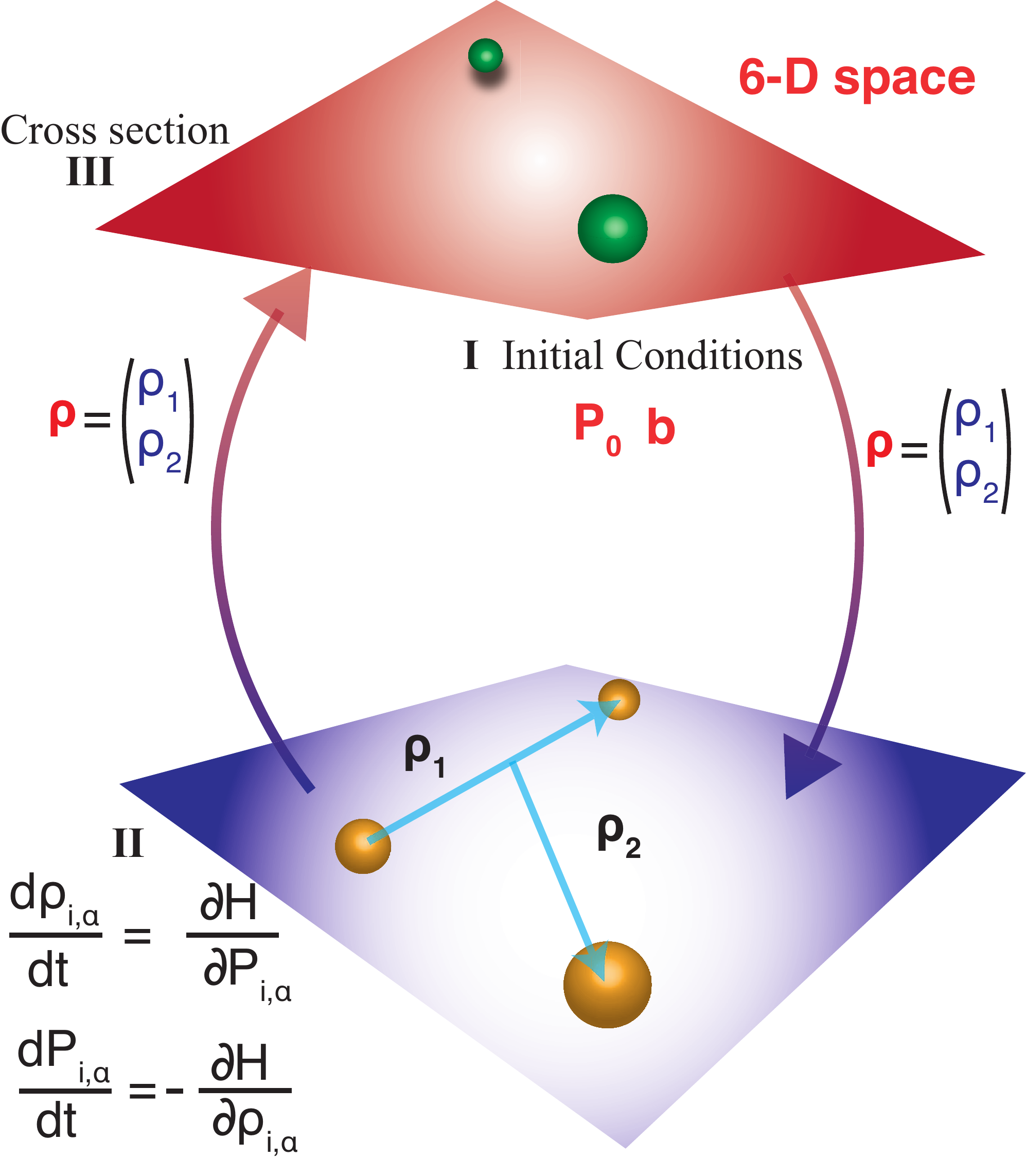}
 \caption{(Color online) Schematic representation of the method 
 developed for treating three-body collisions. We start with the description 
 of the initial conditions in the 6-D space associated to the three-body problem 
 at hand. Then, as indicated in Step I, the initial conditions are transformed into the 
 coordinates associated with the three-body problem in the usual 3D space. Step 
 II represents the solution of the Hamilton's equations of motion in the 3D space. 
 Finally, by means of step III, the results are transformed back into the 6D space, 
 where the cross section is calculated.} 
\label{Class-method}
\end{figure}

In the present approach the mass-weighted 6D vector position $\bm{\rho}$ 
will be employed, in a minor difference from the conventions utilized in Ref.~\cite{JPR-2014}. Then the Hamiltonian is 

\begin{equation}
\label{c-9}
H=\frac{\bm{P}^{2}}{2\mu}+V(\bm{\rho}).
\end{equation}

\noindent
Now that the 
position and the momentum vectors have been defined in this 6D space, the concept 
of impact parameter as the projection of the position vector onto a hyperplane 
perpendicular to the initial momentum is clear. We now implement 
hyperspherical coordinates for the representation of the 6D vectors. 
In particular, it is convenient to implement Avery's definition of the hyperangles~\cite{avery1989} is chosen, where 
all the vectors can be represented by means of their magnitude $r$ 
and five different hyperangles ($\alpha_{i}$, $i=1,2,3,4,5$) as

\begin{equation}
\label{c-10}
\bm{r}=\begin{pmatrix} 
   r_{x_{1}}\\
 r_{x_{2}}\\
r_{x_{3}}\\
r_{x_{4}}\\
r_{x_{5}}\\
 r_{x_{6}} 
\end{pmatrix}
= \begin{pmatrix} 
   r \sin{\alpha_{1}}\sin{\alpha_{2}}\sin{\alpha_{3}}\sin{\alpha_{4}}\sin{\alpha_{5}}\\
 r \cos{\alpha_{1}}\sin{\alpha_{2}}\sin{\alpha_{3}}\sin{\alpha_{4}}\sin{\alpha_{5}}\\
r \cos{\alpha_{2}}\sin{\alpha_{3}}\sin{\alpha_{4}}\sin{\alpha_{5}}\\
r \cos{\alpha_{3}}\sin{\alpha_{4}}\sin{\alpha_{5}}\\
r \cos{\alpha_{4}}\sin{\alpha_{5}}\\
  r\cos{\alpha_{5}}
\end{pmatrix}.
\end{equation}

\noindent
Here the ranges of each angle are $0\le\alpha_{1}\le 2\pi$, $0\le\alpha_{i}\le \pi$, $i=2,3,4,5$. In particular, choosing 
the 3D $z$ axis parallel to $\bm{P}_{2}$, expresses the initial momentum $\bm{P}_{0}$ as 

\begin{equation}
\label{c-11}
\bm{P}_{0}=\begin{pmatrix} 
   P_{0} \sin{\alpha^{P}_{1}}\sin{\alpha^{P}_{2}}\sin{\alpha^{P}_{5}}\\
 P_{0} \cos{\alpha^{P}_{1}}\sin{\alpha^{P}_{2}}\sin{\alpha^{P}_{5}}\\
P_{0} \cos{\alpha^{P}_{2}}\sin{\alpha^{P}_{5}}\\
0\\
0\\
  P_{0}\cos{\alpha^{P}_{5}}
\end{pmatrix},
\end{equation}

\noindent
 where $0\le\alpha^{P}_{1}\le 2\pi$, $0\le\alpha^{P}_{2}\le \pi$ and $0\le\alpha^{P}_{5}\le \pi$. 
 
The impact parameter represents the components of the initial vector position
of the system in the hyperplane perpendicular to the initial momentum of the incoming 
particle, as was introduced previously. Let us define $\tilde{\bm{b}}$ as the 
impact parameter when the 6D vector position is $\bm{\rho}$:

\begin{equation}
\label{c-12}
\tilde{\bm{b}}=\begin{pmatrix} 
   \tilde{b} \sin{\alpha^{\tilde{b}}_{1}} \sin{\alpha^{\tilde{b}}_{2}}\sin{\alpha^{\tilde{b}}_{3}}\sin{\alpha^{\tilde{b}}_{4}}\\
 \tilde{b} \cos{\alpha^{\tilde{b}}_{1}}\sin{\alpha^{\tilde{b}}_{2}}\sin{\alpha^{\tilde{b}}_{3}}\sin{\alpha^{\tilde{b}}_{4}}\\
\tilde{b}\cos{\alpha^{\tilde{b}}_{2}} \sin{\alpha^{\tilde{b}}_{3}}\sin{\alpha^{\tilde{b}}_{4}}\\
\tilde{b}\cos{\alpha^{\tilde{b}}_{3}}\sin{\alpha^{\tilde{b}}_{4}}\\
\tilde{b} \cos{\alpha^{\tilde{b}}_{4}}\\
  0 
\end{pmatrix},
\end{equation}

\noindent
where $0\le\alpha^{\tilde{b}}_{1}\le 2\pi$, $0\le\alpha^{\tilde{b}}_{i}\le \pi$, $i=2,3,4$. Thus, $\tilde{\bm{b}}$ 
is a mass-weighted version of the {\it bare} impact parameter $\bm{b}$. 
These two impact parameters are related by 
$d^{5}{\tilde{\bm{b}}}=(m_{12}^{3}m_{3,12}^{3}/\mu^6)^{1/2}d^{5}{\bm{b}}$, and hence the 
classical cross section is given by

\begin{equation}
\label{c-14}
\sigma_{\rm{process}}(P)=\frac{\int \wp_{\rm{process}}(\tilde{\bm{b}},\bm{P}) d\Omega_{P}^{6}d\Omega_{\tilde{b}}^{5}\tilde{b}^{4}d\tilde{b}}{\left(\frac{m_{12}^3m_{3,12}^3}{\mu^6}\right)^{1/2}\int d\Omega_{P}^{6}},
\end{equation} 

\noindent
where a {\it normalization} mass factor emerges as a consequence of the mass-weighted character 
of the 6D vector position.

 \begin{figure}[h!]
\centering
 \includegraphics[width=9.2 cm]{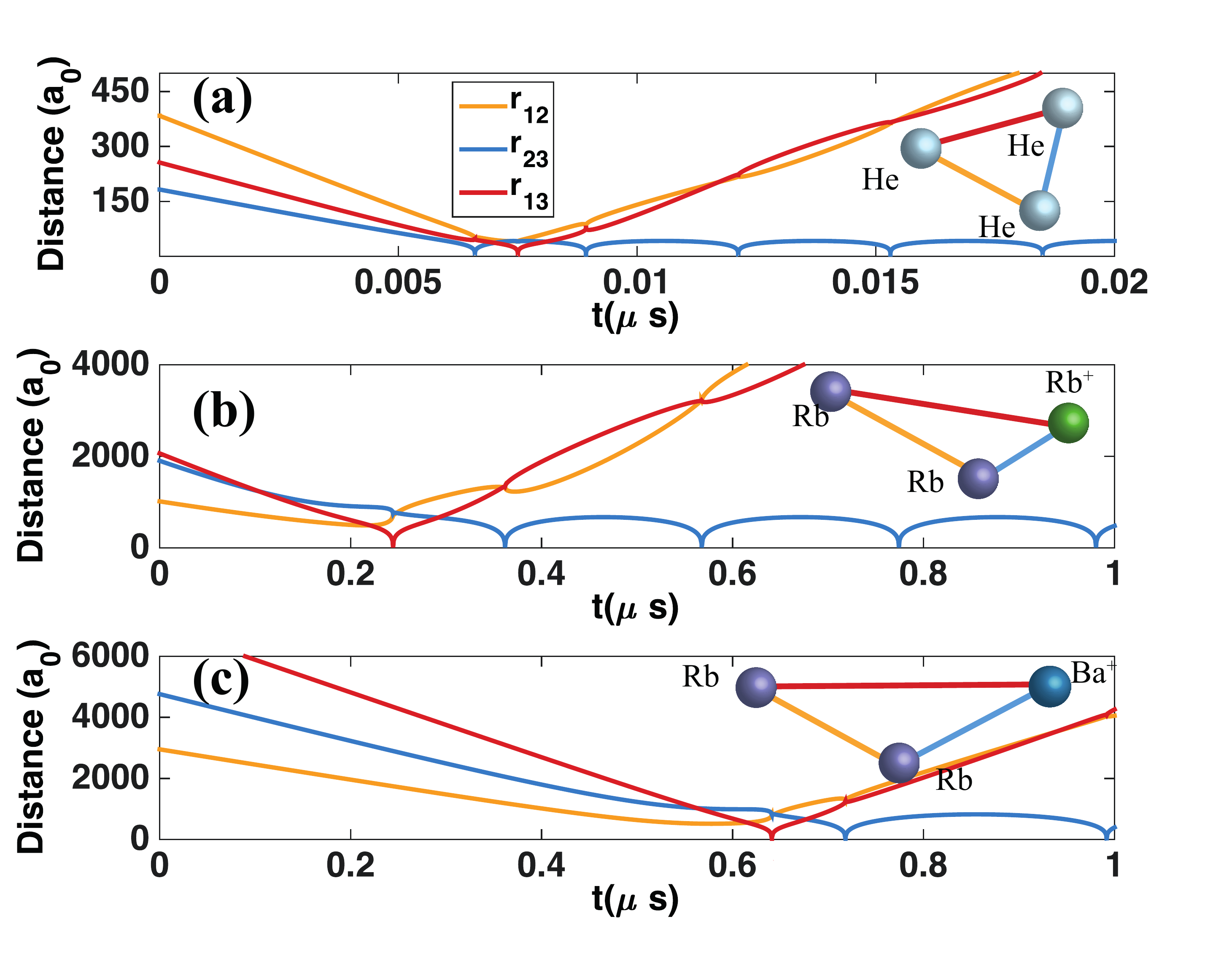}
 \caption{(Color online) Classical trajectories for three-body collisions at a relative collision energy $E= k_B T_{\rm{in}}$ with $T_{\rm{in}}=1$ mK (The Boltzmann constant $k_B$ will usually be omitted in quoting energies in the following.).
 Classical trajectories associated with a three-body recombination event 
 He + He + He $\rightarrow$ He$_{2}$ + He with $b$ = 97 $a_{0}$ , panel (a); 
 $b$ = 1000 $a_{0}$ for Rb + Rb + Rb$^{+}$ $\rightarrow$ Rb$_{2}^{+}$ + Rb 
 in panel (b); and Rb + Rb + Ba$^{+}$ $\rightarrow$ Rb-Ba$^{+}$ + Rb with 
 $b$ = 1000 $a_{0}$ in panel(c).}
 \label{Class-1}
\end{figure}

The initial vector position $|\bm{\rho}_{0}|=R$ is chosen 
in the asymptotic region where the interaction potential is negligible, thus the 
initial momentum satisfies $E=P_{0}^{2}/2\mu$, where $E$ is the incident collision 
kinetic energy. The  hyperangles $\alpha^{P}_{i}$ 
with $i=1,2,5$, and the impact parameter hyperangles $\alpha^{\tilde{b}}_{j}$ with $j=1,2,3,4$, are randomly generated 
 subject to the constrained magnitude of the impact parameter $|\tilde{\bm{b}|}$. The random distribution of those angles must of course be 
chosen consistent with their appropriate probability density function~\cite{JPR-2014}. Exploiting the orthogonality of the initial momentum $\bm{P}_{0}$ and the impact 
parameter $\tilde{\bm{b}}$, the initial vector position is written as
\begin{equation}
\label{c-15}
\bm{\rho}_{0}=\tilde{\bm{b}}-\frac{\sqrt{R^{2}-\tilde{b}^{2}}}{P_{0}}\bm{P}_{0}.
\end{equation}

\noindent
Eq. (\ref{c-15}) generates $\bm{\rho}_{0}$ from $R$, $\tilde{\bm{b}}$ and $\bf{P}_{0}$. 
For a given set of initial conditions $\bm{\rho}_{0}$, $R$, $\bf{P}_{0}$ and $\tilde{b}$, the information 
is transformed into the usual 3D space by means of Eqs.(\ref{c-5}) and (\ref{c-7}), where Hamilton's classical
equations of motion are numerically integrated up to a certain final time \cite{JPR-2014,Numerical_Recipes}. 
Then the coordinates are mapped back into the 6D space, and the classical three-body cross section is
calculated by means of Eq.(\ref{c-14}). This protocol is schematically presented in Fig.~(\ref{Class-method}).

The present approach has been applied to neutral three-body recombination \cite{JPR-2014} and also to neutral-neutral-ion three-body recombination \cite{JPR-2015}. Fig.(\ref{Class-1}) exhibits 
different trajectories associated with recombination events in several atomic systems: He + He + He 
in panel (a), Rb + Rb + Rb$^+$ in panel(b) and Rb + Rb + Ba$^+$ in panel (c). These trajectories have 
been obtained by assuming a pair-wise potential 
$V(\bm{r}_{1},\bm{r}_{2},\bm{r}_{3})=v(r_{12})+v(r_{23})+v(r_{31})$, where $r_{ij}$ are the 
interparticle distances. In particular, for the helium atom-atom interaction the potential of Aziz 
{\it et al.}, designated HFD-B3-FCI1\cite{Aziz-1995} has been employed. The $^{3}\Sigma$ potential of 
\cite{Strauss-2010} for Rb-Rb is employed, and no spin-flip transitions are 
allowed in the theoretical model. The ion-atom interactions are described by the model potential 
$-\alpha_d (1-(r_{m}/r)2)/2r^4$, where $\alpha_d$ denotes the static dipole polarizability of Rb, which is taken as $\alpha_d$ = 320 a.u., and $r_{m}$ represents 
the position of the minimum of the potential. For Rb$^{+}$-Rb, $r_{m}$ is taken from the 
quantum chemistry calculations of \cite{Jraij-2003}, and the information needed for 
Ba$^{+}$-Rb is adapted from \cite{Krych-2011}. For details about the numerical solution 
method, Monte Carlo sampling and convergence see Refs.~\cite{JPR-2014,JPR-2015}.

\subsubsection{Classical three-body recombination for neutrals and ion-neutral-neutral systems}

The hyperspherical classical trajectory method~\cite{JPR-2014,JPR-2015} has been applied to the recombination of three neutrals and to the ion-neutral-neutral recombination process. For a given 
collision energy $E_{k}=P_{0}^2/2\mu$, the average classical
 three-body cross section is given by 

\begin{equation}
\label{c-16}
\sigma_{\rm{rec}}(P_{0})=\frac{\int \wp_{\rm{rec}}(\bm{b},\bm{P}_{0}) d\Omega_{P_{0}}^{6}d\Omega_{b}^{5}\tilde{b}^{4}db}{\int dd\Omega_{b}^{5}\Omega_{P_{0}}^{6}},
\end{equation} 

\noindent
where $d\Omega_{b}^{5}$ and $d\Omega_{P_{0}}^{6}$ stand for the 
differential elements in the hyperangles associated
 with the impact parameter $b$ and the initial momentum $P_{0}$, respectively.
$ \wp_{\rm{rec}}(\bm{b},\bm{P})$ represents the opacity function or 
reaction probability for three-body recombination, that is, the probability that 
the reactants transform into the products of interest for a given set of initial conditions and 
impact parameter. Generally, such a probability shows a stereochemical 
dependence, but the hyperangular degrees of freedom can be averaged out, 
leading to

\begin{equation}
\label{c-17}
 \wp_{\rm{rec}}(b,P_{0})=\frac{\int \wp_{\rm{rec}}(\bm{b},\bm{P}_{0}) d\Omega_{P_{0}}^{6}d\Omega_{b}^{5}}{\int d\Omega_{P_{0}}^{6}}.
\end{equation}

\noindent
This integral is evaluated by Monte Carlo sampling 
Over the different initial conditions and impact parameters. The sampling is performed by 
means of the probability distribution function in each degree of freedom, which can be laborious but is trivially parallelizable. The solution for 
$\wp_{\rm{rec}}(b,P_{0})$ in Eq.~(\ref{c-17}) implies the maximum impact 
parameter that can produce a recombination process for a fixed $P_{0}$, 
denoted as $b_{max}(P_{0})$. In other words, $\wp_{\rm{rec}}(b,P_{0})=0$ for 
$b>b_{max}(P_{0})$. Finally, the three-body recombination cross section can 
be expressed as

\begin{equation}
\label{c-18}
\sigma_{\rm{rec}}(P_{0})=\Omega^{5}_{b}\int_{0}^{b_{max}(P_{0})} \wp_{\rm{rec}}(b,P_{0})b^{4}db,
\end{equation} 

\noindent
where $\Omega^{5}_{b}=8\pi^2/3$ is the total integrated hyperangular solid angle associated with $\bm{b}$ for a collision of 3 particles in 3D. 
This integral is evaluated by means of Monte Carlo importance sampling~\cite{Shui-thesis}.
Next, the energy-dependent three-body rate constant is defined as 

\begin{equation}
\label{c-20}
k_{3}(P_{0})=\frac{P_{0}}{\mu}\sigma_{\rm{rec}}(P_{0}).
\end{equation}

\begin{figure}[h!]
\centering
 \includegraphics[width=7.9 cm]{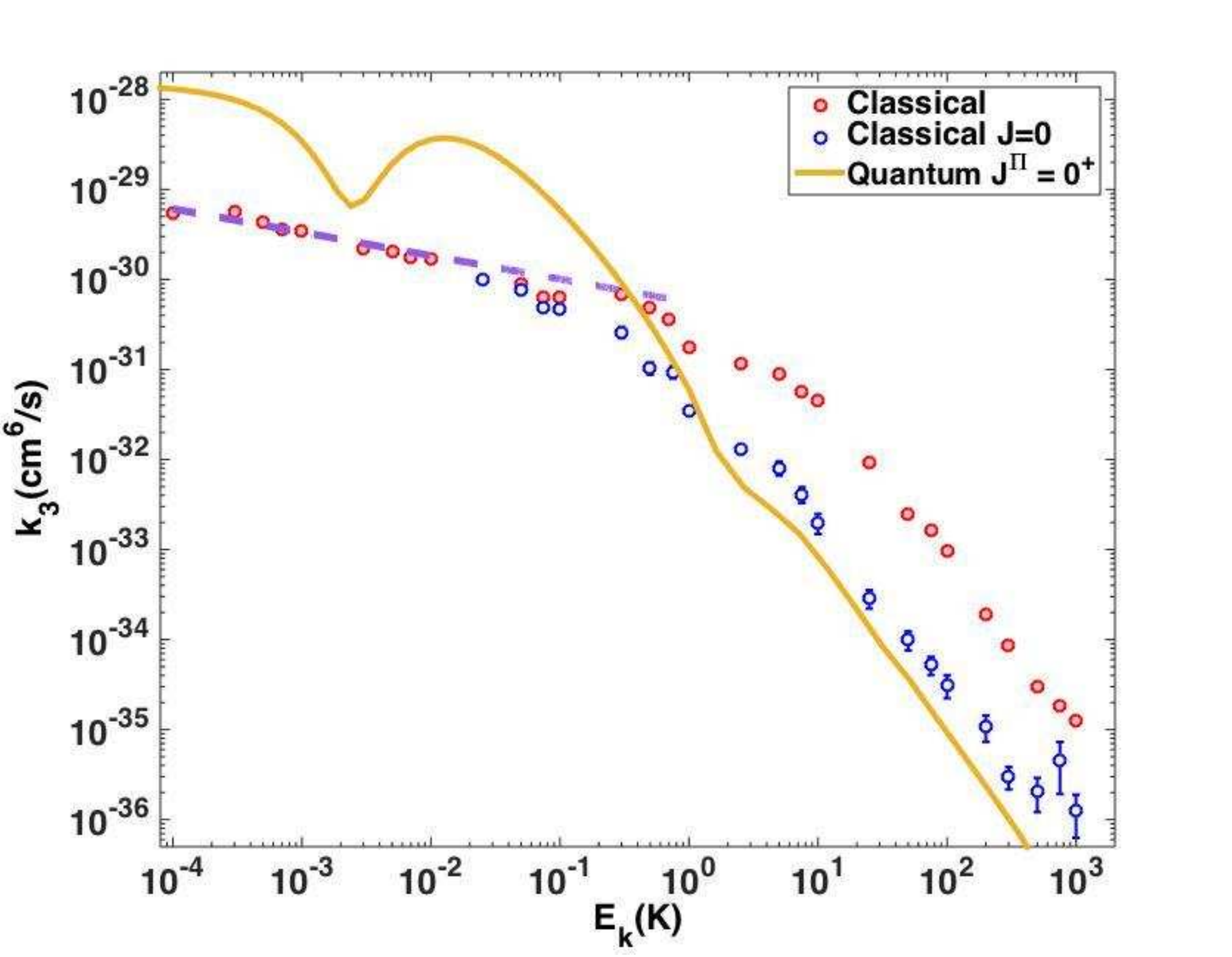}
 \caption{(Color online) Energy dependence of the three-body recombination rate of helium atoms in (cm$^6$/s), 
 {\it i.e.} He + He +He $\rightarrow$ He$_{2}$ + He. Classical trajectory results 
 following the classical treatment in 6D by means 
 of hyperspherical coordinates; red points. The same results but restricted to total angular momentum 
 $J=|\bm{\rho}_{1}\times \bm{P}_{1}+\bm{\rho}_{2}\times \bm{P}_{2}|=0$ shown as 
 the blue circles. The quantum calculation for a fixed angular momentum and 
 parity $J^{\pi}=0^{+}$ is plotted as the solid line. The quantal results show a convergence
 within better than about 15 \% for E = 1000 K regarding the number of channels included and the parameters employed in the calculations.}
 \label{He_recomb}
\end{figure}

The results for the He-He-He classical three-body recombination rate are shown in Fig.(\ref{He_recomb}). 
The quantum mechanical results shown in Fig.~(\ref{He_recomb})
 were obtained using the R-matrix method to solve the coupled hyperradial equations in the adiabatic hyperspherical 
 representation~\cite{lin1995PRep,esry1996PRA,wang2011PRA} to obtain 
the scattering matrix \cite{aymar1996RMP}. Fig.~(\ref{He_recomb}) shows that classical trajectory results for $J$ = 0 are in reasonably good 
agreement with the quantal results at collision energies $\sim$ 1 K, which is 
the same order of magnitude as the van der Waals energy: this serves approximately as the transition 
energy between ultracold physics and thermal physics, as 
was pointed out in Ref.~\cite{JPR-2014}.


\begin{figure}[h!]
\centering
 \includegraphics[width=7.9 cm]{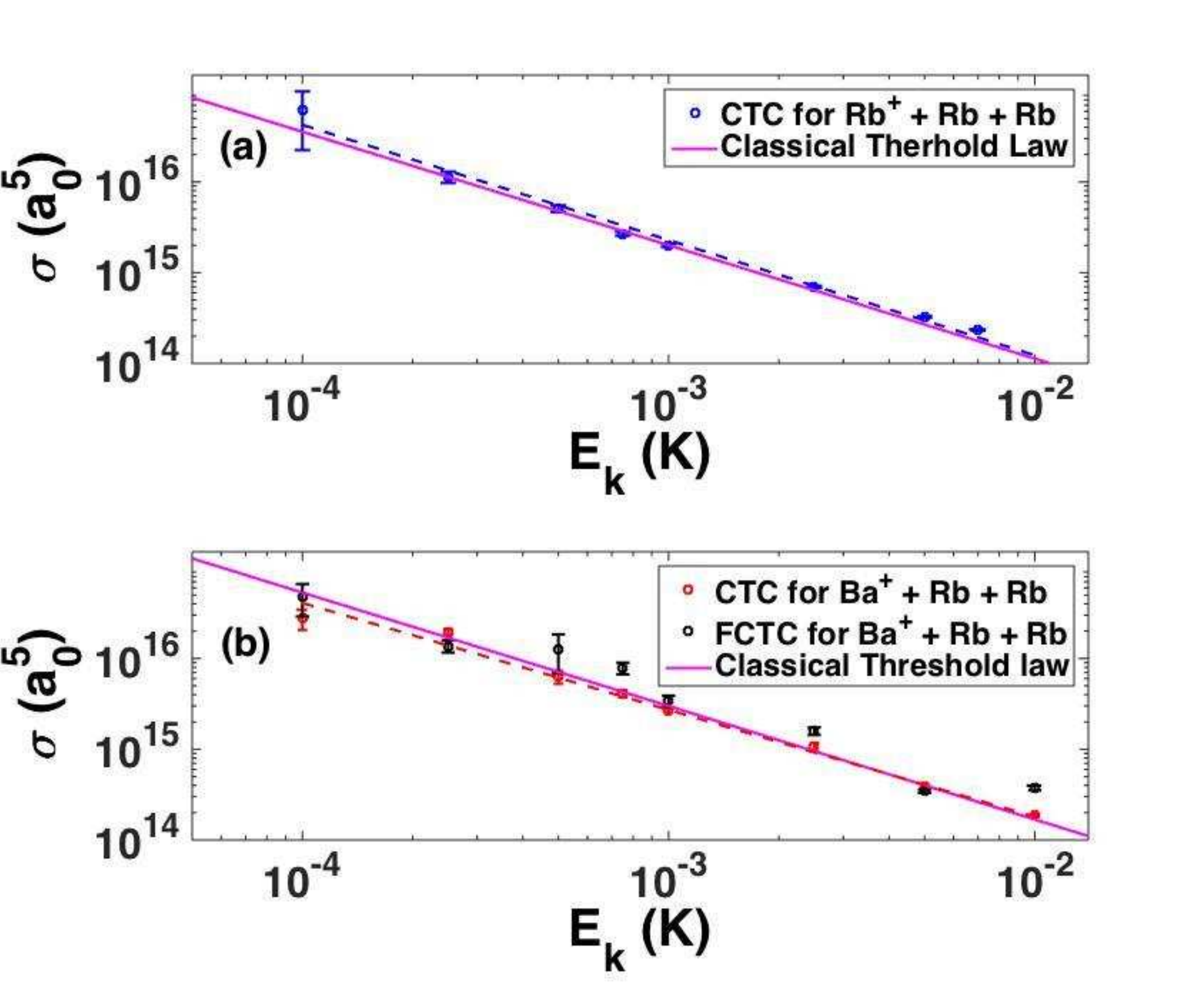}
 \caption{(color online) Three-body recombination cross section 
 (in a$_{0}^{5}$) as a function of the collision energy (in K). Panel (a) 
 $^{87}$Rb$^{+}$ - $^{87}$Rb - $^{87}$Rb ; the circles 
 represent the numerical results by means of CTC whereas the dashed
  line stands for the power-law fit of the points . Panel (b) 
  $^{138}$Ba$^{+}$ - $^{87}$Rb - $^{87}$Rb ; red circles
  represent the numerical results by means of CTC, the black circles denote 
  the results using FCTC (see text for details), the
  dashed line stands for the fit of the obtained CTC results. In both 
  panels, the solid magenta line represents the prediction based on the 
  derived classical threshold law. The fitting function 
  assumed for both systems is $\sigma(E_{k})$ = $\gamma E_{k}^{\beta}$. 
  Figure adapted from Ref.~\cite{JPR-2015}}
 \label{Ion-neutral-neutral}
\end{figure}


The same classical approach has been applied to the study of 
ion-neutral-neutral three body recombination at cold 
temperatures \cite{JPR-2015,Ulm}, which is important 
in ion-neutral hybrid trap experiments. Indeed three-body 
recombination reaction is the main loss mechanism for certain 
ionic species immersed in an ultracold high density 
neutral cloud\cite{harter2012prl,Harter:NaturePhysics:2013,Harter-2014,Ulm}. In particular, 
$^{87}$Rb$^{+}$ - $^{87}$Rb - $^{87}$Rb and 
$^{138}$Ba$^{+}$ - $^{87}$Rb - $^{87}$Rb were studied following 
the hyperspherical classical approach for collision energies ranging
from 100$\mu$K up to 10 mK, and the results are shown in 
Fig.(\ref{Ion-neutral-neutral}). 

The classical trajectory results presented 
in panels (a) and (b) of Fig.(\ref{Ion-neutral-neutral}) have been obtained 
by restricting one of the hyperangles associated with the momentum, guaranteeing 
that 95 \% of the collision energy goes along the vector joining the ion and the 
center of mass of the neutrals. This dynamical constraint is a consequence of 
the typical experimental conditions: the energy of the ion is typically orders of magnitude 
higher than the energy of the neutrals \cite{Harter-2014,Willitsch-2008,Willitsch-2012} because of the trapped ion micromotion.
As for the trajectories shown in Fig.(\ref{Class-1}), the same assumptions about the 
potential energy landscape and the same potentials were employed. 
In panels (a) and (b) of Fig.(\ref{Ion-neutral-neutral}), the 
three-body recombination rate versus collision energy shows a 
power law dependence. The physics behind this numerical observation, including its 
derivation, was explained in Ref.\cite{JPR-2015}, and is summarized below.

\subsubsection{Classical threshold law for three-body recombination: universality in cold chemistry}

In quantum mechanics the existence of threshold laws for elastic and inelastic 
collisions are familiar: the well-known Wigner threshold laws. These threshold
 laws represent the general trend of the cross section for different processes (here, elastic 
 and inelastic collisions) as functions of the collision energy. Analogously, there are also 
 classical threshold laws, such as the famous Langevin cross section~\cite{Langevin} which establishes the behavior of the cross section at low 
 collision energies for two-body ion-neutral collisions. Several years after that, Wannier found the classical
  threshold law for three-body collisions involving charged particles~\cite{wannier1953pr}, implementing 
  a different approach than Langevin developed. \textcolor{black}{Interestingly, in the case of three mixed-charge particles, e.g. two negative electrons escaping from a positive ion,} 
  the exponent in the energy-dependent rate constant is an irrational number \textcolor{black}{which has been 
  experimentally confirmed by measuring the double photoionization of He~\cite{Wiel-1972-JPR,Kossmann-1988-JPR}. \textcolor{black}{The unusual threshold law exponent 1.127... was also verified experimentally for the escape of two electrons from a singly-charged positive ion, as was discussed above in Sec.I ~\cite{Cvejanovic1974jpb,Bryant1982prl}.}  More 
  recently, the classical threshold law for three-body recombination involving neutrals with dominant long range van der Waals attraction, as well as for two
   neutrals and a single ion, have been obtained~\cite{JPR-2014,JPR-2015} following a classical 
  capture model~\cite{Levine}.}

At low collision energies the scattering properties are mainly dominated by the long-range 
tail of the two-body interaction, which here are represented as $V(R)\rightarrow -C_{s}/R^{s}$, 
with $s>2$. We define the maximum impact parameter $\tilde{b}_{\rm{max}}$ as the 
distance where the interaction potential is equal to the collision energy, {\it i.e.}

\begin{equation}
\label{th-1}
E=\frac{C_{s}}{\tilde{b}_{\rm{max}}^{s}}.
\end{equation}

\noindent
This denotes the distance where the motion of the colliding particles starts to deviate from the 
rectilinear uniform trajectory. This distance is the equivalent to the classical capture 
radius employed for the derivation of the Langevin cross section~\cite{Langevin,Levine}, 
but assuming $V(R)=- \alpha_d /2R^{4}$ in that case. In analogy with the classical capture model, it is 
assumed that all the trajectories with $\tilde{b}\le\tilde{b}_{\rm{max}}$ will lead to a three-body 
recombination event, which of course is likely to be an overestimate. The three-body recombination cross section can then be expressed 
as the following (by virtue of Eq.~(\ref{c-16}) )

\begin{eqnarray}
\label{th-2}
\sigma_{\rm{rec}}(E_{k}) &= &\left(\frac{m_{12}^3m_{3,12}^2}{\mu^5}\right)^{-1/2}\frac{8\pi^{2}}{3}\int_{0}^{\tilde{b}_{\rm{max}}(E_{k})}\tilde{b}^{4}d\tilde{b} \nonumber \\ 
&\propto& \tilde{b}^{5}_{\rm{max}}(E_{k}).
\end{eqnarray}

\noindent
 Eqs. (\ref{th-1}), (\ref{th-2}), after incorporating 
the relationship between momentum and energy ($P\propto E_{k}^{1/2}$), yield 

\begin{equation}
\label{th-3}
k_{3}(E_{k})\propto E_{k}^{1/2}\frac{1}{E_{k}^{5/s}}=E_{k}^{\frac{s-10}{2s}}.
\end{equation} 

\noindent
Thus, the neutral three-body recombination rate constant at low collision energies should 
vary with energy in proportion to $k_{3}(E_{k})\propto E_{k}^{1/3}$~\cite{JPR-2014}. Fig.~(\ref{He_recomb}) displays a numerical calculation of 
the three-body recombination rate coefficient for helium, showing that 
at low collision energies $k_{3}(E_{k})$ follows a power law dependence 
as a function of $E_{k}$. A fit of the classical trajectory results to the functional form 
($k_{3}(E_{k})=aE_{k}^{b}$) gives the dashed-purple line. The fitting parameters obtained are
a=(5.89$\pm$3.145 $\times$ 10$^{-31}$) cm$^6$/s and $b=-0.26\pm0.07$, which is 
consistent with the predicted $k_{3}(E_{k})\propto E_{k}^{-1/3}$ behavior~\cite{JPR-2014}.

The preceding derivation has assumed that all of the two-body interactions 
share identical long-range behavior, but an important case to consider is when different particle pairs have different interactions. This case has been explored in the context of determining the threshold law for ion-neutral-neutral three-body recombination~\cite{JPR-2015}. For that system, the two neutral atoms interact through a long-range van der Waals potential 
$V(R)=-C_{6}/R^6$, whereas the two ion-neutral interaction is dominated by the 
charge induced dipole interaction $V(R)=-\alpha_d/2R^4$. A classical capture 
model can be employed in analogy to the neutral three-body 
recombination derivation , but in this case the capture radius is given by 
 
\begin{equation}
\label{in-1}
E=\frac{\alpha_d}{\tilde{b}_{2\rm{max}}^{4}}.
\end{equation}

\noindent
Here it has been assumed that the longer-range attractive ion-neutral interaction 
dominates over the neutral-neutral interaction. Plugging Eq.~(\ref{in-1}) into 
Eq.(\ref{th-2}) the threshold behavior of the ion-neutral-neutral three-body 
recombination cross section is obtained as~\cite{JPR-2015}:

\begin{equation}
\label{in-2}
\sigma_(E_{k})\propto E_{k}^{-5/4}, 
\end{equation} 

\noindent
and the associated rate constant reads as

\begin{equation}
\label{in-3}
k_{3}(E_{k})\propto E_{k}^{-3/4}.
\end{equation} 

Fig.~(\ref{Ion-neutral-neutral}) presents the numerical results for ion-neutral-neutral 
three-body recombination computed classically at low collision energies, as the points 
in both panes of the figure. Also shown is the threshold law given by 
Eq.~(\ref{in-2}) as the magenta solid line.  Power law fits of the numerical results 
are represented by the dashed lines, and the fitting parameters are shown in Table I. 
Fig.~(\ref{Ion-neutral-neutral}) is a numerical confirmation of the predicted threshold law. And the fitted exponents in Table I confirm the validity of the derived classical threshold law.

\begin{table}[h]
\caption{Classical threshold law for the three-body recombination (TBR) cross section. A power law dependence of the TBR cross section as a function of the collision energy is assumed and used as a fitting function 
for the classical trajectory calculations (CTC) numerical results presented in Fig.~(\ref{Ion-neutral-neutral}). The errors quoted for the fitting parameters are associated with a confidence interval of 95 \%. Table adapted from Ref.~\cite{JPR-2015}}
\begin{tabular}{ c c c }
\hline
  System & $ \gamma$ ($a_{0}^{5}$)&$\beta$ (dimensionless) \\ 
  \hline
   $^{87}$Rb$^{+}$ - $^{87}$Rb - $^{87}$Rb & (7.94 $\pm$ 2.72) 10$^{11}$ & -1.178 $\pm$ 0.068 \\
   $^{138}$Ba$^{+}$ - $^{87}$Rb - $^{87}$Rb &  (3.57 $\pm$ 0.07) 10$^{11}$  & -1.269 $\pm$ 0.132  \\
   Classical threshold law & & -1.25 \\ \hline \hline
\end{tabular}
\end{table}

Ion-neutral-neutral collisions play an important role in hybrid trap 
experiments where a high density of neutrals are in presence of a single 
ion or several of them~\cite{Harter-2014,harter2012prl,Harter:NaturePhysics:2013,Ulm}, 
and hence hybrid trap experiments may elucidate the nature of ion-neutral-neutral 
three-body recombination. Indeed, very recently the three-body recombination rate 
for $^{138}$Ba$^{+}$ - $^{87}$Rb - $^{87}$Rb has been experimentally studied~\cite{Ulm}, 
and the results of the experimental three-body recombination rate as a function of the 
micromotion energy $E_{fMM}$ is shown as solid symbols in Fig.~(\ref{K3exp}). In the 
same figure the open symbols stand for the classical trajectory results computed using hyperspherical coordinates 
~\cite{JPR-2014}. The theoretical three-body recombination rate constant presented 
in Fig.~(\ref{K3exp})  is calculated by using the realistic energy distribution of the ion by 
means of a Monte Carlo simulation \cite{Ulm}. Fig.~(\ref{K3exp}) show a good agreement between the classical trajectory calculations in and the 
experimental results, confirming on one hand the validity of the classical Newtonian treatment in 
cold chemistry, and on the other hand, supporting the classical threshold law, Eq.~(\ref{in-3}). 
Apart from the confirmation of the threshold law, this also has important implications in the 
chemistry that occurs after a three-body recombination event in a hybrid trap experiment, since 
the classical results suggest that the dominant product channel will be the formation of shallow 
molecular ions~\cite{JPR-2015,Ulm}. In fact our estimates suggest that classical mechanics should give a reasonable description of the three-body recombination process for Ba$^+$-Rb-Rb down to energies of the order of 100-200nK.

\begin{figure}[h!]
\centering
 \includegraphics[width=7.5 cm]{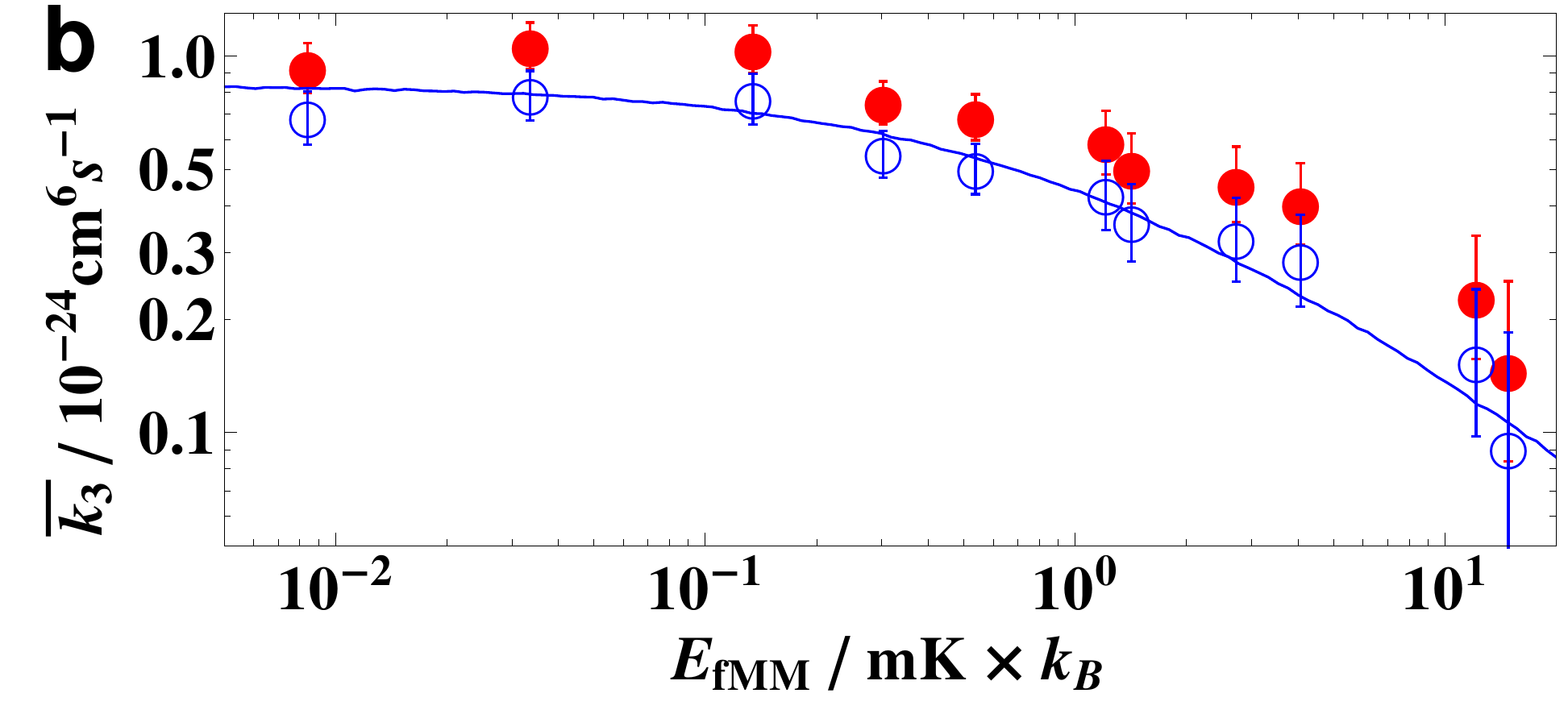}
 \caption{(Color online) The three-body recombination rate for  
 $^{138}$Ba$^{+}$ - $^{87}$Rb - $^{87}$Rb is presented as a function of the controlled 
 micromotion energy. The experimental values are represented by full circles, 
 whereas the theoretical prediction based on classical trajectory calculations 
 are shown as the open circles. Figure adapted from Ref.~\cite{Ulm}.} 
 \label{K3exp}
\end{figure}


\section{Conclusions}
\label{conclusions}
This article reviews developments in 
only a modest subset 
of the many extremely vigorous and dynamic topics in the field of few-body physics.  
\textcolor{black}{Anyone interested in exploring the multi-faceted aspects of this field and 
its interconnections with nuclear physics, chemical physics, and ultracold atomic and 
molecular physics is encouraged to explore the broader literature, and a good start would be 
the following set of review articles: ~\cite{NaidonEndoReview2016, blume2012rpp, wang2013amop,
wang2015AnnRev,rittenhouse2011JPB,braaten2006PRep,nielsen2001PRep,ZinnerJensen2013jpg,
Frederico2012xh,jensen2004RMP,baranov_condensed_2012,Petrov2012,sadeghpour2000JPB,lin1995PRep}.}
The ability to
control interparticle interactions through Fano-Feshbach resonances or confinement-induced
resonances continues to trigger novel experimental efforts, and theoretical progress on many fronts
continues to be rapid as well. This field promises to continue
stimulating new surprises in few-body and many-body physics in the years ahead.  Here is one 
wish list for desirable development of improved understanding in several areas:

\begin{itemize}
{\item (1)} Further insights into the extent of universality for heavy-heavy-light systems with short-range
interactions, including the role of van der Waals forces and the dependence on all parameters including 
the mass ratio.

{\item (2)} Detailed theory and experiment to map out the universality of three-particle systems with all 
masses different, including of course the possible role of van der Waals universality.

{\item (3)} Determination of $N$-body cluster states and recombination rates for both homonuclear
and heteronuclear systems, with $N>3$.  There is a large parameter space to explore here, just for $N=4$, 
for instance.

{\item (4)} Controlled applications of external electromagnetic field dressing of the few-body systems to
suppress or enhance inelastic processes.  Progress in this area could potentially lead to the formation
of a long-lived degenerate Bose gas at infinite two-body scattering length, currently limited by three-body
recombination processes.

{\item (5)} Further development of experiments and theory for mixed-dimension few-body systems.

{\item (6)} Experimental observation of log-periodic {\it energy dependence} of three-body recombination, which has been predicted to be visible for very large scattering lengths.

{\item (7)} In the BCS-BEC crossover problem with a Fermi gas having more than two spin components, it 
will be interesting to see whether macroscopic collapse of the gas is possible when the interaction 
scattering lengths are negative, as was predicted by \cite{blume2008PRAb,rittenhouse2008JPB}.

{\item (8)} The three-body and four-body systems with strong electric and/or magnetic dipolar interactions has received some theoretical attention\cite{wang2011PRL,wang2011PRLb} but little in the way of experimental tests to date, and in view of extensive current interest in polar molecule quantum gases, or strong magnetic dipolar atomic condensates and Fermi gases, far more work is needed from both perspectives.\cite{Kotochigova2014rpp,ZinnerJensen2013jpg,baranov_condensed_2012,wang2015AnnRev} 

{\item (9)}  It should be kept in mind that an accurate description of the two-body scattering length dependence of few-body phenomena hinges critically on having an accurate description of the atom-atom scattering lengths as functions of static and/or oscillating electromagnetic fields.  Theory has improved to the point where a number of alternative techniques can provide this data, when developed in conjunction with experiments, including full close-coupling calculations (CC) \cite{berni2013pra}, the asymptotic bound state model (ABM )\cite{tiecke2010pra}, and variants of multichannel quantum defect theory (MQDT) with or without the additional frame transformation approximation (MQDT-FT) \cite{ruzic2013,Gao2008,burke1998PRLb}.  Ref.\cite{Pires2014pra} and references therein provide comparisons of these different treatments, with application to the recently important heteronuclear system $^6$Li$-^{133}$Cs.  While these theoretical descriptions have been generally successful, extensions and improvements are still desirable in order to gain the fullest possible control of the two-body physics underlying all of the few-body physics addressed in this review.

{\item (10)} Further insights are also desired for systems such as the few-body version of the fractional quantum Hall problem, both in condensed matter systems and in ultracold atomic systems.  Initial studies by \cite{Daily2015PRB,Rittenhouse2016pra,Wooten2016epj} into that subject from the adiabatic hyperspherical perspective suggest that the corresponding 2D $N$-particle Schr\"odinger equation nearly separates in the hyperradial degree of freedom, both for bosons and fermions, as can be deduced from potential energy curves in those references.  Moreover, some of the intriguing degeneracy patterns observed in that problem are deserving of further exploration.
\end{itemize}

\section{Acknowledgments}
Different aspects of this work has been supported over the years by the National Science Foundation, the Department of Energy Office of Science, the AFOSR-MURI program on cold polar molecules, and the Bi-National Science Foundation. CHG specifically thanks many present and former colleagues, especially Doerte Blume, Jose D'Incao, and Brett Esry, for extended collaborations and access to their unpublished results and innumerable discussions.  CHG also acknowledges the Alexander von Humboldt Foundation for supporting an extended stay at the Max-Planck Institute for Nuclear Physics and the University of Heidelberg, where part of the review article was completed.  Many of the extensive revisions following submission were carried out with partial support from the Max-Planck Institute for the Physics of Complex Systems in Dresden.


\begin{thebibliography}{660}%
\makeatletter
\providecommand \@ifxundefined [1]{%
 \@ifx{#1\undefined}
}%
\providecommand \@ifnum [1]{%
 \ifnum #1\expandafter \@firstoftwo
 \else \expandafter \@secondoftwo
 \fi
}%
\providecommand \@ifx [1]{%
 \ifx #1\expandafter \@firstoftwo
 \else \expandafter \@secondoftwo
 \fi
}%
\providecommand \natexlab [1]{#1}%
\providecommand \enquote  [1]{``#1''}%
\providecommand \bibnamefont  [1]{#1}%
\providecommand \bibfnamefont [1]{#1}%
\providecommand \citenamefont [1]{#1}%
\providecommand \href@noop [0]{\@secondoftwo}%
\providecommand \href [0]{\begingroup \@sanitize@url \@href}%
\providecommand \@href[1]{\@@startlink{#1}\@@href}%
\providecommand \@@href[1]{\endgroup#1\@@endlink}%
\providecommand \@sanitize@url [0]{\catcode `\\12\catcode `\$12\catcode
  `\&12\catcode `\#12\catcode `\^12\catcode `\_12\catcode `\%12\relax}%
\providecommand \@@startlink[1]{}%
\providecommand \@@endlink[0]{}%
\providecommand \url  [0]{\begingroup\@sanitize@url \@url }%
\providecommand \@url [1]{\endgroup\@href {#1}{\urlprefix }}%
\providecommand \urlprefix  [0]{URL }%
\providecommand \Eprint [0]{\href }%
\providecommand \doibase [0]{http://dx.doi.org/}%
\providecommand \selectlanguage [0]{\@gobble}%
\providecommand \bibinfo  [0]{\@secondoftwo}%
\providecommand \bibfield  [0]{\@secondoftwo}%
\providecommand \translation [1]{[#1]}%
\providecommand \BibitemOpen [0]{}%
\providecommand \bibitemStop [0]{}%
\providecommand \bibitemNoStop [0]{.\EOS\space}%
\providecommand \EOS [0]{\spacefactor3000\relax}%
\providecommand \BibitemShut  [1]{\csname bibitem#1\endcsname}%
\let\auto@bib@innerbib\@empty
\bibitem [{\citenamefont {Adhikari}\ and\ \citenamefont
  {Fonseca}(1981)}]{AdhikariFonseca1981prd}%
  \BibitemOpen
  \bibfield  {author} {\bibinfo {author} {\bibnamefont {Adhikari},
  \bibfnamefont {Sadhan~K}}, \ and\ \bibinfo {author} {\bibfnamefont
  {Ant\'onio~C.}\ \bibnamefont {Fonseca}}} (\bibinfo {year} {1981}),\ \bibfield
   {title} {\enquote {\bibinfo {title} {Four-body efimov effect in a
  born-oppenheimer model},}\ }\href {\doibase 10.1103/PhysRevD.24.416}
  {\bibfield  {journal} {\bibinfo  {journal} {Phys. Rev. D}\ }\textbf {\bibinfo
  {volume} {24}},\ \bibinfo {pages} {416--425}}\BibitemShut {NoStop}%
\bibitem [{\citenamefont {Adhikari}\ \emph {et~al.}(1995)\citenamefont
  {Adhikari}, \citenamefont {Frederico},\ and\ \citenamefont
  {Goldman}}]{Adhikari1995prl}%
  \BibitemOpen
  \bibfield  {author} {\bibinfo {author} {\bibnamefont {Adhikari},
  \bibfnamefont {Sadhan~K}}, \bibinfo {author} {\bibfnamefont {T.}~\bibnamefont
  {Frederico}}, \ and\ \bibinfo {author} {\bibfnamefont {I.~D.}\ \bibnamefont
  {Goldman}}} (\bibinfo {year} {1995}),\ \bibfield  {title} {\enquote {\bibinfo
  {title} {Perturbative renormalization in quantum few-body problems},}\ }\href
  {\doibase 10.1103/PhysRevLett.74.487} {\bibfield  {journal} {\bibinfo
  {journal} {Phys. Rev. Lett.}\ }\textbf {\bibinfo {volume} {74}},\ \bibinfo
  {pages} {487--491}}\BibitemShut {NoStop}%
\bibitem [{\citenamefont {Akkineni}\ \emph {et~al.}(2007)\citenamefont
  {Akkineni}, \citenamefont {Ceperley},\ and\ \citenamefont
  {Trivedi}}]{Akkineni2007prb}%
  \BibitemOpen
  \bibfield  {author} {\bibinfo {author} {\bibnamefont {Akkineni},
  \bibfnamefont {Vamsi~K}}, \bibinfo {author} {\bibfnamefont {D.~M.}\
  \bibnamefont {Ceperley}}, \ and\ \bibinfo {author} {\bibfnamefont {Nandini}\
  \bibnamefont {Trivedi}}} (\bibinfo {year} {2007}),\ \bibfield  {title}
  {\enquote {\bibinfo {title} {Pairing and superfluid properties of dilute
  fermion gases at unitarity},}\ }\href {\doibase 10.1103/PhysRevB.76.165116}
  {\bibfield  {journal} {\bibinfo  {journal} {Phys. Rev. B}\ }\textbf {\bibinfo
  {volume} {76}},\ \bibinfo {pages} {165116}}\BibitemShut {NoStop}%
\bibitem [{\citenamefont {Alhassid}\ \emph {et~al.}(2008)\citenamefont
  {Alhassid}, \citenamefont {Bertsch},\ and\ \citenamefont
  {Fang}}]{alhassid2008PRL}%
  \BibitemOpen
  \bibfield  {author} {\bibinfo {author} {\bibnamefont {Alhassid},
  \bibfnamefont {Y}}, \bibinfo {author} {\bibfnamefont {G.~F.}\ \bibnamefont
  {Bertsch}}, \ and\ \bibinfo {author} {\bibfnamefont {L.}~\bibnamefont
  {Fang}}} (\bibinfo {year} {2008}),\ \bibfield  {title} {{\selectlanguage
  {English}\enquote {\bibinfo {title} {New effective interaction for the
  trapped {F}ermi gas},}\ }}\href@noop {} {\bibfield  {journal} {\bibinfo
  {journal} {Phys. Rev. Lett.}\ }\textbf {\bibinfo {volume} {100}}~(\bibinfo
  {number} {23}),\ \bibinfo {pages} {230401}}\BibitemShut {NoStop}%
\bibitem [{\citenamefont {{ALICE Collaboration}}(2016)}]{ALICE}%
  \BibitemOpen
  \bibfield  {author} {\bibinfo {author} {\bibnamefont {{ALICE
  Collaboration}},}} (\bibinfo {year} {2016}),\ \bibfield  {title} {\enquote
  {\bibinfo {title} {$^{3}_{\Lambda}${H} and $^{3}_{\bar{\Lambda}}$$\bar{H}$
  production in {Pb}-{Pb} collisions at $\sqrt{s_{NN}}$= 2.76 {TeV}},}\
  }\href@noop {} {\bibfield  {journal} {\bibinfo  {journal} {Phys. Lett. B}\
  }\textbf {\bibinfo {volume} {754}},\ \bibinfo {pages} {360}}\BibitemShut
  {NoStop}%
\bibitem [{\citenamefont {Alt}\ \emph {et~al.}(1967)\citenamefont {Alt},
  \citenamefont {Grassberger},\ and\ \citenamefont {Sandhas}}]{alt1967NPB}%
  \BibitemOpen
  \bibfield  {author} {\bibinfo {author} {\bibnamefont {Alt}, \bibfnamefont
  {E~O}}, \bibinfo {author} {\bibfnamefont {P.}~\bibnamefont {Grassberger}}, \
  and\ \bibinfo {author} {\bibfnamefont {W.}~\bibnamefont {Sandhas}}} (\bibinfo
  {year} {1967}),\ \bibfield  {title} {\enquote {\bibinfo {title} {Reduction of
  the three-particle collision problem to multi-channel two-particle
  {{Li}ppmann-Schwinger} equations},}\ }\href@noop {} {\bibfield  {journal}
  {\bibinfo  {journal} {Nucl. Phys. B}\ }\textbf {\bibinfo {volume}
  {2}}~(\bibinfo {number} {2}),\ \bibinfo {pages} {167 -- 180}}\BibitemShut
  {NoStop}%
\bibitem [{\citenamefont {Alvarez-Rodriguez}\ \emph {et~al.}(2007)\citenamefont
  {Alvarez-Rodriguez}, \citenamefont {Garrido}, \citenamefont {Jensen},
  \citenamefont {Fedorov},\ and\ \citenamefont
  {Fynbo}}]{alvarez-rodriguez2007EPJA}%
  \BibitemOpen
  \bibfield  {author} {\bibinfo {author} {\bibnamefont {Alvarez-Rodriguez},
  \bibfnamefont {R}}, \bibinfo {author} {\bibfnamefont {E.}~\bibnamefont
  {Garrido}}, \bibinfo {author} {\bibfnamefont {A.~S.}\ \bibnamefont {Jensen}},
  \bibinfo {author} {\bibfnamefont {D.~V.}\ \bibnamefont {Fedorov}}, \ and\
  \bibinfo {author} {\bibfnamefont {H.~O.~U.}\ \bibnamefont {Fynbo}}} (\bibinfo
  {year} {2007}),\ \bibfield  {title} {{\selectlanguage {English}\enquote
  {\bibinfo {title} {Structure of low-lying c-12 resonances},}\ }}\href@noop {}
  {\bibfield  {journal} {\bibinfo  {journal} {Euro. Phys. J. A}\ }\textbf
  {\bibinfo {volume} {31}}~(\bibinfo {number} {3}),\ \bibinfo {pages}
  {303--317}}\BibitemShut {NoStop}%
\bibitem [{\citenamefont {Alvarez-Rodriguez}\ \emph {et~al.}(2008)\citenamefont
  {Alvarez-Rodriguez}, \citenamefont {Jensen}, \citenamefont {Garrido},
  \citenamefont {Fedorov},\ and\ \citenamefont
  {Fynbo}}]{alvarez-rodriguez2008PRC}%
  \BibitemOpen
  \bibfield  {author} {\bibinfo {author} {\bibnamefont {Alvarez-Rodriguez},
  \bibfnamefont {R}}, \bibinfo {author} {\bibfnamefont {A.~S.}\ \bibnamefont
  {Jensen}}, \bibinfo {author} {\bibfnamefont {E.}~\bibnamefont {Garrido}},
  \bibinfo {author} {\bibfnamefont {D.~V.}\ \bibnamefont {Fedorov}}, \ and\
  \bibinfo {author} {\bibfnamefont {H.~O.~U.}\ \bibnamefont {Fynbo}}} (\bibinfo
  {year} {2008}),\ \bibfield  {title} {{\selectlanguage {English}\enquote
  {\bibinfo {title} {Momentum distributions of alpha particles from decaying
  low-lying c-12 resonances},}\ }}\href@noop {} {\bibfield  {journal} {\bibinfo
   {journal} {PRC}\ }\textbf {\bibinfo {volume} {77}}~(\bibinfo {number} {6}),\
  \bibinfo {pages} {064305}}\BibitemShut {NoStop}%
\bibitem [{\citenamefont {Amado}\ and\ \citenamefont
  {Greenwood}(1973)}]{Amado1973prd}%
  \BibitemOpen
  \bibfield  {author} {\bibinfo {author} {\bibnamefont {Amado}, \bibfnamefont
  {R~D}}, \ and\ \bibinfo {author} {\bibfnamefont {F.~C.}\ \bibnamefont
  {Greenwood}}} (\bibinfo {year} {1973}),\ \bibfield  {title} {\enquote
  {\bibinfo {title} {There is no {{E}fimov} effect for four or more
  particles},}\ }\href@noop {} {\bibfield  {journal} {\bibinfo  {journal}
  {PRD}\ }\textbf {\bibinfo {volume} {7}}~(\bibinfo {number} {8}),\ \bibinfo
  {pages} {2517--2519}}\BibitemShut {NoStop}%
\bibitem [{\citenamefont {Amorim}\ \emph {et~al.}(1997)\citenamefont {Amorim},
  \citenamefont {Frederico},\ and\ \citenamefont {Tomio}}]{Amorim-1997}%
  \BibitemOpen
  \bibfield  {author} {\bibinfo {author} {\bibnamefont {Amorim}, \bibfnamefont
  {A~E~A}}, \bibinfo {author} {\bibfnamefont {T.}~\bibnamefont {Frederico}}, \
  and\ \bibinfo {author} {\bibfnamefont {L.}~\bibnamefont {Tomio}}} (\bibinfo
  {year} {1997}),\ \bibfield  {title} {\enquote {\bibinfo {title} {{Universal
  aspects of Efimov states and light halo nuclei}},}\ }\href@noop {} {\bibfield
   {journal} {\bibinfo  {journal} {Phys. Rev. C}\ }\textbf {\bibinfo {volume}
  {56}},\ \bibinfo {pages} {R2378}}\BibitemShut {NoStop}%
\bibitem [{\citenamefont {Ancilotto}\ \emph {et~al.}(2015)\citenamefont
  {Ancilotto}, \citenamefont {Rossi}, \citenamefont {Salasnich},\ and\
  \citenamefont {Toigo}}]{Ancilotto2015fbs}%
  \BibitemOpen
  \bibfield  {author} {\bibinfo {author} {\bibnamefont {Ancilotto},
  \bibfnamefont {F}}, \bibinfo {author} {\bibfnamefont {M.}~\bibnamefont
  {Rossi}}, \bibinfo {author} {\bibfnamefont {L.}~\bibnamefont {Salasnich}}, \
  and\ \bibinfo {author} {\bibfnamefont {F.}~\bibnamefont {Toigo}}} (\bibinfo
  {year} {2015}),\ \bibfield  {title} {\enquote {\bibinfo {title} {Quenched
  dynamics of the momentum distribution of the unitary {B}ose gas},}\
  }\href@noop {} {\bibfield  {journal} {\bibinfo  {journal} {Few-Body Systems}\
  }\textbf {\bibinfo {volume} {56}},\ \bibinfo {pages} {801}}\BibitemShut
  {NoStop}%
\bibitem [{\citenamefont {Anderson}\ \emph {et~al.}(1995)\citenamefont
  {Anderson}, \citenamefont {Ensher}, \citenamefont {Matthews}, \citenamefont
  {Wieman},\ and\ \citenamefont {Cornell}}]{anderson1995}%
  \BibitemOpen
  \bibfield  {author} {\bibinfo {author} {\bibnamefont {Anderson},
  \bibfnamefont {M~H}}, \bibinfo {author} {\bibfnamefont {J.~R.}\ \bibnamefont
  {Ensher}}, \bibinfo {author} {\bibfnamefont {M.~R.}\ \bibnamefont
  {Matthews}}, \bibinfo {author} {\bibfnamefont {C.~E.}\ \bibnamefont
  {Wieman}}, \ and\ \bibinfo {author} {\bibfnamefont {E.~A.}\ \bibnamefont
  {Cornell}}} (\bibinfo {year} {1995}),\ \bibfield  {title} {\enquote {\bibinfo
  {title} {Observation of {B}ose-{E}instein condensation in a dilute atomic
  vapor},}\ }\href@noop {} {\bibfield  {journal} {\bibinfo  {journal}
  {Science}\ }\textbf {\bibinfo {volume} {269}}~(\bibinfo {number} {5221}),\
  \bibinfo {pages} {198--201}}\BibitemShut {NoStop}%
\bibitem [{\citenamefont {Andresen}\ \emph {et~al.}(2010)\citenamefont
  {Andresen}, \citenamefont {Ashkezari}, \citenamefont {Baquero-Ruiz},
  \citenamefont {Bertsche}, \citenamefont {Bowe}, \citenamefont {Butler},
  \citenamefont {Cesar}, \citenamefont {Chapman}, \citenamefont {Charlton},
  \citenamefont {Deller}, \citenamefont {Eriksson}, \citenamefont {Fajans},
  \citenamefont {Friesen}, \citenamefont {Fujiwara}, \citenamefont {Gill},
  \citenamefont {Gutierrez}, \citenamefont {Hangst}, \citenamefont {Hardy},
  \citenamefont {Hayden}, \citenamefont {Humphries}, \citenamefont {Hydomako},
  \citenamefont {Jenkins}, \citenamefont {Jonsell}, \citenamefont {Jorgensen},
  \citenamefont {Kurchaninov}, \citenamefont {Madsen}, \citenamefont {Menary},
  \citenamefont {Nolan}, \citenamefont {Olchanski}, \citenamefont {Olin},
  \citenamefont {Povilus}, \citenamefont {Pusa}, \citenamefont {Robicheaux},
  \citenamefont {Sarid}, \citenamefont {el~Nasr}, \citenamefont {Silveira},
  \citenamefont {So}, \citenamefont {Storey}, \citenamefont {Thompson},
  \citenamefont {van~der Werf}, \citenamefont {Wurtele},\ and\ \citenamefont
  {Yamazaki}}]{andresen2010NT}%
  \BibitemOpen
  \bibfield  {author} {\bibinfo {author} {\bibnamefont {Andresen},
  \bibfnamefont {G~B}}, \bibinfo {author} {\bibfnamefont {M.~D.}\ \bibnamefont
  {Ashkezari}}, \bibinfo {author} {\bibfnamefont {M.}~\bibnamefont
  {Baquero-Ruiz}}, \bibinfo {author} {\bibfnamefont {W.}~\bibnamefont
  {Bertsche}}, \bibinfo {author} {\bibfnamefont {P.~D.}\ \bibnamefont {Bowe}},
  \bibinfo {author} {\bibfnamefont {E.}~\bibnamefont {Butler}}, \bibinfo
  {author} {\bibfnamefont {C.~L.}\ \bibnamefont {Cesar}}, \bibinfo {author}
  {\bibfnamefont {S.}~\bibnamefont {Chapman}}, \bibinfo {author} {\bibfnamefont
  {M.}~\bibnamefont {Charlton}}, \bibinfo {author} {\bibfnamefont
  {A.}~\bibnamefont {Deller}}, \bibinfo {author} {\bibfnamefont
  {S.}~\bibnamefont {Eriksson}}, \bibinfo {author} {\bibfnamefont
  {J.}~\bibnamefont {Fajans}}, \bibinfo {author} {\bibfnamefont
  {T.}~\bibnamefont {Friesen}}, \bibinfo {author} {\bibfnamefont {M.~C.}\
  \bibnamefont {Fujiwara}}, \bibinfo {author} {\bibfnamefont {D.~R.}\
  \bibnamefont {Gill}}, \bibinfo {author} {\bibfnamefont {A.}~\bibnamefont
  {Gutierrez}}, \bibinfo {author} {\bibfnamefont {J.~S.}\ \bibnamefont
  {Hangst}}, \bibinfo {author} {\bibfnamefont {W.~N.}\ \bibnamefont {Hardy}},
  \bibinfo {author} {\bibfnamefont {M.~E.}\ \bibnamefont {Hayden}}, \bibinfo
  {author} {\bibfnamefont {A.~J.}\ \bibnamefont {Humphries}}, \bibinfo {author}
  {\bibfnamefont {R.}~\bibnamefont {Hydomako}}, \bibinfo {author}
  {\bibfnamefont {M.~J.}\ \bibnamefont {Jenkins}}, \bibinfo {author}
  {\bibfnamefont {S.}~\bibnamefont {Jonsell}}, \bibinfo {author} {\bibfnamefont
  {L.~V.}\ \bibnamefont {Jorgensen}}, \bibinfo {author} {\bibfnamefont
  {L.}~\bibnamefont {Kurchaninov}}, \bibinfo {author} {\bibfnamefont
  {N.}~\bibnamefont {Madsen}}, \bibinfo {author} {\bibfnamefont
  {S.}~\bibnamefont {Menary}}, \bibinfo {author} {\bibfnamefont
  {P.}~\bibnamefont {Nolan}}, \bibinfo {author} {\bibfnamefont
  {K.}~\bibnamefont {Olchanski}}, \bibinfo {author} {\bibfnamefont
  {A.}~\bibnamefont {Olin}}, \bibinfo {author} {\bibfnamefont {A.}~\bibnamefont
  {Povilus}}, \bibinfo {author} {\bibfnamefont {P.}~\bibnamefont {Pusa}},
  \bibinfo {author} {\bibfnamefont {F.}~\bibnamefont {Robicheaux}}, \bibinfo
  {author} {\bibfnamefont {E.}~\bibnamefont {Sarid}}, \bibinfo {author}
  {\bibfnamefont {S.~Seif}\ \bibnamefont {el~Nasr}}, \bibinfo {author}
  {\bibfnamefont {D.~M.}\ \bibnamefont {Silveira}}, \bibinfo {author}
  {\bibfnamefont {C.}~\bibnamefont {So}}, \bibinfo {author} {\bibfnamefont
  {J.~W.}\ \bibnamefont {Storey}}, \bibinfo {author} {\bibfnamefont {R.~I.}\
  \bibnamefont {Thompson}}, \bibinfo {author} {\bibfnamefont {D.~P.}\
  \bibnamefont {van~der Werf}}, \bibinfo {author} {\bibfnamefont {J.~S.}\
  \bibnamefont {Wurtele}}, \ and\ \bibinfo {author} {\bibfnamefont
  {Y.}~\bibnamefont {Yamazaki}}} (\bibinfo {year} {2010}),\ \bibfield  {title}
  {\enquote {\bibinfo {title} {Trapped antihydrogen},}\ }\href@noop {}
  {\bibfield  {journal} {\bibinfo  {journal} {Nature (London)}\ }\textbf
  {\bibinfo {volume} {468}}~(\bibinfo {number} {7324}),\ \bibinfo {pages}
  {673}}\BibitemShut {NoStop}%
\bibitem [{\citenamefont {Andresen}\ \emph {et~al.}(2011)\citenamefont
  {Andresen}, \citenamefont {Ashkezari}, \citenamefont {Baquero-Ruiz},
  \citenamefont {Bertsche}, \citenamefont {Bowe}, \citenamefont {Butler},
  \citenamefont {Cesar}, \citenamefont {Charlton}, \citenamefont {Deller},
  \citenamefont {Eriksson}, \citenamefont {Fajans}, \citenamefont {Friesen},
  \citenamefont {Fujiwara}, \citenamefont {Gill}, \citenamefont {Gutierrez},
  \citenamefont {Hangst}, \citenamefont {Hardy}, \citenamefont {Hayano},
  \citenamefont {Hayden}, \citenamefont {Humphries}, \citenamefont {Hydomako},
  \citenamefont {Jonsell}, \citenamefont {Kemp}, \citenamefont {Kurchaninov},
  \citenamefont {Madsen}, \citenamefont {Menary}, \citenamefont {Nolan},
  \citenamefont {Olchanski}, \citenamefont {Olin}, \citenamefont {Pusa},
  \citenamefont {Rasmussen}, \citenamefont {Robicheaux}, \citenamefont {Sarid},
  \citenamefont {Silveira}, \citenamefont {So}, \citenamefont {Storey},
  \citenamefont {Thompson}, \citenamefont {van~der Werf}, \citenamefont
  {Wurtele},\ and\ \citenamefont {Yamazaki}}]{andresen2011NP}%
  \BibitemOpen
  \bibfield  {author} {\bibinfo {author} {\bibnamefont {Andresen},
  \bibfnamefont {G~B}}, \bibinfo {author} {\bibfnamefont {M.~D.}\ \bibnamefont
  {Ashkezari}}, \bibinfo {author} {\bibfnamefont {M.}~\bibnamefont
  {Baquero-Ruiz}}, \bibinfo {author} {\bibfnamefont {W.}~\bibnamefont
  {Bertsche}}, \bibinfo {author} {\bibfnamefont {P.~D.}\ \bibnamefont {Bowe}},
  \bibinfo {author} {\bibfnamefont {E.}~\bibnamefont {Butler}}, \bibinfo
  {author} {\bibfnamefont {C.~L.}\ \bibnamefont {Cesar}}, \bibinfo {author}
  {\bibfnamefont {M.}~\bibnamefont {Charlton}}, \bibinfo {author}
  {\bibfnamefont {A.}~\bibnamefont {Deller}}, \bibinfo {author} {\bibfnamefont
  {S.}~\bibnamefont {Eriksson}}, \bibinfo {author} {\bibfnamefont
  {J.}~\bibnamefont {Fajans}}, \bibinfo {author} {\bibfnamefont
  {T.}~\bibnamefont {Friesen}}, \bibinfo {author} {\bibfnamefont {M.~C.}\
  \bibnamefont {Fujiwara}}, \bibinfo {author} {\bibfnamefont {D.~R.}\
  \bibnamefont {Gill}}, \bibinfo {author} {\bibfnamefont {A.}~\bibnamefont
  {Gutierrez}}, \bibinfo {author} {\bibfnamefont {J.~S.}\ \bibnamefont
  {Hangst}}, \bibinfo {author} {\bibfnamefont {W.~N.}\ \bibnamefont {Hardy}},
  \bibinfo {author} {\bibfnamefont {R.~S.}\ \bibnamefont {Hayano}}, \bibinfo
  {author} {\bibfnamefont {M.~E.}\ \bibnamefont {Hayden}}, \bibinfo {author}
  {\bibfnamefont {A.~J.}\ \bibnamefont {Humphries}}, \bibinfo {author}
  {\bibfnamefont {R.}~\bibnamefont {Hydomako}}, \bibinfo {author}
  {\bibfnamefont {S.}~\bibnamefont {Jonsell}}, \bibinfo {author} {\bibfnamefont
  {S.~L.}\ \bibnamefont {Kemp}}, \bibinfo {author} {\bibfnamefont
  {L.}~\bibnamefont {Kurchaninov}}, \bibinfo {author} {\bibfnamefont
  {N.}~\bibnamefont {Madsen}}, \bibinfo {author} {\bibfnamefont
  {S.}~\bibnamefont {Menary}}, \bibinfo {author} {\bibfnamefont
  {P.}~\bibnamefont {Nolan}}, \bibinfo {author} {\bibfnamefont
  {K.}~\bibnamefont {Olchanski}}, \bibinfo {author} {\bibfnamefont
  {A.}~\bibnamefont {Olin}}, \bibinfo {author} {\bibfnamefont {P.}~\bibnamefont
  {Pusa}}, \bibinfo {author} {\bibfnamefont {C.~O.}\ \bibnamefont {Rasmussen}},
  \bibinfo {author} {\bibfnamefont {F.}~\bibnamefont {Robicheaux}}, \bibinfo
  {author} {\bibfnamefont {E.}~\bibnamefont {Sarid}}, \bibinfo {author}
  {\bibfnamefont {D.~M.}\ \bibnamefont {Silveira}}, \bibinfo {author}
  {\bibfnamefont {C.}~\bibnamefont {So}}, \bibinfo {author} {\bibfnamefont
  {J.~W.}\ \bibnamefont {Storey}}, \bibinfo {author} {\bibfnamefont {R.~I.}\
  \bibnamefont {Thompson}}, \bibinfo {author} {\bibfnamefont {D.~P.}\
  \bibnamefont {van~der Werf}}, \bibinfo {author} {\bibfnamefont {J.~S.}\
  \bibnamefont {Wurtele}}, \ and\ \bibinfo {author} {\bibfnamefont
  {Y.}~\bibnamefont {Yamazaki}}} (\bibinfo {year} {2011}),\ \bibfield  {title}
  {\enquote {\bibinfo {title} {Confinement of antihydrogen for 1,000
  seconds},}\ }\href@noop {} {\bibfield  {journal} {\bibinfo  {journal} {Nucl.
  Phys.}\ }\textbf {\bibinfo {volume} {7}}~(\bibinfo {number} {7}),\ \bibinfo
  {pages} {558}}\BibitemShut {NoStop}%
\bibitem [{\citenamefont {Astrakharchik}\ \emph {et~al.}(2005)\citenamefont
  {Astrakharchik}, \citenamefont {Boronat}, \citenamefont {Casulleras},\ and\
  \citenamefont {Giorgini}}]{astrakharchik2005beyond}%
  \BibitemOpen
  \bibfield  {author} {\bibinfo {author} {\bibnamefont {Astrakharchik},
  \bibfnamefont {G~E}}, \bibinfo {author} {\bibfnamefont {J.}~\bibnamefont
  {Boronat}}, \bibinfo {author} {\bibfnamefont {J.}~\bibnamefont {Casulleras}},
  \ and\ \bibinfo {author} {\bibfnamefont {S.}~\bibnamefont {Giorgini}}}
  (\bibinfo {year} {2005}),\ \bibfield  {title} {\enquote {\bibinfo {title}
  {{Beyond the Tonks-Girardeau gas: Strongly correlated regime in
  quasi-one-dimensional Bose gases}},}\ }\href@noop {} {\bibfield  {journal}
  {\bibinfo  {journal} {Phys. Rev. Lett.}\ }\textbf {\bibinfo {volume}
  {95}}~(\bibinfo {number} {19}),\ \bibinfo {pages} {190407}}\BibitemShut
  {NoStop}%
\bibitem [{\citenamefont {Astrakharchik}\ and\ \citenamefont
  {Pitaevskii}(2004)}]{astrakharchik_motion_2004}%
  \BibitemOpen
  \bibfield  {author} {\bibinfo {author} {\bibnamefont {Astrakharchik},
  \bibfnamefont {G~E}}, \ and\ \bibinfo {author} {\bibfnamefont {L.~P.}\
  \bibnamefont {Pitaevskii}}} (\bibinfo {year} {2004}),\ \bibfield  {title}
  {\enquote {\bibinfo {title} {Motion of a heavy impurity through a
  {Bose}-{Einstein} condensate},}\ }\href {\doibase 10.1103/PhysRevA.70.013608}
  {\bibfield  {journal} {\bibinfo  {journal} {Physical Review A}\ }\textbf
  {\bibinfo {volume} {70}}~(\bibinfo {number} {1}),\ \bibinfo {pages}
  {013608}}\BibitemShut {NoStop}%
\bibitem [{\citenamefont {Avery}(1989)}]{avery1989}%
  \BibitemOpen
  \bibfield  {author} {\bibinfo {author} {\bibnamefont {Avery}, \bibfnamefont
  {J}}} (\bibinfo {year} {1989}),\ \href@noop {} {\emph {\bibinfo {title}
  {Hyperspherical Harmonics: Applications in Quantum Theory}}}\ (\bibinfo
  {publisher} {Kluwer Academic Publishers},\ \bibinfo {address} {Norwell,
  MA})\BibitemShut {NoStop}%
\bibitem [{\citenamefont {Aymar}\ \emph {et~al.}(1996)\citenamefont {Aymar},
  \citenamefont {Greene},\ and\ \citenamefont {Luc-Koenig}}]{aymar1996RMP}%
  \BibitemOpen
  \bibfield  {author} {\bibinfo {author} {\bibnamefont {Aymar}, \bibfnamefont
  {M}}, \bibinfo {author} {\bibfnamefont {C.~H.}\ \bibnamefont {Greene}}, \
  and\ \bibinfo {author} {\bibfnamefont {E}~\bibnamefont {Luc-Koenig}}}
  (\bibinfo {year} {1996}),\ \bibfield  {title} {{\selectlanguage
  {English}\enquote {\bibinfo {title} {Multichannel {R}ydberg spectroscopy of
  complex atoms},}\ }}\href@noop {} {\bibfield  {journal} {\bibinfo  {journal}
  {Rev. Mod. Phys.}\ }\textbf {\bibinfo {volume} {68}}~(\bibinfo {number}
  {4}),\ \bibinfo {pages} {1015--1123}}\BibitemShut {NoStop}%
\bibitem [{\citenamefont {Aziz}\ \emph {et~al.}(1995)\citenamefont {Aziz},
  \citenamefont {Janzen},\ and\ \citenamefont {Moldover}}]{Aziz-1995}%
  \BibitemOpen
  \bibfield  {author} {\bibinfo {author} {\bibnamefont {Aziz}, \bibfnamefont
  {R~A}}, \bibinfo {author} {\bibfnamefont {A.~R.}\ \bibnamefont {Janzen}}, \
  and\ \bibinfo {author} {\bibfnamefont {M.~R.}\ \bibnamefont {Moldover}}}
  (\bibinfo {year} {1995}),\ \bibfield  {title} {\enquote {\bibinfo {title}
  {{\em Ab Initio} calculations for helium: A standard for transport property
  measurements},}\ }\href@noop {} {\bibfield  {journal} {\bibinfo  {journal}
  {Phys. Rev. Lett}\ }\textbf {\bibinfo {volume} {74}},\ \bibinfo {pages}
  {1586}}\BibitemShut {NoStop}%
\bibitem [{\citenamefont {Baranov}(2008)}]{baranov_theoretical_2008}%
  \BibitemOpen
  \bibfield  {author} {\bibinfo {author} {\bibnamefont {Baranov}, \bibfnamefont
  {M~A}}} (\bibinfo {year} {2008}),\ \bibfield  {title} {\enquote {\bibinfo
  {title} {Theoretical progress in many-body physics with ultracold dipolar
  gases},}\ }\href {\doibase 10.1016/j.physrep.2008.04.007} {\bibfield
  {journal} {\bibinfo  {journal} {Physics Reports-Review Section of Physics
  Letters}\ }\textbf {\bibinfo {volume} {464}}~(\bibinfo {number} {3}),\
  \bibinfo {pages} {71--111}}\BibitemShut {NoStop}%
\bibitem [{\citenamefont {Baranov}\ \emph {et~al.}(2012)\citenamefont
  {Baranov}, \citenamefont {Dalmonte}, \citenamefont {Pupillo},\ and\
  \citenamefont {Zoller}}]{baranov_condensed_2012}%
  \BibitemOpen
  \bibfield  {author} {\bibinfo {author} {\bibnamefont {Baranov}, \bibfnamefont
  {M~A}}, \bibinfo {author} {\bibfnamefont {M.}~\bibnamefont {Dalmonte}},
  \bibinfo {author} {\bibfnamefont {G.}~\bibnamefont {Pupillo}}, \ and\
  \bibinfo {author} {\bibfnamefont {P.}~\bibnamefont {Zoller}}} (\bibinfo
  {year} {2012}),\ \bibfield  {title} {\enquote {\bibinfo {title} {Condensed
  {Matter} {Theory} of {Dipolar} {Quantum} {Gases}},}\ }\href {\doibase
  10.1021/cr2003568} {\bibfield  {journal} {\bibinfo  {journal} {Chemical
  Reviews}\ }\textbf {\bibinfo {volume} {112}}~(\bibinfo {number} {9}),\
  \bibinfo {pages} {5012--5061}}\BibitemShut {NoStop}%
\bibitem [{\citenamefont {Barletta}\ and\ \citenamefont
  {Kievsky}(2008)}]{barletta2008FBS}%
  \BibitemOpen
  \bibfield  {author} {\bibinfo {author} {\bibnamefont {Barletta},
  \bibfnamefont {P}}, \ and\ \bibinfo {author} {\bibfnamefont {A.}~\bibnamefont
  {Kievsky}}} (\bibinfo {year} {2008}),\ \bibfield  {title} {{\selectlanguage
  {English}\enquote {\bibinfo {title} {Continuum three-body states using the
  hyperspherical adiabatic basis set},}\ }}\href@noop {} {\bibfield  {journal}
  {\bibinfo  {journal} {Few-Body Systems}\ }\textbf {\bibinfo {volume}
  {44}}~(\bibinfo {number} {1-4}),\ \bibinfo {pages} {371--373}}\BibitemShut
  {NoStop}%
\bibitem [{\citenamefont {Barletta}\ and\ \citenamefont
  {Kievsky}(2009)}]{barletta2009FBS}%
  \BibitemOpen
  \bibfield  {author} {\bibinfo {author} {\bibnamefont {Barletta},
  \bibfnamefont {P}}, \ and\ \bibinfo {author} {\bibfnamefont {A.}~\bibnamefont
  {Kievsky}}} (\bibinfo {year} {2009}),\ \bibfield  {title} {{\selectlanguage
  {English}\enquote {\bibinfo {title} {Scattering states of three-body systems
  with the hyperspherical adiabatic method},}\ }}\href@noop {} {\bibfield
  {journal} {\bibinfo  {journal} {Few-Body Systems}\ }\textbf {\bibinfo
  {volume} {45}}~(\bibinfo {number} {2-4}),\ \bibinfo {pages}
  {123--125}}\BibitemShut {NoStop}%
\bibitem [{\citenamefont {Barletta}\ \emph {et~al.}(2009)\citenamefont
  {Barletta}, \citenamefont {Romero-Redondo}, \citenamefont {Kievsky},
  \citenamefont {Viviani},\ and\ \citenamefont {Garrido}}]{barletta2009PRL}%
  \BibitemOpen
  \bibfield  {author} {\bibinfo {author} {\bibnamefont {Barletta},
  \bibfnamefont {P}}, \bibinfo {author} {\bibfnamefont {C.}~\bibnamefont
  {Romero-Redondo}}, \bibinfo {author} {\bibfnamefont {A.}~\bibnamefont
  {Kievsky}}, \bibinfo {author} {\bibfnamefont {M.}~\bibnamefont {Viviani}}, \
  and\ \bibinfo {author} {\bibfnamefont {E.}~\bibnamefont {Garrido}}} (\bibinfo
  {year} {2009}),\ \bibfield  {title} {{\selectlanguage {English}\enquote
  {\bibinfo {title} {Integral relations for three-body continuum states with
  the adiabatic expansion},}\ }}\href@noop {} {\bibfield  {journal} {\bibinfo
  {journal} {Phys. Rev. Lett.}\ }\textbf {\bibinfo {volume} {103}}~(\bibinfo
  {number} {9}),\ \bibinfo {pages} {090402}}\BibitemShut {NoStop}%
\bibitem [{\citenamefont {Barnea}(1999)}]{barnea1999JMP}%
  \BibitemOpen
  \bibfield  {author} {\bibinfo {author} {\bibnamefont {Barnea}, \bibfnamefont
  {N}}} (\bibinfo {year} {1999}),\ \bibfield  {title} {\enquote {\bibinfo
  {title} {A recursive method for the construction of irreducible
  representations of the orthogonal group {O(N)}},}\ }\href@noop {} {\bibfield
  {journal} {\bibinfo  {journal} {J. Math. Phys.}\ }\textbf {\bibinfo {volume}
  {40}},\ \bibinfo {pages} {1011}}\BibitemShut {NoStop}%
\bibitem [{\citenamefont {Barnea}\ and\ \citenamefont
  {Novoselsky}(1997)}]{barnea1997AP}%
  \BibitemOpen
  \bibfield  {author} {\bibinfo {author} {\bibnamefont {Barnea}, \bibfnamefont
  {N}}, \ and\ \bibinfo {author} {\bibfnamefont {A.}~\bibnamefont
  {Novoselsky}}} (\bibinfo {year} {1997}),\ \bibfield  {title} {\enquote
  {\bibinfo {title} {Construction of hyperspherical functions symmetrized with
  respect to the orthogonal and the symmetric groups},}\ }\href@noop {}
  {\bibfield  {journal} {\bibinfo  {journal} {Ann. Phys.}\ }\textbf {\bibinfo
  {volume} {256}}~(\bibinfo {number} {2}),\ \bibinfo {pages} {192}}\BibitemShut
  {NoStop}%
\bibitem [{\citenamefont {Barontini}\ \emph {et~al.}(2009)\citenamefont
  {Barontini}, \citenamefont {Weber}, \citenamefont {Rabatti}, \citenamefont
  {Catani}, \citenamefont {Thalhammer}, \citenamefont {Inguscio},\ and\
  \citenamefont {Minardi}}]{barontini2009PRL}%
  \BibitemOpen
  \bibfield  {author} {\bibinfo {author} {\bibnamefont {Barontini},
  \bibfnamefont {G}}, \bibinfo {author} {\bibfnamefont {C.}~\bibnamefont
  {Weber}}, \bibinfo {author} {\bibfnamefont {F.}~\bibnamefont {Rabatti}},
  \bibinfo {author} {\bibfnamefont {J.}~\bibnamefont {Catani}}, \bibinfo
  {author} {\bibfnamefont {G.}~\bibnamefont {Thalhammer}}, \bibinfo {author}
  {\bibfnamefont {M.}~\bibnamefont {Inguscio}}, \ and\ \bibinfo {author}
  {\bibfnamefont {F.}~\bibnamefont {Minardi}}} (\bibinfo {year} {2009}),\
  \bibfield  {title} {{\selectlanguage {English}\enquote {\bibinfo {title}
  {Observation of heteronuclear atomic {E}fimov resonances},}\ }}\href@noop {}
  {\bibfield  {journal} {\bibinfo  {journal} {Phys. Rev. Lett.}\ }\textbf
  {\bibinfo {volume} {103}}~(\bibinfo {number} {4}),\ \bibinfo {pages}
  {043201}}\BibitemShut {NoStop}%
\bibitem [{\citenamefont {Bartlett}\ \emph {et~al.}(2003)\citenamefont
  {Bartlett}, \citenamefont {Bray}, \citenamefont {Jones}, \citenamefont
  {Stelbovics}, \citenamefont {Kadyrov}, \citenamefont {Bartschat},
  \citenamefont {Steeg}, \citenamefont {Scott},\ and\ \citenamefont
  {Burke}}]{Bartlett2003}%
  \BibitemOpen
  \bibfield  {author} {\bibinfo {author} {\bibnamefont {Bartlett},
  \bibfnamefont {P~L}}, \bibinfo {author} {\bibfnamefont {I.}~\bibnamefont
  {Bray}}, \bibinfo {author} {\bibfnamefont {S.}~\bibnamefont {Jones}},
  \bibinfo {author} {\bibfnamefont {A.~T.}\ \bibnamefont {Stelbovics}},
  \bibinfo {author} {\bibfnamefont {A.~S.}\ \bibnamefont {Kadyrov}}, \bibinfo
  {author} {\bibfnamefont {K.}~\bibnamefont {Bartschat}}, \bibinfo {author}
  {\bibfnamefont {G.~L.~V.}\ \bibnamefont {Steeg}}, \bibinfo {author}
  {\bibfnamefont {M.~P.}\ \bibnamefont {Scott}}, \ and\ \bibinfo {author}
  {\bibfnamefont {P.~G.}\ \bibnamefont {Burke}}} (\bibinfo {year} {2003}),\
  \bibfield  {title} {\enquote {\bibinfo {title} {Unambiguous ionization
  amplitudes for electron-hydrogen scattering},}\ }\href@noop {} {\bibfield
  {journal} {\bibinfo  {journal} {Phys. Rev. A}\ }\textbf {\bibinfo {volume}
  {68}}~(\bibinfo {number} {2}),\ \bibinfo {pages} {020702}}\BibitemShut
  {NoStop}%
\bibitem [{\citenamefont {Bartolo}\ \emph {et~al.}(2013)\citenamefont
  {Bartolo}, \citenamefont {Papoular}, \citenamefont {Barbiero}, \citenamefont
  {Menotti},\ and\ \citenamefont {Recati}}]{bartolo2013dipolar}%
  \BibitemOpen
  \bibfield  {author} {\bibinfo {author} {\bibnamefont {Bartolo}, \bibfnamefont
  {N}}, \bibinfo {author} {\bibfnamefont {D.~J.}\ \bibnamefont {Papoular}},
  \bibinfo {author} {\bibfnamefont {L.}~\bibnamefont {Barbiero}}, \bibinfo
  {author} {\bibfnamefont {C.}~\bibnamefont {Menotti}}, \ and\ \bibinfo
  {author} {\bibfnamefont {A.}~\bibnamefont {Recati}}} (\bibinfo {year}
  {2013}),\ \bibfield  {title} {\enquote {\bibinfo {title} {Dipolar-induced
  resonance for ultracold bosons in a quasi-one-dimensional optical lattice},}\
  }\href@noop {} {\bibfield  {journal} {\bibinfo  {journal} {Phys. Rev. A}\
  }\textbf {\bibinfo {volume} {88}}~(\bibinfo {number} {2}),\ \bibinfo {pages}
  {023603}}\BibitemShut {NoStop}%
\bibitem [{\citenamefont {Basak}\ and\ \citenamefont
  {Cohen}(1979)}]{BasakCohen1979prb}%
  \BibitemOpen
  \bibfield  {author} {\bibinfo {author} {\bibnamefont {Basak}, \bibfnamefont
  {Soumen}}, \ and\ \bibinfo {author} {\bibfnamefont {Morrel~H.}\ \bibnamefont
  {Cohen}}} (\bibinfo {year} {1979}),\ \bibfield  {title} {\enquote {\bibinfo
  {title} {Deformation-potential theory for the mobility of excess electrons in
  liquid argon},}\ }\href {\doibase 10.1103/PhysRevB.20.3404} {\bibfield
  {journal} {\bibinfo  {journal} {Phys. Rev. B}\ }\textbf {\bibinfo {volume}
  {20}},\ \bibinfo {pages} {3404--3414}}\BibitemShut {NoStop}%
\bibitem [{\citenamefont {{Bazak}}\ and\ \citenamefont
  {{Petrov}}(2016)}]{BazakPetrov2016}%
  \BibitemOpen
  \bibfield  {author} {\bibinfo {author} {\bibnamefont {{Bazak}}, \bibfnamefont
  {B}}, \ and\ \bibinfo {author} {\bibfnamefont {D.~S.}\ \bibnamefont
  {{Petrov}}}} (\bibinfo {year} {2016}),\ \bibfield  {title} {\enquote
  {\bibinfo {title} {{Five-body Efimov effect and universal pentamer in
  fermionic mixtures}},}\ }\href@noop {} {\bibfield  {journal} {\bibinfo
  {journal} {ArXiv e-prints}\ }}\Eprint {http://arxiv.org/abs/1612.09341}
  {arXiv:1612.09341 [cond-mat.quant-gas]} \BibitemShut {NoStop}%
\bibitem [{\citenamefont {Bedaque}\ \emph {et~al.}(2000)\citenamefont
  {Bedaque}, \citenamefont {Braaten},\ and\ \citenamefont
  {Hammer}}]{bedaque2000PRL}%
  \BibitemOpen
  \bibfield  {author} {\bibinfo {author} {\bibnamefont {Bedaque}, \bibfnamefont
  {P~F}}, \bibinfo {author} {\bibfnamefont {E.}~\bibnamefont {Braaten}}, \ and\
  \bibinfo {author} {\bibfnamefont {H.{-}W.}\ \bibnamefont {Hammer}}} (\bibinfo
  {year} {2000}),\ \bibfield  {title} {{\selectlanguage {English}\enquote
  {\bibinfo {title} {Three-body recombination in {B}ose gases with large
  scattering length},}\ }}\href@noop {} {\bibfield  {journal} {\bibinfo
  {journal} {Phys. Rev. Lett.}\ }\textbf {\bibinfo {volume} {85}}~(\bibinfo
  {number} {5}),\ \bibinfo {pages} {908--911}}\BibitemShut {NoStop}%
\bibitem [{\citenamefont {Bei-Bing}\ and\ \citenamefont
  {Shao-Long}(2009)}]{bei-bing_polaron_2009}%
  \BibitemOpen
  \bibfield  {author} {\bibinfo {author} {\bibnamefont {Bei-Bing},
  \bibfnamefont {Huang}}, \ and\ \bibinfo {author} {\bibfnamefont {Wan}\
  \bibnamefont {Shao-Long}}} (\bibinfo {year} {2009}),\ \bibfield  {title}
  {\enquote {\bibinfo {title} {Polaron in {Bose}-{Einstein}-{Condensation}
  {System}},}\ }\href@noop {} {\bibfield  {journal} {\bibinfo  {journal}
  {Chinese Physics Letters}\ }\textbf {\bibinfo {volume} {26}}~(\bibinfo
  {number} {8}),\ \bibinfo {pages} {080302}}\BibitemShut {NoStop}%
\bibitem [{\citenamefont {Bellotti}\ \emph {et~al.}(2016)\citenamefont
  {Bellotti}, \citenamefont {Frederico}, \citenamefont {Yamashita},
  \citenamefont {Fedorov}, \citenamefont {Jensen},\ and\ \citenamefont
  {Zinner}}]{BellottiZinner2016}%
  \BibitemOpen
  \bibfield  {author} {\bibinfo {author} {\bibnamefont {Bellotti},
  \bibfnamefont {F~F}}, \bibinfo {author} {\bibfnamefont {T}~\bibnamefont
  {Frederico}}, \bibinfo {author} {\bibfnamefont {M~T}\ \bibnamefont
  {Yamashita}}, \bibinfo {author} {\bibfnamefont {D~V}\ \bibnamefont
  {Fedorov}}, \bibinfo {author} {\bibfnamefont {A~S}\ \bibnamefont {Jensen}}, \
  and\ \bibinfo {author} {\bibfnamefont {N~T}\ \bibnamefont {Zinner}}}
  (\bibinfo {year} {2016}),\ \bibfield  {title} {\enquote {\bibinfo {title}
  {Three-body bound states of two bosonic impurities immersed in a fermi sea in
  2d},}\ }\href {http://stacks.iop.org/1367-2630/18/i=4/a=043023} {\bibfield
  {journal} {\bibinfo  {journal} {New Journal of Physics}\ }\textbf {\bibinfo
  {volume} {18}}~(\bibinfo {number} {4}),\ \bibinfo {pages}
  {043023}}\BibitemShut {NoStop}%
\bibitem [{\citenamefont {Benayoun}\ \emph {et~al.}(1981)\citenamefont
  {Benayoun}, \citenamefont {Cignoux},\ and\ \citenamefont
  {Chauvin}}]{Benayoun-1981}%
  \BibitemOpen
  \bibfield  {author} {\bibinfo {author} {\bibnamefont {Benayoun},
  \bibfnamefont {J~J}}, \bibinfo {author} {\bibfnamefont {C.}~\bibnamefont
  {Cignoux}}, \ and\ \bibinfo {author} {\bibfnamefont {J.}~\bibnamefont
  {Chauvin}}} (\bibinfo {year} {1981}),\ \bibfield  {title} {\enquote {\bibinfo
  {title} {Neutron-deuteron scattering at zero energy with realistic
  nucleon-nucleon interactions},}\ }\href@noop {} {\bibfield  {journal}
  {\bibinfo  {journal} {Phys. Rev. C}\ }\textbf {\bibinfo {volume} {23}},\
  \bibinfo {pages} {1854}}\BibitemShut {NoStop}%
\bibitem [{\citenamefont {Berezinskii}(1971)}]{berezins.vl_destruction_1971}%
  \BibitemOpen
  \bibfield  {author} {\bibinfo {author} {\bibnamefont {Berezinskii},
  \bibfnamefont {VL}}} (\bibinfo {year} {1971}),\ \bibfield  {title} {\enquote
  {\bibinfo {title} {{Destruction of long-range order in one-dimensional and
  2-dimensional systems having a continuous symmetry group 1 - classical
  systems}},}\ }\href@noop {} {\bibfield  {journal} {\bibinfo  {journal}
  {Soviet Physics Jetp-Ussr}\ }\textbf {\bibinfo {volume} {32}}~(\bibinfo
  {number} {3}),\ \bibinfo {pages} {493--\&}}\BibitemShut {NoStop}%
\bibitem [{\citenamefont {Bergeman}\ \emph {et~al.}(2003)\citenamefont
  {Bergeman}, \citenamefont {Moore},\ and\ \citenamefont
  {Olshanii}}]{bergemann2003}%
  \BibitemOpen
  \bibfield  {author} {\bibinfo {author} {\bibnamefont {Bergeman},
  \bibfnamefont {T}}, \bibinfo {author} {\bibfnamefont {M.~G.}\ \bibnamefont
  {Moore}}, \ and\ \bibinfo {author} {\bibfnamefont {M.}~\bibnamefont
  {Olshanii}}} (\bibinfo {year} {2003}),\ \bibfield  {title} {\enquote
  {\bibinfo {title} {Atom-atom scattering under cylindrical harmonic
  confinement: Numerical and analytic studies of the confinement induced
  resonance},}\ }\href@noop {} {\bibfield  {journal} {\bibinfo  {journal}
  {Phys. Rev. Lett.}\ }\textbf {\bibinfo {volume} {91}}~(\bibinfo {number}
  {16}),\ \bibinfo {pages} {163201}}\BibitemShut {NoStop}%
\bibitem [{\citenamefont {Berninger}\ \emph {et~al.}(2011)\citenamefont
  {Berninger}, \citenamefont {Zenesini}, \citenamefont {Huang}, \citenamefont
  {Harm}, \citenamefont {N\"agerl}, \citenamefont {Ferlaino}, \citenamefont
  {Grimm}, \citenamefont {Julienne},\ and\ \citenamefont
  {Hutson}}]{berninger2011PRL}%
  \BibitemOpen
  \bibfield  {author} {\bibinfo {author} {\bibnamefont {Berninger},
  \bibfnamefont {M}}, \bibinfo {author} {\bibfnamefont {A.}~\bibnamefont
  {Zenesini}}, \bibinfo {author} {\bibfnamefont {B.}~\bibnamefont {Huang}},
  \bibinfo {author} {\bibfnamefont {W.}~\bibnamefont {Harm}}, \bibinfo {author}
  {\bibfnamefont {H.-C.}\ \bibnamefont {N\"agerl}}, \bibinfo {author}
  {\bibfnamefont {F.}~\bibnamefont {Ferlaino}}, \bibinfo {author}
  {\bibfnamefont {R.}~\bibnamefont {Grimm}}, \bibinfo {author} {\bibfnamefont
  {P.~S.}\ \bibnamefont {Julienne}}, \ and\ \bibinfo {author} {\bibfnamefont
  {J.~M.}\ \bibnamefont {Hutson}}} (\bibinfo {year} {2011}),\ \bibfield
  {title} {\enquote {\bibinfo {title} {Universality of the three-body parameter
  for {E}fimov states in ultracold cesium},}\ }\href@noop {} {\bibfield
  {journal} {\bibinfo  {journal} {Phys. Rev. Lett.}\ }\textbf {\bibinfo
  {volume} {107}},\ \bibinfo {pages} {120401}}\BibitemShut {NoStop}%
\bibitem [{\citenamefont {Berninger}\ \emph {et~al.}(2013)\citenamefont
  {Berninger}, \citenamefont {Zenesini}, \citenamefont {Huang}, \citenamefont
  {Harm}, \citenamefont {N\"agerl}, \citenamefont {Ferlaino}, \citenamefont
  {Julienne},\ and\ \citenamefont {Hutson}}]{berni2013pra}%
  \BibitemOpen
  \bibfield  {author} {\bibinfo {author} {\bibnamefont {Berninger},
  \bibfnamefont {M}}, \bibinfo {author} {\bibfnamefont {A.}~\bibnamefont
  {Zenesini}}, \bibinfo {author} {\bibfnamefont {B.}~\bibnamefont {Huang}},
  \bibinfo {author} {\bibfnamefont {W.}~\bibnamefont {Harm}}, \bibinfo {author}
  {\bibfnamefont {H.-C.}\ \bibnamefont {N\"agerl}}, \bibinfo {author}
  {\bibfnamefont {R.}~\bibnamefont {Ferlaino}, \bibfnamefont {F.and~Grimm}},
  \bibinfo {author} {\bibfnamefont {P.~S.}\ \bibnamefont {Julienne}}, \ and\
  \bibinfo {author} {\bibfnamefont {J.~M.}\ \bibnamefont {Hutson}}} (\bibinfo
  {year} {2013}),\ \bibfield  {title} {\enquote {\bibinfo {title} {Feshbach
  resonances, weakly bound molecular states, and coupled-channel potentials for
  cesium at high magnetic fields},}\ }\href@noop {} {\bibfield  {journal}
  {\bibinfo  {journal} {Phys. Rev. A}\ }\textbf {\bibinfo {volume} {87}},\
  \bibinfo {pages} {032517}}\BibitemShut {NoStop}%
\bibitem [{\citenamefont {Bloch}\ \emph {et~al.}(2008)\citenamefont {Bloch},
  \citenamefont {Dalibard},\ and\ \citenamefont {Zwerger}}]{blochrmp}%
  \BibitemOpen
  \bibfield  {author} {\bibinfo {author} {\bibnamefont {Bloch}, \bibfnamefont
  {I}}, \bibinfo {author} {\bibfnamefont {J.}~\bibnamefont {Dalibard}}, \ and\
  \bibinfo {author} {\bibfnamefont {W.}~\bibnamefont {Zwerger}}} (\bibinfo
  {year} {2008}),\ \bibfield  {title} {\enquote {\bibinfo {title} {Many-body
  physics with ultracold gases},}\ }\href@noop {} {\bibfield  {journal}
  {\bibinfo  {journal} {Rev. Mod. Phys.}\ }\textbf {\bibinfo {volume} {80}},\
  \bibinfo {pages} {885--964}}\BibitemShut {NoStop}%
\bibitem [{\citenamefont {Bloom}\ \emph {et~al.}(2013)\citenamefont {Bloom},
  \citenamefont {Hu}, \citenamefont {Cumby},\ and\ \citenamefont
  {Jin}}]{bloom2013PRL}%
  \BibitemOpen
  \bibfield  {author} {\bibinfo {author} {\bibnamefont {Bloom}, \bibfnamefont
  {R~S}}, \bibinfo {author} {\bibfnamefont {M.~G.}\ \bibnamefont {Hu}},
  \bibinfo {author} {\bibfnamefont {T.~D.}\ \bibnamefont {Cumby}}, \ and\
  \bibinfo {author} {\bibfnamefont {D.~S.}\ \bibnamefont {Jin}}} (\bibinfo
  {year} {2013}),\ \bibfield  {title} {\enquote {\bibinfo {title} {Tests of
  universal three-body physics in an ultracold {B}ose-{F}ermi mixture},}\
  }\href@noop {} {\bibfield  {journal} {\bibinfo  {journal} {Phys. Rev. Lett.}\
  }\textbf {\bibinfo {volume} {111}}~(\bibinfo {number} {10}),\ \bibinfo
  {pages} {105301}}\BibitemShut {NoStop}%
\bibitem [{\citenamefont {Blumberg}\ \emph {et~al.}(1979)\citenamefont
  {Blumberg}, \citenamefont {Itano},\ and\ \citenamefont
  {Larson}}]{larsonpra1979}%
  \BibitemOpen
  \bibfield  {author} {\bibinfo {author} {\bibnamefont {Blumberg},
  \bibfnamefont {W~A~M}}, \bibinfo {author} {\bibfnamefont {Wayne~M.}\
  \bibnamefont {Itano}}, \ and\ \bibinfo {author} {\bibfnamefont {D.~J.}\
  \bibnamefont {Larson}}} (\bibinfo {year} {1979}),\ \bibfield  {title}
  {\enquote {\bibinfo {title} {Theory of the photodetachment of negative ions
  in a magnetic field},}\ }\href@noop {} {\bibfield  {journal} {\bibinfo
  {journal} {Phys. Rev. A}\ }\textbf {\bibinfo {volume} {19}},\ \bibinfo
  {pages} {139--148}}\BibitemShut {NoStop}%
\bibitem [{\citenamefont {Blume}(2012{\natexlab{a}})}]{blume2012rpp}%
  \BibitemOpen
  \bibfield  {author} {\bibinfo {author} {\bibnamefont {Blume}, \bibfnamefont
  {D}}} (\bibinfo {year} {2012}{\natexlab{a}}),\ \bibfield  {title} {\enquote
  {\bibinfo {title} {Few-body physics with ultracold atomic and molecular
  systems in traps},}\ }\href@noop {} {\bibfield  {journal} {\bibinfo
  {journal} {Rep. Prog. Phys.}\ }\textbf {\bibinfo {volume} {75}}~(\bibinfo
  {number} {4}),\ \bibinfo {pages} {046401}}\BibitemShut {NoStop}%
\bibitem [{\citenamefont {Blume}(2012{\natexlab{b}})}]{Blume2012prl}%
  \BibitemOpen
  \bibfield  {author} {\bibinfo {author} {\bibnamefont {Blume}, \bibfnamefont
  {D}}} (\bibinfo {year} {2012}{\natexlab{b}}),\ \bibfield  {title} {\enquote
  {\bibinfo {title} {Universal four-body states in heavy-light mixtures with a
  positive scattering length},}\ }\href@noop {} {\bibfield  {journal} {\bibinfo
   {journal} {Phys. Rev. Lett.}\ }\textbf {\bibinfo {volume} {109}},\ \bibinfo
  {pages} {230404}}\BibitemShut {NoStop}%
\bibitem [{\citenamefont {Blume}(2014)}]{blume2014pra}%
  \BibitemOpen
  \bibfield  {author} {\bibinfo {author} {\bibnamefont {Blume}, \bibfnamefont
  {D}}} (\bibinfo {year} {2014}),\ \bibfield  {title} {\enquote {\bibinfo
  {title} {Three-body bound states in a harmonic waveguide with cylindrical
  symmetry},}\ }\href@noop {} {\bibfield  {journal} {\bibinfo  {journal} {Phys.
  Rev. A}\ }\textbf {\bibinfo {volume} {89}},\ \bibinfo {pages}
  {053603}}\BibitemShut {NoStop}%
\bibitem [{\citenamefont {Blume}\ and\ \citenamefont
  {Daily}(2009)}]{blume2009PRA}%
  \BibitemOpen
  \bibfield  {author} {\bibinfo {author} {\bibnamefont {Blume}, \bibfnamefont
  {D}}, \ and\ \bibinfo {author} {\bibfnamefont {K.~M.}\ \bibnamefont {Daily}}}
  (\bibinfo {year} {2009}),\ \bibfield  {title} {{\selectlanguage
  {English}\enquote {\bibinfo {title} {Universal relations for a trapped
  four-{F}ermion system with arbitrary s-wave scattering length},}\
  }}\href@noop {} {\bibfield  {journal} {\bibinfo  {journal} {Phys. Rev. A}\
  }\textbf {\bibinfo {volume} {80}}~(\bibinfo {number} {5}),\ \bibinfo {pages}
  {053626}}\BibitemShut {NoStop}%
\bibitem [{\citenamefont {Blume}\ and\ \citenamefont
  {Greene}(2000)}]{blume2000JCP}%
  \BibitemOpen
  \bibfield  {author} {\bibinfo {author} {\bibnamefont {Blume}, \bibfnamefont
  {D}}, \ and\ \bibinfo {author} {\bibfnamefont {C.~H.}\ \bibnamefont
  {Greene}}} (\bibinfo {year} {2000}),\ \bibfield  {title} {\enquote {\bibinfo
  {title} {Monte {C}arlo hyperspherical description of helium cluster excited
  states},}\ }\href@noop {} {\bibfield  {journal} {\bibinfo  {journal} {J.
  Chem. Phys.}\ }\textbf {\bibinfo {volume} {112}}~(\bibinfo {number} {18}),\
  \bibinfo {pages} {8053--8067}}\BibitemShut {NoStop}%
\bibitem [{\citenamefont {Blume}\ \emph {et~al.}(2000)\citenamefont {Blume},
  \citenamefont {Greene},\ and\ \citenamefont {Esry}}]{blume2000JCPc}%
  \BibitemOpen
  \bibfield  {author} {\bibinfo {author} {\bibnamefont {Blume}, \bibfnamefont
  {D}}, \bibinfo {author} {\bibfnamefont {C.~H.}\ \bibnamefont {Greene}}, \
  and\ \bibinfo {author} {\bibfnamefont {B.~D.}\ \bibnamefont {Esry}}}
  (\bibinfo {year} {2000}),\ \bibfield  {title} {\enquote {\bibinfo {title}
  {Comparative study of {H}e$_3$, {N}e$_3$, and {A}r$_3$ using hyperspherical
  coordinates},}\ }\href@noop {} {\bibfield  {journal} {\bibinfo  {journal} {J.
  Chem. Phys.}\ }\textbf {\bibinfo {volume} {113}}~(\bibinfo {number} {6}),\
  \bibinfo {pages} {2145--2158}}\BibitemShut {NoStop}%
\bibitem [{\citenamefont {Blume}\ \emph {et~al.}(2014)\citenamefont {Blume},
  \citenamefont {Greene},\ and\ \citenamefont {Esry}}]{Blume2014jcpErr}%
  \BibitemOpen
  \bibfield  {author} {\bibinfo {author} {\bibnamefont {Blume}, \bibfnamefont
  {D}}, \bibinfo {author} {\bibfnamefont {C.~H.}\ \bibnamefont {Greene}}, \
  and\ \bibinfo {author} {\bibfnamefont {B.~D.}\ \bibnamefont {Esry}}}
  (\bibinfo {year} {2014}),\ \bibfield  {title} {\enquote {\bibinfo {title}
  {Erratum: {Comparative study of He$_3$, Ne$_3$, and Ar$_3$ using
  hyperspherical coordinates (vol 113, pg 2145, 2000)}},}\ }\href@noop {}
  {\bibfield  {journal} {\bibinfo  {journal} {J. Chem. Phys.}\ }\textbf
  {\bibinfo {volume} {141}}~(\bibinfo {number} {069901})}\BibitemShut {NoStop}%
\bibitem [{\citenamefont {Blume}\ and\ \citenamefont
  {Greene}(2002)}]{blume2002EPJD}%
  \BibitemOpen
  \bibfield  {author} {\bibinfo {author} {\bibnamefont {Blume}, \bibfnamefont
  {D}}, \ and\ \bibinfo {author} {\bibfnamefont {CH}~\bibnamefont {Greene}}}
  (\bibinfo {year} {2002}),\ \bibfield  {title} {{\selectlanguage
  {English}\enquote {\bibinfo {title} {Lowest breathing mode of bosonic helium
  clusters},}\ }}\href@noop {} {\bibfield  {journal} {\bibinfo  {journal}
  {Euro. Phys. J. D}\ }\textbf {\bibinfo {volume} {18}}~(\bibinfo {number}
  {1}),\ \bibinfo {pages} {83--86}}\BibitemShut {NoStop}%
\bibitem [{\citenamefont {Blume}\ \emph {et~al.}(2008)\citenamefont {Blume},
  \citenamefont {Rittenhouse}, \citenamefont {von Stecher},\ and\ \citenamefont
  {Greene}}]{blume2008PRAb}%
  \BibitemOpen
  \bibfield  {author} {\bibinfo {author} {\bibnamefont {Blume}, \bibfnamefont
  {D}}, \bibinfo {author} {\bibfnamefont {S.~T.}\ \bibnamefont {Rittenhouse}},
  \bibinfo {author} {\bibfnamefont {J.}~\bibnamefont {von Stecher}}, \ and\
  \bibinfo {author} {\bibfnamefont {C.~H.}\ \bibnamefont {Greene}}} (\bibinfo
  {year} {2008}),\ \bibfield  {title} {{\selectlanguage {English}\enquote
  {\bibinfo {title} {Stability of inhomogeneous multicomponent {F}ermi
  gases},}\ }}\href@noop {} {\bibfield  {journal} {\bibinfo  {journal} {Phys.
  Rev. A}\ }\textbf {\bibinfo {volume} {77}}~(\bibinfo {number} {3}),\ \bibinfo
  {pages} {033627}}\BibitemShut {NoStop}%
\bibitem [{\citenamefont {Blume}\ and\ \citenamefont
  {Yan}({2014})}]{BlumeYan2014prl}%
  \BibitemOpen
  \bibfield  {author} {\bibinfo {author} {\bibnamefont {Blume}, \bibfnamefont
  {D}}, \ and\ \bibinfo {author} {\bibfnamefont {Y.}~\bibnamefont {Yan}}}
  (\bibinfo {year} {{2014}}),\ \bibfield  {title} {\enquote {\bibinfo {title}
  {{Generalized Efimov Scenario for Heavy-Light Mixtures}},}\ }\href@noop {}
  {\bibfield  {journal} {\bibinfo  {journal} {{Phys. Rev. Lett.}}\ }\textbf
  {\bibinfo {volume} {{113}}}~(\bibinfo {number} {{21}})}\BibitemShut {NoStop}%
\bibitem [{\citenamefont {Bohn}\ \emph {et~al.}(1998)\citenamefont {Bohn},
  \citenamefont {Esry},\ and\ \citenamefont {Greene}}]{bohn1998PRA}%
  \BibitemOpen
  \bibfield  {author} {\bibinfo {author} {\bibnamefont {Bohn}, \bibfnamefont
  {J~L}}, \bibinfo {author} {\bibfnamefont {B.~D.}\ \bibnamefont {Esry}}, \
  and\ \bibinfo {author} {\bibfnamefont {C.~H.}\ \bibnamefont {Greene}}}
  (\bibinfo {year} {1998}),\ \bibfield  {title} {{\selectlanguage
  {English}\enquote {\bibinfo {title} {Effective potentials for dilute
  {B}ose-{E}instein condensates},}\ }}\href@noop {} {\bibfield  {journal}
  {\bibinfo  {journal} {Phys. Rev. A}\ }\textbf {\bibinfo {volume}
  {58}}~(\bibinfo {number} {1}),\ \bibinfo {pages} {584--597}}\BibitemShut
  {NoStop}%
\bibitem [{\citenamefont {Bohr}\ \emph {et~al.}({2014})\citenamefont {Bohr},
  \citenamefont {Paolini}, \citenamefont {Forrey}, \citenamefont
  {Balakrishnan},\ and\ \citenamefont {Stancil}}]{Balakrishnan2014JCP}%
  \BibitemOpen
  \bibfield  {author} {\bibinfo {author} {\bibnamefont {Bohr}, \bibfnamefont
  {A}}, \bibinfo {author} {\bibfnamefont {S.}~\bibnamefont {Paolini}}, \bibinfo
  {author} {\bibfnamefont {R.~C.}\ \bibnamefont {Forrey}}, \bibinfo {author}
  {\bibfnamefont {N.}~\bibnamefont {Balakrishnan}}, \ and\ \bibinfo {author}
  {\bibfnamefont {P.~C.}\ \bibnamefont {Stancil}}} (\bibinfo {year} {{2014}}),\
  \bibfield  {title} {\enquote {\bibinfo {title} {{A full-dimensional quantum
  dynamical study of H$_2$+H$_2$ collisions: Coupled-states versus
  close-coupling formulation}},}\ }\href@noop {} {\bibfield  {journal}
  {\bibinfo  {journal} {J. Chem. Phys.}\ }\textbf {\bibinfo {volume}
  {{140}}}~(\bibinfo {number} {{6}})}\BibitemShut {NoStop}%
\bibitem [{\citenamefont {Bongs}\ and\ \citenamefont
  {Sengstock}(2004)}]{bongs2004physics}%
  \BibitemOpen
  \bibfield  {author} {\bibinfo {author} {\bibnamefont {Bongs}, \bibfnamefont
  {K}}, \ and\ \bibinfo {author} {\bibfnamefont {K.}~\bibnamefont {Sengstock}}}
  (\bibinfo {year} {2004}),\ \bibfield  {title} {\enquote {\bibinfo {title}
  {Physics with coherent matter waves},}\ }\href@noop {} {\bibfield  {journal}
  {\bibinfo  {journal} {Rep. Prog. Phys.}\ }\textbf {\bibinfo {volume}
  {67}}~(\bibinfo {number} {6}),\ \bibinfo {pages} {907}}\BibitemShut {NoStop}%
\bibitem [{\citenamefont {Borca}\ \emph {et~al.}(2003)\citenamefont {Borca},
  \citenamefont {Blume},\ and\ \citenamefont {Greene}}]{borca2003NJP}%
  \BibitemOpen
  \bibfield  {author} {\bibinfo {author} {\bibnamefont {Borca}, \bibfnamefont
  {B}}, \bibinfo {author} {\bibfnamefont {D.}~\bibnamefont {Blume}}, \ and\
  \bibinfo {author} {\bibfnamefont {C.~H.}\ \bibnamefont {Greene}}} (\bibinfo
  {year} {2003}),\ \bibfield  {title} {{\selectlanguage {English}\enquote
  {\bibinfo {title} {A two-atom picture of coherent atom-molecule quantum
  beats},}\ }}\href@noop {} {\bibfield  {journal} {\bibinfo  {journal} {New J.
  Phys.}\ }\textbf {\bibinfo {volume} {5}},\ \bibinfo {pages}
  {111}}\BibitemShut {NoStop}%
\bibitem [{\citenamefont {Botero}\ and\ \citenamefont
  {Greene}(1985)}]{botero1985PRA}%
  \BibitemOpen
  \bibfield  {author} {\bibinfo {author} {\bibnamefont {Botero}, \bibfnamefont
  {J}}, \ and\ \bibinfo {author} {\bibfnamefont {C.~H.}\ \bibnamefont
  {Greene}}} (\bibinfo {year} {1985}),\ \bibfield  {title} {{\selectlanguage
  {English}\enquote {\bibinfo {title} {Positronium negative-ion - an adiabatic
  study using hyperspherical coordinates},}\ }}\href@noop {} {\bibfield
  {journal} {\bibinfo  {journal} {Phys. Rev. A}\ }\textbf {\bibinfo {volume}
  {32}}~(\bibinfo {number} {2}),\ \bibinfo {pages} {1249--1251}}\BibitemShut
  {NoStop}%
\bibitem [{\citenamefont {Botero}\ and\ \citenamefont
  {Greene}(1986)}]{botero1986PRL}%
  \BibitemOpen
  \bibfield  {author} {\bibinfo {author} {\bibnamefont {Botero}, \bibfnamefont
  {J}}, \ and\ \bibinfo {author} {\bibfnamefont {C.~H.}\ \bibnamefont
  {Greene}}} (\bibinfo {year} {1986}),\ \bibfield  {title} {{\selectlanguage
  {English}\enquote {\bibinfo {title} {Resonant photodetachment of the
  positronium negative-ion},}\ }}\href@noop {} {\bibfield  {journal} {\bibinfo
  {journal} {Phys. Rev. Lett.}\ }\textbf {\bibinfo {volume} {56}}~(\bibinfo
  {number} {13}),\ \bibinfo {pages} {1366--1369}}\BibitemShut {NoStop}%
\bibitem [{\citenamefont {Braaten}\ and\ \citenamefont
  {Hammer}(2001)}]{braaten2001PRL}%
  \BibitemOpen
  \bibfield  {author} {\bibinfo {author} {\bibnamefont {Braaten}, \bibfnamefont
  {E}}, \ and\ \bibinfo {author} {\bibfnamefont {H.{-}W.}\ \bibnamefont
  {Hammer}}} (\bibinfo {year} {2001}),\ \bibfield  {title} {{\selectlanguage
  {English}\enquote {\bibinfo {title} {Three-body recombination into deep bound
  states in a {B}ose gas with large scattering length},}\ }}\href@noop {}
  {\bibfield  {journal} {\bibinfo  {journal} {Phys. Rev. Lett.}\ }\textbf
  {\bibinfo {volume} {87}}~(\bibinfo {number} {16}),\ \bibinfo {pages}
  {160407}}\BibitemShut {NoStop}%
\bibitem [{\citenamefont {Braaten}\ and\ \citenamefont
  {Hammer}(2003)}]{braaten2003PRA}%
  \BibitemOpen
  \bibfield  {author} {\bibinfo {author} {\bibnamefont {Braaten}, \bibfnamefont
  {E}}, \ and\ \bibinfo {author} {\bibfnamefont {H.{-}W.}\ \bibnamefont
  {Hammer}}} (\bibinfo {year} {2003}),\ \bibfield  {title} {{\selectlanguage
  {English}\enquote {\bibinfo {title} {Universality in the three-body problem
  for $^4${He} atoms},}\ }}\href@noop {} {\bibfield  {journal} {\bibinfo
  {journal} {Phys. Rev. A}\ }\textbf {\bibinfo {volume} {67}}~(\bibinfo
  {number} {4}),\ \bibinfo {pages} {042706}}\BibitemShut {NoStop}%
\bibitem [{\citenamefont {Braaten}\ and\ \citenamefont
  {Hammer}(2006)}]{braaten2006PRep}%
  \BibitemOpen
  \bibfield  {author} {\bibinfo {author} {\bibnamefont {Braaten}, \bibfnamefont
  {E}}, \ and\ \bibinfo {author} {\bibfnamefont {H.{-}W.}\ \bibnamefont
  {Hammer}}} (\bibinfo {year} {2006}),\ \bibfield  {title} {{\selectlanguage
  {English}\enquote {\bibinfo {title} {Universality in few-body systems with
  large scattering length},}\ }}\href@noop {} {\bibfield  {journal} {\bibinfo
  {journal} {Phys. Rep.}\ }\textbf {\bibinfo {volume} {428}}~(\bibinfo {number}
  {5-6}),\ \bibinfo {pages} {259--390}}\BibitemShut {NoStop}%
\bibitem [{\citenamefont {Braaten}\ \emph {et~al.}(2010)\citenamefont
  {Braaten}, \citenamefont {Kang},\ and\ \citenamefont
  {Platter}}]{braaten2010PRL}%
  \BibitemOpen
  \bibfield  {author} {\bibinfo {author} {\bibnamefont {Braaten}, \bibfnamefont
  {E}}, \bibinfo {author} {\bibfnamefont {D.}~\bibnamefont {Kang}}, \ and\
  \bibinfo {author} {\bibfnamefont {L.}~\bibnamefont {Platter}}} (\bibinfo
  {year} {2010}),\ \bibfield  {title} {{\selectlanguage {English}\enquote
  {\bibinfo {title} {Short-time operator product expansion for rf spectroscopy
  of a strongly interacting {F}ermi gas},}\ }}\href@noop {} {\bibfield
  {journal} {\bibinfo  {journal} {Phys. Rev. Lett.}\ }\textbf {\bibinfo
  {volume} {104}}~(\bibinfo {number} {22}),\ \bibinfo {pages}
  {223004}}\BibitemShut {NoStop}%
\bibitem [{\citenamefont {Braaten}\ \emph {et~al.}(2011)\citenamefont
  {Braaten}, \citenamefont {Kang},\ and\ \citenamefont
  {Platter}}]{Braaten-2011}%
  \BibitemOpen
  \bibfield  {author} {\bibinfo {author} {\bibnamefont {Braaten}, \bibfnamefont
  {E}}, \bibinfo {author} {\bibfnamefont {D.}~\bibnamefont {Kang}}, \ and\
  \bibinfo {author} {\bibfnamefont {L.}~\bibnamefont {Platter}}} (\bibinfo
  {year} {2011}),\ \bibfield  {title} {\enquote {\bibinfo {title} {Short-time
  operator product expansion for rf spectroscopy of a strongly interacting
  {F}ermi gas},}\ }\href@noop {} {\bibfield  {journal} {\bibinfo  {journal}
  {Phys. Rev. Lett.}\ }\textbf {\bibinfo {volume} {106}},\ \bibinfo {pages}
  {153005}}\BibitemShut {NoStop}%
\bibitem [{\citenamefont {Braaten}\ and\ \citenamefont
  {Platter}(2009)}]{braaten2009LP}%
  \BibitemOpen
  \bibfield  {author} {\bibinfo {author} {\bibnamefont {Braaten}, \bibfnamefont
  {E}}, \ and\ \bibinfo {author} {\bibfnamefont {L.}~\bibnamefont {Platter}}}
  (\bibinfo {year} {2009}),\ \bibfield  {title} {{\selectlanguage
  {English}\enquote {\bibinfo {title} {Universal relations for the
  strongly-interacting {F}ermi gas},}\ }}\href@noop {} {\ \textbf {\bibinfo
  {volume} {19}}~(\bibinfo {number} {4}),\ \bibinfo {pages}
  {550--553}}\BibitemShut {NoStop}%
\bibitem [{\citenamefont {Bradley}\ \emph {et~al.}(1997)\citenamefont
  {Bradley}, \citenamefont {Sacket},\ and\ \citenamefont
  {Hulet}}]{bradley1997PRL}%
  \BibitemOpen
  \bibfield  {author} {\bibinfo {author} {\bibnamefont {Bradley}, \bibfnamefont
  {C~C}}, \bibinfo {author} {\bibfnamefont {C.~A.}\ \bibnamefont {Sacket}}, \
  and\ \bibinfo {author} {\bibfnamefont {R.~G.}\ \bibnamefont {Hulet}}}
  (\bibinfo {year} {1997}),\ \bibfield  {title} {\enquote {\bibinfo {title}
  {{B}ose-{E}instein condensation of lithium: Observation of limited condensate
  number},}\ }\href@noop {} {\bibfield  {journal} {\bibinfo  {journal} {Phys.
  Rev. Lett.}\ }\textbf {\bibinfo {volume} {78}},\ \bibinfo {pages}
  {985--989}}\BibitemShut {NoStop}%
\bibitem [{\citenamefont {Bradley}\ \emph {et~al.}(1995)\citenamefont
  {Bradley}, \citenamefont {Sackett}, \citenamefont {Tollett},\ and\
  \citenamefont {Hulet}}]{bradley1995}%
  \BibitemOpen
  \bibfield  {author} {\bibinfo {author} {\bibnamefont {Bradley}, \bibfnamefont
  {C~C}}, \bibinfo {author} {\bibfnamefont {C.~A.}\ \bibnamefont {Sackett}},
  \bibinfo {author} {\bibfnamefont {J.~J.}\ \bibnamefont {Tollett}}, \ and\
  \bibinfo {author} {\bibfnamefont {R.~G.}\ \bibnamefont {Hulet}}} (\bibinfo
  {year} {1995}),\ \bibfield  {title} {\enquote {\bibinfo {title} {Evidence of
  {B}ose-{E}instein condensation in an atomic gas with attractive
  interactions},}\ }\href@noop {} {\bibfield  {journal} {\bibinfo  {journal}
  {Phys. Rev. Lett.}\ }\textbf {\bibinfo {volume} {75}}~(\bibinfo {number}
  {9}),\ \bibinfo {pages} {1687}}\BibitemShut {NoStop}%
\bibitem [{\citenamefont {Bray}\ and\ \citenamefont
  {Stelbovics}(1993)}]{bray1993PRL}%
  \BibitemOpen
  \bibfield  {author} {\bibinfo {author} {\bibnamefont {Bray}, \bibfnamefont
  {I}}, \ and\ \bibinfo {author} {\bibfnamefont {A.~T.}\ \bibnamefont
  {Stelbovics}}} (\bibinfo {year} {1993}),\ \bibfield  {title} {\enquote
  {\bibinfo {title} {Calculation of the total ionization cross section and spin
  asymmetry in electron-hydrogen scattering from threshold to 500 {eV}},}\
  }\href@noop {} {\bibfield  {journal} {\bibinfo  {journal} {Phys. Rev. Lett.}\
  }\textbf {\bibinfo {volume} {70}},\ \bibinfo {pages} {746--749}}\BibitemShut
  {NoStop}%
\bibitem [{\citenamefont {Bruch}\ and\ \citenamefont
  {Tjon}(1979)}]{BruchTjon1979pra}%
  \BibitemOpen
  \bibfield  {author} {\bibinfo {author} {\bibnamefont {Bruch}, \bibfnamefont
  {L~W}}, \ and\ \bibinfo {author} {\bibfnamefont {J.~A.}\ \bibnamefont
  {Tjon}}} (\bibinfo {year} {1979}),\ \bibfield  {title} {\enquote {\bibinfo
  {title} {Binding of three identical bosons in two dimensions},}\ }\href
  {\doibase 10.1103/PhysRevA.19.425} {\bibfield  {journal} {\bibinfo  {journal}
  {Phys. Rev. A}\ }\textbf {\bibinfo {volume} {19}},\ \bibinfo {pages}
  {425--432}}\BibitemShut {NoStop}%
\bibitem [{\citenamefont {Bruderer}\ \emph {et~al.}(2008)\citenamefont
  {Bruderer}, \citenamefont {Bao},\ and\ \citenamefont
  {Jaksch}}]{bruderer_self-trapping_2008}%
  \BibitemOpen
  \bibfield  {author} {\bibinfo {author} {\bibnamefont {Bruderer},
  \bibfnamefont {M}}, \bibinfo {author} {\bibfnamefont {W.}~\bibnamefont
  {Bao}}, \ and\ \bibinfo {author} {\bibfnamefont {D.}~\bibnamefont {Jaksch}}}
  (\bibinfo {year} {2008}),\ \bibfield  {title} {\enquote {\bibinfo {title}
  {Self-trapping of impurities in {Bose}-{Einstein} condensates: {Strong}
  attractive and repulsive coupling},}\ }\href {\doibase
  10.1209/0295-5075/82/30004} {\bibfield  {journal} {\bibinfo  {journal}
  {{Europhys. Lett.}}\ }\textbf {\bibinfo {volume} {82}}~(\bibinfo {number}
  {3}),\ \bibinfo {pages} {30004}}\BibitemShut {NoStop}%
\bibitem [{\citenamefont {Bulgac}\ \emph {et~al.}(2006)\citenamefont {Bulgac},
  \citenamefont {Drut},\ and\ \citenamefont {Magierski}}]{bulgac2006PRL}%
  \BibitemOpen
  \bibfield  {author} {\bibinfo {author} {\bibnamefont {Bulgac}, \bibfnamefont
  {A}}, \bibinfo {author} {\bibfnamefont {JE}~\bibnamefont {Drut}}, \ and\
  \bibinfo {author} {\bibfnamefont {P}~\bibnamefont {Magierski}}} (\bibinfo
  {year} {2006}),\ \bibfield  {title} {{\selectlanguage {English}\enquote
  {\bibinfo {title} {Spin 1/2 {F}ermions in the unitary regime: A superfluid of
  a new type},}\ }}\href@noop {} {\bibfield  {journal} {\bibinfo  {journal}
  {Phys. Rev. Lett.}\ }\textbf {\bibinfo {volume} {96}}~(\bibinfo {number}
  {9}),\ \bibinfo {pages} {090404}}\BibitemShut {NoStop}%
\bibitem [{\citenamefont {Burke}\ \emph {et~al.}(1998)\citenamefont {Burke},
  \citenamefont {Greene},\ and\ \citenamefont {Bohn}}]{burke1998PRLb}%
  \BibitemOpen
  \bibfield  {author} {\bibinfo {author} {\bibnamefont {Burke}, \bibfnamefont
  {J~P}}, \bibinfo {author} {\bibfnamefont {C.~H.}\ \bibnamefont {Greene}}, \
  and\ \bibinfo {author} {\bibfnamefont {J.~L.}\ \bibnamefont {Bohn}}}
  (\bibinfo {year} {1998}),\ \bibfield  {title} {{\selectlanguage
  {English}\enquote {\bibinfo {title} {Multichannel cold collisions: Simple
  dependences on energy and magnetic field},}\ }}\href@noop {} {\bibfield
  {journal} {\bibinfo  {journal} {Phys. Rev. Lett.}\ }\textbf {\bibinfo
  {volume} {81}}~(\bibinfo {number} {16}),\ \bibinfo {pages}
  {3355--3358}}\BibitemShut {NoStop}%
\bibitem [{\citenamefont {Burt}\ \emph {et~al.}(1997)\citenamefont {Burt},
  \citenamefont {Ghrist}, \citenamefont {Myatt}, \citenamefont {Holland},
  \citenamefont {Cornell},\ and\ \citenamefont {Wieman}}]{burt1997PRL}%
  \BibitemOpen
  \bibfield  {author} {\bibinfo {author} {\bibnamefont {Burt}, \bibfnamefont
  {E~A}}, \bibinfo {author} {\bibfnamefont {R.~W.}\ \bibnamefont {Ghrist}},
  \bibinfo {author} {\bibfnamefont {C.~J.}\ \bibnamefont {Myatt}}, \bibinfo
  {author} {\bibfnamefont {M.~J.}\ \bibnamefont {Holland}}, \bibinfo {author}
  {\bibfnamefont {E.~A.}\ \bibnamefont {Cornell}}, \ and\ \bibinfo {author}
  {\bibfnamefont {C.~E.}\ \bibnamefont {Wieman}}} (\bibinfo {year} {1997}),\
  \bibfield  {title} {\enquote {\bibinfo {title} {Coherence, correlations, and
  collisions: What one learns about {B}ose-{E}instein condensates from their
  decay},}\ }\href@noop {} {\bibfield  {journal} {\bibinfo  {journal} {Phys.
  Rev. Lett.}\ }\textbf {\bibinfo {volume} {79}},\ \bibinfo {pages}
  {337--340}}\BibitemShut {NoStop}%
\bibitem [{\citenamefont {Campargue}(2001)}]{Campargue}%
  \BibitemOpen
  \bibfield  {author} {\bibinfo {author} {\bibnamefont {Campargue},
  \bibfnamefont {R}}} (\bibinfo {year} {2001}),\ \href@noop {} {\emph {\bibinfo
  {title} {Atomic and Molecular Beams: The state of the Art 2000}}},\ edited
  by\ \bibinfo {editor} {\bibfnamefont {R.}~\bibnamefont {Campargue}}\
  (\bibinfo  {publisher} {Springer},\ \bibinfo {address}
  {Heidelberg})\BibitemShut {NoStop}%
\bibitem [{\citenamefont {Carlson}\ \emph {et~al.}({2015})\citenamefont
  {Carlson}, \citenamefont {Gandolfi}, \citenamefont {Pederiva}, \citenamefont
  {Pieper}, \citenamefont {Schiavilla}, \citenamefont {Schmidt},\ and\
  \citenamefont {Wiringa}}]{Carlson2015rmp}%
  \BibitemOpen
  \bibfield  {author} {\bibinfo {author} {\bibnamefont {Carlson}, \bibfnamefont
  {J}}, \bibinfo {author} {\bibfnamefont {S.}~\bibnamefont {Gandolfi}},
  \bibinfo {author} {\bibfnamefont {F.}~\bibnamefont {Pederiva}}, \bibinfo
  {author} {\bibfnamefont {Steven~C.}\ \bibnamefont {Pieper}}, \bibinfo
  {author} {\bibfnamefont {R.}~\bibnamefont {Schiavilla}}, \bibinfo {author}
  {\bibfnamefont {K.~E.}\ \bibnamefont {Schmidt}}, \ and\ \bibinfo {author}
  {\bibfnamefont {R.~B.}\ \bibnamefont {Wiringa}}} (\bibinfo {year} {{2015}}),\
  \bibfield  {title} {\enquote {\bibinfo {title} {{Quantum Monte Carlo methods
  for nuclear physics}},}\ }\href {\doibase {10.1103/RevModPhys.87.1067}}
  {\bibfield  {journal} {\bibinfo  {journal} {{Reviews of Modern Physics}}\
  }\textbf {\bibinfo {volume} {{87}}}~(\bibinfo {number} {{3}}),\ \bibinfo
  {pages} {{1067--1118}}}\BibitemShut {NoStop}%
\bibitem [{\citenamefont {Casimir}\ and\ \citenamefont
  {Polder}(1948)}]{CasimirPolder1948pr}%
  \BibitemOpen
  \bibfield  {author} {\bibinfo {author} {\bibnamefont {Casimir}, \bibfnamefont
  {H~B~G}}, \ and\ \bibinfo {author} {\bibfnamefont {D.}~\bibnamefont
  {Polder}}} (\bibinfo {year} {1948}),\ \bibfield  {title} {\enquote {\bibinfo
  {title} {The influence of retardation on the london-van der waals forces},}\
  }\href {\doibase 10.1103/PhysRev.73.360} {\bibfield  {journal} {\bibinfo
  {journal} {Phys. Rev.}\ }\textbf {\bibinfo {volume} {73}},\ \bibinfo {pages}
  {360--372}}\BibitemShut {NoStop}%
\bibitem [{\citenamefont {Castin}\ and\ \citenamefont
  {Werner}(2011)}]{Castin-2011}%
  \BibitemOpen
  \bibfield  {author} {\bibinfo {author} {\bibnamefont {Castin}, \bibfnamefont
  {Y}}, \ and\ \bibinfo {author} {\bibfnamefont {F.}~\bibnamefont {Werner}}}
  (\bibinfo {year} {2011}),\ \bibfield  {title} {\enquote {\bibinfo {title}
  {Single-particle momentum distribution of and {E}fimov trimer},}\ }\href@noop
  {} {\bibfield  {journal} {\bibinfo  {journal} {Phys. Rev. A}\ }\textbf
  {\bibinfo {volume} {83}},\ \bibinfo {pages} {063614}}\BibitemShut {NoStop}%
\bibitem [{\citenamefont {Castin}\ \emph {et~al.}({2010})\citenamefont
  {Castin}, \citenamefont {Mora},\ and\ \citenamefont
  {Pricoupenko}}]{CastinMoraPricoupenko2010prl}%
  \BibitemOpen
  \bibfield  {author} {\bibinfo {author} {\bibnamefont {Castin}, \bibfnamefont
  {Yvan}}, \bibinfo {author} {\bibfnamefont {Christophe}\ \bibnamefont {Mora}},
  \ and\ \bibinfo {author} {\bibfnamefont {Ludovic}\ \bibnamefont
  {Pricoupenko}}} (\bibinfo {year} {{2010}}),\ \bibfield  {title} {\enquote
  {\bibinfo {title} {{Four-Body Efimov Effect for Three Fermions and a Lighter
  Particle}},}\ }\href {\doibase {10.1103/PhysRevLett.105.223201}} {\bibfield
  {journal} {\bibinfo  {journal} {{Physical Review Letters}}\ }\textbf
  {\bibinfo {volume} {{105}}}~(\bibinfo {number} {{22}}),\
  {10.1103/PhysRevLett.105.223201}}\BibitemShut {NoStop}%
\bibitem [{\citenamefont {Castin}\ and\ \citenamefont
  {Werner}({2013})}]{CastinWerner2013cjp}%
  \BibitemOpen
  \bibfield  {author} {\bibinfo {author} {\bibnamefont {Castin}, \bibfnamefont
  {Yvan}}, \ and\ \bibinfo {author} {\bibfnamefont {Felix}\ \bibnamefont
  {Werner}}} (\bibinfo {year} {{2013}}),\ \bibfield  {title} {\enquote
  {\bibinfo {title} {{The third coefficient of unitary Bose viral gas}},}\
  }\href@noop {} {\bibfield  {journal} {\bibinfo  {journal} {{Canadian Journal
  of Physics}}\ }\textbf {\bibinfo {volume} {{91}}}~(\bibinfo {number} {{5}}),\
  \bibinfo {pages} {{382--389}}}\BibitemShut {NoStop}%
\bibitem [{\citenamefont {Catani}\ \emph {et~al.}(2009)\citenamefont {Catani},
  \citenamefont {Barontini}, \citenamefont {Lamporesi}, \citenamefont
  {Rabatti}, \citenamefont {Thalhammer}, \citenamefont {Minardi}, \citenamefont
  {Stringari},\ and\ \citenamefont {Inguscio}}]{catani2009prl}%
  \BibitemOpen
  \bibfield  {author} {\bibinfo {author} {\bibnamefont {Catani}, \bibfnamefont
  {J}}, \bibinfo {author} {\bibfnamefont {G.}~\bibnamefont {Barontini}},
  \bibinfo {author} {\bibfnamefont {G.}~\bibnamefont {Lamporesi}}, \bibinfo
  {author} {\bibfnamefont {F.}~\bibnamefont {Rabatti}}, \bibinfo {author}
  {\bibfnamefont {G.}~\bibnamefont {Thalhammer}}, \bibinfo {author}
  {\bibfnamefont {F.}~\bibnamefont {Minardi}}, \bibinfo {author} {\bibfnamefont
  {S.}~\bibnamefont {Stringari}}, \ and\ \bibinfo {author} {\bibfnamefont
  {M.}~\bibnamefont {Inguscio}}} (\bibinfo {year} {2009}),\ \bibfield  {title}
  {\enquote {\bibinfo {title} {Entropy exchange in a mixture of ultracold
  atoms},}\ }\href@noop {} {\bibfield  {journal} {\bibinfo  {journal} {Phys.
  Rev. Lett.}\ }\textbf {\bibinfo {volume} {103}},\ \bibinfo {pages}
  {140401}}\BibitemShut {NoStop}%
\bibitem [{\citenamefont {Catani}\ \emph {et~al.}(2012)\citenamefont {Catani},
  \citenamefont {Lamporesi}, \citenamefont {Naik}, \citenamefont {Gring},
  \citenamefont {Inguscio}, \citenamefont {Minardi}, \citenamefont {Kantian},\
  and\ \citenamefont {Giamarchi}}]{catani_quantum_2012}%
  \BibitemOpen
  \bibfield  {author} {\bibinfo {author} {\bibnamefont {Catani}, \bibfnamefont
  {J}}, \bibinfo {author} {\bibfnamefont {G.}~\bibnamefont {Lamporesi}},
  \bibinfo {author} {\bibfnamefont {D.}~\bibnamefont {Naik}}, \bibinfo {author}
  {\bibfnamefont {M.}~\bibnamefont {Gring}}, \bibinfo {author} {\bibfnamefont
  {M.}~\bibnamefont {Inguscio}}, \bibinfo {author} {\bibfnamefont
  {F.}~\bibnamefont {Minardi}}, \bibinfo {author} {\bibfnamefont
  {A.}~\bibnamefont {Kantian}}, \ and\ \bibinfo {author} {\bibfnamefont
  {T.}~\bibnamefont {Giamarchi}}} (\bibinfo {year} {2012}),\ \bibfield  {title}
  {\enquote {\bibinfo {title} {Quantum dynamics of impurities in a
  one-dimensional {Bose} gas},}\ }\href {\doibase 10.1103/PhysRevA.85.023623}
  {\bibfield  {journal} {\bibinfo  {journal} {Physical Review A}\ }\textbf
  {\bibinfo {volume} {85}}~(\bibinfo {number} {2}),\ \bibinfo {pages}
  {023623}}\BibitemShut {NoStop}%
\bibitem [{\citenamefont {Cavagnero}(1984)}]{cavagnero1984pra}%
  \BibitemOpen
  \bibfield  {author} {\bibinfo {author} {\bibnamefont {Cavagnero},
  \bibfnamefont {M}}} (\bibinfo {year} {1984}),\ \bibfield  {title} {\enquote
  {\bibinfo {title} {Electron correlations in atomic shells: Systematics of
  high-angular-momentum admixtures},}\ }\href@noop {} {\bibfield  {journal}
  {\bibinfo  {journal} {Phys. Rev. A}\ }\textbf {\bibinfo {volume}
  {30}}~(\bibinfo {number} {3}),\ \bibinfo {pages} {1169--1174}}\BibitemShut
  {NoStop}%
\bibitem [{\citenamefont {Cavagnero}(1986)}]{cavagnero1986pra}%
  \BibitemOpen
  \bibfield  {author} {\bibinfo {author} {\bibnamefont {Cavagnero},
  \bibfnamefont {M}}} (\bibinfo {year} {1986}),\ \bibfield  {title} {\enquote
  {\bibinfo {title} {Electron correlations in atomic shells. {II}.
  {A}ntisymmetric basis functions},}\ }\href@noop {} {\bibfield  {journal}
  {\bibinfo  {journal} {Phys. Rev. A}\ }\textbf {\bibinfo {volume}
  {33}}~(\bibinfo {number} {5}),\ \bibinfo {pages} {2877--2886}}\BibitemShut
  {NoStop}%
\bibitem [{\citenamefont {Cazalilla}\ \emph {et~al.}(2011)\citenamefont
  {Cazalilla}, \citenamefont {Citro}, \citenamefont {Giamarchi}, \citenamefont
  {Orignac},\ and\ \citenamefont {Rigol}}]{cazalilla2011one}%
  \BibitemOpen
  \bibfield  {author} {\bibinfo {author} {\bibnamefont {Cazalilla},
  \bibfnamefont {M~A}}, \bibinfo {author} {\bibfnamefont {R.}~\bibnamefont
  {Citro}}, \bibinfo {author} {\bibfnamefont {T.}~\bibnamefont {Giamarchi}},
  \bibinfo {author} {\bibfnamefont {E.}~\bibnamefont {Orignac}}, \ and\
  \bibinfo {author} {\bibfnamefont {M.}~\bibnamefont {Rigol}}} (\bibinfo {year}
  {2011}),\ \bibfield  {title} {\enquote {\bibinfo {title} {One dimensional
  bosons: From condensed matter systems to ultracold gases},}\ }\href@noop {}
  {\bibfield  {journal} {\bibinfo  {journal} {Rev. Mod. Phys.}\ }\textbf
  {\bibinfo {volume} {83}}~(\bibinfo {number} {4}),\ \bibinfo {pages}
  {1405}}\BibitemShut {NoStop}%
\bibitem [{\citenamefont {Chang}\ and\ \citenamefont
  {Bertsch}(2007)}]{chang2007PRA}%
  \BibitemOpen
  \bibfield  {author} {\bibinfo {author} {\bibnamefont {Chang}, \bibfnamefont
  {S~Y}}, \ and\ \bibinfo {author} {\bibfnamefont {G.~F.}\ \bibnamefont
  {Bertsch}}} (\bibinfo {year} {2007}),\ \bibfield  {title} {\enquote {\bibinfo
  {title} {Local-density-functional theory for superfluid {F}ermionic systems:
  The unitary gas},}\ }\href@noop {} {\bibfield  {journal} {\bibinfo  {journal}
  {Phys. Rev. A}\ }\textbf {\bibinfo {volume} {76}}~(\bibinfo {number} {4}),\
  \bibinfo {pages} {021603 (R)}}\BibitemShut {NoStop}%
\bibitem [{\citenamefont {Cheon}\ and\ \citenamefont
  {Shigehara}(1999)}]{Cheon1999prl}%
  \BibitemOpen
  \bibfield  {author} {\bibinfo {author} {\bibnamefont {Cheon}, \bibfnamefont
  {Taksu}}, \ and\ \bibinfo {author} {\bibfnamefont {T.}~\bibnamefont
  {Shigehara}}} (\bibinfo {year} {1999}),\ \bibfield  {title} {\enquote
  {\bibinfo {title} {Fermion-boson duality of one-dimensional quantum particles
  with generalized contact interactions},}\ }\href {\doibase
  10.1103/PhysRevLett.82.2536} {\bibfield  {journal} {\bibinfo  {journal}
  {Phys. Rev. Lett.}\ }\textbf {\bibinfo {volume} {82}},\ \bibinfo {pages}
  {2536--2539}}\BibitemShut {NoStop}%
\bibitem [{\citenamefont {Chevy}\ \emph {et~al.}(2011)\citenamefont {Chevy},
  \citenamefont {Nascimbene}, \citenamefont {Navon}, \citenamefont {Jiang},
  \citenamefont {Lobo},\ and\ \citenamefont
  {Salomon}}]{ChevySalomon2011JPConf}%
  \BibitemOpen
  \bibfield  {author} {\bibinfo {author} {\bibnamefont {Chevy}, \bibfnamefont
  {F}}, \bibinfo {author} {\bibfnamefont {S.}~\bibnamefont {Nascimbene}},
  \bibinfo {author} {\bibfnamefont {N.}~\bibnamefont {Navon}}, \bibinfo
  {author} {\bibfnamefont {K.}~\bibnamefont {Jiang}}, \bibinfo {author}
  {\bibfnamefont {C.}~\bibnamefont {Lobo}}, \ and\ \bibinfo {author}
  {\bibfnamefont {C.}~\bibnamefont {Salomon}}} (\bibinfo {year} {2011}),\
  \bibfield  {title} {\enquote {\bibinfo {title} {{Thermodynamics of the
  unitary Fermi gas}},}\ }\href@noop {} {\bibfield  {journal} {\bibinfo
  {journal} {J. Phys. Conf. Ser.}\ }\textbf {\bibinfo {volume} {264}},\
  \bibinfo {pages} {012012}}\BibitemShut {NoStop}%
\bibitem [{\citenamefont {Child}(1974)}]{Child}%
  \BibitemOpen
  \bibfield  {author} {\bibinfo {author} {\bibnamefont {Child}, \bibfnamefont
  {M~S}}} (\bibinfo {year} {1974}),\ \href@noop {} {\emph {\bibinfo {title}
  {Molecular Collision Theory}}}\ (\bibinfo  {publisher} {Academic Press},\
  \bibinfo {address} {London, England})\BibitemShut {NoStop}%
\bibitem [{\citenamefont {Chin}(2011)}]{chin2011ARX}%
  \BibitemOpen
  \bibfield  {author} {\bibinfo {author} {\bibnamefont {Chin}, \bibfnamefont
  {C}}} (\bibinfo {year} {2011}),\ \bibfield  {title} {\enquote {\bibinfo
  {title} {Universal scaling of {E}fimov resonance positions in cold atom
  systems},}\ }\href@noop {} {\ ,\ \bibinfo {pages}
  {arXiv:1111.1484}}\BibitemShut {NoStop}%
\bibitem [{\citenamefont {Chin}\ \emph {et~al.}(2010)\citenamefont {Chin},
  \citenamefont {Grimm}, \citenamefont {Julienne},\ and\ \citenamefont
  {Tiesinga}}]{chin2010RMP}%
  \BibitemOpen
  \bibfield  {author} {\bibinfo {author} {\bibnamefont {Chin}, \bibfnamefont
  {C}}, \bibinfo {author} {\bibfnamefont {R.}~\bibnamefont {Grimm}}, \bibinfo
  {author} {\bibfnamefont {P.~S.}\ \bibnamefont {Julienne}}, \ and\ \bibinfo
  {author} {\bibfnamefont {E.}~\bibnamefont {Tiesinga}}} (\bibinfo {year}
  {2010}),\ \bibfield  {title} {\enquote {\bibinfo {title} {{F}eshbach
  resonances in ultracold gases},}\ }\href@noop {} {\bibfield  {journal}
  {\bibinfo  {journal} {Rev. Mod. Phys.}\ }\textbf {\bibinfo {volume}
  {82}}~(\bibinfo {number} {2}),\ \bibinfo {pages} {1225--1286}}\BibitemShut
  {NoStop}%
\bibitem [{\citenamefont
  {Christensen-{D}alsgaard}(1984)}]{christensendalsgaard1984PRAb}%
  \BibitemOpen
  \bibfield  {author} {\bibinfo {author} {\bibnamefont
  {Christensen-{D}alsgaard}, \bibfnamefont {B~L}}} (\bibinfo {year} {1984}),\
  \bibfield  {title} {{\selectlanguage {English}\enquote {\bibinfo {title}
  {Combined hyperspherical and close-coupling description of 2-electron
  wave-functions - application to e--{H} elastic-scattering phase-shifts},}\
  }}\href@noop {} {\bibfield  {journal} {\bibinfo  {journal} {Phys. Rev. A}\
  }\textbf {\bibinfo {volume} {29}}~(\bibinfo {number} {4}),\ \bibinfo {pages}
  {2242--2244}}\BibitemShut {NoStop}%
\bibitem [{\citenamefont {Clapp}(1949)}]{CLAPP1949}%
  \BibitemOpen
  \bibfield  {author} {\bibinfo {author} {\bibnamefont {Clapp}, \bibfnamefont
  {R~E}}} (\bibinfo {year} {1949}),\ \bibfield  {title} {\enquote {\bibinfo
  {title} {The binding energy of the triton},}\ }\href@noop {} {\bibfield
  {journal} {\bibinfo  {journal} {Phys. Rev.}\ }\textbf {\bibinfo {volume}
  {76}}~(\bibinfo {number} {6}),\ \bibinfo {pages} {873--874}}\BibitemShut
  {NoStop}%
\bibitem [{\citenamefont {Clare}\ and\ \citenamefont
  {Levinger}(1985)}]{Clare-1985}%
  \BibitemOpen
  \bibfield  {author} {\bibinfo {author} {\bibnamefont {Clare}, \bibfnamefont
  {R~B}}, \ and\ \bibinfo {author} {\bibfnamefont {J.~S.}\ \bibnamefont
  {Levinger}}} (\bibinfo {year} {1985}),\ \bibfield  {title} {\enquote
  {\bibinfo {title} {Hypertriton and hyperspherical harmonics},}\ }\href@noop
  {} {\bibfield  {journal} {\bibinfo  {journal} {Phys. Rev. C}\ }\textbf
  {\bibinfo {volume} {31}},\ \bibinfo {pages} {2303}}\BibitemShut {NoStop}%
\bibitem [{\citenamefont {Clark}(1983)}]{clark1983PRA}%
  \BibitemOpen
  \bibfield  {author} {\bibinfo {author} {\bibnamefont {Clark}, \bibfnamefont
  {C~W}}} (\bibinfo {year} {1983}),\ \bibfield  {title} {\enquote {\bibinfo
  {title} {Low-energy electron-atom scattering in a magnetic field},}\
  }\href@noop {} {\bibfield  {journal} {\bibinfo  {journal} {Phys. Rev. A}\
  }\textbf {\bibinfo {volume} {28}},\ \bibinfo {pages} {83--90}}\BibitemShut
  {NoStop}%
\bibitem [{\citenamefont {Clark}\ and\ \citenamefont
  {Greene}(1980)}]{clark1980PRA}%
  \BibitemOpen
  \bibfield  {author} {\bibinfo {author} {\bibnamefont {Clark}, \bibfnamefont
  {C~W}}, \ and\ \bibinfo {author} {\bibfnamefont {C.~H.}\ \bibnamefont
  {Greene}}} (\bibinfo {year} {1980}),\ \bibfield  {title} {\enquote {\bibinfo
  {title} {Hyperspherical analysis of three-electron dynamics},}\ }\href@noop
  {} {\bibfield  {journal} {\bibinfo  {journal} {Phys. Rev. A}\ }\textbf
  {\bibinfo {volume} {21}}~(\bibinfo {number} {6}),\ \bibinfo {pages}
  {1786--1797}}\BibitemShut {NoStop}%
\bibitem [{\citenamefont {Clary}(1991)}]{Clary1991JCP}%
  \BibitemOpen
  \bibfield  {author} {\bibinfo {author} {\bibnamefont {Clary}, \bibfnamefont
  {D~C}}} (\bibinfo {year} {1991}),\ \bibfield  {title} {\enquote {\bibinfo
  {title} {Quantum reactive scattering of four-atom reactions with nonlinear
  geometry: {OH}+{H}$_2$ $\rightarrow$ {H}$_2${O}+{H}},}\ }\href@noop {}
  {\bibfield  {journal} {\bibinfo  {journal} {J. Chem. Phys.}\ }\textbf
  {\bibinfo {volume} {95}}~(\bibinfo {number} {10}),\ \bibinfo {pages}
  {7298--7310}}\BibitemShut {NoStop}%
\bibitem [{\citenamefont {Cobis}\ \emph {et~al.}(1998)\citenamefont {Cobis},
  \citenamefont {Fedorov},\ and\ \citenamefont {Jensen}}]{cobis1998PRC}%
  \BibitemOpen
  \bibfield  {author} {\bibinfo {author} {\bibnamefont {Cobis}, \bibfnamefont
  {A}}, \bibinfo {author} {\bibfnamefont {DV}~\bibnamefont {Fedorov}}, \ and\
  \bibinfo {author} {\bibfnamefont {AS}~\bibnamefont {Jensen}}} (\bibinfo
  {year} {1998}),\ \bibfield  {title} {{\selectlanguage {English}\enquote
  {\bibinfo {title} {Three-body halos. v. computations of continuum spectra for
  borromean nuclei},}\ }}\href@noop {} {\bibfield  {journal} {\bibinfo
  {journal} {PRC}\ }\textbf {\bibinfo {volume} {58}}~(\bibinfo {number} {3}),\
  \bibinfo {pages} {1403--1421}}\BibitemShut {NoStop}%
\bibitem [{\citenamefont {Cobis}\ \emph {et~al.}(1997)\citenamefont {Cobis},
  \citenamefont {Jensen},\ and\ \citenamefont {Fedorov}}]{Cobis-1997}%
  \BibitemOpen
  \bibfield  {author} {\bibinfo {author} {\bibnamefont {Cobis}, \bibfnamefont
  {A}}, \bibinfo {author} {\bibfnamefont {A.~S.}\ \bibnamefont {Jensen}}, \
  and\ \bibinfo {author} {\bibfnamefont {D.~V.}\ \bibnamefont {Fedorov}}}
  (\bibinfo {year} {1997}),\ \bibfield  {title} {\enquote {\bibinfo {title}
  {The simplest strange three-body halo},}\ }\href@noop {} {\bibfield
  {journal} {\bibinfo  {journal} {J. Phys. G}\ }\textbf {\bibinfo {volume}
  {23}},\ \bibinfo {pages} {401}}\BibitemShut {NoStop}%
\bibitem [{\citenamefont {Combescot}\ \emph {et~al.}(2009)\citenamefont
  {Combescot}, \citenamefont {Alzetto},\ and\ \citenamefont
  {Leyronas}}]{Combescot-2009}%
  \BibitemOpen
  \bibfield  {author} {\bibinfo {author} {\bibnamefont {Combescot},
  \bibfnamefont {R}}, \bibinfo {author} {\bibfnamefont {F.}~\bibnamefont
  {Alzetto}}, \ and\ \bibinfo {author} {\bibfnamefont {X.}~\bibnamefont
  {Leyronas}}} (\bibinfo {year} {2009}),\ \bibfield  {title} {\enquote
  {\bibinfo {title} {Particle distribution tail and related energy formula},}\
  }\href@noop {} {\bibfield  {journal} {\bibinfo  {journal} {Phys. Rev A}\
  }\textbf {\bibinfo {volume} {79}},\ \bibinfo {pages} {053640}}\BibitemShut
  {NoStop}%
\bibitem [{\citenamefont {Corson}\ and\ \citenamefont
  {Bohn}({2016})}]{CorsonBohn2016pra}%
  \BibitemOpen
  \bibfield  {author} {\bibinfo {author} {\bibnamefont {Corson}, \bibfnamefont
  {J~P}}, \ and\ \bibinfo {author} {\bibfnamefont {J.~L.}\ \bibnamefont
  {Bohn}}} (\bibinfo {year} {{2016}}),\ \bibfield  {title} {\enquote {\bibinfo
  {title} {{Ballistic quench-induced correlation waves in ultracold gases}},}\
  }\href@noop {} {\bibfield  {journal} {\bibinfo  {journal} {{Phys. Rev. A}}\
  }\textbf {\bibinfo {volume} {{94}}}~(\bibinfo {number} {{2}})}\BibitemShut
  {NoStop}%
\bibitem [{\citenamefont {C\^ot\'e}(2016)}]{Cote2016adv}%
  \BibitemOpen
  \bibfield  {author} {\bibinfo {author} {\bibnamefont {C\^ot\'e},
  \bibfnamefont {R}}} (\bibinfo {year} {2016}),\ \bibfield  {title} {\enquote
  {\bibinfo {title} {Chapter two - ultracold hybrid atom ion systems},}\
  }\href@noop {} {\bibfield  {journal} {\bibinfo  {journal} {Adv. At. Mol. Opt.
  Phys.}\ }\textbf {\bibinfo {volume} {65}},\ \bibinfo {pages}
  {67}}\BibitemShut {NoStop}%
\bibitem [{\citenamefont {Cowell}\ \emph {et~al.}(2002)\citenamefont {Cowell},
  \citenamefont {Heiselberg}, \citenamefont {Mazets}, \citenamefont {Morales},
  \citenamefont {Pandharipande},\ and\ \citenamefont
  {Pethick}}]{Pethick2002prl}%
  \BibitemOpen
  \bibfield  {author} {\bibinfo {author} {\bibnamefont {Cowell}, \bibfnamefont
  {S}}, \bibinfo {author} {\bibfnamefont {H.}~\bibnamefont {Heiselberg}},
  \bibinfo {author} {\bibfnamefont {I.~E.}\ \bibnamefont {Mazets}}, \bibinfo
  {author} {\bibfnamefont {J.}~\bibnamefont {Morales}}, \bibinfo {author}
  {\bibfnamefont {V.~R.}\ \bibnamefont {Pandharipande}}, \ and\ \bibinfo
  {author} {\bibfnamefont {C.~J.}\ \bibnamefont {Pethick}}} (\bibinfo {year}
  {2002}),\ \bibfield  {title} {\enquote {\bibinfo {title} {Cold {B}ose gases
  with large scattering lengths},}\ }\href@noop {} {\bibfield  {journal}
  {\bibinfo  {journal} {Phys. Rev. Lett.}\ }\textbf {\bibinfo {volume} {88}},\
  \bibinfo {pages} {210403}}\BibitemShut {NoStop}%
\bibitem [{\citenamefont {Crawford}(1988)}]{crawfordpra1988}%
  \BibitemOpen
  \bibfield  {author} {\bibinfo {author} {\bibnamefont {Crawford},
  \bibfnamefont {O~H}}} (\bibinfo {year} {1988}),\ \bibfield  {title} {\enquote
  {\bibinfo {title} {Oscillations in photodetachment cross sections for
  negative ions in magnetic fields},}\ }\href@noop {} {\bibfield  {journal}
  {\bibinfo  {journal} {Phys. Rev. A}\ }\textbf {\bibinfo {volume}
  {37}}~(\bibinfo {number} {7}),\ \bibinfo {pages} {2432--2440}}\BibitemShut
  {NoStop}%
\bibitem [{\citenamefont {Cubizolles}\ \emph {et~al.}(2003)\citenamefont
  {Cubizolles}, \citenamefont {Bourdel}, \citenamefont {Kokkelmans},
  \citenamefont {Shlyapnikov},\ and\ \citenamefont
  {Salomon}}]{cubizolles2003PRL}%
  \BibitemOpen
  \bibfield  {author} {\bibinfo {author} {\bibnamefont {Cubizolles},
  \bibfnamefont {J}}, \bibinfo {author} {\bibfnamefont {T.}~\bibnamefont
  {Bourdel}}, \bibinfo {author} {\bibfnamefont {S.~J. J. M.~F.}\ \bibnamefont
  {Kokkelmans}}, \bibinfo {author} {\bibfnamefont {G.~V.}\ \bibnamefont
  {Shlyapnikov}}, \ and\ \bibinfo {author} {\bibfnamefont {C.}~\bibnamefont
  {Salomon}}} (\bibinfo {year} {2003}),\ \bibfield  {title} {\enquote {\bibinfo
  {title} {Production of long-lived ultracold {Li}$_2$ molecules from a {F}ermi
  gas},}\ }\href@noop {} {\bibfield  {journal} {\bibinfo  {journal} {Phys. Rev.
  Lett.}\ }\textbf {\bibinfo {volume} {91}},\ \bibinfo {pages}
  {240401}}\BibitemShut {NoStop}%
\bibitem [{\citenamefont {Cucchietti}\ and\ \citenamefont
  {Timmermans}(2006)}]{cucchietti_strong-coupling_2006}%
  \BibitemOpen
  \bibfield  {author} {\bibinfo {author} {\bibnamefont {Cucchietti},
  \bibfnamefont {F~M}}, \ and\ \bibinfo {author} {\bibfnamefont
  {E.}~\bibnamefont {Timmermans}}} (\bibinfo {year} {2006}),\ \bibfield
  {title} {\enquote {\bibinfo {title} {Strong-coupling polarons in dilute gas
  {Bose}-{Einstein} condensates},}\ }\href {\doibase
  10.1103/PhysRevLett.96.210401} {\bibfield  {journal} {\bibinfo  {journal}
  {Physical Review Letters}\ }\textbf {\bibinfo {volume} {96}}~(\bibinfo
  {number} {21}),\ \bibinfo {pages} {210401}}\BibitemShut {NoStop}%
\bibitem [{\citenamefont {Cui}(2012)}]{xiaoling2012pra}%
  \BibitemOpen
  \bibfield  {author} {\bibinfo {author} {\bibnamefont {Cui}, \bibfnamefont
  {X}}} (\bibinfo {year} {2012}),\ \bibfield  {title} {\enquote {\bibinfo
  {title} {Mixed-partial-wave scattering with spin-orbit coupling and validity
  of pseudopotentials},}\ }\href@noop {} {\bibfield  {journal} {\bibinfo
  {journal} {Phys. Rev. A}\ }\textbf {\bibinfo {volume} {85}},\ \bibinfo
  {pages} {022705}}\BibitemShut {NoStop}%
\bibitem [{\citenamefont {Cui}\ \emph {et~al.}(2010)\citenamefont {Cui},
  \citenamefont {Wang},\ and\ \citenamefont {Zhou}}]{cui2010}%
  \BibitemOpen
  \bibfield  {author} {\bibinfo {author} {\bibnamefont {Cui}, \bibfnamefont
  {X}}, \bibinfo {author} {\bibfnamefont {Y.}~\bibnamefont {Wang}}, \ and\
  \bibinfo {author} {\bibfnamefont {F.}~\bibnamefont {Zhou}}} (\bibinfo {year}
  {2010}),\ \bibfield  {title} {\enquote {\bibinfo {title} {Resonance
  scattering in optical lattices and molecules: Interband versus intraband
  effects},}\ }\href@noop {} {\bibfield  {journal} {\bibinfo  {journal} {Phys.
  Rev. Lett.}\ }\textbf {\bibinfo {volume} {104}}~(\bibinfo {number} {15}),\
  \bibinfo {pages} {153201}}\BibitemShut {NoStop}%
\bibitem [{\citenamefont {Cvejanovic}\ and\ \citenamefont
  {Read}(1974)}]{Cvejanovic1974jpb}%
  \BibitemOpen
  \bibfield  {author} {\bibinfo {author} {\bibnamefont {Cvejanovic},
  \bibfnamefont {S}}, \ and\ \bibinfo {author} {\bibfnamefont {F~H}\
  \bibnamefont {Read}}} (\bibinfo {year} {1974}),\ \bibfield  {title} {\enquote
  {\bibinfo {title} {Studies of the threshold electron impact ionization of
  helium},}\ }\href {http://stacks.iop.org/0022-3700/7/i=14/a=008} {\bibfield
  {journal} {\bibinfo  {journal} {Journal of Physics B: Atomic and Molecular
  Physics}\ }\textbf {\bibinfo {volume} {7}}~(\bibinfo {number} {14}),\
  \bibinfo {pages} {1841}}\BibitemShut {NoStop}%
\bibitem [{\citenamefont {Daily}\ and\ \citenamefont
  {Greene}(2014)}]{Daily-2014}%
  \BibitemOpen
  \bibfield  {author} {\bibinfo {author} {\bibnamefont {Daily}, \bibfnamefont
  {K~M}}, \ and\ \bibinfo {author} {\bibfnamefont {C.~H.}\ \bibnamefont
  {Greene}}} (\bibinfo {year} {2014}),\ \bibfield  {title} {\enquote {\bibinfo
  {title} {Extension of the correlated {G}aussian hyperspherical method to more
  particles and dimensions},}\ }\href@noop {} {\bibfield  {journal} {\bibinfo
  {journal} {Phys. Rev. A}\ }\textbf {\bibinfo {volume} {89}},\ \bibinfo
  {pages} {012503}}\BibitemShut {NoStop}%
\bibitem [{\citenamefont {Daily}\ \emph
  {et~al.}(2015{\natexlab{a}})\citenamefont {Daily}, \citenamefont {Kievsky},\
  and\ \citenamefont {Greene}}]{Daily-2015}%
  \BibitemOpen
  \bibfield  {author} {\bibinfo {author} {\bibnamefont {Daily}, \bibfnamefont
  {K~M}}, \bibinfo {author} {\bibfnamefont {A.}~\bibnamefont {Kievsky}}, \ and\
  \bibinfo {author} {\bibfnamefont {C.~H.}\ \bibnamefont {Greene}}} (\bibinfo
  {year} {2015}{\natexlab{a}}),\ \bibfield  {title} {\enquote {\bibinfo {title}
  {Adiabatic hyperspherical analysis of realistic nuclear potentials},}\
  }\href@noop {} {\bibfield  {journal} {\bibinfo  {journal} {Few-Body Syst.}\
  }\textbf {\bibinfo {volume} {56}},\ \bibinfo {pages} {753--759}}\BibitemShut
  {NoStop}%
\bibitem [{\citenamefont {Daily}\ \emph
  {et~al.}(2015{\natexlab{b}})\citenamefont {Daily}, \citenamefont {Wooten},\
  and\ \citenamefont {Greene}}]{Daily2015PRB}%
  \BibitemOpen
  \bibfield  {author} {\bibinfo {author} {\bibnamefont {Daily}, \bibfnamefont
  {K~M}}, \bibinfo {author} {\bibfnamefont {R.~E.}\ \bibnamefont {Wooten}}, \
  and\ \bibinfo {author} {\bibfnamefont {C.~H.}\ \bibnamefont {Greene}}}
  (\bibinfo {year} {2015}{\natexlab{b}}),\ \bibfield  {title} {\enquote
  {\bibinfo {title} {Hyperspherical theory of the quantum {H}all effect: The
  role of exceptional degeneracy},}\ }\href@noop {} {\bibfield  {journal}
  {\bibinfo  {journal} {Phys. Rev. B}\ }\textbf {\bibinfo {volume} {92}},\
  \bibinfo {pages} {125427}}\BibitemShut {NoStop}%
\bibitem [{\citenamefont {Dalfovo}\ \emph {et~al.}(1999)\citenamefont
  {Dalfovo}, \citenamefont {Giorgini}, \citenamefont {Pitaevskii},\ and\
  \citenamefont {Stringari}}]{dalfovo1999RMP}%
  \BibitemOpen
  \bibfield  {author} {\bibinfo {author} {\bibnamefont {Dalfovo}, \bibfnamefont
  {F}}, \bibinfo {author} {\bibfnamefont {S.}~\bibnamefont {Giorgini}},
  \bibinfo {author} {\bibfnamefont {L.~P.}\ \bibnamefont {Pitaevskii}}, \ and\
  \bibinfo {author} {\bibfnamefont {S.}~\bibnamefont {Stringari}}} (\bibinfo
  {year} {1999}),\ \bibfield  {title} {\enquote {\bibinfo {title} {Theory of
  {{B}ose-{E}instein} condensation in trapped gases},}\ }\href@noop {}
  {\bibfield  {journal} {\bibinfo  {journal} {Rev. Mod. Phys.}\ }\textbf
  {\bibinfo {volume} {71}}~(\bibinfo {number} {3}),\ \bibinfo {pages}
  {463--512}}\BibitemShut {NoStop}%
\bibitem [{\citenamefont {Davis}\ \emph {et~al.}(1995)\citenamefont {Davis},
  \citenamefont {Mewes}, \citenamefont {Andrews}, \citenamefont {{Van Druten}},
  \citenamefont {Durfee}, \citenamefont {Kurn},\ and\ \citenamefont
  {Ketterle}}]{davis1995}%
  \BibitemOpen
  \bibfield  {author} {\bibinfo {author} {\bibnamefont {Davis}, \bibfnamefont
  {K~B}}, \bibinfo {author} {\bibfnamefont {M-O}\ \bibnamefont {Mewes}},
  \bibinfo {author} {\bibfnamefont {M.~R.~van}\ \bibnamefont {Andrews}},
  \bibinfo {author} {\bibfnamefont {N.~J.}\ \bibnamefont {{Van Druten}}},
  \bibinfo {author} {\bibfnamefont {D.~S.}\ \bibnamefont {Durfee}}, \bibinfo
  {author} {\bibfnamefont {D.~M.}\ \bibnamefont {Kurn}}, \ and\ \bibinfo
  {author} {\bibfnamefont {W.}~\bibnamefont {Ketterle}}} (\bibinfo {year}
  {1995}),\ \bibfield  {title} {\enquote {\bibinfo {title} {{B}ose-{E}instein
  condensation in a gas of sodium atoms},}\ }\href@noop {} {\bibfield
  {journal} {\bibinfo  {journal} {Phys. Rev. Lett.}\ }\textbf {\bibinfo
  {volume} {75}}~(\bibinfo {number} {22}),\ \bibinfo {pages}
  {3969}}\BibitemShut {NoStop}%
\bibitem [{\citenamefont {{de Goey}}\ \emph {et~al.}(1986)\citenamefont {{de
  Goey}}, \citenamefont {{v. d. Berg}}, \citenamefont {Mulders}, \citenamefont
  {Stoof}, \citenamefont {Verhaar},\ and\ \citenamefont
  {Gl\"ockle}}]{Goey-1986}%
  \BibitemOpen
  \bibfield  {author} {\bibinfo {author} {\bibnamefont {{de Goey}},
  \bibfnamefont {L~P~H}}, \bibinfo {author} {\bibfnamefont {T.~H.~M.}\
  \bibnamefont {{v. d. Berg}}}, \bibinfo {author} {\bibfnamefont
  {N.}~\bibnamefont {Mulders}}, \bibinfo {author} {\bibfnamefont {H.~T.~C.}\
  \bibnamefont {Stoof}}, \bibinfo {author} {\bibfnamefont {B.~J.}\ \bibnamefont
  {Verhaar}}, \ and\ \bibinfo {author} {\bibfnamefont {W.}~\bibnamefont
  {Gl\"ockle}}} (\bibinfo {year} {1986}),\ \bibfield  {title} {\enquote
  {\bibinfo {title} {3-body recombination in spin-polarized atomic-hydrogen},}\
  }\href@noop {} {\bibfield  {journal} {\bibinfo  {journal} {Phys. Rev. B}\
  }\textbf {\bibinfo {volume} {34}},\ \bibinfo {pages} {6183}}\BibitemShut
  {NoStop}%
\bibitem [{\citenamefont {Deltuva}(2010)}]{deltuva2010PRAb}%
  \BibitemOpen
  \bibfield  {author} {\bibinfo {author} {\bibnamefont {Deltuva}, \bibfnamefont
  {A}}} (\bibinfo {year} {2010}),\ \bibfield  {title} {\enquote {\bibinfo
  {title} {Efimov physics in bosonic atom-trimer scattering},}\ }\href@noop {}
  {\bibfield  {journal} {\bibinfo  {journal} {Phys. Rev. A}\ }\textbf {\bibinfo
  {volume} {82}},\ \bibinfo {pages} {040701}}\BibitemShut {NoStop}%
\bibitem [{\citenamefont {Deltuva}(2011)}]{deltuva2011PRA}%
  \BibitemOpen
  \bibfield  {author} {\bibinfo {author} {\bibnamefont {Deltuva}, \bibfnamefont
  {A}}} (\bibinfo {year} {2011}),\ \bibfield  {title} {\enquote {\bibinfo
  {title} {Universality in bosonic dimer-dimer scattering},}\ }\href@noop {}
  {\bibfield  {journal} {\bibinfo  {journal} {Phys. Rev. A}\ }\textbf {\bibinfo
  {volume} {84}},\ \bibinfo {pages} {022703}}\BibitemShut {NoStop}%
\bibitem [{\citenamefont {Deltuva}(2012)}]{deltuva2012PRA}%
  \BibitemOpen
  \bibfield  {author} {\bibinfo {author} {\bibnamefont {Deltuva}, \bibfnamefont
  {A}}} (\bibinfo {year} {2012}),\ \bibfield  {title} {\enquote {\bibinfo
  {title} {Momentum-space calculation of four-boson recombination},}\
  }\href@noop {} {\bibfield  {journal} {\bibinfo  {journal} {Phys. Rev. A}\
  }\textbf {\bibinfo {volume} {85}},\ \bibinfo {pages} {012708}}\BibitemShut
  {NoStop}%
\bibitem [{\citenamefont {Delves}(1959)}]{delves1959NP}%
  \BibitemOpen
  \bibfield  {author} {\bibinfo {author} {\bibnamefont {Delves}, \bibfnamefont
  {L~M}}} (\bibinfo {year} {1959}),\ \bibfield  {title} {\enquote {\bibinfo
  {title} {Tertiary and general-order collisions},}\ }\href@noop {} {\bibfield
  {journal} {\bibinfo  {journal} {Nucl. Phys.}\ }\textbf {\bibinfo {volume}
  {9}},\ \bibinfo {pages} {391}}\BibitemShut {NoStop}%
\bibitem [{\citenamefont {Delves}(1960)}]{delves1960NP}%
  \BibitemOpen
  \bibfield  {author} {\bibinfo {author} {\bibnamefont {Delves}, \bibfnamefont
  {L~M}}} (\bibinfo {year} {1960}),\ \bibfield  {title} {\enquote {\bibinfo
  {title} {Tertiary and general-order collisions part {II}},}\ }\href@noop {}
  {\bibfield  {journal} {\bibinfo  {journal} {Nucl. Phys.}\ }\textbf {\bibinfo
  {volume} {20}},\ \bibinfo {pages} {275}}\BibitemShut {NoStop}%
\bibitem [{\citenamefont {DeMarco}\ and\ \citenamefont
  {Jin}(1999)}]{demarco1999Sci}%
  \BibitemOpen
  \bibfield  {author} {\bibinfo {author} {\bibnamefont {DeMarco}, \bibfnamefont
  {B}}, \ and\ \bibinfo {author} {\bibfnamefont {D.~S.}\ \bibnamefont {Jin}}}
  (\bibinfo {year} {1999}),\ \bibfield  {title} {\enquote {\bibinfo {title}
  {Onset of {F}ermi degeneracy in a trapped atomic gas},}\ }\href@noop {}
  {\bibfield  {journal} {\bibinfo  {journal} {Science}\ }\textbf {\bibinfo
  {volume} {285}}~(\bibinfo {number} {5434}),\ \bibinfo {pages}
  {1703}}\BibitemShut {NoStop}%
\bibitem [{\citenamefont {Demkov}\ and\ \citenamefont
  {Drukarev}(1966)}]{demkovjetp1966}%
  \BibitemOpen
  \bibfield  {author} {\bibinfo {author} {\bibnamefont {Demkov}, \bibfnamefont
  {Yu~N}}, \ and\ \bibinfo {author} {\bibfnamefont {G.~F.}\ \bibnamefont
  {Drukarev}}} (\bibinfo {year} {1966}),\ \bibfield  {title} {\enquote
  {\bibinfo {title} {Particle of low binding energy in a magnetic field},}\
  }\href@noop {} {\bibfield  {journal} {\bibinfo  {journal} {Sov. Phys. JETP}\
  }\textbf {\bibinfo {volume} {2}},\ \bibinfo {pages} {182}}\BibitemShut
  {NoStop}%
\bibitem [{\citenamefont {Dickerscheid}\ and\ \citenamefont
  {Stoof}(2005)}]{dickerscheid2005feshbach}%
  \BibitemOpen
  \bibfield  {author} {\bibinfo {author} {\bibnamefont {Dickerscheid},
  \bibfnamefont {D~B~M}}, \ and\ \bibinfo {author} {\bibfnamefont {H.~T.~C.}\
  \bibnamefont {Stoof}}} (\bibinfo {year} {2005}),\ \bibfield  {title}
  {\enquote {\bibinfo {title} {Feshbach molecules in a one-dimensional {F}ermi
  gas},}\ }\href@noop {} {\bibfield  {journal} {\bibinfo  {journal} {Phys. Rev.
  A}\ }\textbf {\bibinfo {volume} {72}}~(\bibinfo {number} {5}),\ \bibinfo
  {pages} {053625}}\BibitemShut {NoStop}%
\bibitem [{\citenamefont {D'Incao}(2003)}]{dincao2003PRA}%
  \BibitemOpen
  \bibfield  {author} {\bibinfo {author} {\bibnamefont {D'Incao}, \bibfnamefont
  {J~P}}} (\bibinfo {year} {2003}),\ \bibfield  {title} {{\selectlanguage
  {English}\enquote {\bibinfo {title} {Hyperspherical angular adiabatic
  separation for three-electron atomic systems},}\ }}\href@noop {} {\bibfield
  {journal} {\bibinfo  {journal} {Phys. Rev. A}\ }\textbf {\bibinfo {volume}
  {67}}~(\bibinfo {number} {2}),\ \bibinfo {pages} {024501}}\BibitemShut
  {NoStop}%
\bibitem [{\citenamefont {D'Incao}\ \emph {et~al.}({2015})\citenamefont
  {D'Incao}, \citenamefont {Anis},\ and\ \citenamefont
  {Esry}}]{DIncaoAnisEsry2015pra}%
  \BibitemOpen
  \bibfield  {author} {\bibinfo {author} {\bibnamefont {D'Incao}, \bibfnamefont
  {J~P}}, \bibinfo {author} {\bibfnamefont {Fatima}\ \bibnamefont {Anis}}, \
  and\ \bibinfo {author} {\bibfnamefont {B.~D.}\ \bibnamefont {Esry}}}
  (\bibinfo {year} {{2015}}),\ \bibfield  {title} {\enquote {\bibinfo {title}
  {{Ultracold three-body recombination in two dimensions}},}\ }\href {\doibase
  {10.1103/PhysRevA.91.062710}} {\bibfield  {journal} {\bibinfo  {journal}
  {{Physical Review A}}\ }\textbf {\bibinfo {volume} {{91}}}~(\bibinfo {number}
  {{6}}),\ {10.1103/PhysRevA.91.062710}}\BibitemShut {NoStop}%
\bibitem [{\citenamefont {D'Incao}\ and\ \citenamefont
  {Esry}(2005)}]{dincao2005PRL}%
  \BibitemOpen
  \bibfield  {author} {\bibinfo {author} {\bibnamefont {D'Incao}, \bibfnamefont
  {J~P}}, \ and\ \bibinfo {author} {\bibfnamefont {B.~D.}\ \bibnamefont
  {Esry}}} (\bibinfo {year} {2005}),\ \bibfield  {title} {{\selectlanguage
  {English}\enquote {\bibinfo {title} {Scattering length scaling laws for
  ultracold three-body collisions},}\ }}\href@noop {} {\bibfield  {journal}
  {\bibinfo  {journal} {Phys. Rev. Lett.}\ }\textbf {\bibinfo {volume}
  {94}}~(\bibinfo {number} {21}),\ \bibinfo {pages} {213201}}\BibitemShut
  {NoStop}%
\bibitem [{\citenamefont {D'Incao}\ and\ \citenamefont
  {Esry}(2006{\natexlab{a}})}]{dincao2006PRAb}%
  \BibitemOpen
  \bibfield  {author} {\bibinfo {author} {\bibnamefont {D'Incao}, \bibfnamefont
  {J~P}}, \ and\ \bibinfo {author} {\bibfnamefont {B.~D.}\ \bibnamefont
  {Esry}}} (\bibinfo {year} {2006}{\natexlab{a}}),\ \bibfield  {title}
  {{\selectlanguage {English}\enquote {\bibinfo {title} {Mass dependence of
  ultracold three-body collision rates},}\ }}\href@noop {} {\bibfield
  {journal} {\bibinfo  {journal} {Phys. Rev. A}\ }\textbf {\bibinfo {volume}
  {73}}~(\bibinfo {number} {3}),\ \bibinfo {pages} {030702}}\BibitemShut
  {NoStop}%
\bibitem [{\citenamefont {D'Incao}\ and\ \citenamefont
  {Esry}({2014})}]{DIncaoEsry2014pra}%
  \BibitemOpen
  \bibfield  {author} {\bibinfo {author} {\bibnamefont {D'Incao}, \bibfnamefont
  {J~P}}, \ and\ \bibinfo {author} {\bibfnamefont {B.~D.}\ \bibnamefont
  {Esry}}} (\bibinfo {year} {{2014}}),\ \bibfield  {title} {\enquote {\bibinfo
  {title} {{Adiabatic hyperspherical representation for the three-body problem
  in two dimensions}},}\ }\href {\doibase {10.1103/PhysRevA.90.042707}}
  {\bibfield  {journal} {\bibinfo  {journal} {{Physical Review A}}\ }\textbf
  {\bibinfo {volume} {{90}}}~(\bibinfo {number} {{4}}),\
  {10.1103/PhysRevA.90.042707}}\BibitemShut {NoStop}%
\bibitem [{\citenamefont {D'Incao}\ \emph {et~al.}(2009)\citenamefont
  {D'Incao}, \citenamefont {von Stecher},\ and\ \citenamefont
  {Greene}}]{dincao2009PRLb}%
  \BibitemOpen
  \bibfield  {author} {\bibinfo {author} {\bibnamefont {D'Incao}, \bibfnamefont
  {J~P}}, \bibinfo {author} {\bibfnamefont {J.}~\bibnamefont {von Stecher}}, \
  and\ \bibinfo {author} {\bibfnamefont {C.~H.}\ \bibnamefont {Greene}}}
  (\bibinfo {year} {2009}),\ \bibfield  {title} {{\selectlanguage
  {English}\enquote {\bibinfo {title} {Universal four-boson states in ultracold
  molecular gases: Resonant effects in dimer-dimer collisions},}\ }}\href@noop
  {} {\bibfield  {journal} {\bibinfo  {journal} {Phys. Rev. Lett.}\ }\textbf
  {\bibinfo {volume} {103}}~(\bibinfo {number} {3}),\ \bibinfo {pages}
  {033004}}\BibitemShut {NoStop}%
\bibitem [{\citenamefont {D'Incao}\ and\ \citenamefont
  {Greene}(2011)}]{dincao2011pra}%
  \BibitemOpen
  \bibfield  {author} {\bibinfo {author} {\bibnamefont {D'Incao}, \bibfnamefont
  {Jos\'e~P}}, \ and\ \bibinfo {author} {\bibfnamefont {Chris~H.}\ \bibnamefont
  {Greene}}} (\bibinfo {year} {2011}),\ \bibfield  {title} {\enquote {\bibinfo
  {title} {Collisional aspects of bosonic and fermionic dipoles in
  quasi-two-dimensional confining geometries},}\ }\href@noop {} {\bibfield
  {journal} {\bibinfo  {journal} {Phys. Rev. A}\ }\textbf {\bibinfo {volume}
  {83}},\ \bibinfo {pages} {030702}}\BibitemShut {NoStop}%
\bibitem [{\citenamefont {D'Incao}\ and\ \citenamefont
  {Esry}(2006{\natexlab{b}})}]{dincao2006PRA}%
  \BibitemOpen
  \bibfield  {author} {\bibinfo {author} {\bibnamefont {D'Incao}, \bibfnamefont
  {JP}}, \ and\ \bibinfo {author} {\bibfnamefont {BD}~\bibnamefont {Esry}}}
  (\bibinfo {year} {2006}{\natexlab{b}}),\ \bibfield  {title} {{\selectlanguage
  {English}\enquote {\bibinfo {title} {Enhancing the observability of the
  {E}fimov effect in ultracold atomic gas mixtures},}\ }}\href@noop {}
  {\bibfield  {journal} {\bibinfo  {journal} {Phys. Rev. A}\ }\textbf {\bibinfo
  {volume} {73}}~(\bibinfo {number} {3}),\ \bibinfo {pages}
  {030703}}\BibitemShut {NoStop}%
\bibitem [{\citenamefont {Domke}\ \emph {et~al.}(1991)\citenamefont {Domke},
  \citenamefont {Xue}, \citenamefont {Puschmann}, \citenamefont {Mandel},
  \citenamefont {Hudson}, \citenamefont {Shirley}, \citenamefont {Kaindl},
  \citenamefont {Greene}, \citenamefont {Sadeghpour},\ and\ \citenamefont
  {Petersen}}]{domke1991PRL}%
  \BibitemOpen
  \bibfield  {author} {\bibinfo {author} {\bibnamefont {Domke}, \bibfnamefont
  {M}}, \bibinfo {author} {\bibfnamefont {C.}~\bibnamefont {Xue}}, \bibinfo
  {author} {\bibfnamefont {A.}~\bibnamefont {Puschmann}}, \bibinfo {author}
  {\bibfnamefont {T.}~\bibnamefont {Mandel}}, \bibinfo {author} {\bibfnamefont
  {E.}~\bibnamefont {Hudson}}, \bibinfo {author} {\bibfnamefont {D.~A.}\
  \bibnamefont {Shirley}}, \bibinfo {author} {\bibfnamefont {G.}~\bibnamefont
  {Kaindl}}, \bibinfo {author} {\bibfnamefont {C.~H.}\ \bibnamefont {Greene}},
  \bibinfo {author} {\bibfnamefont {H.~R.}\ \bibnamefont {Sadeghpour}}, \ and\
  \bibinfo {author} {\bibfnamefont {H.}~\bibnamefont {Petersen}}} (\bibinfo
  {year} {1991}),\ \bibfield  {title} {{\selectlanguage {English}\enquote
  {\bibinfo {title} {Extensive double-excitation states in atomic helium},}\
  }}\href@noop {} {\bibfield  {journal} {\bibinfo  {journal} {Phys. Rev.
  Lett.}\ }\textbf {\bibinfo {volume} {66}}~(\bibinfo {number} {10}),\ \bibinfo
  {pages} {1306--1309}}\BibitemShut {NoStop}%
\bibitem [{\citenamefont {Donahue}\ \emph {et~al.}(1982)\citenamefont
  {Donahue}, \citenamefont {Gram}, \citenamefont {Hynes}, \citenamefont {Hamm},
  \citenamefont {Frost}, \citenamefont {Bryant}, \citenamefont {Butterfield},
  \citenamefont {Clark},\ and\ \citenamefont {Smith}}]{Bryant1982prl}%
  \BibitemOpen
  \bibfield  {author} {\bibinfo {author} {\bibnamefont {Donahue}, \bibfnamefont
  {J~B}}, \bibinfo {author} {\bibfnamefont {P.~A.~M.}\ \bibnamefont {Gram}},
  \bibinfo {author} {\bibfnamefont {M.~V.}\ \bibnamefont {Hynes}}, \bibinfo
  {author} {\bibfnamefont {R.~W.}\ \bibnamefont {Hamm}}, \bibinfo {author}
  {\bibfnamefont {C.~A.}\ \bibnamefont {Frost}}, \bibinfo {author}
  {\bibfnamefont {H.~C.}\ \bibnamefont {Bryant}}, \bibinfo {author}
  {\bibfnamefont {K.~B.}\ \bibnamefont {Butterfield}}, \bibinfo {author}
  {\bibfnamefont {David~A.}\ \bibnamefont {Clark}}, \ and\ \bibinfo {author}
  {\bibfnamefont {W.~W.}\ \bibnamefont {Smith}}} (\bibinfo {year} {1982}),\
  \bibfield  {title} {\enquote {\bibinfo {title} {Observation of two-electron
  photoionization of the ${\mathrm{h}}^{\ensuremath{-}}$ ion near threshold},}\
  }\href {\doibase 10.1103/PhysRevLett.48.1538} {\bibfield  {journal} {\bibinfo
   {journal} {Phys. Rev. Lett.}\ }\textbf {\bibinfo {volume} {48}},\ \bibinfo
  {pages} {1538--1541}}\BibitemShut {NoStop}%
\bibitem [{\citenamefont {Donley}\ \emph {et~al.}(2001)\citenamefont {Donley},
  \citenamefont {Clausen}, \citenamefont {Cornish}, \citenamefont {Roberts},
  \citenamefont {Cornell},\ and\ \citenamefont {Wieman}}]{donley2001NT}%
  \BibitemOpen
  \bibfield  {author} {\bibinfo {author} {\bibnamefont {Donley}, \bibfnamefont
  {E~A}}, \bibinfo {author} {\bibfnamefont {N.~R.}\ \bibnamefont {Clausen}},
  \bibinfo {author} {\bibfnamefont {S.~L.}\ \bibnamefont {Cornish}}, \bibinfo
  {author} {\bibfnamefont {J.~L.}\ \bibnamefont {Roberts}}, \bibinfo {author}
  {\bibfnamefont {E.~A.}\ \bibnamefont {Cornell}}, \ and\ \bibinfo {author}
  {\bibfnamefont {C.~E.}\ \bibnamefont {Wieman}}} (\bibinfo {year} {2001}),\
  \bibfield  {title} {\enquote {\bibinfo {title} {Dynamics of collapsing and
  exploding {B}ose-{E}instein condensates},}\ }\href@noop {} {\bibfield
  {journal} {\bibinfo  {journal} {Nature (London)}\ }\textbf {\bibinfo {volume}
  {412}},\ \bibinfo {pages} {295--299}}\BibitemShut {NoStop}%
\bibitem [{\citenamefont {Donley}\ \emph {et~al.}(2002)\citenamefont {Donley},
  \citenamefont {Clausen}, \citenamefont {Thompson},\ and\ \citenamefont
  {Wieman}}]{donley2002NT}%
  \BibitemOpen
  \bibfield  {author} {\bibinfo {author} {\bibnamefont {Donley}, \bibfnamefont
  {E~A}}, \bibinfo {author} {\bibfnamefont {N.~R.}\ \bibnamefont {Clausen}},
  \bibinfo {author} {\bibfnamefont {S.~T.}\ \bibnamefont {Thompson}}, \ and\
  \bibinfo {author} {\bibfnamefont {C.~E.}\ \bibnamefont {Wieman}}} (\bibinfo
  {year} {2002}),\ \bibfield  {title} {\enquote {\bibinfo {title}
  {Atom-molecule coherence in a {B}ose-{E}instein condensate},}\ }\href@noop {}
  {\bibfield  {journal} {\bibinfo  {journal} {Nature (London)}\ }\textbf
  {\bibinfo {volume} {417}},\ \bibinfo {pages} {529--533}}\BibitemShut
  {NoStop}%
\bibitem [{\citenamefont {Duan}\ \emph {et~al.}(2013)\citenamefont {Duan},
  \citenamefont {You},\ and\ \citenamefont {Gao}}]{duan2013pra}%
  \BibitemOpen
  \bibfield  {author} {\bibinfo {author} {\bibnamefont {Duan}, \bibfnamefont
  {Hao}}, \bibinfo {author} {\bibfnamefont {Li}~\bibnamefont {You}}, \ and\
  \bibinfo {author} {\bibfnamefont {Bo}~\bibnamefont {Gao}}} (\bibinfo {year}
  {2013}),\ \bibfield  {title} {\enquote {\bibinfo {title} {Ultracold
  collisions in the presence of synthetic spin-orbit coupling},}\ }\href@noop
  {} {\bibfield  {journal} {\bibinfo  {journal} {Phys. Rev. A}\ }\textbf
  {\bibinfo {volume} {87}},\ \bibinfo {pages} {052708}}\BibitemShut {NoStop}%
\bibitem [{\citenamefont {Dunjko}\ \emph
  {et~al.}(2011{\natexlab{a}})\citenamefont {Dunjko}, \citenamefont {Moore},
  \citenamefont {Bergeman},\ and\ \citenamefont {Olshanii}}]{dunjko2011}%
  \BibitemOpen
  \bibfield  {author} {\bibinfo {author} {\bibnamefont {Dunjko}, \bibfnamefont
  {V}}, \bibinfo {author} {\bibfnamefont {M.~G.}\ \bibnamefont {Moore}},
  \bibinfo {author} {\bibfnamefont {T.}~\bibnamefont {Bergeman}}, \ and\
  \bibinfo {author} {\bibfnamefont {M.}~\bibnamefont {Olshanii}}} (\bibinfo
  {year} {2011}{\natexlab{a}}),\ \bibfield  {title} {\enquote {\bibinfo {title}
  {Confinement-induced resonances},}\ }\href@noop {} {\bibfield  {journal}
  {\bibinfo  {journal} {Adv. At. Mol. Opt. Phys.}\ }\textbf {\bibinfo {volume}
  {60}},\ \bibinfo {pages} {461--510}}\BibitemShut {NoStop}%
\bibitem [{\citenamefont {Dunjko}\ \emph
  {et~al.}(2011{\natexlab{b}})\citenamefont {Dunjko}, \citenamefont {Moore},
  \citenamefont {Bergeman},\ and\ \citenamefont {Olshanii}}]{Dunjko2011461}%
  \BibitemOpen
  \bibfield  {author} {\bibinfo {author} {\bibnamefont {Dunjko}, \bibfnamefont
  {Vanja}}, \bibinfo {author} {\bibfnamefont {Michael~G.}\ \bibnamefont
  {Moore}}, \bibinfo {author} {\bibfnamefont {Thomas}\ \bibnamefont
  {Bergeman}}, \ and\ \bibinfo {author} {\bibfnamefont {Maxim}\ \bibnamefont
  {Olshanii}}} (\bibinfo {year} {2011}{\natexlab{b}}),\ \bibfield  {title}
  {\enquote {\bibinfo {title} {Chapter 10 - confinement-induced resonances},}\
  }in\ \href {\doibase http://dx.doi.org/10.1016/B978-0-12-385508-4.00010-3}
  {\emph {\bibinfo {booktitle} {Advances in Atomic, Molecular, and Optical
  Physics}}},\ Vol.~\bibinfo {volume} {60},\ \bibinfo {editor} {edited by\
  \bibinfo {editor} {\bibfnamefont {P.R.~Berman}\ \bibnamefont {E.~Arimondo}}\
  and\ \bibinfo {editor} {\bibfnamefont {C.C.}\ \bibnamefont {Lin}}}\ (\bibinfo
   {publisher} {Academic Press})\ pp.\ \bibinfo {pages} {461 --
  510}\BibitemShut {NoStop}%
\bibitem [{\citenamefont {Dyke}\ \emph {et~al.}(2013)\citenamefont {Dyke},
  \citenamefont {Pollack},\ and\ \citenamefont {Hulet}}]{dyke2013PRA}%
  \BibitemOpen
  \bibfield  {author} {\bibinfo {author} {\bibnamefont {Dyke}, \bibfnamefont
  {P}}, \bibinfo {author} {\bibfnamefont {S.~E.}\ \bibnamefont {Pollack}}, \
  and\ \bibinfo {author} {\bibfnamefont {R.~G.}\ \bibnamefont {Hulet}}}
  (\bibinfo {year} {2013}),\ \bibfield  {title} {\enquote {\bibinfo {title}
  {Finite range corrections near a {F}eshbach resonance and their role in the
  {E}fimov effect},}\ }\href@noop {} {\bibfield  {journal} {\bibinfo  {journal}
  {Phys. Rev. A}\ }\textbf {\bibinfo {volume} {88}}~(\bibinfo {number} {2}),\
  \bibinfo {pages} {023625}}\BibitemShut {NoStop}%
\bibitem [{\citenamefont {Efimov}(1970)}]{efimov1970plb}%
  \BibitemOpen
  \bibfield  {author} {\bibinfo {author} {\bibnamefont {Efimov}, \bibfnamefont
  {V}}} (\bibinfo {year} {1970}),\ \bibfield  {title} {\enquote {\bibinfo
  {title} {Energy levels arising from resonant two-body forces in a three-body
  system},}\ }\href@noop {} {\bibfield  {journal} {\bibinfo  {journal} {Phys.
  Lett. B}\ }\textbf {\bibinfo {volume} {33}}~(\bibinfo {number} {8}),\
  \bibinfo {pages} {563 -- 564}}\BibitemShut {NoStop}%
\bibitem [{\citenamefont {Efimov}(1971)}]{efimov1971SJNP}%
  \BibitemOpen
  \bibfield  {author} {\bibinfo {author} {\bibnamefont {Efimov}, \bibfnamefont
  {V}}} (\bibinfo {year} {1971}),\ \bibfield  {title} {\enquote {\bibinfo
  {title} {Weakly-bound states of three resonantly-interacting particles},}\
  }\href@noop {} {\bibfield  {journal} {\bibinfo  {journal} {Sov. J. Nuc.
  Phys.}\ }\textbf {\bibinfo {volume} {12}},\ \bibinfo {pages}
  {589--595}}\BibitemShut {NoStop}%
\bibitem [{\citenamefont {Efimov}(1973{\natexlab{a}})}]{efimov1973npa}%
  \BibitemOpen
  \bibfield  {author} {\bibinfo {author} {\bibnamefont {Efimov}, \bibfnamefont
  {V}}} (\bibinfo {year} {1973}{\natexlab{a}}),\ \bibfield  {title} {\enquote
  {\bibinfo {title} {Energy levels of three resonantly interacting
  particles},}\ }\href@noop {} {\bibfield  {journal} {\bibinfo  {journal}
  {Nucl. Phys. A}\ }\textbf {\bibinfo {volume} {210}}~(\bibinfo {number} {1}),\
  \bibinfo {pages} {157 -- 188}}\BibitemShut {NoStop}%
\bibitem [{\citenamefont {Efimov}(1973{\natexlab{b}})}]{Efimov-1973}%
  \BibitemOpen
  \bibfield  {author} {\bibinfo {author} {\bibnamefont {Efimov}, \bibfnamefont
  {V}}} (\bibinfo {year} {1973}{\natexlab{b}}),\ \bibfield  {title} {\enquote
  {\bibinfo {title} {Energy levels of three resonantly interacting
  particles},}\ }\href@noop {} {\bibfield  {journal} {\bibinfo  {journal} {Nuc.
  Phys. A}\ }\textbf {\bibinfo {volume} {210}},\ \bibinfo {pages}
  {157--188}}\BibitemShut {NoStop}%
\bibitem [{\citenamefont {Efimov}(1979)}]{efimov1979SJNP}%
  \BibitemOpen
  \bibfield  {author} {\bibinfo {author} {\bibnamefont {Efimov}, \bibfnamefont
  {V}}} (\bibinfo {year} {1979}),\ \bibfield  {title} {\enquote {\bibinfo
  {title} {Low-energy properties of three resonantly interacting particles},}\
  }\href@noop {} {\bibfield  {journal} {\bibinfo  {journal} {Sov. J. Nuc.
  Phys.}\ }\textbf {\bibinfo {volume} {29}},\ \bibinfo {pages}
  {546}}\BibitemShut {NoStop}%
\bibitem [{\citenamefont {Efimov}\ and\ \citenamefont
  {Tkachenko}(1988)}]{Efimov-1988}%
  \BibitemOpen
  \bibfield  {author} {\bibinfo {author} {\bibnamefont {Efimov}, \bibfnamefont
  {V}}, \ and\ \bibinfo {author} {\bibfnamefont {E.~G.}\ \bibnamefont
  {Tkachenko}}} (\bibinfo {year} {1988}),\ \bibfield  {title} {\enquote
  {\bibinfo {title} {On the correlation between the triton binding energy and
  the neutron-deuteron doublet scattering length},}\ }\href@noop {} {\bibfield
  {journal} {\bibinfo  {journal} {Few-Body Systems}\ }\textbf {\bibinfo
  {volume} {4}},\ \bibinfo {pages} {71--88}}\BibitemShut {NoStop}%
\bibitem [{\citenamefont {Efremov}\ \emph {et~al.}(2013)\citenamefont
  {Efremov}, \citenamefont {Plimak}, \citenamefont {Ivanov},\ and\
  \citenamefont {Schleich}}]{EfremovSchleich2013PRL}%
  \BibitemOpen
  \bibfield  {author} {\bibinfo {author} {\bibnamefont {Efremov}, \bibfnamefont
  {Maxim~A}}, \bibinfo {author} {\bibfnamefont {Lev}\ \bibnamefont {Plimak}},
  \bibinfo {author} {\bibfnamefont {Misha~Yu.}\ \bibnamefont {Ivanov}}, \ and\
  \bibinfo {author} {\bibfnamefont {Wolfgang~P.}\ \bibnamefont {Schleich}}}
  (\bibinfo {year} {2013}),\ \bibfield  {title} {\enquote {\bibinfo {title}
  {Three-body bound states in atomic mixtures with resonant $p$-wave
  interaction},}\ }\href {\doibase 10.1103/PhysRevLett.111.113201} {\bibfield
  {journal} {\bibinfo  {journal} {Phys. Rev. Lett.}\ }\textbf {\bibinfo
  {volume} {111}},\ \bibinfo {pages} {113201}}\BibitemShut {NoStop}%
\bibitem [{\citenamefont {Eismann}\ \emph {et~al.}({2016})\citenamefont
  {Eismann}, \citenamefont {Khaykovich}, \citenamefont {Laurent}, \citenamefont
  {Ferrier-Barbut}, \citenamefont {Rem}, \citenamefont {Grier}, \citenamefont
  {Delehaye}, \citenamefont {Chevy}, \citenamefont {Salomon}, \citenamefont
  {Ha},\ and\ \citenamefont {Chin}}]{Eismann2016prx}%
  \BibitemOpen
  \bibfield  {author} {\bibinfo {author} {\bibnamefont {Eismann}, \bibfnamefont
  {U}}, \bibinfo {author} {\bibfnamefont {L.}~\bibnamefont {Khaykovich}},
  \bibinfo {author} {\bibfnamefont {S.}~\bibnamefont {Laurent}}, \bibinfo
  {author} {\bibfnamefont {I.}~\bibnamefont {Ferrier-Barbut}}, \bibinfo
  {author} {\bibfnamefont {B.~S.}\ \bibnamefont {Rem}}, \bibinfo {author}
  {\bibfnamefont {A.T.}\ \bibnamefont {Grier}}, \bibinfo {author}
  {\bibfnamefont {M.}~\bibnamefont {Delehaye}}, \bibinfo {author}
  {\bibfnamefont {F.}~\bibnamefont {Chevy}}, \bibinfo {author} {\bibfnamefont
  {C.}~\bibnamefont {Salomon}}, \bibinfo {author} {\bibfnamefont {L.-C.}\
  \bibnamefont {Ha}}, \ and\ \bibinfo {author} {\bibfnamefont {C.}~\bibnamefont
  {Chin}}} (\bibinfo {year} {{2016}}),\ \bibfield  {title} {\enquote {\bibinfo
  {title} {Universal loss dynamics in a unitary {B}ose gas},}\ }\href@noop {}
  {\bibfield  {journal} {\bibinfo  {journal} {Phys. Rev. X}\ }\textbf {\bibinfo
  {volume} {{6}}}~(\bibinfo {number} {{2}})}\BibitemShut {NoStop}%
\bibitem [{\citenamefont {Epelbaum}\ \emph {et~al.}(2009)\citenamefont
  {Epelbaum}, \citenamefont {Hammer},\ and\ \citenamefont
  {Meissner}}]{epelbaum2009RMP}%
  \BibitemOpen
  \bibfield  {author} {\bibinfo {author} {\bibnamefont {Epelbaum},
  \bibfnamefont {E}}, \bibinfo {author} {\bibfnamefont {H.~W.}\ \bibnamefont
  {Hammer}}, \ and\ \bibinfo {author} {\bibfnamefont {Ulf-G}\ \bibnamefont
  {Meissner}}} (\bibinfo {year} {2009}),\ \bibfield  {title} {{\selectlanguage
  {English}\enquote {\bibinfo {title} {Modern theory of nuclear forces},}\
  }}\href@noop {} {\bibfield  {journal} {\bibinfo  {journal} {Rev. Mod. Phys.}\
  }\textbf {\bibinfo {volume} {81}}~(\bibinfo {number} {4}),\ \bibinfo {pages}
  {1773--1825}}\BibitemShut {NoStop}%
\bibitem [{\citenamefont {Ermolova}\ \emph {et~al.}(2014)\citenamefont
  {Ermolova}, \citenamefont {Rusin},\ and\ \citenamefont
  {Sevryuk}}]{Ermolova-2014}%
  \BibitemOpen
  \bibfield  {author} {\bibinfo {author} {\bibnamefont {Ermolova},
  \bibfnamefont {E~V}}, \bibinfo {author} {\bibfnamefont {L.~{Yu}.}\
  \bibnamefont {Rusin}}, \ and\ \bibinfo {author} {\bibfnamefont {M.~B.}\
  \bibnamefont {Sevryuk}}} (\bibinfo {year} {2014}),\ \bibfield  {title}
  {\enquote {\bibinfo {title} {A hard sphere model for direct three-body
  recombination of heavy ions},}\ }\href@noop {} {\bibfield  {journal}
  {\bibinfo  {journal} {Russ. J. Phys. Chem. B}\ }\textbf {\bibinfo {volume}
  {8}},\ \bibinfo {pages} {769}}\BibitemShut {NoStop}%
\bibitem [{\citenamefont {Esry}\ and\ \citenamefont
  {Greene}(1999)}]{esry1999PRAb}%
  \BibitemOpen
  \bibfield  {author} {\bibinfo {author} {\bibnamefont {Esry}, \bibfnamefont
  {B~D}}, \ and\ \bibinfo {author} {\bibfnamefont {C.~H.}\ \bibnamefont
  {Greene}}} (\bibinfo {year} {1999}),\ \bibfield  {title} {{\selectlanguage
  {English}\enquote {\bibinfo {title} {Validity of the shape-independent
  approximation for {B}ose-{E}instein condensates},}\ }}\href@noop {}
  {\bibfield  {journal} {\bibinfo  {journal} {Phys. Rev. A}\ }\textbf {\bibinfo
  {volume} {60}}~(\bibinfo {number} {2}),\ \bibinfo {pages}
  {1451--1462}}\BibitemShut {NoStop}%
\bibitem [{\citenamefont {Esry}\ and\ \citenamefont
  {Greene}(2006)}]{esry2006NT}%
  \BibitemOpen
  \bibfield  {author} {\bibinfo {author} {\bibnamefont {Esry}, \bibfnamefont
  {B~D}}, \ and\ \bibinfo {author} {\bibfnamefont {C.~H.}\ \bibnamefont
  {Greene}}} (\bibinfo {year} {2006}),\ \bibfield  {title} {\enquote {\bibinfo
  {title} {Quantum physics: a m\'enage \`a trois laid bare.}}\ }\href@noop {}
  {\bibfield  {journal} {\bibinfo  {journal} {Nature (London)}\ }\textbf
  {\bibinfo {volume} {440}}~(\bibinfo {number} {7082}),\ \bibinfo {pages}
  {289--290}}\BibitemShut {NoStop}%
\bibitem [{\citenamefont {Esry}\ \emph {et~al.}(1999)\citenamefont {Esry},
  \citenamefont {Greene},\ and\ \citenamefont {Burke}}]{esry1999PRL}%
  \BibitemOpen
  \bibfield  {author} {\bibinfo {author} {\bibnamefont {Esry}, \bibfnamefont
  {B~D}}, \bibinfo {author} {\bibfnamefont {C.~H.}\ \bibnamefont {Greene}}, \
  and\ \bibinfo {author} {\bibfnamefont {J.~P.}\ \bibnamefont {Burke}}}
  (\bibinfo {year} {1999}),\ \bibfield  {title} {{\selectlanguage
  {English}\enquote {\bibinfo {title} {Recombination of three atoms in the
  ultracold limit},}\ }}\href@noop {} {\bibfield  {journal} {\bibinfo
  {journal} {Phys. Rev. Lett.}\ }\textbf {\bibinfo {volume} {83}}~(\bibinfo
  {number} {9}),\ \bibinfo {pages} {1751--1754}}\BibitemShut {NoStop}%
\bibitem [{\citenamefont {Esry}\ \emph
  {et~al.}(1996{\natexlab{a}})\citenamefont {Esry}, \citenamefont {Greene},
  \citenamefont {Zhou},\ and\ \citenamefont {Lin}}]{esry1996JPB}%
  \BibitemOpen
  \bibfield  {author} {\bibinfo {author} {\bibnamefont {Esry}, \bibfnamefont
  {B~D}}, \bibinfo {author} {\bibfnamefont {C.~H.}\ \bibnamefont {Greene}},
  \bibinfo {author} {\bibfnamefont {Y.}~\bibnamefont {Zhou}}, \ and\ \bibinfo
  {author} {\bibfnamefont {C.~D.}\ \bibnamefont {Lin}}} (\bibinfo {year}
  {1996}{\natexlab{a}}),\ \bibfield  {title} {{\selectlanguage
  {English}\enquote {\bibinfo {title} {Role of the scattering length in
  three-boson dynamics and {B}ose-{E}instein condensation},}\ }}\href@noop {}
  {\bibfield  {journal} {\bibinfo  {journal} {J. Phys. B}\ }\textbf {\bibinfo
  {volume} {29}}~(\bibinfo {number} {2}),\ \bibinfo {pages}
  {L51--L57}}\BibitemShut {NoStop}%
\bibitem [{\citenamefont {Esry}\ \emph
  {et~al.}(1996{\natexlab{b}})\citenamefont {Esry}, \citenamefont {Lin},\ and\
  \citenamefont {Greene}}]{esry1996PRA}%
  \BibitemOpen
  \bibfield  {author} {\bibinfo {author} {\bibnamefont {Esry}, \bibfnamefont
  {B~D}}, \bibinfo {author} {\bibfnamefont {C.~D.}\ \bibnamefont {Lin}}, \ and\
  \bibinfo {author} {\bibfnamefont {C.~H.}\ \bibnamefont {Greene}}} (\bibinfo
  {year} {1996}{\natexlab{b}}),\ \bibfield  {title} {{\selectlanguage
  {English}\enquote {\bibinfo {title} {Adiabatic hyperspherical study of the
  helium trimer},}\ }}\href@noop {} {\bibfield  {journal} {\bibinfo  {journal}
  {Phys. Rev. A}\ }\textbf {\bibinfo {volume} {54}}~(\bibinfo {number} {1}),\
  \bibinfo {pages} {394--401}}\BibitemShut {NoStop}%
\bibitem [{\citenamefont {Esry}\ and\ \citenamefont
  {Sadeghpour}(2003)}]{esry2003PRA}%
  \BibitemOpen
  \bibfield  {author} {\bibinfo {author} {\bibnamefont {Esry}, \bibfnamefont
  {B~D}}, \ and\ \bibinfo {author} {\bibfnamefont {H.~R.}\ \bibnamefont
  {Sadeghpour}}} (\bibinfo {year} {2003}),\ \bibfield  {title}
  {{\selectlanguage {English}\enquote {\bibinfo {title} {Ultraslow
  $\bar{p}$--{H} collisions in hyperspherical coordinates: {H}ydrogen and
  protonium channels},}\ }}\href@noop {} {\bibfield  {journal} {\bibinfo
  {journal} {Phys. Rev. A}\ }\textbf {\bibinfo {volume} {67}}~(\bibinfo
  {number} {1}),\ \bibinfo {pages} {012704}}\BibitemShut {NoStop}%
\bibitem [{\citenamefont {Esry}(1997)}]{esry1997PRA}%
  \BibitemOpen
  \bibfield  {author} {\bibinfo {author} {\bibnamefont {Esry}, \bibfnamefont
  {BD}}} (\bibinfo {year} {1997}),\ \bibfield  {title} {{\selectlanguage
  {English}\enquote {\bibinfo {title} {{H}artree-{F}ock theory for
  {B}ose-{E}instein condensates and the inclusion of correlation effects},}\
  }}\href@noop {} {\bibfield  {journal} {\bibinfo  {journal} {Phys. Rev. A}\
  }\textbf {\bibinfo {volume} {55}}~(\bibinfo {number} {2}),\ \bibinfo {pages}
  {1147--1159}}\BibitemShut {NoStop}%
\bibitem [{\citenamefont {Fang}\ and\ \citenamefont
  {Tomusiak}(1977)}]{Fang-1977}%
  \BibitemOpen
  \bibfield  {author} {\bibinfo {author} {\bibnamefont {Fang}, \bibfnamefont
  {K~K}}, \ and\ \bibinfo {author} {\bibfnamefont {E.~L.}\ \bibnamefont
  {Tomusiak}}} (\bibinfo {year} {1977}),\ \bibfield  {title} {\enquote
  {\bibinfo {title} {Three-body model of $^6${Li} usign hyperspherical
  harmonics},}\ }\href@noop {} {\bibfield  {journal} {\bibinfo  {journal}
  {Phys. Rev. C}\ }\textbf {\bibinfo {volume} {16}},\ \bibinfo {pages}
  {2117}}\BibitemShut {NoStop}%
\bibitem [{\citenamefont {Fano}(1970)}]{fano1970}%
  \BibitemOpen
  \bibfield  {author} {\bibinfo {author} {\bibnamefont {Fano}, \bibfnamefont
  {U}}} (\bibinfo {year} {1970}),\ \bibfield  {title} {\enquote {\bibinfo
  {title} {Quantum defect theory of $l$ uncoupling in $\mathrm{H}_2$ as an
  example of channel-interaction treatment},}\ }\href@noop {} {\bibfield
  {journal} {\bibinfo  {journal} {Phys. Rev. A}\ }\textbf {\bibinfo {volume}
  {2}}~(\bibinfo {number} {2}),\ \bibinfo {pages} {353--365}}\BibitemShut
  {NoStop}%
\bibitem [{\citenamefont {Fano}(1976)}]{fano1976PT}%
  \BibitemOpen
  \bibfield  {author} {\bibinfo {author} {\bibnamefont {Fano}, \bibfnamefont
  {U}}} (\bibinfo {year} {1976}),\ \bibfield  {title} {\enquote {\bibinfo
  {title} {Dynamics of electron excitation},}\ }\href@noop {} {\bibfield
  {journal} {\bibinfo  {journal} {Physics Today}\ }\textbf {\bibinfo {volume}
  {29}}~(\bibinfo {number} {9}),\ \bibinfo {pages} {32}}\BibitemShut {NoStop}%
\bibitem [{\citenamefont {Fano}(1981{\natexlab{a}})}]{fano1981}%
  \BibitemOpen
  \bibfield  {author} {\bibinfo {author} {\bibnamefont {Fano}, \bibfnamefont
  {U}}} (\bibinfo {year} {1981}{\natexlab{a}}),\ \bibfield  {title} {\enquote
  {\bibinfo {title} {Stark effect of nonhydrogenic {R}ydberg spectra},}\
  }\href@noop {} {\bibfield  {journal} {\bibinfo  {journal} {Phys. Rev. A}\
  }\textbf {\bibinfo {volume} {24}}~(\bibinfo {number} {1}),\ \bibinfo {pages}
  {619--622}}\BibitemShut {NoStop}%
\bibitem [{\citenamefont {Fano}(1981{\natexlab{b}})}]{fano1981PRA}%
  \BibitemOpen
  \bibfield  {author} {\bibinfo {author} {\bibnamefont {Fano}, \bibfnamefont
  {U}}} (\bibinfo {year} {1981}{\natexlab{b}}),\ \bibfield  {title} {\enquote
  {\bibinfo {title} {Unified treatment of collisions},}\ }\href@noop {}
  {\bibfield  {journal} {\bibinfo  {journal} {Phys. Rev. A}\ }\textbf {\bibinfo
  {volume} {24}}~(\bibinfo {number} {5}),\ \bibinfo {pages}
  {2402--2415}}\BibitemShut {NoStop}%
\bibitem [{\citenamefont {Fano}(1983)}]{fano1983RPP}%
  \BibitemOpen
  \bibfield  {author} {\bibinfo {author} {\bibnamefont {Fano}, \bibfnamefont
  {U}}} (\bibinfo {year} {1983}),\ \bibfield  {title} {\enquote {\bibinfo
  {title} {Correlations of two excited electrons},}\ }\href@noop {} {\bibfield
  {journal} {\bibinfo  {journal} {Rep. Prog. Phys.}\ }\textbf {\bibinfo
  {volume} {46}}~(\bibinfo {number} {2}),\ \bibinfo {pages}
  {97--165}}\BibitemShut {NoStop}%
\bibitem [{\citenamefont {Fano}\ and\ \citenamefont {Rau}(1986)}]{fano1986}%
  \BibitemOpen
  \bibfield  {author} {\bibinfo {author} {\bibnamefont {Fano}, \bibfnamefont
  {U}}, \ and\ \bibinfo {author} {\bibfnamefont {A.R.P.}\ \bibnamefont {Rau}}}
  (\bibinfo {year} {1986}),\ \href@noop {} {\emph {\bibinfo {title} {Atomic
  Collisions and Spectra}}}\ (\bibinfo  {publisher} {Academic Press},\ \bibinfo
  {address} {Orlando, FL})\BibitemShut {NoStop}%
\bibitem [{\citenamefont {Fano}\ and\ \citenamefont
  {Stephens}(1986)}]{FanoStephens1986prb}%
  \BibitemOpen
  \bibfield  {author} {\bibinfo {author} {\bibnamefont {Fano}, \bibfnamefont
  {U}}, \ and\ \bibinfo {author} {\bibfnamefont {J.~A.}\ \bibnamefont
  {Stephens}}} (\bibinfo {year} {1986}),\ \bibfield  {title} {\enquote
  {\bibinfo {title} {Slow electrons in condensed matter},}\ }\href {\doibase
  10.1103/PhysRevB.34.438} {\bibfield  {journal} {\bibinfo  {journal} {Phys.
  Rev. B}\ }\textbf {\bibinfo {volume} {34}},\ \bibinfo {pages}
  {438--441}}\BibitemShut {NoStop}%
\bibitem [{\citenamefont {Fedichev}\ \emph {et~al.}(2004)\citenamefont
  {Fedichev}, \citenamefont {Bijlsma},\ and\ \citenamefont
  {Zoller}}]{fedichev2004}%
  \BibitemOpen
  \bibfield  {author} {\bibinfo {author} {\bibnamefont {Fedichev},
  \bibfnamefont {P~O}}, \bibinfo {author} {\bibfnamefont {M.~J.}\ \bibnamefont
  {Bijlsma}}, \ and\ \bibinfo {author} {\bibfnamefont {P.}~\bibnamefont
  {Zoller}}} (\bibinfo {year} {2004}),\ \bibfield  {title} {\enquote {\bibinfo
  {title} {Extended molecules and geometric scattering resonances in optical
  lattices},}\ }\href@noop {} {\bibfield  {journal} {\bibinfo  {journal} {Phys.
  Rev. Lett.}\ }\textbf {\bibinfo {volume} {92}}~(\bibinfo {number} {8}),\
  \bibinfo {pages} {080401}}\BibitemShut {NoStop}%
\bibitem [{\citenamefont {Fedichev}\ \emph
  {et~al.}(1996{\natexlab{a}})\citenamefont {Fedichev}, \citenamefont {Kagan},
  \citenamefont {Shlyapnikov},\ and\ \citenamefont
  {Walraven}}]{Fedichev1996PRL}%
  \BibitemOpen
  \bibfield  {author} {\bibinfo {author} {\bibnamefont {Fedichev},
  \bibfnamefont {P~O}}, \bibinfo {author} {\bibfnamefont {Yu}~\bibnamefont
  {Kagan}}, \bibinfo {author} {\bibfnamefont {G~V}\ \bibnamefont
  {Shlyapnikov}}, \ and\ \bibinfo {author} {\bibfnamefont {J~T~M}\ \bibnamefont
  {Walraven}}} (\bibinfo {year} {1996}{\natexlab{a}}),\ \bibfield  {title}
  {\enquote {\bibinfo {title} {Influence of nearly resonant light on the
  scattering length in low-temperature atomic gases},}\ }\href@noop {}
  {\bibfield  {journal} {\bibinfo  {journal} {Phys. Rev. Lett.}\ }\textbf
  {\bibinfo {volume} {77}},\ \bibinfo {pages} {2913--2916}}\BibitemShut
  {NoStop}%
\bibitem [{\citenamefont {Fedichev}\ \emph
  {et~al.}(1996{\natexlab{b}})\citenamefont {Fedichev}, \citenamefont
  {Reynolds},\ and\ \citenamefont {Shlyapnikov}}]{fedichev1996PRLb}%
  \BibitemOpen
  \bibfield  {author} {\bibinfo {author} {\bibnamefont {Fedichev},
  \bibfnamefont {P~O}}, \bibinfo {author} {\bibfnamefont {M.~W.}\ \bibnamefont
  {Reynolds}}, \ and\ \bibinfo {author} {\bibfnamefont {G.~V.}\ \bibnamefont
  {Shlyapnikov}}} (\bibinfo {year} {1996}{\natexlab{b}}),\ \bibfield  {title}
  {\enquote {\bibinfo {title} {Three-body recombination of ultracold atoms to a
  weakly bound $s$ level},}\ }\href@noop {} {\bibfield  {journal} {\bibinfo
  {journal} {Phys. Rev. Lett.}\ }\textbf {\bibinfo {volume} {77}},\ \bibinfo
  {pages} {2921--2924}}\BibitemShut {NoStop}%
\bibitem [{\citenamefont {Fedorov}\ and\ \citenamefont
  {Jensen}(2002)}]{Fedorov-2002}%
  \BibitemOpen
  \bibfield  {author} {\bibinfo {author} {\bibnamefont {Fedorov}, \bibfnamefont
  {D~V}}, \ and\ \bibinfo {author} {\bibfnamefont {A.~S.}\ \bibnamefont
  {Jensen}}} (\bibinfo {year} {2002}),\ \bibfield  {title} {\enquote {\bibinfo
  {title} {Regularized zero-range model and an application to the triton and
  the hypertriton},}\ }\href@noop {} {\bibfield  {journal} {\bibinfo  {journal}
  {Nuc. Phys. A}\ }\textbf {\bibinfo {volume} {697}},\ \bibinfo {pages}
  {783--801}}\BibitemShut {NoStop}%
\bibitem [{\citenamefont {Fedorov}\ \emph {et~al.}(1994)\citenamefont
  {Fedorov}, \citenamefont {Jensen},\ and\ \citenamefont
  {Riisager}}]{Fedorov-1994}%
  \BibitemOpen
  \bibfield  {author} {\bibinfo {author} {\bibnamefont {Fedorov}, \bibfnamefont
  {D~V}}, \bibinfo {author} {\bibfnamefont {A.~S.}\ \bibnamefont {Jensen}}, \
  and\ \bibinfo {author} {\bibfnamefont {K.}~\bibnamefont {Riisager}}}
  (\bibinfo {year} {1994}),\ \bibfield  {title} {\enquote {\bibinfo {title}
  {Efimov states in halo nuclei},}\ }\href@noop {} {\bibfield  {journal}
  {\bibinfo  {journal} {Phys. Rev. Lett.}\ }\textbf {\bibinfo {volume} {73}},\
  \bibinfo {pages} {2817}}\BibitemShut {NoStop}%
\bibitem [{\citenamefont {Ferlaino}\ \emph
  {et~al.}(2009{\natexlab{a}})\citenamefont {Ferlaino}, \citenamefont {Knoop},
  \citenamefont {Berninger}, \citenamefont {Harm}, \citenamefont {D'Incao},
  \citenamefont {Naegerl},\ and\ \citenamefont {Grimm}}]{ferlaino2009PRL}%
  \BibitemOpen
  \bibfield  {author} {\bibinfo {author} {\bibnamefont {Ferlaino},
  \bibfnamefont {F}}, \bibinfo {author} {\bibfnamefont {S.}~\bibnamefont
  {Knoop}}, \bibinfo {author} {\bibfnamefont {M.}~\bibnamefont {Berninger}},
  \bibinfo {author} {\bibfnamefont {W.}~\bibnamefont {Harm}}, \bibinfo {author}
  {\bibfnamefont {J.~P.}\ \bibnamefont {D'Incao}}, \bibinfo {author}
  {\bibfnamefont {H.~C.}\ \bibnamefont {Naegerl}}, \ and\ \bibinfo {author}
  {\bibfnamefont {R.}~\bibnamefont {Grimm}}} (\bibinfo {year}
  {2009}{\natexlab{a}}),\ \bibfield  {title} {{\selectlanguage
  {English}\enquote {\bibinfo {title} {Evidence for universal four-body states
  tied to an {E}fimov trimer},}\ }}\href@noop {} {\bibfield  {journal}
  {\bibinfo  {journal} {Phys. Rev. Lett.}\ }\textbf {\bibinfo {volume}
  {102}}~(\bibinfo {number} {14}),\ \bibinfo {pages} {140401}}\BibitemShut
  {NoStop}%
\bibitem [{\citenamefont {Ferlaino}\ \emph
  {et~al.}(2009{\natexlab{b}})\citenamefont {Ferlaino}, \citenamefont {Knoop},
  \citenamefont {Berninger}, \citenamefont {Harm}, \citenamefont {D\'{}Incao},
  \citenamefont {N\"agerl},\ and\ \citenamefont {Grim}}]{Ferlaino-2009}%
  \BibitemOpen
  \bibfield  {author} {\bibinfo {author} {\bibnamefont {Ferlaino},
  \bibfnamefont {F}}, \bibinfo {author} {\bibfnamefont {S.}~\bibnamefont
  {Knoop}}, \bibinfo {author} {\bibfnamefont {M.}~\bibnamefont {Berninger}},
  \bibinfo {author} {\bibfnamefont {W.}~\bibnamefont {Harm}}, \bibinfo {author}
  {\bibfnamefont {J.~P}\ \bibnamefont {D\'{}Incao}}, \bibinfo {author}
  {\bibfnamefont {H.~C.}\ \bibnamefont {N\"agerl}}, \ and\ \bibinfo {author}
  {\bibfnamefont {R.}~\bibnamefont {Grim}}} (\bibinfo {year}
  {2009}{\natexlab{b}}),\ \bibfield  {title} {\enquote {\bibinfo {title}
  {Evidence for universal four-body states tied to an efimov trimer},}\
  }\href@noop {} {\bibfield  {journal} {\bibinfo  {journal} {Phys. Rev. Lett.}\
  }\textbf {\bibinfo {volume} {102}},\ \bibinfo {pages} {140401}}\BibitemShut
  {NoStop}%
\bibitem [{\citenamefont {Ferlaino}\ \emph {et~al.}(2008)\citenamefont
  {Ferlaino}, \citenamefont {Knoop}, \citenamefont {Mark}, \citenamefont
  {Berninger}, \citenamefont {Schoebel}, \citenamefont {Naegerl},\ and\
  \citenamefont {Grimm}}]{ferlaino2008PRL}%
  \BibitemOpen
  \bibfield  {author} {\bibinfo {author} {\bibnamefont {Ferlaino},
  \bibfnamefont {F}}, \bibinfo {author} {\bibfnamefont {S.}~\bibnamefont
  {Knoop}}, \bibinfo {author} {\bibfnamefont {M.}~\bibnamefont {Mark}},
  \bibinfo {author} {\bibfnamefont {M.}~\bibnamefont {Berninger}}, \bibinfo
  {author} {\bibfnamefont {H.}~\bibnamefont {Schoebel}}, \bibinfo {author}
  {\bibfnamefont {H.~C.}\ \bibnamefont {Naegerl}}, \ and\ \bibinfo {author}
  {\bibfnamefont {R.}~\bibnamefont {Grimm}}} (\bibinfo {year} {2008}),\
  \bibfield  {title} {{\selectlanguage {English}\enquote {\bibinfo {title}
  {Collisions between tunable halo dimers: Exploring an elementary four-body
  process with identical bosons},}\ }}\href@noop {} {\bibfield  {journal}
  {\bibinfo  {journal} {Phys. Rev. Lett.}\ }\textbf {\bibinfo {volume}
  {101}}~(\bibinfo {number} {2}),\ \bibinfo {pages} {023201}}\BibitemShut
  {NoStop}%
\bibitem [{\citenamefont {Ferlaino}\ \emph {et~al.}(2011)\citenamefont
  {Ferlaino}, \citenamefont {Zenesini}, \citenamefont {Berninger},
  \citenamefont {Huang}, \citenamefont {N\"agerl},\ and\ \citenamefont
  {Grimm}}]{ferlaino2011FBS}%
  \BibitemOpen
  \bibfield  {author} {\bibinfo {author} {\bibnamefont {Ferlaino},
  \bibfnamefont {F}}, \bibinfo {author} {\bibfnamefont {A.}~\bibnamefont
  {Zenesini}}, \bibinfo {author} {\bibfnamefont {M.}~\bibnamefont {Berninger}},
  \bibinfo {author} {\bibfnamefont {B.}~\bibnamefont {Huang}}, \bibinfo
  {author} {\bibfnamefont {H.-C.}\ \bibnamefont {N\"agerl}}, \ and\ \bibinfo
  {author} {\bibfnamefont {R.}~\bibnamefont {Grimm}}} (\bibinfo {year}
  {2011}),\ \bibfield  {title} {{\selectlanguage {English}\enquote {\bibinfo
  {title} {{E}fimov resonances in ultracold quantum gases},}\ }}\href@noop {}
  {\bibfield  {journal} {\bibinfo  {journal} {Few-Body Systems}\ }\textbf
  {\bibinfo {volume} {51}},\ \bibinfo {pages} {113--133}}\BibitemShut {NoStop}%
\bibitem [{\citenamefont {Fermi}(1934)}]{fermi1934}%
  \BibitemOpen
  \bibfield  {author} {\bibinfo {author} {\bibnamefont {Fermi}, \bibfnamefont
  {E}}} (\bibinfo {year} {1934}),\ \bibfield  {title} {\enquote {\bibinfo
  {title} {Sopra lo spostamento per pressione delle righe elevate delle serie
  spettrali},}\ }\href@noop {} {\bibfield  {journal} {\bibinfo  {journal} {Il
  Nuovo Cimento}\ }\textbf {\bibinfo {volume} {11}}~(\bibinfo {number} {3}),\
  \bibinfo {pages} {157--166}}\BibitemShut {NoStop}%
\bibitem [{\citenamefont {Fetter}\ and\ \citenamefont
  {Walecka}(2003)}]{fetter2003quantum}%
  \BibitemOpen
  \bibfield  {author} {\bibinfo {author} {\bibnamefont {Fetter}, \bibfnamefont
  {AL}}, \ and\ \bibinfo {author} {\bibfnamefont {J.D.}\ \bibnamefont
  {Walecka}}} (\bibinfo {year} {2003}),\ \href
  {https://books.google.de/books?id=0wekf1s83b0C} {\emph {\bibinfo {title}
  {Quantum Theory of Many-particle Systems}}},\ Dover Books on Physics\
  (\bibinfo  {publisher} {Dover Publications})\BibitemShut {NoStop}%
\bibitem [{\citenamefont {Feynman}(1955)}]{Feynman1955PhysRev}%
  \BibitemOpen
  \bibfield  {author} {\bibinfo {author} {\bibnamefont {Feynman}, \bibfnamefont
  {R~P}}} (\bibinfo {year} {1955}),\ \bibfield  {title} {\enquote {\bibinfo
  {title} {Slow electrons in a polar crystal},}\ }\href {\doibase
  10.1103/PhysRev.97.660} {\bibfield  {journal} {\bibinfo  {journal} {Phys.
  Rev.}\ }\textbf {\bibinfo {volume} {97}},\ \bibinfo {pages}
  {660--665}}\BibitemShut {NoStop}%
\bibitem [{\citenamefont {Fink}\ and\ \citenamefont
  {Zoller}(1985)}]{fink1985JPB}%
  \BibitemOpen
  \bibfield  {author} {\bibinfo {author} {\bibnamefont {Fink}, \bibfnamefont
  {M~G~J}}, \ and\ \bibinfo {author} {\bibfnamefont {P.}~\bibnamefont
  {Zoller}}} (\bibinfo {year} {1985}),\ \bibfield  {title} {\enquote {\bibinfo
  {title} {One- and two-photon detachment of negative hydrogen ions: a
  hyperspherical adiabatic approach},}\ }\href@noop {} {\bibfield  {journal}
  {\bibinfo  {journal} {J. Phys. B}\ }\textbf {\bibinfo {volume}
  {18}}~(\bibinfo {number} {12}),\ \bibinfo {pages} {L373}}\BibitemShut
  {NoStop}%
\bibitem [{\citenamefont {Flambaum}\ \emph {et~al.}(1999)\citenamefont
  {Flambaum}, \citenamefont {Gribakin},\ and\ \citenamefont
  {Harabati}}]{flambaum1999pra}%
  \BibitemOpen
  \bibfield  {author} {\bibinfo {author} {\bibnamefont {Flambaum},
  \bibfnamefont {V~V}}, \bibinfo {author} {\bibfnamefont {G.~F.}\ \bibnamefont
  {Gribakin}}, \ and\ \bibinfo {author} {\bibfnamefont {C.}~\bibnamefont
  {Harabati}}} (\bibinfo {year} {1999}),\ \bibfield  {title} {\enquote
  {\bibinfo {title} {Analytical calculation of cold-atom scattering},}\
  }\href@noop {} {\bibfield  {journal} {\bibinfo  {journal} {Phys. Rev. A}\
  }\textbf {\bibinfo {volume} {59}},\ \bibinfo {pages} {1998}}\BibitemShut
  {NoStop}%
\bibitem [{\citenamefont {Fletcher}\ \emph {et~al.}(2013)\citenamefont
  {Fletcher}, \citenamefont {Gaunt}, \citenamefont {Navon}, \citenamefont
  {Smith},\ and\ \citenamefont {Hadzibabic}}]{Hadzibabic2013prl}%
  \BibitemOpen
  \bibfield  {author} {\bibinfo {author} {\bibnamefont {Fletcher},
  \bibfnamefont {R~J}}, \bibinfo {author} {\bibfnamefont {A.~L.}\ \bibnamefont
  {Gaunt}}, \bibinfo {author} {\bibfnamefont {N.}~\bibnamefont {Navon}},
  \bibinfo {author} {\bibfnamefont {R.~P.}\ \bibnamefont {Smith}}, \ and\
  \bibinfo {author} {\bibfnamefont {Z.}~\bibnamefont {Hadzibabic}}} (\bibinfo
  {year} {2013}),\ \bibfield  {title} {\enquote {\bibinfo {title} {Stability of
  a unitary {B}ose gas},}\ }\href@noop {} {\bibfield  {journal} {\bibinfo
  {journal} {Phys. Rev. Lett.}\ }\textbf {\bibinfo {volume} {111}},\ \bibinfo
  {pages} {125303}}\BibitemShut {NoStop}%
\bibitem [{\citenamefont {Flower}\ and\ \citenamefont
  {Harris}(2007)}]{Flower-2007}%
  \BibitemOpen
  \bibfield  {author} {\bibinfo {author} {\bibnamefont {Flower}, \bibfnamefont
  {D~R}}, \ and\ \bibinfo {author} {\bibfnamefont {G.~J.}\ \bibnamefont
  {Harris}}} (\bibinfo {year} {2007}),\ \bibfield  {title} {\enquote {\bibinfo
  {title} {Three-body recombination of hydrogen during primordial star
  formation},}\ }\href@noop {} {\bibfield  {journal} {\bibinfo  {journal}
  {Monthly Notices of the Royal Astronomical Society}\ }\textbf {\bibinfo
  {volume} {377}},\ \bibinfo {pages} {705--710}}\BibitemShut {NoStop}%
\bibitem [{\citenamefont {Fock}(1958)}]{fock1958}%
  \BibitemOpen
  \bibfield  {author} {\bibinfo {author} {\bibnamefont {Fock}, \bibfnamefont
  {V}}} (\bibinfo {year} {1958}),\ \bibfield  {title} {\enquote {\bibinfo
  {title} {Hyperspherical expansion},}\ }\href@noop {} {\bibfield  {journal}
  {\bibinfo  {journal} {K. Norske Vidensk. Selsk. Forhandl.}\ }\textbf
  {\bibinfo {volume} {31}},\ \bibinfo {pages} {138--52}}\BibitemShut {NoStop}%
\bibitem [{\citenamefont {Forrey}(2013)}]{Forrey-2013a}%
  \BibitemOpen
  \bibfield  {author} {\bibinfo {author} {\bibnamefont {Forrey}, \bibfnamefont
  {R~C}}} (\bibinfo {year} {2013}),\ \bibfield  {title} {\enquote {\bibinfo
  {title} {Rate of formation of hydrogen molecules by three-body recombination
  during primordial star formation},}\ }\href@noop {} {\bibfield  {journal}
  {\bibinfo  {journal} {Astrophys. J. Lett.}\ }\textbf {\bibinfo {volume}
  {773}},\ \bibinfo {pages} {L25}}\BibitemShut {NoStop}%
\bibitem [{\citenamefont {Frederico}\ \emph {et~al.}(2012)\citenamefont
  {Frederico}, \citenamefont {Delfino}, \citenamefont {Tomio},\ and\
  \citenamefont {Yamashita}}]{Frederico2012xh}%
  \BibitemOpen
  \bibfield  {author} {\bibinfo {author} {\bibnamefont {Frederico},
  \bibfnamefont {T}}, \bibinfo {author} {\bibfnamefont {A.}~\bibnamefont
  {Delfino}}, \bibinfo {author} {\bibfnamefont {Lauro}\ \bibnamefont {Tomio}},
  \ and\ \bibinfo {author} {\bibfnamefont {M.~T.}\ \bibnamefont {Yamashita}}}
  (\bibinfo {year} {2012}),\ \bibfield  {title} {\enquote {\bibinfo {title}
  {{Universal aspects of light halo nuclei}},}\ }\href {\doibase
  10.1016/j.ppnp.2012.06.001} {\bibfield  {journal} {\bibinfo  {journal} {Prog.
  Part. Nucl. Phys.}\ }\textbf {\bibinfo {volume} {67}},\ \bibinfo {pages}
  {939--994}}\BibitemShut {NoStop}%
\bibitem [{\citenamefont {Friar}\ \emph {et~al.}(1984)\citenamefont {Friar},
  \citenamefont {Gibson}, \citenamefont {Payne},\ and\ \citenamefont
  {R.}}]{Friar-1984}%
  \BibitemOpen
  \bibfield  {author} {\bibinfo {author} {\bibnamefont {Friar}, \bibfnamefont
  {J~L}}, \bibinfo {author} {\bibfnamefont {B.~F.}\ \bibnamefont {Gibson}},
  \bibinfo {author} {\bibfnamefont {G.~L.}\ \bibnamefont {Payne}}, \ and\
  \bibinfo {author} {\bibfnamefont {Chen~C.}\ \bibnamefont {R.}}} (\bibinfo
  {year} {1984}),\ \bibfield  {title} {\enquote {\bibinfo {title}
  {Configuration space {F}addev continuum calculations: {N-d} s-wave scattering
  lengths with tensor-force interactions},}\ }\href@noop {} {\bibfield
  {journal} {\bibinfo  {journal} {Phys. Rev. C}\ }\textbf {\bibinfo {volume}
  {30}},\ \bibinfo {pages} {1121}}\BibitemShut {NoStop}%
\bibitem [{\citenamefont {Fr\"{o}hlich}\ \emph {et~al.}(2011)\citenamefont
  {Fr\"{o}hlich}, \citenamefont {Feld}, \citenamefont {Vogt}, \citenamefont
  {Koschorreck}, \citenamefont {Zwerger},\ and\ \citenamefont
  {K\"{o}hl}}]{froehlich2011}%
  \BibitemOpen
  \bibfield  {author} {\bibinfo {author} {\bibnamefont {Fr\"{o}hlich},
  \bibfnamefont {B}}, \bibinfo {author} {\bibfnamefont {M.}~\bibnamefont
  {Feld}}, \bibinfo {author} {\bibfnamefont {E.}~\bibnamefont {Vogt}}, \bibinfo
  {author} {\bibfnamefont {M.}~\bibnamefont {Koschorreck}}, \bibinfo {author}
  {\bibfnamefont {W.}~\bibnamefont {Zwerger}}, \ and\ \bibinfo {author}
  {\bibfnamefont {M.}~\bibnamefont {K\"{o}hl}}} (\bibinfo {year} {2011}),\
  \bibfield  {title} {\enquote {\bibinfo {title} {Radio-frequency spectroscopy
  of a strongly interacting two-dimensional {F}ermi gas},}\ }\href@noop {}
  {\bibfield  {journal} {\bibinfo  {journal} {Phys. Rev. Lett.}\ }\textbf
  {\bibinfo {volume} {106}}~(\bibinfo {number} {10}),\ \bibinfo {pages}
  {105301}}\BibitemShut {NoStop}%
\bibitem [{\citenamefont {Fr\"ohlich}(1954)}]{Froehlich1954AdvPhys}%
  \BibitemOpen
  \bibfield  {author} {\bibinfo {author} {\bibnamefont {Fr\"ohlich},
  \bibfnamefont {H}}} (\bibinfo {year} {1954}),\ \bibfield  {title} {\enquote
  {\bibinfo {title} {Electrons in lattice fields},}\ }\href {\doibase
  10.1080/00018735400101213} {\bibfield  {journal} {\bibinfo  {journal}
  {Advances in Physics}\ }\textbf {\bibinfo {volume} {3}}~(\bibinfo {number}
  {11}),\ \bibinfo {pages} {325--361}},\ \Eprint
  {http://arxiv.org/abs/http://dx.doi.org/10.1080/00018735400101213}
  {http://dx.doi.org/10.1080/00018735400101213} \BibitemShut {NoStop}%
\bibitem [{\citenamefont {Fujiwara}\ \emph {et~al.}(2008)\citenamefont
  {Fujiwara}, \citenamefont {Suzuki}, \citenamefont {Kohno},\ and\
  \citenamefont {Miyagawa}}]{Fujiwara-2008}%
  \BibitemOpen
  \bibfield  {author} {\bibinfo {author} {\bibnamefont {Fujiwara},
  \bibfnamefont {Y}}, \bibinfo {author} {\bibfnamefont {Y.}~\bibnamefont
  {Suzuki}}, \bibinfo {author} {\bibfnamefont {M.}~\bibnamefont {Kohno}}, \
  and\ \bibinfo {author} {\bibfnamefont {K.}~\bibnamefont {Miyagawa}}}
  (\bibinfo {year} {2008}),\ \bibfield  {title} {\enquote {\bibinfo {title}
  {Addendum to triton and hypertriton binding energies calcualted from
  {SU}$_{6}$ quark-model baryon-baryon interactions},}\ }\href@noop {}
  {\bibfield  {journal} {\bibinfo  {journal} {Phys. Rev. C}\ }\textbf {\bibinfo
  {volume} {77}},\ \bibinfo {pages} {027001}}\BibitemShut {NoStop}%
\bibitem [{\citenamefont {Fukuda}\ \emph {et~al.}(1990)\citenamefont {Fukuda},
  \citenamefont {Ishihara},\ and\ \citenamefont {Hara}}]{FUKUDA1990}%
  \BibitemOpen
  \bibfield  {author} {\bibinfo {author} {\bibnamefont {Fukuda}, \bibfnamefont
  {H}}, \bibinfo {author} {\bibfnamefont {T.}~\bibnamefont {Ishihara}}, \ and\
  \bibinfo {author} {\bibfnamefont {S.}~\bibnamefont {Hara}}} (\bibinfo {year}
  {1990}),\ \bibfield  {title} {\enquote {\bibinfo {title} {Hyperradial
  adiabatic treatment of {d$\mu$+t} collisions at low energies},}\ }\href@noop
  {} {\bibfield  {journal} {\bibinfo  {journal} {Phys. Rev. A}\ }\textbf
  {\bibinfo {volume} {41}}~(\bibinfo {number} {1}),\ \bibinfo {pages}
  {145--149}}\BibitemShut {NoStop}%
\bibitem [{\citenamefont {Gao}(1998)}]{Gao1998a}%
  \BibitemOpen
  \bibfield  {author} {\bibinfo {author} {\bibnamefont {Gao}, \bibfnamefont
  {B}}} (\bibinfo {year} {1998}),\ \bibfield  {title} {\enquote {\bibinfo
  {title} {Quantum-defect theory of atomic collisions and molecular vibration
  spectra},}\ }\href@noop {} {\bibfield  {journal} {\bibinfo  {journal} {Phys.
  Rev. A}\ }\textbf {\bibinfo {volume} {58}}~(\bibinfo {number} {5}),\ \bibinfo
  {pages} {4222--4225}}\BibitemShut {NoStop}%
\bibitem [{\citenamefont {Gao}(2001)}]{Gao2001}%
  \BibitemOpen
  \bibfield  {author} {\bibinfo {author} {\bibnamefont {Gao}, \bibfnamefont
  {B}}} (\bibinfo {year} {2001}),\ \bibfield  {title} {\enquote {\bibinfo
  {title} {Angular-momentum-insensitive quantum-defect theory for diatomic
  systems},}\ }\href@noop {} {\bibfield  {journal} {\bibinfo  {journal} {Phys.
  Rev. A}\ }\textbf {\bibinfo {volume} {64}}~(\bibinfo {number} {1}),\ \bibinfo
  {pages} {010701}}\BibitemShut {NoStop}%
\bibitem [{\citenamefont {Gao}(2008)}]{Gao2008}%
  \BibitemOpen
  \bibfield  {author} {\bibinfo {author} {\bibnamefont {Gao}, \bibfnamefont
  {B}}} (\bibinfo {year} {2008}),\ \bibfield  {title} {\enquote {\bibinfo
  {title} {General form of the quantum-defect theory for $-1/{r}^{\alpha}$ type
  of potentials with $\alpha>2$},}\ }\href@noop {} {\bibfield  {journal}
  {\bibinfo  {journal} {Phys. Rev. A}\ }\textbf {\bibinfo {volume} {78}},\
  \bibinfo {pages} {012702}}\BibitemShut {NoStop}%
\bibitem [{\citenamefont {Gao}\ \emph {et~al.}(2015)\citenamefont {Gao},
  \citenamefont {Wang},\ and\ \citenamefont {Yu}}]{GaoWangYu2015}%
  \BibitemOpen
  \bibfield  {author} {\bibinfo {author} {\bibnamefont {Gao}, \bibfnamefont
  {Chao}}, \bibinfo {author} {\bibfnamefont {Jia}\ \bibnamefont {Wang}}, \ and\
  \bibinfo {author} {\bibfnamefont {Zhenhua}\ \bibnamefont {Yu}}} (\bibinfo
  {year} {2015}),\ \bibfield  {title} {\enquote {\bibinfo {title} {Revealing
  the origin of super efimov states in the hyperspherical formalism},}\ }\href
  {\doibase 10.1103/PhysRevA.92.020504} {\bibfield  {journal} {\bibinfo
  {journal} {Phys. Rev. A}\ }\textbf {\bibinfo {volume} {92}},\ \bibinfo
  {pages} {020504}}\BibitemShut {NoStop}%
\bibitem [{\citenamefont {Gattobigio}\ \emph {et~al.}(2011)\citenamefont
  {Gattobigio}, \citenamefont {Kievsky},\ and\ \citenamefont
  {Viviani}}]{gattobigio2011PRA}%
  \BibitemOpen
  \bibfield  {author} {\bibinfo {author} {\bibnamefont {Gattobigio},
  \bibfnamefont {M}}, \bibinfo {author} {\bibfnamefont {A.}~\bibnamefont
  {Kievsky}}, \ and\ \bibinfo {author} {\bibfnamefont {M.}~\bibnamefont
  {Viviani}}} (\bibinfo {year} {2011}),\ \bibfield  {title} {\enquote {\bibinfo
  {title} {Spectra of helium clusters with up to six atoms using soft-core
  potentials},}\ }\href@noop {} {\bibfield  {journal} {\bibinfo  {journal}
  {Phys. Rev. A}\ }\textbf {\bibinfo {volume} {84}},\ \bibinfo {pages}
  {052503}}\BibitemShut {NoStop}%
\bibitem [{\citenamefont {Gattobigio}\ \emph {et~al.}(2012)\citenamefont
  {Gattobigio}, \citenamefont {Kievsky},\ and\ \citenamefont
  {Viviani}}]{gattobigio2012PRA}%
  \BibitemOpen
  \bibfield  {author} {\bibinfo {author} {\bibnamefont {Gattobigio},
  \bibfnamefont {M}}, \bibinfo {author} {\bibfnamefont {A.}~\bibnamefont
  {Kievsky}}, \ and\ \bibinfo {author} {\bibfnamefont {M.}~\bibnamefont
  {Viviani}}} (\bibinfo {year} {2012}),\ \bibfield  {title} {\enquote {\bibinfo
  {title} {Energy spectra of small bosonic clusters having a large two-body
  scattering length},}\ }\href@noop {} {\bibfield  {journal} {\bibinfo
  {journal} {Phys. Rev. A}\ }\textbf {\bibinfo {volume} {86}},\ \bibinfo
  {pages} {042513}}\BibitemShut {NoStop}%
\bibitem [{\citenamefont {Gharashi}\ \emph {et~al.}(2012)\citenamefont
  {Gharashi}, \citenamefont {Daily},\ and\ \citenamefont
  {Blume}}]{gharashi2012pra}%
  \BibitemOpen
  \bibfield  {author} {\bibinfo {author} {\bibnamefont {Gharashi},
  \bibfnamefont {S~E}}, \bibinfo {author} {\bibfnamefont {K.~M.}\ \bibnamefont
  {Daily}}, \ and\ \bibinfo {author} {\bibfnamefont {D.}~\bibnamefont {Blume}}}
  (\bibinfo {year} {2012}),\ \bibfield  {title} {\enquote {\bibinfo {title}
  {Three $s$-wave-interacting fermions under anisotropic harmonic confinement:
  Dimensional crossover of energetics and virial coefficients},}\ }\href@noop
  {} {\bibfield  {journal} {\bibinfo  {journal} {Phys. Rev. A}\ }\textbf
  {\bibinfo {volume} {86}},\ \bibinfo {pages} {042702}}\BibitemShut {NoStop}%
\bibitem [{\citenamefont {Giamarchi}(2004)}]{giamarchi2004quantum}%
  \BibitemOpen
  \bibfield  {author} {\bibinfo {author} {\bibnamefont {Giamarchi},
  \bibfnamefont {T}}} (\bibinfo {year} {2004}),\ \href@noop {} {\emph {\bibinfo
  {title} {Quantum physics in one dimension}}}\ (\bibinfo  {publisher} {Oxford
  University Press},\ \bibinfo {address} {London})\BibitemShut {NoStop}%
\bibitem [{\citenamefont {Giannakeas}\ \emph {et~al.}(2012)\citenamefont
  {Giannakeas}, \citenamefont {Diakonos},\ and\ \citenamefont
  {Schmelcher}}]{giannakeas2012}%
  \BibitemOpen
  \bibfield  {author} {\bibinfo {author} {\bibnamefont {Giannakeas},
  \bibfnamefont {P}}, \bibinfo {author} {\bibfnamefont {F.~K.}\ \bibnamefont
  {Diakonos}}, \ and\ \bibinfo {author} {\bibfnamefont {P.}~\bibnamefont
  {Schmelcher}}} (\bibinfo {year} {2012}),\ \bibfield  {title} {\enquote
  {\bibinfo {title} {Coupled $\ell$-wave confinement-induced resonances in
  cylindrically symmetric waveguides},}\ }\href@noop {} {\bibfield  {journal}
  {\bibinfo  {journal} {Phys. Rev. A}\ }\textbf {\bibinfo {volume}
  {86}}~(\bibinfo {number} {4}),\ \bibinfo {pages} {042703}}\BibitemShut
  {NoStop}%
\bibitem [{\citenamefont {Giannakeas}\ \emph {et~al.}(2016)\citenamefont
  {Giannakeas}, \citenamefont {Greene},\ and\ \citenamefont
  {Robicheaux}}]{giannakeas2016pra}%
  \BibitemOpen
  \bibfield  {author} {\bibinfo {author} {\bibnamefont {Giannakeas},
  \bibfnamefont {P}}, \bibinfo {author} {\bibfnamefont {C.~H.}\ \bibnamefont
  {Greene}}, \ and\ \bibinfo {author} {\bibfnamefont {F.}~\bibnamefont
  {Robicheaux}}} (\bibinfo {year} {2016}),\ \bibfield  {title} {\enquote
  {\bibinfo {title} {Generalized local-frame-transformation theory for excited
  species in external fields},}\ }\href@noop {} {\bibfield  {journal} {\bibinfo
   {journal} {Phys. Rev. A}\ }\textbf {\bibinfo {volume} {94}},\ \bibinfo
  {pages} {013419}}\BibitemShut {NoStop}%
\bibitem [{\citenamefont {Giannakeas}\ \emph {et~al.}(2013)\citenamefont
  {Giannakeas}, \citenamefont {Melezhik},\ and\ \citenamefont
  {Schmelcher}}]{giannakeas2013prl}%
  \BibitemOpen
  \bibfield  {author} {\bibinfo {author} {\bibnamefont {Giannakeas},
  \bibfnamefont {P}}, \bibinfo {author} {\bibfnamefont {V.~S.}\ \bibnamefont
  {Melezhik}}, \ and\ \bibinfo {author} {\bibfnamefont {P.}~\bibnamefont
  {Schmelcher}}} (\bibinfo {year} {2013}),\ \bibfield  {title} {\enquote
  {\bibinfo {title} {Dipolar confinement-induced resonances of ultracold gases
  in waveguides},}\ }\href@noop {} {\bibfield  {journal} {\bibinfo  {journal}
  {Phys. Rev. Lett.}\ }\textbf {\bibinfo {volume} {111}},\ \bibinfo {pages}
  {183201}}\BibitemShut {NoStop}%
\bibitem [{\citenamefont {Giorgini}\ \emph {et~al.}(2008)\citenamefont
  {Giorgini}, \citenamefont {Pitaevskii},\ and\ \citenamefont
  {Stringari}}]{giorgini2008RMP}%
  \BibitemOpen
  \bibfield  {author} {\bibinfo {author} {\bibnamefont {Giorgini},
  \bibfnamefont {S}}, \bibinfo {author} {\bibfnamefont {L.~P.}\ \bibnamefont
  {Pitaevskii}}, \ and\ \bibinfo {author} {\bibfnamefont {S.}~\bibnamefont
  {Stringari}}} (\bibinfo {year} {2008}),\ \bibfield  {title} {{\selectlanguage
  {English}\enquote {\bibinfo {title} {Theory of ultracold atomic {F}ermi
  gases},}\ }}\href@noop {} {\bibfield  {journal} {\bibinfo  {journal} {Rev.
  Mod. Phys.}\ }\textbf {\bibinfo {volume} {80}}~(\bibinfo {number} {4}),\
  \bibinfo {pages} {1215--1274}}\BibitemShut {NoStop}%
\bibitem [{\citenamefont {Girardeau}(1960)}]{girardeau1960}%
  \BibitemOpen
  \bibfield  {author} {\bibinfo {author} {\bibnamefont {Girardeau},
  \bibfnamefont {M}}} (\bibinfo {year} {1960}),\ \bibfield  {title} {\enquote
  {\bibinfo {title} {Relationship between systems of impenetrable bosons and
  fermions in one dimension},}\ }\href@noop {} {\bibfield  {journal} {\bibinfo
  {journal} {J. Math. Phys.}\ }\textbf {\bibinfo {volume} {1}}~(\bibinfo
  {number} {6}),\ \bibinfo {pages} {516--523}}\BibitemShut {NoStop}%
\bibitem [{\citenamefont {Girardeau}\ and\ \citenamefont
  {Olshanii}(2004)}]{grirardeaupra2004}%
  \BibitemOpen
  \bibfield  {author} {\bibinfo {author} {\bibnamefont {Girardeau},
  \bibfnamefont {M~D}}, \ and\ \bibinfo {author} {\bibfnamefont
  {M.}~\bibnamefont {Olshanii}}} (\bibinfo {year} {2004}),\ \bibfield  {title}
  {\enquote {\bibinfo {title} {Theory of spinor {F}ermi and {B}ose gases in
  tight atom waveguides},}\ }\href@noop {} {\bibfield  {journal} {\bibinfo
  {journal} {Phys. Rev. A}\ }\textbf {\bibinfo {volume} {70}},\ \bibinfo
  {pages} {023608}}\BibitemShut {NoStop}%
\bibitem [{\citenamefont {Gl{\"o}ckle}(2012)}]{glockle2012quantum}%
  \BibitemOpen
  \bibfield  {author} {\bibinfo {author} {\bibnamefont {Gl{\"o}ckle},
  \bibfnamefont {W}}} (\bibinfo {year} {2012}),\ \href@noop {} {\emph {\bibinfo
  {title} {The quantum mechanical few-body problem}}}\ (\bibinfo  {publisher}
  {Springer Science \& Business Media})\BibitemShut {NoStop}%
\bibitem [{\citenamefont {Gogolin}\ \emph {et~al.}(2008)\citenamefont
  {Gogolin}, \citenamefont {Mora},\ and\ \citenamefont
  {Egger}}]{gogolin2008PRL}%
  \BibitemOpen
  \bibfield  {author} {\bibinfo {author} {\bibnamefont {Gogolin}, \bibfnamefont
  {A~O}}, \bibinfo {author} {\bibfnamefont {C.}~\bibnamefont {Mora}}, \ and\
  \bibinfo {author} {\bibfnamefont {R.}~\bibnamefont {Egger}}} (\bibinfo {year}
  {2008}),\ \bibfield  {title} {\enquote {\bibinfo {title} {Analytical solution
  of the bosonic three-body problem},}\ }\href@noop {} {\bibfield  {journal}
  {\bibinfo  {journal} {Phys. Rev. Lett.}\ }\textbf {\bibinfo {volume}
  {100}}~(\bibinfo {number} {14}),\ \bibinfo {pages} {140404}}\BibitemShut
  {NoStop}%
\bibitem [{\citenamefont {Gongleton}(1992)}]{Gongleton-1992}%
  \BibitemOpen
  \bibfield  {author} {\bibinfo {author} {\bibnamefont {Gongleton},
  \bibfnamefont {J~G}}} (\bibinfo {year} {1992}),\ \bibfield  {title} {\enquote
  {\bibinfo {title} {A simple model of the hypertriton},}\ }\href@noop {}
  {\bibfield  {journal} {\bibinfo  {journal} {J. Phys. G}\ }\textbf {\bibinfo
  {volume} {18}},\ \bibinfo {pages} {339}}\BibitemShut {NoStop}%
\bibitem [{\citenamefont {Granger}\ and\ \citenamefont
  {Blume}(2004)}]{granger2004PRL}%
  \BibitemOpen
  \bibfield  {author} {\bibinfo {author} {\bibnamefont {Granger}, \bibfnamefont
  {Brian~E}}, \ and\ \bibinfo {author} {\bibfnamefont {D.}~\bibnamefont
  {Blume}}} (\bibinfo {year} {2004}),\ \bibfield  {title} {\enquote {\bibinfo
  {title} {Tuning the interactions of spin-polarized {F}ermions using
  quasi-one-dimensional confinement},}\ }\href@noop {} {\bibfield  {journal}
  {\bibinfo  {journal} {Phys. Rev. Lett.}\ }\textbf {\bibinfo {volume}
  {92}}~(\bibinfo {number} {13}),\ \bibinfo {pages} {133202}}\BibitemShut
  {NoStop}%
\bibitem [{\citenamefont {Greene}(1981)}]{greene1981PRA}%
  \BibitemOpen
  \bibfield  {author} {\bibinfo {author} {\bibnamefont {Greene}, \bibfnamefont
  {C~H}}} (\bibinfo {year} {1981}),\ \bibfield  {title} {{\selectlanguage
  {English}\enquote {\bibinfo {title} {Doubly excited-states of the
  akaline-earth atoms},}\ }}\href@noop {} {\bibfield  {journal} {\bibinfo
  {journal} {Phys. Rev. A}\ }\textbf {\bibinfo {volume} {23}}~(\bibinfo
  {number} {2}),\ \bibinfo {pages} {661--678}}\BibitemShut {NoStop}%
\bibitem [{\citenamefont {Greene}(1987)}]{greene1987}%
  \BibitemOpen
  \bibfield  {author} {\bibinfo {author} {\bibnamefont {Greene}, \bibfnamefont
  {C~H}}} (\bibinfo {year} {1987}),\ \bibfield  {title} {\enquote {\bibinfo
  {title} {Negative-ion photodetachment in a weak magnetic field},}\
  }\href@noop {} {\bibfield  {journal} {\bibinfo  {journal} {Phys. Rev. A}\
  }\textbf {\bibinfo {volume} {36}}~(\bibinfo {number} {9}),\ \bibinfo {pages}
  {4236--4244}}\BibitemShut {NoStop}%
\bibitem [{\citenamefont {Greene}(2010)}]{greene2010PT}%
  \BibitemOpen
  \bibfield  {author} {\bibinfo {author} {\bibnamefont {Greene}, \bibfnamefont
  {C~H}}} (\bibinfo {year} {2010}),\ \bibfield  {title} {{\selectlanguage
  {English}\enquote {\bibinfo {title} {Universal insights from few-body
  land},}\ }}\href@noop {} {\bibfield  {journal} {\bibinfo  {journal} {Physics
  Today}\ }\textbf {\bibinfo {volume} {63}}~(\bibinfo {number} {3}),\ \bibinfo
  {pages} {40--45}}\BibitemShut {NoStop}%
\bibitem [{\citenamefont {Greene}\ and\ \citenamefont
  {Clark}(1984)}]{greene1984PRAc}%
  \BibitemOpen
  \bibfield  {author} {\bibinfo {author} {\bibnamefont {Greene}, \bibfnamefont
  {C~H}}, \ and\ \bibinfo {author} {\bibfnamefont {C.~W.}\ \bibnamefont
  {Clark}}} (\bibinfo {year} {1984}),\ \bibfield  {title} {\enquote {\bibinfo
  {title} {Adiabatic hyperspherical treatment of lithium $^2${P}$^0$ states},}\
  }\href@noop {} {\bibfield  {journal} {\bibinfo  {journal} {Phys. Rev. A}\
  }\textbf {\bibinfo {volume} {30}}~(\bibinfo {number} {5}),\ \bibinfo {pages}
  {2161}}\BibitemShut {NoStop}%
\bibitem [{\citenamefont {Greene}\ and\ \citenamefont
  {Jungen}(1985)}]{greene1985AMOP}%
  \BibitemOpen
  \bibfield  {author} {\bibinfo {author} {\bibnamefont {Greene}, \bibfnamefont
  {C~H}}, \ and\ \bibinfo {author} {\bibfnamefont {C}~\bibnamefont {Jungen}}}
  (\bibinfo {year} {1985}),\ \bibfield  {title} {{\selectlanguage
  {English}\enquote {\bibinfo {title} {Molecular applications of quantum defect
  theory},}\ }}\href@noop {} {\bibfield  {journal} {\bibinfo  {journal} {Adv.
  At. Mol. Opt. Phys.}\ }\textbf {\bibinfo {volume} {21}},\ \bibinfo {pages}
  {51--121}}\BibitemShut {NoStop}%
\bibitem [{\citenamefont {Greene}\ and\ \citenamefont
  {Rau}(1982)}]{greene1982PRL}%
  \BibitemOpen
  \bibfield  {author} {\bibinfo {author} {\bibnamefont {Greene}, \bibfnamefont
  {C~H}}, \ and\ \bibinfo {author} {\bibfnamefont {A.~R.~P.}\ \bibnamefont
  {Rau}}} (\bibinfo {year} {1982}),\ \bibfield  {title} {{\selectlanguage
  {English}\enquote {\bibinfo {title} {Double escape of 2 electrons at
  threshold - dependence on {L}, {S}, and $\pi$},}\ }}\href@noop {} {\bibfield
  {journal} {\bibinfo  {journal} {Phys. Rev. Lett.}\ }\textbf {\bibinfo
  {volume} {48}}~(\bibinfo {number} {8}),\ \bibinfo {pages}
  {533--537}}\BibitemShut {NoStop}%
\bibitem [{\citenamefont {Greene}\ and\ \citenamefont
  {Rau}(1983)}]{greene1983JPB}%
  \BibitemOpen
  \bibfield  {author} {\bibinfo {author} {\bibnamefont {Greene}, \bibfnamefont
  {C~H}}, \ and\ \bibinfo {author} {\bibfnamefont {A.~R.~P.}\ \bibnamefont
  {Rau}}} (\bibinfo {year} {1983}),\ \bibfield  {title} {{\selectlanguage
  {English}\enquote {\bibinfo {title} {Effect of symmetry on 2-electron escape
  at threshold},}\ }}\href@noop {} {\bibfield  {journal} {\bibinfo  {journal}
  {J. Phys. B}\ }\textbf {\bibinfo {volume} {16}}~(\bibinfo {number} {1}),\
  \bibinfo {pages} {99--106}}\BibitemShut {NoStop}%
\bibitem [{\citenamefont {Greiner}\ \emph {et~al.}(2003)\citenamefont
  {Greiner}, \citenamefont {Regal},\ and\ \citenamefont {Jin}}]{greiner2003NT}%
  \BibitemOpen
  \bibfield  {author} {\bibinfo {author} {\bibnamefont {Greiner}, \bibfnamefont
  {M}}, \bibinfo {author} {\bibfnamefont {C.~A.}\ \bibnamefont {Regal}}, \ and\
  \bibinfo {author} {\bibfnamefont {D.~S.}\ \bibnamefont {Jin}}} (\bibinfo
  {year} {2003}),\ \bibfield  {title} {\enquote {\bibinfo {title} {Emergence of
  a molecular {B}ose-{E}instein condensate from a {F}ermi gas},}\ }\href@noop
  {} {\bibfield  {journal} {\bibinfo  {journal} {Nature (London)}\ }\textbf
  {\bibinfo {volume} {412}},\ \bibinfo {pages} {537--540}}\BibitemShut
  {NoStop}%
\bibitem [{\citenamefont {Gridnev}(2014)}]{Gridnev2014JPA}%
  \BibitemOpen
  \bibfield  {author} {\bibinfo {author} {\bibnamefont {Gridnev}, \bibfnamefont
  {Dmitry~K}}} (\bibinfo {year} {2014}),\ \bibfield  {title} {\enquote
  {\bibinfo {title} {Three resonating fermions in flatland: proof of the super
  efimov effect and the exact discrete spectrum asymptotics},}\ }\href
  {http://stacks.iop.org/1751-8121/47/i=50/a=505204} {\bibfield  {journal}
  {\bibinfo  {journal} {Journal of Physics A: Mathematical and Theoretical}\
  }\textbf {\bibinfo {volume} {47}}~(\bibinfo {number} {50}),\ \bibinfo {pages}
  {505204}}\BibitemShut {NoStop}%
\bibitem [{\citenamefont {Gross}\ \emph {et~al.}(2009)\citenamefont {Gross},
  \citenamefont {Shotan}, \citenamefont {Kokkelmans},\ and\ \citenamefont
  {Khaykovich}}]{gross2009PRL}%
  \BibitemOpen
  \bibfield  {author} {\bibinfo {author} {\bibnamefont {Gross}, \bibfnamefont
  {N}}, \bibinfo {author} {\bibfnamefont {Z.}~\bibnamefont {Shotan}}, \bibinfo
  {author} {\bibfnamefont {S.}~\bibnamefont {Kokkelmans}}, \ and\ \bibinfo
  {author} {\bibfnamefont {L.}~\bibnamefont {Khaykovich}}} (\bibinfo {year}
  {2009}),\ \bibfield  {title} {{\selectlanguage {English}\enquote {\bibinfo
  {title} {Observation of universality in ultracold $^7${Li} three-body
  recombination},}\ }}\href@noop {} {\bibfield  {journal} {\bibinfo  {journal}
  {Phys. Rev. Lett.}\ }\textbf {\bibinfo {volume} {103}}~(\bibinfo {number}
  {16}),\ \bibinfo {pages} {163202}}\BibitemShut {NoStop}%
\bibitem [{\citenamefont {Gross}\ \emph {et~al.}(2010)\citenamefont {Gross},
  \citenamefont {Shotan}, \citenamefont {Kokkelmans},\ and\ \citenamefont
  {Khaykovich}}]{gross2010PRL}%
  \BibitemOpen
  \bibfield  {author} {\bibinfo {author} {\bibnamefont {Gross}, \bibfnamefont
  {N}}, \bibinfo {author} {\bibfnamefont {Z.}~\bibnamefont {Shotan}}, \bibinfo
  {author} {\bibfnamefont {S.}~\bibnamefont {Kokkelmans}}, \ and\ \bibinfo
  {author} {\bibfnamefont {L.}~\bibnamefont {Khaykovich}}} (\bibinfo {year}
  {2010}),\ \bibfield  {title} {\enquote {\bibinfo {title}
  {Nuclear-spin-independent short-range three-body physics in ultracold
  atoms},}\ }\href@noop {} {\bibfield  {journal} {\bibinfo  {journal} {Phys.
  Rev. Lett.}\ }\textbf {\bibinfo {volume} {105}},\ \bibinfo {pages}
  {103203}}\BibitemShut {NoStop}%
\bibitem [{\citenamefont {Gross}\ \emph {et~al.}(2011)\citenamefont {Gross},
  \citenamefont {Shotan}, \citenamefont {Machtey}, \citenamefont {Kokkelmans},\
  and\ \citenamefont {Khaykovich}}]{gross2011CRP}%
  \BibitemOpen
  \bibfield  {author} {\bibinfo {author} {\bibnamefont {Gross}, \bibfnamefont
  {N}}, \bibinfo {author} {\bibfnamefont {Z.}~\bibnamefont {Shotan}}, \bibinfo
  {author} {\bibfnamefont {O.}~\bibnamefont {Machtey}}, \bibinfo {author}
  {\bibfnamefont {S.}~\bibnamefont {Kokkelmans}}, \ and\ \bibinfo {author}
  {\bibfnamefont {L.}~\bibnamefont {Khaykovich}}} (\bibinfo {year} {2011}),\
  \bibfield  {title} {\enquote {\bibinfo {title} {Study of {E}fimov physics in
  two nuclear-spin sublevels of $^7${Li}},}\ }\href@noop {} {\ \textbf
  {\bibinfo {volume} {12}}~(\bibinfo {number} {1}),\ \bibinfo {pages} {4 --
  12}}\BibitemShut {NoStop}%
\bibitem [{\citenamefont {Grozdanov}(1995)}]{grozdanovpra1995}%
  \BibitemOpen
  \bibfield  {author} {\bibinfo {author} {\bibnamefont {Grozdanov},
  \bibfnamefont {T~P}}} (\bibinfo {year} {1995}),\ \bibfield  {title} {\enquote
  {\bibinfo {title} {Photodetachment of an electron bound by a zero-range
  potential in magnetic fields of arbitrary strength},}\ }\href@noop {}
  {\bibfield  {journal} {\bibinfo  {journal} {Phys. Rev. A}\ }\textbf {\bibinfo
  {volume} {51}}~(\bibinfo {number} {1}),\ \bibinfo {pages}
  {607--610}}\BibitemShut {NoStop}%
\bibitem [{\citenamefont {Grusdt}\ \emph {et~al.}(2015)\citenamefont {Grusdt},
  \citenamefont {Shchadilova}, \citenamefont {Rubtsov},\ and\ \citenamefont
  {Demler}}]{grusdt_renormalization_2015}%
  \BibitemOpen
  \bibfield  {author} {\bibinfo {author} {\bibnamefont {Grusdt}, \bibfnamefont
  {F}}, \bibinfo {author} {\bibfnamefont {Y.~E.}\ \bibnamefont {Shchadilova}},
  \bibinfo {author} {\bibfnamefont {A.~N.}\ \bibnamefont {Rubtsov}}, \ and\
  \bibinfo {author} {\bibfnamefont {E.}~\bibnamefont {Demler}}} (\bibinfo
  {year} {2015}),\ \bibfield  {title} {\enquote {\bibinfo {title}
  {Renormalization group approach to the {Frohlich} polaron model: application
  to impurity-{BEC} problem},}\ }\href {\doibase 10.1038/srep12124} {\bibfield
  {journal} {\bibinfo  {journal} {Scientific Reports}\ }\textbf {\bibinfo
  {volume} {5}},\ \bibinfo {pages} {12124}}\BibitemShut {NoStop}%
\bibitem [{\citenamefont {G\"{u}nter}\ \emph {et~al.}(2005)\citenamefont
  {G\"{u}nter}, \citenamefont {St\"{o}ferle}, \citenamefont {Moritz},
  \citenamefont {K\"{o}hl},\ and\ \citenamefont {Esslinger}}]{gunter2005}%
  \BibitemOpen
  \bibfield  {author} {\bibinfo {author} {\bibnamefont {G\"{u}nter},
  \bibfnamefont {K}}, \bibinfo {author} {\bibfnamefont {T.}~\bibnamefont
  {St\"{o}ferle}}, \bibinfo {author} {\bibfnamefont {H.}~\bibnamefont
  {Moritz}}, \bibinfo {author} {\bibfnamefont {M.}~\bibnamefont {K\"{o}hl}}, \
  and\ \bibinfo {author} {\bibfnamefont {T.}~\bibnamefont {Esslinger}}}
  (\bibinfo {year} {2005}),\ \bibfield  {title} {\enquote {\bibinfo {title}
  {p-wave interactions in low-dimensional fermionic gases},}\ }\href@noop {}
  {\bibfield  {journal} {\bibinfo  {journal} {Phys. Rev. Lett.}\ }\textbf
  {\bibinfo {volume} {95}}~(\bibinfo {number} {23}),\ \bibinfo {pages}
  {230401}}\BibitemShut {NoStop}%
\bibitem [{\citenamefont {Hadizadeh}\ \emph {et~al.}(2011)\citenamefont
  {Hadizadeh}, \citenamefont {Yamashita}, \citenamefont {Tomio}, \citenamefont
  {Delfino},\ and\ \citenamefont {Frederico}}]{HadizadehTomio2011prl}%
  \BibitemOpen
  \bibfield  {author} {\bibinfo {author} {\bibnamefont {Hadizadeh},
  \bibfnamefont {M~R}}, \bibinfo {author} {\bibfnamefont {M.~T.}\ \bibnamefont
  {Yamashita}}, \bibinfo {author} {\bibfnamefont {Lauro}\ \bibnamefont
  {Tomio}}, \bibinfo {author} {\bibfnamefont {A.}~\bibnamefont {Delfino}}, \
  and\ \bibinfo {author} {\bibfnamefont {T.}~\bibnamefont {Frederico}}}
  (\bibinfo {year} {2011}),\ \bibfield  {title} {\enquote {\bibinfo {title}
  {Scaling properties of universal tetramers},}\ }\href {\doibase
  10.1103/PhysRevLett.107.135304} {\bibfield  {journal} {\bibinfo  {journal}
  {Phys. Rev. Lett.}\ }\textbf {\bibinfo {volume} {107}},\ \bibinfo {pages}
  {135304}}\BibitemShut {NoStop}%
\bibitem [{\citenamefont {{H{\"a}fner}}\ \emph {et~al.}(2017)\citenamefont
  {{H{\"a}fner}}, \citenamefont {{Ulmanis}}, \citenamefont {{Kuhnle}},
  \citenamefont {{Wang}}, \citenamefont {{Greene}},\ and\ \citenamefont
  {{Weidem{\"u}ller}}}]{Haefner2017arxiv}%
  \BibitemOpen
  \bibfield  {author} {\bibinfo {author} {\bibnamefont {{H{\"a}fner}},
  \bibfnamefont {S}}, \bibinfo {author} {\bibfnamefont {J.}~\bibnamefont
  {{Ulmanis}}}, \bibinfo {author} {\bibfnamefont {E.~D.}\ \bibnamefont
  {{Kuhnle}}}, \bibinfo {author} {\bibfnamefont {Y.}~\bibnamefont {{Wang}}},
  \bibinfo {author} {\bibfnamefont {C.~H.}\ \bibnamefont {{Greene}}}, \ and\
  \bibinfo {author} {\bibfnamefont {M.}~\bibnamefont {{Weidem{\"u}ller}}}}
  (\bibinfo {year} {2017}),\ \bibfield  {title} {\enquote {\bibinfo {title}
  {{Role of the intraspecies scattering length in the Efimov scenario with
  large mass difference}},}\ }\href@noop {} {\bibfield  {journal} {\bibinfo
  {journal} {ArXiv e-prints}\ }}\Eprint {http://arxiv.org/abs/1701.08007}
  {arXiv:1701.08007 [cond-mat.quant-gas]} \BibitemShut {NoStop}%
\bibitem [{\citenamefont {Hagen}\ \emph {et~al.}(2013)\citenamefont {Hagen},
  \citenamefont {Hagen}, \citenamefont {Hammer},\ and\ \citenamefont
  {Platter}}]{Hagen2013prl}%
  \BibitemOpen
  \bibfield  {author} {\bibinfo {author} {\bibnamefont {Hagen}, \bibfnamefont
  {G}}, \bibinfo {author} {\bibfnamefont {P.}~\bibnamefont {Hagen}}, \bibinfo
  {author} {\bibfnamefont {H.-W.}\ \bibnamefont {Hammer}}, \ and\ \bibinfo
  {author} {\bibfnamefont {L.}~\bibnamefont {Platter}}} (\bibinfo {year}
  {2013}),\ \bibfield  {title} {\enquote {\bibinfo {title} {Efimov physics
  around the neutron-rich $^{60}\mathrm{Ca}$ isotope},}\ }\href {\doibase
  10.1103/PhysRevLett.111.132501} {\bibfield  {journal} {\bibinfo  {journal}
  {Phys. Rev. Lett.}\ }\textbf {\bibinfo {volume} {111}},\ \bibinfo {pages}
  {132501}}\BibitemShut {NoStop}%
\bibitem [{\citenamefont {Hall}\ and\ \citenamefont
  {Willitsch}(2012)}]{Hall-2012}%
  \BibitemOpen
  \bibfield  {author} {\bibinfo {author} {\bibnamefont {Hall}, \bibfnamefont
  {F~H~J}}, \ and\ \bibinfo {author} {\bibfnamefont {S.}~\bibnamefont
  {Willitsch}}} (\bibinfo {year} {2012}),\ \bibfield  {title} {\enquote
  {\bibinfo {title} {Millikelvin reactive collisions between sympathetically
  cooled molecular ions and laser-cooled atoms in an ion-atom hybrid trap},}\
  }\href@noop {} {\bibfield  {journal} {\bibinfo  {journal} {Phys. Rev. Lett.}\
  }\textbf {\bibinfo {volume} {109}},\ \bibinfo {pages} {233202}}\BibitemShut
  {NoStop}%
\bibitem [{\citenamefont {Haller}\ \emph {et~al.}(2009)\citenamefont {Haller},
  \citenamefont {Gustavsson}, \citenamefont {Mark}, \citenamefont {Danzl},
  \citenamefont {Hart}, \citenamefont {Pupillo},\ and\ \citenamefont
  {N\"{a}gerl}}]{haller2009}%
  \BibitemOpen
  \bibfield  {author} {\bibinfo {author} {\bibnamefont {Haller}, \bibfnamefont
  {E}}, \bibinfo {author} {\bibfnamefont {M.}~\bibnamefont {Gustavsson}},
  \bibinfo {author} {\bibfnamefont {M.~J.}\ \bibnamefont {Mark}}, \bibinfo
  {author} {\bibfnamefont {J.~G.}\ \bibnamefont {Danzl}}, \bibinfo {author}
  {\bibfnamefont {R.}~\bibnamefont {Hart}}, \bibinfo {author} {\bibfnamefont
  {G.}~\bibnamefont {Pupillo}}, \ and\ \bibinfo {author} {\bibfnamefont
  {H.-C.}\ \bibnamefont {N\"{a}gerl}}} (\bibinfo {year} {2009}),\ \bibfield
  {title} {\enquote {\bibinfo {title} {Realization of an excited, strongly
  correlated quantum gas phase},}\ }\href@noop {} {\bibfield  {journal}
  {\bibinfo  {journal} {Science}\ }\textbf {\bibinfo {volume} {325}}~(\bibinfo
  {number} {5945}),\ \bibinfo {pages} {1224--1227}}\BibitemShut {NoStop}%
\bibitem [{\citenamefont {Haller}\ \emph {et~al.}(2010)\citenamefont {Haller},
  \citenamefont {Mark}, \citenamefont {Hart}, \citenamefont {Danzl},
  \citenamefont {Reichs\"{o}llner}, \citenamefont {Melezhik}, \citenamefont
  {Schmelcher},\ and\ \citenamefont {N\"{a}gerl}}]{Haller2010a}%
  \BibitemOpen
  \bibfield  {author} {\bibinfo {author} {\bibnamefont {Haller}, \bibfnamefont
  {E}}, \bibinfo {author} {\bibfnamefont {M.~J.}\ \bibnamefont {Mark}},
  \bibinfo {author} {\bibfnamefont {R.}~\bibnamefont {Hart}}, \bibinfo {author}
  {\bibfnamefont {J.~G.}\ \bibnamefont {Danzl}}, \bibinfo {author}
  {\bibfnamefont {L.}~\bibnamefont {Reichs\"{o}llner}}, \bibinfo {author}
  {\bibfnamefont {V.~S.}\ \bibnamefont {Melezhik}}, \bibinfo {author}
  {\bibfnamefont {P.}~\bibnamefont {Schmelcher}}, \ and\ \bibinfo {author}
  {\bibfnamefont {H.-C.}\ \bibnamefont {N\"{a}gerl}}} (\bibinfo {year}
  {2010}),\ \bibfield  {title} {\enquote {\bibinfo {title} {Confinement-induced
  resonances in low-dimensional quantum systems},}\ }\href@noop {} {\bibfield
  {journal} {\bibinfo  {journal} {Physical Review Letters}\ }\textbf {\bibinfo
  {volume} {104}}~(\bibinfo {number} {15}),\ \bibinfo {pages}
  {153203}}\BibitemShut {NoStop}%
\bibitem [{\citenamefont {Hammer}\ and\ \citenamefont
  {Platter}(2007)}]{hammer2007EPJAb}%
  \BibitemOpen
  \bibfield  {author} {\bibinfo {author} {\bibnamefont {Hammer}, \bibfnamefont
  {H-W}}, \ and\ \bibinfo {author} {\bibfnamefont {L.}~\bibnamefont {Platter}}}
  (\bibinfo {year} {2007}),\ \bibfield  {title} {{\selectlanguage
  {English}\enquote {\bibinfo {title} {Universal properties of the four-body
  system with large scattering length},}\ }}\href@noop {} {\bibfield  {journal}
  {\bibinfo  {journal} {Euro. Phys. J. A}\ }\textbf {\bibinfo {volume}
  {32}}~(\bibinfo {number} {1}),\ \bibinfo {pages} {113--120}}\BibitemShut
  {NoStop}%
\bibitem [{\citenamefont {Hanna}\ and\ \citenamefont
  {Blume}(2006)}]{hanna2006PRA}%
  \BibitemOpen
  \bibfield  {author} {\bibinfo {author} {\bibnamefont {Hanna}, \bibfnamefont
  {G~J}}, \ and\ \bibinfo {author} {\bibfnamefont {D.}~\bibnamefont {Blume}}}
  (\bibinfo {year} {2006}),\ \bibfield  {title} {{\selectlanguage
  {English}\enquote {\bibinfo {title} {Energetics and structural properties of
  three-dimensional bosonic clusters near threshold},}\ }}\href@noop {}
  {\bibfield  {journal} {\bibinfo  {journal} {Phys. Rev. A}\ }\textbf {\bibinfo
  {volume} {74}}~(\bibinfo {number} {6}),\ \bibinfo {pages}
  {063604}}\BibitemShut {NoStop}%
\bibitem [{\citenamefont {Hanna}\ \emph {et~al.}(2010)\citenamefont {Hanna},
  \citenamefont {Tiesinga},\ and\ \citenamefont {Julienne}}]{Hanna2010njp}%
  \BibitemOpen
  \bibfield  {author} {\bibinfo {author} {\bibnamefont {Hanna}, \bibfnamefont
  {T~M}}, \bibinfo {author} {\bibfnamefont {E.}~\bibnamefont {Tiesinga}}, \
  and\ \bibinfo {author} {\bibfnamefont {P.~S.}\ \bibnamefont {Julienne}}}
  (\bibinfo {year} {2010}),\ \bibfield  {title} {\enquote {\bibinfo {title}
  {Creation and manipulation of {F}eshbach resonances with radiofrequency
  radiation},}\ }\href@noop {} {\bibfield  {journal} {\bibinfo  {journal} {New
  J. Phys.}\ }\textbf {\bibinfo {volume} {12}}~(\bibinfo {number} {8}),\
  \bibinfo {pages} {083031}}\BibitemShut {NoStop}%
\bibitem [{\citenamefont {Hanna}\ \emph {et~al.}(2012)\citenamefont {Hanna},
  \citenamefont {Tiesinga}, \citenamefont {Mitchell},\ and\ \citenamefont
  {Julienne}}]{hanna2012pra}%
  \BibitemOpen
  \bibfield  {author} {\bibinfo {author} {\bibnamefont {Hanna}, \bibfnamefont
  {T~M}}, \bibinfo {author} {\bibfnamefont {E.}~\bibnamefont {Tiesinga}},
  \bibinfo {author} {\bibfnamefont {W.~F.}\ \bibnamefont {Mitchell}}, \ and\
  \bibinfo {author} {\bibfnamefont {P.~S.}\ \bibnamefont {Julienne}}} (\bibinfo
  {year} {2012}),\ \bibfield  {title} {\enquote {\bibinfo {title} {Resonant
  control of polar molecules in individual sites of an optical lattice},}\
  }\href@noop {} {\bibfield  {journal} {\bibinfo  {journal} {Phys. Rev. A}\
  }\textbf {\bibinfo {volume} {85}},\ \bibinfo {pages} {022703}}\BibitemShut
  {NoStop}%
\bibitem [{\citenamefont {Hara}\ \emph {et~al.}(1989)\citenamefont {Hara},
  \citenamefont {Fukuda},\ and\ \citenamefont {Ishihara}}]{HARA1989}%
  \BibitemOpen
  \bibfield  {author} {\bibinfo {author} {\bibnamefont {Hara}, \bibfnamefont
  {S}}, \bibinfo {author} {\bibfnamefont {H.}~\bibnamefont {Fukuda}}, \ and\
  \bibinfo {author} {\bibfnamefont {T.}~\bibnamefont {Ishihara}}} (\bibinfo
  {year} {1989}),\ \bibfield  {title} {\enquote {\bibinfo {title} {Application
  of hyperspheroidal coordinates to {HD}$^+$},}\ }\href@noop {} {\bibfield
  {journal} {\bibinfo  {journal} {Phys. Rev. A}\ }\textbf {\bibinfo {volume}
  {39}}~(\bibinfo {number} {1}),\ \bibinfo {pages} {35--38}}\BibitemShut
  {NoStop}%
\bibitem [{\citenamefont {Hara}\ \emph {et~al.}(1988)\citenamefont {Hara},
  \citenamefont {Fukuda}, \citenamefont {Ishihara},\ and\ \citenamefont
  {Matveenko}}]{hara1988PLA}%
  \BibitemOpen
  \bibfield  {author} {\bibinfo {author} {\bibnamefont {Hara}, \bibfnamefont
  {S}}, \bibinfo {author} {\bibfnamefont {H.}~\bibnamefont {Fukuda}}, \bibinfo
  {author} {\bibfnamefont {T.}~\bibnamefont {Ishihara}}, \ and\ \bibinfo
  {author} {\bibfnamefont {A.~V.}\ \bibnamefont {Matveenko}}} (\bibinfo {year}
  {1988}),\ \bibfield  {title} {{\selectlanguage {English}\enquote {\bibinfo
  {title} {Hyper-radial adiabatic expansion for a muonic molecule dt-$\mu$},}\
  }}\href@noop {} {\bibfield  {journal} {\bibinfo  {journal} {Phys. Lett. A}\
  }\textbf {\bibinfo {volume} {130}}~(\bibinfo {number} {1}),\ \bibinfo {pages}
  {22--25}}\BibitemShut {NoStop}%
\bibitem [{\citenamefont {Harmin}(1982{\natexlab{a}})}]{harmin1982prl}%
  \BibitemOpen
  \bibfield  {author} {\bibinfo {author} {\bibnamefont {Harmin}, \bibfnamefont
  {D~A}}} (\bibinfo {year} {1982}{\natexlab{a}}),\ \bibfield  {title} {\enquote
  {\bibinfo {title} {Theory of the nonhydrogenic {S}tark effect},}\ }\href@noop
  {} {\bibfield  {journal} {\bibinfo  {journal} {Phys. Rev. Lett.}\ }\textbf
  {\bibinfo {volume} {49}}~(\bibinfo {number} {2}),\ \bibinfo {pages}
  {128--131}}\BibitemShut {NoStop}%
\bibitem [{\citenamefont {Harmin}(1982{\natexlab{b}})}]{harmin1982}%
  \BibitemOpen
  \bibfield  {author} {\bibinfo {author} {\bibnamefont {Harmin}, \bibfnamefont
  {D~A}}} (\bibinfo {year} {1982}{\natexlab{b}}),\ \bibfield  {title} {\enquote
  {\bibinfo {title} {Theory of the {S}tark effect},}\ }\href@noop {} {\bibfield
   {journal} {\bibinfo  {journal} {Phys. Rev. A}\ }\textbf {\bibinfo {volume}
  {26}}~(\bibinfo {number} {5}),\ \bibinfo {pages} {2656--2681}}\BibitemShut
  {NoStop}%
\bibitem [{\citenamefont {H\"arter}\ and\ \citenamefont
  {Denschlag}(2014)}]{Harter-2014}%
  \BibitemOpen
  \bibfield  {author} {\bibinfo {author} {\bibnamefont {H\"arter},
  \bibfnamefont {A}}, \ and\ \bibinfo {author} {\bibfnamefont {J.~H.}\
  \bibnamefont {Denschlag}}} (\bibinfo {year} {2014}),\ \bibfield  {title}
  {\enquote {\bibinfo {title} {Cold atom-ion experiments in hybrid traps},}\
  }\href@noop {} {\bibfield  {journal} {\bibinfo  {journal} {Contemporary
  Physics}\ }\textbf {\bibinfo {volume} {55}},\ \bibinfo {pages}
  {33--45}}\BibitemShut {NoStop}%
\bibitem [{\citenamefont {H\"arter}\ \emph {et~al.}(2012)\citenamefont
  {H\"arter}, \citenamefont {Kr\"ukow}, \citenamefont {Brunner}, \citenamefont
  {Schnitzler}, \citenamefont {Schmid},\ and\ \citenamefont
  {Denschlag}}]{harter2012prl}%
  \BibitemOpen
  \bibfield  {author} {\bibinfo {author} {\bibnamefont {H\"arter},
  \bibfnamefont {A}}, \bibinfo {author} {\bibfnamefont {A.}~\bibnamefont
  {Kr\"ukow}}, \bibinfo {author} {\bibfnamefont {A.}~\bibnamefont {Brunner}},
  \bibinfo {author} {\bibfnamefont {W.}~\bibnamefont {Schnitzler}}, \bibinfo
  {author} {\bibfnamefont {S.}~\bibnamefont {Schmid}}, \ and\ \bibinfo {author}
  {\bibfnamefont {J.~H.}\ \bibnamefont {Denschlag}}} (\bibinfo {year} {2012}),\
  \bibfield  {title} {\enquote {\bibinfo {title} {Single ion as a three-body
  reaction center in an ultracold atomic gas},}\ }\href@noop {} {\bibfield
  {journal} {\bibinfo  {journal} {Phys. Rev. Lett.}\ }\textbf {\bibinfo
  {volume} {109}},\ \bibinfo {pages} {123201}}\BibitemShut {NoStop}%
\bibitem [{\citenamefont {H\"{a}rter}\ \emph {et~al.}(2013)\citenamefont
  {H\"{a}rter}, \citenamefont {Kr\"{u}kow}, \citenamefont {Dei\ss},
  \citenamefont {Drews}, \citenamefont {Tiemann},\ and\ \citenamefont
  {Denschlag}}]{Harter:NaturePhysics:2013}%
  \BibitemOpen
  \bibfield  {author} {\bibinfo {author} {\bibnamefont {H\"{a}rter},
  \bibfnamefont {A}}, \bibinfo {author} {\bibfnamefont {A.}~\bibnamefont
  {Kr\"{u}kow}}, \bibinfo {author} {\bibfnamefont {M.}~\bibnamefont {Dei\ss}},
  \bibinfo {author} {\bibfnamefont {B.}~\bibnamefont {Drews}}, \bibinfo
  {author} {\bibfnamefont {E.}~\bibnamefont {Tiemann}}, \ and\ \bibinfo
  {author} {\bibfnamefont {J.~H.}\ \bibnamefont {Denschlag}}} (\bibinfo {year}
  {2013}),\ \bibfield  {title} {\enquote {\bibinfo {title} {Population
  distribution of product states following three-body recombination in an
  ultracold atomic gas},}\ }\href@noop {} {\bibfield  {journal} {\bibinfo
  {journal} {Nat. Phys.}\ }\textbf {\bibinfo {volume} {9}}~(\bibinfo {number}
  {8}),\ \bibinfo {pages} {512}}\BibitemShut {NoStop}%
\bibitem [{\citenamefont {Helfrich}\ \emph {et~al.}(2010)\citenamefont
  {Helfrich}, \citenamefont {Hammer},\ and\ \citenamefont
  {Petrov}}]{helfrich2010PRA}%
  \BibitemOpen
  \bibfield  {author} {\bibinfo {author} {\bibnamefont {Helfrich},
  \bibfnamefont {K}}, \bibinfo {author} {\bibfnamefont {H.{-}W.}\ \bibnamefont
  {Hammer}}, \ and\ \bibinfo {author} {\bibfnamefont {D.~S.}\ \bibnamefont
  {Petrov}}} (\bibinfo {year} {2010}),\ \bibfield  {title} {{\selectlanguage
  {English}\enquote {\bibinfo {title} {Three-body problem in heteronuclear
  mixtures with resonant interspecies interaction},}\ }}\href@noop {}
  {\bibfield  {journal} {\bibinfo  {journal} {Phys. Rev. A}\ }\textbf {\bibinfo
  {volume} {81}}~(\bibinfo {number} {4}),\ \bibinfo {pages}
  {042715}}\BibitemShut {NoStop}%
\bibitem [{\citenamefont {He{\ss}}\ \emph {et~al.}(2015)\citenamefont
  {He{\ss}}, \citenamefont {Giannakeas},\ and\ \citenamefont
  {Schmelcher}}]{hess2015analytical}%
  \BibitemOpen
  \bibfield  {author} {\bibinfo {author} {\bibnamefont {He{\ss}}, \bibfnamefont
  {B}}, \bibinfo {author} {\bibfnamefont {P.}~\bibnamefont {Giannakeas}}, \
  and\ \bibinfo {author} {\bibfnamefont {P.}~\bibnamefont {Schmelcher}}}
  (\bibinfo {year} {2015}),\ \bibfield  {title} {\enquote {\bibinfo {title}
  {Analytical approach to atomic multichannel collisions in tight harmonic
  waveguides},}\ }\href@noop {} {\bibfield  {journal} {\bibinfo  {journal}
  {Phys. Rev. A}\ }\textbf {\bibinfo {volume} {92}}~(\bibinfo {number} {2}),\
  \bibinfo {pages} {022706}}\BibitemShut {NoStop}%
\bibitem [{\citenamefont {Hess}\ \emph {et~al.}(1983)\citenamefont {Hess},
  \citenamefont {Bell}, \citenamefont {Kochanski}, \citenamefont {Cline},
  \citenamefont {Kleppner},\ and\ \citenamefont {Greytak}}]{Hess-1983}%
  \BibitemOpen
  \bibfield  {author} {\bibinfo {author} {\bibnamefont {Hess}, \bibfnamefont
  {H~F}}, \bibinfo {author} {\bibfnamefont {D.~A.}\ \bibnamefont {Bell}},
  \bibinfo {author} {\bibfnamefont {G.~P.}\ \bibnamefont {Kochanski}}, \bibinfo
  {author} {\bibfnamefont {R.~W.}\ \bibnamefont {Cline}}, \bibinfo {author}
  {\bibfnamefont {D.}~\bibnamefont {Kleppner}}, \ and\ \bibinfo {author}
  {\bibfnamefont {T.~J.}\ \bibnamefont {Greytak}}} (\bibinfo {year} {1983}),\
  \bibfield  {title} {\enquote {\bibinfo {title} {Observation of three-body
  recombination in spin-polarized hydrogen},}\ }\href@noop {} {\bibfield
  {journal} {\bibinfo  {journal} {Phys. Rev. Lett.}\ }\textbf {\bibinfo
  {volume} {51}},\ \bibinfo {pages} {483}}\BibitemShut {NoStop}%
\bibitem [{\citenamefont {Hess}\ \emph {et~al.}(1984)\citenamefont {Hess},
  \citenamefont {Bell}, \citenamefont {Kochanski}, \citenamefont {Kleppner},\
  and\ \citenamefont {Greytak}}]{Hess-1984}%
  \BibitemOpen
  \bibfield  {author} {\bibinfo {author} {\bibnamefont {Hess}, \bibfnamefont
  {H~F}}, \bibinfo {author} {\bibfnamefont {D.~A.}\ \bibnamefont {Bell}},
  \bibinfo {author} {\bibfnamefont {G.~P.}\ \bibnamefont {Kochanski}}, \bibinfo
  {author} {\bibfnamefont {D.}~\bibnamefont {Kleppner}}, \ and\ \bibinfo
  {author} {\bibfnamefont {T.~J.}\ \bibnamefont {Greytak}}} (\bibinfo {year}
  {1984}),\ \bibfield  {title} {\enquote {\bibinfo {title} {Temperature and
  magnetic field dependence of three-body recombination in spin-polarized
  hydrogen},}\ }\href@noop {} {\bibfield  {journal} {\bibinfo  {journal} {Phys.
  Rev. Lett.}\ }\textbf {\bibinfo {volume} {52}},\ \bibinfo {pages}
  {1520}}\BibitemShut {NoStop}%
\bibitem [{\citenamefont {Hino}\ and\ \citenamefont
  {Macek}({1996})}]{hino1996PRL}%
  \BibitemOpen
  \bibfield  {author} {\bibinfo {author} {\bibnamefont {Hino}, \bibfnamefont
  {K~I}}, \ and\ \bibinfo {author} {\bibfnamefont {J.~H.}\ \bibnamefont
  {Macek}}} (\bibinfo {year} {{1996}}),\ \bibfield  {title} {\enquote {\bibinfo
  {title} {{Strong induced-dipole-field oscillations of the dt $\mu$ system
  above the t$\mu$(n=2) threshold}},}\ }\href@noop {} {\bibfield  {journal}
  {\bibinfo  {journal} {{Phys. Rev. Lett.}}\ }\textbf {\bibinfo {volume}
  {{77}}}~(\bibinfo {number} {{21}}),\ \bibinfo {pages}
  {{4310--4313}}}\BibitemShut {NoStop}%
\bibitem [{\citenamefont {Hirschfelder}\ \emph {et~al.}(1954)\citenamefont
  {Hirschfelder}, \citenamefont {Curtiss},\ and\ \citenamefont
  {Bird}}]{Molecular_transport}%
  \BibitemOpen
  \bibfield  {author} {\bibinfo {author} {\bibnamefont {Hirschfelder},
  \bibfnamefont {J~O}}, \bibinfo {author} {\bibfnamefont {C.~F.}\ \bibnamefont
  {Curtiss}}, \ and\ \bibinfo {author} {\bibfnamefont {R.~B.}\ \bibnamefont
  {Bird}}} (\bibinfo {year} {1954}),\ \href@noop {} {\emph {\bibinfo {title}
  {Molecular Theory of Transport and Processes in Gases}}}\ (\bibinfo
  {publisher} {Wiley},\ \bibinfo {address} {New York})\BibitemShut {NoStop}%
\bibitem [{\citenamefont {Hiyama}\ and\ \citenamefont
  {Kamimura}(2012)}]{Hiyama-2012}%
  \BibitemOpen
  \bibfield  {author} {\bibinfo {author} {\bibnamefont {Hiyama}, \bibfnamefont
  {E}}, \ and\ \bibinfo {author} {\bibfnamefont {M.}~\bibnamefont {Kamimura}}}
  (\bibinfo {year} {2012}),\ \bibfield  {title} {\enquote {\bibinfo {title}
  {Linear correlations between $^4${He} trimer and tetramer energies calculated
  with various realistic $^4${He} potentials},}\ }\href@noop {} {\bibfield
  {journal} {\bibinfo  {journal} {Phys. Rev A}\ }\textbf {\bibinfo {volume}
  {85}},\ \bibinfo {pages} {062505}}\BibitemShut {NoStop}%
\bibitem [{\citenamefont {Holzmann}\ and\ \citenamefont
  {Castin}(1999)}]{Holzmann1999}%
  \BibitemOpen
  \bibfield  {author} {\bibinfo {author} {\bibnamefont {Holzmann},
  \bibfnamefont {M}}, \ and\ \bibinfo {author} {\bibfnamefont {Y.}~\bibnamefont
  {Castin}}} (\bibinfo {year} {1999}),\ \bibfield  {title} {\enquote {\bibinfo
  {title} {Pair correlation function of an inhomogeneous interacting
  bose-einstein condensate},}\ }\href {\doibase 10.1007/s100530050586}
  {\bibfield  {journal} {\bibinfo  {journal} {The European Physical Journal D -
  Atomic, Molecular, Optical and Plasma Physics}\ }\textbf {\bibinfo {volume}
  {7}}~(\bibinfo {number} {3}),\ \bibinfo {pages} {425--432}}\BibitemShut
  {NoStop}%
\bibitem [{\citenamefont {Houbiers}\ \emph {et~al.}(1997)\citenamefont
  {Houbiers}, \citenamefont {Ferwerda}, \citenamefont {Stoof}, \citenamefont
  {McAlexander}, \citenamefont {Sackett},\ and\ \citenamefont
  {Hulet}}]{houbiers1997PRAb}%
  \BibitemOpen
  \bibfield  {author} {\bibinfo {author} {\bibnamefont {Houbiers},
  \bibfnamefont {M}}, \bibinfo {author} {\bibfnamefont {R.}~\bibnamefont
  {Ferwerda}}, \bibinfo {author} {\bibfnamefont {H.~T.~C}\ \bibnamefont
  {Stoof}}, \bibinfo {author} {\bibfnamefont {W.~I.}\ \bibnamefont
  {McAlexander}}, \bibinfo {author} {\bibfnamefont {C.~A.}\ \bibnamefont
  {Sackett}}, \ and\ \bibinfo {author} {\bibfnamefont {R.~G.}\ \bibnamefont
  {Hulet}}} (\bibinfo {year} {1997}),\ \bibfield  {title} {\enquote {\bibinfo
  {title} {Superfluid state of atomic \textsuperscript{6}{Li} in a magnetic
  trap},}\ }\href@noop {} {\bibfield  {journal} {\bibinfo  {journal} {Phys.
  Rev. A}\ }\textbf {\bibinfo {volume} {56}}~(\bibinfo {number} {6}),\ \bibinfo
  {pages} {4864--4878}}\BibitemShut {NoStop}%
\bibitem [{\citenamefont {Hove}\ \emph {et~al.}({2014})\citenamefont {Hove},
  \citenamefont {Fedorov}, \citenamefont {Fynbo}, \citenamefont {Jensen},
  \citenamefont {Riisager}, \citenamefont {Zinner},\ and\ \citenamefont
  {Garrido}}]{Hove2014prc}%
  \BibitemOpen
  \bibfield  {author} {\bibinfo {author} {\bibnamefont {Hove}, \bibfnamefont
  {D}}, \bibinfo {author} {\bibfnamefont {D.~V.}\ \bibnamefont {Fedorov}},
  \bibinfo {author} {\bibfnamefont {H.~O.~U.}\ \bibnamefont {Fynbo}}, \bibinfo
  {author} {\bibfnamefont {A.~S.}\ \bibnamefont {Jensen}}, \bibinfo {author}
  {\bibfnamefont {K.}~\bibnamefont {Riisager}}, \bibinfo {author}
  {\bibfnamefont {N.~T.}\ \bibnamefont {Zinner}}, \ and\ \bibinfo {author}
  {\bibfnamefont {E.}~\bibnamefont {Garrido}}} (\bibinfo {year} {{2014}}),\
  \bibfield  {title} {\enquote {\bibinfo {title} {{Borromean structures in
  medium-heavy nuclei}},}\ }\href {\doibase {10.1103/PhysRevC.90.064311}}
  {\bibfield  {journal} {\bibinfo  {journal} {{Physical Review C}}\ }\textbf
  {\bibinfo {volume} {{90}}}~(\bibinfo {number} {{6}}),\
  {10.1103/PhysRevC.90.064311}}\BibitemShut {NoStop}%
\bibitem [{\citenamefont {Hove}\ \emph {et~al.}({2016})\citenamefont {Hove},
  \citenamefont {Jensen}, \citenamefont {Fynbo}, \citenamefont {Zinner},
  \citenamefont {Fedorov},\ and\ \citenamefont {Garrido}}]{Hove2016prc}%
  \BibitemOpen
  \bibfield  {author} {\bibinfo {author} {\bibnamefont {Hove}, \bibfnamefont
  {D}}, \bibinfo {author} {\bibfnamefont {A.~S.}\ \bibnamefont {Jensen}},
  \bibinfo {author} {\bibfnamefont {H.~O.~U.}\ \bibnamefont {Fynbo}}, \bibinfo
  {author} {\bibfnamefont {N.~T.}\ \bibnamefont {Zinner}}, \bibinfo {author}
  {\bibfnamefont {D.~V.}\ \bibnamefont {Fedorov}}, \ and\ \bibinfo {author}
  {\bibfnamefont {E.}~\bibnamefont {Garrido}}} (\bibinfo {year} {{2016}}),\
  \bibfield  {title} {\enquote {\bibinfo {title} {{Capture reactions into
  Borromean two-proton systems at r p waiting points}},}\ }\href {\doibase
  {10.1103/PhysRevC.93.024601}} {\bibfield  {journal} {\bibinfo  {journal}
  {{Physical Review C}}\ }\textbf {\bibinfo {volume} {{93}}}~(\bibinfo {number}
  {{2}}),\ {10.1103/PhysRevC.93.024601}}\BibitemShut {NoStop}%
\bibitem [{\citenamefont {Hu}\ \emph {et~al.}(2007)\citenamefont {Hu},
  \citenamefont {Drummond},\ and\ \citenamefont {Liu}}]{hu2007NTP}%
  \BibitemOpen
  \bibfield  {author} {\bibinfo {author} {\bibnamefont {Hu}, \bibfnamefont
  {H}}, \bibinfo {author} {\bibfnamefont {P.~D.}\ \bibnamefont {Drummond}}, \
  and\ \bibinfo {author} {\bibfnamefont {X.-J.}\ \bibnamefont {Liu}}} (\bibinfo
  {year} {2007}),\ \bibfield  {title} {{\selectlanguage {English}\enquote
  {\bibinfo {title} {Universal thermodynamics of strongly interacting {F}ermi
  gases},}\ }}\href@noop {} {\bibfield  {journal} {\bibinfo  {journal} {Nat.
  Phys.}\ }\textbf {\bibinfo {volume} {3}}~(\bibinfo {number} {7}),\ \bibinfo
  {pages} {469--472}}\BibitemShut {NoStop}%
\bibitem [{\citenamefont {Hu}\ \emph {et~al.}(2006)\citenamefont {Hu},
  \citenamefont {Liu},\ and\ \citenamefont {Drummond}}]{hu2006EPL}%
  \BibitemOpen
  \bibfield  {author} {\bibinfo {author} {\bibnamefont {Hu}, \bibfnamefont
  {H}}, \bibinfo {author} {\bibfnamefont {X.-J.}\ \bibnamefont {Liu}}, \ and\
  \bibinfo {author} {\bibfnamefont {P.~D.}\ \bibnamefont {Drummond}}} (\bibinfo
  {year} {2006}),\ \bibfield  {title} {{\selectlanguage {English}\enquote
  {\bibinfo {title} {Equation of state of a superfluid {F}ermi gas in the
  {BCS}-{BEC} crossover},}\ }}\href@noop {} {\bibfield  {journal} {\bibinfo
  {journal} {Europhys. Lett.}\ }\textbf {\bibinfo {volume} {74}}~(\bibinfo
  {number} {4}),\ \bibinfo {pages} {574--580}}\BibitemShut {NoStop}%
\bibitem [{\citenamefont {Hu}\ \emph {et~al.}({2014})\citenamefont {Hu},
  \citenamefont {Bloom}, \citenamefont {Jin},\ and\ \citenamefont
  {Goldwin}}]{HuBloom2014}%
  \BibitemOpen
  \bibfield  {author} {\bibinfo {author} {\bibnamefont {Hu}, \bibfnamefont
  {Ming-Guang}}, \bibinfo {author} {\bibfnamefont {Ruth~S.}\ \bibnamefont
  {Bloom}}, \bibinfo {author} {\bibfnamefont {Deborah~S.}\ \bibnamefont {Jin}},
  \ and\ \bibinfo {author} {\bibfnamefont {Jonathan~M.}\ \bibnamefont
  {Goldwin}}} (\bibinfo {year} {{2014}}),\ \bibfield  {title} {\enquote
  {\bibinfo {title} {{Avalanche-mechanism loss at an atom-molecule Efimov
  resonance}},}\ }\href {\doibase {10.1103/PhysRevA.90.013619}} {\bibfield
  {journal} {\bibinfo  {journal} {{Physical Review A}}\ }\textbf {\bibinfo
  {volume} {{90}}}~(\bibinfo {number} {{1}}),\
  {10.1103/PhysRevA.90.013619}}\BibitemShut {NoStop}%
\bibitem [{\citenamefont {Hu}\ \emph {et~al.}(2016)\citenamefont {Hu},
  \citenamefont {Van~de Graaff}, \citenamefont {Kedar}, \citenamefont {Corson},
  \citenamefont {Cornell},\ and\ \citenamefont {Jin}}]{HuJinCornell2016prl}%
  \BibitemOpen
  \bibfield  {author} {\bibinfo {author} {\bibnamefont {Hu}, \bibfnamefont
  {Ming-Guang}}, \bibinfo {author} {\bibfnamefont {Michael~J.}\ \bibnamefont
  {Van~de Graaff}}, \bibinfo {author} {\bibfnamefont {Dhruv}\ \bibnamefont
  {Kedar}}, \bibinfo {author} {\bibfnamefont {John~P.}\ \bibnamefont {Corson}},
  \bibinfo {author} {\bibfnamefont {Eric~A.}\ \bibnamefont {Cornell}}, \ and\
  \bibinfo {author} {\bibfnamefont {Deborah~S.}\ \bibnamefont {Jin}}} (\bibinfo
  {year} {2016}),\ \bibfield  {title} {\enquote {\bibinfo {title} {Bose
  polarons in the strongly interacting regime},}\ }\href {\doibase
  10.1103/PhysRevLett.117.055301} {\bibfield  {journal} {\bibinfo  {journal}
  {Phys. Rev. Lett.}\ }\textbf {\bibinfo {volume} {117}},\ \bibinfo {pages}
  {055301}}\BibitemShut {NoStop}%
\bibitem [{\citenamefont {Huang}\ \emph
  {et~al.}(2014{\natexlab{a}})\citenamefont {Huang}, \citenamefont {O'Hara},
  \citenamefont {Grimm}, \citenamefont {Hutson},\ and\ \citenamefont
  {Petrov}}]{Huang-2014a}%
  \BibitemOpen
  \bibfield  {author} {\bibinfo {author} {\bibnamefont {Huang}, \bibfnamefont
  {B}}, \bibinfo {author} {\bibfnamefont {K.~M.}\ \bibnamefont {O'Hara}},
  \bibinfo {author} {\bibfnamefont {R.}~\bibnamefont {Grimm}}, \bibinfo
  {author} {\bibfnamefont {J.~M.}\ \bibnamefont {Hutson}}, \ and\ \bibinfo
  {author} {\bibfnamefont {D.~S.}\ \bibnamefont {Petrov}}} (\bibinfo {year}
  {2014}{\natexlab{a}}),\ \bibfield  {title} {\enquote {\bibinfo {title}
  {Three-body parameter for {E}fimov states in $^6${Li}},}\ }\href@noop {}
  {\bibfield  {journal} {\bibinfo  {journal} {Phys. Rev. A}\ }\textbf {\bibinfo
  {volume} {90}},\ \bibinfo {pages} {043636}}\BibitemShut {NoStop}%
\bibitem [{\citenamefont {Huang}\ \emph {et~al.}(2015)\citenamefont {Huang},
  \citenamefont {Sidorenkov},\ and\ \citenamefont {Grimm}}]{Huang-2015}%
  \BibitemOpen
  \bibfield  {author} {\bibinfo {author} {\bibnamefont {Huang}, \bibfnamefont
  {B}}, \bibinfo {author} {\bibfnamefont {L.~A.}\ \bibnamefont {Sidorenkov}}, \
  and\ \bibinfo {author} {\bibfnamefont {R.}~\bibnamefont {Grimm}}} (\bibinfo
  {year} {2015}),\ \href@noop {} {\enquote {\bibinfo {title}
  {Finite-temperature effects on a triatomic {E}fimov resonance in ultracold
  cesium},}\ }\BibitemShut {NoStop}%
\bibitem [{\citenamefont {Huang}\ \emph
  {et~al.}(2014{\natexlab{b}})\citenamefont {Huang}, \citenamefont
  {Sidorenkov}, \citenamefont {Grimm},\ and\ \citenamefont
  {Hutson}}]{Huang-2014b}%
  \BibitemOpen
  \bibfield  {author} {\bibinfo {author} {\bibnamefont {Huang}, \bibfnamefont
  {B}}, \bibinfo {author} {\bibfnamefont {L.~A.}\ \bibnamefont {Sidorenkov}},
  \bibinfo {author} {\bibfnamefont {R.}~\bibnamefont {Grimm}}, \ and\ \bibinfo
  {author} {\bibfnamefont {J.~M.}\ \bibnamefont {Hutson}}} (\bibinfo {year}
  {2014}{\natexlab{b}}),\ \bibfield  {title} {\enquote {\bibinfo {title}
  {Observation of the second triatomic resonance in {E}fimov's scenario},}\
  }\href@noop {} {\bibfield  {journal} {\bibinfo  {journal} {Phys. Rev. Lett.}\
  }\textbf {\bibinfo {volume} {112}},\ \bibinfo {pages} {190401}}\BibitemShut
  {NoStop}%
\bibitem [{\citenamefont {Huckans}\ \emph {et~al.}(2009)\citenamefont
  {Huckans}, \citenamefont {Williams}, \citenamefont {Hazlett}, \citenamefont
  {Sities},\ and\ \citenamefont {O'Hara}}]{Huckans-2009}%
  \BibitemOpen
  \bibfield  {author} {\bibinfo {author} {\bibnamefont {Huckans}, \bibfnamefont
  {J~H}}, \bibinfo {author} {\bibfnamefont {J.~R.}\ \bibnamefont {Williams}},
  \bibinfo {author} {\bibfnamefont {E.~L.}\ \bibnamefont {Hazlett}}, \bibinfo
  {author} {\bibfnamefont {R.~W.}\ \bibnamefont {Sities}}, \ and\ \bibinfo
  {author} {\bibfnamefont {K.~M.}\ \bibnamefont {O'Hara}}} (\bibinfo {year}
  {2009}),\ \bibfield  {title} {\enquote {\bibinfo {title} {Three-body
  recombination in a three-state fermi gas with widely tunable interactions},}\
  }\href@noop {} {\bibfield  {journal} {\bibinfo  {journal} {Phys. Rev. Lett.}\
  }\textbf {\bibinfo {volume} {102}},\ \bibinfo {pages} {165302}}\BibitemShut
  {NoStop}%
\bibitem [{\citenamefont {Hudson}\ \emph {et~al.}({2014})\citenamefont
  {Hudson}, \citenamefont {Braaten}, \citenamefont {Daekyoung},\ and\
  \citenamefont {Platter}}]{Smith2014prl}%
  \BibitemOpen
  \bibfield  {author} {\bibinfo {author} {\bibnamefont {Hudson}, \bibfnamefont
  {S~D}}, \bibinfo {author} {\bibfnamefont {E.}~\bibnamefont {Braaten}},
  \bibinfo {author} {\bibfnamefont {K.}~\bibnamefont {Daekyoung}}, \ and\
  \bibinfo {author} {\bibfnamefont {L.}~\bibnamefont {Platter}}} (\bibinfo
  {year} {{2014}}),\ \bibfield  {title} {\enquote {\bibinfo {title} {Two-body
  and three-body contacts for identical bosons near unitarity},}\ }\href@noop
  {} {\bibfield  {journal} {\bibinfo  {journal} {{Phys. Rev. Lett.}}\ }\textbf
  {\bibinfo {volume} {{112}}}~(\bibinfo {number} {{11}})}\BibitemShut {NoStop}%
\bibitem [{\citenamefont {Hui}\ \emph {et~al.}(2004)\citenamefont {Hui},
  \citenamefont {Minguzzi}, \citenamefont {Liu},\ and\ \citenamefont
  {Tosi}}]{hu2004PRL}%
  \BibitemOpen
  \bibfield  {author} {\bibinfo {author} {\bibnamefont {Hui}, \bibfnamefont
  {H}}, \bibinfo {author} {\bibfnamefont {A.}~\bibnamefont {Minguzzi}},
  \bibinfo {author} {\bibfnamefont {X.-J.}\ \bibnamefont {Liu}}, \ and\
  \bibinfo {author} {\bibfnamefont {M.~P.}\ \bibnamefont {Tosi}}} (\bibinfo
  {year} {2004}),\ \bibfield  {title} {\enquote {\bibinfo {title} {Collective
  modes and ballistic expansion of a {F}ermi gas in the {BCS}-{BEC}
  crossover},}\ }\href@noop {} {\bibfield  {journal} {\bibinfo  {journal}
  {Phys. Rev. Lett.}\ }\textbf {\bibinfo {volume} {93}}~(\bibinfo {number}
  {19}),\ \bibinfo {eid} {190403}}\BibitemShut {NoStop}%
\bibitem [{\citenamefont {Idziaszek}\ and\ \citenamefont
  {Calarco}(2005)}]{idziaszek2005}%
  \BibitemOpen
  \bibfield  {author} {\bibinfo {author} {\bibnamefont {Idziaszek},
  \bibfnamefont {Z}}, \ and\ \bibinfo {author} {\bibfnamefont {T.}~\bibnamefont
  {Calarco}}} (\bibinfo {year} {2005}),\ \bibfield  {title} {\enquote {\bibinfo
  {title} {Two atoms in an anisotropic harmonic trap},}\ }\href@noop {}
  {\bibfield  {journal} {\bibinfo  {journal} {Phys. Rev. A}\ }\textbf {\bibinfo
  {volume} {71}}~(\bibinfo {number} {5}),\ \bibinfo {pages}
  {050701}}\BibitemShut {NoStop}%
\bibitem [{\citenamefont {Idziaszek}\ and\ \citenamefont
  {Calarco}(2006)}]{idziaszek2006}%
  \BibitemOpen
  \bibfield  {author} {\bibinfo {author} {\bibnamefont {Idziaszek},
  \bibfnamefont {Z}}, \ and\ \bibinfo {author} {\bibfnamefont {T.}~\bibnamefont
  {Calarco}}} (\bibinfo {year} {2006}),\ \bibfield  {title} {\enquote {\bibinfo
  {title} {Analytical solutions for the dynamics of two trapped interacting
  ultracold atoms},}\ }\href@noop {} {\bibfield  {journal} {\bibinfo  {journal}
  {Phys. Rev. A}\ }\textbf {\bibinfo {volume} {74}}~(\bibinfo {number} {2}),\
  \bibinfo {pages} {022712}}\BibitemShut {NoStop}%
\bibitem [{\citenamefont {Imambekov}\ \emph {et~al.}(2012)\citenamefont
  {Imambekov}, \citenamefont {Schmidt},\ and\ \citenamefont
  {Glazman}}]{imambekov2012one}%
  \BibitemOpen
  \bibfield  {author} {\bibinfo {author} {\bibnamefont {Imambekov},
  \bibfnamefont {A}}, \bibinfo {author} {\bibfnamefont {T.~L.}\ \bibnamefont
  {Schmidt}}, \ and\ \bibinfo {author} {\bibfnamefont {L.~I.}\ \bibnamefont
  {Glazman}}} (\bibinfo {year} {2012}),\ \bibfield  {title} {\enquote {\bibinfo
  {title} {One-dimensional quantum liquids: {B}eyond the {L}uttinger liquid
  paradigm},}\ }\href@noop {} {\bibfield  {journal} {\bibinfo  {journal} {Rev.
  Mod. Phys.}\ }\textbf {\bibinfo {volume} {84}}~(\bibinfo {number} {3}),\
  \bibinfo {pages} {1253}}\BibitemShut {NoStop}%
\bibitem [{\citenamefont {Inouye}\ \emph {et~al.}(1998)\citenamefont {Inouye},
  \citenamefont {Andrews}, \citenamefont {Stenger}, \citenamefont {Miesner},
  \citenamefont {Stamper-Kurn},\ and\ \citenamefont {Ketterle}}]{inouye1998}%
  \BibitemOpen
  \bibfield  {author} {\bibinfo {author} {\bibnamefont {Inouye}, \bibfnamefont
  {S}}, \bibinfo {author} {\bibfnamefont {M.~R.}\ \bibnamefont {Andrews}},
  \bibinfo {author} {\bibfnamefont {J.}~\bibnamefont {Stenger}}, \bibinfo
  {author} {\bibfnamefont {H.{-}J.}\ \bibnamefont {Miesner}}, \bibinfo {author}
  {\bibfnamefont {D.~M.}\ \bibnamefont {Stamper-Kurn}}, \ and\ \bibinfo
  {author} {\bibfnamefont {W.}~\bibnamefont {Ketterle}}} (\bibinfo {year}
  {1998}),\ \bibfield  {title} {\enquote {\bibinfo {title} {Observation of
  {F}eshbach resonances in a {B}ose-{E}instein condensate},}\ }\href@noop {}
  {\bibfield  {journal} {\bibinfo  {journal} {Nature (London)}\ }\textbf
  {\bibinfo {volume} {392}}~(\bibinfo {number} {6672}),\ \bibinfo {pages}
  {151--154}}\BibitemShut {NoStop}%
\bibitem [{\citenamefont {Jagutzki}(2002)}]{Jagutzki-2002}%
  \BibitemOpen
  \bibfield  {author} {\bibinfo {author} {\bibnamefont {Jagutzki},
  \bibfnamefont {O~{et al}}}} (\bibinfo {year} {2002}),\ \bibfield  {title}
  {\enquote {\bibinfo {title} {Multiple it readout of a microchannel plate
  detector with a three-layer delay-line anode},}\ }\href@noop {} {\bibfield
  {journal} {\bibinfo  {journal} {Nucl. Sci. IEEE Trans.}\ }\textbf {\bibinfo
  {volume} {49}},\ \bibinfo {pages} {2477--2483}}\BibitemShut {NoStop}%
\bibitem [{\citenamefont {Jensen}\ \emph
  {et~al.}(2004{\natexlab{a}})\citenamefont {Jensen}, \citenamefont {Riisager},
  \citenamefont {Fedorov},\ and\ \citenamefont {Garrido}}]{Jensen-2004}%
  \BibitemOpen
  \bibfield  {author} {\bibinfo {author} {\bibnamefont {Jensen}, \bibfnamefont
  {A~S}}, \bibinfo {author} {\bibfnamefont {D.}~\bibnamefont {Riisager}},
  \bibinfo {author} {\bibfnamefont {D.~V.}\ \bibnamefont {Fedorov}}, \ and\
  \bibinfo {author} {\bibfnamefont {E.}~\bibnamefont {Garrido}}} (\bibinfo
  {year} {2004}{\natexlab{a}}),\ \bibfield  {title} {\enquote {\bibinfo {title}
  {Structure and reactions of quantum halos},}\ }\href@noop {} {\bibfield
  {journal} {\bibinfo  {journal} {Rev. Mod. Phys.}\ }\textbf {\bibinfo {volume}
  {76}},\ \bibinfo {pages} {215}}\BibitemShut {NoStop}%
\bibitem [{\citenamefont {Jensen}\ \emph
  {et~al.}(2004{\natexlab{b}})\citenamefont {Jensen}, \citenamefont {Riisager},
  \citenamefont {Fedorov},\ and\ \citenamefont {Garrido}}]{jensen2004RMP}%
  \BibitemOpen
  \bibfield  {author} {\bibinfo {author} {\bibnamefont {Jensen}, \bibfnamefont
  {AS}}, \bibinfo {author} {\bibfnamefont {K}~\bibnamefont {Riisager}},
  \bibinfo {author} {\bibfnamefont {DV}~\bibnamefont {Fedorov}}, \ and\
  \bibinfo {author} {\bibfnamefont {E}~\bibnamefont {Garrido}}} (\bibinfo
  {year} {2004}{\natexlab{b}}),\ \bibfield  {title} {{\selectlanguage
  {English}\enquote {\bibinfo {title} {Structure and reactions of quantum
  halos},}\ }}\href@noop {} {\bibfield  {journal} {\bibinfo  {journal} {RMP}\
  }\textbf {\bibinfo {volume} {76}}~(\bibinfo {number} {1}),\ \bibinfo {pages}
  {215--261}}\BibitemShut {NoStop}%
\bibitem [{\citenamefont {Jeon}\ \emph {et~al.}(2007)\citenamefont {Jeon},
  \citenamefont {Chang},\ and\ \citenamefont {Jain}}]{Jeon2007}%
  \BibitemOpen
  \bibfield  {author} {\bibinfo {author} {\bibnamefont {Jeon}, \bibfnamefont
  {Gun~Sang}}, \bibinfo {author} {\bibfnamefont {Chia-Chen}\ \bibnamefont
  {Chang}}, \ and\ \bibinfo {author} {\bibfnamefont {Jainendra~K.}\
  \bibnamefont {Jain}}} (\bibinfo {year} {2007}),\ \bibfield  {title} {\enquote
  {\bibinfo {title} {Semiconductor quantum dots in high magnetic fields},}\
  }\href {\doibase 10.1140/epjb/e2007-00060-4} {\bibfield  {journal} {\bibinfo
  {journal} {The European Physical Journal B}\ }\textbf {\bibinfo {volume}
  {55}}~(\bibinfo {number} {3}),\ \bibinfo {pages} {271--282}}\BibitemShut
  {NoStop}%
\bibitem [{\citenamefont {Ji}\ \emph {et~al.}(2015)\citenamefont {Ji},
  \citenamefont {Braaten}, \citenamefont {Phillips},\ and\ \citenamefont
  {Platter}}]{platterpra2015}%
  \BibitemOpen
  \bibfield  {author} {\bibinfo {author} {\bibnamefont {Ji}, \bibfnamefont
  {C}}, \bibinfo {author} {\bibfnamefont {E.}~\bibnamefont {Braaten}}, \bibinfo
  {author} {\bibfnamefont {D.~R.}\ \bibnamefont {Phillips}}, \ and\ \bibinfo
  {author} {\bibfnamefont {L.}~\bibnamefont {Platter}}} (\bibinfo {year}
  {2015}),\ \bibfield  {title} {\enquote {\bibinfo {title} {Universal relations
  for range corrections to {E}fimov features},}\ }\href@noop {} {\bibfield
  {journal} {\bibinfo  {journal} {Phys. Rev. A}\ }\textbf {\bibinfo {volume}
  {92}},\ \bibinfo {pages} {030702}}\BibitemShut {NoStop}%
\bibitem [{\citenamefont {Jiang}\ \emph {et~al.}({2014})\citenamefont {Jiang},
  \citenamefont {Liu}, \citenamefont {Semenoff},\ and\ \citenamefont
  {Zhou}}]{JiangZhou2014pra}%
  \BibitemOpen
  \bibfield  {author} {\bibinfo {author} {\bibnamefont {Jiang}, \bibfnamefont
  {S-J}}, \bibinfo {author} {\bibfnamefont {W.-M.}\ \bibnamefont {Liu}},
  \bibinfo {author} {\bibfnamefont {G.~W.}\ \bibnamefont {Semenoff}}, \ and\
  \bibinfo {author} {\bibfnamefont {F.}~\bibnamefont {Zhou}}} (\bibinfo {year}
  {{2014}}),\ \bibfield  {title} {\enquote {\bibinfo {title} {{Universal Bose
  gases near resonance: A rigorous solution}},}\ }\href@noop {} {\bibfield
  {journal} {\bibinfo  {journal} {{Phys. Rev. A}}\ }\textbf {\bibinfo {volume}
  {{89}}}~(\bibinfo {number} {{3}})}\BibitemShut {NoStop}%
\bibitem [{\citenamefont {Jiang}\ \emph {et~al.}({2016})\citenamefont {Jiang},
  \citenamefont {Maki},\ and\ \citenamefont {Zhou}}]{Jiang2016pra}%
  \BibitemOpen
  \bibfield  {author} {\bibinfo {author} {\bibnamefont {Jiang}, \bibfnamefont
  {S-J}}, \bibinfo {author} {\bibfnamefont {J.}~\bibnamefont {Maki}}, \ and\
  \bibinfo {author} {\bibfnamefont {F.}~\bibnamefont {Zhou}}} (\bibinfo {year}
  {{2016}}),\ \bibfield  {title} {\enquote {\bibinfo {title} {{Long-lived
  universal resonant Bose gases}},}\ }\href@noop {} {\bibfield  {journal}
  {\bibinfo  {journal} {{Phys. Rev. A}}\ }\textbf {\bibinfo {volume}
  {{93}}}~(\bibinfo {number} {{4}})}\BibitemShut {NoStop}%
\bibitem [{\citenamefont {{Johansen}}\ \emph {et~al.}(2016)\citenamefont
  {{Johansen}}, \citenamefont {{DeSalvo}}, \citenamefont {{Patel}},\ and\
  \citenamefont {{Chin}}}]{JohansenChin2016arxiv}%
  \BibitemOpen
  \bibfield  {author} {\bibinfo {author} {\bibnamefont {{Johansen}},
  \bibfnamefont {J}}, \bibinfo {author} {\bibfnamefont {B.~J.}\ \bibnamefont
  {{DeSalvo}}}, \bibinfo {author} {\bibfnamefont {K.}~\bibnamefont {{Patel}}},
  \ and\ \bibinfo {author} {\bibfnamefont {C.}~\bibnamefont {{Chin}}}}
  (\bibinfo {year} {2016}),\ \bibfield  {title} {\enquote {\bibinfo {title}
  {{Testing universality of Efimov physics across broad and narrow Feshbach
  resonances}},}\ }\href@noop {} {\bibfield  {journal} {\bibinfo  {journal}
  {ArXiv e-prints}\ }}\Eprint {http://arxiv.org/abs/1612.05169}
  {arXiv:1612.05169 [cond-mat.quant-gas]} \BibitemShut {NoStop}%
\bibitem [{\citenamefont {{Johnson}}(1973)}]{Johnson1973jcp}%
  \BibitemOpen
  \bibfield  {author} {\bibinfo {author} {\bibnamefont {{Johnson}},
  \bibfnamefont {B~R}}} (\bibinfo {year} {1973}),\ \bibfield  {title} {\enquote
  {\bibinfo {title} {{The Multichannel Log-Derivative Method for Scattering
  Calculations}},}\ }\href {\doibase 10.1016/0021-9991(73)90049-1} {\bibfield
  {journal} {\bibinfo  {journal} {Journal of Computational Physics}\ }\textbf
  {\bibinfo {volume} {13}},\ \bibinfo {pages} {445--449}}\BibitemShut {NoStop}%
\bibitem [{\citenamefont {Johnson}(1983)}]{johnson1983JCP}%
  \BibitemOpen
  \bibfield  {author} {\bibinfo {author} {\bibnamefont {Johnson}, \bibfnamefont
  {BR}}} (\bibinfo {year} {1983}),\ \bibfield  {title} {{\selectlanguage
  {English}\enquote {\bibinfo {title} {The quantum dynamics of 3 particles in
  hyperspherical coordinates},}\ }}\href@noop {} {\bibfield  {journal}
  {\bibinfo  {journal} {J. Chem. Phys.}\ }\textbf {\bibinfo {volume}
  {79}}~(\bibinfo {number} {4}),\ \bibinfo {pages} {1916--1925}}\BibitemShut
  {NoStop}%
\bibitem [{\citenamefont {J\o{}rgensen}\ \emph {et~al.}(2016)\citenamefont
  {J\o{}rgensen}, \citenamefont {Wacker}, \citenamefont {Skalmstang},
  \citenamefont {Parish}, \citenamefont {Levinsen}, \citenamefont
  {Christensen}, \citenamefont {Bruun},\ and\ \citenamefont
  {Arlt}}]{JorgensenParishArlt2016prl}%
  \BibitemOpen
  \bibfield  {author} {\bibinfo {author} {\bibnamefont {J\o{}rgensen},
  \bibfnamefont {Nils~B}}, \bibinfo {author} {\bibfnamefont {Lars}\
  \bibnamefont {Wacker}}, \bibinfo {author} {\bibfnamefont {Kristoffer~T.}\
  \bibnamefont {Skalmstang}}, \bibinfo {author} {\bibfnamefont {Meera~M.}\
  \bibnamefont {Parish}}, \bibinfo {author} {\bibfnamefont {Jesper}\
  \bibnamefont {Levinsen}}, \bibinfo {author} {\bibfnamefont {Rasmus~S.}\
  \bibnamefont {Christensen}}, \bibinfo {author} {\bibfnamefont {Georg~M.}\
  \bibnamefont {Bruun}}, \ and\ \bibinfo {author} {\bibfnamefont {Jan~J.}\
  \bibnamefont {Arlt}}} (\bibinfo {year} {2016}),\ \bibfield  {title} {\enquote
  {\bibinfo {title} {Observation of attractive and repulsive polarons in a
  bose-einstein condensate},}\ }\href {\doibase 10.1103/PhysRevLett.117.055302}
  {\bibfield  {journal} {\bibinfo  {journal} {Phys. Rev. Lett.}\ }\textbf
  {\bibinfo {volume} {117}},\ \bibinfo {pages} {055302}}\BibitemShut {NoStop}%
\bibitem [{\citenamefont {Jraij}\ \emph {et~al.}(2003)\citenamefont {Jraij},
  \citenamefont {Allouche}, \citenamefont {Korek},\ and\ \citenamefont
  {Aubert-Fr\'{e}con}}]{Jraij-2003}%
  \BibitemOpen
  \bibfield  {author} {\bibinfo {author} {\bibnamefont {Jraij}, \bibfnamefont
  {A}}, \bibinfo {author} {\bibfnamefont {A.~R.}\ \bibnamefont {Allouche}},
  \bibinfo {author} {\bibfnamefont {M.}~\bibnamefont {Korek}}, \ and\ \bibinfo
  {author} {\bibfnamefont {M.}~\bibnamefont {Aubert-Fr\'{e}con}}} (\bibinfo
  {year} {2003}),\ \bibfield  {title} {\enquote {\bibinfo {title} {Theoretical
  electronic structure of the alkali-dimer cation {Rb}$_2^+$},}\ }\href@noop {}
  {\bibfield  {journal} {\bibinfo  {journal} {Chem. Phys.}\ }\textbf {\bibinfo
  {volume} {290}},\ \bibinfo {pages} {129}}\BibitemShut {NoStop}%
\bibitem [{\citenamefont {Kadomtsev}\ \emph {et~al.}(1987)\citenamefont
  {Kadomtsev}, \citenamefont {Vinitsky},\ and\ \citenamefont
  {Yukajlovic}}]{kadomtsev1987PRA}%
  \BibitemOpen
  \bibfield  {author} {\bibinfo {author} {\bibnamefont {Kadomtsev},
  \bibfnamefont {MB}}, \bibinfo {author} {\bibfnamefont {SI}~\bibnamefont
  {Vinitsky}}, \ and\ \bibinfo {author} {\bibfnamefont {FR}~\bibnamefont
  {Yukajlovic}}} (\bibinfo {year} {1987}),\ \bibfield  {title}
  {{\selectlanguage {English}\enquote {\bibinfo {title} {Adiabatic
  representation for the 3-body problem in the limit of separated atoms in
  appropriate coordinates},}\ }}\href@noop {} {\bibfield  {journal} {\bibinfo
  {journal} {Phys. Rev. A}\ }\textbf {\bibinfo {volume} {36}}~(\bibinfo
  {number} {10}),\ \bibinfo {pages} {4652--4661}}\BibitemShut {NoStop}%
\bibitem [{\citenamefont {Kadyrov}\ \emph {et~al.}(2009)\citenamefont
  {Kadyrov}, \citenamefont {Bray}, \citenamefont {Mukhamedzhanov},\ and\
  \citenamefont {Stelbovics}}]{Kadyrov2009}%
  \BibitemOpen
  \bibfield  {author} {\bibinfo {author} {\bibnamefont {Kadyrov}, \bibfnamefont
  {A~S}}, \bibinfo {author} {\bibfnamefont {I.}~\bibnamefont {Bray}}, \bibinfo
  {author} {\bibfnamefont {A.~M.}\ \bibnamefont {Mukhamedzhanov}}, \ and\
  \bibinfo {author} {\bibfnamefont {A.~T.}\ \bibnamefont {Stelbovics}}}
  (\bibinfo {year} {2009}),\ \bibfield  {title} {\enquote {\bibinfo {title}
  {Surface-integral formulation of scattering theory},}\ }\href@noop {}
  {\bibfield  {journal} {\bibinfo  {journal} {Ann. Phys.}\ }\textbf {\bibinfo
  {volume} {324}}~(\bibinfo {number} {7}),\ \bibinfo {pages}
  {1516--1546}}\BibitemShut {NoStop}%
\bibitem [{\citenamefont {Kagan}\ \emph {et~al.}(1985)\citenamefont {Kagan},
  \citenamefont {Svistunov},\ and\ \citenamefont
  {Shlyapnikov}}]{kagan1985JETPL}%
  \BibitemOpen
  \bibfield  {author} {\bibinfo {author} {\bibnamefont {Kagan}, \bibfnamefont
  {Y}}, \bibinfo {author} {\bibfnamefont {B.~V.}\ \bibnamefont {Svistunov}}, \
  and\ \bibinfo {author} {\bibfnamefont {G.~V.}\ \bibnamefont {Shlyapnikov}}}
  (\bibinfo {year} {1985}),\ \bibfield  {title} {\enquote {\bibinfo {title}
  {Effect of {B}ose condensation on inelastic processes in gases},}\
  }\href@noop {} {\bibfield  {journal} {\bibinfo  {journal} {JETPL}\ }\textbf
  {\bibinfo {volume} {42}}~(\bibinfo {number} {4}),\ \bibinfo {pages}
  {209--212}}\BibitemShut {NoStop}%
\bibitem [{\citenamefont {Kalas}\ and\ \citenamefont
  {Blume}(2006)}]{kalas_interaction-induced_2006}%
  \BibitemOpen
  \bibfield  {author} {\bibinfo {author} {\bibnamefont {Kalas}, \bibfnamefont
  {R~M}}, \ and\ \bibinfo {author} {\bibfnamefont {D.}~\bibnamefont {Blume}}}
  (\bibinfo {year} {2006}),\ \bibfield  {title} {\enquote {\bibinfo {title}
  {Interaction-induced localization of an impurity in a trapped
  {Bose}-{Einstein} condensate},}\ }\href {\doibase 10.1103/PhysRevA.73.043608}
  {\bibfield  {journal} {\bibinfo  {journal} {Physical Review A}\ }\textbf
  {\bibinfo {volume} {73}}~(\bibinfo {number} {4}),\ \bibinfo {pages}
  {043608}}\BibitemShut {NoStop}%
\bibitem [{\citenamefont {Kanjilal}\ and\ \citenamefont
  {Blume}(2004)}]{kanjilalpra2004}%
  \BibitemOpen
  \bibfield  {author} {\bibinfo {author} {\bibnamefont {Kanjilal},
  \bibfnamefont {K}}, \ and\ \bibinfo {author} {\bibfnamefont {D.}~\bibnamefont
  {Blume}}} (\bibinfo {year} {2004}),\ \bibfield  {title} {\enquote {\bibinfo
  {title} {Nondivergent pseudopotential treatment of spin-polarized fermions
  under one- and three-dimensional harmonic confinement},}\ }\href@noop {}
  {\bibfield  {journal} {\bibinfo  {journal} {Phys. Rev. A}\ }\textbf {\bibinfo
  {volume} {70}},\ \bibinfo {pages} {042709}}\BibitemShut {NoStop}%
\bibitem [{\citenamefont {Kanjilal}\ and\ \citenamefont
  {Blume}({2006})}]{KanjilalBlume2006pra}%
  \BibitemOpen
  \bibfield  {author} {\bibinfo {author} {\bibnamefont {Kanjilal},
  \bibfnamefont {K}}, \ and\ \bibinfo {author} {\bibfnamefont {D}~\bibnamefont
  {Blume}}} (\bibinfo {year} {{2006}}),\ \bibfield  {title} {\enquote {\bibinfo
  {title} {{Coupled-channel pseudopotential description of the Feshbach
  resonance in two dimensions}},}\ }\href {\doibase
  {10.1103/PhysRevA.73.060701}} {\bibfield  {journal} {\bibinfo  {journal}
  {{PHYSICAL REVIEW A}}\ }\textbf {\bibinfo {volume} {{73}}}~(\bibinfo {number}
  {{6}}),\ {10.1103/PhysRevA.73.060701}}\BibitemShut {NoStop}%
\bibitem [{\citenamefont {Kanjilal}\ \emph {et~al.}({2007})\citenamefont
  {Kanjilal}, \citenamefont {Bohn},\ and\ \citenamefont
  {Blume}}]{KanjilalBohnBlume2007pra}%
  \BibitemOpen
  \bibfield  {author} {\bibinfo {author} {\bibnamefont {Kanjilal},
  \bibfnamefont {K}}, \bibinfo {author} {\bibfnamefont {John~L.}\ \bibnamefont
  {Bohn}}, \ and\ \bibinfo {author} {\bibfnamefont {D.}~\bibnamefont {Blume}}}
  (\bibinfo {year} {{2007}}),\ \bibfield  {title} {\enquote {\bibinfo {title}
  {{Pseudopotential treatment of two aligned dipoles under external harmonic
  confinement}},}\ }\href {\doibase {10.1103/PhysRevA.75.052705}} {\bibfield
  {journal} {\bibinfo  {journal} {{Physical Review A}}\ }\textbf {\bibinfo
  {volume} {{75}}}~(\bibinfo {number} {{5}}),\
  {10.1103/PhysRevA.75.052705}}\BibitemShut {NoStop}%
\bibitem [{\citenamefont {Karplus}\ \emph {et~al.}(1965)\citenamefont
  {Karplus}, \citenamefont {Porter},\ and\ \citenamefont
  {Sharma}}]{Karplus-1965}%
  \BibitemOpen
  \bibfield  {author} {\bibinfo {author} {\bibnamefont {Karplus}, \bibfnamefont
  {M}}, \bibinfo {author} {\bibfnamefont {R.~N.}\ \bibnamefont {Porter}}, \
  and\ \bibinfo {author} {\bibfnamefont {R.~D.}\ \bibnamefont {Sharma}}}
  (\bibinfo {year} {1965}),\ \bibfield  {title} {\enquote {\bibinfo {title}
  {Exchange reactions with activation energy. {I}. {S}imple barrier potential
  for {(H, H2)}},}\ }\href@noop {} {\bibfield  {journal} {\bibinfo  {journal}
  {J. Chem. Phys.}\ }\textbf {\bibinfo {volume} {43}},\ \bibinfo {pages}
  {3259}}\BibitemShut {NoStop}%
\bibitem [{\citenamefont {Kartavtsev}\ and\ \citenamefont
  {Macek}(2002)}]{kartavtsev2002FBS}%
  \BibitemOpen
  \bibfield  {author} {\bibinfo {author} {\bibnamefont {Kartavtsev},
  \bibfnamefont {O~I}}, \ and\ \bibinfo {author} {\bibfnamefont {J.~H.}\
  \bibnamefont {Macek}}} (\bibinfo {year} {2002}),\ \bibfield  {title}
  {{\selectlanguage {English}\enquote {\bibinfo {title} {Low-energy three-body
  recombination near a {F}eshbach resonance},}\ }}\href@noop {} {\bibfield
  {journal} {\bibinfo  {journal} {Few-Body Systems}\ }\textbf {\bibinfo
  {volume} {31}}~(\bibinfo {number} {2-4}),\ \bibinfo {pages}
  {249--254}}\BibitemShut {NoStop}%
\bibitem [{\citenamefont {Kartavtsev}\ and\ \citenamefont
  {Malykh}(2007)}]{kartavtsev_low-energy_2007}%
  \BibitemOpen
  \bibfield  {author} {\bibinfo {author} {\bibnamefont {Kartavtsev},
  \bibfnamefont {O~I}}, \ and\ \bibinfo {author} {\bibfnamefont {A.~V.}\
  \bibnamefont {Malykh}}} (\bibinfo {year} {2007}),\ \bibfield  {title}
  {\enquote {\bibinfo {title} {Low-energy three-body dynamics in binary quantum
  gases},}\ }\href {\doibase 10.1088/0953-4075/40/7/011} {\bibfield  {journal}
  {\bibinfo  {journal} {Journal of Physics B-Atomic Molecular and Optical
  Physics}\ }\textbf {\bibinfo {volume} {40}}~(\bibinfo {number} {7}),\
  \bibinfo {pages} {1429--1441}}\BibitemShut {NoStop}%
\bibitem [{\citenamefont {Kartavtsev}\ and\ \citenamefont
  {Malykh}(2014)}]{kartavtsev_recent_2014}%
  \BibitemOpen
  \bibfield  {author} {\bibinfo {author} {\bibnamefont {Kartavtsev},
  \bibfnamefont {O~I}}, \ and\ \bibinfo {author} {\bibfnamefont {A.~V.}\
  \bibnamefont {Malykh}}} (\bibinfo {year} {2014}),\ \bibfield  {title}
  {\enquote {\bibinfo {title} {Recent advances in description of few
  two-component fermions},}\ }\href {\doibase 10.1134/S1063778814030120}
  {\bibfield  {journal} {\bibinfo  {journal} {Physics of Atomic Nuclei}\
  }\textbf {\bibinfo {volume} {77}}~(\bibinfo {number} {4}),\ \bibinfo {pages}
  {430--437}}\BibitemShut {NoStop}%
\bibitem [{\citenamefont {Kato}\ and\ \citenamefont
  {Watanabe}(1995)}]{KATO1995}%
  \BibitemOpen
  \bibfield  {author} {\bibinfo {author} {\bibnamefont {Kato}, \bibfnamefont
  {D}}, \ and\ \bibinfo {author} {\bibfnamefont {S.}~\bibnamefont {Watanabe}}}
  (\bibinfo {year} {1995}),\ \bibfield  {title} {\enquote {\bibinfo {title}
  {2-electron correlations in e+{H}$^-$ $\rightarrow$ e+e+p near-threshold},}\
  }\href@noop {} {\bibfield  {journal} {\bibinfo  {journal} {Phys. Rev. Lett.}\
  }\textbf {\bibinfo {volume} {74}}~(\bibinfo {number} {13}),\ \bibinfo {pages}
  {2443--2446}}\BibitemShut {NoStop}%
\bibitem [{\citenamefont {Kato}\ and\ \citenamefont
  {Watanabe}(1997)}]{kato1997PRA}%
  \BibitemOpen
  \bibfield  {author} {\bibinfo {author} {\bibnamefont {Kato}, \bibfnamefont
  {D}}, \ and\ \bibinfo {author} {\bibfnamefont {S}~\bibnamefont {Watanabe}}}
  (\bibinfo {year} {1997}),\ \bibfield  {title} {{\selectlanguage
  {English}\enquote {\bibinfo {title} {Ab initio study of interelectronic
  correlations in electron-impact ionization of hydrogen},}\ }}\href@noop {}
  {\bibfield  {journal} {\bibinfo  {journal} {Phys. Rev. A}\ }\textbf {\bibinfo
  {volume} {56}}~(\bibinfo {number} {5}),\ \bibinfo {pages}
  {3687--3700}}\BibitemShut {NoStop}%
\bibitem [{\citenamefont {Kazansky}\ \emph {et~al.}({2003})\citenamefont
  {Kazansky}, \citenamefont {Selles},\ and\ \citenamefont
  {Malegat}}]{malegat2003PRA}%
  \BibitemOpen
  \bibfield  {author} {\bibinfo {author} {\bibnamefont {Kazansky},
  \bibfnamefont {A~K}}, \bibinfo {author} {\bibfnamefont {P.}~\bibnamefont
  {Selles}}, \ and\ \bibinfo {author} {\bibfnamefont {L}~\bibnamefont
  {Malegat}}} (\bibinfo {year} {{2003}}),\ \bibfield  {title} {\enquote
  {\bibinfo {title} {{Hyperspherical time-dependent method with semiclassical
  outgoing waves for double photoionization of helium}},}\ }\href@noop {}
  {\bibfield  {journal} {\bibinfo  {journal} {Phys. Rev. A}\ }\textbf {\bibinfo
  {volume} {{68}}}~(\bibinfo {number} {{5}})}\BibitemShut {NoStop}%
\bibitem [{\citenamefont {Keck}(1960)}]{Keck-1960}%
  \BibitemOpen
  \bibfield  {author} {\bibinfo {author} {\bibnamefont {Keck}, \bibfnamefont
  {J~C}}} (\bibinfo {year} {1960}),\ \bibfield  {title} {\enquote {\bibinfo
  {title} {Variational theory of chemical reaction rates applied to three-body
  recombination},}\ }\href@noop {} {\bibfield  {journal} {\bibinfo  {journal}
  {J. Chem. Phys}\ }\textbf {\bibinfo {volume} {32}},\ \bibinfo {pages}
  {1035}}\BibitemShut {NoStop}%
\bibitem [{\citenamefont {Keck}(1967)}]{Keck-1967}%
  \BibitemOpen
  \bibfield  {author} {\bibinfo {author} {\bibnamefont {Keck}, \bibfnamefont
  {J~C}}} (\bibinfo {year} {1967}),\ \bibfield  {title} {\enquote {\bibinfo
  {title} {Variational theory of reaction rates},}\ }\href@noop {} {\bibfield
  {journal} {\bibinfo  {journal} {Adv. Chem. Phys.}\ }\textbf {\bibinfo
  {volume} {13}},\ \bibinfo {pages} {85}}\BibitemShut {NoStop}%
\bibitem [{\citenamefont {Kendrick}\ \emph {et~al.}(1999)\citenamefont
  {Kendrick}, \citenamefont {Pack}, \citenamefont {Walker},\ and\ \citenamefont
  {Hayes}}]{kendrick1999JCP}%
  \BibitemOpen
  \bibfield  {author} {\bibinfo {author} {\bibnamefont {Kendrick},
  \bibfnamefont {B~K}}, \bibinfo {author} {\bibfnamefont {R.~T.}\ \bibnamefont
  {Pack}}, \bibinfo {author} {\bibfnamefont {R.~B.}\ \bibnamefont {Walker}}, \
  and\ \bibinfo {author} {\bibfnamefont {E.~F.}\ \bibnamefont {Hayes}}}
  (\bibinfo {year} {1999}),\ \bibfield  {title} {\enquote {\bibinfo {title}
  {Hyperspherical surface functions for nonzero total angular momentum. {I}.
  {E}ckart singularities},}\ }\href@noop {} {\bibfield  {journal} {\bibinfo
  {journal} {J. Chem. Phys.}\ }\textbf {\bibinfo {volume} {110}}~(\bibinfo
  {number} {14}),\ \bibinfo {pages} {6673--6693}}\BibitemShut {NoStop}%
\bibitem [{\citenamefont {Kestner}\ and\ \citenamefont
  {Duan}(2007)}]{kestner2007PRA}%
  \BibitemOpen
  \bibfield  {author} {\bibinfo {author} {\bibnamefont {Kestner}, \bibfnamefont
  {J~P}}, \ and\ \bibinfo {author} {\bibfnamefont {L-M.}\ \bibnamefont {Duan}}}
  (\bibinfo {year} {2007}),\ \bibfield  {title} {{\selectlanguage
  {English}\enquote {\bibinfo {title} {Level crossing in the three-body problem
  for strongly interacting {F}ermions in a harmonic trap},}\ }}\href@noop {}
  {\bibfield  {journal} {\bibinfo  {journal} {Phys. Rev. A}\ }\textbf {\bibinfo
  {volume} {76}}~(\bibinfo {number} {3}),\ \bibinfo {pages}
  {033611}}\BibitemShut {NoStop}%
\bibitem [{\citenamefont {Ketterle}\ and\ \citenamefont
  {Zwierlein}(2008)}]{ketterle2008RIVISTADELNUOVOCIMENTO}%
  \BibitemOpen
  \bibfield  {author} {\bibinfo {author} {\bibnamefont {Ketterle},
  \bibfnamefont {W}}, \ and\ \bibinfo {author} {\bibfnamefont {M.~W.}\
  \bibnamefont {Zwierlein}}} (\bibinfo {year} {2008}),\ \bibfield  {title}
  {\enquote {\bibinfo {title} {Making, probing and understanding ultracold
  {F}ermi gases},}\ }\href@noop {} {\bibfield  {journal} {\bibinfo  {journal}
  {Rivista del Nuovo Cimento}\ }\textbf {\bibinfo {volume} {31}}~(\bibinfo
  {number} {5-6}),\ \bibinfo {pages} {247--422}}\BibitemShut {NoStop}%
\bibitem [{\citenamefont {Kezerashvili}\ \emph {et~al.}(2015)\citenamefont
  {Kezerashvili}, \citenamefont {Tsiklauri},\ and\ \citenamefont
  {Takibayev}}]{Nuclear-clusters}%
  \BibitemOpen
  \bibfield  {author} {\bibinfo {author} {\bibnamefont {Kezerashvili},
  \bibfnamefont {R~Y}}, \bibinfo {author} {\bibfnamefont {S.~M.}\ \bibnamefont
  {Tsiklauri}}, \ and\ \bibinfo {author} {\bibfnamefont {N.~Z.}\ \bibnamefont
  {Takibayev}}} (\bibinfo {year} {2015}),\ \href@noop {} {\enquote {\bibinfo
  {title} {Lighest kaonic nuclear clusters},}\ }\bibinfo {howpublished}
  {arXiv:1510.00478v1}\BibitemShut {NoStop}%
\bibitem [{\citenamefont {Kim}\ \emph {et~al.}(2006)\citenamefont {Kim},
  \citenamefont {Melezhik},\ and\ \citenamefont {Schmelcher}}]{kim2006}%
  \BibitemOpen
  \bibfield  {author} {\bibinfo {author} {\bibnamefont {Kim}, \bibfnamefont
  {J~I}}, \bibinfo {author} {\bibfnamefont {V.~S.}\ \bibnamefont {Melezhik}}, \
  and\ \bibinfo {author} {\bibfnamefont {P.}~\bibnamefont {Schmelcher}}}
  (\bibinfo {year} {2006}),\ \bibfield  {title} {\enquote {\bibinfo {title}
  {Suppression of quantum scattering in strongly confined systems},}\
  }\href@noop {} {\bibfield  {journal} {\bibinfo  {journal} {Phys. Rev. Lett.}\
  }\textbf {\bibinfo {volume} {97}}~(\bibinfo {number} {19}),\ \bibinfo {pages}
  {193203}}\BibitemShut {NoStop}%
\bibitem [{\citenamefont {Kim}\ \emph {et~al.}(2005)\citenamefont {Kim},
  \citenamefont {Schmiedmayer},\ and\ \citenamefont {Schmelcher}}]{kim2005}%
  \BibitemOpen
  \bibfield  {author} {\bibinfo {author} {\bibnamefont {Kim}, \bibfnamefont
  {J~I}}, \bibinfo {author} {\bibfnamefont {J.}~\bibnamefont {Schmiedmayer}}, \
  and\ \bibinfo {author} {\bibfnamefont {P.}~\bibnamefont {Schmelcher}}}
  (\bibinfo {year} {2005}),\ \bibfield  {title} {\enquote {\bibinfo {title}
  {Quantum scattering in quasi-one-dimensional cylindrical confinement},}\
  }\href@noop {} {\bibfield  {journal} {\bibinfo  {journal} {Phys. Rev. A}\
  }\textbf {\bibinfo {volume} {72}}~(\bibinfo {number} {4}),\ \bibinfo {pages}
  {042711}}\BibitemShut {NoStop}%
\bibitem [{\citenamefont {Kim}\ and\ \citenamefont
  {Zubarev}(2002)}]{KimZubarev2002pra}%
  \BibitemOpen
  \bibfield  {author} {\bibinfo {author} {\bibnamefont {Kim}, \bibfnamefont
  {Y~E}}, \ and\ \bibinfo {author} {\bibfnamefont {A.~L.}\ \bibnamefont
  {Zubarev}}} (\bibinfo {year} {2002}),\ \bibfield  {title} {\enquote {\bibinfo
  {title} {Equivalent linear two-body method for {B}ose-{E}instein condensates
  in time-dependent harmonic traps},}\ }\href@noop {} {\bibfield  {journal}
  {\bibinfo  {journal} {Phys. Rev. A}\ }\textbf {\bibinfo {volume} {66}},\
  \bibinfo {pages} {053602}}\BibitemShut {NoStop}%
\bibitem [{\citenamefont {Kim}\ and\ \citenamefont
  {Zubarev}(2000)}]{kim2000JPB}%
  \BibitemOpen
  \bibfield  {author} {\bibinfo {author} {\bibnamefont {Kim}, \bibfnamefont
  {Y~E}}, \ and\ \bibinfo {author} {\bibfnamefont {A.L.}\ \bibnamefont
  {Zubarev}}} (\bibinfo {year} {2000}),\ \bibfield  {title} {\enquote {\bibinfo
  {title} {Equivalent linear two-body method for many-body problems},}\
  }\href@noop {} {\bibfield  {journal} {\bibinfo  {journal} {J. Phys. B}\
  }\textbf {\bibinfo {volume} {33}},\ \bibinfo {pages} {55}}\BibitemShut
  {NoStop}%
\bibitem [{\citenamefont {Kinoshita}\ \emph {et~al.}(2004)\citenamefont
  {Kinoshita}, \citenamefont {Wenger},\ and\ \citenamefont
  {Weiss}}]{kinoshita2004observation}%
  \BibitemOpen
  \bibfield  {author} {\bibinfo {author} {\bibnamefont {Kinoshita},
  \bibfnamefont {T}}, \bibinfo {author} {\bibfnamefont {T.}~\bibnamefont
  {Wenger}}, \ and\ \bibinfo {author} {\bibfnamefont {D.~S}\ \bibnamefont
  {Weiss}}} (\bibinfo {year} {2004}),\ \bibfield  {title} {\enquote {\bibinfo
  {title} {{Observation of a one-dimensional Tonks-Girardeau gas}},}\
  }\href@noop {} {\bibfield  {journal} {\bibinfo  {journal} {Science}\ }\textbf
  {\bibinfo {volume} {305}}~(\bibinfo {number} {5687}),\ \bibinfo {pages}
  {1125--1128}}\BibitemShut {NoStop}%
\bibitem [{\citenamefont {Kira}({2015})}]{Kira2015ncomm}%
  \BibitemOpen
  \bibfield  {author} {\bibinfo {author} {\bibnamefont {Kira}, \bibfnamefont
  {M}}} (\bibinfo {year} {{2015}}),\ \bibfield  {title} {\enquote {\bibinfo
  {title} {{Coherent quantum depletion of an interacting atom condensate}},}\
  }\href@noop {} {\bibfield  {journal} {\bibinfo  {journal} {Nat. Commun.}\
  }\textbf {\bibinfo {volume} {{6}}}}\BibitemShut {NoStop}%
\bibitem [{\citenamefont {Klar}\ and\ \citenamefont
  {Schlecht}(1976)}]{KLAR1976}%
  \BibitemOpen
  \bibfield  {author} {\bibinfo {author} {\bibnamefont {Klar}, \bibfnamefont
  {H}}, \ and\ \bibinfo {author} {\bibfnamefont {W.}~\bibnamefont {Schlecht}}}
  (\bibinfo {year} {1976}),\ \bibfield  {title} {\enquote {\bibinfo {title}
  {Threshold multiple ionization of atoms - energy-dependence for double and
  triple escape},}\ }\href@noop {} {\bibfield  {journal} {\bibinfo  {journal}
  {J. Phys. B}\ }\textbf {\bibinfo {volume} {9}}~(\bibinfo {number} {10}),\
  \bibinfo {pages} {1699--1711}}\BibitemShut {NoStop}%
\bibitem [{\citenamefont {Knirk}(1974)}]{Knirk1974jcp}%
  \BibitemOpen
  \bibfield  {author} {\bibinfo {author} {\bibnamefont {Knirk}, \bibfnamefont
  {Dwayne~L}}} (\bibinfo {year} {1974}),\ \bibfield  {title} {\enquote
  {\bibinfo {title} {Approach to the description of atoms using hyperspherical
  coordinates},}\ }\href {\doibase 10.1063/1.1680808} {\bibfield  {journal}
  {\bibinfo  {journal} {The Journal of Chemical Physics}\ }\textbf {\bibinfo
  {volume} {60}}~(\bibinfo {number} {1}),\ \bibinfo {pages}
  {66--80}}\BibitemShut {NoStop}%
\bibitem [{\citenamefont {Knoop}\ \emph {et~al.}(2010)\citenamefont {Knoop},
  \citenamefont {Ferlaino}, \citenamefont {Berninger}, \citenamefont {Mark},
  \citenamefont {N\"{a}gerl}, \citenamefont {Grimm}, \citenamefont {D'Incao},\
  and\ \citenamefont {Esry}}]{knoop2010PRL}%
  \BibitemOpen
  \bibfield  {author} {\bibinfo {author} {\bibnamefont {Knoop}, \bibfnamefont
  {S}}, \bibinfo {author} {\bibfnamefont {F.}~\bibnamefont {Ferlaino}},
  \bibinfo {author} {\bibfnamefont {M.}~\bibnamefont {Berninger}}, \bibinfo
  {author} {\bibfnamefont {M.}~\bibnamefont {Mark}}, \bibinfo {author}
  {\bibfnamefont {H.{-}C.}\ \bibnamefont {N\"{a}gerl}}, \bibinfo {author}
  {\bibfnamefont {R.}~\bibnamefont {Grimm}}, \bibinfo {author} {\bibfnamefont
  {J.~P.}\ \bibnamefont {D'Incao}}, \ and\ \bibinfo {author} {\bibfnamefont
  {B.~D.}\ \bibnamefont {Esry}}} (\bibinfo {year} {2010}),\ \bibfield  {title}
  {{\selectlanguage {English}\enquote {\bibinfo {title} {Magnetically
  controlled exchange process in an ultracold atom-dimer mixture},}\
  }}\href@noop {} {\bibfield  {journal} {\bibinfo  {journal} {Phys. Rev.
  Lett.}\ }\textbf {\bibinfo {volume} {104}}~(\bibinfo {number} {5}),\ \bibinfo
  {pages} {053201}}\BibitemShut {NoStop}%
\bibitem [{\citenamefont {Knoop}\ \emph {et~al.}(2009)\citenamefont {Knoop},
  \citenamefont {Ferlaino}, \citenamefont {Mark}, \citenamefont {Berninger},
  \citenamefont {Schoebel}, \citenamefont {N\"{a}gerl},\ and\ \citenamefont
  {Grimm}}]{knoop2009NTP}%
  \BibitemOpen
  \bibfield  {author} {\bibinfo {author} {\bibnamefont {Knoop}, \bibfnamefont
  {S}}, \bibinfo {author} {\bibfnamefont {F.}~\bibnamefont {Ferlaino}},
  \bibinfo {author} {\bibfnamefont {M.}~\bibnamefont {Mark}}, \bibinfo {author}
  {\bibfnamefont {M.}~\bibnamefont {Berninger}}, \bibinfo {author}
  {\bibfnamefont {H.}~\bibnamefont {Schoebel}}, \bibinfo {author}
  {\bibfnamefont {H.{-}C.}\ \bibnamefont {N\"{a}gerl}}, \ and\ \bibinfo
  {author} {\bibfnamefont {R.}~\bibnamefont {Grimm}}} (\bibinfo {year}
  {2009}),\ \bibfield  {title} {{\selectlanguage {English}\enquote {\bibinfo
  {title} {Observation of an {E}fimov-like trimer resonance in ultracold
  atom-dimer scattering},}\ }}\href@noop {} {\bibfield  {journal} {\bibinfo
  {journal} {Nat. Phys.}\ }\textbf {\bibinfo {volume} {5}}~(\bibinfo {number}
  {3}),\ \bibinfo {pages} {227--230}}\BibitemShut {NoStop}%
\bibitem [{\citenamefont {K{\"o}hl}\ \emph {et~al.}(2005)\citenamefont
  {K{\"o}hl}, \citenamefont {Moritz}, \citenamefont {St\"{o}ferle},
  \citenamefont {G\"unter},\ and\ \citenamefont
  {Esslinger}}]{kohl2005fermionic}%
  \BibitemOpen
  \bibfield  {author} {\bibinfo {author} {\bibnamefont {K{\"o}hl},
  \bibfnamefont {M}}, \bibinfo {author} {\bibfnamefont {H.}~\bibnamefont
  {Moritz}}, \bibinfo {author} {\bibfnamefont {T.}~\bibnamefont
  {St\"{o}ferle}}, \bibinfo {author} {\bibfnamefont {K.}~\bibnamefont
  {G\"unter}}, \ and\ \bibinfo {author} {\bibfnamefont {T.}~\bibnamefont
  {Esslinger}}} (\bibinfo {year} {2005}),\ \bibfield  {title} {\enquote
  {\bibinfo {title} {Fermionic atoms in a three dimensional optical lattice:
  Observing {F}ermi surfaces, dynamics, and interactions},}\ }\href@noop {}
  {\bibfield  {journal} {\bibinfo  {journal} {Phys. Rev. Lett.}\ }\textbf
  {\bibinfo {volume} {94}}~(\bibinfo {number} {8}),\ \bibinfo {pages}
  {080403}}\BibitemShut {NoStop}%
\bibitem [{\citenamefont {K\"ohl}\ and\ \citenamefont
  {Schaefer}(1983)}]{Kohler-1983}%
  \BibitemOpen
  \bibfield  {author} {\bibinfo {author} {\bibnamefont {K\"ohl}, \bibfnamefont
  {W~E}}, \ and\ \bibinfo {author} {\bibfnamefont {J.}~\bibnamefont
  {Schaefer}}} (\bibinfo {year} {1983}),\ \bibfield  {title} {\enquote
  {\bibinfo {title} {Theoreticall studies of {H}$_2$-{H}$_2$ collisions. vs.
  {Ab} initio calculations of relaxation phenomena in parahydrogen gas},}\
  }\href@noop {} {\bibfield  {journal} {\bibinfo  {journal} {J. Chem. Phys}\
  }\textbf {\bibinfo {volume} {78}},\ \bibinfo {pages} {6602}}\BibitemShut
  {NoStop}%
\bibitem [{\citenamefont {K\"ohler}(2002)}]{kohler2002PRL}%
  \BibitemOpen
  \bibfield  {author} {\bibinfo {author} {\bibnamefont {K\"ohler},
  \bibfnamefont {T}}} (\bibinfo {year} {2002}),\ \bibfield  {title}
  {{\selectlanguage {English}\enquote {\bibinfo {title} {Three-body problem in
  a dilute {B}ose-{E}instein condensate},}\ }}\href@noop {} {\bibfield
  {journal} {\bibinfo  {journal} {Phys. Rev. Lett.}\ }\textbf {\bibinfo
  {volume} {89}}~(\bibinfo {number} {21}),\ \bibinfo {pages}
  {210404}}\BibitemShut {NoStop}%
\bibitem [{\citenamefont {K\"{o}hler}\ \emph {et~al.}(2006)\citenamefont
  {K\"{o}hler}, \citenamefont {G\'{o}ral},\ and\ \citenamefont
  {Julienne}}]{koehler2006}%
  \BibitemOpen
  \bibfield  {author} {\bibinfo {author} {\bibnamefont {K\"{o}hler},
  \bibfnamefont {T}}, \bibinfo {author} {\bibfnamefont {K.}~\bibnamefont
  {G\'{o}ral}}, \ and\ \bibinfo {author} {\bibfnamefont {P.~S.}\ \bibnamefont
  {Julienne}}} (\bibinfo {year} {2006}),\ \bibfield  {title} {\enquote
  {\bibinfo {title} {Production of cold molecules via magnetically tunable
  feshbach resonances},}\ }\href@noop {} {\bibfield  {journal} {\bibinfo
  {journal} {Rev. Mod. Phys.}\ }\textbf {\bibinfo {volume} {78}}~(\bibinfo
  {number} {4}),\ \bibinfo {pages} {1311--1361}}\BibitemShut {NoStop}%
\bibitem [{\citenamefont {Kohstall}\ \emph {et~al.}(2012)\citenamefont
  {Kohstall}, \citenamefont {Zaccanti}, \citenamefont {Jag}, \citenamefont
  {Trenkwalder}, \citenamefont {Massignan}, \citenamefont {Bruun},
  \citenamefont {Schreck},\ and\ \citenamefont
  {Grimm}}]{kohstall_metastability_2012}%
  \BibitemOpen
  \bibfield  {author} {\bibinfo {author} {\bibnamefont {Kohstall},
  \bibfnamefont {C}}, \bibinfo {author} {\bibfnamefont {M.}~\bibnamefont
  {Zaccanti}}, \bibinfo {author} {\bibfnamefont {M.}~\bibnamefont {Jag}},
  \bibinfo {author} {\bibfnamefont {A.}~\bibnamefont {Trenkwalder}}, \bibinfo
  {author} {\bibfnamefont {P.}~\bibnamefont {Massignan}}, \bibinfo {author}
  {\bibfnamefont {G.~M.}\ \bibnamefont {Bruun}}, \bibinfo {author}
  {\bibfnamefont {F.}~\bibnamefont {Schreck}}, \ and\ \bibinfo {author}
  {\bibfnamefont {R.}~\bibnamefont {Grimm}}} (\bibinfo {year} {2012}),\
  \bibfield  {title} {\enquote {\bibinfo {title} {Metastability and coherence
  of repulsive polarons in a strongly interacting {Fermi} mixture},}\ }\href
  {\doibase 10.1038/nature11065} {\bibfield  {journal} {\bibinfo  {journal}
  {Nature}\ }\textbf {\bibinfo {volume} {485}}~(\bibinfo {number} {7400}),\
  \bibinfo {pages} {615--+}}\BibitemShut {NoStop}%
\bibitem [{\citenamefont {Kokoouline}\ \emph {et~al.}(2011)\citenamefont
  {Kokoouline}, \citenamefont {Douguet},\ and\ \citenamefont
  {Greene}}]{Kokoouline2011CPL}%
  \BibitemOpen
  \bibfield  {author} {\bibinfo {author} {\bibnamefont {Kokoouline},
  \bibfnamefont {V}}, \bibinfo {author} {\bibfnamefont {N.}~\bibnamefont
  {Douguet}}, \ and\ \bibinfo {author} {\bibfnamefont {C.~H.}\ \bibnamefont
  {Greene}}} (\bibinfo {year} {2011}),\ \bibfield  {title} {\enquote {\bibinfo
  {title} {Breaking bonds with electrons: Dissociative recombination of
  molecular ions},}\ }\href@noop {} {\bibfield  {journal} {\bibinfo  {journal}
  {Chem. Phys. Lett.}\ }\textbf {\bibinfo {volume} {507}}~(\bibinfo {number}
  {1-3}),\ \bibinfo {pages} {1--10}}\BibitemShut {NoStop}%
\bibitem [{\citenamefont {Kokoouline}\ and\ \citenamefont
  {Greene}(2003)}]{kokoouline2003PRL}%
  \BibitemOpen
  \bibfield  {author} {\bibinfo {author} {\bibnamefont {Kokoouline},
  \bibfnamefont {V}}, \ and\ \bibinfo {author} {\bibfnamefont {C.~H.}\
  \bibnamefont {Greene}}} (\bibinfo {year} {2003}),\ \bibfield  {title}
  {{\selectlanguage {English}\enquote {\bibinfo {title} {Theory of dissociative
  recombination of {D}$_{3h}$ triatomic ions applied to {H}$_3^+$},}\
  }}\href@noop {} {\bibfield  {journal} {\bibinfo  {journal} {Phys. Rev.
  Lett.}\ }\textbf {\bibinfo {volume} {90}}~(\bibinfo {number} {13}),\ \bibinfo
  {pages} {133201}}\BibitemShut {NoStop}%
\bibitem [{\citenamefont {Kokoouline}\ \emph {et~al.}(2001)\citenamefont
  {Kokoouline}, \citenamefont {Greene},\ and\ \citenamefont
  {Esry}}]{kokoouline2001NT}%
  \BibitemOpen
  \bibfield  {author} {\bibinfo {author} {\bibnamefont {Kokoouline},
  \bibfnamefont {V}}, \bibinfo {author} {\bibfnamefont {C.~H.}\ \bibnamefont
  {Greene}}, \ and\ \bibinfo {author} {\bibfnamefont {B.~D.}\ \bibnamefont
  {Esry}}} (\bibinfo {year} {2001}),\ \bibfield  {title} {\enquote {\bibinfo
  {title} {Mechanism for the destruction of {H}$^{+}_{3}$ ions by electron
  impact.}}\ }\href@noop {} {\bibfield  {journal} {\bibinfo  {journal} {Nature
  (London)}\ }\textbf {\bibinfo {volume} {412}}~(\bibinfo {number} {6850}),\
  \bibinfo {pages} {891--894}}\BibitemShut {NoStop}%
\bibitem [{\citenamefont {Kolomeisky}\ and\ \citenamefont
  {Straley}(1996)}]{kolomeisky1996phase}%
  \BibitemOpen
  \bibfield  {author} {\bibinfo {author} {\bibnamefont {Kolomeisky},
  \bibfnamefont {E~B}}, \ and\ \bibinfo {author} {\bibfnamefont {J.~P}\
  \bibnamefont {Straley}}} (\bibinfo {year} {1996}),\ \bibfield  {title}
  {\enquote {\bibinfo {title} {Phase diagram and correlation exponents for
  interacting fermions in one dimension},}\ }\href@noop {} {\bibfield
  {journal} {\bibinfo  {journal} {Rev. Mod. Phys.}\ }\textbf {\bibinfo {volume}
  {68}}~(\bibinfo {number} {1}),\ \bibinfo {pages} {175}}\BibitemShut {NoStop}%
\bibitem [{\citenamefont {Koschorreck}\ \emph {et~al.}(2012)\citenamefont
  {Koschorreck}, \citenamefont {Pertot}, \citenamefont {Vogt}, \citenamefont
  {Froehlich}, \citenamefont {Feld},\ and\ \citenamefont
  {Koehl}}]{koschorreck_attractive_2012}%
  \BibitemOpen
  \bibfield  {author} {\bibinfo {author} {\bibnamefont {Koschorreck},
  \bibfnamefont {Marco}}, \bibinfo {author} {\bibfnamefont {Daniel}\
  \bibnamefont {Pertot}}, \bibinfo {author} {\bibfnamefont {Enrico}\
  \bibnamefont {Vogt}}, \bibinfo {author} {\bibfnamefont {Bernd}\ \bibnamefont
  {Froehlich}}, \bibinfo {author} {\bibfnamefont {Michael}\ \bibnamefont
  {Feld}}, \ and\ \bibinfo {author} {\bibfnamefont {Michael}\ \bibnamefont
  {Koehl}}} (\bibinfo {year} {2012}),\ \bibfield  {title} {\enquote {\bibinfo
  {title} {Attractive and repulsive {Fermi} polarons in two dimensions},}\
  }\href {\doibase 10.1038/nature11151} {\bibfield  {journal} {\bibinfo
  {journal} {Nature}\ }\textbf {\bibinfo {volume} {485}}~(\bibinfo {number}
  {7400}),\ \bibinfo {pages} {619--+}}\BibitemShut {NoStop}%
\bibitem [{\citenamefont {Kossmann}\ \emph {et~al.}(1988)\citenamefont
  {Kossmann}, \citenamefont {Schmidt},\ and\ \citenamefont
  {Andersen}}]{Kossmann-1988-JPR}%
  \BibitemOpen
  \bibfield  {author} {\bibinfo {author} {\bibnamefont {Kossmann},
  \bibfnamefont {H}}, \bibinfo {author} {\bibfnamefont {V.}~\bibnamefont
  {Schmidt}}, \ and\ \bibinfo {author} {\bibfnamefont {T.}~\bibnamefont
  {Andersen}}} (\bibinfo {year} {1988}),\ \bibfield  {title} {\enquote
  {\bibinfo {title} {Test of {W}annier threshold laws:
  {D}ouble{-}photoionization cross section in helium},}\ }\href@noop {}
  {\bibinfo  {journal} {Phys. Rev. Lett.}\ }\BibitemShut {NoStop}%
\bibitem [{\citenamefont {Kosterlitz}\ and\ \citenamefont
  {Thouless}(1973)}]{kosterlitz_ordering_1973}%
  \BibitemOpen
\bibfield  {journal} {  }\bibfield  {author} {\bibinfo {author} {\bibnamefont
  {Kosterlitz}, \bibfnamefont {J~M}}, \ and\ \bibinfo {author} {\bibfnamefont
  {D.~J.}\ \bibnamefont {Thouless}}} (\bibinfo {year} {1973}),\ \bibfield
  {title} {\enquote {\bibinfo {title} {{Ordering}, {Metastability} and
  {Phase}-{Transitions} in 2 {Dimensional} {Systems}},}\ }\href {\doibase
  10.1088/0022-3719/6/7/010} {\bibfield  {journal} {\bibinfo  {journal}
  {Journal of Physics C-Solid State Physics}\ }\textbf {\bibinfo {volume}
  {6}}~(\bibinfo {number} {7}),\ \bibinfo {pages} {1181--1203}}\BibitemShut
  {NoStop}%
\bibitem [{\citenamefont {Kotochigova}({2014})}]{Kotochigova2014rpp}%
  \BibitemOpen
  \bibfield  {author} {\bibinfo {author} {\bibnamefont {Kotochigova},
  \bibfnamefont {Svetlana}}} (\bibinfo {year} {{2014}}),\ \bibfield  {title}
  {\enquote {\bibinfo {title} {{Controlling interactions between highly
  magnetic atoms with Feshbach resonances}},}\ }\href {\doibase
  {10.1088/0034-4885/77/9/093901}} {\bibfield  {journal} {\bibinfo  {journal}
  {{Reports on Progress in Physics}}\ }\textbf {\bibinfo {volume}
  {{77}}}~(\bibinfo {number} {{9}}),\
  {10.1088/0034-4885/77/9/093901}}\BibitemShut {NoStop}%
\bibitem [{\citenamefont {Kraemer}\ \emph {et~al.}(2006)\citenamefont
  {Kraemer}, \citenamefont {Mark}, \citenamefont {Waldburger}, \citenamefont
  {Danzl}, \citenamefont {Chin}, \citenamefont {Engeser}, \citenamefont
  {Lange}, \citenamefont {Pilch}, \citenamefont {Jaakkola}, \citenamefont
  {Nagerl},\ and\ \citenamefont {Grimm}}]{kraemer2006NT}%
  \BibitemOpen
  \bibfield  {author} {\bibinfo {author} {\bibnamefont {Kraemer}, \bibfnamefont
  {T}}, \bibinfo {author} {\bibfnamefont {M.}~\bibnamefont {Mark}}, \bibinfo
  {author} {\bibfnamefont {P.}~\bibnamefont {Waldburger}}, \bibinfo {author}
  {\bibfnamefont {J.~G.}\ \bibnamefont {Danzl}}, \bibinfo {author}
  {\bibfnamefont {C.}~\bibnamefont {Chin}}, \bibinfo {author} {\bibfnamefont
  {B.}~\bibnamefont {Engeser}}, \bibinfo {author} {\bibfnamefont {A.~D.}\
  \bibnamefont {Lange}}, \bibinfo {author} {\bibfnamefont {K.}~\bibnamefont
  {Pilch}}, \bibinfo {author} {\bibfnamefont {A.}~\bibnamefont {Jaakkola}},
  \bibinfo {author} {\bibfnamefont {H.{-}C.}\ \bibnamefont {Nagerl}}, \ and\
  \bibinfo {author} {\bibfnamefont {R.}~\bibnamefont {Grimm}}} (\bibinfo {year}
  {2006}),\ \bibfield  {title} {{\selectlanguage {English}\enquote {\bibinfo
  {title} {Evidence for {E}fimov quantum states in an ultracold gas of caesium
  atoms},}\ }}\href@noop {} {\bibfield  {journal} {\bibinfo  {journal} {Nature
  (London)}\ }\textbf {\bibinfo {volume} {440}}~(\bibinfo {number} {7082}),\
  \bibinfo {pages} {315--318}}\BibitemShut {NoStop}%
\bibitem [{\citenamefont {Kristensen}\ and\ \citenamefont
  {Pricoupenko}(2015)}]{kristensen2015ultracold}%
  \BibitemOpen
  \bibfield  {author} {\bibinfo {author} {\bibnamefont {Kristensen},
  \bibfnamefont {T}}, \ and\ \bibinfo {author} {\bibfnamefont {L.}~\bibnamefont
  {Pricoupenko}}} (\bibinfo {year} {2015}),\ \bibfield  {title} {\enquote
  {\bibinfo {title} {Ultracold-atom collisions in atomic waveguides: A
  two-channel analysis},}\ }\href@noop {} {\bibfield  {journal} {\bibinfo
  {journal} {Phys. Rev. A}\ }\textbf {\bibinfo {volume} {91}}~(\bibinfo
  {number} {4}),\ \bibinfo {pages} {042703}}\BibitemShut {NoStop}%
\bibitem [{\citenamefont {Kr\"oger}\ and\ \citenamefont
  {Perne}(1980)}]{KrogerPerne1980prc}%
  \BibitemOpen
  \bibfield  {author} {\bibinfo {author} {\bibnamefont {Kr\"oger},
  \bibfnamefont {H}}, \ and\ \bibinfo {author} {\bibfnamefont {R.}~\bibnamefont
  {Perne}}} (\bibinfo {year} {1980}),\ \bibfield  {title} {\enquote {\bibinfo
  {title} {Efimov effect in the four-body case},}\ }\href {\doibase
  10.1103/PhysRevC.22.21} {\bibfield  {journal} {\bibinfo  {journal} {Phys.
  Rev. C}\ }\textbf {\bibinfo {volume} {22}},\ \bibinfo {pages}
  {21--27}}\BibitemShut {NoStop}%
\bibitem [{\citenamefont {Kr\"{u}kow}\ \emph {et~al.}(2016)\citenamefont
  {Kr\"{u}kow}, \citenamefont {Mohammadi}, \citenamefont {H\"arter},
  \citenamefont {{Hecker Denschlag}}, \citenamefont {P\'{e}rez-R\'{i}os},\ and\
  \citenamefont {Greene}}]{Ulm}%
  \BibitemOpen
  \bibfield  {author} {\bibinfo {author} {\bibnamefont {Kr\"{u}kow},
  \bibfnamefont {A}}, \bibinfo {author} {\bibfnamefont {A.}~\bibnamefont
  {Mohammadi}}, \bibinfo {author} {\bibfnamefont {A.}~\bibnamefont {H\"arter}},
  \bibinfo {author} {\bibfnamefont {J.}~\bibnamefont {{Hecker Denschlag}}},
  \bibinfo {author} {\bibfnamefont {J.}~\bibnamefont {P\'{e}rez-R\'{i}os}}, \
  and\ \bibinfo {author} {\bibfnamefont {C.~H.}\ \bibnamefont {Greene}}}
  (\bibinfo {year} {2016}),\ \bibfield  {title} {\enquote {\bibinfo {title}
  {Energy scaling of cold atom-atom-ion three-body recombination},}\
  }\href@noop {} {\bibfield  {journal} {\bibinfo  {journal} {Phys. Rev. Lett.}\
  }\textbf {\bibinfo {volume} {116}},\ \bibinfo {pages} {193201}}\BibitemShut
  {NoStop}%
\bibitem [{\citenamefont {Krych}\ \emph {et~al.}(2011)\citenamefont {Krych},
  \citenamefont {Skomorowski}, \citenamefont {Pawlowski}, \citenamefont
  {Moszynski},\ and\ \citenamefont {Idziaszek}}]{Krych-2011}%
  \BibitemOpen
  \bibfield  {author} {\bibinfo {author} {\bibnamefont {Krych}, \bibfnamefont
  {M}}, \bibinfo {author} {\bibfnamefont {W.}~\bibnamefont {Skomorowski}},
  \bibinfo {author} {\bibfnamefont {F.}~\bibnamefont {Pawlowski}}, \bibinfo
  {author} {\bibfnamefont {R.}~\bibnamefont {Moszynski}}, \ and\ \bibinfo
  {author} {\bibfnamefont {Z.}~\bibnamefont {Idziaszek}}} (\bibinfo {year}
  {2011}),\ \bibfield  {title} {\enquote {\bibinfo {title} {Sympathetic cooling
  of the {Ba}$^+$ ion by collisions with ultracold {Rb} atoms: Theoretical
  prospects},}\ }\href@noop {} {\bibfield  {journal} {\bibinfo  {journal}
  {Phys. Rev A}\ }\textbf {\bibinfo {volume} {83}},\ \bibinfo {pages}
  {032723}}\BibitemShut {NoStop}%
\bibitem [{\citenamefont {Ku}\ \emph {et~al.}({2012})\citenamefont {Ku},
  \citenamefont {Sommer}, \citenamefont {Cheuk},\ and\ \citenamefont
  {Zwierlein}}]{KuZwierlein2012science}%
  \BibitemOpen
  \bibfield  {author} {\bibinfo {author} {\bibnamefont {Ku}, \bibfnamefont
  {M~J~H}}, \bibinfo {author} {\bibfnamefont {A.~T.}\ \bibnamefont {Sommer}},
  \bibinfo {author} {\bibfnamefont {L.~W.}\ \bibnamefont {Cheuk}}, \ and\
  \bibinfo {author} {\bibfnamefont {M.~W.}\ \bibnamefont {Zwierlein}}}
  (\bibinfo {year} {{2012}}),\ \bibfield  {title} {\enquote {\bibinfo {title}
  {{Revealing the Superfluid Lambda Transition in the Universal Thermodynamics
  of a Unitary Fermi Gas}},}\ }\href@noop {} {\bibfield  {journal} {\bibinfo
  {journal} {Science}\ }\textbf {\bibinfo {volume} {{335}}}~(\bibinfo {number}
  {{6068}}),\ \bibinfo {pages} {{563--567}}}\BibitemShut {NoStop}%
\bibitem [{\citenamefont {Kunitski}\ \emph {et~al.}(2015)\citenamefont
  {Kunitski}, \citenamefont {Zeller}, \citenamefont {Voigtsberger},
  \citenamefont {Kalinin}, \citenamefont {Schmidt}, \citenamefont
  {Sch\"offler}, \citenamefont {Czasch}, \citenamefont {Sch\"{o}llkopf},
  \citenamefont {Grisenti}, \citenamefont {Jahnke}, \citenamefont {Blume},\
  and\ \citenamefont {D\"orner}}]{Kunitski-2015}%
  \BibitemOpen
  \bibfield  {author} {\bibinfo {author} {\bibnamefont {Kunitski},
  \bibfnamefont {M}}, \bibinfo {author} {\bibfnamefont {S.}~\bibnamefont
  {Zeller}}, \bibinfo {author} {\bibfnamefont {J.}~\bibnamefont
  {Voigtsberger}}, \bibinfo {author} {\bibfnamefont {A.}~\bibnamefont
  {Kalinin}}, \bibinfo {author} {\bibfnamefont {L.~P.~H.}\ \bibnamefont
  {Schmidt}}, \bibinfo {author} {\bibfnamefont {M.}~\bibnamefont
  {Sch\"offler}}, \bibinfo {author} {\bibfnamefont {A.}~\bibnamefont {Czasch}},
  \bibinfo {author} {\bibfnamefont {W.}~\bibnamefont {Sch\"{o}llkopf}},
  \bibinfo {author} {\bibfnamefont {R.~E.}\ \bibnamefont {Grisenti}}, \bibinfo
  {author} {\bibfnamefont {T.}~\bibnamefont {Jahnke}}, \bibinfo {author}
  {\bibfnamefont {D.}~\bibnamefont {Blume}}, \ and\ \bibinfo {author}
  {\bibfnamefont {R.}~\bibnamefont {D\"orner}}} (\bibinfo {year} {2015}),\
  \bibfield  {title} {\enquote {\bibinfo {title} {{Observation of the Efimov
  state of the helium trimer}},}\ }\href@noop {} {\bibfield  {journal}
  {\bibinfo  {journal} {Science}\ }\textbf {\bibinfo {volume} {348}},\ \bibinfo
  {pages} {551--555}}\BibitemShut {NoStop}%
\bibitem [{\citenamefont {Kuppermann}\ and\ \citenamefont
  {Hipes}(1986)}]{kuppermannJCP1986}%
  \BibitemOpen
  \bibfield  {author} {\bibinfo {author} {\bibnamefont {Kuppermann},
  \bibfnamefont {A}}, \ and\ \bibinfo {author} {\bibfnamefont {P.~G.}\
  \bibnamefont {Hipes}}} (\bibinfo {year} {1986}),\ \bibfield  {title}
  {\enquote {\bibinfo {title} {3-{D}imensional quantum-mechanical reactive
  scattering using symmetrized hyperspherical coordinates},}\ }\href@noop {}
  {\bibfield  {journal} {\bibinfo  {journal} {J. Chem. Phys.}\ }\textbf
  {\bibinfo {volume} {84}}~(\bibinfo {number} {10}),\ \bibinfo {pages}
  {5962--5963}}\BibitemShut {NoStop}%
\bibitem [{\citenamefont {Kushibe}\ \emph {et~al.}(2004)\citenamefont
  {Kushibe}, \citenamefont {Mutou}, \citenamefont {Morishita}, \citenamefont
  {Watanabe},\ and\ \citenamefont {Matsuzawa}}]{kushibe2004PRA}%
  \BibitemOpen
  \bibfield  {author} {\bibinfo {author} {\bibnamefont {Kushibe}, \bibfnamefont
  {D}}, \bibinfo {author} {\bibfnamefont {M}~\bibnamefont {Mutou}}, \bibinfo
  {author} {\bibfnamefont {T}~\bibnamefont {Morishita}}, \bibinfo {author}
  {\bibfnamefont {S}~\bibnamefont {Watanabe}}, \ and\ \bibinfo {author}
  {\bibfnamefont {M}~\bibnamefont {Matsuzawa}}} (\bibinfo {year} {2004}),\
  \bibfield  {title} {{\selectlanguage {English}\enquote {\bibinfo {title}
  {Aspects of hyperspherical adiabaticity in an atomic-gas {B}ose-{E}instein
  condensate},}\ }}\href@noop {} {\bibfield  {journal} {\bibinfo  {journal}
  {Phys. Rev. A}\ }\textbf {\bibinfo {volume} {70}}~(\bibinfo {number} {6}),\
  \bibinfo {pages} {063617}}\BibitemShut {NoStop}%
\bibitem [{\citenamefont {Lamporesi}\ \emph {et~al.}(2010)\citenamefont
  {Lamporesi}, \citenamefont {Catani}, \citenamefont {Barontini}, \citenamefont
  {Nishida}, \citenamefont {Inguscio},\ and\ \citenamefont
  {Minardi}}]{lamporesi2010}%
  \BibitemOpen
  \bibfield  {author} {\bibinfo {author} {\bibnamefont {Lamporesi},
  \bibfnamefont {G}}, \bibinfo {author} {\bibfnamefont {J.}~\bibnamefont
  {Catani}}, \bibinfo {author} {\bibfnamefont {G.}~\bibnamefont {Barontini}},
  \bibinfo {author} {\bibfnamefont {Y.}~\bibnamefont {Nishida}}, \bibinfo
  {author} {\bibfnamefont {M.}~\bibnamefont {Inguscio}}, \ and\ \bibinfo
  {author} {\bibfnamefont {F.}~\bibnamefont {Minardi}}} (\bibinfo {year}
  {2010}),\ \bibfield  {title} {\enquote {\bibinfo {title} {Scattering in mixed
  dimensions with ultracold gases},}\ }\href@noop {} {\bibfield  {journal}
  {\bibinfo  {journal} {Phys. Rev. Lett.}\ }\textbf {\bibinfo {volume}
  {104}}~(\bibinfo {number} {15}),\ \bibinfo {pages} {153202}}\BibitemShut
  {NoStop}%
\bibitem [{\citenamefont {Landau}\ and\ \citenamefont
  {Lifshitz}(1976)}]{Landau-mechanics}%
  \BibitemOpen
  \bibfield  {author} {\bibinfo {author} {\bibnamefont {Landau}, \bibfnamefont
  {L~D}}, \ and\ \bibinfo {author} {\bibfnamefont {E.~M.}\ \bibnamefont
  {Lifshitz}}} (\bibinfo {year} {1976}),\ \href@noop {} {\emph {\bibinfo
  {title} {Mechanics}}}\ (\bibinfo  {publisher} {Elsevier
  Butterworth-Heinemann},\ \bibinfo {address} {Burlington, MA})\BibitemShut
  {NoStop}%
\bibitem [{\citenamefont {Landau}\ and\ \citenamefont
  {Lifshitz}(1997)}]{Landau1997}%
  \BibitemOpen
  \bibfield  {author} {\bibinfo {author} {\bibnamefont {Landau}, \bibfnamefont
  {LD}}, \ and\ \bibinfo {author} {\bibfnamefont {EM}~\bibnamefont {Lifshitz}}}
  (\bibinfo {year} {1997}),\ \href@noop {} {\emph {\bibinfo {title} {Quantum
  Mechanics: Non-Relativistic Theory}}}\ (\bibinfo  {publisher}
  {Butterworth-Heinemann})\BibitemShut {NoStop}%
\bibitem [{\citenamefont {Langevin}(1905)}]{Langevin}%
  \BibitemOpen
  \bibfield  {author} {\bibinfo {author} {\bibnamefont {Langevin},
  \bibfnamefont {P}}} (\bibinfo {year} {1905}),\ \bibfield  {title} {\enquote
  {\bibinfo {title} {Une formule fodnamentale de th\'{e}orie cin\'{e}tique},}\
  }\href@noop {} {\bibfield  {journal} {\bibinfo  {journal} {C. R. Acad. Sci.}\
  }\textbf {\bibinfo {volume} {140}},\ \bibinfo {pages} {35}}\BibitemShut
  {NoStop}%
\bibitem [{\citenamefont {Larson}\ and\ \citenamefont
  {Stoneman}(1985)}]{larsonpra1985}%
  \BibitemOpen
  \bibfield  {author} {\bibinfo {author} {\bibnamefont {Larson}, \bibfnamefont
  {D~J}}, \ and\ \bibinfo {author} {\bibfnamefont {R.}~\bibnamefont
  {Stoneman}}} (\bibinfo {year} {1985}),\ \bibfield  {title} {\enquote
  {\bibinfo {title} {Photodetachment of atomic negative ions near threshold in
  a magnetic field},}\ }\href@noop {} {\bibfield  {journal} {\bibinfo
  {journal} {Phys. Rev. A}\ }\textbf {\bibinfo {volume} {31}},\ \bibinfo
  {pages} {2210--2214}}\BibitemShut {NoStop}%
\bibitem [{\citenamefont {Lattimer}\ and\ \citenamefont
  {Prakash}(2004)}]{Lattimer-2004}%
  \BibitemOpen
  \bibfield  {author} {\bibinfo {author} {\bibnamefont {Lattimer},
  \bibfnamefont {J~M}}, \ and\ \bibinfo {author} {\bibfnamefont
  {M.}~\bibnamefont {Prakash}}} (\bibinfo {year} {2004}),\ \bibfield  {title}
  {\enquote {\bibinfo {title} {The physics of neutron stars},}\ }\href@noop {}
  {\bibfield  {journal} {\bibinfo  {journal} {Science}\ }\textbf {\bibinfo
  {volume} {304}},\ \bibinfo {pages} {536--542}}\BibitemShut {NoStop}%
\bibitem [{\citenamefont {Laughlin}(1983)}]{Laughlin1983prb}%
  \BibitemOpen
  \bibfield  {author} {\bibinfo {author} {\bibnamefont {Laughlin},
  \bibfnamefont {R~B}}} (\bibinfo {year} {1983}),\ \bibfield  {title} {\enquote
  {\bibinfo {title} {Quantized motion of three two-dimensional electrons in a
  strong magnetic field},}\ }\href {\doibase 10.1103/PhysRevB.27.3383}
  {\bibfield  {journal} {\bibinfo  {journal} {Phys. Rev. B}\ }\textbf {\bibinfo
  {volume} {27}},\ \bibinfo {pages} {3383--3389}}\BibitemShut {NoStop}%
\bibitem [{\citenamefont {Launay}\ and\ \citenamefont
  {Dourneuf}(1990)}]{Launay1990CPL}%
  \BibitemOpen
  \bibfield  {author} {\bibinfo {author} {\bibnamefont {Launay}, \bibfnamefont
  {J~M}}, \ and\ \bibinfo {author} {\bibfnamefont {M.~Le}\ \bibnamefont
  {Dourneuf}}} (\bibinfo {year} {1990}),\ \bibfield  {title} {\enquote
  {\bibinfo {title} {Quantum-mechanical calculation of integral cross sections
  for the reaction {F+H$_2$($v=0, j=0$)$\rightarrow$FH($v' j'$)+H} by the
  hyperspherical method},}\ }\href@noop {} {\bibfield  {journal} {\bibinfo
  {journal} {Chem. Phys. Lett.}\ }\textbf {\bibinfo {volume} {169}}~(\bibinfo
  {number} {6}),\ \bibinfo {pages} {473 -- 481}}\BibitemShut {NoStop}%
\bibitem [{\citenamefont {Laurent}\ \emph {et~al.}({2014})\citenamefont
  {Laurent}, \citenamefont {Leyronas},\ and\ \citenamefont
  {Chevy}}]{Laurent2014prl}%
  \BibitemOpen
  \bibfield  {author} {\bibinfo {author} {\bibnamefont {Laurent}, \bibfnamefont
  {S}}, \bibinfo {author} {\bibfnamefont {X.}~\bibnamefont {Leyronas}}, \ and\
  \bibinfo {author} {\bibfnamefont {F.}~\bibnamefont {Chevy}}} (\bibinfo {year}
  {{2014}}),\ \bibfield  {title} {\enquote {\bibinfo {title} {Momentum
  distribution of a dilute unitary {B}ose gas with three-body losses},}\
  }\href@noop {} {\bibfield  {journal} {\bibinfo  {journal} {{Phys. Rev.
  Lett.}}\ }\textbf {\bibinfo {volume} {{113}}}~(\bibinfo {number}
  {{22}})}\BibitemShut {NoStop}%
\bibitem [{\citenamefont {Lazauskas}\ and\ \citenamefont
  {Carbonell}(2006)}]{lazauskas2006PRA}%
  \BibitemOpen
  \bibfield  {author} {\bibinfo {author} {\bibnamefont {Lazauskas},
  \bibfnamefont {R}}, \ and\ \bibinfo {author} {\bibfnamefont {J}~\bibnamefont
  {Carbonell}}} (\bibinfo {year} {2006}),\ \bibfield  {title} {{\selectlanguage
  {English}\enquote {\bibinfo {title} {Description of $^4${He} tetramer bound
  and scattering states},}\ }}\href@noop {} {\bibfield  {journal} {\bibinfo
  {journal} {Phys. Rev. A}\ }\textbf {\bibinfo {volume} {73}}~(\bibinfo
  {number} {6}),\ \bibinfo {pages} {062717}}\BibitemShut {NoStop}%
\bibitem [{\citenamefont {LeBlanc}\ and\ \citenamefont
  {Thywissen}(2007)}]{leblancpra2007}%
  \BibitemOpen
  \bibfield  {author} {\bibinfo {author} {\bibnamefont {LeBlanc}, \bibfnamefont
  {L~J}}, \ and\ \bibinfo {author} {\bibfnamefont {J.~H.}\ \bibnamefont
  {Thywissen}}} (\bibinfo {year} {2007}),\ \bibfield  {title} {\enquote
  {\bibinfo {title} {Species-specific optical lattices},}\ }\href@noop {}
  {\bibfield  {journal} {\bibinfo  {journal} {Phys. Rev. A}\ }\textbf {\bibinfo
  {volume} {75}},\ \bibinfo {pages} {053612}}\BibitemShut {NoStop}%
\bibitem [{\citenamefont {Lee}\ \emph {et~al.}(2007)\citenamefont {Lee},
  \citenamefont {K\"ohler},\ and\ \citenamefont {Julienne}}]{lee2007PRA}%
  \BibitemOpen
  \bibfield  {author} {\bibinfo {author} {\bibnamefont {Lee}, \bibfnamefont
  {M~D}}, \bibinfo {author} {\bibfnamefont {T.}~\bibnamefont {K\"ohler}}, \
  and\ \bibinfo {author} {\bibfnamefont {P.~S.}\ \bibnamefont {Julienne}}}
  (\bibinfo {year} {2007}),\ \bibfield  {title} {{\selectlanguage
  {English}\enquote {\bibinfo {title} {Excited {T}homas-{E}fimov levels in
  ultracold gases},}\ }}\href@noop {} {\bibfield  {journal} {\bibinfo
  {journal} {Phys. Rev. A}\ }\textbf {\bibinfo {volume} {76}}~(\bibinfo
  {number} {1}),\ \bibinfo {pages} {012720}}\BibitemShut {NoStop}%
\bibitem [{\citenamefont {Lee}\ and\ \citenamefont
  {Lee}(2010)}]{LeeLee2010pra}%
  \BibitemOpen
  \bibfield  {author} {\bibinfo {author} {\bibnamefont {Lee}, \bibfnamefont
  {Yu-Li}}, \ and\ \bibinfo {author} {\bibfnamefont {Yu-Wen}\ \bibnamefont
  {Lee}}} (\bibinfo {year} {2010}),\ \bibfield  {title} {\enquote {\bibinfo
  {title} {Universality and stability for a dilute {B}ose gas with a {F}eshbach
  resonance},}\ }\href@noop {} {\bibfield  {journal} {\bibinfo  {journal}
  {Phys. Rev. A}\ }\textbf {\bibinfo {volume} {81}},\ \bibinfo {pages}
  {063613}}\BibitemShut {NoStop}%
\bibitem [{\citenamefont {Lemeshko}(2017)}]{Lemeshko2017prl}%
  \BibitemOpen
  \bibfield  {author} {\bibinfo {author} {\bibnamefont {Lemeshko},
  \bibfnamefont {Mikhail}}} (\bibinfo {year} {2017}),\ \bibfield  {title}
  {\enquote {\bibinfo {title} {Quasiparticle approach to molecules interacting
  with quantum solvents},}\ }\href {\doibase 10.1103/PhysRevLett.118.095301}
  {\bibfield  {journal} {\bibinfo  {journal} {Phys. Rev. Lett.}\ }\textbf
  {\bibinfo {volume} {118}},\ \bibinfo {pages} {095301}}\BibitemShut {NoStop}%
\bibitem [{\citenamefont {Lepage}(1989)}]{lepage1989}%
  \BibitemOpen
  \bibfield  {author} {\bibinfo {author} {\bibnamefont {Lepage}, \bibfnamefont
  {G~P}}} (\bibinfo {year} {1989}),\ \href@noop {} {\emph {\bibinfo {title}
  {From Actions to Answers (TASI-89)}}},\ edited by\ \bibinfo {editor}
  {\bibfnamefont {T.}~\bibnamefont {DeGrand}}\ and\ \bibinfo {editor}
  {\bibfnamefont {D.}~\bibnamefont {Toussaint}}\ (\bibinfo  {publisher} {World
  Scientific, Singapore})\ \Eprint {http://arxiv.org/abs/G.~P.~Lepage
  nucl-th/9706029} {G.~P.~Lepage nucl-th/9706029} \BibitemShut {NoStop}%
\bibitem [{\citenamefont {Lepage}(1997)}]{lepage1997ARX}%
  \BibitemOpen
  \bibfield  {author} {\bibinfo {author} {\bibnamefont {Lepage}, \bibfnamefont
  {G~P}}} (\bibinfo {year} {1997}),\ \bibfield  {title} {\enquote {\bibinfo
  {title} {How to renormalize the {S}chr\"{o}dinger equation},}\ }\href@noop {}
  {\bibinfo  {journal} {arXiv}\ ,\ \bibinfo {pages}
  {nucl--th/9706029}}\BibitemShut {NoStop}%
\bibitem [{\citenamefont {Lepetit}\ \emph {et~al.}(1990)\citenamefont
  {Lepetit}, \citenamefont {Peng},\ and\ \citenamefont
  {Kuppermann}}]{Lepetit-1990}%
  \BibitemOpen
\bibfield  {journal} {  }\bibfield  {author} {\bibinfo {author} {\bibnamefont
  {Lepetit}, \bibfnamefont {B}}, \bibinfo {author} {\bibfnamefont
  {Z.}~\bibnamefont {Peng}}, \ and\ \bibinfo {author} {\bibfnamefont
  {A.}~\bibnamefont {Kuppermann}}} (\bibinfo {year} {1990}),\ \bibfield
  {title} {\enquote {\bibinfo {title} {Calculation of bound rovibrational
  states on teh first electronically excited state of the {H}$_3$ system},}\
  }\href@noop {} {\bibfield  {journal} {\bibinfo  {journal} {Chem. Phys.
  Lett.}\ }\textbf {\bibinfo {volume} {166}},\ \bibinfo {pages}
  {572}}\BibitemShut {NoStop}%
\bibitem [{\citenamefont {Levine}\ and\ \citenamefont
  {Bernstein}(1987)}]{Levine}%
  \BibitemOpen
  \bibfield  {author} {\bibinfo {author} {\bibnamefont {Levine}, \bibfnamefont
  {R~D}}, \ and\ \bibinfo {author} {\bibfnamefont {R.~B.}\ \bibnamefont
  {Bernstein}}} (\bibinfo {year} {1987}),\ \href@noop {} {\emph {\bibinfo
  {title} {Molecular Reaction Dynamics and Chemical Reactivity}}}\ (\bibinfo
  {publisher} {Oxford University Press},\ \bibinfo {address} {New
  York})\BibitemShut {NoStop}%
\bibitem [{\citenamefont {Levinger}(1974)}]{Levinger}%
  \BibitemOpen
  \bibfield  {author} {\bibinfo {author} {\bibnamefont {Levinger},
  \bibfnamefont {J~S}}} (\bibinfo {year} {1974}),\ \enquote {\bibinfo {title}
  {The two and three body problem},}\ \ (\bibinfo  {publisher} {Springer},\
  \bibinfo {address} {Berlin})\ p.~\bibinfo {pages} {88}\BibitemShut {NoStop}%
\bibitem [{\citenamefont {Levinsen}\ \emph {et~al.}(2014)\citenamefont
  {Levinsen}, \citenamefont {Massignan},\ and\ \citenamefont
  {Parish}}]{Levinsen2014}%
  \BibitemOpen
  \bibfield  {author} {\bibinfo {author} {\bibnamefont {Levinsen},
  \bibfnamefont {Jesper}}, \bibinfo {author} {\bibfnamefont {Pietro}\
  \bibnamefont {Massignan}}, \ and\ \bibinfo {author} {\bibfnamefont
  {Meera~M.}\ \bibnamefont {Parish}}} (\bibinfo {year} {2014}),\ \bibfield
  {title} {\enquote {\bibinfo {title} {Efimov trimers under strong
  confinement},}\ }\href@noop {} {\bibfield  {journal} {\bibinfo  {journal}
  {Physical Review X}\ }\textbf {\bibinfo {volume} {4}}~(\bibinfo {number}
  {3}),\ \bibinfo {pages} {031020}}\BibitemShut {NoStop}%
\bibitem [{\citenamefont {Levinsen}\ \emph {et~al.}(2015)\citenamefont
  {Levinsen}, \citenamefont {Parish},\ and\ \citenamefont
  {Bruun}}]{levinsen_impurity_2015}%
  \BibitemOpen
  \bibfield  {author} {\bibinfo {author} {\bibnamefont {Levinsen},
  \bibfnamefont {Jesper}}, \bibinfo {author} {\bibfnamefont {Meera~M.}\
  \bibnamefont {Parish}}, \ and\ \bibinfo {author} {\bibfnamefont {Georg~M.}\
  \bibnamefont {Bruun}}} (\bibinfo {year} {2015}),\ \bibfield  {title}
  {\enquote {\bibinfo {title} {Impurity in a {Bose}-{Einstein} {Condensate} and
  the {Efimov} {Effect}},}\ }\href {\doibase 10.1103/PhysRevLett.115.125302}
  {\bibfield  {journal} {\bibinfo  {journal} {Physical Review Letters}\
  }\textbf {\bibinfo {volume} {115}}~(\bibinfo {number} {12}),\ \bibinfo
  {pages} {125302}}\BibitemShut {NoStop}%
\bibitem [{\citenamefont {Lewenstein}\ \emph {et~al.}(2012)\citenamefont
  {Lewenstein}, \citenamefont {Sanpera},\ and\ \citenamefont
  {Ahufinger}}]{lewenstein2012ultracold}%
  \BibitemOpen
  \bibfield  {author} {\bibinfo {author} {\bibnamefont {Lewenstein},
  \bibfnamefont {M}}, \bibinfo {author} {\bibfnamefont {A.}~\bibnamefont
  {Sanpera}}, \ and\ \bibinfo {author} {\bibfnamefont {V.}~\bibnamefont
  {Ahufinger}}} (\bibinfo {year} {2012}),\ \href@noop {} {\emph {\bibinfo
  {title} {Ultracold Atoms in Optical Lattices: Simulating quantum many-body
  systems}}}\ (\bibinfo  {publisher} {Oxford University Press},\ \bibinfo
  {address} {London})\BibitemShut {NoStop}%
\bibitem [{\citenamefont {Lewenstein}\ \emph {et~al.}(2007)\citenamefont
  {Lewenstein}, \citenamefont {Sanpera}, \citenamefont {Ahufinger},
  \citenamefont {Damski}, \citenamefont {Sen},\ and\ \citenamefont
  {Sen}}]{lewenstein2007ultracold}%
  \BibitemOpen
  \bibfield  {author} {\bibinfo {author} {\bibnamefont {Lewenstein},
  \bibfnamefont {M}}, \bibinfo {author} {\bibfnamefont {A.}~\bibnamefont
  {Sanpera}}, \bibinfo {author} {\bibfnamefont {V.}~\bibnamefont {Ahufinger}},
  \bibinfo {author} {\bibfnamefont {B.}~\bibnamefont {Damski}}, \bibinfo
  {author} {\bibfnamefont {A.}~\bibnamefont {Sen}}, \ and\ \bibinfo {author}
  {\bibfnamefont {U.}~\bibnamefont {Sen}}} (\bibinfo {year} {2007}),\ \bibfield
   {title} {\enquote {\bibinfo {title} {Ultracold atomic gases in optical
  lattices: mimicking condensed matter physics and beyond},}\ }\href@noop {}
  {\bibfield  {journal} {\bibinfo  {journal} {Adv. Phys.}\ }\textbf {\bibinfo
  {volume} {56}}~(\bibinfo {number} {2}),\ \bibinfo {pages}
  {243--379}}\BibitemShut {NoStop}%
\bibitem [{\citenamefont {Lewerenz}(1997)}]{lewerenz1997JCP}%
  \BibitemOpen
  \bibfield  {author} {\bibinfo {author} {\bibnamefont {Lewerenz},
  \bibfnamefont {M}}} (\bibinfo {year} {1997}),\ \bibfield  {title}
  {{\selectlanguage {English}\enquote {\bibinfo {title} {Structure and
  energetics of small helium clusters: Quantum simulations using a recent
  perturbational pair potential},}\ }}\href@noop {} {\bibfield  {journal}
  {\bibinfo  {journal} {J. Chem. Phys.}\ }\textbf {\bibinfo {volume}
  {106}}~(\bibinfo {number} {11}),\ \bibinfo {pages} {4596--4603}}\BibitemShut
  {NoStop}%
\bibitem [{\citenamefont {Leyton}\ \emph {et~al.}(2014)\citenamefont {Leyton},
  \citenamefont {Roghani}, \citenamefont {Peano},\ and\ \citenamefont
  {Thorwart}}]{leyton2014photon}%
  \BibitemOpen
  \bibfield  {author} {\bibinfo {author} {\bibnamefont {Leyton}, \bibfnamefont
  {V}}, \bibinfo {author} {\bibfnamefont {M.}~\bibnamefont {Roghani}}, \bibinfo
  {author} {\bibfnamefont {V.}~\bibnamefont {Peano}}, \ and\ \bibinfo {author}
  {\bibfnamefont {M.}~\bibnamefont {Thorwart}}} (\bibinfo {year} {2014}),\
  \bibfield  {title} {\enquote {\bibinfo {title} {Photon-assisted
  confinement-induced resonances for ultracold atoms},}\ }\href@noop {}
  {\bibfield  {journal} {\bibinfo  {journal} {Phys. Rev. Lett.}\ }\textbf
  {\bibinfo {volume} {112}}~(\bibinfo {number} {23}),\ \bibinfo {pages}
  {233201}}\BibitemShut {NoStop}%
\bibitem [{\citenamefont {Li}\ and\ \citenamefont
  {Das~Sarma}(2014)}]{li_variational_2014}%
  \BibitemOpen
  \bibfield  {author} {\bibinfo {author} {\bibnamefont {Li}, \bibfnamefont
  {Weiran}}, \ and\ \bibinfo {author} {\bibfnamefont {S.}~\bibnamefont
  {Das~Sarma}}} (\bibinfo {year} {2014}),\ \bibfield  {title} {\enquote
  {\bibinfo {title} {Variational study of polarons in {Bose}-{Einstein}
  condensates},}\ }\href {\doibase 10.1103/PhysRevA.90.013618} {\bibfield
  {journal} {\bibinfo  {journal} {Physical Review A}\ }\textbf {\bibinfo
  {volume} {90}}~(\bibinfo {number} {1}),\ \bibinfo {pages}
  {013618}}\BibitemShut {NoStop}%
\bibitem [{\citenamefont {Lieb}\ \emph {et~al.}(2004)\citenamefont {Lieb},
  \citenamefont {Seiringer},\ and\ \citenamefont {Yngvason}}]{lieb2004one}%
  \BibitemOpen
  \bibfield  {author} {\bibinfo {author} {\bibnamefont {Lieb}, \bibfnamefont
  {E~H}}, \bibinfo {author} {\bibfnamefont {R.}~\bibnamefont {Seiringer}}, \
  and\ \bibinfo {author} {\bibfnamefont {J.}~\bibnamefont {Yngvason}}}
  (\bibinfo {year} {2004}),\ \bibfield  {title} {\enquote {\bibinfo {title}
  {One-dimensional behavior of dilute, trapped bose gases},}\ }\href@noop {}
  {\bibfield  {journal} {\bibinfo  {journal} {Commun. Math. Phys.}\ }\textbf
  {\bibinfo {volume} {244}}~(\bibinfo {number} {2}),\ \bibinfo {pages}
  {347--393}}\BibitemShut {NoStop}%
\bibitem [{\citenamefont {Lin}(1986)}]{lin1986AMOP}%
  \BibitemOpen
  \bibfield  {author} {\bibinfo {author} {\bibnamefont {Lin}, \bibfnamefont
  {C~D}}} (\bibinfo {year} {1986}),\ \bibfield  {title} {{\selectlanguage
  {English}\enquote {\bibinfo {title} {Doubly excited-states, including new
  classification schemes},}\ }}\href@noop {} {\bibfield  {journal} {\bibinfo
  {journal} {Adv. At. Mol. Opt. Phys.}\ }\textbf {\bibinfo {volume} {22}},\
  \bibinfo {pages} {77--142}}\BibitemShut {NoStop}%
\bibitem [{\citenamefont {Lin}(1995)}]{lin1995PRep}%
  \BibitemOpen
  \bibfield  {author} {\bibinfo {author} {\bibnamefont {Lin}, \bibfnamefont
  {C~D}}} (\bibinfo {year} {1995}),\ \bibfield  {title} {\enquote {\bibinfo
  {title} {Hyperspherical coordinate approach to atomic and other {C}oulombic
  three-body systems},}\ }\href@noop {} {\bibfield  {journal} {\bibinfo
  {journal} {Phys. Rep.}\ }\textbf {\bibinfo {volume} {257}}~(\bibinfo {number}
  {1}),\ \bibinfo {pages} {1--83}}\BibitemShut {NoStop}%
\bibitem [{\citenamefont {Lin}\ and\ \citenamefont
  {Morishita}(2000)}]{lin2000PHYSICSESSAYS}%
  \BibitemOpen
  \bibfield  {author} {\bibinfo {author} {\bibnamefont {Lin}, \bibfnamefont
  {C~D}}, \ and\ \bibinfo {author} {\bibfnamefont {T.}~\bibnamefont
  {Morishita}}} (\bibinfo {year} {2000}),\ \bibfield  {title} {{\selectlanguage
  {English}\enquote {\bibinfo {title} {Few-body problems: The hyperspherical
  way},}\ }}\href@noop {} {\bibfield  {journal} {\bibinfo  {journal} {Physics
  Essays}\ }\textbf {\bibinfo {volume} {13}}~(\bibinfo {number} {2-3, Sp. Iss.
  SI}),\ \bibinfo {pages} {367--376}}\BibitemShut {NoStop}%
\bibitem [{\citenamefont {Liu}\ \emph {et~al.}({2009})\citenamefont {Liu},
  \citenamefont {Hu},\ and\ \citenamefont {Drummond}}]{LiuDrummond2009prl}%
  \BibitemOpen
  \bibfield  {author} {\bibinfo {author} {\bibnamefont {Liu}, \bibfnamefont
  {Xia-Ji}}, \bibinfo {author} {\bibfnamefont {Hui}\ \bibnamefont {Hu}}, \ and\
  \bibinfo {author} {\bibfnamefont {Peter~D.}\ \bibnamefont {Drummond}}}
  (\bibinfo {year} {{2009}}),\ \bibfield  {title} {\enquote {\bibinfo {title}
  {{Virial Expansion for a Strongly Correlated Fermi Gas}},}\ }\href {\doibase
  {10.1103/PhysRevLett.102.160401}} {\bibfield  {journal} {\bibinfo  {journal}
  {{Physical Review Letters}}\ }\textbf {\bibinfo {volume} {{102}}}~(\bibinfo
  {number} {{16}}),\ {10.1103/PhysRevLett.102.160401}}\BibitemShut {NoStop}%
\bibitem [{\citenamefont {Loftus}\ \emph {et~al.}(2002)\citenamefont {Loftus},
  \citenamefont {Regal}, \citenamefont {Ticknor}, \citenamefont {Bohn},\ and\
  \citenamefont {Jin}}]{loftus2002PRL}%
  \BibitemOpen
  \bibfield  {author} {\bibinfo {author} {\bibnamefont {Loftus}, \bibfnamefont
  {T}}, \bibinfo {author} {\bibfnamefont {C.~A.}\ \bibnamefont {Regal}},
  \bibinfo {author} {\bibfnamefont {C.}~\bibnamefont {Ticknor}}, \bibinfo
  {author} {\bibfnamefont {J.~L.}\ \bibnamefont {Bohn}}, \ and\ \bibinfo
  {author} {\bibfnamefont {D.~S.}\ \bibnamefont {Jin}}} (\bibinfo {year}
  {2002}),\ \bibfield  {title} {\enquote {\bibinfo {title} {Resonant control of
  elastic collisions in an optically trapped {F}ermi gas of atoms},}\
  }\href@noop {} {\bibfield  {journal} {\bibinfo  {journal} {Phys. Rev. Lett.}\
  }\textbf {\bibinfo {volume} {88}},\ \bibinfo {pages} {173201}}\BibitemShut
  {NoStop}%
\bibitem [{\citenamefont {Lompe}\ \emph
  {et~al.}(2010{\natexlab{a}})\citenamefont {Lompe}, \citenamefont
  {Ottenstein}, \citenamefont {Serwane}, \citenamefont {Viering}, \citenamefont
  {Wenz}, \citenamefont {Z\"urn},\ and\ \citenamefont {Jochim}}]{lompe2010PRL}%
  \BibitemOpen
  \bibfield  {author} {\bibinfo {author} {\bibnamefont {Lompe}, \bibfnamefont
  {T}}, \bibinfo {author} {\bibfnamefont {T.~B.}\ \bibnamefont {Ottenstein}},
  \bibinfo {author} {\bibfnamefont {F.}~\bibnamefont {Serwane}}, \bibinfo
  {author} {\bibfnamefont {K.}~\bibnamefont {Viering}}, \bibinfo {author}
  {\bibfnamefont {A.~N.}\ \bibnamefont {Wenz}}, \bibinfo {author}
  {\bibfnamefont {G.}~\bibnamefont {Z\"urn}}, \ and\ \bibinfo {author}
  {\bibfnamefont {S.}~\bibnamefont {Jochim}}} (\bibinfo {year}
  {2010}{\natexlab{a}}),\ \bibfield  {title} {\enquote {\bibinfo {title}
  {Atom-dimer scattering in a three-component fermi gas},}\ }\href@noop {}
  {\bibfield  {journal} {\bibinfo  {journal} {Phys. Rev. Lett.}\ }\textbf
  {\bibinfo {volume} {105}},\ \bibinfo {pages} {103201}}\BibitemShut {NoStop}%
\bibitem [{\citenamefont {Lompe}\ \emph
  {et~al.}(2010{\natexlab{b}})\citenamefont {Lompe}, \citenamefont {Ottestein},
  \citenamefont {Serwane}, \citenamefont {Wenz}, \citenamefont {Z\"urn},\ and\
  \citenamefont {Joachim}}]{Lompe-2010b}%
  \BibitemOpen
  \bibfield  {author} {\bibinfo {author} {\bibnamefont {Lompe}, \bibfnamefont
  {T}}, \bibinfo {author} {\bibfnamefont {T.~B.}\ \bibnamefont {Ottestein}},
  \bibinfo {author} {\bibfnamefont {F.}~\bibnamefont {Serwane}}, \bibinfo
  {author} {\bibfnamefont {A.~N.}\ \bibnamefont {Wenz}}, \bibinfo {author}
  {\bibfnamefont {G.}~\bibnamefont {Z\"urn}}, \ and\ \bibinfo {author}
  {\bibfnamefont {S.}~\bibnamefont {Joachim}}} (\bibinfo {year}
  {2010}{\natexlab{b}}),\ \bibfield  {title} {\enquote {\bibinfo {title}
  {Radio-frequency association of efimov trimers},}\ }\href@noop {} {\bibfield
  {journal} {\bibinfo  {journal} {Science}\ }\textbf {\bibinfo {volume}
  {330}},\ \bibinfo {pages} {940--944}}\BibitemShut {NoStop}%
\bibitem [{\citenamefont {Lonardoni}\ \emph {et~al.}(2014)\citenamefont
  {Lonardoni}, \citenamefont {Pederiva},\ and\ \citenamefont
  {Gandolfi}}]{Lonardoni-2014}%
  \BibitemOpen
  \bibfield  {author} {\bibinfo {author} {\bibnamefont {Lonardoni},
  \bibfnamefont {D}}, \bibinfo {author} {\bibfnamefont {F.}~\bibnamefont
  {Pederiva}}, \ and\ \bibinfo {author} {\bibfnamefont {S.}~\bibnamefont
  {Gandolfi}}} (\bibinfo {year} {2014}),\ \bibfield  {title} {\enquote
  {\bibinfo {title} {From hypernuclei to the inner core of neutron stars: A
  quantum {M}onte {C}arlo study},}\ }\href@noop {} {\bibfield  {journal}
  {\bibinfo  {journal} {J. Phys. Conf. Ser.}\ }\textbf {\bibinfo {volume}
  {529}},\ \bibinfo {pages} {012012}}\BibitemShut {NoStop}%
\bibitem [{\citenamefont {Macek}(1968)}]{macek1968JPB}%
  \BibitemOpen
  \bibfield  {author} {\bibinfo {author} {\bibnamefont {Macek}, \bibfnamefont
  {J}}} (\bibinfo {year} {1968}),\ \bibfield  {title} {\enquote {\bibinfo
  {title} {Properties of autoionizing states of {He}},}\ }\href@noop {}
  {\bibfield  {journal} {\bibinfo  {journal} {J. Phys. B}\ }\textbf {\bibinfo
  {volume} {1}}~(\bibinfo {number} {5}),\ \bibinfo {pages} {831}}\BibitemShut
  {NoStop}%
\bibitem [{\citenamefont {Macek}(1986)}]{macek1986ZPD}%
  \BibitemOpen
  \bibfield  {author} {\bibinfo {author} {\bibnamefont {Macek}, \bibfnamefont
  {J}}} (\bibinfo {year} {1986}),\ \bibfield  {title} {\enquote {\bibinfo
  {title} {Loosely bound states of three particles},}\ }\href@noop {}
  {\bibfield  {journal} {\bibinfo  {journal} {Z. Phys. D}\ }\textbf {\bibinfo
  {volume} {3}}~(\bibinfo {number} {1}),\ \bibinfo {pages} {31}}\BibitemShut
  {NoStop}%
\bibitem [{\citenamefont {Macek}\ and\ \citenamefont
  {Jerjian}(1986)}]{macek1986PRA}%
  \BibitemOpen
  \bibfield  {author} {\bibinfo {author} {\bibnamefont {Macek}, \bibfnamefont
  {J}}, \ and\ \bibinfo {author} {\bibfnamefont {K.~A.}\ \bibnamefont
  {Jerjian}}} (\bibinfo {year} {1986}),\ \bibfield  {title} {{\selectlanguage
  {English}\enquote {\bibinfo {title} {Adiabatic hyperspherical treatment of
  {HD}$^+$},}\ }}\href@noop {} {\bibfield  {journal} {\bibinfo  {journal}
  {Phys. Rev. A}\ }\textbf {\bibinfo {volume} {33}}~(\bibinfo {number} {1}),\
  \bibinfo {pages} {233--241}}\BibitemShut {NoStop}%
\bibitem [{\citenamefont {Macek}(2002)}]{macek2002FBS}%
  \BibitemOpen
  \bibfield  {author} {\bibinfo {author} {\bibnamefont {Macek}, \bibfnamefont
  {J~H}}} (\bibinfo {year} {2002}),\ \bibfield  {title} {{\selectlanguage
  {English}\enquote {\bibinfo {title} {Multichannel zero-range potentials in
  the hyperspherical theory of three-body dynamics},}\ }}\href@noop {}
  {\bibfield  {journal} {\bibinfo  {journal} {Few-Body Systems}\ }\textbf
  {\bibinfo {volume} {31}}~(\bibinfo {number} {2-4}),\ \bibinfo {pages}
  {241--248}}\BibitemShut {NoStop}%
\bibitem [{\citenamefont {Macek}\ \emph {et~al.}(2005)\citenamefont {Macek},
  \citenamefont {Ovchinnikov},\ and\ \citenamefont {Gasaneo}}]{macek2005PRA}%
  \BibitemOpen
  \bibfield  {author} {\bibinfo {author} {\bibnamefont {Macek}, \bibfnamefont
  {J~H}}, \bibinfo {author} {\bibfnamefont {S.}~\bibnamefont {Ovchinnikov}}, \
  and\ \bibinfo {author} {\bibfnamefont {G.}~\bibnamefont {Gasaneo}}} (\bibinfo
  {year} {2005}),\ \bibfield  {title} {{\selectlanguage {English}\enquote
  {\bibinfo {title} {Solution for boson-diboson elastic scattering at zero
  energy in the shape-independent model},}\ }}\href@noop {} {\bibfield
  {journal} {\bibinfo  {journal} {Phys. Rev. A}\ }\textbf {\bibinfo {volume}
  {72}}~(\bibinfo {number} {3}),\ \bibinfo {pages} {032709}}\BibitemShut
  {NoStop}%
\bibitem [{\citenamefont {Macek}\ \emph {et~al.}(2006)\citenamefont {Macek},
  \citenamefont {Yu~Ovchinnikov},\ and\ \citenamefont
  {Gasaneo}}]{macek2006PRA}%
  \BibitemOpen
  \bibfield  {author} {\bibinfo {author} {\bibnamefont {Macek}, \bibfnamefont
  {J~H}}, \bibinfo {author} {\bibfnamefont {S.}~\bibnamefont {Yu~Ovchinnikov}},
  \ and\ \bibinfo {author} {\bibfnamefont {G.}~\bibnamefont {Gasaneo}}}
  (\bibinfo {year} {2006}),\ \bibfield  {title} {\enquote {\bibinfo {title}
  {Exact solution for three particles interacting via zero-range potentials},}\
  }\href@noop {} {\bibfield  {journal} {\bibinfo  {journal} {Phys. Rev. A}\
  }\textbf {\bibinfo {volume} {73}},\ \bibinfo {pages} {032704}}\BibitemShut
  {NoStop}%
\bibitem [{\citenamefont {Machtey}\ \emph
  {et~al.}(2012{\natexlab{a}})\citenamefont {Machtey}, \citenamefont
  {Kessler},\ and\ \citenamefont {Khaykovich}}]{machtey2012PRL}%
  \BibitemOpen
  \bibfield  {author} {\bibinfo {author} {\bibnamefont {Machtey}, \bibfnamefont
  {O}}, \bibinfo {author} {\bibfnamefont {D.~A.}\ \bibnamefont {Kessler}}, \
  and\ \bibinfo {author} {\bibfnamefont {L.}~\bibnamefont {Khaykovich}}}
  (\bibinfo {year} {2012}{\natexlab{a}}),\ \bibfield  {title} {\enquote
  {\bibinfo {title} {Universal dimer in a collisionally opaque medium:
  Experimental observables and {E}fimov resonances},}\ }\href@noop {}
  {\bibfield  {journal} {\bibinfo  {journal} {Phys. Rev. Lett.}\ }\textbf
  {\bibinfo {volume} {108}},\ \bibinfo {pages} {130403}}\BibitemShut {NoStop}%
\bibitem [{\citenamefont {Machtey}\ \emph
  {et~al.}(2012{\natexlab{b}})\citenamefont {Machtey}, \citenamefont {Shotan},
  \citenamefont {Gross},\ and\ \citenamefont {Khaykovich}}]{machtey2012PRLb}%
  \BibitemOpen
  \bibfield  {author} {\bibinfo {author} {\bibnamefont {Machtey}, \bibfnamefont
  {O}}, \bibinfo {author} {\bibfnamefont {Z.}~\bibnamefont {Shotan}}, \bibinfo
  {author} {\bibfnamefont {N.}~\bibnamefont {Gross}}, \ and\ \bibinfo {author}
  {\bibfnamefont {L.}~\bibnamefont {Khaykovich}}} (\bibinfo {year}
  {2012}{\natexlab{b}}),\ \bibfield  {title} {\enquote {\bibinfo {title}
  {Association of {E}fimov trimers from a three-atom continuum},}\ }\href@noop
  {} {\bibfield  {journal} {\bibinfo  {journal} {Phys. Rev. Lett.}\ }\textbf
  {\bibinfo {volume} {108}},\ \bibinfo {pages} {210406}}\BibitemShut {NoStop}%
\bibitem [{\citenamefont {Maier}\ \emph {et~al.}(2015)\citenamefont {Maier},
  \citenamefont {Eisele}, \citenamefont {Tiemann},\ and\ \citenamefont
  {Zimmermann}}]{Maier-2015}%
  \BibitemOpen
  \bibfield  {author} {\bibinfo {author} {\bibnamefont {Maier}, \bibfnamefont
  {R~A~W}}, \bibinfo {author} {\bibfnamefont {M.}~\bibnamefont {Eisele}},
  \bibinfo {author} {\bibfnamefont {E.}~\bibnamefont {Tiemann}}, \ and\
  \bibinfo {author} {\bibfnamefont {C.}~\bibnamefont {Zimmermann}}} (\bibinfo
  {year} {2015}),\ \bibfield  {title} {\enquote {\bibinfo {title} {Efimov
  resonance and three-body parameter in a lithium-rubidium mixture},}\
  }\href@noop {} {\bibfield  {journal} {\bibinfo  {journal} {Phys. Rev. Lett.}\
  }\textbf {\bibinfo {volume} {115}},\ \bibinfo {pages} {043201}}\BibitemShut
  {NoStop}%
\bibitem [{\citenamefont {Makotyn}\ \emph {et~al.}(2014)\citenamefont
  {Makotyn}, \citenamefont {Klauss}, \citenamefont {Goldberger}, \citenamefont
  {Cornell},\ and\ \citenamefont {Jin}}]{Makotyn:NaturePhysics:2014}%
  \BibitemOpen
  \bibfield  {author} {\bibinfo {author} {\bibnamefont {Makotyn}, \bibfnamefont
  {P}}, \bibinfo {author} {\bibfnamefont {C.~E. .~E.}\ \bibnamefont {Klauss}},
  \bibinfo {author} {\bibfnamefont {D.~L. .~L.}\ \bibnamefont {Goldberger}},
  \bibinfo {author} {\bibfnamefont {E.~A.}\ \bibnamefont {Cornell}}, \ and\
  \bibinfo {author} {\bibfnamefont {D.~S.}\ \bibnamefont {Jin}}} (\bibinfo
  {year} {2014}),\ \bibfield  {title} {\enquote {\bibinfo {title} {Universal
  dynamics of a degenerate unitary {B}ose gas},}\ }\href@noop {} {\bibfield
  {journal} {\bibinfo  {journal} {Nat. Phys.}\ }\textbf {\bibinfo {volume}
  {10}}~(\bibinfo {number} {2}),\ \bibinfo {pages} {116--119}}\BibitemShut
  {NoStop}%
\bibitem [{\citenamefont {Malegat}({2003})}]{malegat2003PS}%
  \BibitemOpen
  \bibfield  {author} {\bibinfo {author} {\bibnamefont {Malegat}, \bibfnamefont
  {L}}} (\bibinfo {year} {{2003}}),\ \bibfield  {title} {\enquote {\bibinfo
  {title} {{On the theoretical treatment of the double electronic
  continuum}},}\ }\href@noop {} {\bibfield  {journal} {\bibinfo  {journal}
  {Physica Scripta}\ }\textbf {\bibinfo {volume} {{68}}}~(\bibinfo {number}
  {{6}}),\ \bibinfo {pages} {{C113--C117}}}\BibitemShut {NoStop}%
\bibitem [{\citenamefont {Malegat}({2004})}]{malegat2004PSb}%
  \BibitemOpen
  \bibfield  {author} {\bibinfo {author} {\bibnamefont {Malegat}, \bibfnamefont
  {L}}} (\bibinfo {year} {{2004}}),\ \bibfield  {title} {\enquote {\bibinfo
  {title} {{(e, 2e) and ($\gamma$, 2e) processes: Open and closed
  questions}},}\ }\href@noop {} {\bibfield  {journal} {\bibinfo  {journal}
  {Physica Scripta}\ }\textbf {\bibinfo {volume} {{T110}}},\ \bibinfo {pages}
  {{83--89}}}\BibitemShut {NoStop}%
\bibitem [{\citenamefont {Manolopoulos}\ \emph {et~al.}(1993)\citenamefont
  {Manolopoulos}, \citenamefont {Jamieson},\ and\ \citenamefont
  {Pradhan}}]{MANOLOPOULOS1993}%
  \BibitemOpen
  \bibfield  {author} {\bibinfo {author} {\bibnamefont {Manolopoulos},
  \bibfnamefont {David~E}}, \bibinfo {author} {\bibfnamefont {Michael~J.}\
  \bibnamefont {Jamieson}}, \ and\ \bibinfo {author} {\bibfnamefont {Atul~D.}\
  \bibnamefont {Pradhan}}} (\bibinfo {year} {1993}),\ \bibfield  {title}
  {\enquote {\bibinfo {title} {Johnson's log derivative algorithm rederived},}\
  }\href {\doibase http://dx.doi.org/10.1006/jcph.1993.1062} {\bibfield
  {journal} {\bibinfo  {journal} {Journal of Computational Physics}\ }\textbf
  {\bibinfo {volume} {105}}~(\bibinfo {number} {1}),\ \bibinfo {pages} {169 --
  172}}\BibitemShut {NoStop}%
\bibitem [{\citenamefont {Mansbach}\ and\ \citenamefont
  {Keck}(1969)}]{mansbach1969pr}%
  \BibitemOpen
  \bibfield  {author} {\bibinfo {author} {\bibnamefont {Mansbach},
  \bibfnamefont {P}}, \ and\ \bibinfo {author} {\bibfnamefont {J.}~\bibnamefont
  {Keck}}} (\bibinfo {year} {1969}),\ \bibfield  {title} {\enquote {\bibinfo
  {title} {{Monte Carlo} trajectory calculations of atomic excitation and
  ionization by thermal electrons},}\ }\href@noop {} {\bibfield  {journal}
  {\bibinfo  {journal} {Phys. Rev.}\ }\textbf {\bibinfo {volume} {181}},\
  \bibinfo {pages} {275--289}}\BibitemShut {NoStop}%
\bibitem [{\citenamefont {Maslov}\ and\ \citenamefont
  {Fedoriuk}(1981)}]{Maslov}%
  \BibitemOpen
  \bibfield  {author} {\bibinfo {author} {\bibnamefont {Maslov}, \bibfnamefont
  {V~P}}, \ and\ \bibinfo {author} {\bibfnamefont {M.~V.}\ \bibnamefont
  {Fedoriuk}}} (\bibinfo {year} {1981}),\ \href@noop {} {\emph {\bibinfo
  {title} {Semi-classical approximation in quantum mechanics}}}\ (\bibinfo
  {publisher} {D. Reidel publishing company},\ \bibinfo {address} {London,
  England})\BibitemShut {NoStop}%
\bibitem [{\citenamefont {Mason}\ and\ \citenamefont
  {Monchick}(1962)}]{Mason-1962}%
  \BibitemOpen
  \bibfield  {author} {\bibinfo {author} {\bibnamefont {Mason}, \bibfnamefont
  {E~A}}, \ and\ \bibinfo {author} {\bibfnamefont {L.}~\bibnamefont
  {Monchick}}} (\bibinfo {year} {1962}),\ \bibfield  {title} {\enquote
  {\bibinfo {title} {Heat conductivity of polyatomic and polar gases},}\
  }\href@noop {} {\bibfield  {journal} {\bibinfo  {journal} {J. Chem. Phys.}\
  }\textbf {\bibinfo {volume} {36}},\ \bibinfo {pages} {1622}}\BibitemShut
  {NoStop}%
\bibitem [{\citenamefont {Massignan}\ and\ \citenamefont
  {Castin}(2006)}]{massigan2006pra}%
  \BibitemOpen
  \bibfield  {author} {\bibinfo {author} {\bibnamefont {Massignan},
  \bibfnamefont {P}}, \ and\ \bibinfo {author} {\bibfnamefont {Y.}~\bibnamefont
  {Castin}}} (\bibinfo {year} {2006}),\ \bibfield  {title} {\enquote {\bibinfo
  {title} {Three-dimensional strong localization of matter waves by scattering
  from atoms in a lattice with a confinement-induced resonance},}\ }\href@noop
  {} {\bibfield  {journal} {\bibinfo  {journal} {Phys. Rev. A}\ }\textbf
  {\bibinfo {volume} {74}},\ \bibinfo {pages} {013616}}\BibitemShut {NoStop}%
\bibitem [{\citenamefont {McCourt}\ \emph {et~al.}(1991)\citenamefont
  {McCourt}, \citenamefont {Beenaker}, \citenamefont {K\"ohler},\ and\
  \citenamefont {Kuscer}}]{McCourt}%
  \BibitemOpen
  \bibfield  {author} {\bibinfo {author} {\bibnamefont {McCourt}, \bibfnamefont
  {F~R~W}}, \bibinfo {author} {\bibfnamefont {J.~J.~M.}\ \bibnamefont
  {Beenaker}}, \bibinfo {author} {\bibfnamefont {W.~E.}\ \bibnamefont
  {K\"ohler}}, \ and\ \bibinfo {author} {\bibfnamefont {I.}~\bibnamefont
  {Kuscer}}} (\bibinfo {year} {1991}),\ \href@noop {} {\emph {\bibinfo {title}
  {Nonequilibrium Phenomena in Polyatomic Gases}}}\ (\bibinfo  {publisher}
  {Oxford University Press},\ \bibinfo {address} {Clarendon,
  Oxfort})\BibitemShut {NoStop}%
\bibitem [{\citenamefont {McCurdy}\ \emph {et~al.}(2004)\citenamefont
  {McCurdy}, \citenamefont {Baertschy},\ and\ \citenamefont
  {Rescigno}}]{McCurdy2004JPB}%
  \BibitemOpen
  \bibfield  {author} {\bibinfo {author} {\bibnamefont {McCurdy}, \bibfnamefont
  {C~W}}, \bibinfo {author} {\bibfnamefont {M}~\bibnamefont {Baertschy}}, \
  and\ \bibinfo {author} {\bibfnamefont {T~N}\ \bibnamefont {Rescigno}}}
  (\bibinfo {year} {2004}),\ \bibfield  {title} {\enquote {\bibinfo {title}
  {Solving the three-body {C}oulomb breakup problem using exterior complex
  scaling},}\ }\href@noop {} {\bibfield  {journal} {\bibinfo  {journal} {J.
  Phys. B}\ }\textbf {\bibinfo {volume} {37}}~(\bibinfo {number} {17}),\
  \bibinfo {pages} {R137}}\BibitemShut {NoStop}%
\bibitem [{\citenamefont {McCurdy}\ \emph {et~al.}(1997)\citenamefont
  {McCurdy}, \citenamefont {Rescigno},\ and\ \citenamefont
  {Byrum}}]{McCurdy1997}%
  \BibitemOpen
  \bibfield  {author} {\bibinfo {author} {\bibnamefont {McCurdy}, \bibfnamefont
  {C~W}}, \bibinfo {author} {\bibfnamefont {T.~N.}\ \bibnamefont {Rescigno}}, \
  and\ \bibinfo {author} {\bibfnamefont {D.}~\bibnamefont {Byrum}}} (\bibinfo
  {year} {1997}),\ \bibfield  {title} {\enquote {\bibinfo {title} {Approach to
  electron-impact ionization that avoids the three-body {C}oulomb asymptotic
  form},}\ }\href@noop {} {\bibfield  {journal} {\bibinfo  {journal} {Phys.
  Rev. A}\ }\textbf {\bibinfo {volume} {56}}~(\bibinfo {number} {3}),\ \bibinfo
  {pages} {1958--1969}}\BibitemShut {NoStop}%
\bibitem [{\citenamefont {McGuire}(1964)}]{mcguire1964JMP}%
  \BibitemOpen
  \bibfield  {author} {\bibinfo {author} {\bibnamefont {McGuire}, \bibfnamefont
  {J~B}}} (\bibinfo {year} {1964}),\ \bibfield  {title} {\enquote {\bibinfo
  {title} {Study of exactly soluble one-dimensional n-body problems},}\
  }\href@noop {} {\bibfield  {journal} {\bibinfo  {journal} {JMP}\ }\textbf
  {\bibinfo {volume} {5}}~(\bibinfo {number} {5}),\ \bibinfo {pages}
  {622--636}}\BibitemShut {NoStop}%
\bibitem [{\citenamefont {McKay}\ and\ \citenamefont
  {DeMarco}(2011)}]{mckay2011cooling}%
  \BibitemOpen
  \bibfield  {author} {\bibinfo {author} {\bibnamefont {McKay}, \bibfnamefont
  {DC}}, \ and\ \bibinfo {author} {\bibfnamefont {B}~\bibnamefont {DeMarco}}}
  (\bibinfo {year} {2011}),\ \bibfield  {title} {\enquote {\bibinfo {title}
  {Cooling in strongly correlated optical lattices: prospects and
  challenges},}\ }\href@noop {} {\bibfield  {journal} {\bibinfo  {journal}
  {Rep. Prog. Phys.}\ }\textbf {\bibinfo {volume} {74}}~(\bibinfo {number}
  {5}),\ \bibinfo {pages} {054401}}\BibitemShut {NoStop}%
\bibitem [{\citenamefont {Mehta}\ \emph {et~al.}(2007)\citenamefont {Mehta},
  \citenamefont {Esry},\ and\ \citenamefont {Greene}}]{mehta2007PRA}%
  \BibitemOpen
  \bibfield  {author} {\bibinfo {author} {\bibnamefont {Mehta}, \bibfnamefont
  {N~P}}, \bibinfo {author} {\bibfnamefont {B.~D.}\ \bibnamefont {Esry}}, \
  and\ \bibinfo {author} {\bibfnamefont {C.~H.}\ \bibnamefont {Greene}}}
  (\bibinfo {year} {2007}),\ \bibfield  {title} {{\selectlanguage
  {English}\enquote {\bibinfo {title} {Three-body recombination in one
  dimension},}\ }}\href@noop {} {\bibfield  {journal} {\bibinfo  {journal}
  {Phys. Rev. A}\ }\textbf {\bibinfo {volume} {76}}~(\bibinfo {number} {2}),\
  \bibinfo {pages} {022711}}\BibitemShut {NoStop}%
\bibitem [{\citenamefont {Mehta}\ \emph {et~al.}(2009)\citenamefont {Mehta},
  \citenamefont {Rittenhouse}, \citenamefont {D'Incao}, \citenamefont {von
  Stecher},\ and\ \citenamefont {Greene}}]{mehta2009PRL}%
  \BibitemOpen
  \bibfield  {author} {\bibinfo {author} {\bibnamefont {Mehta}, \bibfnamefont
  {N~P}}, \bibinfo {author} {\bibfnamefont {S.~T.}\ \bibnamefont
  {Rittenhouse}}, \bibinfo {author} {\bibfnamefont {J.~P.}\ \bibnamefont
  {D'Incao}}, \bibinfo {author} {\bibfnamefont {J.}~\bibnamefont {von
  Stecher}}, \ and\ \bibinfo {author} {\bibfnamefont {C.~H.}\ \bibnamefont
  {Greene}}} (\bibinfo {year} {2009}),\ \bibfield  {title} {{\selectlanguage
  {English}\enquote {\bibinfo {title} {General theoretical description of
  ${N}$-body recombination},}\ }}\href@noop {} {\bibfield  {journal} {\bibinfo
  {journal} {Phys. Rev. Lett.}\ }\textbf {\bibinfo {volume} {103}}~(\bibinfo
  {number} {15}),\ \bibinfo {pages} {153201}}\BibitemShut {NoStop}%
\bibitem [{\citenamefont {Mehta}\ \emph {et~al.}(2008)\citenamefont {Mehta},
  \citenamefont {Rittenhouse}, \citenamefont {D'Incao},\ and\ \citenamefont
  {Greene}}]{mehta2008PRA}%
  \BibitemOpen
  \bibfield  {author} {\bibinfo {author} {\bibnamefont {Mehta}, \bibfnamefont
  {N~P}}, \bibinfo {author} {\bibfnamefont {Seth~T.}\ \bibnamefont
  {Rittenhouse}}, \bibinfo {author} {\bibfnamefont {J.~P.}\ \bibnamefont
  {D'Incao}}, \ and\ \bibinfo {author} {\bibfnamefont {C.~H.}\ \bibnamefont
  {Greene}}} (\bibinfo {year} {2008}),\ \bibfield  {title} {\enquote {\bibinfo
  {title} {{E}fimov states embedded in the three-body continuum},}\ }\href@noop
  {} {\bibfield  {journal} {\bibinfo  {journal} {Phys. Rev. A}\ }\textbf
  {\bibinfo {volume} {78}},\ \bibinfo {pages} {020701}}\BibitemShut {NoStop}%
\bibitem [{\citenamefont {Mehta}\ and\ \citenamefont
  {Shepard}(2005)}]{mehta2005PRA}%
  \BibitemOpen
  \bibfield  {author} {\bibinfo {author} {\bibnamefont {Mehta}, \bibfnamefont
  {NP}}, \ and\ \bibinfo {author} {\bibfnamefont {JR}~\bibnamefont {Shepard}}}
  (\bibinfo {year} {2005}),\ \bibfield  {title} {\enquote {\bibinfo {title}
  {Three bosons in one dimension with short-range interactions: Zero-range
  potentials},}\ }\href@noop {} {\bibfield  {journal} {\bibinfo  {journal}
  {Phys. Rev. A}\ }\textbf {\bibinfo {volume} {72}}~(\bibinfo {number} {3}),\
  \bibinfo {pages} {032728}}\BibitemShut {NoStop}%
\bibitem [{\citenamefont {Melezhik}\ and\ \citenamefont
  {Schmelcher}(2009)}]{melezhik2009}%
  \BibitemOpen
  \bibfield  {author} {\bibinfo {author} {\bibnamefont {Melezhik},
  \bibfnamefont {V~S}}, \ and\ \bibinfo {author} {\bibfnamefont
  {P.}~\bibnamefont {Schmelcher}}} (\bibinfo {year} {2009}),\ \bibfield
  {title} {\enquote {\bibinfo {title} {Quantum dynamics of resonant molecule
  formation in waveguides},}\ }\href@noop {} {\bibfield  {journal} {\bibinfo
  {journal} {New J. Phys.}\ }\textbf {\bibinfo {volume} {11}}~(\bibinfo
  {number} {7}),\ \bibinfo {pages} {073031}}\BibitemShut {NoStop}%
\bibitem [{\citenamefont {Melezhik}\ and\ \citenamefont
  {Schmelcher}(2011)}]{melezhik2011}%
  \BibitemOpen
  \bibfield  {author} {\bibinfo {author} {\bibnamefont {Melezhik},
  \bibfnamefont {V~S}}, \ and\ \bibinfo {author} {\bibfnamefont
  {P.}~\bibnamefont {Schmelcher}}} (\bibinfo {year} {2011}),\ \bibfield
  {title} {\enquote {\bibinfo {title} {Multichannel effects near
  confinement-induced resonances in harmonic waveguides},}\ }\href@noop {}
  {\bibfield  {journal} {\bibinfo  {journal} {Phys. Rev. A}\ }\textbf {\bibinfo
  {volume} {84}}~(\bibinfo {number} {4}),\ \bibinfo {pages}
  {042712}}\BibitemShut {NoStop}%
\bibitem [{\citenamefont {Meyer}\ and\ \citenamefont
  {Greene}(1994)}]{meyer1994PRA}%
  \BibitemOpen
  \bibfield  {author} {\bibinfo {author} {\bibnamefont {Meyer}, \bibfnamefont
  {K~W}}, \ and\ \bibinfo {author} {\bibfnamefont {C.~H.}\ \bibnamefont
  {Greene}}} (\bibinfo {year} {1994}),\ \bibfield  {title} {{\selectlanguage
  {English}\enquote {\bibinfo {title} {Double photoionization of helium using
  {R}{-}matrix methods},}\ }}\href@noop {} {\bibfield  {journal} {\bibinfo
  {journal} {Phys. Rev. A}\ }\textbf {\bibinfo {volume} {50}}~(\bibinfo
  {number} {5}),\ \bibinfo {pages} {R3573--R3576}}\BibitemShut {NoStop}%
\bibitem [{\citenamefont {Meyer}\ \emph {et~al.}(1997)\citenamefont {Meyer},
  \citenamefont {Greene},\ and\ \citenamefont {Esry}}]{meyer1997PRL}%
  \BibitemOpen
  \bibfield  {author} {\bibinfo {author} {\bibnamefont {Meyer}, \bibfnamefont
  {K~W}}, \bibinfo {author} {\bibfnamefont {C.~H.}\ \bibnamefont {Greene}}, \
  and\ \bibinfo {author} {\bibfnamefont {B.~D.}\ \bibnamefont {Esry}}}
  (\bibinfo {year} {1997}),\ \bibfield  {title} {{\selectlanguage
  {English}\enquote {\bibinfo {title} {Two-electron photoejection of {He} and
  {H}$^-$},}\ }}\href@noop {} {\bibfield  {journal} {\bibinfo  {journal} {Phys.
  Rev. Lett.}\ }\textbf {\bibinfo {volume} {78}}~(\bibinfo {number} {26}),\
  \bibinfo {pages} {4902--4905}}\BibitemShut {NoStop}%
\bibitem [{\citenamefont {Michaud}\ and\ \citenamefont
  {Sanche}({1987})}]{MichaudSanche1987pra}%
  \BibitemOpen
  \bibfield  {author} {\bibinfo {author} {\bibnamefont {Michaud}, \bibfnamefont
  {M}}, \ and\ \bibinfo {author} {\bibfnamefont {L}~\bibnamefont {Sanche}}}
  (\bibinfo {year} {{1987}}),\ \bibfield  {title} {\enquote {\bibinfo {title}
  {{Absolute Vibrational-Excitation Cross-Sections for Slow-Electron (1-18 eV)
  Scattering in Solid H$_2$O}},}\ }\href {\doibase {10.1103/PhysRevA.36.4684}}
  {\bibfield  {journal} {\bibinfo  {journal} {{Physical Review A}}\ }\textbf
  {\bibinfo {volume} {{36}}}~(\bibinfo {number} {{10}}),\ \bibinfo {pages}
  {{4684--4699}}}\BibitemShut {NoStop}%
\bibitem [{\citenamefont {Micheli}\ \emph {et~al.}(2010)\citenamefont
  {Micheli}, \citenamefont {Idziaszek}, \citenamefont {Pupillo}, \citenamefont
  {Baranov}, \citenamefont {Zoller},\ and\ \citenamefont
  {Julienne}}]{micheli2010universal}%
  \BibitemOpen
  \bibfield  {author} {\bibinfo {author} {\bibnamefont {Micheli}, \bibfnamefont
  {A}}, \bibinfo {author} {\bibfnamefont {Z.}~\bibnamefont {Idziaszek}},
  \bibinfo {author} {\bibfnamefont {G.}~\bibnamefont {Pupillo}}, \bibinfo
  {author} {\bibfnamefont {M.~A.}\ \bibnamefont {Baranov}}, \bibinfo {author}
  {\bibfnamefont {P.}~\bibnamefont {Zoller}}, \ and\ \bibinfo {author}
  {\bibfnamefont {P.~S}\ \bibnamefont {Julienne}}} (\bibinfo {year} {2010}),\
  \bibfield  {title} {\enquote {\bibinfo {title} {Universal rates for reactive
  ultracold polar molecules in reduced dimensions},}\ }\href@noop {} {\bibfield
   {journal} {\bibinfo  {journal} {Phys. Rev. Lett.}\ }\textbf {\bibinfo
  {volume} {105}}~(\bibinfo {number} {7}),\ \bibinfo {pages}
  {073202}}\BibitemShut {NoStop}%
\bibitem [{\citenamefont {Mies}(1984)}]{Mies-1984}%
  \BibitemOpen
  \bibfield  {author} {\bibinfo {author} {\bibnamefont {Mies}, \bibfnamefont
  {F~H}}} (\bibinfo {year} {1984}),\ \bibfield  {title} {\enquote {\bibinfo
  {title} {A multichannel quantum defect analysis of diatomic predissociation
  and inelastic atomic scattering},}\ }\href@noop {} {\bibfield  {journal}
  {\bibinfo  {journal} {J. Chem. Phys}\ }\textbf {\bibinfo {volume} {80}},\
  \bibinfo {pages} {2514}}\BibitemShut {NoStop}%
\bibitem [{\citenamefont {Mies}\ and\ \citenamefont
  {Raoult}(2000)}]{Mies-2000}%
  \BibitemOpen
  \bibfield  {author} {\bibinfo {author} {\bibnamefont {Mies}, \bibfnamefont
  {F~H}}, \ and\ \bibinfo {author} {\bibfnamefont {M}~\bibnamefont {Raoult}}}
  (\bibinfo {year} {2000}),\ \bibfield  {title} {\enquote {\bibinfo {title}
  {Analysis of threshold effects in ultracold atomic collisions},}\ }\href@noop
  {} {\bibfield  {journal} {\bibinfo  {journal} {Phys. Rev A}\ }\textbf
  {\bibinfo {volume} {62}},\ \bibinfo {pages} {012708}}\BibitemShut {NoStop}%
\bibitem [{\citenamefont {Mikkelsen}\ \emph {et~al.}({2015})\citenamefont
  {Mikkelsen}, \citenamefont {Jensen}, \citenamefont {Fedorov},\ and\
  \citenamefont {Zinner}}]{Mikkelsen2015jpb}%
  \BibitemOpen
  \bibfield  {author} {\bibinfo {author} {\bibnamefont {Mikkelsen},
  \bibfnamefont {M}}, \bibinfo {author} {\bibfnamefont {A.~S.}\ \bibnamefont
  {Jensen}}, \bibinfo {author} {\bibfnamefont {D.~V.}\ \bibnamefont {Fedorov}},
  \ and\ \bibinfo {author} {\bibfnamefont {N.~T.}\ \bibnamefont {Zinner}}}
  (\bibinfo {year} {{2015}}),\ \bibfield  {title} {\enquote {\bibinfo {title}
  {{Three-body recombination of two-component cold atomic gases into deep
  dimers in an optical model}},}\ }\href {\doibase
  {10.1088/0953-4075/48/8/085301}} {\bibfield  {journal} {\bibinfo  {journal}
  {{J. Phys. B}}\ }\textbf {\bibinfo {volume} {{48}}}~(\bibinfo {number}
  {{8}}),\ {10.1088/0953-4075/48/8/085301}}\BibitemShut {NoStop}%
\bibitem [{\citenamefont {Mitroy}\ \emph {et~al.}({2013})\citenamefont
  {Mitroy}, \citenamefont {Bubin}, \citenamefont {Horiuchi}, \citenamefont
  {Suzuki}, \citenamefont {Adamowicz}, \citenamefont {Cencek}, \citenamefont
  {Szalewicz}, \citenamefont {Komasa}, \citenamefont {Blume},\ and\
  \citenamefont {Varga}}]{Mitroy2013rmp}%
  \BibitemOpen
  \bibfield  {author} {\bibinfo {author} {\bibnamefont {Mitroy}, \bibfnamefont
  {Jim}}, \bibinfo {author} {\bibfnamefont {Sergiy}\ \bibnamefont {Bubin}},
  \bibinfo {author} {\bibfnamefont {Wataru}\ \bibnamefont {Horiuchi}}, \bibinfo
  {author} {\bibfnamefont {Yasuyuki}\ \bibnamefont {Suzuki}}, \bibinfo {author}
  {\bibfnamefont {Ludwik}\ \bibnamefont {Adamowicz}}, \bibinfo {author}
  {\bibfnamefont {Wojciech}\ \bibnamefont {Cencek}}, \bibinfo {author}
  {\bibfnamefont {Krzysztof}\ \bibnamefont {Szalewicz}}, \bibinfo {author}
  {\bibfnamefont {Jacek}\ \bibnamefont {Komasa}}, \bibinfo {author}
  {\bibfnamefont {D.}~\bibnamefont {Blume}}, \ and\ \bibinfo {author}
  {\bibfnamefont {Kalman}\ \bibnamefont {Varga}}} (\bibinfo {year} {{2013}}),\
  \bibfield  {title} {\enquote {\bibinfo {title} {{Theory and application of
  explicitly correlated Gaussians}},}\ }\href {\doibase
  {10.1103/RevModPhys.85.693}} {\bibfield  {journal} {\bibinfo  {journal}
  {{Reviews of Modern Physics}}\ }\textbf {\bibinfo {volume} {{85}}}~(\bibinfo
  {number} {{2}}),\ \bibinfo {pages} {{693--749}}}\BibitemShut {NoStop}%
\bibitem [{\citenamefont {Moerdijk}\ \emph {et~al.}({1996})\citenamefont
  {Moerdijk}, \citenamefont {Boesten},\ and\ \citenamefont
  {Verhaar}}]{Moerdijk1996PRA}%
  \BibitemOpen
  \bibfield  {author} {\bibinfo {author} {\bibnamefont {Moerdijk},
  \bibfnamefont {A~J}}, \bibinfo {author} {\bibfnamefont {H.~M. J.~M.}\
  \bibnamefont {Boesten}}, \ and\ \bibinfo {author} {\bibfnamefont {B.~J.}\
  \bibnamefont {Verhaar}}} (\bibinfo {year} {{1996}}),\ \bibfield  {title}
  {\enquote {\bibinfo {title} {{Decay of trapped ultracold alkali atoms by
  recombination}},}\ }\href@noop {} {\bibfield  {journal} {\bibinfo  {journal}
  {{Phys. Rev. A}}\ }\textbf {\bibinfo {volume} {{53}}}~(\bibinfo {number}
  {{2}}),\ \bibinfo {pages} {{916--920}}}\BibitemShut {NoStop}%
\bibitem [{\citenamefont {Moerdijk}\ and\ \citenamefont
  {Verhaar}({1996})}]{Moerdijk1996PRA2}%
  \BibitemOpen
  \bibfield  {author} {\bibinfo {author} {\bibnamefont {Moerdijk},
  \bibfnamefont {A~J}}, \ and\ \bibinfo {author} {\bibfnamefont {B.~J.}\
  \bibnamefont {Verhaar}}} (\bibinfo {year} {{1996}}),\ \bibfield  {title}
  {\enquote {\bibinfo {title} {{Collisional two- and three-body decay rates of
  dilute quantum gases at ultralow temperatures}},}\ }\href@noop {} {\bibfield
  {journal} {\bibinfo  {journal} {{Phys. Rev. A}}\ }\textbf {\bibinfo {volume}
  {{53}}}~(\bibinfo {number} {{1}}),\ \bibinfo {pages}
  {{R19--R22}}}\BibitemShut {NoStop}%
\bibitem [{\citenamefont {Montero}\ and\ \citenamefont
  {P\'{e}rez-R\'{i}os}(2014)}]{Montero-2014}%
  \BibitemOpen
  \bibfield  {author} {\bibinfo {author} {\bibnamefont {Montero}, \bibfnamefont
  {S}}, \ and\ \bibinfo {author} {\bibfnamefont {J.}~\bibnamefont
  {P\'{e}rez-R\'{i}os}}} (\bibinfo {year} {2014}),\ \bibfield  {title}
  {\enquote {\bibinfo {title} {Rotational relaxation in molecular hydrogen and
  deuterium: Theory versus acoustic experiments},}\ }\href@noop {} {\bibfield
  {journal} {\bibinfo  {journal} {J. Chem. Phys}\ }\textbf {\bibinfo {volume}
  {141}},\ \bibinfo {pages} {114301}}\BibitemShut {NoStop}%
\bibitem [{\citenamefont {Moore}\ \emph {et~al.}(2004)\citenamefont {Moore},
  \citenamefont {Bergeman},\ and\ \citenamefont {Olshanii}}]{moore2004}%
  \BibitemOpen
  \bibfield  {author} {\bibinfo {author} {\bibnamefont {Moore}, \bibfnamefont
  {M~G}}, \bibinfo {author} {\bibfnamefont {T.}~\bibnamefont {Bergeman}}, \
  and\ \bibinfo {author} {\bibfnamefont {M.}~\bibnamefont {Olshanii}}}
  (\bibinfo {year} {2004}),\ \bibfield  {title} {\enquote {\bibinfo {title}
  {Scattering in tight atom waveguides},}\ }in\ \href@noop {} {\emph {\bibinfo
  {booktitle} {Journal de Physique IV (Proceedings)}}},\ Vol.\ \bibinfo
  {volume} {116}\ (\bibinfo {organization} {EDP sciences})\ pp.\ \bibinfo
  {pages} {69--86}\BibitemShut {NoStop}%
\bibitem [{\citenamefont {Mora}\ \emph
  {et~al.}(2005{\natexlab{a}})\citenamefont {Mora}, \citenamefont {Egger},\
  and\ \citenamefont {Gogolin}}]{mora2005PRA}%
  \BibitemOpen
  \bibfield  {author} {\bibinfo {author} {\bibnamefont {Mora}, \bibfnamefont
  {C}}, \bibinfo {author} {\bibfnamefont {R.}~\bibnamefont {Egger}}, \ and\
  \bibinfo {author} {\bibfnamefont {A.~O.}\ \bibnamefont {Gogolin}}} (\bibinfo
  {year} {2005}{\natexlab{a}}),\ \bibfield  {title} {{\selectlanguage
  {English}\enquote {\bibinfo {title} {Three-body problem for ultracold atoms
  in quasi-one-dimensional traps},}\ }}\href@noop {} {\bibfield  {journal}
  {\bibinfo  {journal} {Phys. Rev. A}\ }\textbf {\bibinfo {volume}
  {71}}~(\bibinfo {number} {5}),\ \bibinfo {pages} {052705}}\BibitemShut
  {NoStop}%
\bibitem [{\citenamefont {Mora}\ \emph {et~al.}(2004)\citenamefont {Mora},
  \citenamefont {Egger}, \citenamefont {Gogolin},\ and\ \citenamefont
  {Komnik}}]{mora2004PRL}%
  \BibitemOpen
  \bibfield  {author} {\bibinfo {author} {\bibnamefont {Mora}, \bibfnamefont
  {C}}, \bibinfo {author} {\bibfnamefont {R.}~\bibnamefont {Egger}}, \bibinfo
  {author} {\bibfnamefont {A.~O.}\ \bibnamefont {Gogolin}}, \ and\ \bibinfo
  {author} {\bibfnamefont {A.}~\bibnamefont {Komnik}}} (\bibinfo {year}
  {2004}),\ \bibfield  {title} {{\selectlanguage {English}\enquote {\bibinfo
  {title} {Atom-dimer scattering for confined ultracold {F}ermion gases},}\
  }}\href@noop {} {\bibfield  {journal} {\bibinfo  {journal} {Phys. Rev.
  Lett.}\ }\textbf {\bibinfo {volume} {93}}~(\bibinfo {number} {17}),\ \bibinfo
  {pages} {170403}}\BibitemShut {NoStop}%
\bibitem [{\citenamefont {Mora}\ \emph
  {et~al.}(2005{\natexlab{b}})\citenamefont {Mora}, \citenamefont {Komnik},
  \citenamefont {Egger},\ and\ \citenamefont {Gogolin}}]{mora2005PRL}%
  \BibitemOpen
  \bibfield  {author} {\bibinfo {author} {\bibnamefont {Mora}, \bibfnamefont
  {C}}, \bibinfo {author} {\bibfnamefont {A.}~\bibnamefont {Komnik}}, \bibinfo
  {author} {\bibfnamefont {R.}~\bibnamefont {Egger}}, \ and\ \bibinfo {author}
  {\bibfnamefont {A.~O.}\ \bibnamefont {Gogolin}}} (\bibinfo {year}
  {2005}{\natexlab{b}}),\ \bibfield  {title} {{\selectlanguage
  {English}\enquote {\bibinfo {title} {Four-body problem and {BEC}-{BCS}
  crossover in a quasi-one-dimensional cold {F}ermion gas},}\ }}\href@noop {}
  {\bibfield  {journal} {\bibinfo  {journal} {Phys. Rev. Lett.}\ }\textbf
  {\bibinfo {volume} {95}}~(\bibinfo {number} {8}),\ \bibinfo {pages}
  {080403}}\BibitemShut {NoStop}%
\bibitem [{\citenamefont {Mora}\ \emph {et~al.}({2011})\citenamefont {Mora},
  \citenamefont {Gogolin},\ and\ \citenamefont {Egger}}]{MoraGogolinEgger2011}%
  \BibitemOpen
  \bibfield  {author} {\bibinfo {author} {\bibnamefont {Mora}, \bibfnamefont
  {Christophe}}, \bibinfo {author} {\bibfnamefont {Alexander~O.}\ \bibnamefont
  {Gogolin}}, \ and\ \bibinfo {author} {\bibfnamefont {Reinhold}\ \bibnamefont
  {Egger}}} (\bibinfo {year} {{2011}}),\ \bibfield  {title} {\enquote {\bibinfo
  {title} {{Exact solution of the three-boson problem at vanishing energy}},}\
  }\href {\doibase {10.1016/j.crhy.2010.11.002}} {\bibfield  {journal}
  {\bibinfo  {journal} {{Comptes Rendus Physique}}\ }\textbf {\bibinfo {volume}
  {{12}}}~(\bibinfo {number} {{1}}),\ \bibinfo {pages} {{27--38}}}\BibitemShut
  {NoStop}%
\bibitem [{\citenamefont {Morishita}\ and\ \citenamefont
  {Lin}(1998)}]{morishita1998PRA}%
  \BibitemOpen
  \bibfield  {author} {\bibinfo {author} {\bibnamefont {Morishita},
  \bibfnamefont {T}}, \ and\ \bibinfo {author} {\bibfnamefont {C.D.}\
  \bibnamefont {Lin}}} (\bibinfo {year} {1998}),\ \bibfield  {title} {\enquote
  {\bibinfo {title} {Hyperspherical analysis of doubly and triply excited
  states of {Li}},}\ }\href@noop {} {\bibfield  {journal} {\bibinfo  {journal}
  {Phys. Rev. A}\ }\textbf {\bibinfo {volume} {57}}~(\bibinfo {number} {6}),\
  \bibinfo {pages} {4268--4274}}\BibitemShut {NoStop}%
\bibitem [{\citenamefont {Morishita}\ and\ \citenamefont
  {Lin}(1999)}]{morishita1999PRA}%
  \BibitemOpen
  \bibfield  {author} {\bibinfo {author} {\bibnamefont {Morishita},
  \bibfnamefont {T}}, \ and\ \bibinfo {author} {\bibfnamefont {CD}~\bibnamefont
  {Lin}}} (\bibinfo {year} {1999}),\ \bibfield  {title} {{\selectlanguage
  {English}\enquote {\bibinfo {title} {Comprehensive analysis of electron
  correlations in three-electron atoms},}\ }}\href@noop {} {\bibfield
  {journal} {\bibinfo  {journal} {Phys. Rev. A}\ }\textbf {\bibinfo {volume}
  {59}}~(\bibinfo {number} {3}),\ \bibinfo {pages} {1835--1843}}\BibitemShut
  {NoStop}%
\bibitem [{\citenamefont {Morishita}\ and\ \citenamefont
  {Lin}(2005)}]{morishita2005PRA}%
  \BibitemOpen
  \bibfield  {author} {\bibinfo {author} {\bibnamefont {Morishita},
  \bibfnamefont {T}}, \ and\ \bibinfo {author} {\bibfnamefont {C.D.}\
  \bibnamefont {Lin}}} (\bibinfo {year} {2005}),\ \bibfield  {title} {\enquote
  {\bibinfo {title} {Hyperspherical analysis of radial correlations in
  four-electron atoms},}\ }\href@noop {} {\bibfield  {journal} {\bibinfo
  {journal} {Phys. Rev. A}\ }\textbf {\bibinfo {volume} {71}}~(\bibinfo
  {number} {1}),\ \bibinfo {pages} {012504}}\BibitemShut {NoStop}%
\bibitem [{\citenamefont {Morishita}\ \emph {et~al.}(1997)\citenamefont
  {Morishita}, \citenamefont {Tolstikhin}, \citenamefont {Watanabe},\ and\
  \citenamefont {Matsuzawa}}]{morishita1997PRA}%
  \BibitemOpen
  \bibfield  {author} {\bibinfo {author} {\bibnamefont {Morishita},
  \bibfnamefont {T}}, \bibinfo {author} {\bibfnamefont {O.~I.}\ \bibnamefont
  {Tolstikhin}}, \bibinfo {author} {\bibfnamefont {S.}~\bibnamefont
  {Watanabe}}, \ and\ \bibinfo {author} {\bibfnamefont {M.}~\bibnamefont
  {Matsuzawa}}} (\bibinfo {year} {1997}),\ \bibfield  {title} {{\selectlanguage
  {English}\enquote {\bibinfo {title} {Hyperspherical hierarchy of
  three-electron radial excitations},}\ }}\href@noop {} {\bibfield  {journal}
  {\bibinfo  {journal} {Phys. Rev. A}\ }\textbf {\bibinfo {volume}
  {56}}~(\bibinfo {number} {5}),\ \bibinfo {pages} {3559--3568}}\BibitemShut
  {NoStop}%
\bibitem [{\citenamefont {Moritz}\ \emph {et~al.}(2005)\citenamefont {Moritz},
  \citenamefont {St\"{o}ferle}, \citenamefont {G\"{u}nter}, \citenamefont
  {K\"{o}hl},\ and\ \citenamefont {Esslinger}}]{moritz2005}%
  \BibitemOpen
  \bibfield  {author} {\bibinfo {author} {\bibnamefont {Moritz}, \bibfnamefont
  {H}}, \bibinfo {author} {\bibfnamefont {T.}~\bibnamefont {St\"{o}ferle}},
  \bibinfo {author} {\bibfnamefont {K.}~\bibnamefont {G\"{u}nter}}, \bibinfo
  {author} {\bibfnamefont {M.}~\bibnamefont {K\"{o}hl}}, \ and\ \bibinfo
  {author} {\bibfnamefont {T.}~\bibnamefont {Esslinger}}} (\bibinfo {year}
  {2005}),\ \bibfield  {title} {\enquote {\bibinfo {title} {Confinement induced
  molecules in a {1D} {F}ermi gas},}\ }\href@noop {} {\bibfield  {journal}
  {\bibinfo  {journal} {Phys. Rev. Lett.}\ }\textbf {\bibinfo {volume}
  {94}}~(\bibinfo {number} {21}),\ \bibinfo {pages} {210401}}\BibitemShut
  {NoStop}%
\bibitem [{\citenamefont {Moroz}(2014)}]{Moroz2014}%
  \BibitemOpen
  \bibfield  {author} {\bibinfo {author} {\bibnamefont {Moroz}, \bibfnamefont
  {Sand~Nishida, Y}}} (\bibinfo {year} {2014}),\ \bibfield  {title} {\enquote
  {\bibinfo {title} {Super {E}fimov effect for mass-imbalanced systems},}\
  }\href@noop {} {\bibfield  {journal} {\bibinfo  {journal} {Phys. Rev. A.}\
  }\textbf {\bibinfo {volume} {90}}~(\bibinfo {number} {6}),\ \bibinfo {pages}
  {063631}}\BibitemShut {NoStop}%
\bibitem [{\citenamefont {Morse}\ and\ \citenamefont
  {Feshbach}(1953)}]{morse1953}%
  \BibitemOpen
  \bibfield  {author} {\bibinfo {author} {\bibnamefont {Morse}, \bibfnamefont
  {P~M}}, \ and\ \bibinfo {author} {\bibfnamefont {H.~A.}\ \bibnamefont
  {Feshbach}}} (\bibinfo {year} {1953}),\ \href@noop {} {\emph {\bibinfo
  {title} {Methods of Theoretical Physics}}},\ \bibinfo {edition} {1st}\ ed.\
  (\bibinfo  {publisher} {McGraw-Hill},\ \bibinfo {address} {New
  York})\BibitemShut {NoStop}%
\bibitem [{\citenamefont {{Naidon}}(2016)}]{Naidon2016arxiv}%
  \BibitemOpen
  \bibfield  {author} {\bibinfo {author} {\bibnamefont {{Naidon}},
  \bibfnamefont {P}}} (\bibinfo {year} {2016}),\ \bibfield  {title} {\enquote
  {\bibinfo {title} {{Two impurities in a Bose-Einstein condensate: from Yukawa
  to Efimov attracted polarons}},}\ }\href@noop {} {\bibfield  {journal}
  {\bibinfo  {journal} {ArXiv e-prints}\ }}\Eprint
  {http://arxiv.org/abs/1607.04507} {arXiv:1607.04507 [cond-mat.quant-gas]}
  \BibitemShut {NoStop}%
\bibitem [{\citenamefont {{Naidon}}\ and\ \citenamefont
  {{Endo}}(2016)}]{NaidonEndoReview2016}%
  \BibitemOpen
  \bibfield  {author} {\bibinfo {author} {\bibnamefont {{Naidon}},
  \bibfnamefont {P}}, \ and\ \bibinfo {author} {\bibfnamefont {S.}~\bibnamefont
  {{Endo}}}} (\bibinfo {year} {2016}),\ \bibfield  {title} {\enquote {\bibinfo
  {title} {{Efimov Physics: a review}},}\ }\href@noop {} {\bibfield  {journal}
  {\bibinfo  {journal} {ArXiv e-prints}\ }}\Eprint
  {http://arxiv.org/abs/1610.09805} {arXiv:1610.09805 [quant-ph]} \BibitemShut
  {NoStop}%
\bibitem [{\citenamefont {Naidon}\ \emph
  {et~al.}(2014{\natexlab{a}})\citenamefont {Naidon}, \citenamefont {Endo},\
  and\ \citenamefont {Ueda}}]{naidon2014prl}%
  \BibitemOpen
  \bibfield  {author} {\bibinfo {author} {\bibnamefont {Naidon}, \bibfnamefont
  {P}}, \bibinfo {author} {\bibfnamefont {S.}~\bibnamefont {Endo}}, \ and\
  \bibinfo {author} {\bibfnamefont {M.}~\bibnamefont {Ueda}}} (\bibinfo {year}
  {2014}{\natexlab{a}}),\ \bibfield  {title} {\enquote {\bibinfo {title}
  {Microscopic origin and universality classes of the {E}fimov three-body
  parameter},}\ }\href@noop {} {\bibfield  {journal} {\bibinfo  {journal}
  {Phys. Rev. Lett.}\ }\textbf {\bibinfo {volume} {112}},\ \bibinfo {pages}
  {105301}}\BibitemShut {NoStop}%
\bibitem [{\citenamefont {Naidon}\ \emph
  {et~al.}(2014{\natexlab{b}})\citenamefont {Naidon}, \citenamefont {Endo},\
  and\ \citenamefont {Ueda}}]{naidon2014PRA}%
  \BibitemOpen
  \bibfield  {author} {\bibinfo {author} {\bibnamefont {Naidon}, \bibfnamefont
  {P}}, \bibinfo {author} {\bibfnamefont {S.}~\bibnamefont {Endo}}, \ and\
  \bibinfo {author} {\bibfnamefont {M.}~\bibnamefont {Ueda}}} (\bibinfo {year}
  {2014}{\natexlab{b}}),\ \bibfield  {title} {\enquote {\bibinfo {title}
  {Physical origin of the universal three-body parameter in atomic {E}fimov
  physics},}\ }\href@noop {} {\bibfield  {journal} {\bibinfo  {journal} {Phys.
  Rev. A}\ }\textbf {\bibinfo {volume} {90}},\ \bibinfo {pages}
  {022106}}\BibitemShut {NoStop}%
\bibitem [{\citenamefont {Naidon}\ \emph {et~al.}(2012)\citenamefont {Naidon},
  \citenamefont {Hiyama},\ and\ \citenamefont {Ueda}}]{naidon2012PRA}%
  \BibitemOpen
  \bibfield  {author} {\bibinfo {author} {\bibnamefont {Naidon}, \bibfnamefont
  {P}}, \bibinfo {author} {\bibfnamefont {E.}~\bibnamefont {Hiyama}}, \ and\
  \bibinfo {author} {\bibfnamefont {M.}~\bibnamefont {Ueda}}} (\bibinfo {year}
  {2012}),\ \bibfield  {title} {\enquote {\bibinfo {title} {Universality and
  the three-body parameter of ${}^{4}${He} trimers},}\ }\href@noop {}
  {\bibfield  {journal} {\bibinfo  {journal} {Phys. Rev. A}\ }\textbf {\bibinfo
  {volume} {86}},\ \bibinfo {pages} {012502}}\BibitemShut {NoStop}%
\bibitem [{\citenamefont {Nakajima}\ \emph {et~al.}(2010)\citenamefont
  {Nakajima}, \citenamefont {Horikoshi}, \citenamefont {Mukaiyama},
  \citenamefont {Naidon},\ and\ \citenamefont {Ueda}}]{nakajima2010PRL}%
  \BibitemOpen
  \bibfield  {author} {\bibinfo {author} {\bibnamefont {Nakajima},
  \bibfnamefont {S}}, \bibinfo {author} {\bibfnamefont {M.}~\bibnamefont
  {Horikoshi}}, \bibinfo {author} {\bibfnamefont {T.}~\bibnamefont
  {Mukaiyama}}, \bibinfo {author} {\bibfnamefont {P.}~\bibnamefont {Naidon}}, \
  and\ \bibinfo {author} {\bibfnamefont {M.}~\bibnamefont {Ueda}}} (\bibinfo
  {year} {2010}),\ \bibfield  {title} {{\selectlanguage {English}\enquote
  {\bibinfo {title} {Nonuniversal {E}fimov atom-dimer resonances in a
  three-component mixture of $^6${Li}},}\ }}\href@noop {} {\bibfield  {journal}
  {\bibinfo  {journal} {Phys. Rev. Lett.}\ }\textbf {\bibinfo {volume}
  {105}}~(\bibinfo {number} {2}),\ \bibinfo {pages} {023201}}\BibitemShut
  {NoStop}%
\bibitem [{\citenamefont {Nakajima}\ \emph
  {et~al.}(2011{\natexlab{a}})\citenamefont {Nakajima}, \citenamefont
  {Horikoshi}, \citenamefont {Mukaiyama}, \citenamefont {Naidon},\ and\
  \citenamefont {Ueda}}]{nakajima2011PRL}%
  \BibitemOpen
  \bibfield  {author} {\bibinfo {author} {\bibnamefont {Nakajima},
  \bibfnamefont {S}}, \bibinfo {author} {\bibfnamefont {M.}~\bibnamefont
  {Horikoshi}}, \bibinfo {author} {\bibfnamefont {T.}~\bibnamefont
  {Mukaiyama}}, \bibinfo {author} {\bibfnamefont {P.}~\bibnamefont {Naidon}}, \
  and\ \bibinfo {author} {\bibfnamefont {M.}~\bibnamefont {Ueda}}} (\bibinfo
  {year} {2011}{\natexlab{a}}),\ \bibfield  {title} {\enquote {\bibinfo {title}
  {Measurement of an {E}fimov trimer binding energy in a three-component
  mixture of $^{6}${Li}},}\ }\href@noop {} {\bibfield  {journal} {\bibinfo
  {journal} {Phys. Rev. Lett.}\ }\textbf {\bibinfo {volume} {106}},\ \bibinfo
  {pages} {143201}}\BibitemShut {NoStop}%
\bibitem [{\citenamefont {Nakajima}\ \emph
  {et~al.}(2011{\natexlab{b}})\citenamefont {Nakajima}, \citenamefont
  {Horikoshi}, \citenamefont {Mukaiyama}, \citenamefont {Naidon},\ and\
  \citenamefont {Ueda}}]{Nakajima-2011}%
  \BibitemOpen
  \bibfield  {author} {\bibinfo {author} {\bibnamefont {Nakajima},
  \bibfnamefont {S}}, \bibinfo {author} {\bibfnamefont {M.}~\bibnamefont
  {Horikoshi}}, \bibinfo {author} {\bibfnamefont {T.}~\bibnamefont
  {Mukaiyama}}, \bibinfo {author} {\bibfnamefont {P.}~\bibnamefont {Naidon}}, \
  and\ \bibinfo {author} {\bibfnamefont {M.}~\bibnamefont {Ueda}}} (\bibinfo
  {year} {2011}{\natexlab{b}}),\ \bibfield  {title} {\enquote {\bibinfo {title}
  {Measurement of an {E}fimov trimer binding energy in a three-component
  mixture of $^{6}$li},}\ }\href@noop {} {\bibfield  {journal} {\bibinfo
  {journal} {Phys. Rev. Lett.}\ }\textbf {\bibinfo {volume} {106}},\ \bibinfo
  {pages} {143201}}\BibitemShut {NoStop}%
\bibitem [{\citenamefont {Nascimbene}\ \emph {et~al.}({2010})\citenamefont
  {Nascimbene}, \citenamefont {Navon}, \citenamefont {Jiang}, \citenamefont
  {Chevy},\ and\ \citenamefont {Salomon}}]{NascimbeneSalomon2010nature}%
  \BibitemOpen
  \bibfield  {author} {\bibinfo {author} {\bibnamefont {Nascimbene},
  \bibfnamefont {S}}, \bibinfo {author} {\bibfnamefont {N.}~\bibnamefont
  {Navon}}, \bibinfo {author} {\bibfnamefont {K.~J.}\ \bibnamefont {Jiang}},
  \bibinfo {author} {\bibfnamefont {F.}~\bibnamefont {Chevy}}, \ and\ \bibinfo
  {author} {\bibfnamefont {C.}~\bibnamefont {Salomon}}} (\bibinfo {year}
  {{2010}}),\ \bibfield  {title} {\enquote {\bibinfo {title} {{Exploring the
  thermodynamics of a universal Fermi gas}},}\ }\href {\doibase
  {10.1038/nature08814}} {\bibfield  {journal} {\bibinfo  {journal} {{Nature}}\
  }\textbf {\bibinfo {volume} {{463}}}~(\bibinfo {number} {{7284}}),\ \bibinfo
  {pages} {{1057--U73}}}\BibitemShut {NoStop}%
\bibitem [{\citenamefont {Naus}\ and\ \citenamefont
  {Tjon}({1987})}]{NausTjon1987fbs}%
  \BibitemOpen
  \bibfield  {author} {\bibinfo {author} {\bibnamefont {Naus}, \bibfnamefont
  {HWL}}, \ and\ \bibinfo {author} {\bibfnamefont {JA}~\bibnamefont {Tjon}}}
  (\bibinfo {year} {{1987}}),\ \bibfield  {title} {\enquote {\bibinfo {title}
  {{The Efimov Effect in a 4-Body System}},}\ }\href {\doibase
  {10.1007/BF01080835}} {\bibfield  {journal} {\bibinfo  {journal} {{Few-Body
  Systems}}\ }\textbf {\bibinfo {volume} {{2}}}~(\bibinfo {number} {{3}}),\
  \bibinfo {pages} {{121--126}}}\BibitemShut {NoStop}%
\bibitem [{\citenamefont {Neuhauser}\ \emph {et~al.}({1991})\citenamefont
  {Neuhauser}, \citenamefont {Judson}, \citenamefont {Jaffe}, \citenamefont
  {Baer},\ and\ \citenamefont {Kouri}}]{Neuhauser1991CPL}%
  \BibitemOpen
  \bibfield  {author} {\bibinfo {author} {\bibnamefont {Neuhauser},
  \bibfnamefont {D}}, \bibinfo {author} {\bibfnamefont {R.~S.}\ \bibnamefont
  {Judson}}, \bibinfo {author} {\bibfnamefont {R.~L.}\ \bibnamefont {Jaffe}},
  \bibinfo {author} {\bibfnamefont {M.}~\bibnamefont {Baer}}, \ and\ \bibinfo
  {author} {\bibfnamefont {D.~J.}\ \bibnamefont {Kouri}}} (\bibinfo {year}
  {{1991}}),\ \bibfield  {title} {\enquote {\bibinfo {title} {Total integral
  reactive cross-sections for {F} + {H}$_2$ $\rightarrow$ {HF} + {H}:
  comparison of converged quantum, quasi-classical trajectory and experimental
  results},}\ }\href@noop {} {\bibfield  {journal} {\bibinfo  {journal} {Chem.
  Phys. Lett.}\ }\textbf {\bibinfo {volume} {{176}}}~(\bibinfo {number}
  {{6}}),\ \bibinfo {pages} {{546--550}}}\BibitemShut {NoStop}%
\bibitem [{\citenamefont {Neumark}\ \emph {et~al.}({1985})\citenamefont
  {Neumark}, \citenamefont {Wodtke}, \citenamefont {Robinson}, \citenamefont
  {Hayden},\ and\ \citenamefont {Lee}}]{Neumark1985JCP}%
  \BibitemOpen
  \bibfield  {author} {\bibinfo {author} {\bibnamefont {Neumark}, \bibfnamefont
  {D~M}}, \bibinfo {author} {\bibfnamefont {A.~M.}\ \bibnamefont {Wodtke}},
  \bibinfo {author} {\bibfnamefont {G.~N.}\ \bibnamefont {Robinson}}, \bibinfo
  {author} {\bibfnamefont {C.~C.}\ \bibnamefont {Hayden}}, \ and\ \bibinfo
  {author} {\bibfnamefont {Y.~T.}\ \bibnamefont {Lee}}} (\bibinfo {year}
  {{1985}}),\ \bibfield  {title} {\enquote {\bibinfo {title} {Molecular-beam
  studies of the {F} +{H}$_2$ reaction},}\ }\href@noop {} {\bibfield  {journal}
  {\bibinfo  {journal} {J. Chem. Phys.}\ }\textbf {\bibinfo {volume}
  {{82}}}~(\bibinfo {number} {{7}}),\ \bibinfo {pages}
  {{3045--3066}}}\BibitemShut {NoStop}%
\bibitem [{\citenamefont {Ngampruetikorn}\ \emph {et~al.}({2013})\citenamefont
  {Ngampruetikorn}, \citenamefont {Parish},\ and\ \citenamefont
  {Levinsen}}]{Ngampruetikorn2013epl}%
  \BibitemOpen
  \bibfield  {author} {\bibinfo {author} {\bibnamefont {Ngampruetikorn},
  \bibfnamefont {Vudtiwat}}, \bibinfo {author} {\bibfnamefont {Meera~M.}\
  \bibnamefont {Parish}}, \ and\ \bibinfo {author} {\bibfnamefont {Jesper}\
  \bibnamefont {Levinsen}}} (\bibinfo {year} {{2013}}),\ \bibfield  {title}
  {\enquote {\bibinfo {title} {{Three-body problem in a two-dimensional Fermi
  gas}},}\ }\href {\doibase {10.1209/0295-5075/102/13001}} {\bibfield
  {journal} {\bibinfo  {journal} {{EPL}}\ }\textbf {\bibinfo {volume}
  {{102}}}~(\bibinfo {number} {{1}}),\ \bibinfo {pages} {{13001}}}\BibitemShut
  {NoStop}%
\bibitem [{\citenamefont {Nicholson}(2012)}]{Nicholson2012prl}%
  \BibitemOpen
  \bibfield  {author} {\bibinfo {author} {\bibnamefont {Nicholson},
  \bibfnamefont {Amy~N}}} (\bibinfo {year} {2012}),\ \bibfield  {title}
  {\enquote {\bibinfo {title} {$n$},}\ }\href {\doibase
  10.1103/PhysRevLett.109.073003} {\bibfield  {journal} {\bibinfo  {journal}
  {Phys. Rev. Lett.}\ }\textbf {\bibinfo {volume} {109}},\ \bibinfo {pages}
  {073003}}\BibitemShut {NoStop}%
\bibitem [{\citenamefont {Nielsen}\ \emph {et~al.}(2001)\citenamefont
  {Nielsen}, \citenamefont {Fedorov}, \citenamefont {Jensen},\ and\
  \citenamefont {Garrido}}]{nielsen2001PRep}%
  \BibitemOpen
  \bibfield  {author} {\bibinfo {author} {\bibnamefont {Nielsen}, \bibfnamefont
  {E}}, \bibinfo {author} {\bibfnamefont {DV}~\bibnamefont {Fedorov}}, \bibinfo
  {author} {\bibfnamefont {AS}~\bibnamefont {Jensen}}, \ and\ \bibinfo {author}
  {\bibfnamefont {E}~\bibnamefont {Garrido}}} (\bibinfo {year} {2001}),\
  \bibfield  {title} {{\selectlanguage {English}\enquote {\bibinfo {title} {The
  three-body problem with short-range interactions},}\ }}\href@noop {}
  {\bibfield  {journal} {\bibinfo  {journal} {Phys. Rep.}\ }\textbf {\bibinfo
  {volume} {347}}~(\bibinfo {number} {5}),\ \bibinfo {pages}
  {373--459}}\BibitemShut {NoStop}%
\bibitem [{\citenamefont {Nielsen}\ and\ \citenamefont
  {Macek}(1999)}]{nielsen1999PRLb}%
  \BibitemOpen
  \bibfield  {author} {\bibinfo {author} {\bibnamefont {Nielsen}, \bibfnamefont
  {Esben}}, \ and\ \bibinfo {author} {\bibfnamefont {J.~H.}\ \bibnamefont
  {Macek}}} (\bibinfo {year} {1999}),\ \bibfield  {title} {\enquote {\bibinfo
  {title} {Low-energy recombination of identical bosons by three-body
  collisions},}\ }\href@noop {} {\bibfield  {journal} {\bibinfo  {journal}
  {Phys. Rev. Lett.}\ }\textbf {\bibinfo {volume} {83}}~(\bibinfo {number}
  {8}),\ \bibinfo {pages} {1566--1569}}\BibitemShut {NoStop}%
\bibitem [{\citenamefont {Nikitin}(1970)}]{nikitin1970}%
  \BibitemOpen
  \bibfield  {author} {\bibinfo {author} {\bibnamefont {Nikitin}, \bibfnamefont
  {E~E}}} (\bibinfo {year} {1970}),\ \href@noop {} {\emph {\bibinfo {title}
  {Teoriya Elementarnykh Atomnomolekulyarnykh Protsessov v Gazakh (Theory of
  Elementary Atom-molecule Processes in Gases)}}}\ (\bibinfo  {publisher} {Izd.
  Khimiya})\BibitemShut {NoStop}%
\bibitem [{\citenamefont {{Nishida}}(2017)}]{Nishida2017arxiv}%
  \BibitemOpen
  \bibfield  {author} {\bibinfo {author} {\bibnamefont {{Nishida}},
  \bibfnamefont {Y}}} (\bibinfo {year} {2017}),\ \bibfield  {title} {\enquote
  {\bibinfo {title} {{Semi-super Efimov effect of two-dimensional bosons at a
  three-body resonance}},}\ }\href@noop {} {\bibfield  {journal} {\bibinfo
  {journal} {ArXiv e-prints}\ }}\Eprint {http://arxiv.org/abs/1702.07532}
  {arXiv:1702.07532 [cond-mat.quant-gas]} \BibitemShut {NoStop}%
\bibitem [{\citenamefont {Nishida}\ \emph {et~al.}(2013)\citenamefont
  {Nishida}, \citenamefont {Moroz},\ and\ \citenamefont {Dam}}]{Nishida2013}%
  \BibitemOpen
  \bibfield  {author} {\bibinfo {author} {\bibnamefont {Nishida}, \bibfnamefont
  {Y}}, \bibinfo {author} {\bibfnamefont {S.}~\bibnamefont {Moroz}}, \ and\
  \bibinfo {author} {\bibfnamefont {T.~S.}\ \bibnamefont {Dam}}} (\bibinfo
  {year} {2013}),\ \bibfield  {title} {\enquote {\bibinfo {title} {Super
  {E}fimov effect of resonantly interacting fermions in two dimensions},}\
  }\href@noop {} {\bibfield  {journal} {\bibinfo  {journal} {Phys. Rev. Lett.}\
  }\textbf {\bibinfo {volume} {110}}~(\bibinfo {number} {23}),\ \bibinfo
  {pages} {235301}}\BibitemShut {NoStop}%
\bibitem [{\citenamefont {Nishida}\ and\ \citenamefont
  {Tan}(2008)}]{nishida2008universal}%
  \BibitemOpen
  \bibfield  {author} {\bibinfo {author} {\bibnamefont {Nishida}, \bibfnamefont
  {Y}}, \ and\ \bibinfo {author} {\bibfnamefont {S.}~\bibnamefont {Tan}}}
  (\bibinfo {year} {2008}),\ \bibfield  {title} {\enquote {\bibinfo {title}
  {Universal {F}ermi gases in mixed dimensions},}\ }\href@noop {} {\bibfield
  {journal} {\bibinfo  {journal} {Phys. Rev. Lett.}\ }\textbf {\bibinfo
  {volume} {101}}~(\bibinfo {number} {17}),\ \bibinfo {pages}
  {170401}}\BibitemShut {NoStop}%
\bibitem [{\citenamefont {Nishida}\ and\ \citenamefont
  {Tan}(2010)}]{nishida2010}%
  \BibitemOpen
  \bibfield  {author} {\bibinfo {author} {\bibnamefont {Nishida}, \bibfnamefont
  {Y}}, \ and\ \bibinfo {author} {\bibfnamefont {S.}~\bibnamefont {Tan}}}
  (\bibinfo {year} {2010}),\ \bibfield  {title} {\enquote {\bibinfo {title}
  {Confinement-induced $p$-wave resonances from $s$-wave interactions},}\
  }\href@noop {} {\bibfield  {journal} {\bibinfo  {journal} {Phys. Rev. A}\
  }\textbf {\bibinfo {volume} {82}}~(\bibinfo {number} {6}),\ \bibinfo {pages}
  {062713}}\BibitemShut {NoStop}%
\bibitem [{\citenamefont {Nishida}\ and\ \citenamefont
  {Tan}(2011)}]{nishida2011FBS}%
  \BibitemOpen
  \bibfield  {author} {\bibinfo {author} {\bibnamefont {Nishida}, \bibfnamefont
  {Yusuke}}, \ and\ \bibinfo {author} {\bibfnamefont {Shina}\ \bibnamefont
  {Tan}}} (\bibinfo {year} {2011}),\ \bibfield  {title} {{\selectlanguage
  {English}\enquote {\bibinfo {title} {Liberating {E}fimov physics from three
  dimensions},}\ }}\href@noop {} {\bibfield  {journal} {\bibinfo  {journal}
  {Few-Body Systems}\ }\textbf {\bibinfo {volume} {51}},\ \bibinfo {pages}
  {191--206}}\BibitemShut {NoStop}%
\bibitem [{\citenamefont {Nogga}\ \emph {et~al.}(2000)\citenamefont {Nogga},
  \citenamefont {Kamada},\ and\ \citenamefont {Gl\"ockle}}]{Nogga-2000}%
  \BibitemOpen
  \bibfield  {author} {\bibinfo {author} {\bibnamefont {Nogga}, \bibfnamefont
  {A}}, \bibinfo {author} {\bibfnamefont {H.}~\bibnamefont {Kamada}}, \ and\
  \bibinfo {author} {\bibfnamefont {W.}~\bibnamefont {Gl\"ockle}}} (\bibinfo
  {year} {2000}),\ \bibfield  {title} {\enquote {\bibinfo {title} {Modern
  nuclear forces predictions for the $\alpha$ particle},}\ }\href@noop {}
  {\bibfield  {journal} {\bibinfo  {journal} {Phys. Rev. Lett.}\ }\textbf
  {\bibinfo {volume} {85}},\ \bibinfo {pages} {944}}\BibitemShut {NoStop}%
\bibitem [{\citenamefont {Ohsaki}\ and\ \citenamefont
  {Nakamura}(1990)}]{ohsaki1990PRep}%
  \BibitemOpen
  \bibfield  {author} {\bibinfo {author} {\bibnamefont {Ohsaki}, \bibfnamefont
  {A}}, \ and\ \bibinfo {author} {\bibfnamefont {H.}~\bibnamefont {Nakamura}}}
  (\bibinfo {year} {1990}),\ \bibfield  {title} {\enquote {\bibinfo {title}
  {Hyperspherical coordinate approach to atom-diatom chemical reactions in the
  sudden and adiabatic approximations},}\ }\href@noop {} {\bibfield  {journal}
  {\bibinfo  {journal} {Phys. Rep.}\ }\textbf {\bibinfo {volume}
  {187}}~(\bibinfo {number} {1}),\ \bibinfo {pages} {1 -- 62}}\BibitemShut
  {NoStop}%
\bibitem [{\citenamefont {Olshanii}(1998)}]{olshanii1998}%
  \BibitemOpen
  \bibfield  {author} {\bibinfo {author} {\bibnamefont {Olshanii},
  \bibfnamefont {M}}} (\bibinfo {year} {1998}),\ \bibfield  {title} {\enquote
  {\bibinfo {title} {Atomic scattering in the presence of an external
  confinement and a gas of impenetrable bosons},}\ }\href@noop {} {\bibfield
  {journal} {\bibinfo  {journal} {Phys. Rev. Lett.}\ }\textbf {\bibinfo
  {volume} {81}}~(\bibinfo {number} {5}),\ \bibinfo {pages}
  {938--941}}\BibitemShut {NoStop}%
\bibitem [{\citenamefont {Ottenstein}\ \emph {et~al.}(2008)\citenamefont
  {Ottenstein}, \citenamefont {Lompe}, \citenamefont {Kohnen}, \citenamefont
  {Wenz},\ and\ \citenamefont {Jochim}}]{ottenstein2008PRL}%
  \BibitemOpen
  \bibfield  {author} {\bibinfo {author} {\bibnamefont {Ottenstein},
  \bibfnamefont {T~B}}, \bibinfo {author} {\bibfnamefont {T.}~\bibnamefont
  {Lompe}}, \bibinfo {author} {\bibfnamefont {M.}~\bibnamefont {Kohnen}},
  \bibinfo {author} {\bibfnamefont {A.~N.}\ \bibnamefont {Wenz}}, \ and\
  \bibinfo {author} {\bibfnamefont {S.}~\bibnamefont {Jochim}}} (\bibinfo
  {year} {2008}),\ \bibfield  {title} {{\selectlanguage {English}\enquote
  {\bibinfo {title} {Collisional stability of a three-component degenerate
  {F}ermi gas},}\ }}\href@noop {} {\bibfield  {journal} {\bibinfo  {journal}
  {Phys. Rev. Lett.}\ }\textbf {\bibinfo {volume} {101}}~(\bibinfo {number}
  {20}),\ \bibinfo {pages} {203202}}\BibitemShut {NoStop}%
\bibitem [{\citenamefont {Owens}\ \emph {et~al.}(2016)\citenamefont {Owens},
  \citenamefont {Xie},\ and\ \citenamefont {Hutson}}]{Hutson2016rf}%
  \BibitemOpen
  \bibfield  {author} {\bibinfo {author} {\bibnamefont {Owens}, \bibfnamefont
  {D, J}}, \bibinfo {author} {\bibfnamefont {T.}~\bibnamefont {Xie}}, \ and\
  \bibinfo {author} {\bibfnamefont {J.~M.}\ \bibnamefont {Hutson}}} (\bibinfo
  {year} {2016}),\ \bibfield  {title} {\enquote {\bibinfo {title} {Creating
  {F}eshbach resonances for ultracold molecule formation with radio-frequency
  fields},}\ }\href@noop {} {\bibfield  {journal} {\bibinfo  {journal} {Phys.
  Rev. A}\ }\textbf {\bibinfo {volume} {94}},\ \bibinfo {pages}
  {023619}}\BibitemShut {NoStop}%
\bibitem [{\citenamefont {Pack}\ and\ \citenamefont
  {Parker}(1987)}]{pack1987JCP}%
  \BibitemOpen
  \bibfield  {author} {\bibinfo {author} {\bibnamefont {Pack}, \bibfnamefont
  {R~T}}, \ and\ \bibinfo {author} {\bibfnamefont {G.~A.}\ \bibnamefont
  {Parker}}} (\bibinfo {year} {1987}),\ \bibfield  {title} {\enquote {\bibinfo
  {title} {Quantum reactive scattering in three dimensions using hyperspherical
  ({APH}) coordinates},}\ }\href@noop {} {\bibfield  {journal} {\bibinfo
  {journal} {J. Chem. Phys.}\ }\textbf {\bibinfo {volume} {87}},\ \bibinfo
  {pages} {3888--3921}}\BibitemShut {NoStop}%
\bibitem [{\citenamefont {Pack}\ and\ \citenamefont
  {Parker}(1989)}]{pack1989JCP}%
  \BibitemOpen
  \bibfield  {author} {\bibinfo {author} {\bibnamefont {Pack}, \bibfnamefont
  {R~T}}, \ and\ \bibinfo {author} {\bibfnamefont {G.~A.}\ \bibnamefont
  {Parker}}} (\bibinfo {year} {1989}),\ \bibfield  {title} {\enquote {\bibinfo
  {title} {Quantum reactive scattering in three dimensions using hyperspherical
  {(APH)} coordinates. {III}. small theta behavior and corrigenda},}\
  }\href@noop {} {\bibfield  {journal} {\bibinfo  {journal} {J. Chem. Phys.}\
  }\textbf {\bibinfo {volume} {90}}~(\bibinfo {number} {7}),\ \bibinfo {pages}
  {3511--3519}}\BibitemShut {NoStop}%
\bibitem [{\citenamefont {Paredes}\ \emph {et~al.}(2004)\citenamefont
  {Paredes}, \citenamefont {Widera}, \citenamefont {Murg}, \citenamefont
  {Mandel}, \citenamefont {F\"{o}lling}, \citenamefont {Cirac}, \citenamefont
  {Shlyapnikov}, \citenamefont {H\"{a}nsch},\ and\ \citenamefont
  {Bloch}}]{paredes2004tonks}%
  \BibitemOpen
  \bibfield  {author} {\bibinfo {author} {\bibnamefont {Paredes}, \bibfnamefont
  {B}}, \bibinfo {author} {\bibfnamefont {A.}~\bibnamefont {Widera}}, \bibinfo
  {author} {\bibfnamefont {V.}~\bibnamefont {Murg}}, \bibinfo {author}
  {\bibfnamefont {O.}~\bibnamefont {Mandel}}, \bibinfo {author} {\bibfnamefont
  {S.}~\bibnamefont {F\"{o}lling}}, \bibinfo {author} {\bibfnamefont {J.~I.}\
  \bibnamefont {Cirac}}, \bibinfo {author} {\bibfnamefont {G.~V.}\ \bibnamefont
  {Shlyapnikov}}, \bibinfo {author} {\bibfnamefont {T.~W.}\ \bibnamefont
  {H\"{a}nsch}}, \ and\ \bibinfo {author} {\bibfnamefont {I.}~\bibnamefont
  {Bloch}}} (\bibinfo {year} {2004}),\ \bibfield  {title} {\enquote {\bibinfo
  {title} {{Tonks-Girardeau gas of ultracold atoms in an optical lattice}},}\
  }\href@noop {} {\bibfield  {journal} {\bibinfo  {journal} {Nature (London)}\
  }\textbf {\bibinfo {volume} {429}}~(\bibinfo {number} {6989}),\ \bibinfo
  {pages} {277--281}}\BibitemShut {NoStop}%
\bibitem [{\citenamefont {Patton}\ \emph {et~al.}(2003)\citenamefont {Patton},
  \citenamefont {Langbein},\ and\ \citenamefont {Woggon}}]{Patton2003prb}%
  \BibitemOpen
  \bibfield  {author} {\bibinfo {author} {\bibnamefont {Patton}, \bibfnamefont
  {B}}, \bibinfo {author} {\bibfnamefont {W.}~\bibnamefont {Langbein}}, \ and\
  \bibinfo {author} {\bibfnamefont {U.}~\bibnamefont {Woggon}}} (\bibinfo
  {year} {2003}),\ \bibfield  {title} {\enquote {\bibinfo {title} {Trion,
  biexciton, and exciton dynamics in single self-assembled cdse quantum
  dots},}\ }\href {\doibase 10.1103/PhysRevB.68.125316} {\bibfield  {journal}
  {\bibinfo  {journal} {Phys. Rev. B}\ }\textbf {\bibinfo {volume} {68}},\
  \bibinfo {pages} {125316}}\BibitemShut {NoStop}%
\bibitem [{\citenamefont {Peano}\ \emph {et~al.}(2005)\citenamefont {Peano},
  \citenamefont {Thorwart}, \citenamefont {Mora},\ and\ \citenamefont
  {Egger}}]{peano2005}%
  \BibitemOpen
  \bibfield  {author} {\bibinfo {author} {\bibnamefont {Peano}, \bibfnamefont
  {V}}, \bibinfo {author} {\bibfnamefont {M.}~\bibnamefont {Thorwart}},
  \bibinfo {author} {\bibfnamefont {C.}~\bibnamefont {Mora}}, \ and\ \bibinfo
  {author} {\bibfnamefont {R.}~\bibnamefont {Egger}}} (\bibinfo {year}
  {2005}),\ \bibfield  {title} {\enquote {\bibinfo {title} {Confinement-induced
  resonances for a two-component ultracold atom gas in arbitrary
  quasi-one-dimensional traps},}\ }\href@noop {} {\bibfield  {journal}
  {\bibinfo  {journal} {New J. Phys.}\ }\textbf {\bibinfo {volume}
  {7}}~(\bibinfo {number} {1}),\ \bibinfo {pages} {192}}\BibitemShut {NoStop}%
\bibitem [{\citenamefont {Peng}\ \emph {et~al.}(2010)\citenamefont {Peng},
  \citenamefont {Bohloul}, \citenamefont {Liu}, \citenamefont {Hu},\ and\
  \citenamefont {Drummond}}]{peng2010}%
  \BibitemOpen
  \bibfield  {author} {\bibinfo {author} {\bibnamefont {Peng}, \bibfnamefont
  {S{-}G}}, \bibinfo {author} {\bibfnamefont {S.}~\bibnamefont {Bohloul}},
  \bibinfo {author} {\bibfnamefont {X.-J. .~J.}\ \bibnamefont {Liu}}, \bibinfo
  {author} {\bibfnamefont {H.}~\bibnamefont {Hu}}, \ and\ \bibinfo {author}
  {\bibfnamefont {P.~D.}\ \bibnamefont {Drummond}}} (\bibinfo {year} {2010}),\
  \bibfield  {title} {\enquote {\bibinfo {title} {Confinement-induced resonance
  in quasi-one-dimensional systems under transversely anisotropic
  confinement},}\ }\href@noop {} {\bibfield  {journal} {\bibinfo  {journal}
  {Phys. Rev. A}\ }\textbf {\bibinfo {volume} {82}}~(\bibinfo {number} {6}),\
  \bibinfo {pages} {063633}}\BibitemShut {NoStop}%
\bibitem [{\citenamefont {Peng}\ \emph {et~al.}(2011)\citenamefont {Peng},
  \citenamefont {Hu}, \citenamefont {Liu},\ and\ \citenamefont
  {Drummond}}]{peng2011pra}%
  \BibitemOpen
  \bibfield  {author} {\bibinfo {author} {\bibnamefont {Peng}, \bibfnamefont
  {S{-}G}}, \bibinfo {author} {\bibfnamefont {H.}~\bibnamefont {Hu}}, \bibinfo
  {author} {\bibfnamefont {X.-J.}\ \bibnamefont {Liu}}, \ and\ \bibinfo
  {author} {\bibfnamefont {P.~D.}\ \bibnamefont {Drummond}}} (\bibinfo {year}
  {2011}),\ \bibfield  {title} {\enquote {\bibinfo {title} {Confinement-induced
  resonances in anharmonic waveguides},}\ }\href@noop {} {\bibfield  {journal}
  {\bibinfo  {journal} {Phys. Rev. A}\ }\textbf {\bibinfo {volume} {84}},\
  \bibinfo {pages} {043619}}\BibitemShut {NoStop}%
\bibitem [{\citenamefont {Peng}\ \emph {et~al.}(2012)\citenamefont {Peng},
  \citenamefont {Hu}, \citenamefont {Liu},\ and\ \citenamefont
  {Jiang}}]{peng2012two}%
  \BibitemOpen
  \bibfield  {author} {\bibinfo {author} {\bibnamefont {Peng}, \bibfnamefont
  {S-G}}, \bibinfo {author} {\bibfnamefont {H.}~\bibnamefont {Hu}}, \bibinfo
  {author} {\bibfnamefont {X.-J.}\ \bibnamefont {Liu}}, \ and\ \bibinfo
  {author} {\bibfnamefont {K.}~\bibnamefont {Jiang}}} (\bibinfo {year}
  {2012}),\ \bibfield  {title} {\enquote {\bibinfo {title} {Two-channel-model
  description of confinement-induced {F}eshbach molecules},}\ }\href@noop {}
  {\bibfield  {journal} {\bibinfo  {journal} {Phys. Rev. A}\ }\textbf {\bibinfo
  {volume} {86}}~(\bibinfo {number} {3}),\ \bibinfo {pages}
  {033601}}\BibitemShut {NoStop}%
\bibitem [{\citenamefont {Peng}\ \emph {et~al.}(2014)\citenamefont {Peng},
  \citenamefont {Tan},\ and\ \citenamefont {Jiang}}]{peng2014prl}%
  \BibitemOpen
  \bibfield  {author} {\bibinfo {author} {\bibnamefont {Peng}, \bibfnamefont
  {S{-}G}}, \bibinfo {author} {\bibfnamefont {S.}~\bibnamefont {Tan}}, \ and\
  \bibinfo {author} {\bibfnamefont {K.}~\bibnamefont {Jiang}}} (\bibinfo {year}
  {2014}),\ \bibfield  {title} {\enquote {\bibinfo {title} {Manipulation of
  $p$-wave scattering of cold atoms in low dimensions using the magnetic field
  vector},}\ }\href@noop {} {\bibfield  {journal} {\bibinfo  {journal} {Phys.
  Rev. Lett.}\ }\textbf {\bibinfo {volume} {112}},\ \bibinfo {pages}
  {250401}}\BibitemShut {NoStop}%
\bibitem [{\citenamefont {P\'{e}rez-R\'{i}os}\ and\ \citenamefont
  {Greene}(2015)}]{JPR-2015}%
  \BibitemOpen
  \bibfield  {author} {\bibinfo {author} {\bibnamefont {P\'{e}rez-R\'{i}os},
  \bibfnamefont {J}}, \ and\ \bibinfo {author} {\bibfnamefont {C.~H.}\
  \bibnamefont {Greene}}} (\bibinfo {year} {2015}),\ \bibfield  {title}
  {\enquote {\bibinfo {title} {Communication: Classical threhold law for
  ion-neutral-neutral three-body recombination},}\ }\href@noop {} {\bibfield
  {journal} {\bibinfo  {journal} {J. Chem. Phys.}\ }\textbf {\bibinfo {volume}
  {143}},\ \bibinfo {pages} {041105}}\BibitemShut {NoStop}%
\bibitem [{\citenamefont {P\'{e}rez-R\'{i}os}\ \emph
  {et~al.}(2014)\citenamefont {P\'{e}rez-R\'{i}os}, \citenamefont {Ragole},
  \citenamefont {Wang},\ and\ \citenamefont {Greene}}]{JPR-2014}%
  \BibitemOpen
  \bibfield  {author} {\bibinfo {author} {\bibnamefont {P\'{e}rez-R\'{i}os},
  \bibfnamefont {J}}, \bibinfo {author} {\bibfnamefont {S.}~\bibnamefont
  {Ragole}}, \bibinfo {author} {\bibfnamefont {J.}~\bibnamefont {Wang}}, \ and\
  \bibinfo {author} {\bibfnamefont {C.~H.}\ \bibnamefont {Greene}}} (\bibinfo
  {year} {2014}),\ \bibfield  {title} {\enquote {\bibinfo {title} {Comparison
  of classical and quantal calculations of helium three-body recombination},}\
  }\href@noop {} {\bibfield  {journal} {\bibinfo  {journal} {J. Chem. Phys.}\
  }\textbf {\bibinfo {volume} {140}},\ \bibinfo {pages} {044307}}\BibitemShut
  {NoStop}%
\bibitem [{\citenamefont {Peterkop}(1971)}]{PETERKOP1971}%
  \BibitemOpen
  \bibfield  {author} {\bibinfo {author} {\bibnamefont {Peterkop},
  \bibfnamefont {R}}} (\bibinfo {year} {1971}),\ \bibfield  {title} {\enquote
  {\bibinfo {title} {{WKB} approximation and threshold law for electron-atom
  ionization},}\ }\href@noop {} {\bibfield  {journal} {\bibinfo  {journal} {J.
  Phys. B}\ }\textbf {\bibinfo {volume} {4}}~(\bibinfo {number} {4}),\ \bibinfo
  {pages} {513--\&}}\BibitemShut {NoStop}%
\bibitem [{\citenamefont {Peterkop}(1983)}]{PETERKOP1983}%
  \BibitemOpen
  \bibfield  {author} {\bibinfo {author} {\bibnamefont {Peterkop},
  \bibfnamefont {R}}} (\bibinfo {year} {1983}),\ \bibfield  {title} {\enquote
  {\bibinfo {title} {On the derivation of the ionization threshold law},}\
  }\href@noop {} {\bibfield  {journal} {\bibinfo  {journal} {J. Phys. B}\
  }\textbf {\bibinfo {volume} {16}}~(\bibinfo {number} {19}),\ \bibinfo {pages}
  {L587--L593}}\BibitemShut {NoStop}%
\bibitem [{\citenamefont {Petrignani}\ \emph {et~al.}(2011)\citenamefont
  {Petrignani}, \citenamefont {Altevogt}, \citenamefont {Berg}, \citenamefont
  {Bing}, \citenamefont {Grieser}, \citenamefont {Hoffmann}, \citenamefont
  {Jordon-Thaden}, \citenamefont {Krantz}, \citenamefont {Mendes},
  \citenamefont {Novotny}, \citenamefont {Orlov}, \citenamefont {Repnow},
  \citenamefont {Sorg}, \citenamefont {St\"utzel}, \citenamefont {Wolf},
  \citenamefont {Buhr}, \citenamefont {Kreckel}, \citenamefont {Kokoouline},\
  and\ \citenamefont {Greene}}]{petrignani2011}%
  \BibitemOpen
  \bibfield  {author} {\bibinfo {author} {\bibnamefont {Petrignani},
  \bibfnamefont {A}}, \bibinfo {author} {\bibfnamefont {S.}~\bibnamefont
  {Altevogt}}, \bibinfo {author} {\bibfnamefont {M.~H.}\ \bibnamefont {Berg}},
  \bibinfo {author} {\bibfnamefont {D.}~\bibnamefont {Bing}}, \bibinfo {author}
  {\bibfnamefont {M.}~\bibnamefont {Grieser}}, \bibinfo {author} {\bibfnamefont
  {J.}~\bibnamefont {Hoffmann}}, \bibinfo {author} {\bibfnamefont
  {B.}~\bibnamefont {Jordon-Thaden}}, \bibinfo {author} {\bibfnamefont
  {C.}~\bibnamefont {Krantz}}, \bibinfo {author} {\bibfnamefont {M.~B.}\
  \bibnamefont {Mendes}}, \bibinfo {author} {\bibfnamefont {S.}~\bibnamefont
  {Novotny}}, \bibinfo {author} {\bibfnamefont {D.~A.}\ \bibnamefont {Orlov}},
  \bibinfo {author} {\bibfnamefont {R.}~\bibnamefont {Repnow}}, \bibinfo
  {author} {\bibfnamefont {T.}~\bibnamefont {Sorg}}, \bibinfo {author}
  {\bibfnamefont {J.}~\bibnamefont {St\"utzel}}, \bibinfo {author}
  {\bibfnamefont {A.}~\bibnamefont {Wolf}}, \bibinfo {author} {\bibfnamefont
  {H.}~\bibnamefont {Buhr}}, \bibinfo {author} {\bibfnamefont {H.}~\bibnamefont
  {Kreckel}}, \bibinfo {author} {\bibfnamefont {V.}~\bibnamefont {Kokoouline}},
  \ and\ \bibinfo {author} {\bibfnamefont {C.~H.}\ \bibnamefont {Greene}}}
  (\bibinfo {year} {2011}),\ \bibfield  {title} {\enquote {\bibinfo {title}
  {Resonant structure of low-energy {H}$_3^+$ dissociative recombination},}\
  }\href@noop {} {\bibfield  {journal} {\bibinfo  {journal} {Phys. Rev. A}\
  }\textbf {\bibinfo {volume} {83}},\ \bibinfo {pages} {032711}}\BibitemShut
  {NoStop}%
\bibitem [{\citenamefont {Petrov}(2004)}]{Petrov-2004}%
  \BibitemOpen
  \bibfield  {author} {\bibinfo {author} {\bibnamefont {Petrov}, \bibfnamefont
  {D~S}}} (\bibinfo {year} {2004}),\ \bibfield  {title} {\enquote {\bibinfo
  {title} {Three-boson problem near a narrow {F}eshbach resonance},}\
  }\href@noop {} {\bibfield  {journal} {\bibinfo  {journal} {Phys. Rev. Lett.}\
  }\textbf {\bibinfo {volume} {93}},\ \bibinfo {pages} {143201}}\BibitemShut
  {NoStop}%
\bibitem [{\citenamefont {Petrov}(2012)}]{Petrov2012}%
  \BibitemOpen
  \bibfield  {author} {\bibinfo {author} {\bibnamefont {Petrov}, \bibfnamefont
  {D~S}}} (\bibinfo {year} {2012}),\ \bibfield  {title} {\enquote {\bibinfo
  {title} {The few-atom problem},}\ }\href@noop {} {\bibinfo  {journal}
  {arXiv}\ ,\ \bibinfo {pages} {1206.5752}}\BibitemShut {NoStop}%
\bibitem [{\citenamefont {Petrov}\ \emph
  {et~al.}(2000{\natexlab{a}})\citenamefont {Petrov}, \citenamefont
  {Holzmann},\ and\ \citenamefont {Shlyapnikov}}]{petrov2000prl}%
  \BibitemOpen
\bibfield  {journal} {  }\bibfield  {author} {\bibinfo {author} {\bibnamefont
  {Petrov}, \bibfnamefont {D~S}}, \bibinfo {author} {\bibfnamefont
  {M.}~\bibnamefont {Holzmann}}, \ and\ \bibinfo {author} {\bibfnamefont
  {G.~V.}\ \bibnamefont {Shlyapnikov}}} (\bibinfo {year}
  {2000}{\natexlab{a}}),\ \bibfield  {title} {\enquote {\bibinfo {title}
  {Bose-{E}instein condensation in quasi-{2D} trapped gases},}\ }\href@noop {}
  {\bibfield  {journal} {\bibinfo  {journal} {Phys. Rev. Lett.}\ }\textbf
  {\bibinfo {volume} {84}},\ \bibinfo {pages} {2551--2555}}\BibitemShut
  {NoStop}%
\bibitem [{\citenamefont {Petrov}\ \emph
  {et~al.}(2000{\natexlab{b}})\citenamefont {Petrov}, \citenamefont
  {Holzmann},\ and\ \citenamefont
  {Shlyapnikov}}]{PetrovHolzmannShlyapnikov2000prl}%
  \BibitemOpen
  \bibfield  {author} {\bibinfo {author} {\bibnamefont {Petrov}, \bibfnamefont
  {D~S}}, \bibinfo {author} {\bibfnamefont {M.}~\bibnamefont {Holzmann}}, \
  and\ \bibinfo {author} {\bibfnamefont {G.~V.}\ \bibnamefont {Shlyapnikov}}}
  (\bibinfo {year} {2000}{\natexlab{b}}),\ \bibfield  {title} {\enquote
  {\bibinfo {title} {Bose-einstein condensation in quasi-2d trapped gases},}\
  }\href {\doibase 10.1103/PhysRevLett.84.2551} {\bibfield  {journal} {\bibinfo
   {journal} {Phys. Rev. Lett.}\ }\textbf {\bibinfo {volume} {84}},\ \bibinfo
  {pages} {2551--2555}}\BibitemShut {NoStop}%
\bibitem [{\citenamefont {Petrov}\ \emph
  {et~al.}(2005{\natexlab{a}})\citenamefont {Petrov}, \citenamefont {Salomon},\
  and\ \citenamefont {Shlyapnikov}}]{petrov2005JPB}%
  \BibitemOpen
  \bibfield  {author} {\bibinfo {author} {\bibnamefont {Petrov}, \bibfnamefont
  {D~S}}, \bibinfo {author} {\bibfnamefont {C.}~\bibnamefont {Salomon}}, \ and\
  \bibinfo {author} {\bibfnamefont {G.~V.}\ \bibnamefont {Shlyapnikov}}}
  (\bibinfo {year} {2005}{\natexlab{a}}),\ \bibfield  {title} {{\selectlanguage
  {English}\enquote {\bibinfo {title} {Diatomic molecules in ultracold {F}ermi
  gases - novel composite bosons},}\ }}\href@noop {} {\bibfield  {journal}
  {\bibinfo  {journal} {J. Phys. B}\ }\textbf {\bibinfo {volume}
  {38}}~(\bibinfo {number} {9, Sp. Iss. SI}),\ \bibinfo {pages}
  {S645--S660}}\BibitemShut {NoStop}%
\bibitem [{\citenamefont {Petrov}\ \emph
  {et~al.}(2005{\natexlab{b}})\citenamefont {Petrov}, \citenamefont {Salomon},\
  and\ \citenamefont {Shlyapnikov}}]{petrov2005PRA}%
  \BibitemOpen
  \bibfield  {author} {\bibinfo {author} {\bibnamefont {Petrov}, \bibfnamefont
  {D~S}}, \bibinfo {author} {\bibfnamefont {C.}~\bibnamefont {Salomon}}, \ and\
  \bibinfo {author} {\bibfnamefont {G.~V.}\ \bibnamefont {Shlyapnikov}}}
  (\bibinfo {year} {2005}{\natexlab{b}}),\ \bibfield  {title} {{\selectlanguage
  {English}\enquote {\bibinfo {title} {Scattering properties of weakly bound
  dimers of {F}ermionic atoms},}\ }}\href@noop {} {\bibfield  {journal}
  {\bibinfo  {journal} {Phys. Rev. A}\ }\textbf {\bibinfo {volume}
  {71}}~(\bibinfo {number} {1}),\ \bibinfo {pages} {012708}}\BibitemShut
  {NoStop}%
\bibitem [{\citenamefont {Petrov}\ and\ \citenamefont
  {Shlyapnikov}(2001)}]{petrov2001}%
  \BibitemOpen
  \bibfield  {author} {\bibinfo {author} {\bibnamefont {Petrov}, \bibfnamefont
  {D~S}}, \ and\ \bibinfo {author} {\bibfnamefont {G.~V.}\ \bibnamefont
  {Shlyapnikov}}} (\bibinfo {year} {2001}),\ \bibfield  {title} {\enquote
  {\bibinfo {title} {Interatomic collisions in a tightly confined {B}ose
  gas},}\ }\href@noop {} {\bibfield  {journal} {\bibinfo  {journal} {Phys. Rev.
  A}\ }\textbf {\bibinfo {volume} {64}}~(\bibinfo {number} {1}),\ \bibinfo
  {pages} {012706}}\BibitemShut {NoStop}%
\bibitem [{\citenamefont {Petrov}\ and\ \citenamefont
  {Werner}(2015)}]{Petrov2015}%
  \BibitemOpen
  \bibfield  {author} {\bibinfo {author} {\bibnamefont {Petrov}, \bibfnamefont
  {D~S}}, \ and\ \bibinfo {author} {\bibfnamefont {F.}~\bibnamefont {Werner}}}
  (\bibinfo {year} {2015}),\ \bibfield  {title} {\enquote {\bibinfo {title}
  {Three-body recombination in heteronuclear mixtures at finite temperature},}\
  }\href@noop {} {\bibfield  {journal} {\bibinfo  {journal} {Phys. Rev. A}\
  }\textbf {\bibinfo {volume} {92}},\ \bibinfo {pages} {022704}}\BibitemShut
  {NoStop}%
\bibitem [{\citenamefont {Phillips}(1968)}]{Phillips-1968}%
  \BibitemOpen
  \bibfield  {author} {\bibinfo {author} {\bibnamefont {Phillips},
  \bibfnamefont {A~C}}} (\bibinfo {year} {1968}),\ \bibfield  {title} {\enquote
  {\bibinfo {title} {Consistency of the low-energy three-nucleon observables
  and the separable interaction model},}\ }\href@noop {} {\bibfield  {journal}
  {\bibinfo  {journal} {Nuc. Phys. A}\ }\textbf {\bibinfo {volume} {107}},\
  \bibinfo {pages} {209}}\BibitemShut {NoStop}%
\bibitem [{\citenamefont {Piatecki}\ and\ \citenamefont
  {Krauth}({2014})}]{Krauth2014ncomm}%
  \BibitemOpen
  \bibfield  {author} {\bibinfo {author} {\bibnamefont {Piatecki},
  \bibfnamefont {S}}, \ and\ \bibinfo {author} {\bibfnamefont {W.}~\bibnamefont
  {Krauth}}} (\bibinfo {year} {{2014}}),\ \bibfield  {title} {\enquote
  {\bibinfo {title} {{Efimov-driven phase transitions of the unitary Bose
  gas}},}\ }\href@noop {} {\bibfield  {journal} {\bibinfo  {journal} {Nat.
  Commun.}\ }\textbf {\bibinfo {volume} {{5}}},\ \bibinfo {pages}
  {3503}}\BibitemShut {NoStop}%
\bibitem [{\citenamefont {Pindzola}\ and\ \citenamefont
  {Robicheaux}(1998)}]{pindzola1998PRA}%
  \BibitemOpen
  \bibfield  {author} {\bibinfo {author} {\bibnamefont {Pindzola},
  \bibfnamefont {M~S}}, \ and\ \bibinfo {author} {\bibfnamefont
  {F.}~\bibnamefont {Robicheaux}}} (\bibinfo {year} {1998}),\ \bibfield
  {title} {\enquote {\bibinfo {title} {Correlated photoionization processes in
  {H}$^-$},}\ }\href@noop {} {\bibfield  {journal} {\bibinfo  {journal} {Phys.
  Rev. A}\ }\textbf {\bibinfo {volume} {58}},\ \bibinfo {pages}
  {4229--4231}}\BibitemShut {NoStop}%
\bibitem [{\citenamefont {Pindzola}\ \emph {et~al.}(2007)\citenamefont
  {Pindzola}, \citenamefont {Robicheaux}, \citenamefont {Loch}, \citenamefont
  {Berengut}, \citenamefont {Topcu}, \citenamefont {Colgan}, \citenamefont
  {Foster}, \citenamefont {Griffin}, \citenamefont {Ballance}, \citenamefont
  {Schultz}, \citenamefont {Minami}, \citenamefont {Badnell}, \citenamefont
  {Witthoeft}, \citenamefont {Plante}, \citenamefont {Mitnik}, \citenamefont
  {Ludlow},\ and\ \citenamefont {Kleiman}}]{pindzola2007JPB}%
  \BibitemOpen
  \bibfield  {author} {\bibinfo {author} {\bibnamefont {Pindzola},
  \bibfnamefont {M~S}}, \bibinfo {author} {\bibfnamefont {F.}~\bibnamefont
  {Robicheaux}}, \bibinfo {author} {\bibfnamefont {S.~D.}\ \bibnamefont
  {Loch}}, \bibinfo {author} {\bibfnamefont {J.~C.}\ \bibnamefont {Berengut}},
  \bibinfo {author} {\bibfnamefont {T.}~\bibnamefont {Topcu}}, \bibinfo
  {author} {\bibfnamefont {J.}~\bibnamefont {Colgan}}, \bibinfo {author}
  {\bibfnamefont {M.}~\bibnamefont {Foster}}, \bibinfo {author} {\bibfnamefont
  {D.~C.}\ \bibnamefont {Griffin}}, \bibinfo {author} {\bibfnamefont {C.~P.}\
  \bibnamefont {Ballance}}, \bibinfo {author} {\bibfnamefont {D.~R.}\
  \bibnamefont {Schultz}}, \bibinfo {author} {\bibfnamefont {T.}~\bibnamefont
  {Minami}}, \bibinfo {author} {\bibfnamefont {N.~R.}\ \bibnamefont {Badnell}},
  \bibinfo {author} {\bibfnamefont {M.~C.}\ \bibnamefont {Witthoeft}}, \bibinfo
  {author} {\bibfnamefont {D.~R.}\ \bibnamefont {Plante}}, \bibinfo {author}
  {\bibfnamefont {D.~M.}\ \bibnamefont {Mitnik}}, \bibinfo {author}
  {\bibfnamefont {J.~A.}\ \bibnamefont {Ludlow}}, \ and\ \bibinfo {author}
  {\bibfnamefont {U.}~\bibnamefont {Kleiman}}} (\bibinfo {year} {2007}),\
  \bibfield  {title} {\enquote {\bibinfo {title} {The time-dependent
  close-coupling method for atomic and molecular collision processes},}\
  }\href@noop {} {\bibfield  {journal} {\bibinfo  {journal} {J. Phys. B}\
  }\textbf {\bibinfo {volume} {40}}~(\bibinfo {number} {7}),\ \bibinfo {pages}
  {R39}}\BibitemShut {NoStop}%
\bibitem [{\citenamefont {Pires}\ \emph
  {et~al.}(2014{\natexlab{a}})\citenamefont {Pires}, \citenamefont {Repp},
  \citenamefont {Ulmanis}, \citenamefont {Kuhnle}, \citenamefont
  {Weidem\"uller}, \citenamefont {Tiecke}, \citenamefont {Greene},
  \citenamefont {Ruzic}, \citenamefont {Bohn},\ and\ \citenamefont
  {Tiemann}}]{Pires2014pra}%
  \BibitemOpen
  \bibfield  {author} {\bibinfo {author} {\bibnamefont {Pires}, \bibfnamefont
  {R}}, \bibinfo {author} {\bibfnamefont {M.}~\bibnamefont {Repp}}, \bibinfo
  {author} {\bibfnamefont {J.}~\bibnamefont {Ulmanis}}, \bibinfo {author}
  {\bibfnamefont {E.~D.}\ \bibnamefont {Kuhnle}}, \bibinfo {author}
  {\bibfnamefont {M.}~\bibnamefont {Weidem\"uller}}, \bibinfo {author}
  {\bibfnamefont {T.~G.}\ \bibnamefont {Tiecke}}, \bibinfo {author}
  {\bibfnamefont {C.~H.}\ \bibnamefont {Greene}}, \bibinfo {author}
  {\bibfnamefont {B.~P.}\ \bibnamefont {Ruzic}}, \bibinfo {author}
  {\bibfnamefont {J.~L.}\ \bibnamefont {Bohn}}, \ and\ \bibinfo {author}
  {\bibfnamefont {E.}~\bibnamefont {Tiemann}}} (\bibinfo {year}
  {2014}{\natexlab{a}}),\ \bibfield  {title} {\enquote {\bibinfo {title}
  {Analyzing {F}eshbach resonances: A {$^{6}\mathrm{Li}$-${}^{133}\mathrm{Cs}$}
  case study},}\ }\href@noop {} {\bibfield  {journal} {\bibinfo  {journal}
  {Phys. Rev. A}\ }\textbf {\bibinfo {volume} {90}},\ \bibinfo {pages}
  {012710}}\BibitemShut {NoStop}%
\bibitem [{\citenamefont {Pires}\ \emph
  {et~al.}(2014{\natexlab{b}})\citenamefont {Pires}, \citenamefont {Ulmanis},
  \citenamefont {H\"afner}, \citenamefont {M.}, \citenamefont {Arias},
  \citenamefont {Kuhnle},\ and\ \citenamefont {Weidem\"uller}}]{Pires-2014}%
  \BibitemOpen
  \bibfield  {author} {\bibinfo {author} {\bibnamefont {Pires}, \bibfnamefont
  {R}}, \bibinfo {author} {\bibfnamefont {J.}~\bibnamefont {Ulmanis}}, \bibinfo
  {author} {\bibfnamefont {S.}~\bibnamefont {H\"afner}}, \bibinfo {author}
  {\bibfnamefont {Repp.}\ \bibnamefont {M.}}, \bibinfo {author} {\bibfnamefont
  {A.}~\bibnamefont {Arias}}, \bibinfo {author} {\bibfnamefont {E.~D.}\
  \bibnamefont {Kuhnle}}, \ and\ \bibinfo {author} {\bibfnamefont
  {M.}~\bibnamefont {Weidem\"uller}}} (\bibinfo {year} {2014}{\natexlab{b}}),\
  \bibfield  {title} {\enquote {\bibinfo {title} {Observation of {E}fimov
  resonances in a mixture with extreme mass imbalance},}\ }\href@noop {}
  {\bibfield  {journal} {\bibinfo  {journal} {Phys. Rev. Lett.}\ }\textbf
  {\bibinfo {volume} {112}},\ \bibinfo {pages} {250404}}\BibitemShut {NoStop}%
\bibitem [{\citenamefont {Platter}\ \emph {et~al.}(2005)\citenamefont
  {Platter}, \citenamefont {Hammer},\ and\ \citenamefont
  {Mei$\beta$ner}}]{Platter-2005}%
  \BibitemOpen
  \bibfield  {author} {\bibinfo {author} {\bibnamefont {Platter}, \bibfnamefont
  {L}}, \bibinfo {author} {\bibfnamefont {H.{-}W.}\ \bibnamefont {Hammer}}, \
  and\ \bibinfo {author} {\bibfnamefont {U.{-}G.}\ \bibnamefont
  {Mei$\beta$ner}}} (\bibinfo {year} {2005}),\ \bibfield  {title} {\enquote
  {\bibinfo {title} {On the correlation between the binding energies of the
  triton and the $\alpha$-particle},}\ }\href@noop {} {\bibfield  {journal}
  {\bibinfo  {journal} {Phys. Lett. B}\ }\textbf {\bibinfo {volume} {607}},\
  \bibinfo {pages} {254--258}}\BibitemShut {NoStop}%
\bibitem [{\citenamefont {Platter}\ \emph {et~al.}(2004)\citenamefont
  {Platter}, \citenamefont {Hammer},\ and\ \citenamefont
  {Meissner}}]{platter2004PRA}%
  \BibitemOpen
  \bibfield  {author} {\bibinfo {author} {\bibnamefont {Platter}, \bibfnamefont
  {L}}, \bibinfo {author} {\bibfnamefont {H.{-}W.}\ \bibnamefont {Hammer}}, \
  and\ \bibinfo {author} {\bibfnamefont {U.{-}G}\ \bibnamefont {Meissner}}}
  (\bibinfo {year} {2004}),\ \bibfield  {title} {{\selectlanguage
  {English}\enquote {\bibinfo {title} {Four-boson system with short-range
  interactions},}\ }}\href@noop {} {\bibfield  {journal} {\bibinfo  {journal}
  {Phys. Rev. A}\ }\textbf {\bibinfo {volume} {70}}~(\bibinfo {number} {5}),\
  \bibinfo {pages} {052101}}\BibitemShut {NoStop}%
\bibitem [{\citenamefont {Pohl}\ \emph {et~al.}(2008)\citenamefont {Pohl},
  \citenamefont {Vrinceanu},\ and\ \citenamefont {Sadeghpour}}]{pohl2008prl}%
  \BibitemOpen
  \bibfield  {author} {\bibinfo {author} {\bibnamefont {Pohl}, \bibfnamefont
  {T}}, \bibinfo {author} {\bibfnamefont {D.}~\bibnamefont {Vrinceanu}}, \ and\
  \bibinfo {author} {\bibfnamefont {H.~R.}\ \bibnamefont {Sadeghpour}}}
  (\bibinfo {year} {2008}),\ \bibfield  {title} {\enquote {\bibinfo {title}
  {Rydberg atom formation in ultracold plasmas: Small energy transfer with
  large consequences},}\ }\href@noop {} {\bibfield  {journal} {\bibinfo
  {journal} {Phys. Rev. Lett.}\ }\textbf {\bibinfo {volume} {100}},\ \bibinfo
  {pages} {223201}}\BibitemShut {NoStop}%
\bibitem [{\citenamefont {Pollack}\ \emph {et~al.}(2009)\citenamefont
  {Pollack}, \citenamefont {Dries},\ and\ \citenamefont
  {Hulet}}]{Pollack-2009}%
  \BibitemOpen
  \bibfield  {author} {\bibinfo {author} {\bibnamefont {Pollack}, \bibfnamefont
  {S~E}}, \bibinfo {author} {\bibfnamefont {D.}~\bibnamefont {Dries}}, \ and\
  \bibinfo {author} {\bibfnamefont {R.~G.}\ \bibnamefont {Hulet}}} (\bibinfo
  {year} {2009}),\ \bibfield  {title} {\enquote {\bibinfo {title} {Universality
  in three- and four-body bound states of ultracold atoms},}\ }\href@noop {}
  {\bibfield  {journal} {\bibinfo  {journal} {Science}\ }\textbf {\bibinfo
  {volume} {326}},\ \bibinfo {pages} {1683--1685}}\BibitemShut {NoStop}%
\bibitem [{\citenamefont {Press}\ \emph {et~al.}(1986)\citenamefont {Press},
  \citenamefont {Teukolsky}, \citenamefont {Vetterling},\ and\ \citenamefont
  {Flannery}}]{Numerical_Recipes}%
  \BibitemOpen
  \bibfield  {author} {\bibinfo {author} {\bibnamefont {Press}, \bibfnamefont
  {W~H}}, \bibinfo {author} {\bibfnamefont {S.~A.}\ \bibnamefont {Teukolsky}},
  \bibinfo {author} {\bibfnamefont {W.~T.}\ \bibnamefont {Vetterling}}, \ and\
  \bibinfo {author} {\bibfnamefont {B.~P.}\ \bibnamefont {Flannery}}} (\bibinfo
  {year} {1986}),\ \href@noop {} {\emph {\bibinfo {title} {Numerical Recipes in
  Fotran 77}}}\ (\bibinfo  {publisher} {Cambridge University Press},\ \bibinfo
  {address} {Cambridge, England})\BibitemShut {NoStop}%
\bibitem [{\citenamefont {Pricoupenko}(2008)}]{pricoupenko2008prl}%
  \BibitemOpen
  \bibfield  {author} {\bibinfo {author} {\bibnamefont {Pricoupenko},
  \bibfnamefont {L}}} (\bibinfo {year} {2008}),\ \bibfield  {title} {\enquote
  {\bibinfo {title} {Resonant scattering of ultracold atoms in low
  dimensions},}\ }\href@noop {} {\bibfield  {journal} {\bibinfo  {journal}
  {Phys. Rev. Lett.}\ }\textbf {\bibinfo {volume} {100}},\ \bibinfo {pages}
  {170404}}\BibitemShut {NoStop}%
\bibitem [{\citenamefont {Pricoupenko}(2011)}]{Pricoupenk2011pra}%
  \BibitemOpen
  \bibfield  {author} {\bibinfo {author} {\bibnamefont {Pricoupenko},
  \bibfnamefont {Ludovic}}} (\bibinfo {year} {2011}),\ \bibfield  {title}
  {\enquote {\bibinfo {title} {Isotropic contact forces in arbitrary
  representation: Heterogeneous few-body problems and low dimensions},}\ }\href
  {\doibase 10.1103/PhysRevA.83.062711} {\bibfield  {journal} {\bibinfo
  {journal} {Phys. Rev. A}\ }\textbf {\bibinfo {volume} {83}},\ \bibinfo
  {pages} {062711}}\BibitemShut {NoStop}%
\bibitem [{\citenamefont {Radzihovsky}\ \emph {et~al.}(2008)\citenamefont
  {Radzihovsky}, \citenamefont {Weichman},\ and\ \citenamefont
  {Park}}]{Radzihovsky20082376}%
  \BibitemOpen
  \bibfield  {author} {\bibinfo {author} {\bibnamefont {Radzihovsky},
  \bibfnamefont {L}}, \bibinfo {author} {\bibfnamefont {P.~B.}\ \bibnamefont
  {Weichman}}, \ and\ \bibinfo {author} {\bibfnamefont {J.~I.}\ \bibnamefont
  {Park}}} (\bibinfo {year} {2008}),\ \bibfield  {title} {\enquote {\bibinfo
  {title} {Superfluidity and phase transitions in a resonant bose gas},}\
  }\href@noop {} {\bibfield  {journal} {\bibinfo  {journal} {Ann. Phys.}\
  }\textbf {\bibinfo {volume} {323}}~(\bibinfo {number} {10}),\ \bibinfo
  {pages} {2376 -- 2451}}\BibitemShut {NoStop}%
\bibitem [{\citenamefont {Rakshit}\ and\ \citenamefont
  {Blume}({2012})}]{RakshitBlume2012pra}%
  \BibitemOpen
  \bibfield  {author} {\bibinfo {author} {\bibnamefont {Rakshit}, \bibfnamefont
  {D}}, \ and\ \bibinfo {author} {\bibfnamefont {D.}~\bibnamefont {Blume}}}
  (\bibinfo {year} {{2012}}),\ \bibfield  {title} {\enquote {\bibinfo {title}
  {{Hyperspherical explicitly correlated Gaussian approach for few-body systems
  with finite angular momentum}},}\ }\href {\doibase
  {10.1103/PhysRevA.86.062513}} {\bibfield  {journal} {\bibinfo  {journal}
  {{Physical Review A}}\ }\textbf {\bibinfo {volume} {{86}}}~(\bibinfo {number}
  {{6}}),\ {10.1103/PhysRevA.86.062513}}\BibitemShut {NoStop}%
\bibitem [{\citenamefont {Rancon}\ and\ \citenamefont
  {Levin}({2014})}]{Levin2014pra}%
  \BibitemOpen
  \bibfield  {author} {\bibinfo {author} {\bibnamefont {Rancon}, \bibfnamefont
  {A}}, \ and\ \bibinfo {author} {\bibfnamefont {K.}~\bibnamefont {Levin}}}
  (\bibinfo {year} {{2014}}),\ \bibfield  {title} {\enquote {\bibinfo {title}
  {{Equilibrating dynamics in quenched Bose gases: Characterizing multiple time
  regimes}},}\ }\href@noop {} {\bibfield  {journal} {\bibinfo  {journal}
  {{Phys. Rev. A}}\ }\textbf {\bibinfo {volume} {{90}}}~(\bibinfo {number}
  {{2}})}\BibitemShut {NoStop}%
\bibitem [{\citenamefont {Rath}\ and\ \citenamefont
  {Schmidt}(2013)}]{rath_field-theoretical_2013}%
  \BibitemOpen
  \bibfield  {author} {\bibinfo {author} {\bibnamefont {Rath}, \bibfnamefont
  {Steffen~Patrick}}, \ and\ \bibinfo {author} {\bibfnamefont {Richard}\
  \bibnamefont {Schmidt}}} (\bibinfo {year} {2013}),\ \bibfield  {title}
  {\enquote {\bibinfo {title} {Field-theoretical study of the {Bose}
  polaron},}\ }\href {\doibase 10.1103/PhysRevA.88.053632} {\bibfield
  {journal} {\bibinfo  {journal} {Physical Review A}\ }\textbf {\bibinfo
  {volume} {88}}~(\bibinfo {number} {5}),\ \bibinfo {pages}
  {053632}}\BibitemShut {NoStop}%
\bibitem [{\citenamefont {Rau}(1971)}]{RAU1971}%
  \BibitemOpen
  \bibfield  {author} {\bibinfo {author} {\bibnamefont {Rau}, \bibfnamefont
  {A~R~P}}} (\bibinfo {year} {1971}),\ \bibfield  {title} {\enquote {\bibinfo
  {title} {2 electrons in a {C}oulomb potential - double-continuum wave
  functions and threshold law for electron-atom ionization},}\ }\href@noop {}
  {\bibfield  {journal} {\bibinfo  {journal} {Phys. Rev. A}\ }\textbf {\bibinfo
  {volume} {4}}~(\bibinfo {number} {1}),\ \bibinfo {pages}
  {207--\&}}\BibitemShut {NoStop}%
\bibitem [{\citenamefont {Rau}(1984)}]{rau1984PRep}%
  \BibitemOpen
  \bibfield  {author} {\bibinfo {author} {\bibnamefont {Rau}, \bibfnamefont
  {A~R~P}}} (\bibinfo {year} {1984}),\ \bibfield  {title} {{\selectlanguage
  {English}\enquote {\bibinfo {title} {The {W}annier theory for 2 electrons
  escaping from a positive-ion},}\ }}\href@noop {} {\bibfield  {journal}
  {\bibinfo  {journal} {Phys. Rep.}\ }\textbf {\bibinfo {volume}
  {110}}~(\bibinfo {number} {5-6}),\ \bibinfo {pages} {369--387}}\BibitemShut
  {NoStop}%
\bibitem [{\citenamefont {RAU}({1992})}]{Rau1992science}%
  \BibitemOpen
  \bibfield  {author} {\bibinfo {author} {\bibnamefont {RAU}, \bibfnamefont
  {ARP}}} (\bibinfo {year} {{1992}}),\ \bibfield  {title} {\enquote {\bibinfo
  {title} {{Excitation and Decay of Correlated Atomic States}},}\ }\href
  {\doibase {10.1126/science.258.5087.1444}} {\bibfield  {journal} {\bibinfo
  {journal} {{SCIENCE}}\ }\textbf {\bibinfo {volume} {{258}}}~(\bibinfo
  {number} {{5087}}),\ \bibinfo {pages} {{1444--1451}}}\BibitemShut {NoStop}%
\bibitem [{\citenamefont {Read}(1984)}]{READ1984}%
  \BibitemOpen
  \bibfield  {author} {\bibinfo {author} {\bibnamefont {Read}, \bibfnamefont
  {F~H}}} (\bibinfo {year} {1984}),\ \bibfield  {title} {\enquote {\bibinfo
  {title} {Extensions of the {W}annier theory for near-threshold excitation and
  ionization of atoms by electron-impact},}\ }\href@noop {} {\bibfield
  {journal} {\bibinfo  {journal} {J. Phys. B}\ }\textbf {\bibinfo {volume}
  {17}}~(\bibinfo {number} {19}),\ \bibinfo {pages} {3965--3986}}\BibitemShut
  {NoStop}%
\bibitem [{\citenamefont {Regal}\ \emph {et~al.}(2005)\citenamefont {Regal},
  \citenamefont {Greiner}, \citenamefont {Giorgini}, \citenamefont {Holland},\
  and\ \citenamefont {Jin}}]{regal2005PRL}%
  \BibitemOpen
  \bibfield  {author} {\bibinfo {author} {\bibnamefont {Regal}, \bibfnamefont
  {C~A}}, \bibinfo {author} {\bibfnamefont {M.}~\bibnamefont {Greiner}},
  \bibinfo {author} {\bibfnamefont {S.}~\bibnamefont {Giorgini}}, \bibinfo
  {author} {\bibfnamefont {M.}~\bibnamefont {Holland}}, \ and\ \bibinfo
  {author} {\bibfnamefont {D.~S.}\ \bibnamefont {Jin}}} (\bibinfo {year}
  {2005}),\ \bibfield  {title} {\enquote {\bibinfo {title} {Momentum
  distribution of a {F}ermi gas of atoms in the {BCS}-{BEC} crossover},}\
  }\href@noop {} {\bibfield  {journal} {\bibinfo  {journal} {Phys. Rev. Lett.}\
  }\textbf {\bibinfo {volume} {95}}~(\bibinfo {number} {25}),\ \bibinfo {pages}
  {250404}}\BibitemShut {NoStop}%
\bibitem [{\citenamefont {Regal}\ \emph
  {et~al.}(2004{\natexlab{a}})\citenamefont {Regal}, \citenamefont {Greiner},\
  and\ \citenamefont {Jin}}]{regal2004PRL}%
  \BibitemOpen
  \bibfield  {author} {\bibinfo {author} {\bibnamefont {Regal}, \bibfnamefont
  {C~A}}, \bibinfo {author} {\bibfnamefont {M.}~\bibnamefont {Greiner}}, \ and\
  \bibinfo {author} {\bibfnamefont {D.~S.}\ \bibnamefont {Jin}}} (\bibinfo
  {year} {2004}{\natexlab{a}}),\ \bibfield  {title} {\enquote {\bibinfo {title}
  {{Li}fetime of molecule-atom mixtures near a {F}eshbach resonance in
  $^{40}${K}},}\ }\href@noop {} {\bibfield  {journal} {\bibinfo  {journal}
  {Phys. Rev. Lett.}\ }\textbf {\bibinfo {volume} {92}},\ \bibinfo {pages}
  {083201}}\BibitemShut {NoStop}%
\bibitem [{\citenamefont {Regal}\ \emph
  {et~al.}(2004{\natexlab{b}})\citenamefont {Regal}, \citenamefont {Greiner},\
  and\ \citenamefont {Jin}}]{regal2004PRLb}%
  \BibitemOpen
  \bibfield  {author} {\bibinfo {author} {\bibnamefont {Regal}, \bibfnamefont
  {C~A}}, \bibinfo {author} {\bibfnamefont {M.}~\bibnamefont {Greiner}}, \ and\
  \bibinfo {author} {\bibfnamefont {D.~S.}\ \bibnamefont {Jin}}} (\bibinfo
  {year} {2004}{\natexlab{b}}),\ \bibfield  {title} {\enquote {\bibinfo {title}
  {Observation of resonance condensation of {F}ermionic atom pairs},}\
  }\href@noop {} {\bibfield  {journal} {\bibinfo  {journal} {Phys. Rev. Lett.}\
  }\textbf {\bibinfo {volume} {92}}~(\bibinfo {number} {4}),\ \bibinfo {pages}
  {040403}}\BibitemShut {NoStop}%
\bibitem [{\citenamefont {Regal}\ and\ \citenamefont {Jin}(2007)}]{regal2007}%
  \BibitemOpen
  \bibfield  {author} {\bibinfo {author} {\bibnamefont {Regal}, \bibfnamefont
  {C~A}}, \ and\ \bibinfo {author} {\bibfnamefont {D.~S.}\ \bibnamefont {Jin}}}
  (\bibinfo {year} {2007}),\ \bibfield  {title} {\enquote {\bibinfo {title}
  {Experimental realization of the {BCS}-{BEC} crossover with a {F}ermi gas of
  atoms},}\ }\href@noop {} {\bibfield  {journal} {\bibinfo  {journal} {Adv. At.
  Mol. Opt. Phys.}\ }\textbf {\bibinfo {volume} {54}},\ \bibinfo {pages}
  {1--79}}\BibitemShut {NoStop}%
\bibitem [{\citenamefont {Regal}\ \emph {et~al.}(2003)\citenamefont {Regal},
  \citenamefont {Ticknor}, \citenamefont {Bohn},\ and\ \citenamefont
  {Jin}}]{regal2003PRLb}%
  \BibitemOpen
  \bibfield  {author} {\bibinfo {author} {\bibnamefont {Regal}, \bibfnamefont
  {C~A}}, \bibinfo {author} {\bibfnamefont {C.}~\bibnamefont {Ticknor}},
  \bibinfo {author} {\bibfnamefont {J.~L.}\ \bibnamefont {Bohn}}, \ and\
  \bibinfo {author} {\bibfnamefont {D.~S.}\ \bibnamefont {Jin}}} (\bibinfo
  {year} {2003}),\ \bibfield  {title} {\enquote {\bibinfo {title} {Tuning
  $p$-wave interactions in an ultracold {F}ermi gas of atoms},}\ }\href@noop {}
  {\bibfield  {journal} {\bibinfo  {journal} {Phys. Rev. Lett.}\ }\textbf
  {\bibinfo {volume} {90}},\ \bibinfo {pages} {053201}}\BibitemShut {NoStop}%
\bibitem [{\citenamefont {Rem}\ \emph {et~al.}(2013)\citenamefont {Rem},
  \citenamefont {Grier}, \citenamefont {Ferrier-Barbut}, \citenamefont
  {Eismann}, \citenamefont {Langen}, \citenamefont {Navon}, \citenamefont
  {Khaykovich}, \citenamefont {Werner}, \citenamefont {Petrov}, \citenamefont
  {Chevy},\ and\ \citenamefont {Salomon}}]{Salomon2013prl}%
  \BibitemOpen
  \bibfield  {author} {\bibinfo {author} {\bibnamefont {Rem}, \bibfnamefont
  {B~S}}, \bibinfo {author} {\bibfnamefont {A.~T.}\ \bibnamefont {Grier}},
  \bibinfo {author} {\bibfnamefont {I.}~\bibnamefont {Ferrier-Barbut}},
  \bibinfo {author} {\bibfnamefont {U.}~\bibnamefont {Eismann}}, \bibinfo
  {author} {\bibfnamefont {T.}~\bibnamefont {Langen}}, \bibinfo {author}
  {\bibfnamefont {N.}~\bibnamefont {Navon}}, \bibinfo {author} {\bibfnamefont
  {L.}~\bibnamefont {Khaykovich}}, \bibinfo {author} {\bibfnamefont
  {F.}~\bibnamefont {Werner}}, \bibinfo {author} {\bibfnamefont {D.~S.}\
  \bibnamefont {Petrov}}, \bibinfo {author} {\bibfnamefont {F.}~\bibnamefont
  {Chevy}}, \ and\ \bibinfo {author} {\bibfnamefont {C.}~\bibnamefont
  {Salomon}}} (\bibinfo {year} {2013}),\ \bibfield  {title} {\enquote {\bibinfo
  {title} {Lifetime of the {B}ose gas with resonant interactions},}\
  }\href@noop {} {\bibfield  {journal} {\bibinfo  {journal} {Phys. Rev. Lett.}\
  }\textbf {\bibinfo {volume} {110}},\ \bibinfo {pages} {163202}}\BibitemShut
  {NoStop}%
\bibitem [{\citenamefont {Fabre~de~la
  Ripelle}(1993)}]{FabreDeLaRipelle1993FBS}%
  \BibitemOpen
  \bibfield  {author} {\bibinfo {author} {\bibnamefont {Fabre~de~la Ripelle},
  \bibfnamefont {M}}} (\bibinfo {year} {1993}),\ \bibfield  {title} {\enquote
  {\bibinfo {title} {Green function and scattering amplitudes in
  many-dimensional space},}\ }\href@noop {} {\bibfield  {journal} {\bibinfo
  {journal} {Few-Body Systems}\ }\textbf {\bibinfo {volume} {14}}~(\bibinfo
  {number} {1}),\ \bibinfo {pages} {1}}\BibitemShut {NoStop}%
\bibitem [{\citenamefont {Rittenhouse}\ \emph {et~al.}(2006)\citenamefont
  {Rittenhouse}, \citenamefont {Cavagnero}, \citenamefont {{von Stecher}},\
  and\ \citenamefont {Greene}}]{rittenhouse2006FBS}%
  \BibitemOpen
  \bibfield  {author} {\bibinfo {author} {\bibnamefont {Rittenhouse},
  \bibfnamefont {S~T}}, \bibinfo {author} {\bibfnamefont {M.~J.}\ \bibnamefont
  {Cavagnero}}, \bibinfo {author} {\bibfnamefont {J.}~\bibnamefont {{von
  Stecher}}}, \ and\ \bibinfo {author} {\bibfnamefont {C.~H.}\ \bibnamefont
  {Greene}}} (\bibinfo {year} {2006}),\ \bibfield  {title} {{\selectlanguage
  {English}\enquote {\bibinfo {title} {A hyperspherical variational approach to
  the {N}-{F}ermion problem},}\ }}\href@noop {} {\bibfield  {journal} {\bibinfo
   {journal} {Few-Body Systems}\ }\textbf {\bibinfo {volume} {38}}~(\bibinfo
  {number} {2-4}),\ \bibinfo {pages} {85--90}}\BibitemShut {NoStop}%
\bibitem [{\citenamefont {Rittenhouse}\ and\ \citenamefont
  {Greene}(2008)}]{rittenhouse2008JPB}%
  \BibitemOpen
  \bibfield  {author} {\bibinfo {author} {\bibnamefont {Rittenhouse},
  \bibfnamefont {S~T}}, \ and\ \bibinfo {author} {\bibfnamefont {C.~H.}\
  \bibnamefont {Greene}}} (\bibinfo {year} {2008}),\ \bibfield  {title}
  {{\selectlanguage {English}\enquote {\bibinfo {title} {The degenerate {F}ermi
  gas with density-dependent interactions in the large-n limit under the
  {K}-harmonic approximation},}\ }}\href@noop {} {\bibfield  {journal}
  {\bibinfo  {journal} {J. Phys. B}\ }\textbf {\bibinfo {volume}
  {41}}~(\bibinfo {number} {20}),\ \bibinfo {pages} {205302}}\BibitemShut
  {NoStop}%
\bibitem [{\citenamefont {Rittenhouse}\ \emph
  {et~al.}(2011{\natexlab{a}})\citenamefont {Rittenhouse}, \citenamefont {von
  Stecher}, \citenamefont {D\'{}Incao}, \citenamefont {Mehta},\ and\
  \citenamefont {Greene}}]{Rittenhouse-2011JPB}%
  \BibitemOpen
  \bibfield  {author} {\bibinfo {author} {\bibnamefont {Rittenhouse},
  \bibfnamefont {S~T}}, \bibinfo {author} {\bibfnamefont {J.}~\bibnamefont {von
  Stecher}}, \bibinfo {author} {\bibfnamefont {J.~P.}\ \bibnamefont
  {D\'{}Incao}}, \bibinfo {author} {\bibfnamefont {N.~P.}\ \bibnamefont
  {Mehta}}, \ and\ \bibinfo {author} {\bibfnamefont {C.~H.}\ \bibnamefont
  {Greene}}} (\bibinfo {year} {2011}{\natexlab{a}}),\ \href@noop {} {\bibfield
  {journal} {\bibinfo  {journal} {J. Phys. B}\ }\textbf {\bibinfo {volume}
  {44}},\ \bibinfo {pages} {172001}}\BibitemShut {NoStop}%
\bibitem [{\citenamefont {Rittenhouse}\ \emph
  {et~al.}(2011{\natexlab{b}})\citenamefont {Rittenhouse}, \citenamefont {von
  Stecher}, \citenamefont {D'Incao}, \citenamefont {Mehta},\ and\ \citenamefont
  {Greene}}]{rittenhouse2011JPB}%
  \BibitemOpen
  \bibfield  {author} {\bibinfo {author} {\bibnamefont {Rittenhouse},
  \bibfnamefont {S~T}}, \bibinfo {author} {\bibfnamefont {J.}~\bibnamefont {von
  Stecher}}, \bibinfo {author} {\bibfnamefont {J.~P.}\ \bibnamefont {D'Incao}},
  \bibinfo {author} {\bibfnamefont {N.~P.}\ \bibnamefont {Mehta}}, \ and\
  \bibinfo {author} {\bibfnamefont {C.~H.}\ \bibnamefont {Greene}}} (\bibinfo
  {year} {2011}{\natexlab{b}}),\ \bibfield  {title} {{\selectlanguage
  {English}\enquote {\bibinfo {title} {The hyperspherical four-fermion
  problem},}\ }}\href@noop {} {\bibfield  {journal} {\bibinfo  {journal} {J.
  Phys. B}\ }\textbf {\bibinfo {volume} {44}}~(\bibinfo {number} {17}),\
  \bibinfo {pages} {172001}}\BibitemShut {NoStop}%
\bibitem [{\citenamefont {Rittenhouse}\ \emph {et~al.}(2016)\citenamefont
  {Rittenhouse}, \citenamefont {Wray},\ and\ \citenamefont
  {Johnson}}]{Rittenhouse2016pra}%
  \BibitemOpen
  \bibfield  {author} {\bibinfo {author} {\bibnamefont {Rittenhouse},
  \bibfnamefont {S~T}}, \bibinfo {author} {\bibfnamefont {A.}~\bibnamefont
  {Wray}}, \ and\ \bibinfo {author} {\bibfnamefont {B.~L.}\ \bibnamefont
  {Johnson}}} (\bibinfo {year} {2016}),\ \bibfield  {title} {\enquote {\bibinfo
  {title} {Hyperspherical approach to a three-boson problem in two dimensions
  with a magnetic field},}\ }\href@noop {} {\bibfield  {journal} {\bibinfo
  {journal} {Phys. Rev. A}\ }\textbf {\bibinfo {volume} {93}},\ \bibinfo
  {pages} {012511}}\BibitemShut {NoStop}%
\bibitem [{\citenamefont {Robicheaux}(2006)}]{Francis-2006}%
  \BibitemOpen
  \bibfield  {author} {\bibinfo {author} {\bibnamefont {Robicheaux},
  \bibfnamefont {F}}} (\bibinfo {year} {2006}),\ \bibfield  {title} {\enquote
  {\bibinfo {title} {Three-body recombination for electrons in a strong
  magnetic field: {M}agnetic moment},}\ }\href@noop {} {\bibfield  {journal}
  {\bibinfo  {journal} {Phys. Rev A}\ }\textbf {\bibinfo {volume} {73}},\
  \bibinfo {pages} {033401}}\BibitemShut {NoStop}%
\bibitem [{\citenamefont {Robicheaux}(2007)}]{robicheaux2007JPB}%
  \BibitemOpen
  \bibfield  {author} {\bibinfo {author} {\bibnamefont {Robicheaux},
  \bibfnamefont {F}}} (\bibinfo {year} {2007}),\ \bibfield  {title} {\enquote
  {\bibinfo {title} {Three-body recombination with mixed sign light
  particles},}\ }\href@noop {} {\bibfield  {journal} {\bibinfo  {journal} {J.
  Phys. B}\ }\textbf {\bibinfo {volume} {40}}~(\bibinfo {number} {2}),\
  \bibinfo {pages} {271--280}}\BibitemShut {NoStop}%
\bibitem [{\citenamefont {Robicheaux}\ \emph {et~al.}(2015)\citenamefont
  {Robicheaux}, \citenamefont {Giannakeas},\ and\ \citenamefont
  {Greene}}]{robicheauxschwinger2015}%
  \BibitemOpen
  \bibfield  {author} {\bibinfo {author} {\bibnamefont {Robicheaux},
  \bibfnamefont {F}}, \bibinfo {author} {\bibfnamefont {P.}~\bibnamefont
  {Giannakeas}}, \ and\ \bibinfo {author} {\bibfnamefont {C.~H.}\ \bibnamefont
  {Greene}}} (\bibinfo {year} {2015}),\ \bibfield  {title} {\enquote {\bibinfo
  {title} {Schwinger-variational-principle theory of collisions in the presence
  of multiple potentials},}\ }\href@noop {} {\bibfield  {journal} {\bibinfo
  {journal} {Phys. Rev. A}\ }\textbf {\bibinfo {volume} {92}},\ \bibinfo
  {pages} {022711}}\BibitemShut {NoStop}%
\bibitem [{\citenamefont {Robicheaux}\ \emph {et~al.}(2010)\citenamefont
  {Robicheaux}, \citenamefont {Loch}, \citenamefont {Pindzola},\ and\
  \citenamefont {Ballance}}]{robicheaux2010PRL}%
  \BibitemOpen
  \bibfield  {author} {\bibinfo {author} {\bibnamefont {Robicheaux},
  \bibfnamefont {F}}, \bibinfo {author} {\bibfnamefont {S.~D.}\ \bibnamefont
  {Loch}}, \bibinfo {author} {\bibfnamefont {M.~S.}\ \bibnamefont {Pindzola}},
  \ and\ \bibinfo {author} {\bibfnamefont {C.~P.}\ \bibnamefont {Ballance}}}
  (\bibinfo {year} {2010}),\ \bibfield  {title} {\enquote {\bibinfo {title}
  {Contribution of near threshold states to recombination in plasmas},}\
  }\href@noop {} {\bibfield  {journal} {\bibinfo  {journal} {Phys. Rev. Lett.}\
  }\textbf {\bibinfo {volume} {105}}~(\bibinfo {number} {23}),\ \bibinfo
  {pages} {233201}}\BibitemShut {NoStop}%
\bibitem [{\citenamefont {Robicheaux}\ \emph {et~al.}(1997)\citenamefont
  {Robicheaux}, \citenamefont {Pindzola},\ and\ \citenamefont
  {Plante}}]{Robicheaux1997}%
  \BibitemOpen
  \bibfield  {author} {\bibinfo {author} {\bibnamefont {Robicheaux},
  \bibfnamefont {F}}, \bibinfo {author} {\bibfnamefont {M.~S.}\ \bibnamefont
  {Pindzola}}, \ and\ \bibinfo {author} {\bibfnamefont {D.~R.}\ \bibnamefont
  {Plante}}} (\bibinfo {year} {1997}),\ \bibfield  {title} {\enquote {\bibinfo
  {title} {Time-dependent quantal calculations for {L}=0 models of the
  electron-impact ionization of hydrogen near threshold},}\ }\href@noop {}
  {\bibfield  {journal} {\bibinfo  {journal} {Physical Review A}\ }\textbf
  {\bibinfo {volume} {55}}~(\bibinfo {number} {5}),\ \bibinfo {pages}
  {3573--3579}}\BibitemShut {NoStop}%
\bibitem [{\citenamefont {Rodberg}\ and\ \citenamefont
  {Thaler}(1970)}]{Rodberg1970}%
  \BibitemOpen
  \bibfield  {author} {\bibinfo {author} {\bibnamefont {Rodberg}, \bibfnamefont
  {L~S}}, \ and\ \bibinfo {author} {\bibfnamefont {R.~M.}\ \bibnamefont
  {Thaler}}} (\bibinfo {year} {1970}),\ \href@noop {} {\emph {\bibinfo {title}
  {Introduction to the Quantum Theory of Scattering (Pure and Applied Physics,
  A Series of Monographs and Textbooks, Volume 26)}}}\ (\bibinfo  {publisher}
  {Academic Press})\BibitemShut {NoStop}%
\bibitem [{\citenamefont {Roy}\ \emph {et~al.}(2013)\citenamefont {Roy},
  \citenamefont {Landini}, \citenamefont {Trenkwalder}, \citenamefont
  {Semeghini}, \citenamefont {Spagnolli}, \citenamefont {Simoni}, \citenamefont
  {Fattori}, \citenamefont {Inguscio},\ and\ \citenamefont
  {Modugno}}]{roy2013prl}%
  \BibitemOpen
  \bibfield  {author} {\bibinfo {author} {\bibnamefont {Roy}, \bibfnamefont
  {S}}, \bibinfo {author} {\bibfnamefont {M.}~\bibnamefont {Landini}}, \bibinfo
  {author} {\bibfnamefont {A.}~\bibnamefont {Trenkwalder}}, \bibinfo {author}
  {\bibfnamefont {G.}~\bibnamefont {Semeghini}}, \bibinfo {author}
  {\bibfnamefont {G.}~\bibnamefont {Spagnolli}}, \bibinfo {author}
  {\bibfnamefont {A.}~\bibnamefont {Simoni}}, \bibinfo {author} {\bibfnamefont
  {M.}~\bibnamefont {Fattori}}, \bibinfo {author} {\bibfnamefont
  {M.}~\bibnamefont {Inguscio}}, \ and\ \bibinfo {author} {\bibfnamefont
  {G.}~\bibnamefont {Modugno}}} (\bibinfo {year} {2013}),\ \bibfield  {title}
  {\enquote {\bibinfo {title} {Test of the universality of the three-body
  {E}fimov parameter at narrow {F}eshbach resonances},}\ }\href@noop {}
  {\bibfield  {journal} {\bibinfo  {journal} {Phys. Rev. Lett.}\ }\textbf
  {\bibinfo {volume} {111}},\ \bibinfo {pages} {053202}}\BibitemShut {NoStop}%
\bibitem [{\citenamefont {Ruzic}\ \emph {et~al.}(2013)\citenamefont {Ruzic},
  \citenamefont {Greene},\ and\ \citenamefont {Bohn}}]{ruzic2013}%
  \BibitemOpen
  \bibfield  {author} {\bibinfo {author} {\bibnamefont {Ruzic}, \bibfnamefont
  {B}}, \bibinfo {author} {\bibfnamefont {C.~H.}\ \bibnamefont {Greene}}, \
  and\ \bibinfo {author} {\bibfnamefont {J.}~\bibnamefont {Bohn}}} (\bibinfo
  {year} {2013}),\ \bibfield  {title} {\enquote {\bibinfo {title} {Quantum
  defect theory for high-partial-wave cold collisions},}\ }\href@noop {}
  {\bibfield  {journal} {\bibinfo  {journal} {Phys. Rev. A}\ }\textbf {\bibinfo
  {volume} {87}}~(\bibinfo {number} {3}),\ \bibinfo {pages}
  {032706}}\BibitemShut {NoStop}%
\bibitem [{\citenamefont {{S\'{a} de Melo}}\ \emph {et~al.}(1993)\citenamefont
  {{S\'{a} de Melo}}, \citenamefont {Randeria},\ and\ \citenamefont
  {Engelbrecht}}]{demelo1993PRL}%
  \BibitemOpen
  \bibfield  {author} {\bibinfo {author} {\bibnamefont {{S\'{a} de Melo}},
  \bibfnamefont {C~A~R}}, \bibinfo {author} {\bibfnamefont {M.}~\bibnamefont
  {Randeria}}, \ and\ \bibinfo {author} {\bibfnamefont {J.~R.}\ \bibnamefont
  {Engelbrecht}}} (\bibinfo {year} {1993}),\ \bibfield  {title} {\enquote
  {\bibinfo {title} {Crossover from {{BCS}} to {B}ose superconductivity:
  Transition temperature and time-dependent {G}inzburg-{L}andau theory},}\
  }\href@noop {} {\bibfield  {journal} {\bibinfo  {journal} {Phys. Rev. Lett.}\
  }\textbf {\bibinfo {volume} {71}},\ \bibinfo {pages} {3202}}\BibitemShut
  {NoStop}%
\bibitem [{\citenamefont {Sadeghpour}\ and\ \citenamefont
  {Greene}(1990)}]{sadeghpour1990PRL}%
  \BibitemOpen
  \bibfield  {author} {\bibinfo {author} {\bibnamefont {Sadeghpour},
  \bibfnamefont {H~R}}, \ and\ \bibinfo {author} {\bibfnamefont {C.~H.}\
  \bibnamefont {Greene}}} (\bibinfo {year} {1990}),\ \bibfield  {title}
  {{\selectlanguage {English}\enquote {\bibinfo {title} {Dominant
  photodetachment channels in {H}$^-$},}\ }}\href@noop {} {\bibfield  {journal}
  {\bibinfo  {journal} {Phys. Rev. Lett.}\ }\textbf {\bibinfo {volume}
  {65}}~(\bibinfo {number} {3}),\ \bibinfo {pages} {313--316}}\BibitemShut
  {NoStop}%
\bibitem [{\citenamefont {Sadeghpour}\ \emph {et~al.}(2000)\citenamefont
  {Sadeghpour}, \citenamefont {Bohn}, \citenamefont {Cavagnero}, \citenamefont
  {Esry}, \citenamefont {Fabrikant}, \citenamefont {Macek},\ and\ \citenamefont
  {Rau}}]{sadeghpour2000JPB}%
  \BibitemOpen
  \bibfield  {author} {\bibinfo {author} {\bibnamefont {Sadeghpour},
  \bibfnamefont {HR}}, \bibinfo {author} {\bibfnamefont {JL}~\bibnamefont
  {Bohn}}, \bibinfo {author} {\bibfnamefont {MJ}~\bibnamefont {Cavagnero}},
  \bibinfo {author} {\bibfnamefont {BD}~\bibnamefont {Esry}}, \bibinfo {author}
  {\bibfnamefont {II}~\bibnamefont {Fabrikant}}, \bibinfo {author}
  {\bibfnamefont {JH}~\bibnamefont {Macek}}, \ and\ \bibinfo {author}
  {\bibfnamefont {ARP}\ \bibnamefont {Rau}}} (\bibinfo {year} {2000}),\
  \bibfield  {title} {{\selectlanguage {English}\enquote {\bibinfo {title}
  {Collisions near threshold in atomic and molecular physics},}\ }}\href@noop
  {} {\bibfield  {journal} {\bibinfo  {journal} {J. Phys. B}\ }\textbf
  {\bibinfo {volume} {33}}~(\bibinfo {number} {5}),\ \bibinfo {pages}
  {R93--R140}}\BibitemShut {NoStop}%
\bibitem [{\citenamefont {Saeidian}\ \emph {et~al.}(2008)\citenamefont
  {Saeidian}, \citenamefont {Melezhik},\ and\ \citenamefont
  {Schmelcher}}]{saeidian2008}%
  \BibitemOpen
  \bibfield  {author} {\bibinfo {author} {\bibnamefont {Saeidian},
  \bibfnamefont {S}}, \bibinfo {author} {\bibfnamefont {V.~S.}\ \bibnamefont
  {Melezhik}}, \ and\ \bibinfo {author} {\bibfnamefont {P.}~\bibnamefont
  {Schmelcher}}} (\bibinfo {year} {2008}),\ \bibfield  {title} {\enquote
  {\bibinfo {title} {Multichannel atomic scattering and confinement-induced
  resonances in waveguides},}\ }\href@noop {} {\bibfield  {journal} {\bibinfo
  {journal} {Phys. Rev. A}\ }\textbf {\bibinfo {volume} {77}}~(\bibinfo
  {number} {4}),\ \bibinfo {pages} {042721}}\BibitemShut {NoStop}%
\bibitem [{\citenamefont {Saeidian}\ \emph {et~al.}(2012)\citenamefont
  {Saeidian}, \citenamefont {Melezhik},\ and\ \citenamefont
  {Schmelcher}}]{saeidian2012}%
  \BibitemOpen
  \bibfield  {author} {\bibinfo {author} {\bibnamefont {Saeidian},
  \bibfnamefont {S}}, \bibinfo {author} {\bibfnamefont {V.~S.}\ \bibnamefont
  {Melezhik}}, \ and\ \bibinfo {author} {\bibfnamefont {P.}~\bibnamefont
  {Schmelcher}}} (\bibinfo {year} {2012}),\ \bibfield  {title} {\enquote
  {\bibinfo {title} {Shifts and widths of {F}eshbach resonances in atomic
  waveguides},}\ }\href@noop {} {\bibfield  {journal} {\bibinfo  {journal}
  {Phys. Rev. A}\ }\textbf {\bibinfo {volume} {86}}~(\bibinfo {number} {6}),\
  \bibinfo {pages} {062713}}\BibitemShut {NoStop}%
\bibitem [{\citenamefont {Saeidian}\ \emph {et~al.}(2015)\citenamefont
  {Saeidian}, \citenamefont {Melezhik},\ and\ \citenamefont
  {Schmelcher}}]{saeidian2015shifts}%
  \BibitemOpen
  \bibfield  {author} {\bibinfo {author} {\bibnamefont {Saeidian},
  \bibfnamefont {S}}, \bibinfo {author} {\bibfnamefont {V.~S}\ \bibnamefont
  {Melezhik}}, \ and\ \bibinfo {author} {\bibfnamefont {P.}~\bibnamefont
  {Schmelcher}}} (\bibinfo {year} {2015}),\ \bibfield  {title} {\enquote
  {\bibinfo {title} {Shifts and widths of $p$-wave confinement induced
  resonances in atomic waveguides},}\ }\href@noop {} {\bibfield  {journal}
  {\bibinfo  {journal} {J. Phys. B}\ }\textbf {\bibinfo {volume}
  {48}}~(\bibinfo {number} {15}),\ \bibinfo {pages} {155301}}\BibitemShut
  {NoStop}%
\bibitem [{\citenamefont {Safavi-Naini}\ \emph {et~al.}({2013})\citenamefont
  {Safavi-Naini}, \citenamefont {Rittenhouse}, \citenamefont {Blume},\ and\
  \citenamefont {Sadeghpour}}]{Safavi-Naini2013pra}%
  \BibitemOpen
  \bibfield  {author} {\bibinfo {author} {\bibnamefont {Safavi-Naini},
  \bibfnamefont {A}}, \bibinfo {author} {\bibfnamefont {Seth~T.}\ \bibnamefont
  {Rittenhouse}}, \bibinfo {author} {\bibfnamefont {D.}~\bibnamefont {Blume}},
  \ and\ \bibinfo {author} {\bibfnamefont {H.~R.}\ \bibnamefont {Sadeghpour}}}
  (\bibinfo {year} {{2013}}),\ \bibfield  {title} {\enquote {\bibinfo {title}
  {{Nonuniversal bound states of two identical heavy fermions and one light
  particle}},}\ }\href {\doibase {10.1103/PhysRevA.87.032713}} {\bibfield
  {journal} {\bibinfo  {journal} {{Physical Review A}}\ }\textbf {\bibinfo
  {volume} {{87}}}~(\bibinfo {number} {{3}}),\
  {10.1103/PhysRevA.87.032713}}\BibitemShut {NoStop}%
\bibitem [{\citenamefont {Sagi}\ \emph {et~al.}(2013)\citenamefont {Sagi},
  \citenamefont {Drake}, \citenamefont {Paudel}, \citenamefont {Chapurin},\
  and\ \citenamefont {Jin}}]{SagiJin2013}%
  \BibitemOpen
  \bibfield  {author} {\bibinfo {author} {\bibnamefont {Sagi}, \bibfnamefont
  {Y}}, \bibinfo {author} {\bibfnamefont {T.~E.}\ \bibnamefont {Drake}},
  \bibinfo {author} {\bibfnamefont {R.}~\bibnamefont {Paudel}}, \bibinfo
  {author} {\bibfnamefont {R.}~\bibnamefont {Chapurin}}, \ and\ \bibinfo
  {author} {\bibfnamefont {D.~S.}\ \bibnamefont {Jin}}} (\bibinfo {year}
  {2013}),\ \bibfield  {title} {\enquote {\bibinfo {title} {Probing local
  quantities in a strongly interacting {F}ermi gas},}\ }\href@noop {}
  {\bibfield  {journal} {\bibinfo  {journal} {Journal of Physics: Conference
  Series}\ }\textbf {\bibinfo {volume} {467}}~(\bibinfo {number} {1}),\
  \bibinfo {pages} {012010}}\BibitemShut {NoStop}%
\bibitem [{\citenamefont {Sagi}\ \emph {et~al.}(2012)\citenamefont {Sagi},
  \citenamefont {Drake}, \citenamefont {Paudel},\ and\ \citenamefont
  {Jin}}]{SagiJin2012prl}%
  \BibitemOpen
  \bibfield  {author} {\bibinfo {author} {\bibnamefont {Sagi}, \bibfnamefont
  {Y}}, \bibinfo {author} {\bibfnamefont {T.~E.}\ \bibnamefont {Drake}},
  \bibinfo {author} {\bibfnamefont {R.}~\bibnamefont {Paudel}}, \ and\ \bibinfo
  {author} {\bibfnamefont {D.~S.}\ \bibnamefont {Jin}}} (\bibinfo {year}
  {2012}),\ \bibfield  {title} {\enquote {\bibinfo {title} {Measurement of the
  homogeneous contact of a unitary {F}ermi gas},}\ }\href@noop {} {\bibfield
  {journal} {\bibinfo  {journal} {Phys. Rev. Lett.}\ }\textbf {\bibinfo
  {volume} {109}},\ \bibinfo {pages} {220402}}\BibitemShut {NoStop}%
\bibitem [{\citenamefont {Sala}\ \emph {et~al.}(2012)\citenamefont {Sala},
  \citenamefont {Schneider},\ and\ \citenamefont {Saenz}}]{sala2012}%
  \BibitemOpen
  \bibfield  {author} {\bibinfo {author} {\bibnamefont {Sala}, \bibfnamefont
  {S}}, \bibinfo {author} {\bibfnamefont {P.~I.}\ \bibnamefont {Schneider}}, \
  and\ \bibinfo {author} {\bibfnamefont {A.}~\bibnamefont {Saenz}}} (\bibinfo
  {year} {2012}),\ \bibfield  {title} {\enquote {\bibinfo {title} {Inelastic
  confinement-induced resonances in low-dimensional quantum systems},}\
  }\href@noop {} {\bibfield  {journal} {\bibinfo  {journal} {Phys. Rev. Lett.}\
  }\textbf {\bibinfo {volume} {109}}~(\bibinfo {number} {7}),\ \bibinfo {pages}
  {073201}}\BibitemShut {NoStop}%
\bibitem [{\citenamefont {Sala}\ \emph {et~al.}(2013)\citenamefont {Sala},
  \citenamefont {Z\"{u}rn}, \citenamefont {Lompe}, \citenamefont {Wenz},
  \citenamefont {Murmann}, \citenamefont {Serwane}, \citenamefont {Jochim},\
  and\ \citenamefont {Saenz}}]{sala2013}%
  \BibitemOpen
  \bibfield  {author} {\bibinfo {author} {\bibnamefont {Sala}, \bibfnamefont
  {S}}, \bibinfo {author} {\bibfnamefont {G.}~\bibnamefont {Z\"{u}rn}},
  \bibinfo {author} {\bibfnamefont {T.}~\bibnamefont {Lompe}}, \bibinfo
  {author} {\bibfnamefont {A.~N. .~N.}\ \bibnamefont {Wenz}}, \bibinfo {author}
  {\bibfnamefont {S.}~\bibnamefont {Murmann}}, \bibinfo {author} {\bibfnamefont
  {F.}~\bibnamefont {Serwane}}, \bibinfo {author} {\bibfnamefont
  {S.}~\bibnamefont {Jochim}}, \ and\ \bibinfo {author} {\bibfnamefont
  {A.}~\bibnamefont {Saenz}}} (\bibinfo {year} {2013}),\ \bibfield  {title}
  {\enquote {\bibinfo {title} {Coherent molecule formation in anharmonic
  potentials near confinement-induced resonances},}\ }\href@noop {} {\bibfield
  {journal} {\bibinfo  {journal} {Phys. Rev. Lett.}\ }\textbf {\bibinfo
  {volume} {110}}~(\bibinfo {number} {20}),\ \bibinfo {pages}
  {203202}}\BibitemShut {NoStop}%
\bibitem [{\citenamefont {Scelle}\ \emph {et~al.}(2013)\citenamefont {Scelle},
  \citenamefont {Rentrop}, \citenamefont {Trautmann}, \citenamefont
  {Schuster},\ and\ \citenamefont {Oberthaler}}]{scelle_motional_2013}%
  \BibitemOpen
  \bibfield  {author} {\bibinfo {author} {\bibnamefont {Scelle}, \bibfnamefont
  {R}}, \bibinfo {author} {\bibfnamefont {T.}~\bibnamefont {Rentrop}}, \bibinfo
  {author} {\bibfnamefont {A.}~\bibnamefont {Trautmann}}, \bibinfo {author}
  {\bibfnamefont {T.}~\bibnamefont {Schuster}}, \ and\ \bibinfo {author}
  {\bibfnamefont {M.~K.}\ \bibnamefont {Oberthaler}}} (\bibinfo {year}
  {2013}),\ \bibfield  {title} {\enquote {\bibinfo {title} {Motional
  {Coherence} of {Fermions} {Immersed} in a {Bose} {Gas}},}\ }\href {\doibase
  10.1103/PhysRevLett.111.070401} {\bibfield  {journal} {\bibinfo  {journal}
  {Physical Review Letters}\ }\textbf {\bibinfo {volume} {111}}~(\bibinfo
  {number} {7}),\ \bibinfo {pages} {070401}}\BibitemShut {NoStop}%
\bibitem [{\citenamefont {Schakel}(2010)}]{Schakel-2010}%
  \BibitemOpen
  \bibfield  {author} {\bibinfo {author} {\bibnamefont {Schakel}, \bibfnamefont
  {A~M~J}}} (\bibinfo {year} {2010}),\ \href@noop {} {\enquote {\bibinfo
  {title} {Tan relations in dilute {B}ose gases},}\ }\bibinfo {howpublished}
  {arXiv:1007.3452v1}\BibitemShut {NoStop}%
\bibitem [{\citenamefont {Schirotzek}\ \emph {et~al.}(2009)\citenamefont
  {Schirotzek}, \citenamefont {Wu}, \citenamefont {Sommer},\ and\ \citenamefont
  {Zwierlein}}]{schirotzek_observation_2009}%
  \BibitemOpen
  \bibfield  {author} {\bibinfo {author} {\bibnamefont {Schirotzek},
  \bibfnamefont {Andre}}, \bibinfo {author} {\bibfnamefont {Cheng-Hsun}\
  \bibnamefont {Wu}}, \bibinfo {author} {\bibfnamefont {Ariel}\ \bibnamefont
  {Sommer}}, \ and\ \bibinfo {author} {\bibfnamefont {Martin~W.}\ \bibnamefont
  {Zwierlein}}} (\bibinfo {year} {2009}),\ \bibfield  {title} {\enquote
  {\bibinfo {title} {Observation of {Fermi} {Polarons} in a {Tunable} {Fermi}
  {Liquid} of {Ultracold} {Atoms}},}\ }\href {\doibase
  10.1103/PhysRevLett.102.230402} {\bibfield  {journal} {\bibinfo  {journal}
  {Physical Review Letters}\ }\textbf {\bibinfo {volume} {102}}~(\bibinfo
  {number} {23}),\ \bibinfo {pages} {230402}}\BibitemShut {NoStop}%
\bibitem [{\citenamefont {Schmidt}\ \emph {et~al.}(2012)\citenamefont
  {Schmidt}, \citenamefont {Rath},\ and\ \citenamefont
  {Zwerger}}]{schmidt2012EPJB}%
  \BibitemOpen
  \bibfield  {author} {\bibinfo {author} {\bibnamefont {Schmidt}, \bibfnamefont
  {R}}, \bibinfo {author} {\bibfnamefont {S.P.}\ \bibnamefont {Rath}}, \ and\
  \bibinfo {author} {\bibfnamefont {W.}~\bibnamefont {Zwerger}}} (\bibinfo
  {year} {2012}),\ \bibfield  {title} {{\selectlanguage {English}\enquote
  {\bibinfo {title} {{E}fimov physics beyond universality},}\ }}\href@noop {}
  {\bibfield  {journal} {\bibinfo  {journal} {EPJB}\ }\textbf {\bibinfo
  {volume} {85}},\ \bibinfo {pages} {386}}\BibitemShut {NoStop}%
\bibitem [{\citenamefont {Schmidt}\ and\ \citenamefont
  {Lemeshko}(2016)}]{SchmidtLemeshko2016prx}%
  \BibitemOpen
  \bibfield  {author} {\bibinfo {author} {\bibnamefont {Schmidt}, \bibfnamefont
  {Richard}}, \ and\ \bibinfo {author} {\bibfnamefont {Mikhail}\ \bibnamefont
  {Lemeshko}}} (\bibinfo {year} {2016}),\ \bibfield  {title} {\enquote
  {\bibinfo {title} {Deformation of a quantum many-particle system by a
  rotating impurity},}\ }\href {\doibase 10.1103/PhysRevX.6.011012} {\bibfield
  {journal} {\bibinfo  {journal} {Phys. Rev. X}\ }\textbf {\bibinfo {volume}
  {6}},\ \bibinfo {pages} {011012}}\BibitemShut {NoStop}%
\bibitem [{\citenamefont {Sch\"{o}llkopf}\ and\ \citenamefont
  {Toennies}(1994)}]{He-trimer}%
  \BibitemOpen
  \bibfield  {author} {\bibinfo {author} {\bibnamefont {Sch\"{o}llkopf},
  \bibfnamefont {W}}, \ and\ \bibinfo {author} {\bibfnamefont {J.~P.}\
  \bibnamefont {Toennies}}} (\bibinfo {year} {1994}),\ \bibfield  {title}
  {\enquote {\bibinfo {title} {Nondestructive mass selection of small
  van-der-{W}aals clusters},}\ }\href@noop {} {\bibfield  {journal} {\bibinfo
  {journal} {Science}\ }\textbf {\bibinfo {volume} {266}},\ \bibinfo {pages}
  {1345--1348}}\BibitemShut {NoStop}%
\bibitem [{\citenamefont {Sch\"{o}llkopf}\ and\ \citenamefont
  {Toennies}(1996)}]{He-trimerb}%
  \BibitemOpen
  \bibfield  {author} {\bibinfo {author} {\bibnamefont {Sch\"{o}llkopf},
  \bibfnamefont {W}}, \ and\ \bibinfo {author} {\bibfnamefont {J.~P.}\
  \bibnamefont {Toennies}}} (\bibinfo {year} {1996}),\ \bibfield  {title}
  {\enquote {\bibinfo {title} {The nondestructive detection of the helium dimer
  and trimer},}\ }\href@noop {} {\bibfield  {journal} {\bibinfo  {journal} {J.
  Chem. Phys.}\ }\textbf {\bibinfo {volume} {104}},\ \bibinfo {pages}
  {1155}}\BibitemShut {NoStop}%
\bibitem [{\citenamefont {Schunck}\ \emph {et~al.}(2005)\citenamefont
  {Schunck}, \citenamefont {Zwierlein}, \citenamefont {Stan}, \citenamefont
  {Raupach},\ and\ \citenamefont {Ketterle}}]{schunck2005PRA}%
  \BibitemOpen
  \bibfield  {author} {\bibinfo {author} {\bibnamefont {Schunck}, \bibfnamefont
  {C~H}}, \bibinfo {author} {\bibfnamefont {M.~W.}\ \bibnamefont {Zwierlein}},
  \bibinfo {author} {\bibfnamefont {C.~A.}\ \bibnamefont {Stan}}, \bibinfo
  {author} {\bibfnamefont {S.~M.~F.}\ \bibnamefont {Raupach}}, \ and\ \bibinfo
  {author} {\bibfnamefont {W.}~\bibnamefont {Ketterle}}} (\bibinfo {year}
  {2005}),\ \bibfield  {title} {\enquote {\bibinfo {title} {{F}eshbach
  resonances in {F}ermionic $^6${Li}},}\ }\href@noop {} {\bibfield  {journal}
  {\bibinfo  {journal} {Phys. Rev. A}\ }\textbf {\bibinfo {volume} {71}},\
  \bibinfo {pages} {045601}}\BibitemShut {NoStop}%
\bibitem [{\citenamefont {Seaton}(1983)}]{seaton1983rpp}%
  \BibitemOpen
  \bibfield  {author} {\bibinfo {author} {\bibnamefont {Seaton}, \bibfnamefont
  {M~J}}} (\bibinfo {year} {1983}),\ \bibfield  {title} {\enquote {\bibinfo
  {title} {Quantum defect theory},}\ }\href@noop {} {\bibfield  {journal}
  {\bibinfo  {journal} {Rep. Prog. Phys.}\ }\textbf {\bibinfo {volume}
  {46}}~(\bibinfo {number} {2}),\ \bibinfo {pages} {167--257}}\BibitemShut
  {NoStop}%
\bibitem [{\citenamefont {Selles}\ \emph {et~al.}({2004})\citenamefont
  {Selles}, \citenamefont {Malegat}, \citenamefont {Huetz}, \citenamefont
  {Kazansky}, \citenamefont {Collins}, \citenamefont {Seccombe},\ and\
  \citenamefont {Reddish}}]{malegat2004PRA}%
  \BibitemOpen
  \bibfield  {author} {\bibinfo {author} {\bibnamefont {Selles}, \bibfnamefont
  {P}}, \bibinfo {author} {\bibfnamefont {L.}~\bibnamefont {Malegat}}, \bibinfo
  {author} {\bibfnamefont {A.}~\bibnamefont {Huetz}}, \bibinfo {author}
  {\bibfnamefont {A.~K.}\ \bibnamefont {Kazansky}}, \bibinfo {author}
  {\bibfnamefont {S.~A.}\ \bibnamefont {Collins}}, \bibinfo {author}
  {\bibfnamefont {D.~P.}\ \bibnamefont {Seccombe}}, \ and\ \bibinfo {author}
  {\bibfnamefont {T.~J.}\ \bibnamefont {Reddish}}} (\bibinfo {year} {{2004}}),\
  \bibfield  {title} {\enquote {\bibinfo {title} {{Convergence of the method of
  the hyperspherical R matrix with semiclassical outgoing waves}},}\
  }\href@noop {} {\bibfield  {journal} {\bibinfo  {journal} {Phys. Rev. A}\
  }\textbf {\bibinfo {volume} {{69}}}~(\bibinfo {number} {{5}})}\BibitemShut
  {NoStop}%
\bibitem [{\citenamefont {Selles}\ \emph {et~al.}(1987)\citenamefont {Selles},
  \citenamefont {Mazeau},\ and\ \citenamefont {Huetz}}]{SELLES1987}%
  \BibitemOpen
  \bibfield  {author} {\bibinfo {author} {\bibnamefont {Selles}, \bibfnamefont
  {P}}, \bibinfo {author} {\bibfnamefont {J.}~\bibnamefont {Mazeau}}, \ and\
  \bibinfo {author} {\bibfnamefont {A.}~\bibnamefont {Huetz}}} (\bibinfo {year}
  {1987}),\ \bibfield  {title} {\enquote {\bibinfo {title} {Wannier theory for
  {P}$^o$ and {D}$^e$ states of 2 electrons},}\ }\href@noop {} {\bibfield
  {journal} {\bibinfo  {journal} {J. Phys. B}\ }\textbf {\bibinfo {volume}
  {20}}~(\bibinfo {number} {19}),\ \bibinfo {pages} {5183--5193}}\BibitemShut
  {NoStop}%
\bibitem [{\citenamefont {Shepard}(2007)}]{shepard2007PRA}%
  \BibitemOpen
  \bibfield  {author} {\bibinfo {author} {\bibnamefont {Shepard}, \bibfnamefont
  {J~R}}} (\bibinfo {year} {2007}),\ \bibfield  {title} {{\selectlanguage
  {English}\enquote {\bibinfo {title} {Calculations of recombination rates for
  cold $^4${He} atoms from atom-dimer phase shifts and determination of
  universal scaling functions},}\ }}\href@noop {} {\bibfield  {journal}
  {\bibinfo  {journal} {Phys. Rev. A}\ }\textbf {\bibinfo {volume}
  {75}}~(\bibinfo {number} {6}),\ \bibinfo {pages} {062713}}\BibitemShut
  {NoStop}%
\bibitem [{\citenamefont {Shi}\ and\ \citenamefont {Yi}(2014)}]{tao2014pra}%
  \BibitemOpen
  \bibfield  {author} {\bibinfo {author} {\bibnamefont {Shi}, \bibfnamefont
  {T}}, \ and\ \bibinfo {author} {\bibfnamefont {S.}~\bibnamefont {Yi}}}
  (\bibinfo {year} {2014}),\ \bibfield  {title} {\enquote {\bibinfo {title}
  {Observing dipolar confinement-induced resonances in quasi-one-dimensional
  atomic gases},}\ }\href@noop {} {\bibfield  {journal} {\bibinfo  {journal}
  {Phys. Rev. A}\ }\textbf {\bibinfo {volume} {90}},\ \bibinfo {pages}
  {042710}}\BibitemShut {NoStop}%
\bibitem [{\citenamefont {Shotan}\ \emph {et~al.}({2014})\citenamefont
  {Shotan}, \citenamefont {Machtey}, \citenamefont {Kokkelmans},\ and\
  \citenamefont {Khaykovich}}]{Shotan2014prl}%
  \BibitemOpen
  \bibfield  {author} {\bibinfo {author} {\bibnamefont {Shotan}, \bibfnamefont
  {Z}}, \bibinfo {author} {\bibfnamefont {O.}~\bibnamefont {Machtey}}, \bibinfo
  {author} {\bibfnamefont {S.}~\bibnamefont {Kokkelmans}}, \ and\ \bibinfo
  {author} {\bibfnamefont {L.}~\bibnamefont {Khaykovich}}} (\bibinfo {year}
  {{2014}}),\ \bibfield  {title} {\enquote {\bibinfo {title} {Three-body
  recombination at vanishing scattering lengths in an ultracold {B}ose gas},}\
  }\href@noop {} {\bibfield  {journal} {\bibinfo  {journal} {{Phys. Rev.
  Lett.}}\ }\textbf {\bibinfo {volume} {{113}}}~(\bibinfo {number}
  {{5}})}\BibitemShut {NoStop}%
\bibitem [{\citenamefont {Shui}(1972)}]{Shui-thesis}%
  \BibitemOpen
  \bibfield  {author} {\bibinfo {author} {\bibnamefont {Shui}, \bibfnamefont
  {V~H}}} (\bibinfo {year} {1972}),\ \emph {\bibinfo {title} {Thermal
  dissociation and recombination of hydrogen according to the reactions
  {H}$_{2}$ + {H}$\rightarrow$ {H} +{H} + {H}}},\ \href@noop {} {Ph.D. thesis}\
  (\bibinfo  {school} {Deparment of Mechanical Engineering MIT})\BibitemShut
  {NoStop}%
\bibitem [{\citenamefont {Shui}(1973)}]{Shui-1973}%
  \BibitemOpen
  \bibfield  {author} {\bibinfo {author} {\bibnamefont {Shui}, \bibfnamefont
  {V~H}}} (\bibinfo {year} {1973}),\ \bibfield  {title} {\enquote {\bibinfo
  {title} {Thermal dissociation and recombination of hydrogen according to the
  reactions {H}$_2$+{H}$\rightarrow$ {H}+{H}+{H}},}\ }\href@noop {} {\bibfield
  {journal} {\bibinfo  {journal} {J. Chem. Phys.}\ }\textbf {\bibinfo {volume}
  {58}},\ \bibinfo {pages} {4868}}\BibitemShut {NoStop}%
\bibitem [{\citenamefont {Shui}\ \emph {et~al.}(1970)\citenamefont {Shui},
  \citenamefont {Appleton},\ and\ \citenamefont {Keck}}]{Shui-1970}%
  \BibitemOpen
  \bibfield  {author} {\bibinfo {author} {\bibnamefont {Shui}, \bibfnamefont
  {V~H}}, \bibinfo {author} {\bibfnamefont {J.~P.}\ \bibnamefont {Appleton}}, \
  and\ \bibinfo {author} {\bibfnamefont {J.~C.}\ \bibnamefont {Keck}}}
  (\bibinfo {year} {1970}),\ \bibfield  {title} {\enquote {\bibinfo {title}
  {Three-body recombination and dissociation of nitrogen: {A} comparison
  between theory and experiment},}\ }\href@noop {} {\bibfield  {journal}
  {\bibinfo  {journal} {J. Chem. Phys.}\ }\textbf {\bibinfo {volume} {53}},\
  \bibinfo {pages} {2547}}\BibitemShut {NoStop}%
\bibitem [{\citenamefont {Simoni}\ \emph {et~al.}(2015)\citenamefont {Simoni},
  \citenamefont {Srinivasan}, \citenamefont {Launay}, \citenamefont
  {Jachymski}, \citenamefont {Idziaszek},\ and\ \citenamefont
  {Julienne}}]{simoni2015polar}%
  \BibitemOpen
  \bibfield  {author} {\bibinfo {author} {\bibnamefont {Simoni}, \bibfnamefont
  {A}}, \bibinfo {author} {\bibfnamefont {S.}~\bibnamefont {Srinivasan}},
  \bibinfo {author} {\bibfnamefont {J.~M.}\ \bibnamefont {Launay}}, \bibinfo
  {author} {\bibfnamefont {K.}~\bibnamefont {Jachymski}}, \bibinfo {author}
  {\bibfnamefont {Z.}~\bibnamefont {Idziaszek}}, \ and\ \bibinfo {author}
  {\bibfnamefont {P.~S.}\ \bibnamefont {Julienne}}} (\bibinfo {year} {2015}),\
  \bibfield  {title} {\enquote {\bibinfo {title} {Polar molecule reactive
  collisions in quasi-1d systems},}\ }\href@noop {} {\bibfield  {journal}
  {\bibinfo  {journal} {New. J. Phys.}\ }\textbf {\bibinfo {volume}
  {17}}~(\bibinfo {number} {1}),\ \bibinfo {pages} {013020}}\BibitemShut
  {NoStop}%
\bibitem [{\citenamefont {Sinha}\ and\ \citenamefont
  {Santos}(2007)}]{sinha2007}%
  \BibitemOpen
  \bibfield  {author} {\bibinfo {author} {\bibnamefont {Sinha}, \bibfnamefont
  {S}}, \ and\ \bibinfo {author} {\bibfnamefont {L.}~\bibnamefont {Santos}}}
  (\bibinfo {year} {2007}),\ \bibfield  {title} {\enquote {\bibinfo {title}
  {Cold dipolar gases in quasi-one-dimensional geometries},}\ }\href@noop {}
  {\bibfield  {journal} {\bibinfo  {journal} {Phys. Rev. Lett.}\ }\textbf
  {\bibinfo {volume} {99}}~(\bibinfo {number} {14}),\ \bibinfo {pages}
  {140406}}\BibitemShut {NoStop}%
\bibitem [{\citenamefont {Skorniakov}\ and\ \citenamefont
  {{Ter-Martirosian}}(1957)}]{skorniakov1957JETP}%
  \BibitemOpen
  \bibfield  {author} {\bibinfo {author} {\bibnamefont {Skorniakov},
  \bibfnamefont {G~V}}, \ and\ \bibinfo {author} {\bibfnamefont {K.~A.}\
  \bibnamefont {{Ter-Martirosian}}}} (\bibinfo {year} {1957}),\ \bibfield
  {title} {\enquote {\bibinfo {title} {Three body problem for short range
  forces. {I}. scattering of low energy neutrons by deuterons},}\ }\href@noop
  {} {\bibfield  {journal} {\bibinfo  {journal} {Sov. Phys. JETP}\ }\textbf
  {\bibinfo {volume} {4}}}\BibitemShut {NoStop}%
\bibitem [{\citenamefont {Smirnov}\ and\ \citenamefont
  {Shitikova}(1977)}]{smirnov1977Sov.J.Part.Nucl.}%
  \BibitemOpen
  \bibfield  {author} {\bibinfo {author} {\bibnamefont {Smirnov}, \bibfnamefont
  {Y~F}}, \ and\ \bibinfo {author} {\bibfnamefont {K.~V.}\ \bibnamefont
  {Shitikova}}} (\bibinfo {year} {1977}),\ \bibfield  {title} {\enquote
  {\bibinfo {title} {Method of {K} harmonics and the shell model},}\
  }\href@noop {} {\bibfield  {journal} {\bibinfo  {journal} {Sov. J. Part.
  Nucl.}\ }\textbf {\bibinfo {volume} {8}},\ \bibinfo {pages} {44}}\BibitemShut
  {NoStop}%
\bibitem [{\citenamefont {Smith}(1960)}]{Smith-1960}%
  \BibitemOpen
  \bibfield  {author} {\bibinfo {author} {\bibnamefont {Smith}, \bibfnamefont
  {F~T}}} (\bibinfo {year} {1960}),\ \bibfield  {title} {\enquote {\bibinfo
  {title} {Generalized angular momentum in many-body collisions},}\ }\href@noop
  {} {\bibfield  {journal} {\bibinfo  {journal} {Phys. Rev.}\ }\textbf
  {\bibinfo {volume} {120}},\ \bibinfo {pages} {1058}}\BibitemShut {NoStop}%
\bibitem [{\citenamefont {Smith}(1962)}]{Smith-1962}%
  \BibitemOpen
  \bibfield  {author} {\bibinfo {author} {\bibnamefont {Smith}, \bibfnamefont
  {F~T}}} (\bibinfo {year} {1962}),\ \bibfield  {title} {\enquote {\bibinfo
  {title} {Three-body collision rates in atomic recombination reactions},}\
  }\href@noop {} {\bibfield  {journal} {\bibinfo  {journal} {Discuss. Faraday
  Soc.}\ }\textbf {\bibinfo {volume} {33}},\ \bibinfo {pages}
  {183}}\BibitemShut {NoStop}%
\bibitem [{\citenamefont {Snider}(1960)}]{Snider-1960}%
  \BibitemOpen
  \bibfield  {author} {\bibinfo {author} {\bibnamefont {Snider}, \bibfnamefont
  {R~F}}} (\bibinfo {year} {1960}),\ \bibfield  {title} {\enquote {\bibinfo
  {title} {Quantum-mechanical modified {B}oltzmann equation for degenerate
  internal states},}\ }\href@noop {} {\bibfield  {journal} {\bibinfo  {journal}
  {J. Chem. Phys}\ }\textbf {\bibinfo {volume} {32}},\ \bibinfo {pages}
  {1051}}\BibitemShut {NoStop}%
\bibitem [{\citenamefont {S\"{o}ding}\ \emph {et~al.}(1999)\citenamefont
  {S\"{o}ding}, \citenamefont {{Gu{\'e}ry-Odelin}}, \citenamefont {Desbiolles},
  \citenamefont {Chevy}, \citenamefont {Inamori},\ and\ \citenamefont
  {Dalibard}}]{Dalibard1999}%
  \BibitemOpen
  \bibfield  {author} {\bibinfo {author} {\bibnamefont {S\"{o}ding},
  \bibfnamefont {J}}, \bibinfo {author} {\bibfnamefont {D.}~\bibnamefont
  {{Gu{\'e}ry-Odelin}}}, \bibinfo {author} {\bibfnamefont {P.}~\bibnamefont
  {Desbiolles}}, \bibinfo {author} {\bibfnamefont {F.}~\bibnamefont {Chevy}},
  \bibinfo {author} {\bibfnamefont {H.}~\bibnamefont {Inamori}}, \ and\
  \bibinfo {author} {\bibfnamefont {J.}~\bibnamefont {Dalibard}}} (\bibinfo
  {year} {1999}),\ \bibfield  {title} {\enquote {\bibinfo {title} {Three-body
  decay of a rubidium {B}ose-{E}instein condensate},}\ }\href@noop {}
  {\bibfield  {journal} {\bibinfo  {journal} {Appl. Phys. B}\ }\textbf
  {\bibinfo {volume} {69}}~(\bibinfo {number} {4}),\ \bibinfo {pages}
  {257--261}}\BibitemShut {NoStop}%
\bibitem [{\citenamefont {Sogo}\ \emph {et~al.}(2005)\citenamefont {Sogo},
  \citenamefont {Sorensen}, \citenamefont {Jensen},\ and\ \citenamefont
  {Fedorov}}]{sogo2005EPL}%
  \BibitemOpen
  \bibfield  {author} {\bibinfo {author} {\bibnamefont {Sogo}, \bibfnamefont
  {T}}, \bibinfo {author} {\bibfnamefont {O.}~\bibnamefont {Sorensen}},
  \bibinfo {author} {\bibfnamefont {A.~S.}\ \bibnamefont {Jensen}}, \ and\
  \bibinfo {author} {\bibfnamefont {D.~V.}\ \bibnamefont {Fedorov}}} (\bibinfo
  {year} {2005}),\ \bibfield  {title} {{\selectlanguage {English}\enquote
  {\bibinfo {title} {Semi-analytic solution to the {N}-boson problem with
  zero-range interactions},}\ }}\href@noop {} {\bibfield  {journal} {\bibinfo
  {journal} {EPL}\ }\textbf {\bibinfo {volume} {69}}~(\bibinfo {number} {5}),\
  \bibinfo {pages} {732--738}}\BibitemShut {NoStop}%
\bibitem [{\citenamefont {Sorensen}\ \emph {et~al.}(2004)\citenamefont
  {Sorensen}, \citenamefont {Fedorov},\ and\ \citenamefont
  {Jensen}}]{FedorovJensen2004becJPB}%
  \BibitemOpen
  \bibfield  {author} {\bibinfo {author} {\bibnamefont {Sorensen},
  \bibfnamefont {O}}, \bibinfo {author} {\bibfnamefont {D.~V.}\ \bibnamefont
  {Fedorov}}, \ and\ \bibinfo {author} {\bibfnamefont {A.~S.}\ \bibnamefont
  {Jensen}}} (\bibinfo {year} {2004}),\ \bibfield  {title} {\enquote {\bibinfo
  {title} {Structure of boson systems beyond the mean field},}\ }\href@noop {}
  {\bibfield  {journal} {\bibinfo  {journal} {J. Phys. B.}\ }\textbf {\bibinfo
  {volume} {37}}~(\bibinfo {number} {1}),\ \bibinfo {pages} {93}}\BibitemShut
  {NoStop}%
\bibitem [{\citenamefont {S\o{}rensen}\ \emph {et~al.}(2012)\citenamefont
  {S\o{}rensen}, \citenamefont {Fedorov}, \citenamefont {Jensen},\ and\
  \citenamefont {Zinner}}]{sorensen2012PRA}%
  \BibitemOpen
  \bibfield  {author} {\bibinfo {author} {\bibnamefont {S\o{}rensen},
  \bibfnamefont {P~K}}, \bibinfo {author} {\bibfnamefont {D.~V.}\ \bibnamefont
  {Fedorov}}, \bibinfo {author} {\bibfnamefont {A.~S.}\ \bibnamefont {Jensen}},
  \ and\ \bibinfo {author} {\bibfnamefont {N.~T.}\ \bibnamefont {Zinner}}}
  (\bibinfo {year} {2012}),\ \bibfield  {title} {\enquote {\bibinfo {title}
  {{E}fimov physics and the three-body parameter within a two-channel
  framework},}\ }\href@noop {} {\bibfield  {journal} {\bibinfo  {journal}
  {Phys. Rev. A}\ }\textbf {\bibinfo {volume} {86}},\ \bibinfo {pages}
  {052516}}\BibitemShut {NoStop}%
\bibitem [{\citenamefont {Spethmann}\ \emph {et~al.}(2012)\citenamefont
  {Spethmann}, \citenamefont {Kindermann}, \citenamefont {John}, \citenamefont
  {Weber}, \citenamefont {Meschede},\ and\ \citenamefont
  {Widera}}]{spethmann_dynamics_2012}%
  \BibitemOpen
  \bibfield  {author} {\bibinfo {author} {\bibnamefont {Spethmann},
  \bibfnamefont {Nicolas}}, \bibinfo {author} {\bibfnamefont {Farina}\
  \bibnamefont {Kindermann}}, \bibinfo {author} {\bibfnamefont {Shincy}\
  \bibnamefont {John}}, \bibinfo {author} {\bibfnamefont {Claudia}\
  \bibnamefont {Weber}}, \bibinfo {author} {\bibfnamefont {Dieter}\
  \bibnamefont {Meschede}}, \ and\ \bibinfo {author} {\bibfnamefont {Artur}\
  \bibnamefont {Widera}}} (\bibinfo {year} {2012}),\ \bibfield  {title}
  {\enquote {\bibinfo {title} {Dynamics of {Single} {Neutral} {Impurity}
  {Atoms} {Immersed} in an {Ultracold} {Gas}},}\ }\href {\doibase
  10.1103/PhysRevLett.109.235301} {\bibfield  {journal} {\bibinfo  {journal}
  {Physical Review Letters}\ }\textbf {\bibinfo {volume} {109}}~(\bibinfo
  {number} {23}),\ \bibinfo {pages} {235301}}\BibitemShut {NoStop}%
\bibitem [{\citenamefont {{STAR Collaboration}}(2010)}]{STAR}%
  \BibitemOpen
  \bibfield  {author} {\bibinfo {author} {\bibnamefont {{STAR
  Collaboration}},}} (\bibinfo {year} {2010}),\ \bibfield  {title} {\enquote
  {\bibinfo {title} {Observation of an antimatter hypernucleus},}\ }\href@noop
  {} {\bibfield  {journal} {\bibinfo  {journal} {Science}\ }\textbf {\bibinfo
  {volume} {428}},\ \bibinfo {pages} {58--62}}\BibitemShut {NoStop}%
\bibitem [{\citenamefont {von Stecher}(2010)}]{stecher2010JPB}%
  \BibitemOpen
  \bibfield  {author} {\bibinfo {author} {\bibnamefont {von Stecher},
  \bibfnamefont {J}}} (\bibinfo {year} {2010}),\ \bibfield  {title}
  {{\selectlanguage {English}\enquote {\bibinfo {title} {Weakly bound cluster
  states of {E}fimov character},}\ }}\href@noop {} {\bibfield  {journal}
  {\bibinfo  {journal} {J. Phys. B}\ }\textbf {\bibinfo {volume}
  {43}}~(\bibinfo {number} {10}),\ \bibinfo {pages} {101002}}\BibitemShut
  {NoStop}%
\bibitem [{\citenamefont {von Stecher}(2011)}]{stecher2011PRL}%
  \BibitemOpen
  \bibfield  {author} {\bibinfo {author} {\bibnamefont {von Stecher},
  \bibfnamefont {J}}} (\bibinfo {year} {2011}),\ \bibfield  {title} {\enquote
  {\bibinfo {title} {Five- and six-body resonances tied to an {E}fimov
  trimer},}\ }\href@noop {} {\bibfield  {journal} {\bibinfo  {journal} {Phys.
  Rev. Lett.}\ }\textbf {\bibinfo {volume} {107}},\ \bibinfo {pages}
  {200402}}\BibitemShut {NoStop}%
\bibitem [{\citenamefont {von Stecher}\ \emph {et~al.}(2009)\citenamefont {von
  Stecher}, \citenamefont {D'Incao},\ and\ \citenamefont
  {Greene}}]{stecher2009NTP}%
  \BibitemOpen
  \bibfield  {author} {\bibinfo {author} {\bibnamefont {von Stecher},
  \bibfnamefont {J}}, \bibinfo {author} {\bibfnamefont {J.~P.}\ \bibnamefont
  {D'Incao}}, \ and\ \bibinfo {author} {\bibfnamefont {C.~H.}\ \bibnamefont
  {Greene}}} (\bibinfo {year} {2009}),\ \bibfield  {title} {{\selectlanguage
  {English}\enquote {\bibinfo {title} {Signatures of universal four-body
  phenomena and their relation to the {E}fimov effect},}\ }}\href@noop {}
  {\bibfield  {journal} {\bibinfo  {journal} {Nat. Phys.}\ }\textbf {\bibinfo
  {volume} {5}}~(\bibinfo {number} {6}),\ \bibinfo {pages}
  {417--421}}\BibitemShut {NoStop}%
\bibitem [{\citenamefont {von Stecher}\ and\ \citenamefont
  {Greene}(2009)}]{stecher2009PRA}%
  \BibitemOpen
  \bibfield  {author} {\bibinfo {author} {\bibnamefont {von Stecher},
  \bibfnamefont {J}}, \ and\ \bibinfo {author} {\bibfnamefont {C.~H.}\
  \bibnamefont {Greene}}} (\bibinfo {year} {2009}),\ \bibfield  {title}
  {{\selectlanguage {English}\enquote {\bibinfo {title} {Correlated {G}aussian
  hyperspherical method for few-body systems},}\ }}\href@noop {} {\bibfield
  {journal} {\bibinfo  {journal} {Phys. Rev. A}\ }\textbf {\bibinfo {volume}
  {80}}~(\bibinfo {number} {2}),\ \bibinfo {pages} {022504}}\BibitemShut
  {NoStop}%
\bibitem [{\citenamefont {Stephens}\ and\ \citenamefont
  {Fano}(1988)}]{StephensFano1988pra}%
  \BibitemOpen
  \bibfield  {author} {\bibinfo {author} {\bibnamefont {Stephens},
  \bibfnamefont {J~A}}, \ and\ \bibinfo {author} {\bibfnamefont
  {U.}~\bibnamefont {Fano}}} (\bibinfo {year} {1988}),\ \bibfield  {title}
  {\enquote {\bibinfo {title} {Slow electrons in condensed matter: The large
  polaron},}\ }\href {\doibase 10.1103/PhysRevA.38.3372} {\bibfield  {journal}
  {\bibinfo  {journal} {Phys. Rev. A}\ }\textbf {\bibinfo {volume} {38}},\
  \bibinfo {pages} {3372--3376}}\BibitemShut {NoStop}%
\bibitem [{\citenamefont {Stormer}\ \emph {et~al.}(1999)\citenamefont
  {Stormer}, \citenamefont {Tsui},\ and\ \citenamefont
  {Gossard}}]{Stormer1999rmp}%
  \BibitemOpen
  \bibfield  {author} {\bibinfo {author} {\bibnamefont {Stormer}, \bibfnamefont
  {Horst~L}}, \bibinfo {author} {\bibfnamefont {Daniel~C.}\ \bibnamefont
  {Tsui}}, \ and\ \bibinfo {author} {\bibfnamefont {Arthur~C.}\ \bibnamefont
  {Gossard}}} (\bibinfo {year} {1999}),\ \bibfield  {title} {\enquote {\bibinfo
  {title} {The fractional quantum hall effect},}\ }\href {\doibase
  10.1103/RevModPhys.71.S298} {\bibfield  {journal} {\bibinfo  {journal} {Rev.
  Mod. Phys.}\ }\textbf {\bibinfo {volume} {71}},\ \bibinfo {pages}
  {S298--S305}}\BibitemShut {NoStop}%
\bibitem [{\citenamefont {Strauss}\ \emph {et~al.}(2010)\citenamefont
  {Strauss}, \citenamefont {Takekoshi}, \citenamefont {Winker}, \citenamefont
  {Grimm},\ and\ \citenamefont {Denschlag}}]{Strauss-2010}%
  \BibitemOpen
  \bibfield  {author} {\bibinfo {author} {\bibnamefont {Strauss}, \bibfnamefont
  {C}}, \bibinfo {author} {\bibfnamefont {T.}~\bibnamefont {Takekoshi}},
  \bibinfo {author} {\bibfnamefont {K.}~\bibnamefont {Winker}}, \bibinfo
  {author} {\bibfnamefont {R.}~\bibnamefont {Grimm}}, \ and\ \bibinfo {author}
  {\bibfnamefont {J.~H.}\ \bibnamefont {Denschlag}}} (\bibinfo {year} {2010}),\
  \bibfield  {title} {\enquote {\bibinfo {title} {Hyperfine, rotational, and
  vibrational structure of the a$^{3}$$\sigma$$_{u}^{+}$ state of
  $^{87}${Rb}$_2$},}\ }\href@noop {} {\bibfield  {journal} {\bibinfo  {journal}
  {Phys. Rev A}\ }\textbf {\bibinfo {volume} {82}},\ \bibinfo {pages}
  {052514}}\BibitemShut {NoStop}%
\bibitem [{\citenamefont {Strecker}\ \emph {et~al.}(2003)\citenamefont
  {Strecker}, \citenamefont {Partridge},\ and\ \citenamefont
  {Hulet}}]{strecker2003PRL}%
  \BibitemOpen
  \bibfield  {author} {\bibinfo {author} {\bibnamefont {Strecker},
  \bibfnamefont {K~E}}, \bibinfo {author} {\bibfnamefont {G.~B.}\ \bibnamefont
  {Partridge}}, \ and\ \bibinfo {author} {\bibfnamefont {R.~G.}\ \bibnamefont
  {Hulet}}} (\bibinfo {year} {2003}),\ \bibfield  {title} {\enquote {\bibinfo
  {title} {Conversion of an atomic {F}ermi gas to a long-lived molecular {B}ose
  gas},}\ }\href@noop {} {\bibfield  {journal} {\bibinfo  {journal} {Phys. Rev.
  Lett.}\ }\textbf {\bibinfo {volume} {91}},\ \bibinfo {pages}
  {080406}}\BibitemShut {NoStop}%
\bibitem [{\citenamefont {Suno}\ and\ \citenamefont
  {Esry}(2008)}]{suno2008PRA}%
  \BibitemOpen
  \bibfield  {author} {\bibinfo {author} {\bibnamefont {Suno}, \bibfnamefont
  {H}}, \ and\ \bibinfo {author} {\bibfnamefont {B.~D.}\ \bibnamefont {Esry}}}
  (\bibinfo {year} {2008}),\ \bibfield  {title} {{\selectlanguage
  {English}\enquote {\bibinfo {title} {Adiabatic hyperspherical study of
  triatomic helium systems},}\ }}\href@noop {} {\bibfield  {journal} {\bibinfo
  {journal} {Phys. Rev. A}\ }\textbf {\bibinfo {volume} {78}}~(\bibinfo
  {number} {6, Part A}),\ \bibinfo {pages} {062701}}\BibitemShut {NoStop}%
\bibitem [{\citenamefont {Suno}\ \emph
  {et~al.}(2003{\natexlab{a}})\citenamefont {Suno}, \citenamefont {Esry},\ and\
  \citenamefont {Greene}}]{suno2003PRL}%
  \BibitemOpen
  \bibfield  {author} {\bibinfo {author} {\bibnamefont {Suno}, \bibfnamefont
  {H}}, \bibinfo {author} {\bibfnamefont {B.~D.}\ \bibnamefont {Esry}}, \ and\
  \bibinfo {author} {\bibfnamefont {C.~H.}\ \bibnamefont {Greene}}} (\bibinfo
  {year} {2003}{\natexlab{a}}),\ \bibfield  {title} {{\selectlanguage
  {English}\enquote {\bibinfo {title} {Recombination of three ultracold
  {F}ermionic atoms},}\ }}\href@noop {} {\bibfield  {journal} {\bibinfo
  {journal} {Phys. Rev. Lett.}\ }\textbf {\bibinfo {volume} {90}}~(\bibinfo
  {number} {5}),\ \bibinfo {pages} {053202}}\BibitemShut {NoStop}%
\bibitem [{\citenamefont {Suno}\ \emph
  {et~al.}(2003{\natexlab{b}})\citenamefont {Suno}, \citenamefont {Esry},\ and\
  \citenamefont {Greene}}]{suno2003NJP}%
  \BibitemOpen
  \bibfield  {author} {\bibinfo {author} {\bibnamefont {Suno}, \bibfnamefont
  {H}}, \bibinfo {author} {\bibfnamefont {B.~D.}\ \bibnamefont {Esry}}, \ and\
  \bibinfo {author} {\bibfnamefont {C.~H.}\ \bibnamefont {Greene}}} (\bibinfo
  {year} {2003}{\natexlab{b}}),\ \bibfield  {title} {{\selectlanguage
  {English}\enquote {\bibinfo {title} {Three-body recombination of cold
  {F}ermionic atoms},}\ }}\href@noop {} {\bibfield  {journal} {\bibinfo
  {journal} {New J. Phys.}\ }\textbf {\bibinfo {volume} {5}},\ \bibinfo {pages}
  {53}}\BibitemShut {NoStop}%
\bibitem [{\citenamefont {Suno}\ \emph {et~al.}(2002)\citenamefont {Suno},
  \citenamefont {Esry}, \citenamefont {Greene},\ and\ \citenamefont
  {Burke}}]{suno2002PRA}%
  \BibitemOpen
  \bibfield  {author} {\bibinfo {author} {\bibnamefont {Suno}, \bibfnamefont
  {H}}, \bibinfo {author} {\bibfnamefont {B.~D.}\ \bibnamefont {Esry}},
  \bibinfo {author} {\bibfnamefont {C.~H.}\ \bibnamefont {Greene}}, \ and\
  \bibinfo {author} {\bibfnamefont {J.~P.}\ \bibnamefont {Burke}}} (\bibinfo
  {year} {2002}),\ \bibfield  {title} {{\selectlanguage {English}\enquote
  {\bibinfo {title} {Three-body recombination of cold helium atoms},}\
  }}\href@noop {} {\bibfield  {journal} {\bibinfo  {journal} {Phys. Rev. A}\
  }\textbf {\bibinfo {volume} {65}},\ \bibinfo {pages} {042725}}\BibitemShut
  {NoStop}%
\bibitem [{\citenamefont {Suno}\ \emph {et~al.}(2015)\citenamefont {Suno},
  \citenamefont {Suzuki},\ and\ \citenamefont {Descouvemont}}]{suno2015PRC}%
  \BibitemOpen
  \bibfield  {author} {\bibinfo {author} {\bibnamefont {Suno}, \bibfnamefont
  {H}}, \bibinfo {author} {\bibfnamefont {Y.}~\bibnamefont {Suzuki}}, \ and\
  \bibinfo {author} {\bibfnamefont {P.}~\bibnamefont {Descouvemont}}} (\bibinfo
  {year} {2015}),\ \bibfield  {title} {\enquote {\bibinfo {title}
  {$\mathrm{Triple}-\ensuremath{\alpha}$ continuum structure and hoyle
  resonance of $^{12}\mathrm{C}$ using the hyperspherical slow variable
  discretization},}\ }\href@noop {} {\bibfield  {journal} {\bibinfo  {journal}
  {Phys. Rev. C}\ }\textbf {\bibinfo {volume} {91}},\ \bibinfo {pages}
  {014004}}\BibitemShut {NoStop}%
\bibitem [{\citenamefont {Suzuki}\ and\ \citenamefont
  {Varga}(1998)}]{suzuki1998}%
  \BibitemOpen
  \bibfield  {author} {\bibinfo {author} {\bibnamefont {Suzuki}, \bibfnamefont
  {Y}}, \ and\ \bibinfo {author} {\bibfnamefont {K.}~\bibnamefont {Varga}}}
  (\bibinfo {year} {1998}),\ \href@noop {} {\emph {\bibinfo {title} {Stochastic
  Variational Approach to Quantum-Mechanical Few-Body Problems}}}\ (\bibinfo
  {publisher} {Springer-Verlag, Berlin})\BibitemShut {NoStop}%
\bibitem [{\citenamefont {Sykes}\ \emph {et~al.}({2014})\citenamefont {Sykes},
  \citenamefont {Corson}, \citenamefont {D'Incao}, \citenamefont {Koller},
  \citenamefont {Greene}, \citenamefont {Rey}, \citenamefont {Hazzard},\ and\
  \citenamefont {Bohn}}]{Sykes2014pra}%
  \BibitemOpen
  \bibfield  {author} {\bibinfo {author} {\bibnamefont {Sykes}, \bibfnamefont
  {A~G}}, \bibinfo {author} {\bibfnamefont {J.~P.}\ \bibnamefont {Corson}},
  \bibinfo {author} {\bibfnamefont {J.~P.}\ \bibnamefont {D'Incao}}, \bibinfo
  {author} {\bibfnamefont {A.~P.}\ \bibnamefont {Koller}}, \bibinfo {author}
  {\bibfnamefont {C.~H.}\ \bibnamefont {Greene}}, \bibinfo {author}
  {\bibfnamefont {A.~M.}\ \bibnamefont {Rey}}, \bibinfo {author} {\bibfnamefont
  {K.~R.~A.}\ \bibnamefont {Hazzard}}, \ and\ \bibinfo {author} {\bibfnamefont
  {J.~L.}\ \bibnamefont {Bohn}}} (\bibinfo {year} {{2014}}),\ \bibfield
  {title} {\enquote {\bibinfo {title} {{Quenching to unitarity: Quantum
  dynamics in a three-dimensional Bose gas}},}\ }\href@noop {} {\bibfield
  {journal} {\bibinfo  {journal} {{Phys. Rev. A}}\ }\textbf {\bibinfo {volume}
  {{89}}}~(\bibinfo {number} {{2}})}\BibitemShut {NoStop}%
\bibitem [{\citenamefont {Tan}(2008{\natexlab{a}})}]{tan2008AP}%
  \BibitemOpen
  \bibfield  {author} {\bibinfo {author} {\bibnamefont {Tan}, \bibfnamefont
  {S}}} (\bibinfo {year} {2008}{\natexlab{a}}),\ \bibfield  {title} {\enquote
  {\bibinfo {title} {Energetics of a strongly correlated {F}ermi gas},}\
  }\href@noop {} {\bibfield  {journal} {\bibinfo  {journal} {Ann. Phys.}\
  }\textbf {\bibinfo {volume} {323}}~(\bibinfo {number} {12}),\ \bibinfo
  {pages} {2952 -- 2970}}\BibitemShut {NoStop}%
\bibitem [{\citenamefont {Tan}(2008{\natexlab{b}})}]{tan2008APb}%
  \BibitemOpen
  \bibfield  {author} {\bibinfo {author} {\bibnamefont {Tan}, \bibfnamefont
  {S}}} (\bibinfo {year} {2008}{\natexlab{b}}),\ \bibfield  {title} {\enquote
  {\bibinfo {title} {Generalized virial theorem and pressure relation for a
  strongly correlated {F}ermi gas},}\ }\href@noop {} {\bibfield  {journal}
  {\bibinfo  {journal} {Ann. Phys.}\ }\textbf {\bibinfo {volume}
  {323}}~(\bibinfo {number} {12}),\ \bibinfo {pages} {2987 --
  2990}}\BibitemShut {NoStop}%
\bibitem [{\citenamefont {Tan}(2008{\natexlab{c}})}]{tan2008APc}%
  \BibitemOpen
  \bibfield  {author} {\bibinfo {author} {\bibnamefont {Tan}, \bibfnamefont
  {S}}} (\bibinfo {year} {2008}{\natexlab{c}}),\ \bibfield  {title} {\enquote
  {\bibinfo {title} {Large momentum part of a strongly correlated {F}ermi
  gas},}\ }\href@noop {} {\bibfield  {journal} {\bibinfo  {journal} {Ann.
  Phys.}\ }\textbf {\bibinfo {volume} {323}}~(\bibinfo {number} {12}),\
  \bibinfo {pages} {2971 -- 2986}}\BibitemShut {NoStop}%
\bibitem [{\citenamefont {Tang}\ \emph {et~al.}(1992)\citenamefont {Tang},
  \citenamefont {Watanabe}, \citenamefont {Matsuzawa},\ and\ \citenamefont
  {Lin}}]{tang1992PRL}%
  \BibitemOpen
  \bibfield  {author} {\bibinfo {author} {\bibnamefont {Tang}, \bibfnamefont
  {J~Z}}, \bibinfo {author} {\bibfnamefont {S.}~\bibnamefont {Watanabe}},
  \bibinfo {author} {\bibfnamefont {M.}~\bibnamefont {Matsuzawa}}, \ and\
  \bibinfo {author} {\bibfnamefont {C.~D.}\ \bibnamefont {Lin}}} (\bibinfo
  {year} {1992}),\ \bibfield  {title} {{\selectlanguage {English}\enquote
  {\bibinfo {title} {Evidence of an excited angular-correlation mode in
  high-lying {He}},}\ }}\href@noop {} {\bibfield  {journal} {\bibinfo
  {journal} {Phys. Rev. Lett.}\ }\textbf {\bibinfo {volume} {69}}~(\bibinfo
  {number} {11}),\ \bibinfo {pages} {1633--1635}}\BibitemShut {NoStop}%
\bibitem [{\citenamefont {Tanihata}(1996)}]{Tanihata-1996}%
  \BibitemOpen
  \bibfield  {author} {\bibinfo {author} {\bibnamefont {Tanihata},
  \bibfnamefont {I}}} (\bibinfo {year} {1996}),\ \bibfield  {title} {\enquote
  {\bibinfo {title} {Neutron halo nuclei},}\ }\href@noop {} {\bibfield
  {journal} {\bibinfo  {journal} {J. Phy. G: Nucl. Part. Phys.}\ }\textbf
  {\bibinfo {volume} {22}},\ \bibinfo {pages} {157--198}}\BibitemShut {NoStop}%
\bibitem [{\citenamefont {Tanner}\ \emph {et~al.}(2000)\citenamefont {Tanner},
  \citenamefont {Richter},\ and\ \citenamefont
  {Rost}}]{TannerRichterRost2000rmp}%
  \BibitemOpen
  \bibfield  {author} {\bibinfo {author} {\bibnamefont {Tanner}, \bibfnamefont
  {Gregor}}, \bibinfo {author} {\bibfnamefont {Klaus}\ \bibnamefont {Richter}},
  \ and\ \bibinfo {author} {\bibfnamefont {Jan-Michael}\ \bibnamefont {Rost}}}
  (\bibinfo {year} {2000}),\ \bibfield  {title} {\enquote {\bibinfo {title}
  {The theory of two-electron atoms: between ground state and complete
  fragmentation},}\ }\href {\doibase 10.1103/RevModPhys.72.497} {\bibfield
  {journal} {\bibinfo  {journal} {Rev. Mod. Phys.}\ }\textbf {\bibinfo {volume}
  {72}},\ \bibinfo {pages} {497--544}}\BibitemShut {NoStop}%
\bibitem [{\citenamefont {Taylor}(1972)}]{taylor1972}%
  \BibitemOpen
  \bibfield  {author} {\bibinfo {author} {\bibnamefont {Taylor}, \bibfnamefont
  {J~R}}} (\bibinfo {year} {1972}),\ \href@noop {} {\emph {\bibinfo {title}
  {Scattering theory}}}\ (\bibinfo  {publisher} {John Wiley and Sons, Inc., New
  York})\BibitemShut {NoStop}%
\bibitem [{\citenamefont {Tempere}\ \emph {et~al.}(2009)\citenamefont
  {Tempere}, \citenamefont {Casteels}, \citenamefont {Oberthaler},
  \citenamefont {Knoop}, \citenamefont {Timmermans},\ and\ \citenamefont
  {Devreese}}]{tempere_feynman_2009}%
  \BibitemOpen
  \bibfield  {author} {\bibinfo {author} {\bibnamefont {Tempere}, \bibfnamefont
  {J}}, \bibinfo {author} {\bibfnamefont {W.}~\bibnamefont {Casteels}},
  \bibinfo {author} {\bibfnamefont {M.~K.}\ \bibnamefont {Oberthaler}},
  \bibinfo {author} {\bibfnamefont {S.}~\bibnamefont {Knoop}}, \bibinfo
  {author} {\bibfnamefont {E.}~\bibnamefont {Timmermans}}, \ and\ \bibinfo
  {author} {\bibfnamefont {J.~T.}\ \bibnamefont {Devreese}}} (\bibinfo {year}
  {2009}),\ \bibfield  {title} {\enquote {\bibinfo {title} {Feynman
  path-integral treatment of the {BEC}-impurity polaron},}\ }\href {\doibase
  10.1103/PhysRevB.80.184504} {\bibfield  {journal} {\bibinfo  {journal}
  {Physical Review B}\ }\textbf {\bibinfo {volume} {80}}~(\bibinfo {number}
  {18}),\ \bibinfo {pages} {184504}}\BibitemShut {NoStop}%
\bibitem [{\citenamefont {Thomas}(1935)}]{thomas1935pr}%
  \BibitemOpen
  \bibfield  {author} {\bibinfo {author} {\bibnamefont {Thomas}, \bibfnamefont
  {L~H}}} (\bibinfo {year} {1935}),\ \bibfield  {title} {\enquote {\bibinfo
  {title} {The interaction between a neutron and a proton and the structure of
  $^3${H}},}\ }\href@noop {} {\bibfield  {journal} {\bibinfo  {journal} {Phys.
  Rev.}\ }\textbf {\bibinfo {volume} {47}}~(\bibinfo {number} {12}),\ \bibinfo
  {pages} {903--909}}\BibitemShut {NoStop}%
\bibitem [{\citenamefont {Thouless}\ \emph {et~al.}(1982)\citenamefont
  {Thouless}, \citenamefont {Kohmoto}, \citenamefont {Nightingale},\ and\
  \citenamefont {Dennijs}}]{thouless_quantized_1982}%
  \BibitemOpen
  \bibfield  {author} {\bibinfo {author} {\bibnamefont {Thouless},
  \bibfnamefont {DJ}}, \bibinfo {author} {\bibfnamefont {M.}~\bibnamefont
  {Kohmoto}}, \bibinfo {author} {\bibfnamefont {MP}~\bibnamefont
  {Nightingale}}, \ and\ \bibinfo {author} {\bibfnamefont {M.}~\bibnamefont
  {Dennijs}}} (\bibinfo {year} {1982}),\ \bibfield  {title} {\enquote {\bibinfo
  {title} {{Quantized Hall conductance in a two-dimensional periodic
  potential}},}\ }\href {\doibase 10.1103/PhysRevLett.49.405} {\bibfield
  {journal} {\bibinfo  {journal} {Physical Review Letters}\ }\textbf {\bibinfo
  {volume} {49}}~(\bibinfo {number} {6}),\ \bibinfo {pages}
  {405--408}}\BibitemShut {NoStop}%
\bibitem [{\citenamefont {Ticknor}\ \emph {et~al.}(2004)\citenamefont
  {Ticknor}, \citenamefont {Regal}, \citenamefont {Jin},\ and\ \citenamefont
  {Bohn}}]{ticknor2004multiplet}%
  \BibitemOpen
  \bibfield  {author} {\bibinfo {author} {\bibnamefont {Ticknor}, \bibfnamefont
  {C}}, \bibinfo {author} {\bibfnamefont {C.~A.}\ \bibnamefont {Regal}},
  \bibinfo {author} {\bibfnamefont {D.~S.}\ \bibnamefont {Jin}}, \ and\
  \bibinfo {author} {\bibfnamefont {J.~L.}\ \bibnamefont {Bohn}}} (\bibinfo
  {year} {2004}),\ \bibfield  {title} {\enquote {\bibinfo {title} {Multiplet
  structure of {F}eshbach resonances in nonzero partial waves},}\ }\href@noop
  {} {\bibfield  {journal} {\bibinfo  {journal} {Phys. Rev. A}\ }\textbf
  {\bibinfo {volume} {69}}~(\bibinfo {number} {4}),\ \bibinfo {pages}
  {042712}}\BibitemShut {NoStop}%
\bibitem [{\citenamefont {Tiecke}\ \emph {et~al.}(2010)\citenamefont {Tiecke},
  \citenamefont {Goosen}, \citenamefont {Walraven},\ and\ \citenamefont
  {Kokkelmans}}]{tiecke2010pra}%
  \BibitemOpen
  \bibfield  {author} {\bibinfo {author} {\bibnamefont {Tiecke}, \bibfnamefont
  {T~G}}, \bibinfo {author} {\bibfnamefont {M.~R.}\ \bibnamefont {Goosen}},
  \bibinfo {author} {\bibfnamefont {J.~T.~M.}\ \bibnamefont {Walraven}}, \ and\
  \bibinfo {author} {\bibfnamefont {S.~J. J. M.~F.}\ \bibnamefont
  {Kokkelmans}}} (\bibinfo {year} {2010}),\ \bibfield  {title} {\enquote
  {\bibinfo {title} {Asymptotic-bound-state model for {F}eshbach resonances},}\
  }\href@noop {} {\bibfield  {journal} {\bibinfo  {journal} {Phys. Rev. A}\
  }\textbf {\bibinfo {volume} {82}},\ \bibinfo {pages} {042712}}\BibitemShut
  {NoStop}%
\bibitem [{\citenamefont {Tjon}(1975)}]{Tjon-1975}%
  \BibitemOpen
  \bibfield  {author} {\bibinfo {author} {\bibnamefont {Tjon}, \bibfnamefont
  {J~A}}} (\bibinfo {year} {1975}),\ \bibfield  {title} {\enquote {\bibinfo
  {title} {Bound states of $^{4}${He} with local interactions},}\ }\href@noop
  {} {\bibfield  {journal} {\bibinfo  {journal} {Phys. Lett. B}\ }\textbf
  {\bibinfo {volume} {56}},\ \bibinfo {pages} {217}}\BibitemShut {NoStop}%
\bibitem [{\citenamefont {Tolra}\ \emph {et~al.}(2004)\citenamefont {Tolra},
  \citenamefont {O'Hara}, \citenamefont {Huckans}, \citenamefont {Phillips},
  \citenamefont {Rolston},\ and\ \citenamefont
  {Porto}}]{TolraPhillips1D2004prl}%
  \BibitemOpen
  \bibfield  {author} {\bibinfo {author} {\bibnamefont {Tolra}, \bibfnamefont
  {B~Laburthe}}, \bibinfo {author} {\bibfnamefont {K.~M.}\ \bibnamefont
  {O'Hara}}, \bibinfo {author} {\bibfnamefont {J.~H.}\ \bibnamefont {Huckans}},
  \bibinfo {author} {\bibfnamefont {W.~D.}\ \bibnamefont {Phillips}}, \bibinfo
  {author} {\bibfnamefont {S.~L.}\ \bibnamefont {Rolston}}, \ and\ \bibinfo
  {author} {\bibfnamefont {J.~V.}\ \bibnamefont {Porto}}} (\bibinfo {year}
  {2004}),\ \bibfield  {title} {\enquote {\bibinfo {title} {Observation of
  reduced three-body recombination in a correlated 1d degenerate bose gas},}\
  }\href {\doibase 10.1103/PhysRevLett.92.190401} {\bibfield  {journal}
  {\bibinfo  {journal} {Phys. Rev. Lett.}\ }\textbf {\bibinfo {volume} {92}},\
  \bibinfo {pages} {190401}}\BibitemShut {NoStop}%
\bibitem [{\citenamefont {Tolstikhin}\ \emph {et~al.}(1996)\citenamefont
  {Tolstikhin}, \citenamefont {Watanabe},\ and\ \citenamefont
  {Matsuzawa}}]{tolstikhin1996JPB}%
  \BibitemOpen
  \bibfield  {author} {\bibinfo {author} {\bibnamefont {Tolstikhin},
  \bibfnamefont {OI}}, \bibinfo {author} {\bibfnamefont {S}~\bibnamefont
  {Watanabe}}, \ and\ \bibinfo {author} {\bibfnamefont {M}~\bibnamefont
  {Matsuzawa}}} (\bibinfo {year} {1996}),\ \bibfield  {title} {{\selectlanguage
  {English}\enquote {\bibinfo {title} {`slow' variable discretization: A novel
  approach for hamiltonians allowing adiabatic separation of variables},}\
  }}\href@noop {} {\bibfield  {journal} {\bibinfo  {journal} {J. Phys. B}\
  }\textbf {\bibinfo {volume} {29}}~(\bibinfo {number} {11}),\ \bibinfo {pages}
  {L389--L395}}\BibitemShut {NoStop}%
\bibitem [{\citenamefont {Tonks}(1936)}]{tonks1936}%
  \BibitemOpen
  \bibfield  {author} {\bibinfo {author} {\bibnamefont {Tonks}, \bibfnamefont
  {L}}} (\bibinfo {year} {1936}),\ \bibfield  {title} {\enquote {\bibinfo
  {title} {The complete equation of state of one, two and three-dimensional
  gases of hard elastic spheres},}\ }\href@noop {} {\bibfield  {journal}
  {\bibinfo  {journal} {Phys. Rev.}\ }\textbf {\bibinfo {volume}
  {50}}~(\bibinfo {number} {10}),\ \bibinfo {pages} {955--963}}\BibitemShut
  {NoStop}%
\bibitem [{\citenamefont {Truhlar}\ and\ \citenamefont
  {Muckerman}(1975)}]{Truhlar-1975}%
  \BibitemOpen
  \bibfield  {author} {\bibinfo {author} {\bibnamefont {Truhlar}, \bibfnamefont
  {D~G}}, \ and\ \bibinfo {author} {\bibfnamefont {J.~T.}\ \bibnamefont
  {Muckerman}}} (\bibinfo {year} {1975}),\ \bibfield  {title} {\enquote
  {\bibinfo {title} {Reactive scattering cross sections: Quasiclassical and
  semiclassical methods},}\ }in\ \href@noop {} {\emph {\bibinfo {booktitle}
  {Atom-Molecule Collision Theory: A Guide for the Experimentalist}}},\
  \bibinfo {editor} {edited by\ \bibinfo {editor} {\bibfnamefont {R.~B.}\
  \bibnamefont {Bernstein}}},\ Chap.~\bibinfo {chapter} {16}\ (\bibinfo
  {publisher} {Plenum Press, New York})\ pp.\ \bibinfo {pages}
  {505--566}\BibitemShut {NoStop}%
\bibitem [{\citenamefont {Tscherbul}\ \emph {et~al.}(2010)\citenamefont
  {Tscherbul}, \citenamefont {Calarco}, \citenamefont {Lesanovsky},
  \citenamefont {Krems}, \citenamefont {Dalgarno},\ and\ \citenamefont
  {Schmiedmayer}}]{Tscherbul2010pra}%
  \BibitemOpen
  \bibfield  {author} {\bibinfo {author} {\bibnamefont {Tscherbul},
  \bibfnamefont {T~V}}, \bibinfo {author} {\bibfnamefont {T.}~\bibnamefont
  {Calarco}}, \bibinfo {author} {\bibfnamefont {I.}~\bibnamefont {Lesanovsky}},
  \bibinfo {author} {\bibfnamefont {R.~V.}\ \bibnamefont {Krems}}, \bibinfo
  {author} {\bibfnamefont {A.}~\bibnamefont {Dalgarno}}, \ and\ \bibinfo
  {author} {\bibfnamefont {J.}~\bibnamefont {Schmiedmayer}}} (\bibinfo {year}
  {2010}),\ \bibfield  {title} {\enquote {\bibinfo {title} {rf-field-induced
  {F}eshbach resonances},}\ }\href@noop {} {\bibfield  {journal} {\bibinfo
  {journal} {Phys. Rev. A}\ }\textbf {\bibinfo {volume} {81}},\ \bibinfo
  {pages} {050701}}\BibitemShut {NoStop}%
\bibitem [{\citenamefont {Tung}\ \emph {et~al.}(2014)\citenamefont {Tung},
  \citenamefont {Jim\'enez-Garc\'ia}, \citenamefont {Johansen}, \citenamefont
  {Parker},\ and\ \citenamefont {Chin}}]{Tung-2014}%
  \BibitemOpen
  \bibfield  {author} {\bibinfo {author} {\bibnamefont {Tung}, \bibfnamefont
  {S~K}}, \bibinfo {author} {\bibfnamefont {K.}~\bibnamefont
  {Jim\'enez-Garc\'ia}}, \bibinfo {author} {\bibfnamefont {J.}~\bibnamefont
  {Johansen}}, \bibinfo {author} {\bibfnamefont {C.~V.}\ \bibnamefont
  {Parker}}, \ and\ \bibinfo {author} {\bibfnamefont {C.}~\bibnamefont {Chin}}}
  (\bibinfo {year} {2014}),\ \bibfield  {title} {\enquote {\bibinfo {title}
  {Geometric scaling of {E}fimov states in a $^6${L}i-$^{133}${C}s},}\
  }\href@noop {} {\bibfield  {journal} {\bibinfo  {journal} {Phys. Rev. Lett.}\
  }\textbf {\bibinfo {volume} {113}},\ \bibinfo {pages} {240402}}\BibitemShut
  {NoStop}%
\bibitem [{\citenamefont {Ullrich}(2003)}]{Ullrich-2003}%
  \BibitemOpen
  \bibfield  {author} {\bibinfo {author} {\bibnamefont {Ullrich}, \bibfnamefont
  {J~{et al}}}} (\bibinfo {year} {2003}),\ \bibfield  {title} {\enquote
  {\bibinfo {title} {Recoil-ion and electron momentum spectroscopy:
  {R}eaction-microscopes},}\ }\href@noop {} {\bibfield  {journal} {\bibinfo
  {journal} {Rep. Prog. Phys.}\ }\textbf {\bibinfo {volume} {66}},\ \bibinfo
  {pages} {1463--1545}}\BibitemShut {NoStop}%
\bibitem [{\citenamefont {Ulmanis}\ \emph
  {et~al.}(2016{\natexlab{a}})\citenamefont {Ulmanis}, \citenamefont
  {H\"afner}, \citenamefont {Pires}, \citenamefont {Kuhnle}, \citenamefont
  {Wang}, \citenamefont {Greene},\ and\ \citenamefont
  {Weidem\"uller}}]{Ulmanis2016prl}%
  \BibitemOpen
  \bibfield  {author} {\bibinfo {author} {\bibnamefont {Ulmanis}, \bibfnamefont
  {J}}, \bibinfo {author} {\bibfnamefont {S.}~\bibnamefont {H\"afner}},
  \bibinfo {author} {\bibfnamefont {R.}~\bibnamefont {Pires}}, \bibinfo
  {author} {\bibfnamefont {E.~D.}\ \bibnamefont {Kuhnle}}, \bibinfo {author}
  {\bibfnamefont {Y.}~\bibnamefont {Wang}}, \bibinfo {author} {\bibfnamefont
  {C.~H.}\ \bibnamefont {Greene}}, \ and\ \bibinfo {author} {\bibfnamefont
  {M.}~\bibnamefont {Weidem\"uller}}} (\bibinfo {year} {2016}{\natexlab{a}}),\
  \bibfield  {title} {\enquote {\bibinfo {title} {Heteronuclear {E}fimov
  scenario with positive intraspecies scattering length},}\ }\href@noop {}
  {\bibfield  {journal} {\bibinfo  {journal} {Phys. Rev. Lett.}\ }\textbf
  {\bibinfo {volume} {117}},\ \bibinfo {pages} {153201}}\BibitemShut {NoStop}%
\bibitem [{\citenamefont {Ulmanis}\ \emph {et~al.}(2015)\citenamefont
  {Ulmanis}, \citenamefont {H\"{a}fner}, \citenamefont {Pires}, \citenamefont
  {Weidem\"uller},\ and\ \citenamefont {Timeann}}]{Ulmanis-2015}%
  \BibitemOpen
  \bibfield  {author} {\bibinfo {author} {\bibnamefont {Ulmanis}, \bibfnamefont
  {J}}, \bibinfo {author} {\bibfnamefont {S.}~\bibnamefont {H\"{a}fner}},
  \bibinfo {author} {\bibfnamefont {R.}~\bibnamefont {Pires}}, \bibinfo
  {author} {\bibfnamefont {M.}~\bibnamefont {Weidem\"uller}}, \ and\ \bibinfo
  {author} {\bibfnamefont {E.}~\bibnamefont {Timeann}}} (\bibinfo {year}
  {2015}),\ \bibfield  {title} {\enquote {\bibinfo {title} {Universality of
  weakly bound dimers and {E}fimov trimers close to {Li}-{Cs} {F}eshbach
  resonances},}\ }\href@noop {} {\bibfield  {journal} {\bibinfo  {journal} {New
  J. Phys.}\ }\textbf {\bibinfo {volume} {17}},\ \bibinfo {pages}
  {05509}}\BibitemShut {NoStop}%
\bibitem [{\citenamefont {Ulmanis}\ \emph
  {et~al.}(2016{\natexlab{b}})\citenamefont {Ulmanis}, \citenamefont
  {H\"afner}, \citenamefont {Pires}, \citenamefont {Werner}, \citenamefont
  {Petrov}, \citenamefont {Weidem\"uller},\ and\ \citenamefont
  {Kuhnle}}]{Ulmanis-2015b}%
  \BibitemOpen
  \bibfield  {author} {\bibinfo {author} {\bibnamefont {Ulmanis}, \bibfnamefont
  {J}}, \bibinfo {author} {\bibfnamefont {S.}~\bibnamefont {H\"afner}},
  \bibinfo {author} {\bibfnamefont {R.}~\bibnamefont {Pires}}, \bibinfo
  {author} {\bibfnamefont {F.}~\bibnamefont {Werner}}, \bibinfo {author}
  {\bibfnamefont {D.~S.}\ \bibnamefont {Petrov}}, \bibinfo {author}
  {\bibfnamefont {M.}~\bibnamefont {Weidem\"uller}}, \ and\ \bibinfo {author}
  {\bibfnamefont {E.~D.}\ \bibnamefont {Kuhnle}}} (\bibinfo {year}
  {2016}{\natexlab{b}}),\ \bibfield  {title} {\enquote {\bibinfo {title}
  {Universal three-body recombination and {E}fimov resonances in an ultracold
  {L}i-{C}s mixture},}\ }\href@noop {} {\bibfield  {journal} {\bibinfo
  {journal} {Phys. Rev. A.}\ }\textbf {\bibinfo {volume} {93}},\ \bibinfo
  {pages} {022707}}\BibitemShut {NoStop}%
\bibitem [{\citenamefont {Valliers}\ \emph {et~al.}(1976)\citenamefont
  {Valliers}, \citenamefont {Das},\ and\ \citenamefont
  {Coelho}}]{Valliers-1976}%
  \BibitemOpen
  \bibfield  {author} {\bibinfo {author} {\bibnamefont {Valliers},
  \bibfnamefont {M}}, \bibinfo {author} {\bibfnamefont {T.~K.}\ \bibnamefont
  {Das}}, \ and\ \bibinfo {author} {\bibfnamefont {H.~T.}\ \bibnamefont
  {Coelho}}} (\bibinfo {year} {1976}),\ \bibfield  {title} {\enquote {\bibinfo
  {title} {Generalized {N}-body harmonic oscillator basis funcitons for the
  hyperspherical approach},}\ }\href@noop {} {\bibfield  {journal} {\bibinfo
  {journal} {Nuc. Phys. A}\ }\textbf {\bibinfo {volume} {257}},\ \bibinfo
  {pages} {389}}\BibitemShut {NoStop}%
\bibitem [{\citenamefont {{Van der Wiel}}(1972)}]{Wiel-1972-JPR}%
  \BibitemOpen
  \bibfield  {author} {\bibinfo {author} {\bibnamefont {{Van der Wiel}},
  \bibfnamefont {M~J}}} (\bibinfo {year} {1972}),\ \bibfield  {title} {\enquote
  {\bibinfo {title} {Threshold behavior of double photo{-}ionisation in
  {H}e},}\ }\href@noop {} {\bibfield  {journal} {\bibinfo  {journal} {Phys.
  Lett. A}\ }\textbf {\bibinfo {volume} {41}},\ \bibinfo {pages}
  {389}}\BibitemShut {NoStop}%
\bibitem [{\citenamefont {Verma}\ and\ \citenamefont
  {Sural}(1979)}]{Verma-1979}%
  \BibitemOpen
  \bibfield  {author} {\bibinfo {author} {\bibnamefont {Verma}, \bibfnamefont
  {S~P}}, \ and\ \bibinfo {author} {\bibfnamefont {D.~P.}\ \bibnamefont
  {Sural}}} (\bibinfo {year} {1979}),\ \bibfield  {title} {\enquote {\bibinfo
  {title} {Hyperspherical harmonics method for the hypertriton and
  $^{9}_{\Lambda}${Be}},}\ }\href@noop {} {\bibfield  {journal} {\bibinfo
  {journal} {Phys. Rev. C}\ }\textbf {\bibinfo {volume} {20}},\ \bibinfo
  {pages} {781}}\BibitemShut {NoStop}%
\bibitem [{\citenamefont {Verma}\ and\ \citenamefont
  {Sural}(1982)}]{Verma-1982}%
  \BibitemOpen
  \bibfield  {author} {\bibinfo {author} {\bibnamefont {Verma}, \bibfnamefont
  {S~P}}, \ and\ \bibinfo {author} {\bibfnamefont {D.~P.}\ \bibnamefont
  {Sural}}} (\bibinfo {year} {1982}),\ \bibfield  {title} {\enquote {\bibinfo
  {title} {Hyperspherical harmonics caluclaitons for $^{3}_{\Lambda}${H} with
  realisitic n-p potential},}\ }\href@noop {} {\bibfield  {journal} {\bibinfo
  {journal} {J. Phy. G: Nucl. Part. Phys.}\ }\textbf {\bibinfo {volume} {8}},\
  \bibinfo {pages} {73--81}}\BibitemShut {NoStop}%
\bibitem [{\citenamefont {Vida\~{n}a}(2013)}]{Vidana-2013}%
  \BibitemOpen
  \bibfield  {author} {\bibinfo {author} {\bibnamefont {Vida\~{n}a},
  \bibfnamefont {I}}} (\bibinfo {year} {2013}),\ \bibfield  {title} {\enquote
  {\bibinfo {title} {Hyperons and neutrons stars},}\ }\href@noop {} {\bibfield
  {journal} {\bibinfo  {journal} {Nuc. Phys. A}\ }\textbf {\bibinfo {volume}
  {914}},\ \bibinfo {pages} {367--376}}\BibitemShut {NoStop}%
\bibitem [{\citenamefont {Volosniev}\ \emph {et~al.}(2014)\citenamefont
  {Volosniev}, \citenamefont {Fedorov}, \citenamefont {Jensen},\ and\
  \citenamefont {Zinner}}]{Volosniev2014JPB}%
  \BibitemOpen
  \bibfield  {author} {\bibinfo {author} {\bibnamefont {Volosniev},
  \bibfnamefont {A~G}}, \bibinfo {author} {\bibfnamefont {D~V}\ \bibnamefont
  {Fedorov}}, \bibinfo {author} {\bibfnamefont {A~S}\ \bibnamefont {Jensen}}, \
  and\ \bibinfo {author} {\bibfnamefont {N~T}\ \bibnamefont {Zinner}}}
  (\bibinfo {year} {2014}),\ \bibfield  {title} {\enquote {\bibinfo {title}
  {Borromean ground state of fermions in two dimensions},}\ }\href
  {http://stacks.iop.org/0953-4075/47/i=18/a=185302} {\bibfield  {journal}
  {\bibinfo  {journal} {Journal of Physics B: Atomic, Molecular and Optical
  Physics}\ }\textbf {\bibinfo {volume} {47}}~(\bibinfo {number} {18}),\
  \bibinfo {pages} {185302}}\BibitemShut {NoStop}%
\bibitem [{\citenamefont {Wacker}\ \emph {et~al.}(2016)\citenamefont {Wacker},
  \citenamefont {J\o{}rgensen}, \citenamefont {Birkmose}, \citenamefont
  {Winter}, \citenamefont {Mikkelsen}, \citenamefont {Sherson}, \citenamefont
  {Zinner},\ and\ \citenamefont {Arlt}}]{Arlt2016}%
  \BibitemOpen
  \bibfield  {author} {\bibinfo {author} {\bibnamefont {Wacker}, \bibfnamefont
  {L~J}}, \bibinfo {author} {\bibfnamefont {N.~B.}\ \bibnamefont
  {J\o{}rgensen}}, \bibinfo {author} {\bibfnamefont {D.}~\bibnamefont
  {Birkmose}}, \bibinfo {author} {\bibfnamefont {N.}~\bibnamefont {Winter}},
  \bibinfo {author} {\bibfnamefont {M.}~\bibnamefont {Mikkelsen}}, \bibinfo
  {author} {\bibfnamefont {J.}~\bibnamefont {Sherson}}, \bibinfo {author}
  {\bibfnamefont {N.}~\bibnamefont {Zinner}}, \ and\ \bibinfo {author}
  {\bibfnamefont {J.~J.}\ \bibnamefont {Arlt}}} (\bibinfo {year} {2016}),\
  \bibfield  {title} {\enquote {\bibinfo {title} {Universal three-body physics
  in ultracold {KR}b mixtures},}\ }\href {\doibase
  10.1103/PhysRevLett.117.163201} {\bibfield  {journal} {\bibinfo  {journal}
  {Phys. Rev. Lett.}\ }\textbf {\bibinfo {volume} {117}},\ \bibinfo {pages}
  {163201}}\BibitemShut {NoStop}%
\bibitem [{\citenamefont {{Wang}}\ \emph {et~al.}(2016)\citenamefont {{Wang}},
  \citenamefont {{Ye}}, \citenamefont {{Guo}}, \citenamefont {{Blume}},\ and\
  \citenamefont {{Wang}}}]{WangBlumeWang2016}%
  \BibitemOpen
  \bibfield  {author} {\bibinfo {author} {\bibnamefont {{Wang}}, \bibfnamefont
  {F}}, \bibinfo {author} {\bibfnamefont {X.}~\bibnamefont {{Ye}}}, \bibinfo
  {author} {\bibfnamefont {M.}~\bibnamefont {{Guo}}}, \bibinfo {author}
  {\bibfnamefont {D.}~\bibnamefont {{Blume}}}, \ and\ \bibinfo {author}
  {\bibfnamefont {D.}~\bibnamefont {{Wang}}}} (\bibinfo {year} {2016}),\
  \bibfield  {title} {\enquote {\bibinfo {title} {{Exploring Few-Body Processes
  with an Ultracold Light-Heavy Bose-Bose Mixture}},}\ }\href@noop {}
  {\bibfield  {journal} {\bibinfo  {journal} {ArXiv e-prints}\ }}\Eprint
  {http://arxiv.org/abs/1611.03198} {arXiv:1611.03198 [cond-mat.quant-gas]}
  \BibitemShut {NoStop}%
\bibitem [{\citenamefont {Wang}\ \emph
  {et~al.}(2012{\natexlab{a}})\citenamefont {Wang}, \citenamefont {D'Incao},
  \citenamefont {Esry},\ and\ \citenamefont {Greene}}]{wang2012PRL}%
  \BibitemOpen
  \bibfield  {author} {\bibinfo {author} {\bibnamefont {Wang}, \bibfnamefont
  {J}}, \bibinfo {author} {\bibfnamefont {J.~P.}\ \bibnamefont {D'Incao}},
  \bibinfo {author} {\bibfnamefont {B.~D.}\ \bibnamefont {Esry}}, \ and\
  \bibinfo {author} {\bibfnamefont {C.~H.}\ \bibnamefont {Greene}}} (\bibinfo
  {year} {2012}{\natexlab{a}}),\ \bibfield  {title} {\enquote {\bibinfo {title}
  {Origin of the three-body parameter universality in {E}fimov physics},}\
  }\href@noop {} {\bibfield  {journal} {\bibinfo  {journal} {Phys. Rev. Lett.}\
  }\textbf {\bibinfo {volume} {108}},\ \bibinfo {pages} {263001}}\BibitemShut
  {NoStop}%
\bibitem [{\citenamefont {Wang}\ \emph
  {et~al.}(2011{\natexlab{a}})\citenamefont {Wang}, \citenamefont {D'Incao},\
  and\ \citenamefont {Greene}}]{wang2011PRA}%
  \BibitemOpen
  \bibfield  {author} {\bibinfo {author} {\bibnamefont {Wang}, \bibfnamefont
  {J}}, \bibinfo {author} {\bibfnamefont {J.~P.}\ \bibnamefont {D'Incao}}, \
  and\ \bibinfo {author} {\bibfnamefont {C.~H.}\ \bibnamefont {Greene}}}
  (\bibinfo {year} {2011}{\natexlab{a}}),\ \bibfield  {title} {{\selectlanguage
  {English}\enquote {\bibinfo {title} {Numerical study of three-body
  recombination for systems with many bound states},}\ }}\href@noop {}
  {\bibfield  {journal} {\bibinfo  {journal} {Phys. Rev. A}\ }\textbf {\bibinfo
  {volume} {84}}~(\bibinfo {number} {5}),\ \bibinfo {pages}
  {052721}}\BibitemShut {NoStop}%
\bibitem [{\citenamefont {Wang}\ \emph
  {et~al.}(2012{\natexlab{b}})\citenamefont {Wang}, \citenamefont {D\'{}Incao},
  \citenamefont {Wang},\ and\ \citenamefont {Greene}}]{Wang-2012}%
  \BibitemOpen
  \bibfield  {author} {\bibinfo {author} {\bibnamefont {Wang}, \bibfnamefont
  {J}}, \bibinfo {author} {\bibfnamefont {J.~P.}\ \bibnamefont {D\'{}Incao}},
  \bibinfo {author} {\bibfnamefont {Y.}~\bibnamefont {Wang}}, \ and\ \bibinfo
  {author} {\bibfnamefont {C.~H.}\ \bibnamefont {Greene}}} (\bibinfo {year}
  {2012}{\natexlab{b}}),\ \bibfield  {title} {\enquote {\bibinfo {title}
  {Universal three-body recombination via resonant $d$-wave interactions},}\
  }\href@noop {} {\bibfield  {journal} {\bibinfo  {journal} {Phys. Rev. A}\
  }\textbf {\bibinfo {volume} {86}},\ \bibinfo {pages} {062511}}\BibitemShut
  {NoStop}%
\bibitem [{\citenamefont {Wang}\ and\ \citenamefont {Greene}(2015)}]{susocpra}%
  \BibitemOpen
  \bibfield  {author} {\bibinfo {author} {\bibnamefont {Wang}, \bibfnamefont
  {S-J}}, \ and\ \bibinfo {author} {\bibfnamefont {C.~H.}\ \bibnamefont
  {Greene}}} (\bibinfo {year} {2015}),\ \bibfield  {title} {\enquote {\bibinfo
  {title} {General formalism for ultracold scattering with isotropic spin-orbit
  coupling},}\ }\href@noop {} {\bibfield  {journal} {\bibinfo  {journal} {Phys.
  Rev. A}\ }\textbf {\bibinfo {volume} {91}},\ \bibinfo {pages}
  {022706}}\BibitemShut {NoStop}%
\bibitem [{\citenamefont {Wang}\ \emph {et~al.}(2013)\citenamefont {Wang},
  \citenamefont {D'Incao},\ and\ \citenamefont {Esry}}]{wang2013amop}%
  \BibitemOpen
  \bibfield  {author} {\bibinfo {author} {\bibnamefont {Wang}, \bibfnamefont
  {Y}}, \bibinfo {author} {\bibfnamefont {J.~P.}\ \bibnamefont {D'Incao}}, \
  and\ \bibinfo {author} {\bibfnamefont {B.~D.}\ \bibnamefont {Esry}}}
  (\bibinfo {year} {2013}),\ \bibfield  {title} {\enquote {\bibinfo {title}
  {{U}ltracold few-body systems},}\ }\href@noop {} {\bibfield  {journal}
  {\bibinfo  {journal} {Adv. At. Mol. Opt. Phys.}\ }\textbf {\bibinfo {volume}
  {62}},\ \bibinfo {pages} {1 -- 115}}\BibitemShut {NoStop}%
\bibitem [{\citenamefont {Wang}\ \emph
  {et~al.}(2011{\natexlab{b}})\citenamefont {Wang}, \citenamefont {D'Incao},\
  and\ \citenamefont {Greene}}]{wang2011PRL}%
  \BibitemOpen
  \bibfield  {author} {\bibinfo {author} {\bibnamefont {Wang}, \bibfnamefont
  {Y}}, \bibinfo {author} {\bibfnamefont {J.~P.}\ \bibnamefont {D'Incao}}, \
  and\ \bibinfo {author} {\bibfnamefont {C.~H.}\ \bibnamefont {Greene}}}
  (\bibinfo {year} {2011}{\natexlab{b}}),\ \bibfield  {title} {\enquote
  {\bibinfo {title} {{E}fimov effect for three interacting bosonic dipoles},}\
  }\href@noop {} {\bibfield  {journal} {\bibinfo  {journal} {Phys. Rev. Lett.}\
  }\textbf {\bibinfo {volume} {106}},\ \bibinfo {pages} {233201}}\BibitemShut
  {NoStop}%
\bibitem [{\citenamefont {Wang}\ \emph
  {et~al.}(2011{\natexlab{c}})\citenamefont {Wang}, \citenamefont {D'Incao},\
  and\ \citenamefont {Greene}}]{wang2011PRLb}%
  \BibitemOpen
  \bibfield  {author} {\bibinfo {author} {\bibnamefont {Wang}, \bibfnamefont
  {Y}}, \bibinfo {author} {\bibfnamefont {J.~P.}\ \bibnamefont {D'Incao}}, \
  and\ \bibinfo {author} {\bibfnamefont {C.~H.}\ \bibnamefont {Greene}}}
  (\bibinfo {year} {2011}{\natexlab{c}}),\ \bibfield  {title} {\enquote
  {\bibinfo {title} {Universal three-body physics for {F}ermionic dipoles},}\
  }\href@noop {} {\bibfield  {journal} {\bibinfo  {journal} {Phys. Rev. Lett.}\
  }\textbf {\bibinfo {volume} {107}},\ \bibinfo {pages} {233201}}\BibitemShut
  {NoStop}%
\bibitem [{\citenamefont {Wang}\ and\ \citenamefont
  {Esry}(2009)}]{wang2009PRL}%
  \BibitemOpen
  \bibfield  {author} {\bibinfo {author} {\bibnamefont {Wang}, \bibfnamefont
  {Y}}, \ and\ \bibinfo {author} {\bibfnamefont {B.~D.}\ \bibnamefont {Esry}}}
  (\bibinfo {year} {2009}),\ \bibfield  {title} {{\selectlanguage
  {English}\enquote {\bibinfo {title} {{E}fimov trimer formation via ultracold
  four-body recombination},}\ }}\href@noop {} {\bibfield  {journal} {\bibinfo
  {journal} {Phys. Rev. Lett.}\ }\textbf {\bibinfo {volume} {102}}~(\bibinfo
  {number} {13}),\ \bibinfo {pages} {133201}}\BibitemShut {NoStop}%
\bibitem [{\citenamefont {Wang}\ and\ \citenamefont
  {Julienne}(2014)}]{wang2014natphys}%
  \BibitemOpen
  \bibfield  {author} {\bibinfo {author} {\bibnamefont {Wang}, \bibfnamefont
  {Y}}, \ and\ \bibinfo {author} {\bibfnamefont {P.~S.}\ \bibnamefont
  {Julienne}}} (\bibinfo {year} {2014}),\ \bibfield  {title} {\enquote
  {\bibinfo {title} {Universal van der waals physics for three cold atoms near
  feshbach resonances},}\ }\href@noop {} {\bibfield  {journal} {\bibinfo
  {journal} {Nat. Phys.}\ }\textbf {\bibinfo {volume} {10}}~(\bibinfo {number}
  {10}),\ \bibinfo {pages} {768--773}}\BibitemShut {NoStop}%
\bibitem [{\citenamefont {Wang}\ \emph
  {et~al.}(2015{\natexlab{a}})\citenamefont {Wang}, \citenamefont {Julienne},\
  and\ \citenamefont {Greene}}]{wang2015AnnRev}%
  \BibitemOpen
  \bibfield  {author} {\bibinfo {author} {\bibnamefont {Wang}, \bibfnamefont
  {Y}}, \bibinfo {author} {\bibfnamefont {P.~S.}\ \bibnamefont {Julienne}}, \
  and\ \bibinfo {author} {\bibfnamefont {C.~H.}\ \bibnamefont {Greene}}}
  (\bibinfo {year} {2015}{\natexlab{a}}),\ \enquote {\bibinfo {title} {Few-body
  physics of ultracold atoms and molecules with long-range interactions},}\ in\
  \href@noop {} {\emph {\bibinfo {booktitle} {Annual Review of Cold Atoms and
  Molecules}}},\ Chap.~\bibinfo {chapter} {2}\ (\bibinfo  {publisher} {World
  Scientific})\ pp.\ \bibinfo {pages} {77--134}\BibitemShut {NoStop}%
\bibitem [{\citenamefont {Wang}\ \emph
  {et~al.}(2012{\natexlab{c}})\citenamefont {Wang}, \citenamefont {Laing},
  \citenamefont {von Stecher},\ and\ \citenamefont {Esry}}]{wang2012PRLc}%
  \BibitemOpen
  \bibfield  {author} {\bibinfo {author} {\bibnamefont {Wang}, \bibfnamefont
  {Y}}, \bibinfo {author} {\bibfnamefont {W.~B.}\ \bibnamefont {Laing}},
  \bibinfo {author} {\bibfnamefont {J.}~\bibnamefont {von Stecher}}, \ and\
  \bibinfo {author} {\bibfnamefont {B.~D.}\ \bibnamefont {Esry}}} (\bibinfo
  {year} {2012}{\natexlab{c}}),\ \bibfield  {title} {\enquote {\bibinfo {title}
  {Efimov physics in heteronuclear four-body systems},}\ }\href@noop {}
  {\bibfield  {journal} {\bibinfo  {journal} {Phys. Rev. Lett.}\ }\textbf
  {\bibinfo {volume} {108}},\ \bibinfo {pages} {073201}}\BibitemShut {NoStop}%
\bibitem [{\citenamefont {Wang}\ \emph
  {et~al.}(2012{\natexlab{d}})\citenamefont {Wang}, \citenamefont {Wang},
  \citenamefont {D'Incao},\ and\ \citenamefont {Greene}}]{Wang-2012b}%
  \BibitemOpen
  \bibfield  {author} {\bibinfo {author} {\bibnamefont {Wang}, \bibfnamefont
  {Y}}, \bibinfo {author} {\bibfnamefont {J.}~\bibnamefont {Wang}}, \bibinfo
  {author} {\bibfnamefont {J.~P.}\ \bibnamefont {D'Incao}}, \ and\ \bibinfo
  {author} {\bibfnamefont {C.~H.}\ \bibnamefont {Greene}}} (\bibinfo {year}
  {2012}{\natexlab{d}}),\ \bibfield  {title} {\enquote {\bibinfo {title}
  {Universal three-body parameter in heteronuclear atomic systems},}\
  }\href@noop {} {\bibfield  {journal} {\bibinfo  {journal} {Phys. Rev. Lett.}\
  }\textbf {\bibinfo {volume} {109}},\ \bibinfo {pages} {243201}}\BibitemShut
  {NoStop}%
\bibitem [{\citenamefont {Wang}\ \emph
  {et~al.}(2015{\natexlab{b}})\citenamefont {Wang}, \citenamefont {Wang},
  \citenamefont {D'Incao},\ and\ \citenamefont {Greene}}]{Wang-2015erratum}%
  \BibitemOpen
  \bibfield  {author} {\bibinfo {author} {\bibnamefont {Wang}, \bibfnamefont
  {Y}}, \bibinfo {author} {\bibfnamefont {J.}~\bibnamefont {Wang}}, \bibinfo
  {author} {\bibfnamefont {J.~P.}\ \bibnamefont {D'Incao}}, \ and\ \bibinfo
  {author} {\bibfnamefont {C.~H.}\ \bibnamefont {Greene}}} (\bibinfo {year}
  {2015}{\natexlab{b}}),\ \bibfield  {title} {\enquote {\bibinfo {title}
  {Erratum: Universal three-body parameter in heteronuclear atomic systems
  [{P}hys. {R}ev. {L}ett. \textbf{109} , 243201 (2012)]},}\ }\href@noop {}
  {\bibfield  {journal} {\bibinfo  {journal} {Phys. Rev. Lett.}\ }\textbf
  {\bibinfo {volume} {115}},\ \bibinfo {pages} {069901}}\BibitemShut {NoStop}%
\bibitem [{\citenamefont {Wang-Chang}\ \emph {et~al.}(1964)\citenamefont
  {Wang-Chang}, \citenamefont {Uhlenbeck},\ and\ \citenamefont
  {deBoer}}]{Wang-Chang}%
  \BibitemOpen
  \bibfield  {author} {\bibinfo {author} {\bibnamefont {Wang-Chang},
  \bibfnamefont {C~S}}, \bibinfo {author} {\bibfnamefont {G.~E.}\ \bibnamefont
  {Uhlenbeck}}, \ and\ \bibinfo {author} {\bibfnamefont {J.}~\bibnamefont
  {deBoer}}} (\bibinfo {year} {1964}),\ \href@noop {} {\emph {\bibinfo {title}
  {Studies in Statistical Mechanics}}},\ edited by\ \bibinfo {editor}
  {\bibfnamefont {J.}~\bibnamefont {deBoer}}\ and\ \bibinfo {editor}
  {\bibfnamefont {G.~E.}\ \bibnamefont {Uhlenbeck}},\ Vol.~\bibinfo {volume}
  {2}\ (\bibinfo  {publisher} {North-Holland},\ \bibinfo {address}
  {Amsterdam})\BibitemShut {NoStop}%
\bibitem [{\citenamefont {Wannier}(1953)}]{wannier1953pr}%
  \BibitemOpen
  \bibfield  {author} {\bibinfo {author} {\bibnamefont {Wannier}, \bibfnamefont
  {G~H}}} (\bibinfo {year} {1953}),\ \bibfield  {title} {\enquote {\bibinfo
  {title} {The threshold law for single ionization of atoms or ions by
  electrons},}\ }\href@noop {} {\bibfield  {journal} {\bibinfo  {journal}
  {Phys. Rev.}\ }\textbf {\bibinfo {volume} {90}},\ \bibinfo {pages}
  {817--825}}\BibitemShut {NoStop}%
\bibitem [{\citenamefont {Watanabe}(1982)}]{watanabe1982PRA}%
  \BibitemOpen
  \bibfield  {author} {\bibinfo {author} {\bibnamefont {Watanabe},
  \bibfnamefont {S}}} (\bibinfo {year} {1982}),\ \bibfield  {title} {\enquote
  {\bibinfo {title} {Doubly excited states of the helium negative ion},}\
  }\href@noop {} {\bibfield  {journal} {\bibinfo  {journal} {Phys. Rev. A}\
  }\textbf {\bibinfo {volume} {25}},\ \bibinfo {pages}
  {2074--2098}}\BibitemShut {NoStop}%
\bibitem [{\citenamefont {Watanabe}(1991)}]{watanabe1991JPB}%
  \BibitemOpen
  \bibfield  {author} {\bibinfo {author} {\bibnamefont {Watanabe},
  \bibfnamefont {S}}} (\bibinfo {year} {1991}),\ \bibfield  {title}
  {{\selectlanguage {English}\enquote {\bibinfo {title} {The adiabatic
  expansion and {W}annier's ionization threshold law},}\ }}\href@noop {}
  {\bibfield  {journal} {\bibinfo  {journal} {J. Phys. B}\ }\textbf {\bibinfo
  {volume} {24}}~(\bibinfo {number} {3}),\ \bibinfo {pages}
  {L39--L44}}\BibitemShut {NoStop}%
\bibitem [{\citenamefont {Watanabe}\ and\ \citenamefont
  {Komine}(1989)}]{WatanabeEfimov1989}%
  \BibitemOpen
  \bibfield  {author} {\bibinfo {author} {\bibnamefont {Watanabe},
  \bibfnamefont {S}}, \ and\ \bibinfo {author} {\bibfnamefont {K.}~\bibnamefont
  {Komine}}} (\bibinfo {year} {1989}),\ \bibfield  {title} {\enquote {\bibinfo
  {title} {Efimov states of the helium trimer using the hyperspherical
  method},}\ }\href@noop {} {\bibfield  {journal} {\bibinfo  {journal} {{Atomic
  Collision Research in Japan - Progress Report}}\ }\textbf {\bibinfo {volume}
  {15}},\ \bibinfo {pages} {60--62}}\BibitemShut {NoStop}%
\bibitem [{\citenamefont {Watson}\ and\ \citenamefont
  {McKinney}(1999)}]{Watson1999pra}%
  \BibitemOpen
  \bibfield  {author} {\bibinfo {author} {\bibnamefont {Watson}, \bibfnamefont
  {D~K}}, \ and\ \bibinfo {author} {\bibfnamefont {B.~A.}\ \bibnamefont
  {McKinney}}} (\bibinfo {year} {1999}),\ \bibfield  {title} {\enquote
  {\bibinfo {title} {Improved large-{N} limit for {B}ose-{E}instein condensates
  from perturbation theory},}\ }\href@noop {} {\bibfield  {journal} {\bibinfo
  {journal} {Phys. Rev. A}\ }\textbf {\bibinfo {volume} {59}},\ \bibinfo
  {pages} {4091--4094}}\BibitemShut {NoStop}%
\bibitem [{\citenamefont {Weber}\ \emph {et~al.}(2007)\citenamefont {Weber},
  \citenamefont {Negreiros}, \citenamefont {Rosenfield},\ and\ \citenamefont
  {Cuadrad}}]{Weber-2007}%
  \BibitemOpen
  \bibfield  {author} {\bibinfo {author} {\bibnamefont {Weber}, \bibfnamefont
  {F}}, \bibinfo {author} {\bibfnamefont {R.}~\bibnamefont {Negreiros}},
  \bibinfo {author} {\bibfnamefont {P.}~\bibnamefont {Rosenfield}}, \ and\
  \bibinfo {author} {\bibfnamefont {A.~T.~I.}\ \bibnamefont {Cuadrad}}}
  (\bibinfo {year} {2007}),\ \bibfield  {title} {\enquote {\bibinfo {title}
  {Neutron star interiors and the equation of state of ultradense matter},}\
  }\href@noop {} {\bibfield  {journal} {\bibinfo  {journal} {AIP Conf. Proc.}\
  }\textbf {\bibinfo {volume} {892}},\ \bibinfo {pages} {515--517}}\BibitemShut
  {NoStop}%
\bibitem [{\citenamefont {Weber}\ \emph {et~al.}(2003)\citenamefont {Weber},
  \citenamefont {Herbig}, \citenamefont {Mark}, \citenamefont {N\"{a}gerl},\
  and\ \citenamefont {Grimm}}]{weber2003PRL}%
  \BibitemOpen
  \bibfield  {author} {\bibinfo {author} {\bibnamefont {Weber}, \bibfnamefont
  {T}}, \bibinfo {author} {\bibfnamefont {J.}~\bibnamefont {Herbig}}, \bibinfo
  {author} {\bibfnamefont {M.}~\bibnamefont {Mark}}, \bibinfo {author}
  {\bibfnamefont {H.{-}C.}\ \bibnamefont {N\"{a}gerl}}, \ and\ \bibinfo
  {author} {\bibfnamefont {R.}~\bibnamefont {Grimm}}} (\bibinfo {year}
  {2003}),\ \bibfield  {title} {{\selectlanguage {English}\enquote {\bibinfo
  {title} {Three-body recombination at large scattering lengths in an ultracold
  atomic gas},}\ }}\href@noop {} {\bibfield  {journal} {\bibinfo  {journal}
  {Phys. Rev. Lett.}\ }\textbf {\bibinfo {volume} {91}}~(\bibinfo {number}
  {12}),\ \bibinfo {pages} {123201}}\BibitemShut {NoStop}%
\bibitem [{\citenamefont {Weinberg}(1979)}]{Weinberg-1979}%
  \BibitemOpen
  \bibfield  {author} {\bibinfo {author} {\bibnamefont {Weinberg},
  \bibfnamefont {S}}} (\bibinfo {year} {1979}),\ \bibfield  {title} {\enquote
  {\bibinfo {title} {Phenomenological {L}agrangians},}\ }\href@noop {}
  {\bibfield  {journal} {\bibinfo  {journal} {Physica A}\ }\textbf {\bibinfo
  {volume} {96}},\ \bibinfo {pages} {327}}\BibitemShut {NoStop}%
\bibitem [{\citenamefont {Weinberg}(1990)}]{Weinberg-1990}%
  \BibitemOpen
  \bibfield  {author} {\bibinfo {author} {\bibnamefont {Weinberg},
  \bibfnamefont {S}}} (\bibinfo {year} {1990}),\ \bibfield  {title} {\enquote
  {\bibinfo {title} {Nuclear forces from chiral {L}agrangians},}\ }\href@noop
  {} {\bibfield  {journal} {\bibinfo  {journal} {Phys. Lett. B}\ }\textbf
  {\bibinfo {volume} {251}},\ \bibinfo {pages} {288}}\BibitemShut {NoStop}%
\bibitem [{\citenamefont {Weinberg}(1991)}]{Weinberg-1991}%
  \BibitemOpen
  \bibfield  {author} {\bibinfo {author} {\bibnamefont {Weinberg},
  \bibfnamefont {S}}} (\bibinfo {year} {1991}),\ \bibfield  {title} {\enquote
  {\bibinfo {title} {Effective chiral {L}agrangians for nulceon-pion
  interactions and nuclear forces},}\ }\href@noop {} {\bibfield  {journal}
  {\bibinfo  {journal} {Nuc. Phys. B}\ }\textbf {\bibinfo {volume} {363}},\
  \bibinfo {pages} {3}}\BibitemShut {NoStop}%
\bibitem [{\citenamefont {Wenz}\ \emph {et~al.}(2009)\citenamefont {Wenz},
  \citenamefont {Lompe}, \citenamefont {Ottenstein}, \citenamefont {Serwane},
  \citenamefont {Z\"urn},\ and\ \citenamefont {Jochim}}]{wenz2009PRA}%
  \BibitemOpen
  \bibfield  {author} {\bibinfo {author} {\bibnamefont {Wenz}, \bibfnamefont
  {A~N}}, \bibinfo {author} {\bibfnamefont {T.}~\bibnamefont {Lompe}}, \bibinfo
  {author} {\bibfnamefont {T.~B.}\ \bibnamefont {Ottenstein}}, \bibinfo
  {author} {\bibfnamefont {F.}~\bibnamefont {Serwane}}, \bibinfo {author}
  {\bibnamefont {Z\"urn}}, \ and\ \bibinfo {author} {\bibfnamefont
  {S.}~\bibnamefont {Jochim}}} (\bibinfo {year} {2009}),\ \bibfield  {title}
  {{\selectlanguage {English}\enquote {\bibinfo {title} {Universal trimer in a
  three-component {F}ermi gas},}\ }}\href@noop {} {\bibfield  {journal}
  {\bibinfo  {journal} {Phys. Rev. A}\ }\textbf {\bibinfo {volume}
  {80}}~(\bibinfo {number} {4}),\ \bibinfo {pages} {040702}}\BibitemShut
  {NoStop}%
\bibitem [{\citenamefont {Werner}\ \emph {et~al.}({2009})\citenamefont
  {Werner}, \citenamefont {Tarruell},\ and\ \citenamefont
  {Castin}}]{WernerTarruellCastin2009epjb}%
  \BibitemOpen
  \bibfield  {author} {\bibinfo {author} {\bibnamefont {Werner}, \bibfnamefont
  {F}}, \bibinfo {author} {\bibfnamefont {L.}~\bibnamefont {Tarruell}}, \ and\
  \bibinfo {author} {\bibfnamefont {Y.}~\bibnamefont {Castin}}} (\bibinfo
  {year} {{2009}}),\ \bibfield  {title} {\enquote {\bibinfo {title} {{Number of
  closed-channel molecules in the BEC-BCS crossover}},}\ }\href {\doibase
  {10.1140/epjb/e2009-00040-8}} {\bibfield  {journal} {\bibinfo  {journal}
  {{European Physical Journal B}}\ }\textbf {\bibinfo {volume}
  {{68}}}~(\bibinfo {number} {{3}}),\ \bibinfo {pages}
  {{401--415}}}\BibitemShut {NoStop}%
\bibitem [{\citenamefont {Werner}\ and\ \citenamefont
  {Castin}(2006)}]{werner2006PRL}%
  \BibitemOpen
  \bibfield  {author} {\bibinfo {author} {\bibnamefont {Werner}, \bibfnamefont
  {F\'elix}}, \ and\ \bibinfo {author} {\bibfnamefont {Yvan}\ \bibnamefont
  {Castin}}} (\bibinfo {year} {2006}),\ \bibfield  {title} {\enquote {\bibinfo
  {title} {Unitary quantum three-body problem in a harmonic trap},}\
  }\href@noop {} {\bibfield  {journal} {\bibinfo  {journal} {Phys. Rev. Lett.}\
  }\textbf {\bibinfo {volume} {97}},\ \bibinfo {pages} {150401}}\BibitemShut
  {NoStop}%
\bibitem [{\citenamefont {Whittaker}(1937)}]{Whittaker-1937}%
  \BibitemOpen
  \bibfield  {author} {\bibinfo {author} {\bibnamefont {Whittaker},
  \bibfnamefont {E~T}}} (\bibinfo {year} {1937}),\ \href@noop {} {\emph
  {\bibinfo {title} {A Trataise on the Analytical Dynamics of Particles and
  Rigid Bodies}}}\ (\bibinfo  {publisher} {Cambridge University Press},\
  \bibinfo {address} {Cambridge, England})\BibitemShut {NoStop}%
\bibitem [{\citenamefont {Whitten}\ and\ \citenamefont
  {Smith}(1968)}]{whitten1968JMP}%
  \BibitemOpen
  \bibfield  {author} {\bibinfo {author} {\bibnamefont {Whitten}, \bibfnamefont
  {R~C}}, \ and\ \bibinfo {author} {\bibfnamefont {F.~T.}\ \bibnamefont
  {Smith}}} (\bibinfo {year} {1968}),\ \bibfield  {title} {\enquote {\bibinfo
  {title} {Symmetric representation for three-body problems. {II}. motion in
  space},}\ }\href@noop {} {\bibfield  {journal} {\bibinfo  {journal} {J. Math.
  Phys.}\ }\textbf {\bibinfo {volume} {9}}~(\bibinfo {number} {7}),\ \bibinfo
  {pages} {1103--1113}}\BibitemShut {NoStop}%
\bibitem [{\citenamefont {Wigner}(1937)}]{Wigner}%
  \BibitemOpen
  \bibfield  {author} {\bibinfo {author} {\bibnamefont {Wigner}, \bibfnamefont
  {E~P}}} (\bibinfo {year} {1937}),\ \bibfield  {title} {\enquote {\bibinfo
  {title} {Calculation of the rate of elementary association reactions},}\
  }\href@noop {} {\bibfield  {journal} {\bibinfo  {journal} {J. Chem. Phys}\
  }\textbf {\bibinfo {volume} {5}},\ \bibinfo {pages} {720}}\BibitemShut
  {NoStop}%
\bibitem [{\citenamefont {Wild}\ \emph {et~al.}(2012)\citenamefont {Wild},
  \citenamefont {Makotyn}, \citenamefont {Pino}, \citenamefont {Cornell},\ and\
  \citenamefont {Jin}}]{wild2012PRL}%
  \BibitemOpen
  \bibfield  {author} {\bibinfo {author} {\bibnamefont {Wild}, \bibfnamefont
  {R~J}}, \bibinfo {author} {\bibfnamefont {P.}~\bibnamefont {Makotyn}},
  \bibinfo {author} {\bibfnamefont {J.~M.}\ \bibnamefont {Pino}}, \bibinfo
  {author} {\bibfnamefont {E.~A.}\ \bibnamefont {Cornell}}, \ and\ \bibinfo
  {author} {\bibfnamefont {D.~S.}\ \bibnamefont {Jin}}} (\bibinfo {year}
  {2012}),\ \bibfield  {title} {\enquote {\bibinfo {title} {Measurements of
  {T}an's contact in an atomic {B}ose-{E}instein condensate},}\ }\href@noop {}
  {\bibfield  {journal} {\bibinfo  {journal} {Phys. Rev. Lett.}\ }\textbf
  {\bibinfo {volume} {108}},\ \bibinfo {pages} {145305}}\BibitemShut {NoStop}%
\bibitem [{\citenamefont {Williams}\ \emph {et~al.}(2009)\citenamefont
  {Williams}, \citenamefont {Hazlett}, \citenamefont {Huckans}, \citenamefont
  {Stites}, \citenamefont {Zhang},\ and\ \citenamefont
  {O'Hara}}]{Williams-2009}%
  \BibitemOpen
  \bibfield  {author} {\bibinfo {author} {\bibnamefont {Williams},
  \bibfnamefont {J~R}}, \bibinfo {author} {\bibfnamefont {E.~L.}\ \bibnamefont
  {Hazlett}}, \bibinfo {author} {\bibfnamefont {J.~H.}\ \bibnamefont
  {Huckans}}, \bibinfo {author} {\bibfnamefont {R.~W.}\ \bibnamefont {Stites}},
  \bibinfo {author} {\bibfnamefont {Y.}~\bibnamefont {Zhang}}, \ and\ \bibinfo
  {author} {\bibfnamefont {K.~M.}\ \bibnamefont {O'Hara}}} (\bibinfo {year}
  {2009}),\ \bibfield  {title} {\enquote {\bibinfo {title} {Evidence for and
  excited-state {E}fimov trimer in a three-component fermi gas},}\ }\href@noop
  {} {\bibfield  {journal} {\bibinfo  {journal} {Phys. Rev. Lett.}\ }\textbf
  {\bibinfo {volume} {103}},\ \bibinfo {pages} {130404}}\BibitemShut {NoStop}%
\bibitem [{\citenamefont {Williams}\ \emph {et~al.}(2012)\citenamefont
  {Williams}, \citenamefont {LeBlanc}, \citenamefont {Jim{\'e}nez-Garc{\'\i}a},
  \citenamefont {Beeler}, \citenamefont {Perry}, \citenamefont {Phillips},\
  and\ \citenamefont {Spielman}}]{williams2012science}%
  \BibitemOpen
  \bibfield  {author} {\bibinfo {author} {\bibnamefont {Williams},
  \bibfnamefont {R~A}}, \bibinfo {author} {\bibfnamefont {L.~J.}\ \bibnamefont
  {LeBlanc}}, \bibinfo {author} {\bibfnamefont {K.}~\bibnamefont
  {Jim{\'e}nez-Garc{\'\i}a}}, \bibinfo {author} {\bibfnamefont {M.~C.}\
  \bibnamefont {Beeler}}, \bibinfo {author} {\bibfnamefont {A.~R.}\
  \bibnamefont {Perry}}, \bibinfo {author} {\bibfnamefont {W.~D.}\ \bibnamefont
  {Phillips}}, \ and\ \bibinfo {author} {\bibfnamefont {I.~B.}\ \bibnamefont
  {Spielman}}} (\bibinfo {year} {2012}),\ \bibfield  {title} {\enquote
  {\bibinfo {title} {Synthetic partial waves in ultracold atomic collisions},}\
  }\href@noop {} {\bibfield  {journal} {\bibinfo  {journal} {Science}\ }\textbf
  {\bibinfo {volume} {335}}~(\bibinfo {number} {6066}),\ \bibinfo {pages}
  {314--317}}\BibitemShut {NoStop}%
\bibitem [{\citenamefont {Willitsch}(2012)}]{Willitsch-2012}%
  \BibitemOpen
  \bibfield  {author} {\bibinfo {author} {\bibnamefont {Willitsch},
  \bibfnamefont {S}}} (\bibinfo {year} {2012}),\ \bibfield  {title} {\enquote
  {\bibinfo {title} {Coulomb-crystallised molecular ions in traps: methods,
  applications, prospects},}\ }\href@noop {} {\bibfield  {journal} {\bibinfo
  {journal} {Int. Rev. Phys. Chem.}\ }\textbf {\bibinfo {volume} {31}},\
  \bibinfo {pages} {175}}\BibitemShut {NoStop}%
\bibitem [{\citenamefont {Willitsch}\ \emph {et~al.}(2008)\citenamefont
  {Willitsch}, \citenamefont {Bell}, \citenamefont {Gingell},\ and\
  \citenamefont {Softley}}]{Willitsch-2008}%
  \BibitemOpen
  \bibfield  {author} {\bibinfo {author} {\bibnamefont {Willitsch},
  \bibfnamefont {S}}, \bibinfo {author} {\bibfnamefont {M.}~\bibnamefont
  {Bell}}, \bibinfo {author} {\bibfnamefont {A.}~\bibnamefont {Gingell}}, \
  and\ \bibinfo {author} {\bibfnamefont {T.~P.}\ \bibnamefont {Softley}}}
  (\bibinfo {year} {2008}),\ \bibfield  {title} {\enquote {\bibinfo {title}
  {Chemical applications of laser- and sympathetically cooled ions in traps},}\
  }\href@noop {} {\bibfield  {journal} {\bibinfo  {journal} {Phys. Chem. Chem.
  Phys.}\ }\textbf {\bibinfo {volume} {10}},\ \bibinfo {pages}
  {7200}}\BibitemShut {NoStop}%
\bibitem [{\citenamefont {Wilson}(1971)}]{Wilson-1971}%
  \BibitemOpen
  \bibfield  {author} {\bibinfo {author} {\bibnamefont {Wilson}, \bibfnamefont
  {K~G}}} (\bibinfo {year} {1971}),\ \bibfield  {title} {\enquote {\bibinfo
  {title} {Renormalization group and strong interactions},}\ }\href@noop {}
  {\bibfield  {journal} {\bibinfo  {journal} {Phys. Rev. D}\ }\textbf {\bibinfo
  {volume} {3}},\ \bibinfo {pages} {1818}}\BibitemShut {NoStop}%
\bibitem [{\citenamefont {Wilson}(1983)}]{Wilson-review}%
  \BibitemOpen
  \bibfield  {author} {\bibinfo {author} {\bibnamefont {Wilson}, \bibfnamefont
  {K~G}}} (\bibinfo {year} {1983}),\ \bibfield  {title} {\enquote {\bibinfo
  {title} {The renormalization group and critical phenomena},}\ }\href@noop {}
  {\bibfield  {journal} {\bibinfo  {journal} {Rev. Mod. Phys.}\ }\textbf
  {\bibinfo {volume} {55}},\ \bibinfo {pages} {583}}\BibitemShut {NoStop}%
\bibitem [{\citenamefont {Wilson}\ and\ \citenamefont
  {Kogut}(1974)}]{Wilson-1974}%
  \BibitemOpen
  \bibfield  {author} {\bibinfo {author} {\bibnamefont {Wilson}, \bibfnamefont
  {K~G}}, \ and\ \bibinfo {author} {\bibfnamefont {J.}~\bibnamefont {Kogut}}}
  (\bibinfo {year} {1974}),\ \bibfield  {title} {\enquote {\bibinfo {title}
  {The renormalization group and the $\epsilon$ expansion},}\ }\href@noop {}
  {\bibfield  {journal} {\bibinfo  {journal} {Phys. Rep.}\ }\textbf {\bibinfo
  {volume} {12}},\ \bibinfo {pages} {75--200}}\BibitemShut {NoStop}%
\bibitem [{\citenamefont {W\'odkiewicz}(1991)}]{Wodkiewicz1991}%
  \BibitemOpen
  \bibfield  {author} {\bibinfo {author} {\bibnamefont {W\'odkiewicz},
  \bibfnamefont {K}}} (\bibinfo {year} {1991}),\ \bibfield  {title} {\enquote
  {\bibinfo {title} {Fermi pseudopotential in arbitrary dimensions},}\ }\href
  {\doibase 10.1103/PhysRevA.43.68} {\bibfield  {journal} {\bibinfo  {journal}
  {Phys. Rev. A}\ }\textbf {\bibinfo {volume} {43}},\ \bibinfo {pages}
  {68--76}}\BibitemShut {NoStop}%
\bibitem [{\citenamefont {Wong}\ \emph {et~al.}(1988)\citenamefont {Wong},
  \citenamefont {Rau},\ and\ \citenamefont {Greene}}]{wong1988}%
  \BibitemOpen
  \bibfield  {author} {\bibinfo {author} {\bibnamefont {Wong}, \bibfnamefont
  {Hin-Yiu}}, \bibinfo {author} {\bibfnamefont {A.~R.~P.}\ \bibnamefont {Rau}},
  \ and\ \bibinfo {author} {\bibfnamefont {C.~H.}\ \bibnamefont {Greene}}}
  (\bibinfo {year} {1988}),\ \bibfield  {title} {\enquote {\bibinfo {title}
  {Negative-ion photodetachment in an electric field},}\ }\href@noop {}
  {\bibfield  {journal} {\bibinfo  {journal} {Phys. Rev. A}\ }\textbf {\bibinfo
  {volume} {37}},\ \bibinfo {pages} {2393--2403}}\BibitemShut {NoStop}%
\bibitem [{\citenamefont {Wooten}\ \emph {et~al.}(2016)\citenamefont {Wooten},
  \citenamefont {Daily},\ and\ \citenamefont {Greene}}]{Wooten2016epj}%
  \BibitemOpen
  \bibfield  {author} {\bibinfo {author} {\bibnamefont {Wooten}, \bibfnamefont
  {R~E}}, \bibinfo {author} {\bibfnamefont {K.~M.}\ \bibnamefont {Daily}}, \
  and\ \bibinfo {author} {\bibfnamefont {C.~H.}\ \bibnamefont {Greene}}}
  (\bibinfo {year} {2016}),\ \bibfield  {title} {\enquote {\bibinfo {title}
  {Few-body, hyperspherical treatment of the quantum {H}all effect},}\
  }\href@noop {} {\bibfield  {journal} {\bibinfo  {journal} {EPJ Web of
  Conferences}\ }\textbf {\bibinfo {volume} {113}},\ \bibinfo {pages}
  {02006}}\BibitemShut {NoStop}%
\bibitem [{\citenamefont {Wooten}\ \emph {et~al.}(2017)\citenamefont {Wooten},
  \citenamefont {Yan},\ and\ \citenamefont {Greene}}]{Wooten2017prb}%
  \BibitemOpen
  \bibfield  {author} {\bibinfo {author} {\bibnamefont {Wooten}, \bibfnamefont
  {R~E}}, \bibinfo {author} {\bibfnamefont {B.}~\bibnamefont {Yan}}, \ and\
  \bibinfo {author} {\bibfnamefont {Chris~H.}\ \bibnamefont {Greene}}}
  (\bibinfo {year} {2017}),\ \bibfield  {title} {\enquote {\bibinfo {title}
  {Few-body collective excitations beyond {K}ohn's theorem in quantum hall
  systems},}\ }\href {\doibase 10.1103/PhysRevB.95.035150} {\bibfield
  {journal} {\bibinfo  {journal} {Phys. Rev. B}\ }\textbf {\bibinfo {volume}
  {95}},\ \bibinfo {pages} {035150}}\BibitemShut {NoStop}%
\bibitem [{\citenamefont {Yamaguchi}(1954)}]{yamaguchi1954PR}%
  \BibitemOpen
  \bibfield  {author} {\bibinfo {author} {\bibnamefont {Yamaguchi},
  \bibfnamefont {Y}}} (\bibinfo {year} {1954}),\ \bibfield  {title} {\enquote
  {\bibinfo {title} {Two-nucleon problem when the potential is nonlocal but
  separable. {I}},}\ }\href@noop {} {\bibfield  {journal} {\bibinfo  {journal}
  {Phys. Rev.}\ }\textbf {\bibinfo {volume} {95}}~(\bibinfo {number} {6}),\
  \bibinfo {pages} {1628--1634}}\BibitemShut {NoStop}%
\bibitem [{\citenamefont {Yamashita}\ \emph {et~al.}(2015)\citenamefont
  {Yamashita}, \citenamefont {Bellotti}, \citenamefont {Frederico},
  \citenamefont {Fedorov}, \citenamefont {Jensen},\ and\ \citenamefont
  {Zinner}}]{Yamashita2015jpb}%
  \BibitemOpen
  \bibfield  {author} {\bibinfo {author} {\bibnamefont {Yamashita},
  \bibfnamefont {M~T}}, \bibinfo {author} {\bibfnamefont {F~F}\ \bibnamefont
  {Bellotti}}, \bibinfo {author} {\bibfnamefont {T}~\bibnamefont {Frederico}},
  \bibinfo {author} {\bibfnamefont {D~V}\ \bibnamefont {Fedorov}}, \bibinfo
  {author} {\bibfnamefont {A~S}\ \bibnamefont {Jensen}}, \ and\ \bibinfo
  {author} {\bibfnamefont {N~T}\ \bibnamefont {Zinner}}} (\bibinfo {year}
  {2015}),\ \bibfield  {title} {\enquote {\bibinfo {title} {Weakly bound states
  of two- and three-boson systems in the crossover from two to three
  dimensions},}\ }\href {http://stacks.iop.org/0953-4075/48/i=2/a=025302}
  {\bibfield  {journal} {\bibinfo  {journal} {Journal of Physics B: Atomic,
  Molecular and Optical Physics}\ }\textbf {\bibinfo {volume} {48}}~(\bibinfo
  {number} {2}),\ \bibinfo {pages} {025302}}\BibitemShut {NoStop}%
\bibitem [{\citenamefont {Yamashita}\ \emph {et~al.}(2010)\citenamefont
  {Yamashita}, \citenamefont {Fedorov},\ and\ \citenamefont
  {Jensen}}]{yamashita2010pra}%
  \BibitemOpen
  \bibfield  {author} {\bibinfo {author} {\bibnamefont {Yamashita},
  \bibfnamefont {M~T}}, \bibinfo {author} {\bibfnamefont {D.~V.}\ \bibnamefont
  {Fedorov}}, \ and\ \bibinfo {author} {\bibfnamefont {A.~S.}\ \bibnamefont
  {Jensen}}} (\bibinfo {year} {2010}),\ \bibfield  {title} {{\selectlanguage
  {English}\enquote {\bibinfo {title} {Universality of brunnian (n-body
  borromean) four- and five-body systems},}\ }}\href@noop {} {\bibfield
  {journal} {\bibinfo  {journal} {Phys. Rev. A}\ }\textbf {\bibinfo {volume}
  {81}}~(\bibinfo {number} {6}),\ \bibinfo {pages} {063607}}\BibitemShut
  {NoStop}%
\bibitem [{\citenamefont {Yamashita}\ \emph {et~al.}(2006)\citenamefont
  {Yamashita}, \citenamefont {Tomio}, \citenamefont {Delfino},\ and\
  \citenamefont {Frederico}}]{Yamashita2006epl}%
  \BibitemOpen
  \bibfield  {author} {\bibinfo {author} {\bibnamefont {Yamashita},
  \bibfnamefont {M~T}}, \bibinfo {author} {\bibfnamefont {Lauro}\ \bibnamefont
  {Tomio}}, \bibinfo {author} {\bibfnamefont {A.}~\bibnamefont {Delfino}}, \
  and\ \bibinfo {author} {\bibfnamefont {T.}~\bibnamefont {Frederico}}}
  (\bibinfo {year} {2006}),\ \bibfield  {title} {{\selectlanguage
  {English}\enquote {\bibinfo {title} {Four-boson scale near a {F}eshbach
  resonance},}\ }}\href@noop {} {\bibfield  {journal} {\bibinfo  {journal}
  {EPL}\ }\textbf {\bibinfo {volume} {75}}~(\bibinfo {number} {4}),\ \bibinfo
  {pages} {555--561}}\BibitemShut {NoStop}%
\bibitem [{\citenamefont {Yan}\ and\ \citenamefont
  {Blume}({2013})}]{YanBlume2013pra}%
  \BibitemOpen
  \bibfield  {author} {\bibinfo {author} {\bibnamefont {Yan}, \bibfnamefont
  {Y}}, \ and\ \bibinfo {author} {\bibfnamefont {D.}~\bibnamefont {Blume}}}
  (\bibinfo {year} {{2013}}),\ \bibfield  {title} {\enquote {\bibinfo {title}
  {Harmonically trapped {F}ermi gas: {T}emperature dependence of the {T}an
  contact},}\ }\href@noop {} {\bibfield  {journal} {\bibinfo  {journal} {{Phys.
  Rev. A}}\ }\textbf {\bibinfo {volume} {{88}}}~(\bibinfo {number}
  {{2}})}\BibitemShut {NoStop}%
\bibitem [{\citenamefont {Yan}\ and\ \citenamefont
  {Blume}({2015})}]{YanBlume2015pra}%
  \BibitemOpen
  \bibfield  {author} {\bibinfo {author} {\bibnamefont {Yan}, \bibfnamefont
  {Y}}, \ and\ \bibinfo {author} {\bibfnamefont {D.}~\bibnamefont {Blume}}}
  (\bibinfo {year} {{2015}}),\ \bibfield  {title} {\enquote {\bibinfo {title}
  {Energy and structural properties of {N}-boson clusters attached to
  three-body {E}fimov states: {T}wo-body zero-range interactions and the role
  of the three-body regulator},}\ }\href@noop {} {\bibfield  {journal}
  {\bibinfo  {journal} {{Phys. Rev. A}}\ }\textbf {\bibinfo {volume}
  {{92}}}~(\bibinfo {number} {{3}})}\BibitemShut {NoStop}%
\bibitem [{\citenamefont {Yan}\ and\ \citenamefont
  {Blume}({2016})}]{YanBlume2016prl}%
  \BibitemOpen
  \bibfield  {author} {\bibinfo {author} {\bibnamefont {Yan}, \bibfnamefont
  {Y}}, \ and\ \bibinfo {author} {\bibfnamefont {D.}~\bibnamefont {Blume}}}
  (\bibinfo {year} {{2016}}),\ \bibfield  {title} {\enquote {\bibinfo {title}
  {Path-integral {M}onte {C}arlo determination of the fourth-order {V}irial
  coefficient for a unitary two-component {F}ermi gas with zero-range
  interactions},}\ }\href@noop {} {\bibfield  {journal} {\bibinfo  {journal}
  {{Phys. Rev. Lett.}}\ }\textbf {\bibinfo {volume} {{116}}}~(\bibinfo {number}
  {{23}})}\BibitemShut {NoStop}%
\bibitem [{\citenamefont {Yin}\ and\ \citenamefont
  {Radzihovsky}({2016})}]{radzihovsky2016pra}%
  \BibitemOpen
  \bibfield  {author} {\bibinfo {author} {\bibnamefont {Yin}, \bibfnamefont
  {X}}, \ and\ \bibinfo {author} {\bibfnamefont {L.}~\bibnamefont
  {Radzihovsky}}} (\bibinfo {year} {{2016}}),\ \bibfield  {title} {\enquote
  {\bibinfo {title} {{Postquench dynamics and prethermalization in a resonant
  Bose gas}},}\ }\href@noop {} {\bibfield  {journal} {\bibinfo  {journal}
  {{Phys. Rev. A}}\ }\textbf {\bibinfo {volume} {{93}}}~(\bibinfo {number}
  {{3}})}\BibitemShut {NoStop}%
\bibitem [{\citenamefont {Yin}\ and\ \citenamefont
  {Blume}({2015})}]{YinBlume2015pra}%
  \BibitemOpen
  \bibfield  {author} {\bibinfo {author} {\bibnamefont {Yin}, \bibfnamefont
  {X~Y}}, \ and\ \bibinfo {author} {\bibfnamefont {D.}~\bibnamefont {Blume}}}
  (\bibinfo {year} {{2015}}),\ \bibfield  {title} {\enquote {\bibinfo {title}
  {{Trapped unitary two-component Fermi gases with up to ten particles}},}\
  }\href@noop {} {\bibfield  {journal} {\bibinfo  {journal} {{Phys. Rev. A}}\
  }\textbf {\bibinfo {volume} {{92}}}~(\bibinfo {number} {{1}})}\BibitemShut
  {NoStop}%
\bibitem [{\citenamefont {Yurovsky}(2005)}]{yurovsky2005feshbach}%
  \BibitemOpen
  \bibfield  {author} {\bibinfo {author} {\bibnamefont {Yurovsky},
  \bibfnamefont {V~A}}} (\bibinfo {year} {2005}),\ \bibfield  {title} {\enquote
  {\bibinfo {title} {Feshbach resonance scattering under cylindrical harmonic
  confinement},}\ }\href@noop {} {\bibfield  {journal} {\bibinfo  {journal}
  {Phys. Rev. A}\ }\textbf {\bibinfo {volume} {71}}~(\bibinfo {number} {1}),\
  \bibinfo {pages} {012709}}\BibitemShut {NoStop}%
\bibitem [{\citenamefont {Yurovsky}(2006)}]{yurovsky2006properties}%
  \BibitemOpen
  \bibfield  {author} {\bibinfo {author} {\bibnamefont {Yurovsky},
  \bibfnamefont {V~A}}} (\bibinfo {year} {2006}),\ \bibfield  {title} {\enquote
  {\bibinfo {title} {Properties of quasi-one-dimensional molecules with
  {F}eshbach-resonance interaction},}\ }\href@noop {} {\bibfield  {journal}
  {\bibinfo  {journal} {Phys. Rev. A}\ }\textbf {\bibinfo {volume}
  {73}}~(\bibinfo {number} {5}),\ \bibinfo {pages} {052709}}\BibitemShut
  {NoStop}%
\bibitem [{\citenamefont {Yurovsky}\ \emph {et~al.}(2008)\citenamefont
  {Yurovsky}, \citenamefont {Olshanii},\ and\ \citenamefont
  {Weiss}}]{Yurovsky2008a}%
  \BibitemOpen
  \bibfield  {author} {\bibinfo {author} {\bibnamefont {Yurovsky},
  \bibfnamefont {V~A}}, \bibinfo {author} {\bibfnamefont {M.}~\bibnamefont
  {Olshanii}}, \ and\ \bibinfo {author} {\bibfnamefont {D.~S.}\ \bibnamefont
  {Weiss}}} (\bibinfo {year} {2008}),\ \bibfield  {title} {\enquote {\bibinfo
  {title} {Collisions, correlations, and integrability in atom waveguides},}\
  }\href@noop {} {\bibfield  {journal} {\bibinfo  {journal} {Adv. At. Mol. Opt.
  Phys.}\ }\textbf {\bibinfo {volume} {55}},\ \bibinfo {pages} {61 --
  138}}\BibitemShut {NoStop}%
\bibitem [{\citenamefont {Zaccanti}\ \emph {et~al.}(2009)\citenamefont
  {Zaccanti}, \citenamefont {Deissler}, \citenamefont {D'Errico}, \citenamefont
  {Fattori}, \citenamefont {Jona-Lasinio}, \citenamefont {Mueller},
  \citenamefont {Roati}, \citenamefont {Inguscio},\ and\ \citenamefont
  {Modugno}}]{zaccanti2009NTP}%
  \BibitemOpen
  \bibfield  {author} {\bibinfo {author} {\bibnamefont {Zaccanti},
  \bibfnamefont {M}}, \bibinfo {author} {\bibfnamefont {B.}~\bibnamefont
  {Deissler}}, \bibinfo {author} {\bibfnamefont {C.}~\bibnamefont {D'Errico}},
  \bibinfo {author} {\bibfnamefont {M.}~\bibnamefont {Fattori}}, \bibinfo
  {author} {\bibfnamefont {M.}~\bibnamefont {Jona-Lasinio}}, \bibinfo {author}
  {\bibfnamefont {S.}~\bibnamefont {Mueller}}, \bibinfo {author} {\bibfnamefont
  {G.}~\bibnamefont {Roati}}, \bibinfo {author} {\bibfnamefont
  {M.}~\bibnamefont {Inguscio}}, \ and\ \bibinfo {author} {\bibfnamefont
  {G.}~\bibnamefont {Modugno}}} (\bibinfo {year} {2009}),\ \bibfield  {title}
  {{\selectlanguage {English}\enquote {\bibinfo {title} {Observation of an
  {E}fimov spectrum in an atomic system},}\ }}\href@noop {} {\bibfield
  {journal} {\bibinfo  {journal} {Nat. Phys.}\ }\textbf {\bibinfo {volume}
  {5}}~(\bibinfo {number} {8}),\ \bibinfo {pages} {586--591}}\BibitemShut
  {NoStop}%
\bibitem [{\citenamefont {Zenesini}\ \emph {et~al.}(2013)\citenamefont
  {Zenesini}, \citenamefont {Huang}, \citenamefont {Berninger}, \citenamefont
  {Besler}, \citenamefont {N\"agerl}, \citenamefont {Ferlaino}, \citenamefont
  {Grimm}, \citenamefont {Greene},\ and\ \citenamefont {von
  Stecher}}]{zenesini2013NJP}%
  \BibitemOpen
  \bibfield  {author} {\bibinfo {author} {\bibnamefont {Zenesini},
  \bibfnamefont {A}}, \bibinfo {author} {\bibfnamefont {B.}~\bibnamefont
  {Huang}}, \bibinfo {author} {\bibfnamefont {M.}~\bibnamefont {Berninger}},
  \bibinfo {author} {\bibfnamefont {S.}~\bibnamefont {Besler}}, \bibinfo
  {author} {\bibfnamefont {H.-C.}\ \bibnamefont {N\"agerl}}, \bibinfo {author}
  {\bibfnamefont {F.}~\bibnamefont {Ferlaino}}, \bibinfo {author}
  {\bibfnamefont {R.}~\bibnamefont {Grimm}}, \bibinfo {author} {\bibfnamefont
  {C.~H.}\ \bibnamefont {Greene}}, \ and\ \bibinfo {author} {\bibfnamefont
  {J.}~\bibnamefont {von Stecher}}} (\bibinfo {year} {2013}),\ \bibfield
  {title} {\enquote {\bibinfo {title} {Resonant five-body recombination in an
  ultracold gas of bosonic atoms},}\ }\href@noop {} {\bibfield  {journal}
  {\bibinfo  {journal} {New J. Phys.}\ }\textbf {\bibinfo {volume} {15}},\
  \bibinfo {pages} {043040}}\BibitemShut {NoStop}%
\bibitem [{\citenamefont {Zhai}(2015)}]{HuiZhai2015rpp}%
  \BibitemOpen
  \bibfield  {author} {\bibinfo {author} {\bibnamefont {Zhai}, \bibfnamefont
  {Hui}}} (\bibinfo {year} {2015}),\ \bibfield  {title} {\enquote {\bibinfo
  {title} {Degenerate quantum gases with spin-orbit coupling: a review},}\
  }\href {http://stacks.iop.org/0034-4885/78/i=2/a=026001} {\bibfield
  {journal} {\bibinfo  {journal} {Reports on Progress in Physics}\ }\textbf
  {\bibinfo {volume} {78}}~(\bibinfo {number} {2}),\ \bibinfo {pages}
  {026001}}\BibitemShut {NoStop}%
\bibitem [{\citenamefont {Zhang}\ and\ \citenamefont
  {Greene}(2013)}]{zhang2013}%
  \BibitemOpen
  \bibfield  {author} {\bibinfo {author} {\bibnamefont {Zhang}, \bibfnamefont
  {C}}, \ and\ \bibinfo {author} {\bibfnamefont {C.~H.}\ \bibnamefont
  {Greene}}} (\bibinfo {year} {2013}),\ \bibfield  {title} {\enquote {\bibinfo
  {title} {Quasi-one-dimensional scattering with general transverse
  two-dimensional confinement},}\ }\href@noop {} {\bibfield  {journal}
  {\bibinfo  {journal} {Phys. Rev. A}\ }\textbf {\bibinfo {volume}
  {88}}~(\bibinfo {number} {1}),\ \bibinfo {pages} {12715}}\BibitemShut
  {NoStop}%
\bibitem [{\citenamefont {Zhang}\ \emph {et~al.}(2012)\citenamefont {Zhang},
  \citenamefont {Zhang},\ and\ \citenamefont {Zhang}}]{Zhang2012pra2d}%
  \BibitemOpen
  \bibfield  {author} {\bibinfo {author} {\bibnamefont {Zhang}, \bibfnamefont
  {P}}, \bibinfo {author} {\bibfnamefont {L.}~\bibnamefont {Zhang}}, \ and\
  \bibinfo {author} {\bibfnamefont {W.}~\bibnamefont {Zhang}}} (\bibinfo {year}
  {2012}),\ \bibfield  {title} {\enquote {\bibinfo {title} {Interatomic
  collisions in two-dimensional and quasi-two-dimensional confinements with
  spin-orbit coupling},}\ }\href@noop {} {\bibfield  {journal} {\bibinfo
  {journal} {Phys. Rev. A}\ }\textbf {\bibinfo {volume} {86}},\ \bibinfo
  {pages} {042707}}\BibitemShut {NoStop}%
\bibitem [{\citenamefont {Zhang}\ and\ \citenamefont
  {Zhang}(2013)}]{zhang2013effective}%
  \BibitemOpen
  \bibfield  {author} {\bibinfo {author} {\bibnamefont {Zhang}, \bibfnamefont
  {R}}, \ and\ \bibinfo {author} {\bibfnamefont {W.}~\bibnamefont {Zhang}}}
  (\bibinfo {year} {2013}),\ \bibfield  {title} {\enquote {\bibinfo {title}
  {Effective {Hamiltonians} for quasi-one-dimensional {Fermi} gases with
  spin-orbit coupling},}\ }\href@noop {} {\bibfield  {journal} {\bibinfo
  {journal} {Phys. Rev. A}\ }\textbf {\bibinfo {volume} {88}}~(\bibinfo
  {number} {5}),\ \bibinfo {pages} {053605}}\BibitemShut {NoStop}%
\bibitem [{\citenamefont {Zhang}\ and\ \citenamefont
  {Zhang}(2011)}]{zhangwei2011pra}%
  \BibitemOpen
  \bibfield  {author} {\bibinfo {author} {\bibnamefont {Zhang}, \bibfnamefont
  {W}}, \ and\ \bibinfo {author} {\bibfnamefont {P.}~\bibnamefont {Zhang}}}
  (\bibinfo {year} {2011}),\ \bibfield  {title} {\enquote {\bibinfo {title}
  {Confinement-induced resonances in quasi-one-dimensional traps with
  transverse anisotropy},}\ }\href@noop {} {\bibfield  {journal} {\bibinfo
  {journal} {Phys. Rev. A}\ }\textbf {\bibinfo {volume} {83}},\ \bibinfo
  {pages} {053615}}\BibitemShut {NoStop}%
\bibitem [{\citenamefont {Zhang}\ \emph {et~al.}(2014)\citenamefont {Zhang},
  \citenamefont {Song},\ and\ \citenamefont {Liu}}]{soc2014}%
  \BibitemOpen
  \bibfield  {author} {\bibinfo {author} {\bibnamefont {Zhang}, \bibfnamefont
  {Y-C}}, \bibinfo {author} {\bibfnamefont {S.-W.}\ \bibnamefont {Song}}, \
  and\ \bibinfo {author} {\bibfnamefont {W.-M.}\ \bibnamefont {Liu}}} (\bibinfo
  {year} {2014}),\ \bibfield  {title} {\enquote {\bibinfo {title} {The
  confinement induced resonance in spin-orbit coupled cold atoms with {R}aman
  coupling},}\ }\href@noop {} {\bibfield  {journal} {\bibinfo  {journal} {Sci.
  Rep.}\ }\textbf {\bibinfo {volume} {4}}}\BibitemShut {NoStop}%
\bibitem [{\citenamefont {Zhdanov}(2002)}]{Zhdanov}%
  \BibitemOpen
  \bibfield  {author} {\bibinfo {author} {\bibnamefont {Zhdanov}, \bibfnamefont
  {V~M}}} (\bibinfo {year} {2002}),\ \href@noop {} {\emph {\bibinfo {title}
  {Transport Processes in Multicomponent Plasma}}}\ (\bibinfo  {publisher}
  {Taylor and Francis},\ \bibinfo {address} {London})\BibitemShut {NoStop}%
\bibitem [{\citenamefont {Zhu}\ \emph {et~al.}(2001)\citenamefont {Zhu},
  \citenamefont {Teranishi},\ and\ \citenamefont {Nakamura}}]{zhu2001}%
  \BibitemOpen
  \bibfield  {author} {\bibinfo {author} {\bibnamefont {Zhu}, \bibfnamefont
  {C~Y}}, \bibinfo {author} {\bibfnamefont {Y.}~\bibnamefont {Teranishi}}, \
  and\ \bibinfo {author} {\bibfnamefont {H.}~\bibnamefont {Nakamura}}}
  (\bibinfo {year} {2001}),\ \bibfield  {title} {\enquote {\bibinfo {title}
  {Nonadiabatic transitions due to curve crossings: Complete solutions of the
  {L}andau-{Z}ener-{S}t\"{u}ckelberg problems and their applications},}\
  }\href@noop {} {\bibfield  {journal} {\bibinfo  {journal} {Adv. Chem. Phys.}\
  }\textbf {\bibinfo {volume} {117}},\ \bibinfo {pages} {127--233}}\BibitemShut
  {NoStop}%
\bibitem [{\citenamefont {Zhu}(2013)}]{Zhu-2013}%
  \BibitemOpen
  \bibfield  {author} {\bibinfo {author} {\bibnamefont {Zhu}, \bibfnamefont
  {Y}}} (\bibinfo {year} {2013}),\ \bibfield  {title} {\enquote {\bibinfo
  {title} {Beam energy scan on hypertriton production and lifetime measurement
  at {RHIC} {STAR}},}\ }\href@noop {} {\bibfield  {journal} {\bibinfo
  {journal} {Nuc. Phys. A}\ }\textbf {\bibinfo {volume} {904-905}},\ \bibinfo
  {pages} {551c--554c}}\BibitemShut {NoStop}%
\bibitem [{\citenamefont {Zhukov}(1993)}]{Zhukov-1993}%
  \BibitemOpen
  \bibfield  {author} {\bibinfo {author} {\bibnamefont {Zhukov}, \bibfnamefont
  {M~V}}} (\bibinfo {year} {1993}),\ \bibfield  {title} {\enquote {\bibinfo
  {title} {Bound state properties of borromean halo nuclei: $^{6}${He} and
  $^{11}${Li}},}\ }\href@noop {} {\bibfield  {journal} {\bibinfo  {journal}
  {Phys. Rep.}\ }\textbf {\bibinfo {volume} {231}},\ \bibinfo {pages}
  {151}}\BibitemShut {NoStop}%
\bibitem [{\citenamefont {Zinner}(2012)}]{Zinner2012jpa}%
  \BibitemOpen
  \bibfield  {author} {\bibinfo {author} {\bibnamefont {Zinner}, \bibfnamefont
  {N~T}}} (\bibinfo {year} {2012}),\ \bibfield  {title} {\enquote {\bibinfo
  {title} {Universal two-body spectra of ultracold harmonically trapped atoms
  in two and three dimensions},}\ }\href
  {http://stacks.iop.org/1751-8121/45/i=20/a=205302} {\bibfield  {journal}
  {\bibinfo  {journal} {Journal of Physics A: Mathematical and Theoretical}\
  }\textbf {\bibinfo {volume} {45}}~(\bibinfo {number} {20}),\ \bibinfo {pages}
  {205302}}\BibitemShut {NoStop}%
\bibitem [{\citenamefont {Zinner}\ and\ \citenamefont
  {Jensen}(2013)}]{ZinnerJensen2013jpg}%
  \BibitemOpen
  \bibfield  {author} {\bibinfo {author} {\bibnamefont {Zinner}, \bibfnamefont
  {N~T}}, \ and\ \bibinfo {author} {\bibfnamefont {A~S}\ \bibnamefont
  {Jensen}}} (\bibinfo {year} {2013}),\ \bibfield  {title} {\enquote {\bibinfo
  {title} {Comparing and contrasting nuclei and cold atomic gases},}\ }\href
  {http://stacks.iop.org/0954-3899/40/i=5/a=053101} {\bibfield  {journal}
  {\bibinfo  {journal} {Journal of Physics G: Nuclear and Particle Physics}\
  }\textbf {\bibinfo {volume} {40}}~(\bibinfo {number} {5}),\ \bibinfo {pages}
  {053101}}\BibitemShut {NoStop}%
\bibitem [{\citenamefont {Zinner}\ \emph {et~al.}(2009)\citenamefont {Zinner},
  \citenamefont {M\o{}lmer}, \citenamefont {\"Ozen}, \citenamefont {Dean},\
  and\ \citenamefont {Langanke}}]{ZinnerMolmer2009pra}%
  \BibitemOpen
  \bibfield  {author} {\bibinfo {author} {\bibnamefont {Zinner}, \bibfnamefont
  {N~T}}, \bibinfo {author} {\bibfnamefont {K.}~\bibnamefont {M\o{}lmer}},
  \bibinfo {author} {\bibfnamefont {C.}~\bibnamefont {\"Ozen}}, \bibinfo
  {author} {\bibfnamefont {D.~J.}\ \bibnamefont {Dean}}, \ and\ \bibinfo
  {author} {\bibfnamefont {K.}~\bibnamefont {Langanke}}} (\bibinfo {year}
  {2009}),\ \bibfield  {title} {\enquote {\bibinfo {title} {Shell-model {M}onte
  {C}arlo simulations of the {BCS-BEC} crossover in few-fermion systems},}\
  }\href {\doibase 10.1103/PhysRevA.80.013613} {\bibfield  {journal} {\bibinfo
  {journal} {Phys. Rev. A}\ }\textbf {\bibinfo {volume} {80}},\ \bibinfo
  {pages} {013613}}\BibitemShut {NoStop}%
\bibitem [{\citenamefont {Zucrow}\ and\ \citenamefont
  {Hoffman}(1976)}]{Zucrow}%
  \BibitemOpen
  \bibfield  {author} {\bibinfo {author} {\bibnamefont {Zucrow}, \bibfnamefont
  {M~J}}, \ and\ \bibinfo {author} {\bibfnamefont {J.~D.}\ \bibnamefont
  {Hoffman}}} (\bibinfo {year} {1976}),\ \href@noop {} {\emph {\bibinfo {title}
  {Gas dynamics}}}\ (\bibinfo  {publisher} {Wiley},\ \bibinfo {address} {Nw
  York})\BibitemShut {NoStop}%
\bibitem [{\citenamefont {Zwierlein}\ \emph {et~al.}(2006)\citenamefont
  {Zwierlein}, \citenamefont {Schirotzek}, \citenamefont {Christian},\ and\
  \citenamefont {Ketterle}}]{zwierlein2006Sci}%
  \BibitemOpen
  \bibfield  {author} {\bibinfo {author} {\bibnamefont {Zwierlein},
  \bibfnamefont {M~W}}, \bibinfo {author} {\bibfnamefont {A.}~\bibnamefont
  {Schirotzek}}, \bibinfo {author} {\bibfnamefont {H.~S.}\ \bibnamefont
  {Christian}}, \ and\ \bibinfo {author} {\bibfnamefont {W.}~\bibnamefont
  {Ketterle}}} (\bibinfo {year} {2006}),\ \bibfield  {title} {\enquote
  {\bibinfo {title} {{F}ermionic superfluidity with imbalanced spin
  populations},}\ }\href@noop {} {\bibfield  {journal} {\bibinfo  {journal}
  {Science}\ }\textbf {\bibinfo {volume} {311}},\ \bibinfo {pages}
  {492--496}}\BibitemShut {NoStop}%
\bibitem [{\citenamefont {Zwierlein}\ \emph {et~al.}(2004)\citenamefont
  {Zwierlein}, \citenamefont {Stan}, \citenamefont {Schunck}, \citenamefont
  {Raupach}, \citenamefont {Kerman},\ and\ \citenamefont
  {Ketterle}}]{zwierlein2004PRL}%
  \BibitemOpen
  \bibfield  {author} {\bibinfo {author} {\bibnamefont {Zwierlein},
  \bibfnamefont {M~W}}, \bibinfo {author} {\bibfnamefont {C.~A.}\ \bibnamefont
  {Stan}}, \bibinfo {author} {\bibfnamefont {C.~H.}\ \bibnamefont {Schunck}},
  \bibinfo {author} {\bibfnamefont {S.~M.~F.}\ \bibnamefont {Raupach}},
  \bibinfo {author} {\bibfnamefont {A.~J.}\ \bibnamefont {Kerman}}, \ and\
  \bibinfo {author} {\bibfnamefont {W.}~\bibnamefont {Ketterle}}} (\bibinfo
  {year} {2004}),\ \bibfield  {title} {\enquote {\bibinfo {title} {Condensation
  of pairs of {F}ermionic atoms near a {F}eshbach resonance},}\ }\href@noop {}
  {\bibfield  {journal} {\bibinfo  {journal} {Phys. Rev. Lett.}\ }\textbf
  {\bibinfo {volume} {92}},\ \bibinfo {pages} {120403}}\BibitemShut {NoStop}%
\end{thebibliography}


%

\end{document}